\documentclass[english,11pt,b5paper,twoside]{book}
\usepackage{amsmath,amsfonts,amssymb,array,xspace,times,multirow,tabularx,textcomp,graphicx,pifont,gensymb}
\usepackage[usenames,dvipsnames]{color}
\usepackage{txfonts}
\usepackage[T1]{fontenc}
\RequirePackage[english]{babel}
\usepackage{authordateJL}
\usepackage{epsfig}
\usepackage[bf,it]{caption}
\usepackage[caption=false]{subfig}
\usepackage{url}
\usepackage{enumerate}
\usepackage{fancybox}
\usepackage[toc,page]{appendix}
\usepackage{lscape}
\usepackage{vmargin}
\usepackage{float}
\usepackage[official]{eurosym}
\usepackage{fancyhdr}             % en-t�tes
\usepackage[english]{minitoc}
\usepackage[Glenn2]{fncychap}
\usepackage[colorlinks=true]{hyperref}
\usepackage[figure,table]{hypcap}
\usepackage{bookmark}
\hypersetup{urlcolor=NavyBlue,linkcolor=Blue,citecolor=Blue,colorlinks=true}
\ChTitleVar{\bfseries\huge\rm}
\ChNameVar{ \bfseries\Large\sf}
\ChNumVar{\fontsize{50}{52}\usefont{OT1}{ptm}{b}{n}\selectfont}

\setpapersize{B5}
\setmarginsrb{16mm}{16mm}{16mm}{23mm}{10pt}{5mm}{10pt}{5mm}

\newif\ifCUPmtlplainloaded \CUPmtlplainloadedfalse

\makeatletter
\renewcommand{\fps@figure}{htbp!}
\renewcommand{\fps@table}{htbp!}

\makeatother

\DeclareGraphicsExtensions{.eps,.eps.gz,.ps,.fig,.jpg,.png}

\graphicspath{{./CHAP1-RAPPELS/FIGURES/}{./CHAP2-PERSEE/FIGURES/}{./CHAP3-INTEG-DEV_BANC/FIGURES/}{./CHAP4-OPTIM/FIGURES/}{./CHAP5-PERFO_NULLING/FIGURES/}{./CHAP6-EXTRAPOL/FIGURES/}{./ANNEXES/FIGURES/}}%{./vracs/}}

\newcommand{\tir}[1]{\mbox{---}~#1~\mbox{---}}
\newcommand{\cre}[1]{{\small\textit{---~Credit: #1}}}
\newcommand{\hs}{\hspace{3pt}}
\newcommand{\hsb}{\hspace{7.5pt}}

\newcommand{\pe}{\mbox{PERS\'EE}\xspace}
\newcommand{\peg}{\mbox{P\'EGASE}\xspace}
\newcommand{\sce}{\mbox{SCE\lowercase{x}AO}\xspace}
\newcommand{\Ifn}{nulling interferometry\xspace}

\newcommand{\ifns}{nulling interferometers\xspace}

\newcommand{\ac}{autocollimation\xspace}

\newcommand{\mum}{\mbox{{\usefont{U}{eur}{m}{n}{\char22}}m}\xspace}
\newcommand{\bmum}{\mbox{{\usefont{U}{eur}{b}{n}{\char22}}\textbf{m}}\xspace}
\newcommand{\imum}{\mbox{{\usefont{U}{eur}{m}{n}{\char22}}m$^{-1}$}\xspace}
\newcommand{\bimum}{\mbox{{\usefont{U}{eur}{b}{n}{\char22}}\textbf{m}$\boldsymbol{^{-1}}$}\xspace}
\newcommand{\muk}{\mbox{{\usefont{U}{eur}{m}{n}{\char22}}K}\xspace}

\newcommand{\FIG}[3]{\includegraphics[width=#1\linewidth,draft=#2]{#3-eps-converted-to.pdf}}
\newcommand{\FIGA}[4]{\includegraphics[angle=#1,width=#2\linewidth,draft=#3]{#4-eps-converted-to.pdf}}
\newcommand{\FIGH}[3]{\includegraphics[height=#1cm,draft=#2]{#3-eps-converted-to.pdf}}
\newcommand{\FIGWH}[4]{\includegraphics[width=#1\linewidth,height=#2\linewidth,draft=#3]{#4-eps-converted-to.pdf}}

\let\oldsqrt\sqrt
\def\sqrt{\mathpalette\DHLhksqrt}
\def\DHLhksqrt#1#2{%
\setbox0=\hbox{$#1\oldsqrt{#2\,}$}\dimen0=\ht0
\advance\dimen0-0.2\ht0
\setbox2=\hbox{\vrule height\ht0 depth -\dimen0}%
{\box0\lower0.4pt\box2}}

\newcommand{\BV}{\mathcal{V}}

\newcommand{\I}{\mathcal{I}}
\newcommand{\J}{\mathcal{J}}
\newcommand{\BH}{\mathcal{H}}
\newcommand{\K}{\mathcal{K}}

\newcommand{\GC}[1]{\left[#1\right]}
\newcommand{\GP}[1]{\left(#1\right)}
\newcommand{\GN}[1]{\left|#1\right|}

\newcommand{\moy}[2][]{\left\langle#2\right\rangle_{#1}}
\newcommand{\der}[2][]{\dfrac{\mathrm d#2}{\mathrm d#1}}
\newcommand{\dd}{\:\mathrm d}
\newcommand{\TF}[1]{\operatorname{TF}\GC{#1}}
\newcommand{\erf}[1]{\operatorname{erf}\GP{#1}}
\newcommand{\wrap}[1]{\operatorname{wrap}\GP{#1}}
\newcommand{\esp}{\operatorname{E}}
\renewcommand{\mod}{\operatorname{mod}\ }
\newcommand{\e}[1]{10^{#1}}
\newcommand{\E}[1]{\times10^{#1}}
\newcommand{\V}[1]{{\boldsymbol{\mathrm{#1}}}} % Vect discret

\newcommand{\maa}{\mathrm{a}}
\newcommand{\mar}{\mathrm{Ar}}
\newcommand{\mbf}{\mathrm{CL}}
\newcommand{\mbo}{\mathrm{OL}}
\newcommand{\mc}{\mathrm{c}}
\newcommand{\mcam}{\mathrm{cam}}
\newcommand{\mcor}{\mathrm{cor}}
\newcommand{\md}{\mathrm{d}}

\newcommand{\mdyn}{\mathrm{dyn}}
\newcommand{\mech}{\mathrm{samp}}
\newcommand{\mexp}{\mathrm{exp}}
\newcommand{\mf}{\mathrm{F}}
\newcommand{\mfras}{\mathrm{FRAS}}
\newcommand{\mfwhm}{\mathrm{FWHM}}
\newcommand{\mfs}{\mathrm{FS}}
\newcommand{\mgeo}{\mathrm{geo}}
\newcommand{\mi}{\mathrm{i}}
\newcommand{\minj}{\mathrm{inj}}
\newcommand{\mint}{\mathrm{int}}
\newcommand{\mj}{\mathrm{j}}
\newcommand{\ml}{\mathrm{L}}
\newcommand{\mlqg}{\mathrm{LQG}}
\newcommand{\mmax}{\mathrm{max}}
\newcommand{\mmin}{\mathrm{min}}
\newcommand{\mmoy}{\mathrm{mean}}
\newcommand{\mni}{\mathrm{I}}
\newcommand{\mnii}{\mathrm{II}}
\newcommand{\mniii}{\mathrm{III}}
\newcommand{\mniv}{\mathrm{IV}}
\newcommand{\mpl}{\mathrm{p}}
\newcommand{\mper}{\mathrm{per}}
\newcommand{\mph}{\mathrm{ph}}
\newcommand{\mpol}{\mathrm{pol}}
\newcommand{\mps}{\mathrm{SP}}
\newcommand{\mr}{\mathrm{r}}
\newcommand{\mref}{\mathrm{ref}}
\newcommand{\mrot}{\mathrm{rot}}
\newcommand{\ms}{\mathrm{s}}
\newcommand{\msta}{\mathrm{stat}}
\newcommand{\mtot}{\mathrm{tot}}

\newcommand{\muu}{\mathrm{u}}
\newcommand{\mv}{\mathrm{v}}
\newcommand{\mvib}{\mathrm{vib}}
\newcommand{\my}{\mathrm{y}}

\newcommand{\Tc}{T=100\mathrm{s}}
\newcommand{\Ts}{T=7\mathrm{h}}
\newcommand{\Td}{T=10\mathrm{h}}
\newcommand{\tauc}{\tau=100\mathrm{s}}
\newcommand{\tauu}{\tau=1\mathrm{s}}
\newcommand{\taud}{\tau=2.58\mathrm{ms}}

\newcommand{\mss}{\mathsf{s}}
\newcommand{\mpp}{\mathsf{p}}
\newcommand{\msp}{{\mss\text{-}\mpp}}
\newcommand{\ma}{\mathsf{a}}
\newcommand{\mb}{\mathsf{b}}

\dominitoc

\title{Characterization of the Stabilized Nulling Interferometry Testbed PERS\'EE}

\author{Julien Lozi}

\begin{document}
%\ttfamily
\begin{titlepage}
  \begin{center}
    {\Large \'Ecole Doctorale d'Astronomie \& Astrophysique \\
      d'\^Ile-de-France}\\
    \vspace{0.5cm}
    CNES --- Centre National d'\'Etudes Spatiales\\
    Onera --- The French Aerospace Lab\\
    \vspace{1cm}

    {\underline {{\Large {PhD THESIS}}}}\\
    \vspace{0.5cm}
 
    defended to obtain the title of\\
    
    \vspace{0.5cm}
    {\large PhD IN SCIENCES OF\\
    	UNIVERSIT\'E PARIS-SUD~XI}\\
    \vspace{0.5cm}
    \large Specialty Astronomy \& Astrophysics\\
    \vspace{1cm}

    by\\
    \vspace{0.1cm}
    {\textbf{\large{M. Julien LOZI}}}\\
    \vspace{1cm}

    {\LARGE {CHARACTERIZATION OF THE}}\\
    \vspace{0.2cm}
    {\LARGE {STABILIZED NULLING INTERFEROMETRY}}\\
    \vspace{0.2cm} {\LARGE {TESTBED PERS\'EE}}

    \vspace{1cm}

    % \textbf{Th�se dirig�e par Marc OLLIVIER et Fr�d�ric CASSAING,}\\
    % \vspace{0.5cm}

    defended on March 12, 2012 in front of the thesis commity composed of:\\
    \vspace{1cm}
    \begin{tabular}{ll}
      {\textbf{Prof.~Pierre Chavel}}              & President of the jury \\
      {\textbf{Dr.~Olivier Guyon}}                & Thesis Reader \\
      {\textbf{Dr.~Fabien Malbet}}               & Thesis Reader \\
      {\textbf{Dr.~Francoise Delplancke}} & Thesis Referee \\
      {\textbf{Prof.~Marc Ollivier}}              & PhD Advisor \\
      {\textbf{Dr.~Frederic Cassaing}}      & PhD Co-Advisor \\
      {\textbf{M.~Jean-Michel LeDuigou}} & Invited \\
    \end{tabular}\\

  \end{center}
\end{titlepage}

%%% Local Variables: 
%%% mode: latex
%%% TeX-master: "these_JL.tex"
%%% End: 

\thispagestyle{empty}

\pagestyle{fancy}
\renewcommand{\headrulewidth}{1.2pt}
%\changepage{0cm}{0cm}{0cm}{0cm}{0cm}{+3cm}{0cm}{0cm}{0cm}
%Red�finir plain page style (pour les pages de titre)
\fancypagestyle{plain}{
  \fancyhf{}% supprime les en-t�tes et pieds de pages d�finis par d�faut
  \renewcommand{\headrulewidth}{0pt} % supprime le trait
}

\renewcommand{\arraystretch}{1.}
\renewcommand{\labelitemi}{\Pisymbol{pzd}{114}}
  
  %en-t�tes
 \fancyhf{}
 \fancyhead[LE,RO]{\thepage}

\frontmatter

\thispagestyle{empty}
\vspace*{\fill}
\begin{flushright}
{\it To Fr\'ed\'eric Cassaing}
\end{flushright}
\vspace*{\fill}

%%% Local Variables: 
%%% mode: latex
%%% TeX-master: "these_JL"
%%% End: 

\cleardoublepage
\fancyhead{}
\fancyhead[LE,RO]{\thepage}
\fancyhead[RE,LO]{\small ABSTRACT}
\chapter*{Abstract}
\label{sec-abstract}
%\addstarredchapter{Abstract}

There are two problems with the observation of exoplanets: the contrast between the planet and the star and their very low separation. One technique solving these problems is nulling interferometry: two pupils are recombined to make a destructive interference on the star, and their base is adjusted to create a constructive interference on the planet. However, to ensure a sufficient extinction of the star, the optical path difference between the beams must be around the nanometer, and the pointing must be better than one hundredth of Airy disk, despite the external disturbances.

To validate the critical points of such a space mission, a laboratory demonstrator, PERS\'EE, was defined by a consortium led by CNES, including IAS, LESIA, ONERA, OCA and Thales Alenia Space and integrated in Meudon Observatory. This bench simulates the entire space mission (interferometer and nanometric cophasing system). Its goal is to deliver and maintain an extinction of $\e{-4}$ stable at better than $\e{-5}$ over a few hours in the presence of typical injected disturbances.

My thesis work consisted in integrating the bench in successive stages and to develop calibration procedures. This helped me to characterize the critical elements separately before grouping them. After having implemented the control loops of the cophasing system, their precise analysis helped me to reduce down to $0.3$~nm rms the residual OPD, and $0.4$~\% of the Airy disk the residual tip/tilt, despite disturbances of tens of nanometers, consisting of several tens of vibrational frequencies between 1 and 100~Hz. This has been achieved by the implementation of a linear quadratic Gaussian controller, parameterized by the preliminary measurement of the disturbance to minimize. Thanks to these excellent results, I obtained on the band $[1.65$--$2.45]$~\mum a record null rate of $8.8\E{-6}$ stabilized at $9\E{-7}$ over a few hours, a decade better than the original specifications. An extrapolation of these results to the case of a space mission shows that the expected performance is achievable if the available flux is sufficiently important. With telescopes of 40~cm and a control frequency around 100~Hz, stars brighter than magnitude 9 should be observable.

%%% Local Variables: 
%%% mode: latex
%%% TeX-master: "these_JL"
%%% End: 

\cleardoublepage
\fancyhead{}
\fancyhead[LE,RO]{\thepage}
\renewcommand{\chaptermark}[1]{\markboth{\MakeUppercase{\chaptername\ \thechapter.\ #1}}{}}
\fancyhead[RE]{\small\leftmark}
\fancyhead[LO]{\small\rightmark}
\tableofcontents

\cleardoublepage\fancyhead{}
\fancyhead[LE,RO]{\thepage}
\fancyhead[RE,LO]{\small INTRODUCTION}
\thispagestyle{empty}
\chapter*{Introduction}
\label{intro}
\addstarredchapter{Introduction}
%\markboth{INTRODUCTION}{INTRODUCTION}
%\addcontentsline{toc}{chapter}{\protect Introduction}

\thispagestyle{empty} 

With the first detection of an exoplanet orbiting a main-sequence star in 1995 \cite{Mayor95}, the young field of exoplanetology has experienced rapid growth. Since then, more than 700~extrasolar planets have been detected (2012)\footnote{\url{http://www.exoplanet.eu/catalog.php}}, with highly diverse characteristics. Consequently, categories of planets distinct from those in the Solar System have been defined: hot Jupiters, or Pegasids—Jupiter-sized planets orbiting very close to their host star; super-Earths—rocky planets more massive than Earth; and potential ocean planets, whose surface is entirely liquid. These new discoveries have refined models of planetary evolution following their formation.

Among all these detected planets, very few have been directly imaged. Indeed, collecting photons from the planet while minimizing those from the star is extremely complex due to the intrinsic properties of the star–planet pair. The first challenge is the angular separation between the two bodies: resolving a planet next to a star requires a high angular resolution from the instrument, which is directly tied to its physical dimensions. To overcome this difficulty, two main families of instruments stand out: single-pupil telescopes—equipped with a single or segmented mirror forming a single aperture—and interferometers, consisting of multiple recombined pupils designed to generate interference patterns. The second challenge is the high contrast between the star and the planet. Indeed, the star emits far more photons than its planet, with the ratio between the two quantities depending strongly on the observational spectral band. It is therefore necessary to suppress the stellar flux as much as possible to observe the exoplanet. Here again, two families of instruments stand out to achieve this extinction. The first, installed on monolithic telescopes, comprises coronagraphs. An image of the planet is acquired while the star is masked, either through the use of a physical mask or via more complex methods. The second, utilizing interferometers, creates a destructive interference fringe that extinguishes the light from the star.

In both instrument categories, high optical quality is essential. This is why it took over a decade of design work and testbeds before pointing the first coronagraphs and \ifns toward the sky. Because space environments offer much more favorable conditions for observing Earth-like planets, the next step consists of placing instruments on satellites; however, technological and budgetary constraints have significantly delayed the launch of such missions. In particular, all space-based \Ifn projects have been postponed due to a lack of technology readiness. To pave the way for these future missions, it is crucial to validate these methods using testbeds that increasingly simulate actual space conditions with high fidelity.

It was precisely with the goal of simulating the \Ifn \peg space mission that the \pe (\peg Experiment for Research and Stabilization of Extreme Extinction) testbed was created in 2005. Funded by CNES and the Île-de-France region, it was designed by a consortium of experienced institutions in the field: IAS, Paris Observatory–Meudon, ONERA, Observatoire de la Côte d'Azur, and Thales Alenia Space. This testbed uniquely simulates both the optical train of \peg and the disturbance conditions of the mission. My PhD work consisted of integrating the testbed inside a cleanroom at the Meudon Observatory, characterizing and analyzing the interactions between its various components, achieving and stabilizing maximum fringe contrast even under simulated \peg disturbances, and finally extrapolating the results to aid in sizing a mission like \peg.

In Chapter~\ref{chap-rappels}, I will review the background of exoplanet detection, providing a description of these planets alongside an overview of the history of astronomical instrumentation that led to precision instruments such as coronagraphs and \ifns. Chapter~\ref{chap-persee} presents a description of the \pe testbed, as well as the methods used to achieve beam cophasing (i.e., the stabilization of the testbed). Next, Chapter~\ref{sec-integr-devel} outlines the integration methodology of the bench and the modifications I implemented on the initial design following identified performance limitations. In this chapter, I will also describe a method I implemented to finely characterize the testbed. In Chapter~\ref{sec-optimisation-boucles}, I will analyze the control loops that perform the cophasing of the two interferometer arms, demonstrating the advantage of using an optimized controller to mitigate injected vibrations on the testbed. Furthermore, I will attempt to apply this controller to a ground-based telescope coronagraph, the \sce (Subaru Coronagraphic Extreme Adaptive Optics) instrument. I will then present in Chapter~\ref{sec-performance-nulling} the testbed's performance in suppressing stellar flux in a stable manner over long durations. This analysis is also carried out under simulated typical \peg disturbances. Finally, in Chapter~\ref{sec-extrapolation}, I will extrapolate the results obtained on the bench to the case of a space mission such as \peg, drawing initial conclusions regarding its feasibility and sizing.

%%% Local Variables: 
%%% mode: latex
%%% TeX-master: "these_JL"
%%% End:
\thispagestyle{empty}
\cleardoublepage

\mainmatter

\renewcommand{\arraystretch}{1.25}

\fancyhead{}
\fancyhead[LE,RO]{\thepage}
\renewcommand{\chaptermark}[1]{\markboth{\MakeUppercase{\chaptername\ \thechapter.\ #1}}{}}
\fancyhead[RE]{\small\leftmark}
\fancyhead[LO]{\small\rightmark}
\thispagestyle{empty}

\setcounter{chapter}{0}

%§§§§§§§§§§§§§§§§§§§§§§§§§§§§§§§§§§§§§§§§§§§§§§§§§§§§§§§§§§§§§§§§§§§§§§§§§§§§§§§
\chapter{Exoplanet Observation}
\label{chap-rappels}
 
\begin{flushright}
 \begin{minipage}{12cm} {\small \textit{A gas giant was in view, seen at an
       angle that allowed most of it to be sunlit. About it, there curved a
       broad and brilliant ring of material, tipped so as to catch the
       sunlight on the side being viewed. It was brighter than the planet
       itself and along it, one third of the way in toward the planet, was a
       narrow, dividing line. Trevize threw in a request for maximum
       enhancement and the ring became ringlets, narrow and concentric,
       glittering in the sunlight. Only a portion of the ring system was
       visible on the viewscreen and the planet itself had moved on: A
       further direction from Trevize and one corner of the screen marked
       itself off and showed, within itself, a miniature of the planet and
       rings under lesser magnification. "Is that sort of thing common?"
       asked Bliss, awed. "No," said Trevize. "Almost every gas giant has
       rings of debris, but they tend to be faint and narrow. I once saw one
       in which the rings were narrow, but quite bright. But I never saw
       anything like this; or heard of it, either."}}

   \raggedleft{{\small Isaac Asimov, Foundation and Earth (1986)}}
 \end{minipage}
\end{flushright}

\minitoc

\bigskip

%§§§§§§§§§§§§§§§§§§§§§§§§§§§§§§§§§§§§§§§§§§§§§§§§§§§§§§§§§§§§§§§§§§§§§§§§§§§§§§§
\section{Exoplanet Detection and Observation}
\label{sec-detection-observation}

%¤¤¤¤¤¤¤¤¤¤¤¤¤¤¤¤¤¤¤¤¤¤¤¤¤¤¤¤¤¤¤¤¤¤¤¤¤¤¤¤¤¤¤¤¤¤¤¤¤¤¤¤¤¤¤¤¤¤¤¤¤¤¤¤¤¤¤¤¤¤¤¤¤¤¤¤¤¤¤
\subsection{What is an Exoplanet?}
\label{sec-exoplanete}

%-------------------------------------------------------------------------------
\subsubsection{A Millennia-Old Quest}
\label{sec-quete-plus}

The concept of planets outside the Solar System is not new. Since Ancient Greece, astronomers and philosophers have pondered the possibility of worlds similar to ours around other stars and the existence of extraterrestrial life. In 1584, Giordano Bruno, a proponent of the Copernican theory, hypothesized an infinite universe whose stars were suns with their own inhabited planets in his work \emph{De l'Infinito Universo et Mondi}. During the 17th century, the idea was revisited multiple times by Fontenelle in his \emph{Entretiens sur la pluralit\'e des mondes} and by Huygens in his essay \emph{Cosmotheoros}. The latter even proposed observing these new worlds using astronomical instruments. The search intensified during the 19th century, notably with the observation of orbital irregularities of the binary star 70 Ophiuchi, described in 1855 by Captain Jacob at the Madras Observatory; these observations were actually erroneous. The existence of other planets, sometimes inhabited, was no longer even questioned by major 20th-century science fiction authors like Isaac Asimov or Frank Herbert, who, through seminal works such as \emph{Foundation} and \emph{Dune}, speculated on a vast number of planets populating the galaxy. However, it was not until 1995 that a planet was finally discovered around a Sun-like star \cite{Mayor95}, initiating the conquest of new worlds.

%-------------------------------------------------------------------------------
\subsubsection{Definition}
\label{sec-definition}

First, it is important to define the goal of this research. The terms "exoplanet" or "extrasolar planet" designate any planet orbiting a star other than the Sun. Some planets not orbiting any star are also suspected following the discovery of isolated brown dwarfs, referred to as "free-floating planets". The definition of an exoplanet is therefore intrinsically tied to the definition of a planet.

Starting in 2003, the IAU\footnote{International Astronomical Union} attempted to redefine the term "planet" after the discovery of objects larger than Pluto in the Solar System. The definition established in 2006 is as follows \cite{Binzel06}:
\begin{quotation}
  A "planet" is a celestial body that: (a) is in orbit around the Sun, (b) has sufficient mass for its self-gravity to overcome rigid body forces so that it assumes a hydrostatic equilibrium (nearly round) shape, and (c) has cleared the neighborhood around its orbit.
\end{quotation}

However, while valid for planets in the Solar System, this definition sets no upper mass limit for exoplanets and omits free-floating planets. Consequently, the IAU defined extrasolar planets as follows\footnote{\url{http://www.dtm.ciw.edu/boss/definition.html}}:
\begin{quotation}
  \noindent
  \begin{enumerate}
  \item Objects with true masses below the limiting mass for thermonuclear fusion of deuterium (currently calculated to be 13~Jupiter masses for objects of solar metallicity) that orbit stars or stellar remnants are "planets" (regardless of how they formed). The minimum mass required for an extrasolar object to be considered a planet should be the same as that used in the Solar System;
  \item "Sub-stellar" objects with true masses above the limiting mass for thermonuclear fusion of deuterium are "brown dwarfs", regardless of how they formed or where they are located;
  \item Free-floating objects in young star clusters with masses below the limiting mass for thermonuclear fusion of deuterium are not "planets", but are "sub-brown dwarfs" (or whatever name is most appropriate).
  \end{enumerate}
\end{quotation}

As shown, the official definition is not completely finalized, especially since certain objects with masses below 13~M$_\mj$ can undergo deuterium fusion, and this boundary has not been observed in current measurements. There is thus a grey area between the definition of an exoplanet and that of a brown dwarf, which can only be clarified by increasing the number of observations.

%-------------------------------------------------------------------------------
\subsubsection{An Increasing Number of Discoveries}
\label{sec-decouvertes-plus}

Since the discovery of the first exoplanet around a solar-type star in 1995, the number of detected planets has grown continuously, reaching 708 as of December 7, 2011. Figure~\ref{fig-planetes-vs-annee} shows the number of exoplanets detected per year. The histogram begins in 1989 because several planets were detected around pulsars or identified via \emph{a posteriori} processing of archival data.

\begin{figure} \centering
  \FIG{.7}{false}{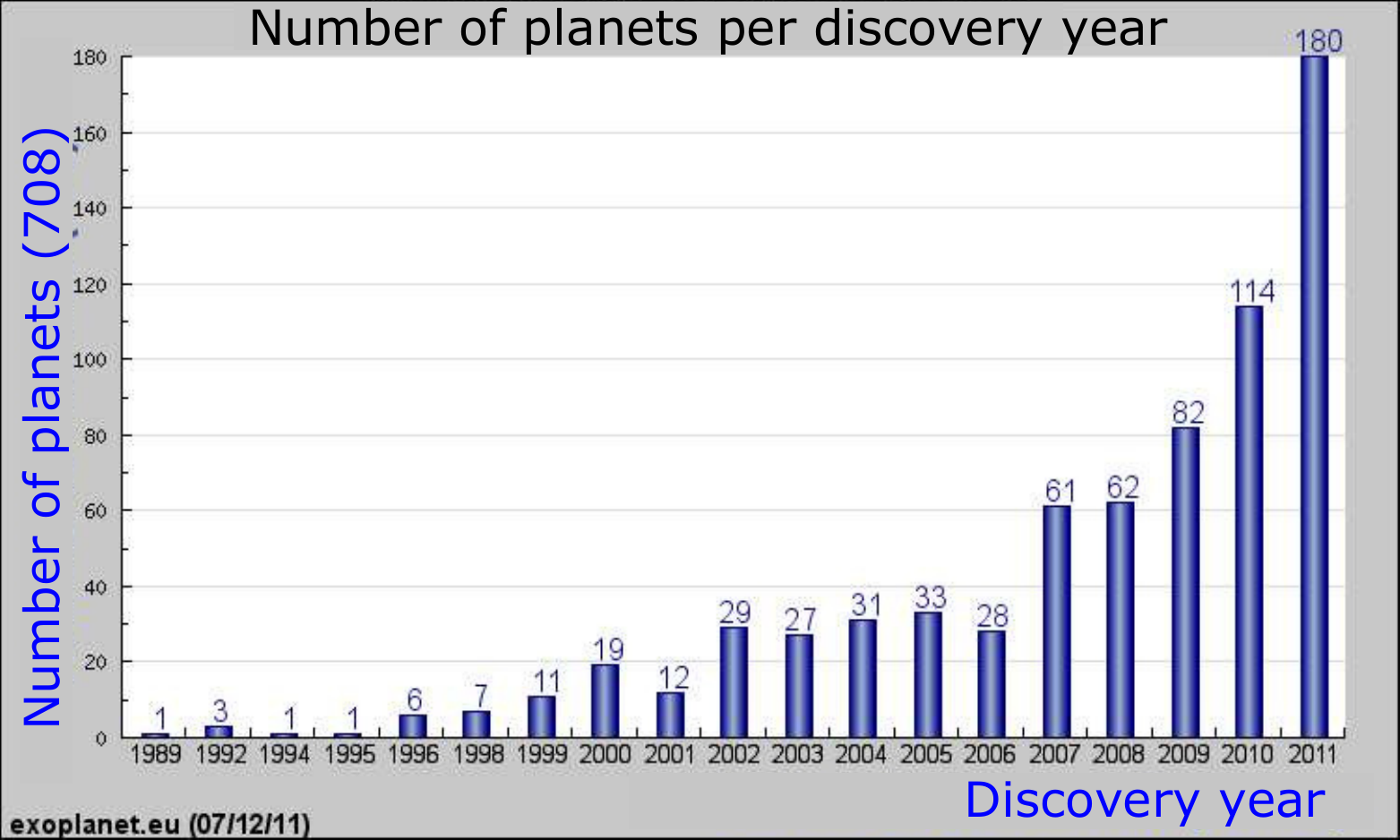}
  \caption{Number of detected exoplanets per year.}
  \label{fig-planetes-vs-annee}
\end{figure}

Examining the mass distribution of detected planets (Figure~\ref{fig-massep-vs-annee}), we see that the range of masses is broadening, particularly toward lower values—planets of smaller and smaller mass are being detected. Earth mass ($3\E{-3}$~M$_\mj$) was reached only very recently \tir{announced on December 20, 2011, Kepler-20e and Kepler-20f, whose masses are estimated to be less than 3 Earth masses with diameters roughly equal to Earth (to be confirmed) \cite{Fressin11}, excluding the exceptional 1992 case of a rocky body detected around a pulsar} due to method sensitivity limits. Furthermore, planets with masses significantly higher than Jupiter are observed, and only 14\% are in multiple-planet systems, demonstrating that the Solar System model is not entirely standard. Jupiter's mass remains a useful benchmark for gas giants, as 74\% of detected planets have masses under 3~Jupiter masses.

\begin{figure} \centering
  \FIG{.7}{false}{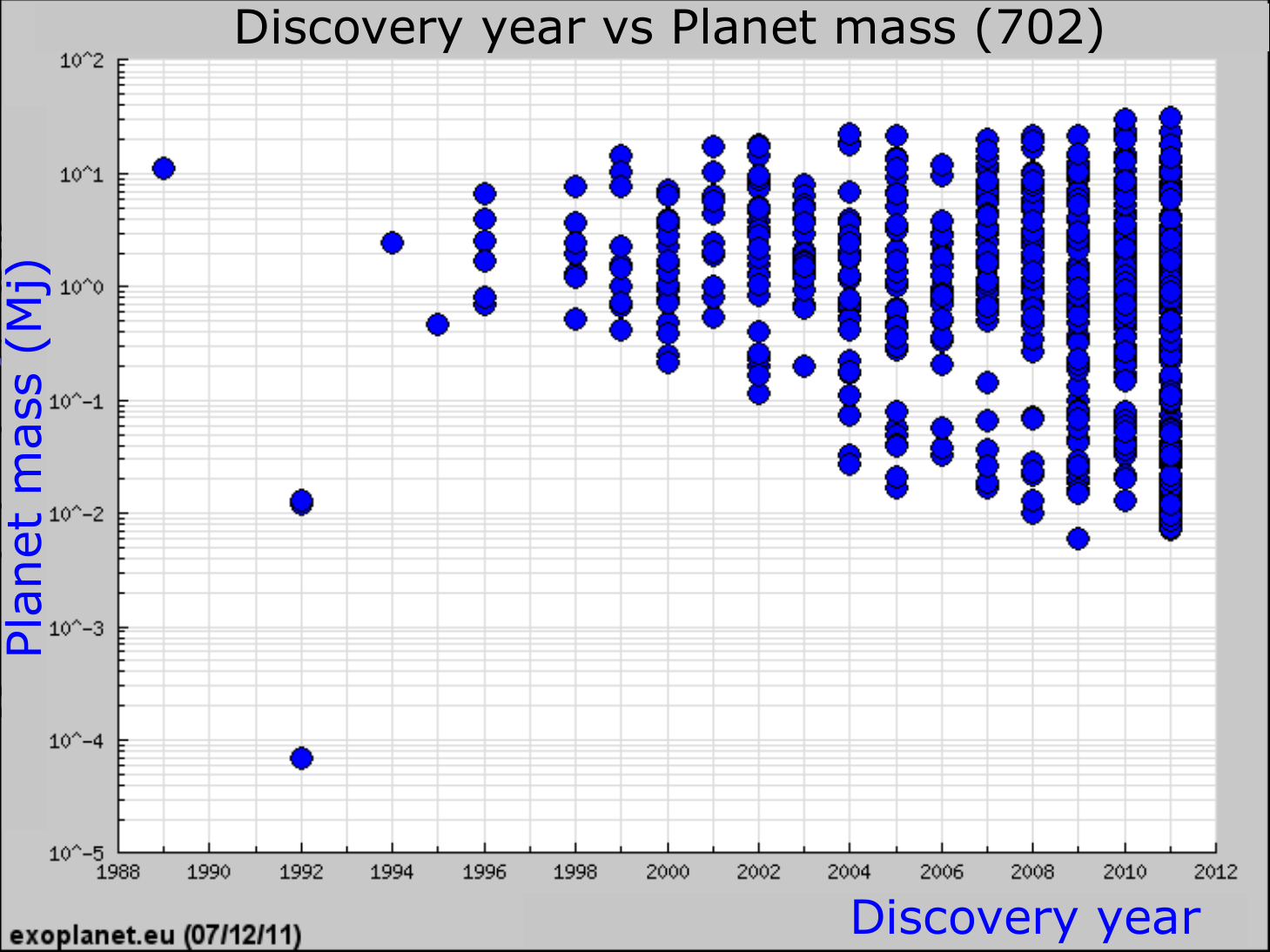}
  \caption{Mass of extrasolar planets versus discovery year.}
  \label{fig-massep-vs-annee}
\end{figure}

The growing count of lower-mass planet detections stems from the development of numerous detection and observation techniques, summarized in Figure~\ref{fig-detection}.

\begin{figure} \centering
  \FIG{1.}{false}{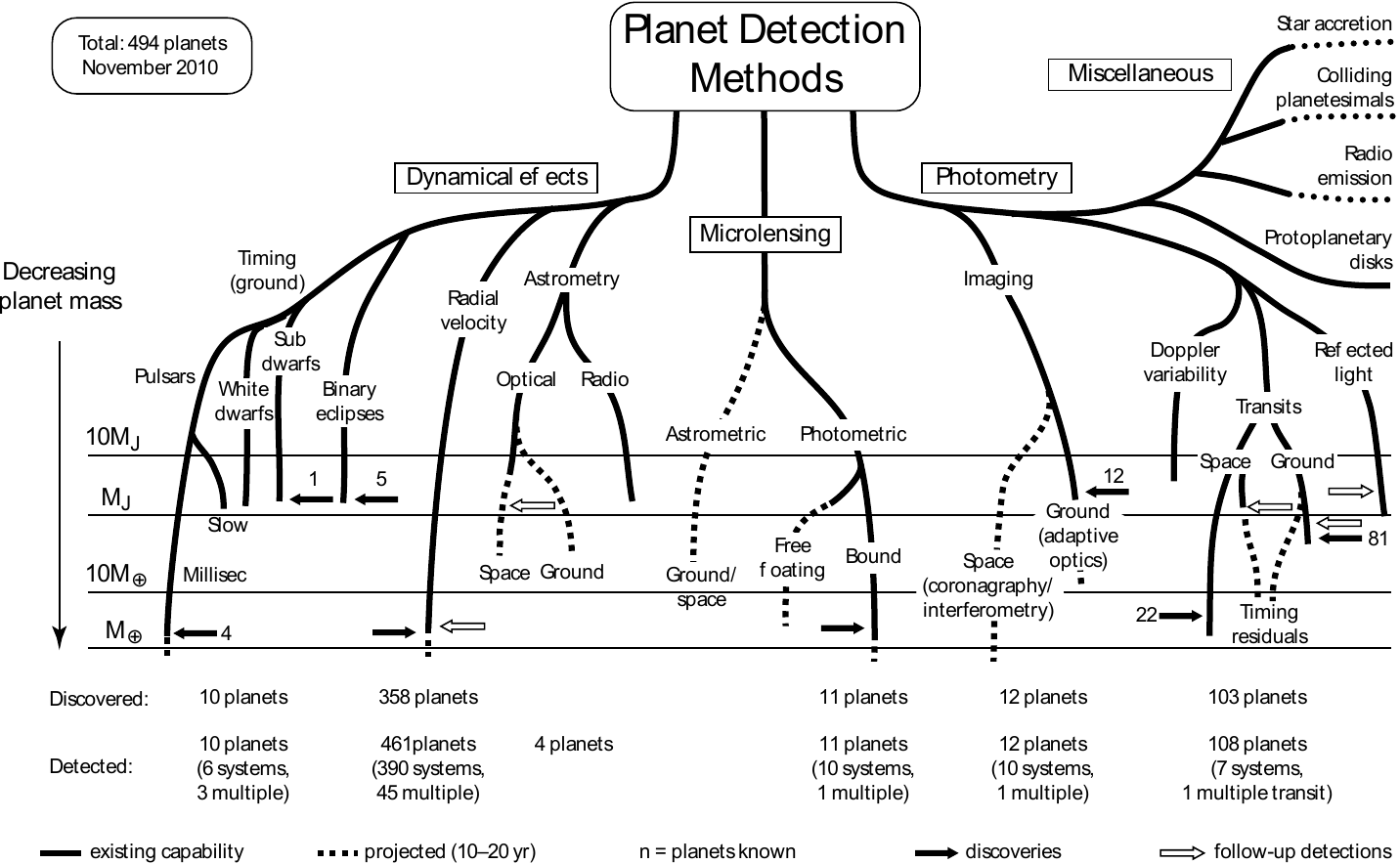}
  \caption[Different exoplanet detection methods.]{Different exoplanet detection methods \cre{M. Perryman (11/12/2010)}.}
  \label{fig-detection}
\end{figure}

Over the past decade, most PhD theses on exoplanets have provided descriptions of these techniques to varying degrees of exhaustiveness; I will spare the reader this recap by referring to the theses of my predecessors on the \pe testbed \cite{Houairi09b,Jacquinod10}. Nevertheless, several major approaches can be distinguished based on their yields:
\begin{itemize}
\item Methods studying the dynamical effects of the planet on its host star. These do not allow direct observation of exoplanets and are thus indirect methods. They currently yield the highest number of detections, primarily through radial velocity measurements;
\item Gravitational microlensing methods, which monitor the flux of a star amplified by the passage of a planetary system across the line of sight from Earth. These transient phenomena are strictly non-repeatable, yielding only the star-to-planet mass ratio after event modeling;
\item Methods studying stellar or planetary photometry—excluding microlensing, which is also photometric. Some are indirect, detecting the planet within the star's flux (e.g., the transit method), while others are direct imaging methods. The latter will ultimately enable atmospheric characterization but remain the most challenging to implement.
\end{itemize}

%-------------------------------------------------------------------------------
\subsubsection{Hot Jupiters: A New Class of Planets}
\label{sec-jupiters-chauds}

The first exoplanet detected in 1995, 51 Pegasi b, possessed unusual characteristics compared to expectations built on the Solar System. Although about half the mass of Jupiter, it orbits its host star 100 times closer, at a distance of 0.05~AU\footnote{Astronomical Unit, $1\text{~AU} \simeq 1.5\E{8}$~km}. Positioned so close to its star, its estimated equilibrium temperature reaches 1265~K.

The rapid increase in discoveries of gas giants orbiting close to their stars prompted a fundamental shift in post-formation planetary evolution models, particularly regarding the speed of planet formation and subsequent orbital migration. These planets were subsequently named "hot Jupiters" or "Pegasids" after this first discovery.

%¤¤¤¤¤¤¤¤¤¤¤¤¤¤¤¤¤¤¤¤¤¤¤¤¤¤¤¤¤¤¤¤¤¤¤¤¤¤¤¤¤¤¤¤¤¤¤¤¤¤¤¤¤¤¤¤¤¤¤¤¤¤¤¤¤¤¤¤¤¤¤¤¤¤¤¤¤¤¤
\subsection{A Challenging Observation}
\label{sec-observation-difficile}

Most methods currently in use provide only indirect detection. However, understanding planet formation and searching for potential biosignatures requires directly analyzing planetary light. Among direct techniques, imaging has enabled the study of a small subset of exoplanets.

Direct observation is hindered by several major issues:
\begin{itemize}
\item The small angular separation between the star and the planet, requiring high instrument resolution to separate them;
\item The high contrast between the star and the planet, as the stellar flux is overwhelmingly dominant;
\item The exoplanetary environment, which can introduce parasitic stray flux that degrades observations.
\end{itemize}

%-------------------------------------------------------------------------------
\subsubsection{The Angular Separation Problem}
\label{sec-probleme-separation}

Angular separation is a key parameter in exoplanet observation. By analogy with the Solar System, observing an Earth-like planet orbiting at 1~AU around a star located at 1~parsec corresponds to an angular separation of 1~arcsecond (arcsec)—a parsec being defined as the distance yielding a parallax of one arcsecond for an observer on Earth at 1~AU. However, target stars are typically observed at distances around 10~parsecs, with the closest star (Proxima Centauri) at 1.3~parsecs. The angular separation for the same system at 10~parsecs is therefore 0.1~arcsec.

For gas giant planets located typically between 5 and 30~AU, the angular separation ranges from 0.5 to 3~arcsec. Conversely, hot Jupiters orbit extremely close to their star, typically between 0.01 and 0.1~AU; at 10~parsecs, this yields tiny angular separations between 1 and 10~milliarcseconds (mas). Figure~\ref{fig-angle_vs_massep} plots angular separation against estimated planet mass for directly imaged exoplanets.

\begin{figure} \centering
  \FIG{0.7}{false}{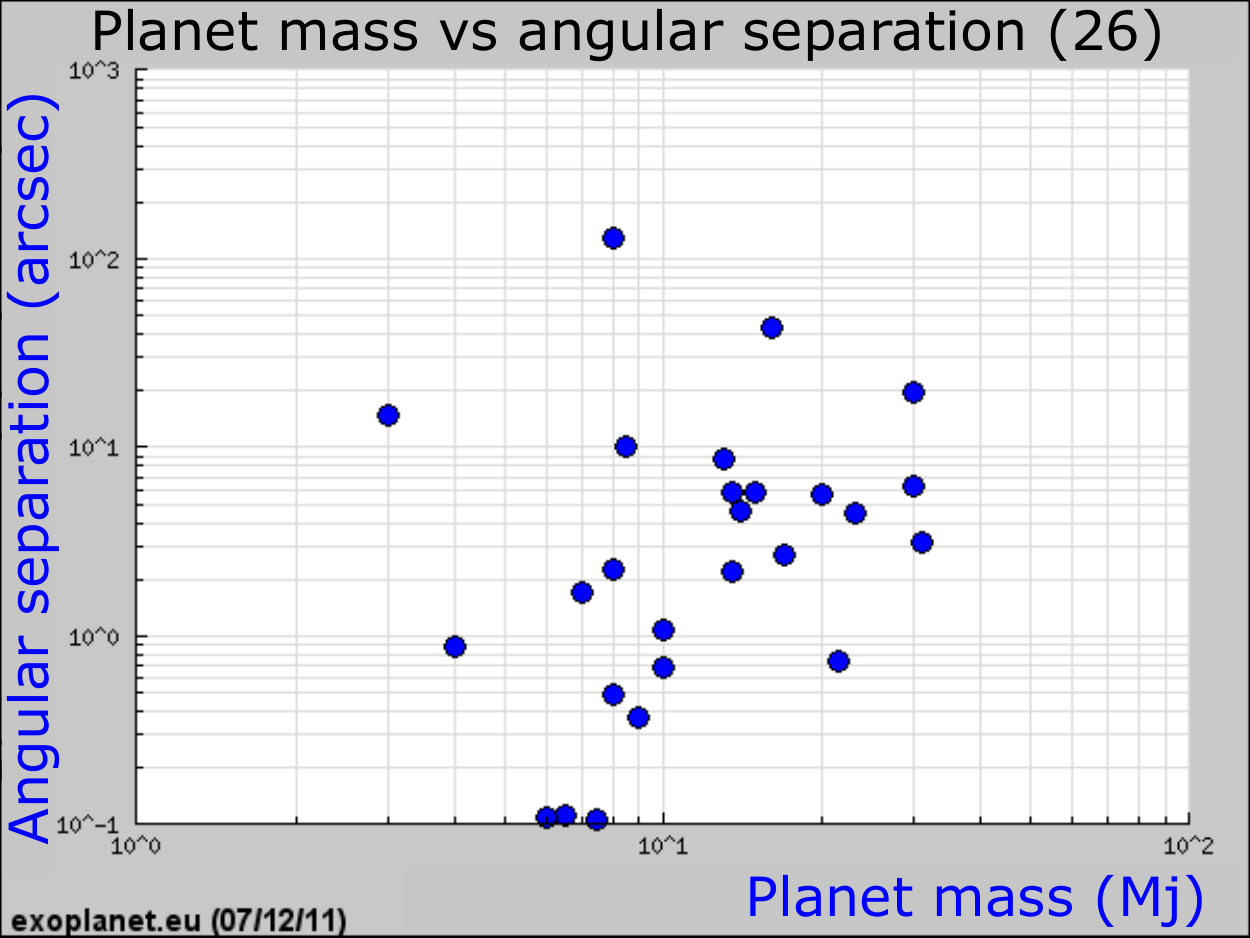}
  \caption[Angular separation versus mass for imaged planets.]{Angular separation versus mass for planets observed via direct imaging.}
  \label{fig-angle_vs_massep}
\end{figure}

As seen in this figure, direct imaging begins detecting planets at separations around 100~mas, but most sit at separations larger than 1~arcsec. Directly imaging hot Jupiters remains beyond the reach of current instruments. Imaged planets typically range between 3 and 30~Jupiter masses; observing Earth-like planets with current imaging instruments is still far off.

\begin{figure} \centering
  \FIG{0.7}{false}{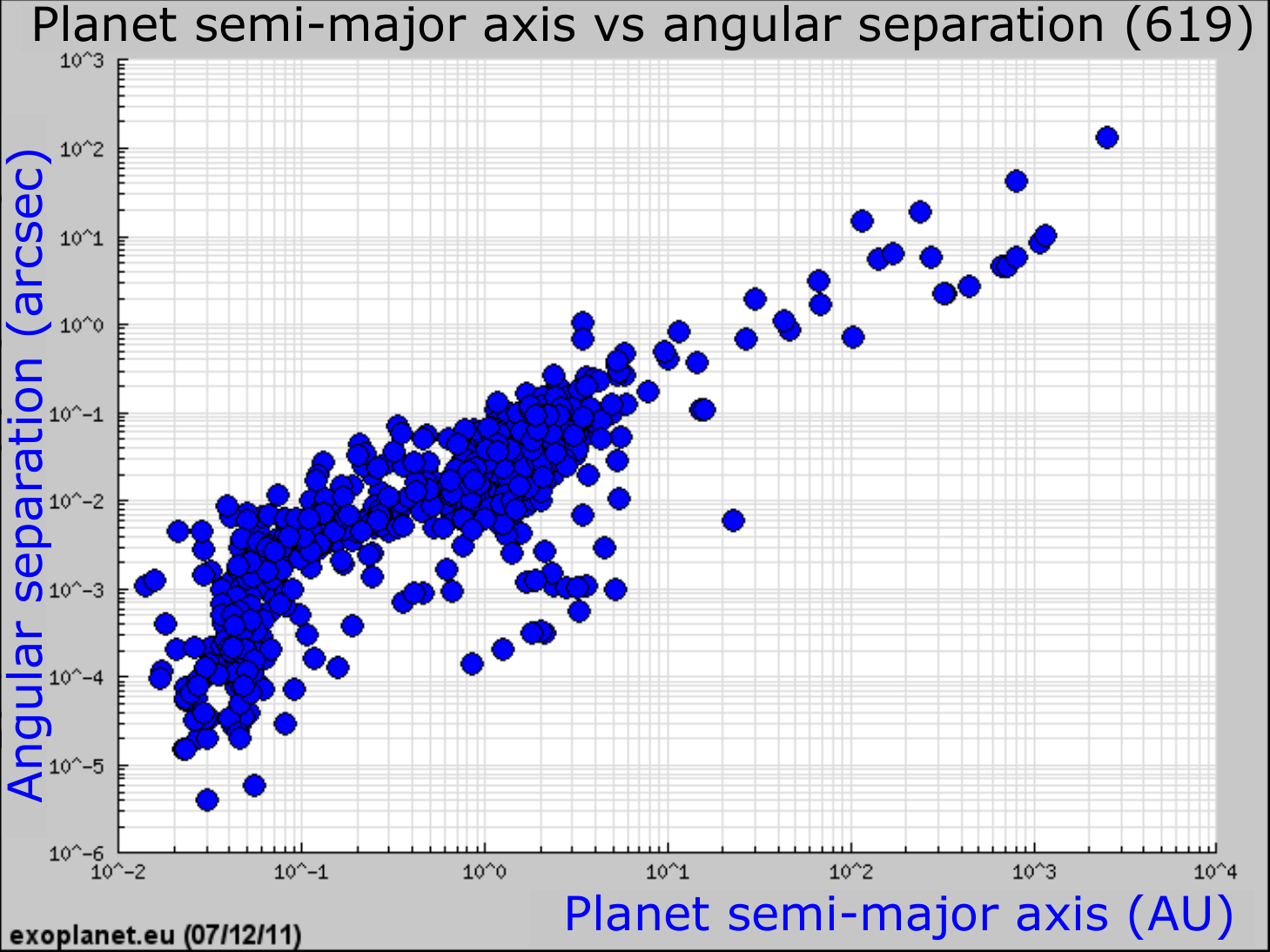}
  \caption{Angular separation versus semi-major axis of the planet's orbit.}
  \label{fig-angle_vs_axe}
\end{figure}

Plotting angular separation against semi-major axis (Figure~\ref{fig-angle_vs_axe}) highlights the gap between direct imaging and indirect techniques. Direct imaging targets wide-orbit planets \tir{between 10 and 1000~AU}, whereas indirect techniques probe planets with orbits mainly between 0.01 and 10~AU. Most observed hot Jupiters, with semi-major axes between 0.01 and 0.1~AU, have angular separations between $\e{-6}$ and $\e{-2}$~arcsec. Novel imaging techniques are therefore needed to directly analyze their atmospheres and understand their formation.

%-------------------------------------------------------------------------------
\subsubsection{The Contrast Problem}
\label{sec-probleme-contraste}

Angular separation is not the only obstacle to direct observation. Planets have been imaged at 0.1~arcsec from their host stars—a separation sufficient to resolve an Earth analogue at 10~parsecs. The second hurdle is the extreme flux contrast between the planet and its star. This contrast depends strongly on the chosen spectral band, which targets specific atmospheric chemical species such as molecular oxygen. Figure~\ref{fig-spectre-terre} illustrates this contrast problem by comparing the spectra of the Earth and the Sun. The bands of interest correspond either to regions where the planet reflects maximum stellar light or where it emits significant thermal radiation.

\begin{figure} \centering
  \FIG{0.5}{false}{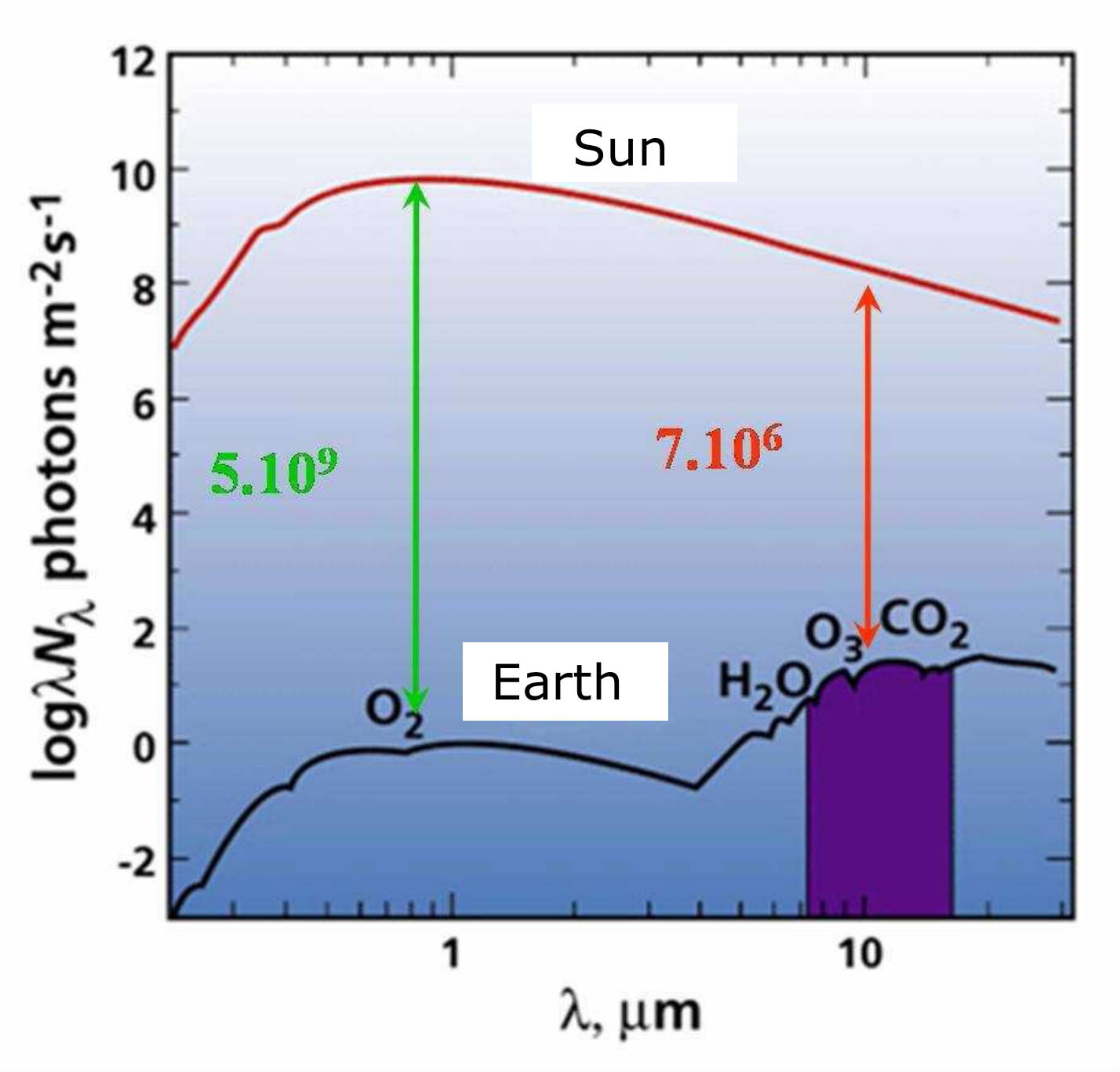}
  \caption{Comparison between the spectra of the Earth and the Sun.}
  \label{fig-spectre-terre}
\end{figure}

Figure~\ref{fig-spectre-terre} displays the two local maxima in Earth's spectrum: the first, in the visible domain, coincides with the solar peak, giving an extreme contrast of $5\E{9}$. Around 10~\mum, however, thermal emission from the planet peaks while solar emission declines, lowering the contrast to $7\E{6}$—a value that remains high. A tradeoff exists: instrument resolution is proportional to wavelength; operating at a wavelength 10 times longer degrades angular resolution by a factor of 10. On the other hand, wavefront errors and stability constraints are relaxed by a factor of 10. Operating at 10~\mum involves different technological challenges, such as cryogenic cooling and mid-infrared single-mode fiber development.

For a gas giant like Jupiter, the visible contrast is more favorable ($5\E{8}$) owing to its larger radius. However, being colder, its thermal emission at 10~\mum is weaker than that of a warmer Earth-sized planet, yielding a contrast of $5\E{7}$.

For hot Jupiters (assuming Jupiter's size and mass at 0.05~AU), contrast ratios are much more favorable in both the visible ($4\E{4}$) and mid-infrared at 10~\mum ($\e{3}$). Observing these targets requires significantly higher angular resolution, however.

%-------------------------------------------------------------------------------
\subsubsection{The Exoplanetary Environment: A Source of Difficulty}
\label{sec-envir-exopl}

Interplanetary space in the Solar System is not empty. It contains a dust disk composed of cometary and asteroidal grains aligned near the ecliptic plane, extending well beyond Earth's orbit. Warmed and illuminated by the Sun, this dust is visible from Earth as zodiacal light under dark skies. With a temperature around 300~K at 1~AU, its thermal emission peaks near 10~\mum, and its integrated flux across the disk is 300 times brighter than Earth. Information regarding exozodiacal dust disks (exozodis) remains sparse; if extrasolar systems feature dust levels comparable to or higher than the Solar System, exozodiacal emission can swamp the weak planetary signal alongside stellar light leakage.

%-------------------------------------------------------------------------------
\subsubsection{Summary of Observational Challenges}
\label{sec-synthese-contraste}

Instruments must address these three challenges to observe exoplanets, whether hot Jupiters requiring high angular resolution under moderate contrast, or Earth-like planets at 10~parsecs requiring extreme contrast mitigation at modest angular separations.

Furthermore, these observations must be carried out despite potential background limits set by exozodiacal dust emission.

%§§§§§§§§§§§§§§§§§§§§§§§§§§§§§§§§§§§§§§§§§§§§§§§§§§§§§§§§§§§§§§§§§§§§§§§§§§§§§§§
\section{Historical Evolution of High Angular Resolution}
\label{sec--evolution}

%¤¤¤¤¤¤¤¤¤¤¤¤¤¤¤¤¤¤¤¤¤¤¤¤¤¤¤¤¤¤¤¤¤¤¤¤¤¤¤¤¤¤¤¤¤¤¤¤¤¤¤¤¤¤¤¤¤¤¤¤¤¤¤¤¤¤¤¤¤¤¤¤¤¤¤¤¤¤¤
\subsection{What is Angular Resolution?}
\label{sec-resolution}

%-------------------------------------------------------------------------------
\subsubsection{The Importance of Pupil Size\dots}
\label{sec-importance-taille}

Humans have observed stars since antiquity; astronomy is considered the oldest science. Civilizations studied seasonal cycles, lunar phases, constellations, and complex planetary motions. Early astronomical sites appeared at Nabta Playa in Egypt during the 5th millennium BCE and at Stonehenge $\sim$2000 years later, serving both religious and practical agricultural functions. Modern astronomy emerged when Greek mathematics enabled theoretical modeling of celestial motion. Early astronomers distinguished fixed stars from wandering planets (from the Greek "planetes asteres", meaning wandering star). The naked-eye planets \tir{Mercury, Venus, Mars, Jupiter, and Saturn} have been known since prehistory. Until the early 17th century, astronomers relied solely on naked-eye observation.

\begin{table} \centering
  \caption[Maximum angular diameter of Solar System planets.]{Maximum angular diameter of naked-eye Solar System planets.}
  \medskip
  \begin{tabular}{*{5}{>{\centering}m{2cm}}}
    \hline\hline Mercury & Venus & Mars & Jupiter & Saturn \tabularnewline
    \hline 12.4~arcsec & 65.3~arcsec & 25.6~arcsec & 49.8~arcsec & 20.8~arcsec
    \tabularnewline
    \hline\hline
  \end{tabular}
  \label{tab-diam}
\end{table}

In scotopic vision, the theoretical resolution limit of the human eye is approximately 30~arcsec. Comparing this to Table~\ref{tab-diam}, the nearest planets sit close to this resolution limit. Resolving surface details was virtually impossible before the 17th century, though Mayan astronomers accurately tracked the phases of Venus at the edge of human visual acuity.

%-------------------------------------------------------------------------------
\subsubsection{\dots But Size Alone is Not Enough!}
\label{sec-dots-mais}

By the end of the 16th century, early spyglasses were invented for military surveillance. In 1609, Galileo turned these instruments skyward, demonstrating his spyglass on August 21, 1609, to the Venetian Senate. His early objective lenses ranged from 30 to 70~mm in diameter but suffered from severe glass inhomogeneity and crude surface polishing. Galileo reported that only a few out of 60 lenses were usable; optical aberrations—rather than pupil diameter—severely limited performance.

Despite these poor optics, Galileo discovered Jupiter's four main moons on January 7, 1610. On July 25, 1610, he observed Saturn's rings, describing them successively as "ears," "handles," or "two companions helping old Saturn on his way" (Figure~\ref{fig-obs-saturne}).

\begin{figure} \centering
  \FIG{0.5}{false}{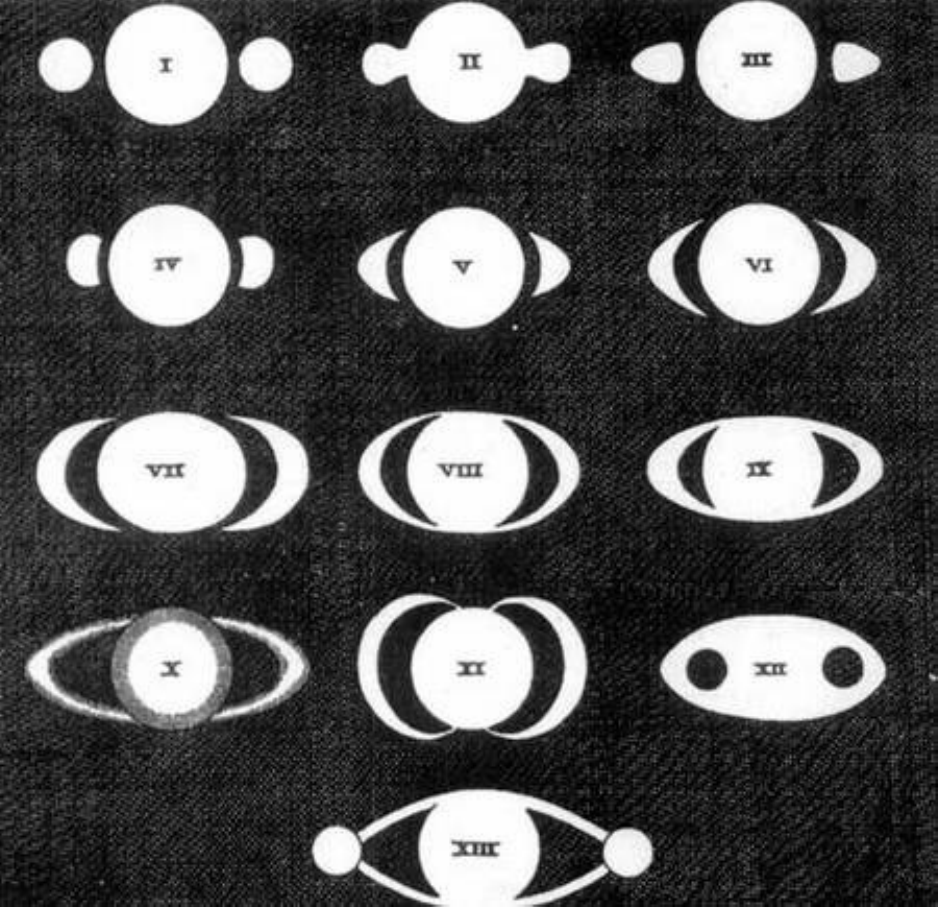}
  \caption[Early drawings of Saturn.]{Early drawings of Saturn from the early 17th century showing its changing appearance, unexplained until 1656.
    \textsc{i}. Galileo, 1610. \textsc{ii}. Scheiner, 1614. \textsc{iii}. Riccioli, 1640. \textsc{iv}–\textsc{vii}. Hevelius, 1640–1650.
    \textsc{viii}, \textsc{ix}. Riccioli, 1648, 1650. \textsc{x}. Eustachio Divini, 1647. \textsc{xi}. Fontana, 1648. \textsc{xii}. Gassendi, 1645.
    \textsc{xiii}. Riccioli, 1630 \cre{DR}.}
  \label{fig-obs-saturne}
\end{figure}

Huygens identified their true nature in 1656: "It is surrounded by a thin, flat ring, nowhere touching the planet, and inclined to the ecliptic". In September 1610, Galileo also tracked the phases of Venus.

Instrument quality improved markedly before the end of the 17th century as we will see later in Section~\ref{sec-mesure-front}. Chromatic aberration remained the primary limitation of simple lenses, prompting astronomers to build telescopes with extremely long focal lengths to mitigate chromatic dispersion.

%¤¤¤¤¤¤¤¤¤¤¤¤¤¤¤¤¤¤¤¤¤¤¤¤¤¤¤¤¤¤¤¤¤¤¤¤¤¤¤¤¤¤¤¤¤¤¤¤¤¤¤¤¤¤¤¤¤¤¤¤¤¤¤¤¤¤¤¤¤¤¤¤¤¤¤¤¤¤¤
\subsection{Cassini's Lenses and Planetary Observation}
\label{sec-lentilles-cassini}

In 1669, Jean-Dominique Cassini was invited to France by Colbert to join the newly formed Académie des Sciences. In 1671, with funding from Louis XIV, he helped establish the Paris Observatory. Cassini brought a 17-foot focal length lens (5.5~m) crafted by Giuseppe Campani, a famous Roman optician. Made before 1665, this lens enabled Cassini to discover transits/eclipses of Jovian satellites and the Great Red Spot. In 1665 and 1666, he derived the rotation periods of Jupiter and Mars, respectively \cite{Cassini1705}. Colbert commissioned Campani to supply his finest lenses and build longer focal length objectives. In December 1672, Campani delivered a 34-foot objective (11~m), followed later by lenses of 80, 90, 100, and 136 feet. The 34-foot lens was mounted in an 11-meter tube adjusted with ropes and pulleys, visible in the background of Cassini's portrait of Fig.~\ref{fig-cassini}.

\begin{figure} \centering
  \FIG{0.5}{false}{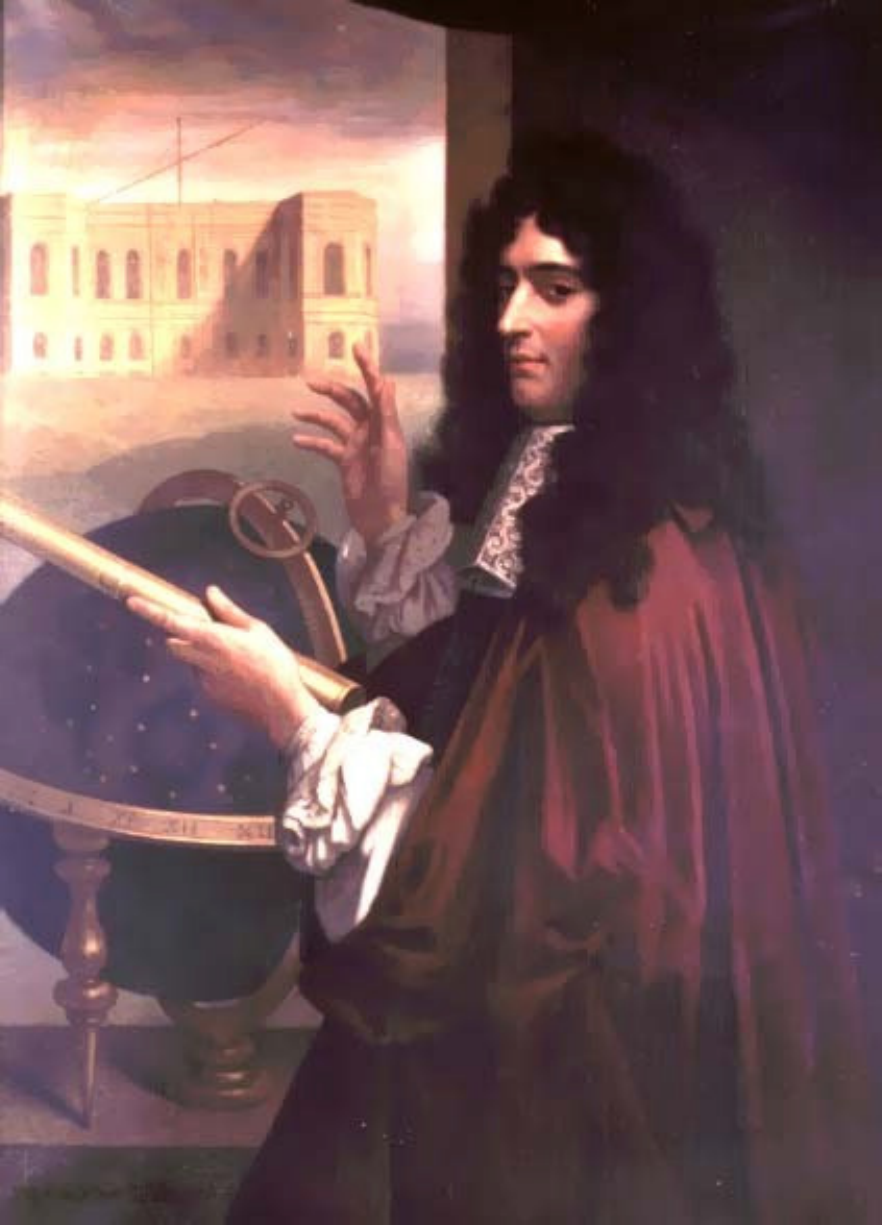}
  \caption[Portrait of Jean-Dominique Cassini.]{Portrait of Jean-Dominique Cassini showing the south facade of the Paris Observatory in the background, with the 34-foot telescope mounted on its roof.}
  \label{fig-cassini}
\end{figure}

Using this telescope in 1675, Cassini discovered the main gap in Saturn's rings, now known as the Cassini Division:

\begin{quotation}
  Deinde latitudo Annuli dividebatur bifariam, Lineâ obscurâ apparenter Ellipticâ reverâ Circulari quasi in duos annulos concentricos, quorum interior exteriori lucidior erat. Hanc phasim statim post emersionem Saturni è Solaribus radiis per totum annum usque ad ejus Immersionem conspexi; primùo quidem, Telescopie Pedum 35, deinde minori, Pedum 20. Ejus delineationem, utcumque rudem, properante calamo hic adjeci.

  Then the breadth of the Ring was divided in two by a dark line, apparently elliptical but actually circular, as if into two concentric rings, the inner being brighter than the outer. I observed this phase immediately after Saturn emerged from the Sun's rays throughout the year until its immersion; first with the 35-foot telescope, then with a smaller 20-foot one. I have added here a sketch of it, rough as it may be, drawn in haste.
\end{quotation}

\emph{An Extract of Signor Cassini's Letter concerning a Spot lately seen in the Sun, together with a remarkable Observation of Saturn, made by the same}, Philosophical Transactions, \textbf{Vol.~XI}, September 25, 1676, p.~690.

At the time of observation, Saturn's ring system spanned a maximum diameter of 38.5~arcsec, while the Cassini Division measured at most 0.65~arcsec across its ansae. For a 108~mm diameter objective like the 34-foot lens, the theoretical diffraction limit is $\sim$1~arcsec. Accounting for chromatic aberration, optical figure errors, and atmospheric turbulence, historians long questioned whether Cassini truly resolved the division.

Other objectives featured immense focal lengths up to 50~m, requiring mounting solutions beyond pulley tubes. Figure~\ref{fig-obs-paris} illustrates an observation night at the Paris Observatory around 1685.

\begin{figure} \centering
  \FIG{0.7}{false}{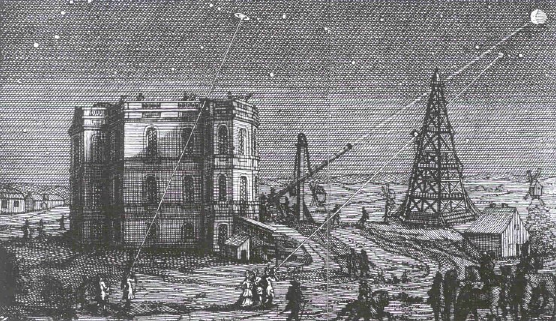}
  \caption{Observation night at the Paris Observatory, circa 1685.}
  \label{fig-obs-paris}
\end{figure}

The engraving shows an astronomer using a sub-100-foot lens on the Observatory roof (left), the 34-foot telescope (center), and an aerial telescope setup suspended from the Marly Tower (right)—a wooden military tower relocated at the request of Louis XIV's minister, Louvois.

These historical lenses are preserved at the Paris Observatory. At the request of the observatory curators, an optical engineering group was formed at LESIA\footnote{Laboratoire d'Études Spatiales et d'Instrumentation en Astrophysique} (Paris Observatory–Meudon), in which I participated, to characterize these poorly documented historic optics. We initially evaluated five lenses, including the famous 34-foot objective. Results are detailed below and will be published in a peer-reviewed journal and displayed at the 2012 exhibition commemorating the tercentenary of Cassini's death.

%-------------------------------------------------------------------------------
\subsubsection{Measurement of Lens Characteristics}
\label{sec-mesure-caracteristiques}

The Paris Observatory holds over 30 historic objective lenses from Cassini's era. Little documentation exists beyond approximate focal lengths; precise focal lengths, glass quality, and chromatic dispersion curves remained unknown. These metrics are vital to evaluating Cassini's visual acuity and confirming whether he could resolve Saturn's division. We measured the five largest wooden-framed objective lenses (Figure~\ref{fig-obj-C40}).

\begin{figure} \centering
  \subfloat[Objective lenses used by Cassini \cre{J. Counil / Bibliothèque de l'Observatoire de Paris}.]{\label{fig-objectifs}
    \FIG{0.8}{false}{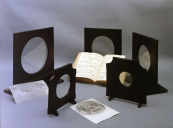}}\\
  \subfloat[Campani's engraving on the 34-foot lens.]{
    \label{fig-C40} \FIG{0.8}{false}{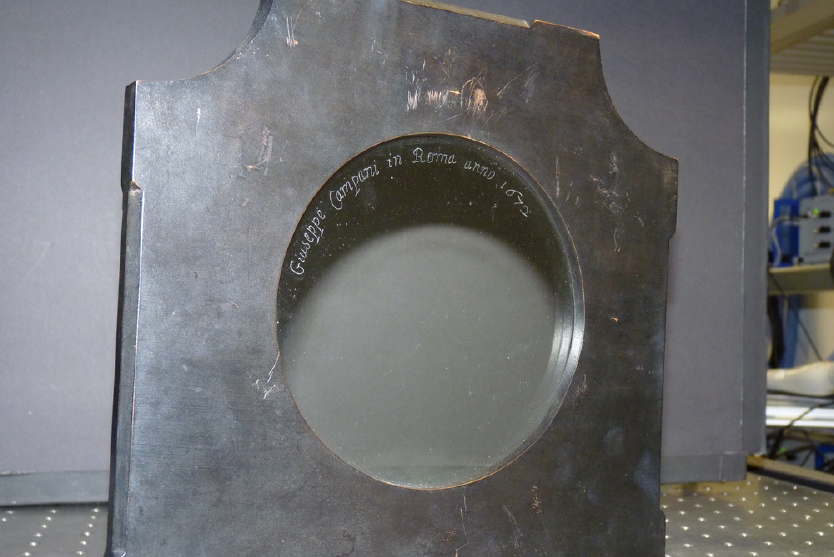}}
  \caption[Cassini's objective lenses.]{Cassini's objective lenses, at least two of which were crafted by Campani in Rome.}
  \label{fig-obj-C40}
\end{figure}

Figure~\ref{fig-objectifs} shows the five lenses; Figure~\ref{fig-C40} details the inscription on the 34-foot lens: \emph{Giuseppe Campani in Roma anno 1672\/}. Lenses are designated by inventory number. In Figure~\ref{fig-objectifs}, bottom left to right: Nos. 40 (the 34-foot lens) and 44; top left to right: Nos. 42, 43, and 41. All exceed 135~mm in diameter, though some are vignetted by wooden frames or paper diaphragms to reduce peripheral aberrations. Only Nos. 40 and 41 bear Campani's signature.

Mechanical properties—clear aperture diameter $D$, center thickness $e$, and surface radii of curvature—were measured first using calipers and a mechanical feeler gauge. Radii of curvature were determined using a spherometer, which measures sagitta $h$ over a known radius $r$ relative to a flat optical reference:
\begin{equation}
  R = \frac{r^2+h^2}{2h}.
\end{equation}

Table~\ref{tab-car-lentilles} summarizes the measured parameters. Focal lengths were measured using a 532~nm laser. Given strong chromatic dispersion, the focus shifts by several centimeters across wavelengths (e.g., a 50~mm focus shift between 633~nm red and green light for lens No. 40).

\begin{table} \centering
  \caption{Summary of experimental measurements for each lens.}
  \medskip
  \begin{tabular}{cccccc}
    \hline\hline
    Inventory No. & 40 & 41 & 42 & 43 & 44 \\
    \hline
    $D$ [mm] & 137 & 181 & 239 & 183 & 84 \\
    \hline
    $r$ [mm] & 45 & 75 & 75 & 75 & 35 \\
    $h_1$ [\mum{}] & 85 & 75 & 22 & $-15$ & $-64$ \\
    $h_2$ [\mum{}] & 85 & 61 & 105 & 120 & 245 \\
    \hline
    $R_1$ [m] & $11.9$ & $37.5$ & $128$ & $-188$ & $-9.57$ \\
    $R_2$ [m] & $11.9$ & $46.1$ & $26.8$ & $23.4$ & $2.50$ \\
    $e$ [mm] & $6.25$ & --- & --- & --- & --- \\
    \hline
    $f_\text{@ 532~nm}$ [m] & $10.9$ & $40.2$ & $47.3$ & $48,5$ & $6.30$ \\
    \hline\hline
  \end{tabular}
  \label{tab-car-lentilles}
\end{table}

Not all thicknesses $e$ were measured directly to protect fragile wooden mounts. All lenses are remarkably thin, with aspect ratios $D/e \approx 20$ (compared to modern lenses where $D/e < 10$). Minimizing glass thickness helped historical opticians reduce wavefront distortions caused by glass inhomogeneities.

%-------------------------------------------------------------------------------
\subsubsection{Wavefront Characterization}
\label{sec-mesure-front}

Evaluating visual performance requires measuring wavefront aberrations introduced by the optics. We used a Zygo Fizeau interferometer combined with a high-precision spherical reference mirror placed with its center of curvature at the focal point of the lens. The Zygo outputs a collimated monochromatic beam; rays passing through the test lens reflect off the spherical mirror back along their incident path. The returning wavefront recombines with a flat reference to generate pupil interference fringes, yielding the double-pass wavefront error (divided by two for single-pass equivalent).

While standard for refractive testing, the long focal lengths involved would require up to 50~m of propagation space for certain lenses! Over such distances, air turbulence degrades measurements. To solve this, I proposed adding a high-quality convergent doublet ($f = 750$~mm) directly after the test lens. This shortened focal distances to under one meter for all setups. The known spherical aberration introduced by using the reference doublet away from its infinite-conjugate design point was subtracted during post-processing.

Point Spread Functions (PSFs) were calculated from the measured wavefront maps. Results for lens No. 40 (used to discover the Cassini Division) are shown in Figure~\ref{fig-res-40}.

\begin{figure} \centering
  \FIG{0.7}{false}{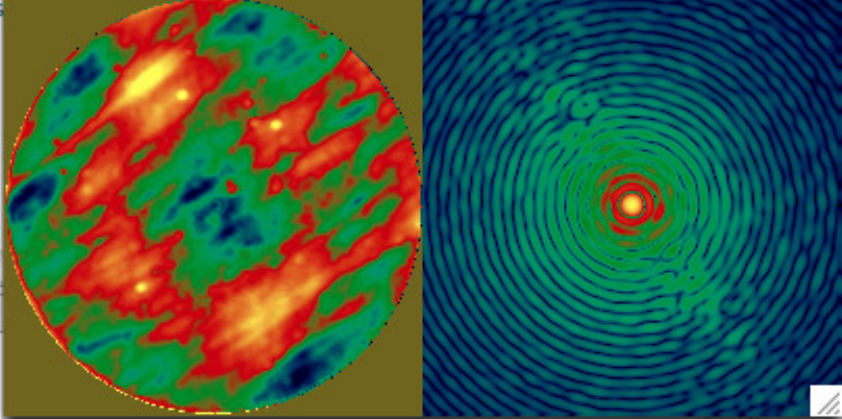}
  \caption{Measured wavefront map and calculated PSF for lens No. 40.}
  \label{fig-res-40}
\end{figure}

The lens exhibits high optical quality: peak-to-valley error is $0.325\;\lambda$, with an RMS error of $\boldsymbol{0.049\;\lambda}$. Diagonal striae in the wavefront map indicate refractive index variations from incomplete glass stirring during manufacturing.

Optical quality is quantified by the Strehl ratio $S$ (ratio of peak central intensity to that of an ideal diffraction-limited system). For lens No. 40, the calculated Strehl ratio is $\boldsymbol{S = 0.83}$, demonstrating excellent image quality.

Figure~\ref{fig-res-obj} shows wavefronts and PSFs for lenses Nos. 41, 43, and 44. Lenses 41–43 exceeded the clear aperture of our setup, requiring sub-aperture stitching across translated positions. Stitching failed for lens No. 42 (omitted). Partial pupil coverage accounts for the custom pupil outlines in Figures~\ref{fig-res-41} and \ref{fig-res-43}.

\begin{figure} \centering
  \subfloat[Wavefront and PSF for lens No. 41.]{\label{fig-res-41}
    \FIG{0.7}{false}{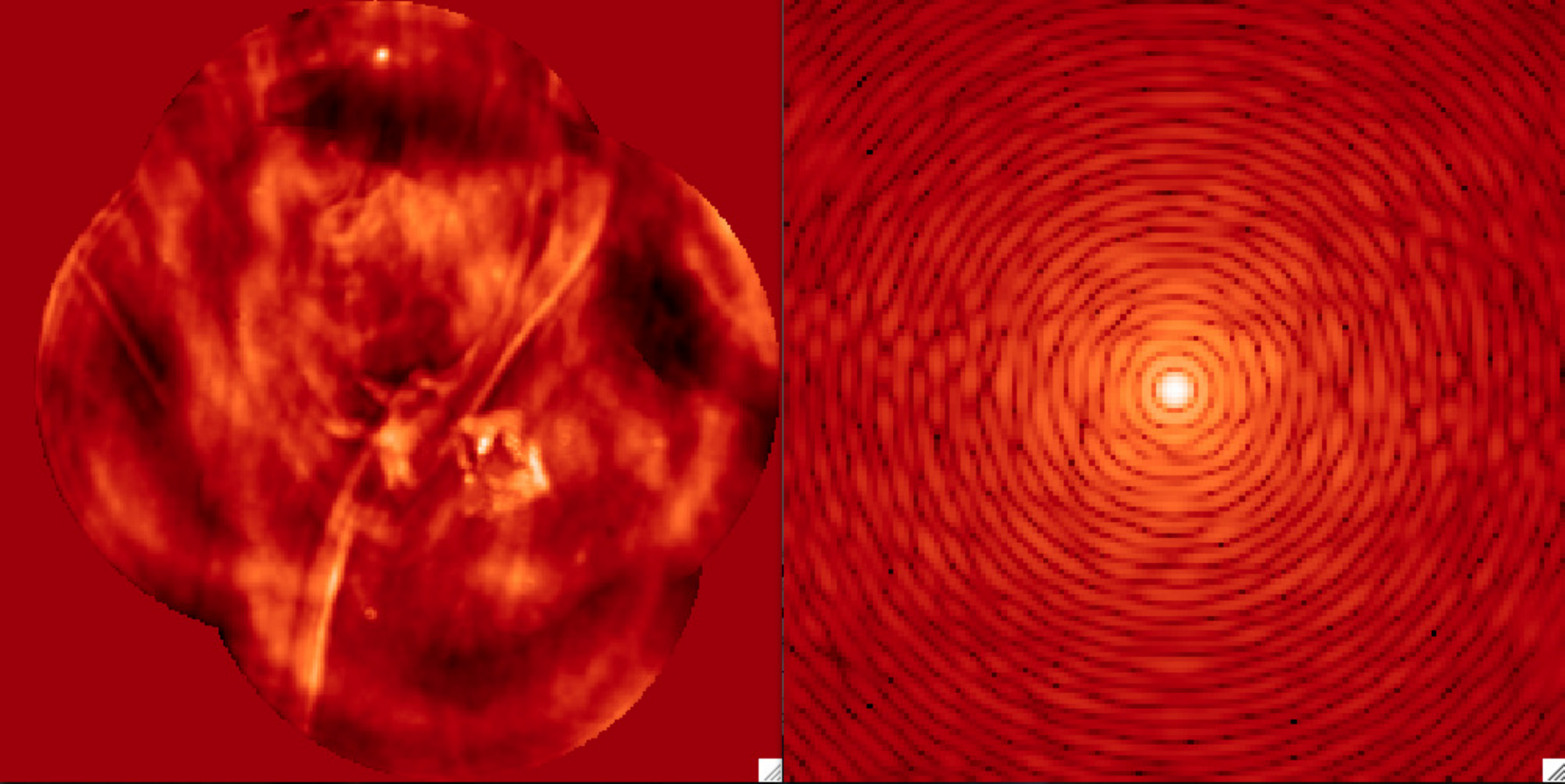}}\\
  \subfloat[Wavefront and PSF for lens No. 43.]{\label{fig-res-43}
    \FIG{0.7}{false}{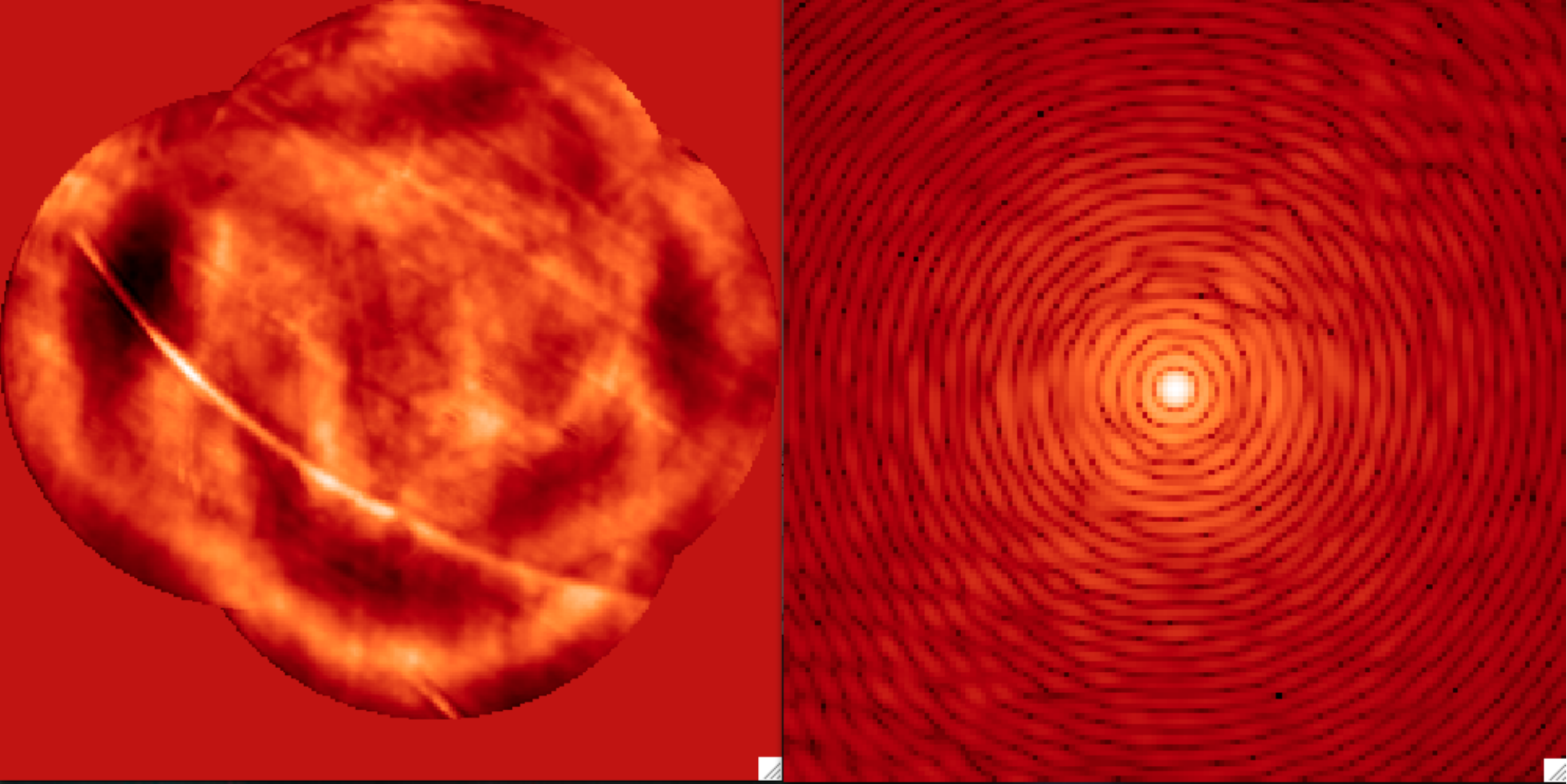}}\\
  \subfloat[Wavefront and PSF for lens No. 44.]{\label{fig-res-44}
    \FIG{0.44}{false}{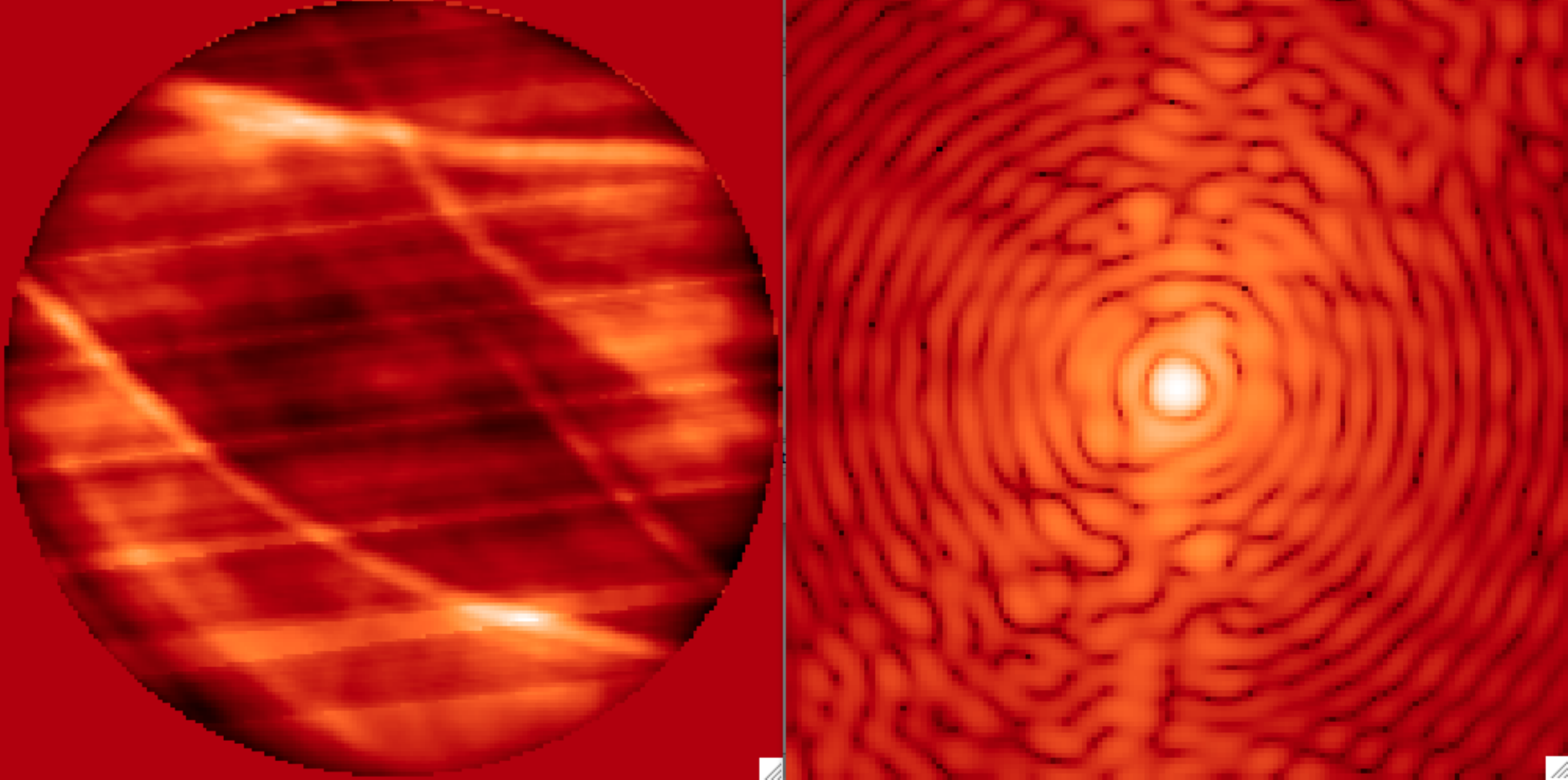}}\\
  \caption{Measured wavefront maps and associated PSFs for lenses Nos. 41, 43, and 44.}
  \label{fig-res-obj}
\end{figure}

Lenses Nos. 41 and 43 demonstrate exceptional performance for their era, achieving RMS wavefront errors of $\boldsymbol{0.027\;\lambda}$ ($S = \boldsymbol{0.94}$). Glass striae are again visible.

Lens No. 44 exhibited lower quality (RMS error $\boldsymbol{0.07\;\lambda}$, $S = \boldsymbol{0.67}$). It is heavily vignetted by an attached card diaphragm, indicating stronger edge aberrations.

%-------------------------------------------------------------------------------
\subsubsection{What Did Cassini See?}
\label{sec-que-voyait}

Wavefronts were measured in monochromatic light at 633~nm. The primary limitation of non-achromatic aerial telescopes is chromatic dispersion, which worsens at shorter focal ratios. Pending direct dispersion measurements, simulations were conducted assuming BK7 glass properties.

Eyepiece aberrations were assumed negligible for the simulations shown in Figure~\ref{fig-pla-obj}. Simulations combine computed PSFs, glass chromatic dispersion, and human eye spectral sensitivity for Saturn, Jupiter, and Mars as aligned in early 1675.

\begin{figure} \centering
  \subfloat[Planets through objective No. 40.]{\label{fig-pla-40}
    \FIG{0.75}{false}{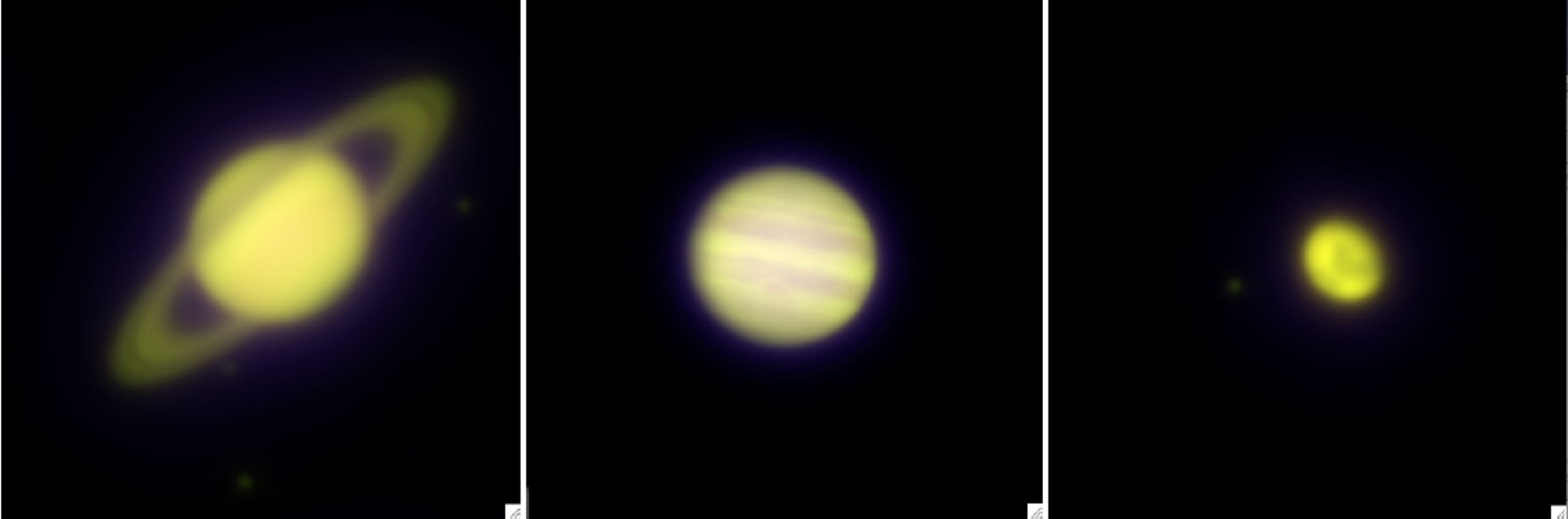}}\\
  \subfloat[Planets through objective No. 41.]{\label{fig-pla-41}
    \FIG{0.75}{false}{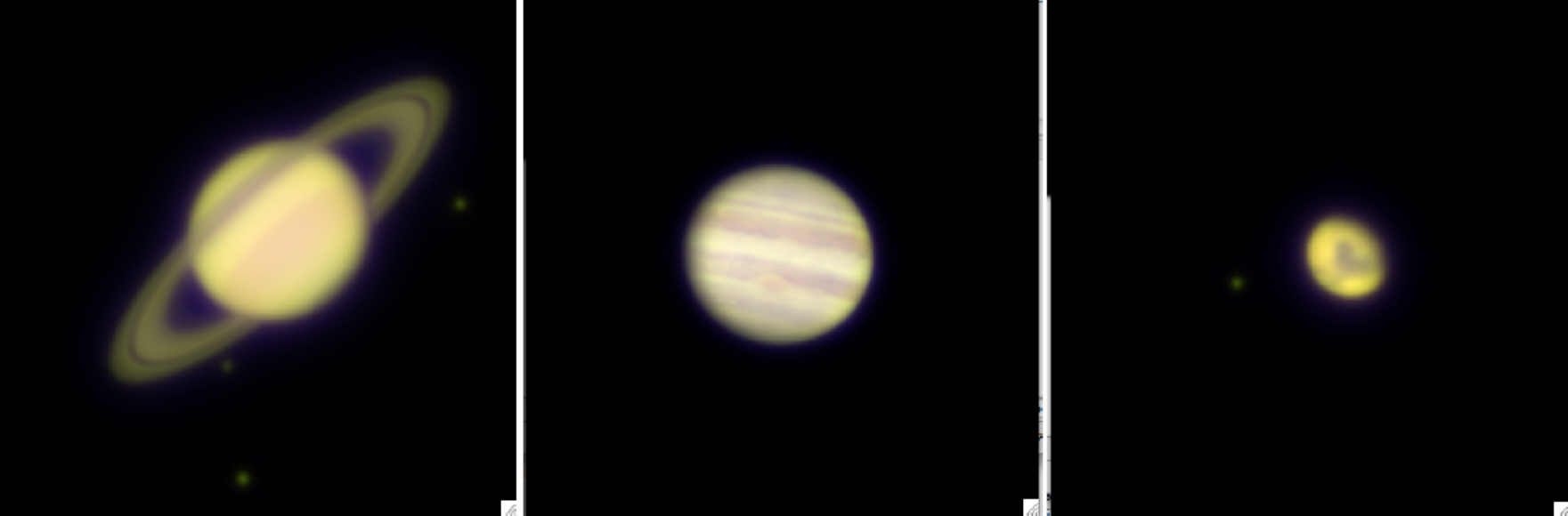}}\\
  \subfloat[Planets through objective No. 43.]{\label{fig-pla-43}
    \FIG{0.75}{false}{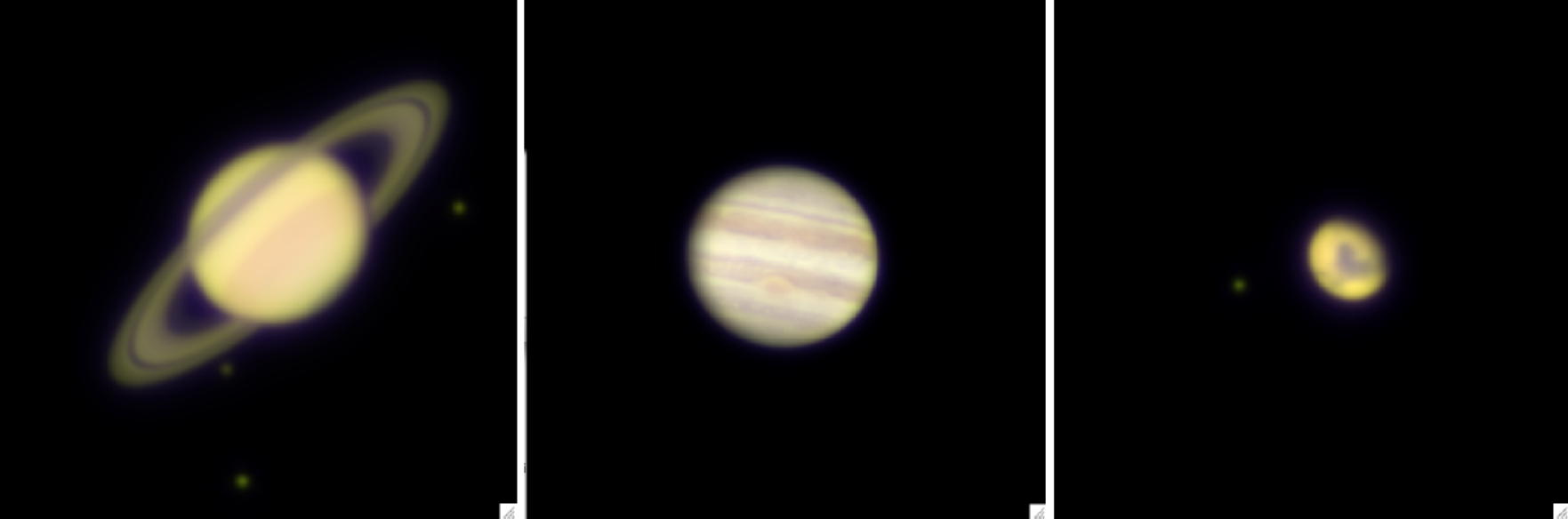}}\\
  \subfloat[Planets through objective No. 44.]{\label{fig-pla-44}
    \FIG{0.75}{false}{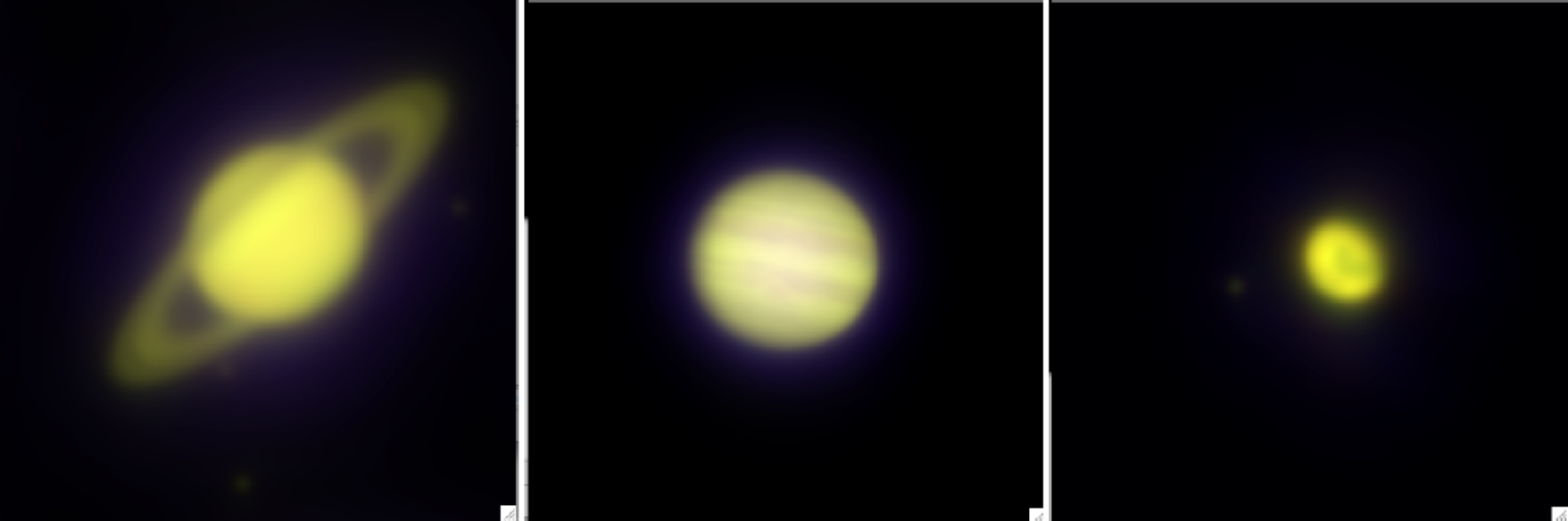}}\\
  \caption{Saturn, Jupiter, and Mars simulated through objectives Nos. 40, 41, 43, and 44.}
  \label{fig-pla-obj}
\end{figure}

Moons are not rendered to photometric scale. Chromatic defocus in the blue channel is prominent when focused at 550~nm (peak human visual sensitivity), particularly for shorter focal lengths (Nos. 40 and 44). Nevertheless, Saturn's ring division, Jupiter's Great Red Spot, and Martian dark features remain discernible across simulations.

Standard resolution criteria are difficult to apply strictly to human visual observation. Furthermore, these simulations omit atmospheric seeing. However, the human brain integrates visual imagery over ~6-second intervals, selectively retaining moments of steady seeing—analogous to modern "lucky imaging" techniques \cite{Fried78}. Cassini likely experienced moments of excellent seeing throughout a full year of tracking Saturn, allowing him to resolve the division. His discovery was genuine, unlike Schiaparelli's Martian "canals" or the fictitious Venusian moon Neith twice reported by Cassini himself.

%¤¤¤¤¤¤¤¤¤¤¤¤¤¤¤¤¤¤¤¤¤¤¤¤¤¤¤¤¤¤¤¤¤¤¤¤¤¤¤¤¤¤¤¤¤¤¤¤¤¤¤¤¤¤¤¤¤¤¤¤¤¤¤¤¤¤¤¤¤¤¤¤¤¤¤¤¤¤¤
\subsection{Ever-Larger Telescopes}
\label{sec-telescopes-plus}

%-------------------------------------------------------------------------------
\subsubsection{Refractors: Cumbersome Instruments}
\label{sec-lunettes-instruments}

Following Cassini, astronomers sought larger apertures primarily to boost light-gathering power and observe fainter targets. In 1758, John Dollond invented the achromatic doublet, reducing focal ratios from $f/100$–$f/600$ (for simple lenses) down to $f/10$–$f/20$ while maintaining comparable color correction. This dramatically reduced telescope lengths.

\begin{figure} \centering
  \subfloat[Meudon Great Refractor.]{\label{fig-lun-meudon}
    \FIGH{9}{false}{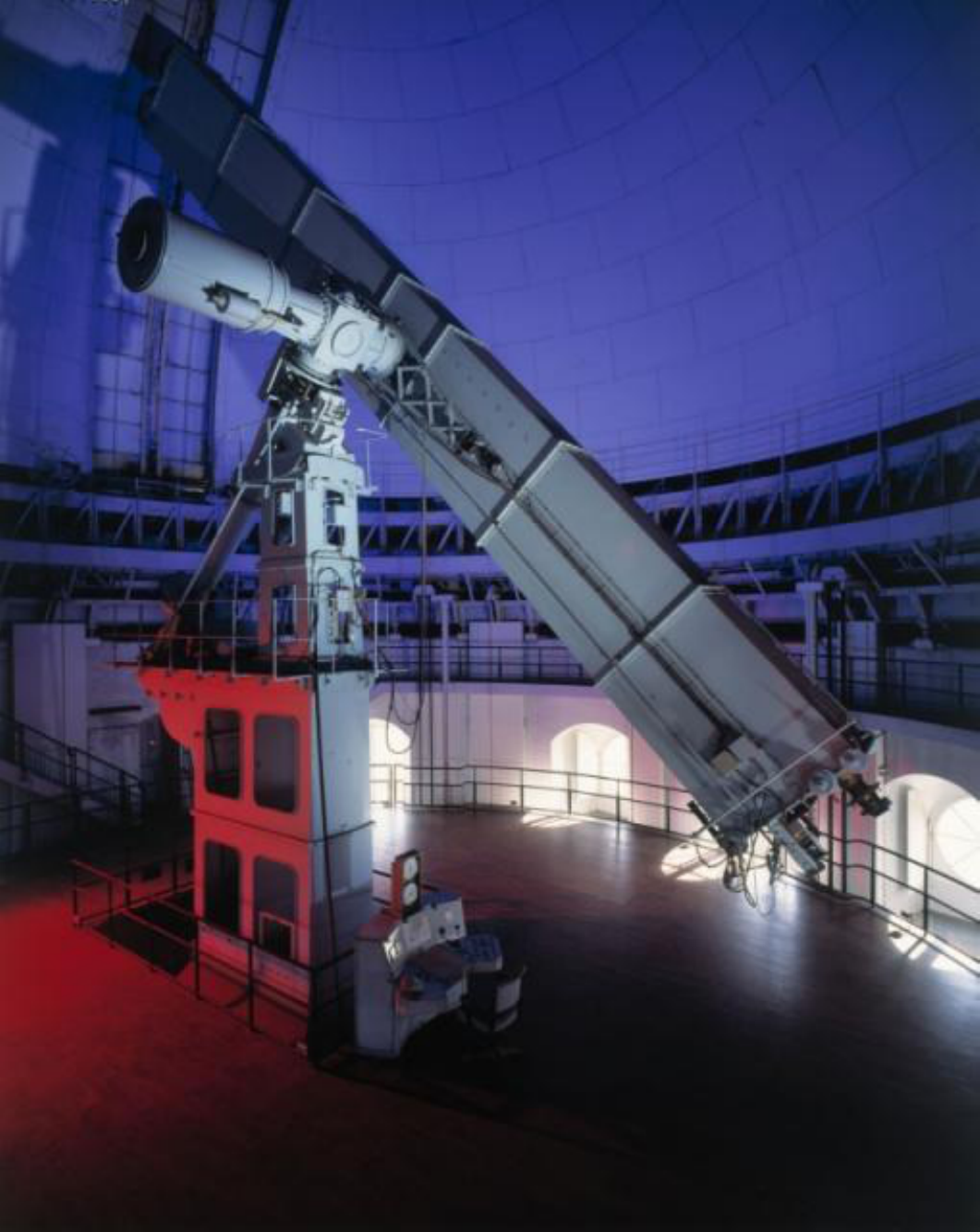}} \hspace{10pt}
  \subfloat[Lens of the 1900 Paris Exhibition Great Refractor.]{\label{fig-lun-1900}
    \FIGH{9}{false}{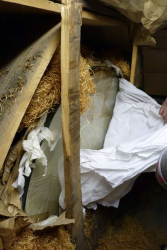}}
  \caption{Europe's largest historic refracting telescopes.}
  \label{fig-lunettes}
\end{figure}

Refractors up to 1~m in diameter were constructed before 1900. Europe's largest operational refractor sits at the Meudon Observatory (double objective of 62 and 83~cm diameter, $f = 12$~m; Figure~\ref{fig-lun-meudon}). The largest lens doublet ever built was for the Great Paris Exhibition Telescope of 1900 (1.25~m diameter, 57~m focal length). Immovable due to its size, it used a 2-meter Foucault siderostat mirror. The project failed optical tests and was dismantled; its optics reside in the Paris Observatory cellars (Figure~\ref{fig-lun-1900}). This marked the limit of large refractors. Lenses supported only along their perimeter sag and deform under gravity when tilted, distorting optical figures. Consequently, reflective optics (mirrors) took center stage.

%-------------------------------------------------------------------------------
\subsubsection{The Evolution of Reflecting Telescopes}
\label{sec-evolution-telescopes}

In parallel, reflecting telescopes evolved initially using speculum metal mirrors up to 1.8~m in size, though these suffered from gravitational sag and thermal distortion.

In 1857, Foucault introduced silvering on glass substrates, initiating the era of modern glass mirrors. Around this period, Herschel (1828) described stellar diffraction rings, and Airy (1835) established their mathematical formulation. Rayleigh later formulated his criterion for the minimum resolvable angular separation $\theta$ between two point sources of equal brightness, for circular apertures:
\begin{equation}
  \theta \simeq \sin\theta = 1.22\frac{\lambda}{D}.
\end{equation}

In this case, we consider the two stars of equal brightness, with one star on the first zero of the second star. Alternative metrics exist; for instance, the Sparrow criterion defines resolution at the threshold where we do not see a line between two equaly bright Airy patterns (roughly half the Rayleigh limit).

Alternatively, resolution is frequently evaluated via spatial cutoff frequency (treating the aperture as a spatial low-pass filter):
\begin{equation}
  \theta = \frac{\lambda}{D}.
\end{equation}

We adopt this definition ($\lambda/D$) hereafter. In all cases, larger apertures $D$ yield finer angular resolution and collect light proportional to aperture area ($D^2$).

Figure~\ref{fig-tailles-telescopes} traces aperture scaling over time. During the 20th century, primary mirrors scaled from 1 to 10~m, boosting angular resolution tenfold and collecting power 100-fold.

\begin{figure} \centering
  \FIG{0.9}{false}{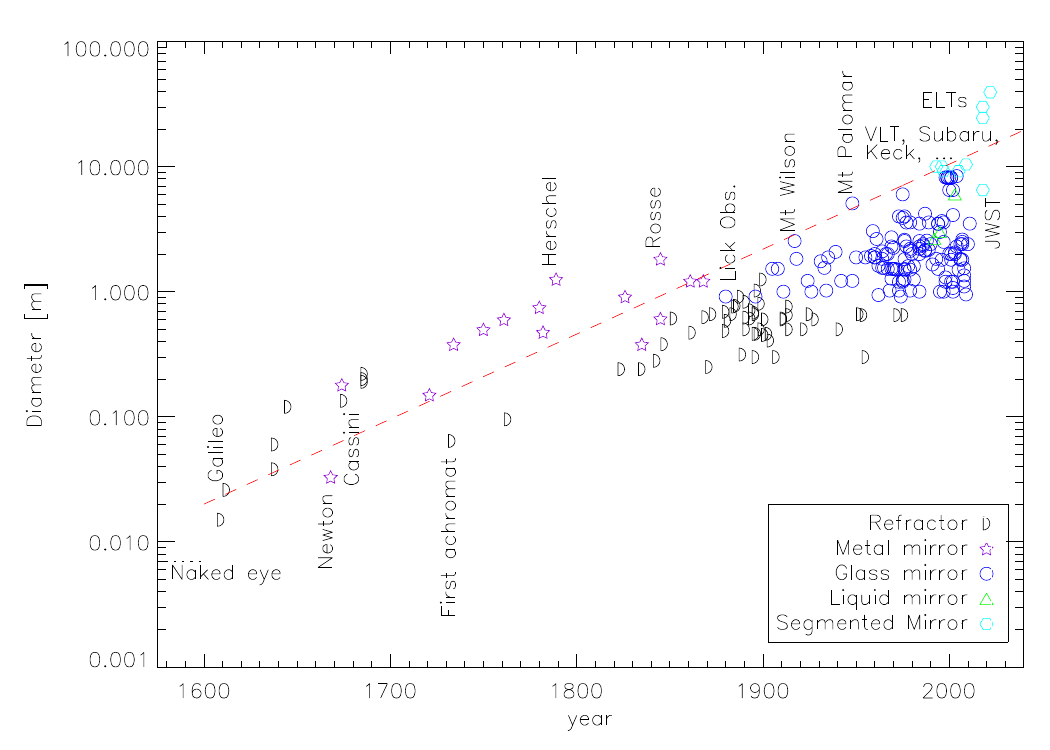}
  \caption[Evolution of refractor and reflector sizes over time.]{Evolution of refractor and reflector sizes over time.}
  \label{fig-tailles-telescopes}
\end{figure}

Monolithic glass mirrors reach a physical limit near 8.4~m, as seen in the Gemini telescopes, Very Large Telescope (VLT) UTs, The two honeycombed mirrors of the Large Binocular Telescope (LBT), and the Subaru telescope. Segmented mirror designs bypassed this limit, yielding ~10-meter apertures in the Keck Observatory and Gran Telescopio Canarias (GTC). Future Extremely Large Telescopes (ELTs) will scale apertures further: 24.5~m for the Giant Magellan Telescope (GMT), 30~m for the Thirty Meter Telescope (TMT), and 39~m for the European Extremely Large Telescope (E-ELT) (Figure~\ref{fig-telescopes}).

\begin{figure} \centering
  \subfloat[Subaru Telescope \cre{NAOJ}.]{\label{fig-subaru}
    \FIGH{5.1}{false}{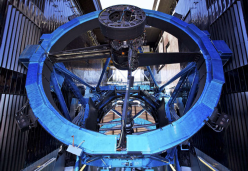}}\hfill
  \subfloat[VLT Telescopes \cre{ESO / H. H. Heyer}.]{\label{fig-vlt}
    \FIGH{5.1}{false}{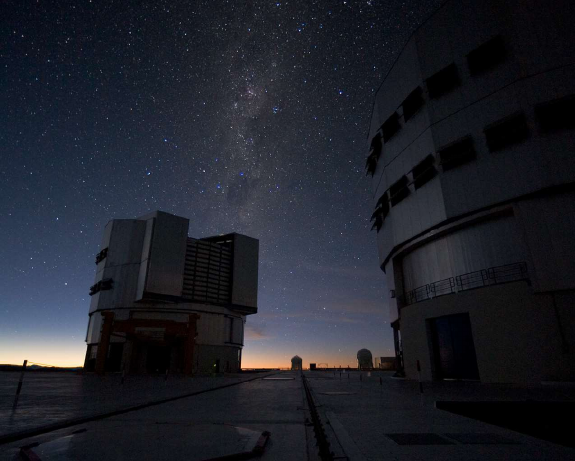}}\\
  \subfloat[TMT \cre{TMT Observatory Corporation}.]{\label{fig-tmt}
    \FIGH{4.4}{false}{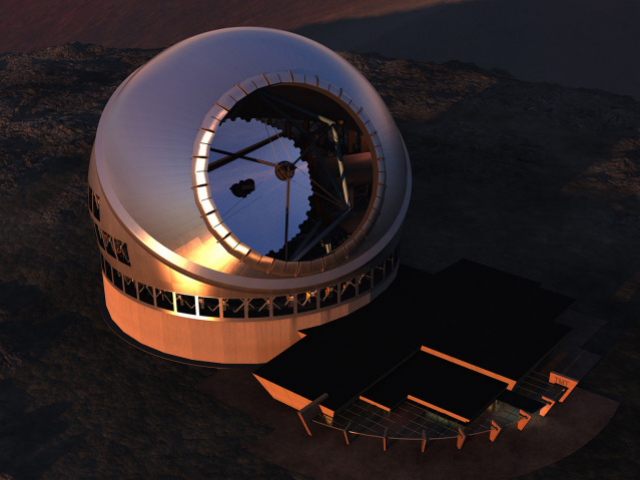}}\hfill \subfloat[E-ELT
  \cre{ESO}.]{\label{fig-eelt}
    \FIGH{4.4}{false}{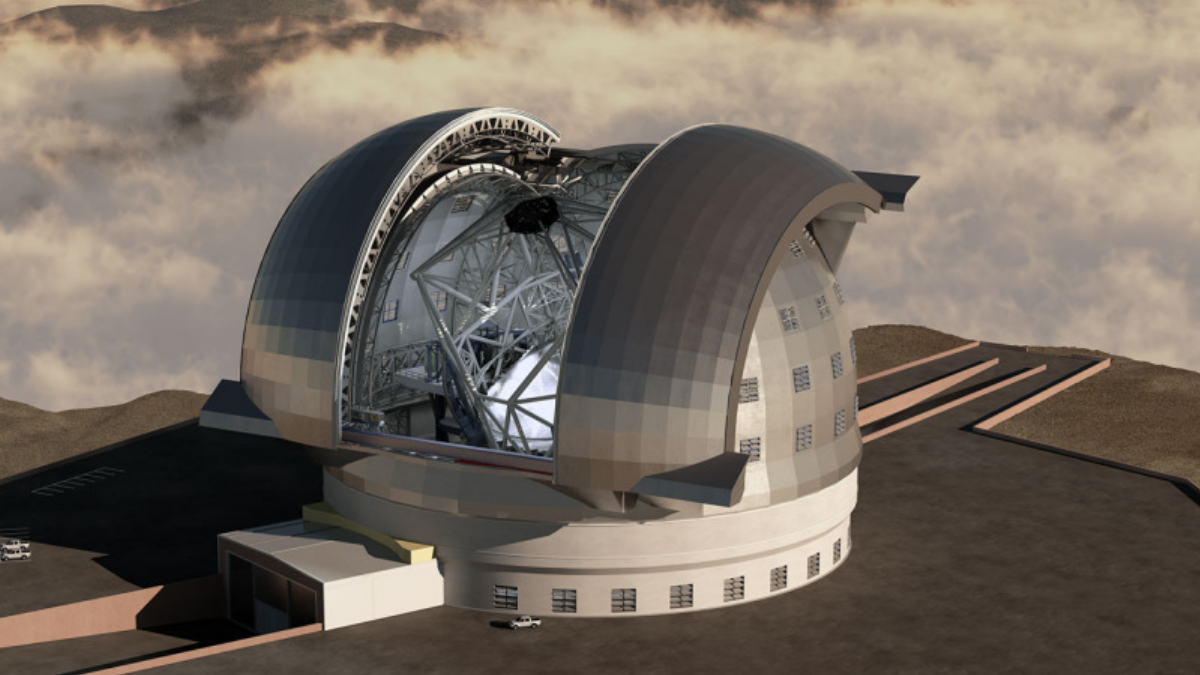}}
  \caption[Current and future ground-based telescopes.]{Current large telescopes (monolithic apertures) and upcoming Extremely Large Telescopes (segmented apertures).}
  \label{fig-telescopes}
\end{figure}

%¤¤¤¤¤¤¤¤¤¤¤¤¤¤¤¤¤¤¤¤¤¤¤¤¤¤¤¤¤¤¤¤¤¤¤¤¤¤¤¤¤¤¤¤¤¤¤¤¤¤¤¤¤¤¤¤¤¤¤¤¤¤¤¤¤¤¤¤¤¤¤¤¤¤¤¤¤¤¤
\subsection{Observing Exoplanets with a Single Telescope}
\label{sec-observer-exoplanetes}

%-------------------------------------------------------------------------------
\subsubsection{Adaptive Optics: Unlocking Telescope Potential}
\label{sec-optique-adaptative}

Ground-based optical resolution remains limited by atmospheric turbulence. Index fluctuations across turbulent atmospheric layers distort incoming plane wavefronts, causing dynamic image blurring. Uncorrected resolution is limited by the spatial coherence scale of turbulence, known as the Fried parameter $r_0$. Scaling as $\lambda^{6/5}$, $r_0$ is $\sim 10$~cm in the visible and $\sim 40$~cm in the near-infrared. Consequently, an uncorrected 8-meter visible light telescope yields the spatial resolution of a 10~cm backyard instrument. Adaptive Optics (AO) compensates for atmospheric phase distortions in real time (Figure~\ref{fig-boucle-oa}).

\begin{figure} \centering
  \FIG{0.7}{false}{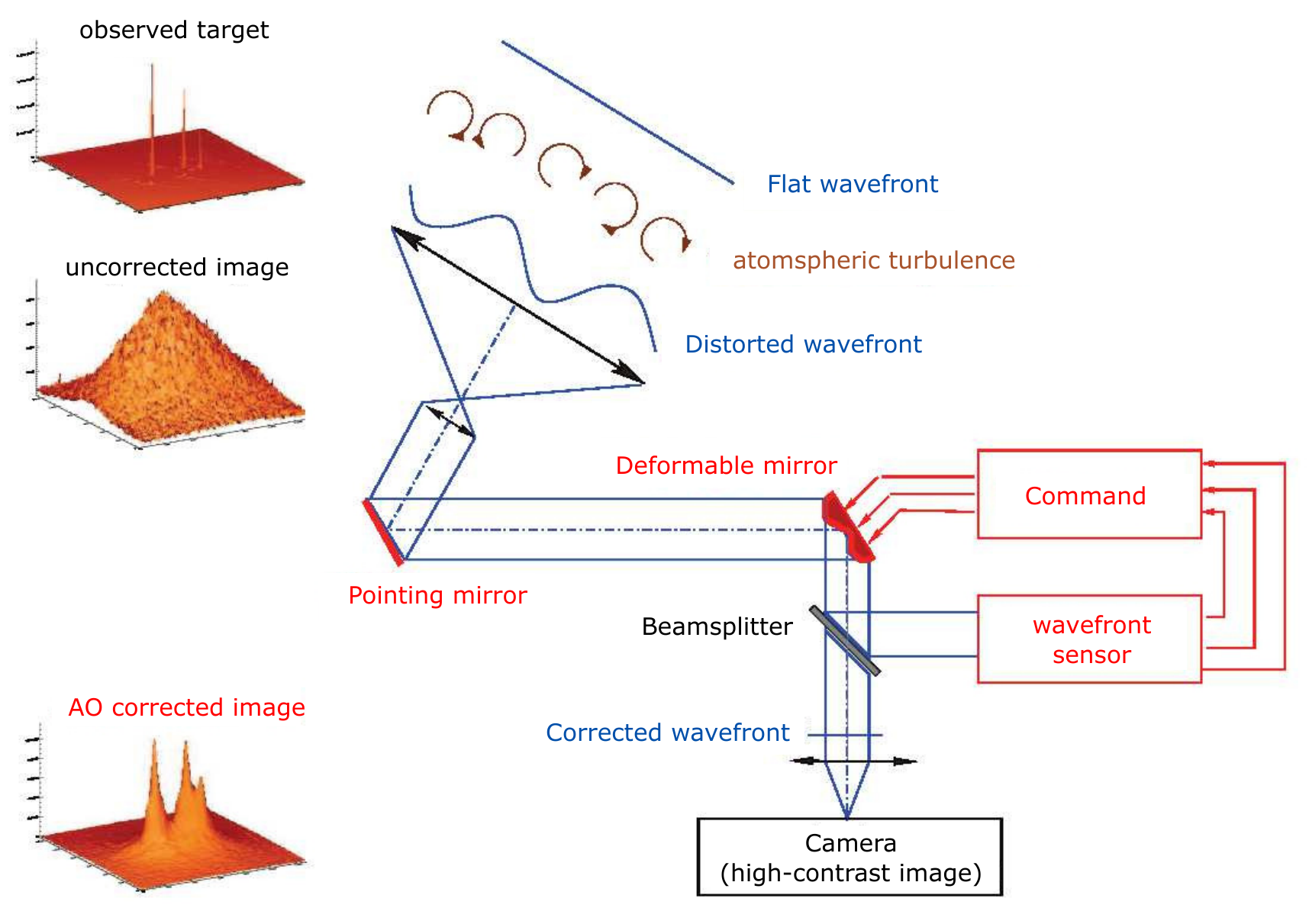}
  \caption{Principle of Adaptive Optics.}
  \label{fig-boucle-oa}
\end{figure}

An AO system operates as a closed feedback control loop:
\begin{itemize}
\item A wavefront sensor measures phase aberrations introduced by the atmosphere;
\item A real-time computer computes corrective commands;
\item A deformable mirror (driven by piezoelectric actuators, electromagnetic actuators, or other technologies) alters its profile to flatten the aberrated wavefront, which is re-measured by the sensor.
\end{itemize}

Correction quality depends on sensor sensitivity, actuator density, loop execution frequency, and control algorithms. For high-contrast exoplanet imaging over narrow fields, a single deformable mirror conjugated to on-axis turbulence provides sufficient correction. When closed-loop AO operates efficiently, diffraction-limited resolution ($\lambda/D$) is achieved, concentrating light into a central Airy core surrounded by faint speckles and diffraction rings.

Without AO, large ground-based telescopes yield Strehl ratios of only ~1\%. First- and second-generation AO systems achieve Strehl ratios around 0.6 (e.g., 0.5 for MACAO\footnote{Multi Application Curvature Adaptive Optics} and 0.6 for NAOS-CONICA\footnote{Nasmyth Adaptive Optics System} at the VLT, up to 0.7 for AO188 \cite{Hayano10} at Subaru). AO enabled the first direct image of an exoplanet in 2004 using VLT/NACO (Figure~\ref{fig-exo-vlt}) \cite{Chauvin04}.

\begin{figure} \centering
  \FIG{0.5}{false}{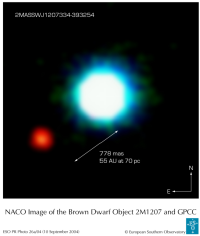}
  \caption[First direct image of an exoplanet.]{First direct image of an exoplanet, obtained at the VLT in 2004 \cre{ESO}.}
  \label{fig-exo-vlt}
\end{figure}

In that system (2M1207b), the 5-Jupiter-mass planet orbits a brown dwarf star at a wide separation under a modest flux contrast ($\sim 10^2$). Detecting closer planets at extreme contrasts requires far higher wavefront correction. Halo speckles from uncorrected phase errors degrade achievable contrast. This motivated eXtreme Adaptive Optics (XAO), which scales actuator counts by an order of magnitude and boosts loop rates to achieve ultra-high Strehl ratios over narrow fields around target stars.

LBT's FLAO\footnote{First Light Adaptive Optics} system demonstrated operational Strehl ratios exceeding 0.9 \cite{Esposito11}. Other XAO systems include SAXO\footnote{SPHERE Adaptive optics for eXoplanet Observation} \cite{Fusco06} for SPHERE\footnote{Spectro-Polarimetric High-contrast Exoplanet REsearch} on the VLT \cite{Beuzit06}, GPI\footnote{Gemini Planet Imager} for Gemini \cite{Macintosh06}, and SCExAO\footnote{Subaru Coronagraphic Extreme Adaptive Optics} on Subaru \cite{Martinache09}, to which I contributed.

%-------------------------------------------------------------------------------
\subsubsection{Coronagraphy: Suppressing Stellar Light}
\label{sec-coronographie}

Diffraction-limited imaging alone is insufficient to resolve faint exoplanets. Contrast ratios for Sun-Jupiter analogues reach $10^{-7}$ in the visible and $10^{-6}$ in the thermal infrared ($10^{-4}$ and $10^{-3}$ for hot Jupiters). Without stellar suppression, planet signals remain buried under stellar diffraction rings and speckle halos. High-contrast imaging uses coronagraphy to suppress on-axis stellar light while transmitting off-axis planet photons (Figure~\ref{fig-corono}).

\begin{figure} \centering
  \FIG{0.9}{false}{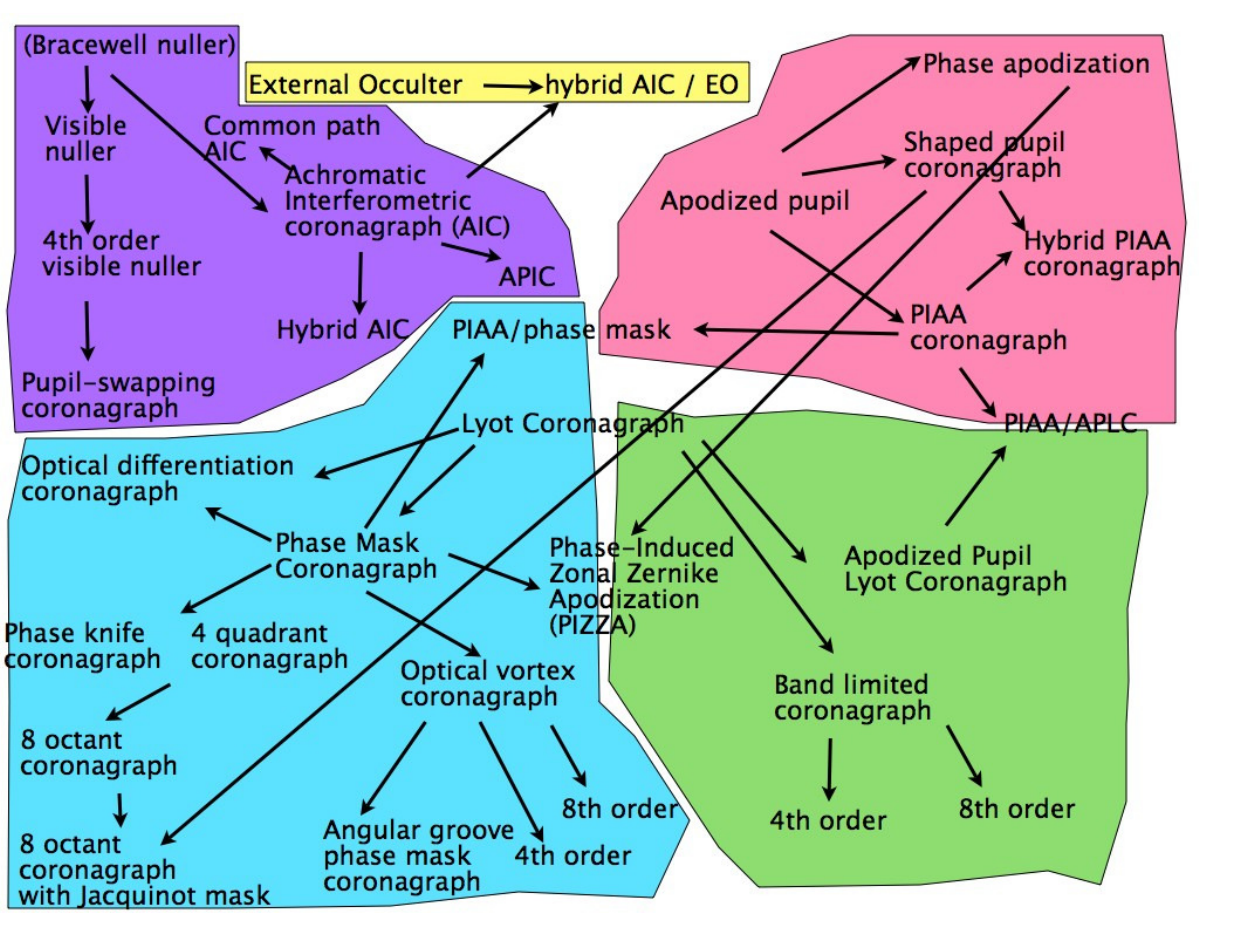}
  \caption[High-contrast imaging techniques.]{Overview of high-contrast imaging techniques \cre{O. Guyon}.}
  \label{fig-corono}
\end{figure}

Monolithic aperture techniques stem from Bernard Lyot's 1930s solar coronagraph \cite{Lyot33} built at Paris Observatory in Meudon. Modern stellar coronagraphs are categorized into four primary families (see Fig.~\ref{fig-corono}:
\begin{itemize}
\item Pupil-plane interferometric coronagraphs (top left, purple), creating destructive interference across the pupil;
\item Image-plane interferometric coronagraphs (bottom left, blue), inducing destructive interference in a intermediate focal plane before a Lyot stop in the pupil blocks the diffracted light;
\item Amplitude-mask coronagraphs (bottom right, green), utilizing pupil apodization without changing the phase, they too use a Lyot stop;
\item Phase-induced amplitude apodization coronagraphs (top right, pink), remapping pupil phase to shape intensity profiles prior to focal masking.
\end{itemize}

XAO systems integrate with advanced coronagraphs. With their modularity, SPHERE and SCExAO will serve as testbeds for comparing multi-coronagraph configurations.

\begin{figure} \centering
  \FIG{0.7}{false}{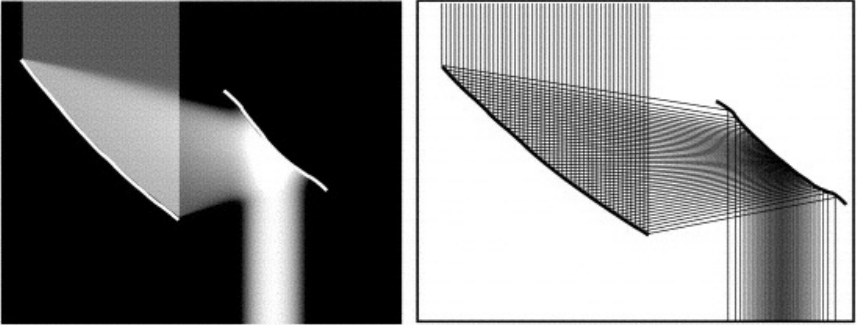}
  \caption[Phase Induced Amplitude Apodization (PIAA) concept.]{Phase Induced Amplitude Apodization (PIAA) concept \cre{O. Guyon}.}
  \label{fig-PIAA}
\end{figure}

One technique enabling inner working angles down to $1\,\lambda/D$ is the Phase Induced Amplitude Apodization Coronagraph (PIAAC) developed by O. Guyon \cite{Guyon03}, implemented on SCExAO. Custom aspheric mirrors remap uniform input beams into Gaussian-like intensity profiles without light loss, suppressing diffraction rings close to the core. PIAA delivers theoretical contrasts of $10^{-6}$ at $1\,\lambda/D$ on SCExAO, and up to $10^{-10}$ for space applications.

%§§§§§§§§§§§§§§§§§§§§§§§§§§§§§§§§§§§§§§§§§§§§§§§§§§§§§§§§§§§§§§§§§§§§§§§§§§§§§§§
\section{Interferometry for Exoplanet Observation}
\label{sec-interf-observ}

%¤¤¤¤¤¤¤¤¤¤¤¤¤¤¤¤¤¤¤¤¤¤¤¤¤¤¤¤¤¤¤¤¤¤¤¤¤¤¤¤¤¤¤¤¤¤¤¤¤¤¤¤¤¤¤¤¤¤¤¤¤¤¤¤¤¤¤¤¤¤¤¤¤¤¤¤¤¤¤
\subsection{Evolution of Optical Interferometry}
\label{sec--evolution-1}

%-------------------------------------------------------------------------------
\subsubsection{Two-Wave Interference}
\label{sec-interferences-deux}

Thomas Young demonstrated the wave nature of light in 1801 using his double-slit experiment (Figure~\ref{fig-franges-young}).

\begin{figure} \centering
  \FIGWH{0.3}{0.28}{false}{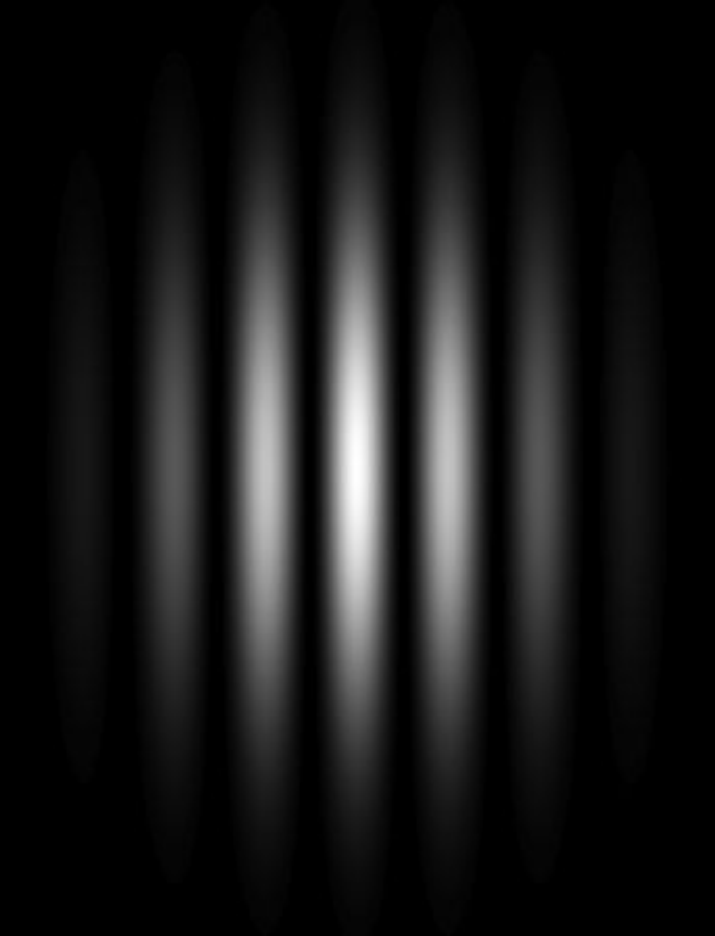}
  \caption{Interference pattern obtained in Young's double-slit experiment.}
  \label{fig-franges-young}
\end{figure}

Interferometers fall into two main classes: wavefront-splitting (e.g., Young's double slits) and amplitude-splitting (e.g., Michelson interferometer). The first one combines different parts of the same wavefront, while the second one splits the same incident wavefront before recombination.  Some coronagraphs presented above use amplitude-splitting interferometry, while long-baseline astronomical interferometers split wavefronts by combining light collected from separate sub-pupils (Figure~\ref{fig-interfero}).

\begin{figure} \centering
  \FIG{0.7}{false}{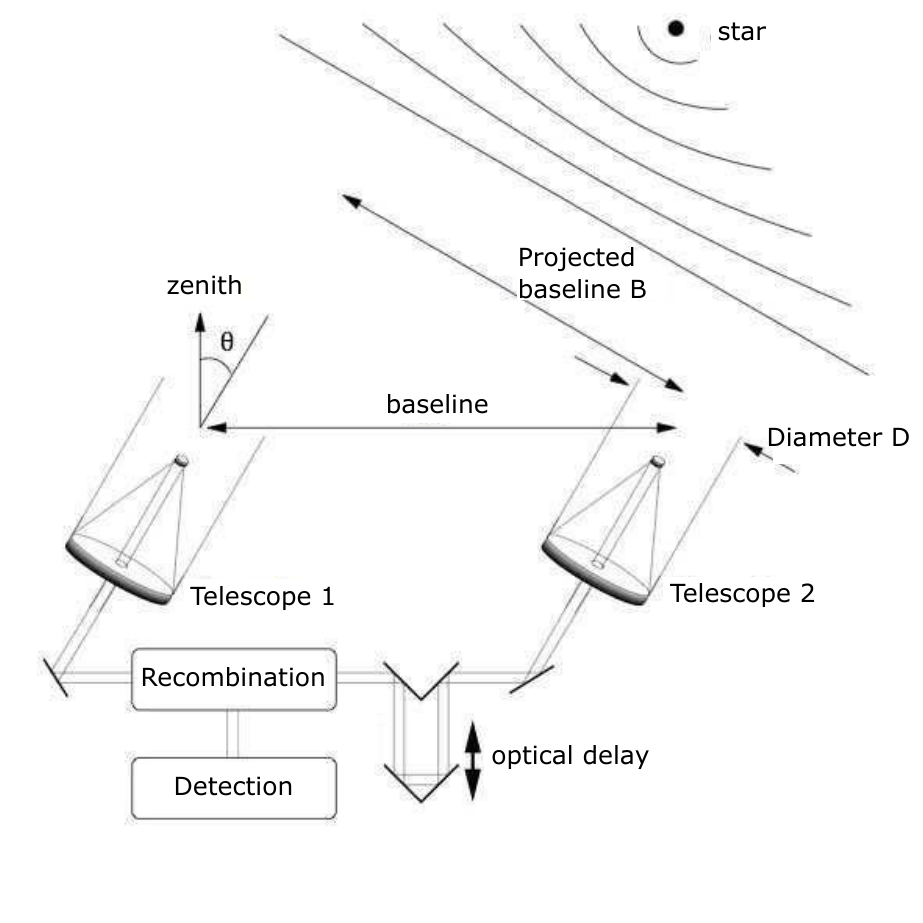}
  \caption{Principle of optical stellar interferometry.}
  \label{fig-interfero}
\end{figure}

For a point source, two-beam interference of intensities $I_1$ and $I_2$ as a function of Optical Path Difference (OPD) $\delta$ is given by:
\begin{equation}
  I(\delta) = I_1 + I_2 + 2\sqrt{I_1I_2}|\gamma(\delta)|
  \cos\GP{\phi_\gamma+2\pi\sigma\delta}.
  \label{eq-interf}
\end{equation}

OPD is dynamically tuned using optical Delay Lines (DL) consisting of mobile retroreflecting mirrors (Figure~\ref{fig-interfero}).

We adopt hereafter the wavenumber $\sigma = 1/\lambda$ rather than wavelength $\lambda$, as phase and delay equations scale linearly with $\sigma$.

In Equation~\eqref{eq-interf}, $\gamma$ denotes the complex degree of temporal coherence, given by the Fourier Transform (FT) of the source spectral energy distribution $S(\sigma)$ \cite{Goodman85}:
\begin{equation}
  \gamma(\delta) = |\gamma(\delta)|e^{i\phi_\gamma} = \TF{S}(\delta).
\end{equation} 

Narrower source spectra yield longer temporal coherence lengths $L_\mc$ over which fringes can be observed:
\begin{equation}
  L_\mc = \frac{\lambda^2}{\Delta\lambda} = \frac1{\Delta\sigma}.
\end{equation}

Coherence lengths range from meters/kilometers for lasers down to tens of micrometers for broadband stellar light.

This relationship between coherence length and spectrum can be use to our advantage. Scanning OPD linearly thanks to delay lines yields Fourier Transform Spectrometry (FTS) datasets \cite{Connes75}.

%-------------------------------------------------------------------------------
\subsubsection{Early Stellar Interferometry}
\label{sec-debuts-interferometrie}

Fizeau and Stephan attempted the first stellar diameter measurements in 1868 \cite{Fizeau68}, though sub-meter baselines proved insufficient. Michelson validated the technique in 1890 on Jupiter's Galilean moons. In 1920, Michelson and Pease mounted a 4-meter periscope beam on the 2.5-meter Hooker Telescope at Mt. Wilson (Figure~\ref{fig-interf-1920}), measuring the stellar diameter of Betelgeuse \cite{Michelson20}.

\begin{figure} \centering
  \FIG{0.7}{false}{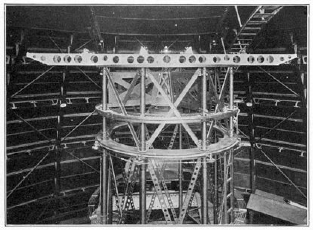}
  \caption[Michelson interferometer on the Hooker Telescope.]{Michelson 6-meter beam interferometer mounted on the 2.5-meter Hooker Telescope (1920).}
  \label{fig-interf-1920}
\end{figure}

Radio astronomy advanced rapidly in the late 1930s due to relaxed structural tolerances at longer wavelengths. Optical interferometry regained momentum in the 1970s through Antoine Labeyrie's I2T\footnote{Interféromètre à 2 Télescopes} \cite{Labeyrie75} and GI2T\footnote{Grand Interféromètre à 2 Télescopes} \cite{Labeyrie86,Mourard94} on the Calern plateau. Since then, interferometry continued to grow, both in number of telescope and baseline $B$, i.e. the distance between pupils.

%-------------------------------------------------------------------------------
\subsubsection{Resolution and Image Reconstruction}
\label{sec-res-interf}

An interferometer's spatial resolution is defined by the angular spacing where fringe patterns from two point sources shift into anti-phase (half-fringe spacing):
\begin{equation}
  \theta = \frac{\lambda}{2B}.
\label{eq-res-interf}
\end{equation}

Interferometers sample visibility parameters in the Fourier domain rather than capturing direct images. Image reconstruction relies on the Van Cittert-Zernike theorem:
\begin{quote}
  For a quasi-monochromatic, spatially incoherent source, the complex degree of spatial coherence equals the normalized Fourier transform of the source brightness distribution.
\end{quote}

The complex coherence factor $\mu(u,v)$ is defined as \cite{Goodman85}:
\begin{equation}
  \mu(u,v) = \frac{\moy[S]{\esp(x,y)\esp^*(x+\lambda u,y+\lambda
      v)}}{\moy[S]{|\esp(x,y)|^2}},
\end{equation}
where $u$ and $v$ are spatial frequency coordinates in the pupil plane. Thus $\mu$ is related to the object's intensity $I$ with:
\begin{equation}
  \mu(u,v)=\frac{\TF{I}(u,v)}{\TF{I}(0,0)}.
\end{equation}

Reconstructing an image requires inverse Fourier transforming $\mu(u,v)$ sampled across dense spatial frequencies. Earth's rotation sweeps fixed baselines through the $(u,v)$ plane over time. Re-configurable arrays with multiple telescopes further improve spatial frequency coverage.

\begin{figure} \centering
  \FIG{0.5}{false}{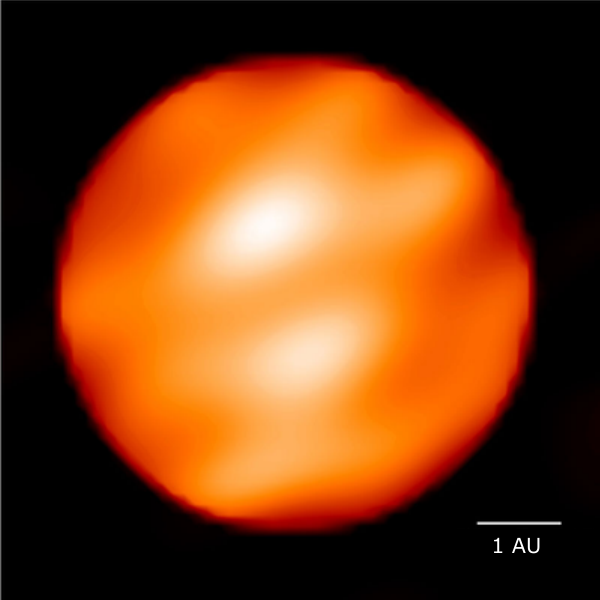}
  \caption[Interferometric image of Betelgeuse.]{Surface image of Betelgeuse at $\lambda = 1.64$~\mum obtained with IOTA, resolving 9~mas surface features across a 55~mas field \cre{Haubois/Perrin (LESIA, Observatoire de Paris)}.}
  \label{fig-betelgeuse}
\end{figure}

Figure~\ref{fig-betelgeuse} displays an image of Betelgeuse reconstructed from IOTA\footnote{Infrared Optical Telescope Array} data \cite{Haubois09} using algorithms such as MIRA (\'Eric Thi\'ebaut) and WISARD (Serge Meimon \& Laurent Mugnier, ONERA) \cite{Meimon05}.

%-------------------------------------------------------------------------------
\subsubsection{Ground-Based Interferometers Worldwide}
\label{sec-interferometres-au}

\begin{table} \small \centering
  \caption{Operational and planned long-baseline optical interferometers.}
  \renewcommand{\arraystretch}{1.0}
  \begin{minipage}{14cm}
    \begin{tabular}{@{}>{\centering}m{1.45cm}@{\hsb}>{\centering}m{2.65cm}
        @{\hsb}>{\centering}m{2.65cm}@{\hsb}c@{\hsb}c@{\hsb}c@{\hsb}c@{\hsb}
        c@{\hsb}c@{}}
      \hline\hline
      Name & Operator & Location & Start & $N_D$ & $D$ [m] & $N_{B}$ & $B$ [m]
      & $\lambda$ [\mum{}] \\

      \hline ISI\footnote{Infrared Spatial Interferometer} & University of
      \mbox{California} & Mt. Wilson, \mbox{CA, USA} & 1988 & 3 & 1.65 & 3 & 70
      & 8--13 \\

      \hline SUSI\footnote{Sydney University Stellar Interferometer} & Sydney
      Institute for Astronomy & Narrabri, NSW, Australia & 1991 & 2 & 0.12 & 10
      & 5-160 & 0.43--0.95 \\

      \hline NOI\footnote{Navy (Prototype) Optical Interferometer} (NPOI) &
      USNO\footnote{United States Naval Observatory} & Anderson Mesa, AZ, USA &
      1994 & 6 & 0.12 & 435 & 2--437 & 0.45--0.85 \\

      \hline CHARA\footnote{Center for High Angular Resolution Astronomy} Array
      & CHARA & Mt. Wilson, \mbox{CA, USA} & 1999 & 6 & 1 & 15 & 34--331 &
      0.5--2.4 \\

      \hline KI\footnote{Keck Interferometer}, end & \multirow{2}{*}{JPL} &
      Mauna Kea, & 2001 & \multirow{2}{*}{2} & \multirow{2}{*}{10} &
      \multirow{2}{*}{2} & \multirow{2}{*}{85} & 2--2.4 \\
      mid-2012 & & HI, USA & 2004 & & & &  & 8--13 \\

      \hline MIRA-I.2\footnote{Mitaka InfraRed and optical Array} &
      NAOJ\footnote{National Astronomical Observatory of Japan} & Tokyo, Japan &
      2002 & 2 & 0.3 & 2 & 30 & 0.6--1 \\

      \hline VLTI\footnote{Very Large Telescope Interferometer} &
      \multirow{2}{*}{ESO\footnote{European Southern Observatory}} & Cerro
      Paranal, & 2002 & 2 & \multirow{2}{*}{8.2} & \multirow{2}{*}{6} &
      \multirow{2}{*}{47--130} & 8--13 \\
      VIMA & & Chile & 2004 & 3 & & & & 1.1--2.4 \\

      \hline VLTI & \multirow{2}{*}{ESO} & Cerro Paranal, &
      \multirow{2}{*}{2005} & 2 & \multirow{2}{*}{1.8} & \multirow{2}{*}{248} &
      \multirow{2}{*}{8--202} & 8--13 \\
      VISA & & Chile & & 3 & & & & 1.1--2.4 \\

      \hline OHANA\footnote{Optical Hawaiian Array for Nanoradian Astronomy}
      (tests) & Mauna Kea Observatories & Mauna Kea, \mbox{HI, USA} & 2005 & 7 &
      3--10 & 7 & 800 & 0.5--2.4 \\

      \hline LBTI\footnote{Large Binocular Telescope Interferometer} &
      Consortium LBT & Mt. Graham, \mbox{AZ, USA} & 2011 & 2 & 8.4 & 2 & 23 &
      1--2.4 \\

      \hline MROI\footnote{Magdalena Ridge Observatory Interferometer} & New
      Mexico Tech & Magdalena Ridge, NM, USA & 2013 & 10 & 1.5 & 10 & 340 &
      0.6--2.4 \\
      \hline\hline
    \end{tabular}
  \end{minipage}
  \label{tab-interferometres}
\end{table}

Table~\ref{tab-interferometres} lists long-baseline optical interferometers operating or under construction in 2011. Large fixed-baseline facilities (KI, VLTI/VIMA) target faint objects, while multi-element arrays (NPOI, VLTI/VISA) prioritize imaging synthesis with smaller telescopes.

%¤¤¤¤¤¤¤¤¤¤¤¤¤¤¤¤¤¤¤¤¤¤¤¤¤¤¤¤¤¤¤¤¤¤¤¤¤¤¤¤¤¤¤¤¤¤¤¤¤¤¤¤¤¤¤¤¤¤¤¤¤¤¤¤¤¤¤¤¤¤¤¤¤¤¤¤¤¤¤
\subsection{Principle of Nulling Interferometry}
\label{sec-principe-interferometrie}

Nulling interferometry suppresses stellar light by producing destructive interference on-axis. Ronald N. Bracewell (1978) proposed a two-telescope nulling arrangement to observe exoplanets (Figure~\ref{fig-interf-bracewell}) \cite{Bracewell78}.

\begin{figure} \centering
  \FIG{0.7}{false}{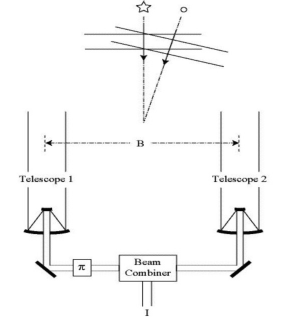}
  \caption{Principle of Bracewell's nulling interferometer.}
  \label{fig-interf-bracewell}
\end{figure}

Two collector apertures separated by baseline $B$ observe a target star-planet system separated by angle $\theta$. Introducing an achromatic $\pi$ phase shift between interferometer arms converts the zero-order central fringe from constructive to destructive interference. On-axis stellar light is canceled, while off-axis planet light located on a constructive fringe maximum is transmitted. Ideal baseline length satisfies:
\begin{equation}
  B=\frac{\lambda}{2\theta}.
\end{equation}

A two-aperture nuller projects a sinusoidal transmission map onto the sky rather than forming a direct image. Figure~\ref{fig-trans-nuller} illustrates a transmission map for a system separated by $0.1$~arcsec observed at $10$~\mum.

\begin{figure} \centering
  \FIG{0.5}{false}{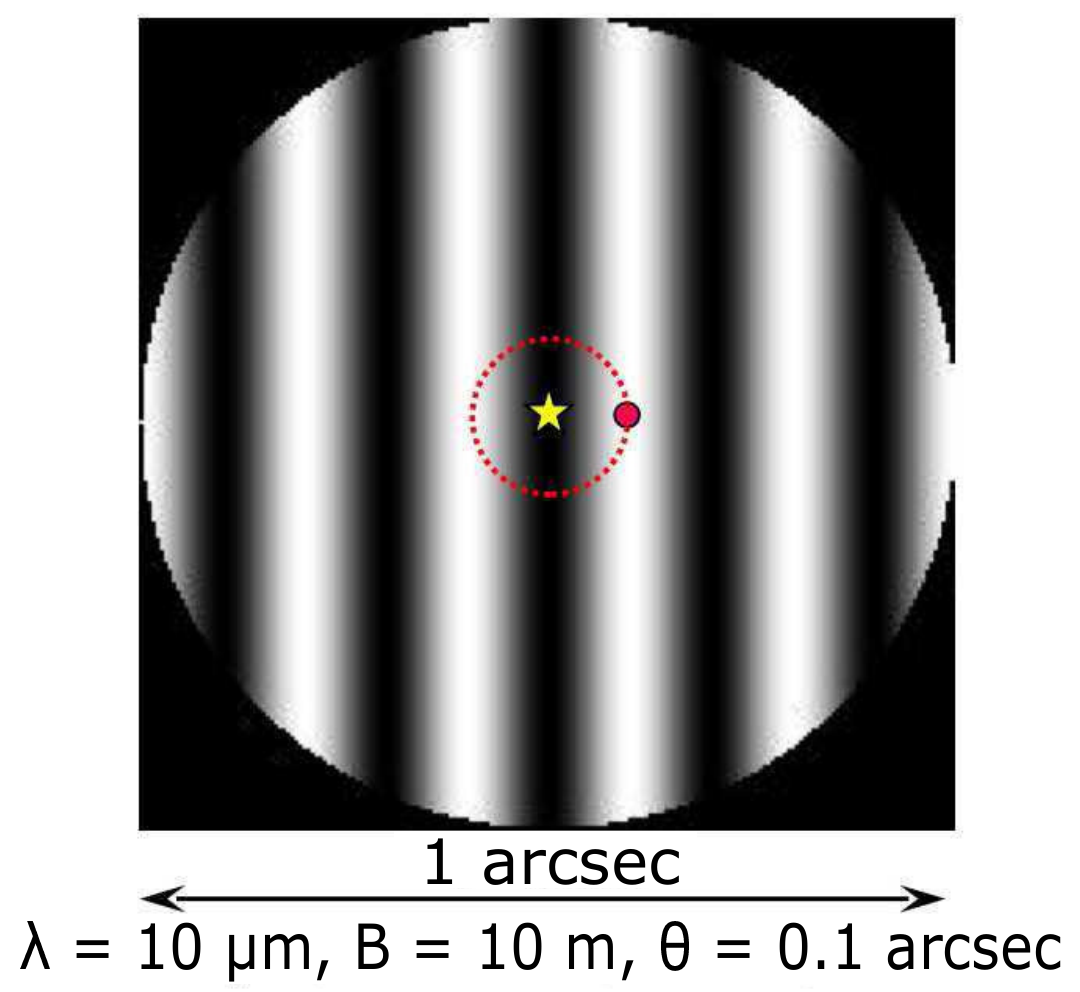}
  \caption[Transmission map of a Bracewell interferometer.]{Transmission map of a $B=10$~m Bracewell interferometer observing at $\lambda=10$~\mum for a $0.1$~arcsec binary separation.}
  \label{fig-trans-nuller}
\end{figure}

Rotating the array baseline modulates off-axis planetary signals between dark and bright fringes, distinguishing point-like planetary signals from continuous exozodiacal backgrounds and stationary residual stellar leakage.

%¤¤¤¤¤¤¤¤¤¤¤¤¤¤¤¤¤¤¤¤¤¤¤¤¤¤¤¤¤¤¤¤¤¤¤¤¤¤¤¤¤¤¤¤¤¤¤¤¤¤¤¤¤¤¤¤¤¤¤¤¤¤¤¤¤¤¤¤¤¤¤¤¤¤¤¤¤¤¤
\subsection{Null Depth and Limiting Factors}
\label{sec-taux-extinction}

Nulling performance is quantified by the null depth $N$ and rejection ratio $\rho$:
\begin{equation}
  N = \frac{I_\mmin}{I_\mmax} \qquad \text{and} \qquad \rho = \frac{1}{N} =
  \frac{I_\mmax}{I_\mmin},
  \label{eq-taux-extinction}
\end{equation}
where $I_\mmin$ is the residual transmitted flux at destructive interference and $I_\mmax$ is constructive peak flux.

Achieving ideal cancellation ($N=0$) is impossible because real stars subtend finite angular diameters $\theta_\star$. Unsuppressed stellar light passing through the transmission map constitutes geometric stellar leakage.

Instrumental aberrations further degrade performance. While higher-order spatial modes can be filtered using single-mode optical fibers, low-order modes—namely differential path delay (piston) and tip-tilt tracking errors—modulate fiber injection efficiencies and drive instrumental leakage.

E. Serabyn (2000) formulated a independent error budget for null depth $N$ \cite{Serabyn00}:
\begin{equation}
  N = N_\star+N_\phi+N_\varepsilon+N_\mpol,
  \label{eq-systeme-nulling}
\end{equation}
where:
\begin{itemize}
\item $N_\star$ is stellar geometric leakage due to finite angular size;
\item $N_\phi$ represents phase errors across the interferometer arms;
\item $N_\varepsilon$ represents photometric power imbalances;
\item $N_\mpol$ represents polarization mismatches.
\end{itemize}

%-------------------------------------------------------------------------------
\subsubsection{Stellar Leakage}
\label{sec-residu-stellaire}

Stellar leakage depends on stellar angular diameter $\theta_\star$, baseline $B$, and source spectrum $S(\sigma)$ \cite{Serabyn00}:
\begin{equation}
  N_\star = \frac{\pi^2}{16}(B\theta_\star)^2 \int{S(\sigma)\sigma^2\dd\sigma}
  \simeq \frac{\pi^2}{16}(\sigma_\mmax B\theta_\star)^2.
\label{eq-N-star}
\end{equation}

Quadratic scaling ($\theta_\star^2$) arises from parabolic approximations near the fringe minimum in two-telescope systems. Multi-aperture configurations (e.g., Angel 4-telescope cross array) generate 4th-order ($\theta_\star^4$) central transmission profiles, substantially suppressing stellar leakage \cite{Mennesson97,Absil01}.

%-------------------------------------------------------------------------------
\subsubsection{Phase Perturbations}
\label{sec-perturbations-phase}

Phase errors originate from two mechanisms:
\begin{itemize}
\item OPD fluctuations driving operational state away from zero-path destructive interference;
\item Chromatic dispersion causing deviations from exact $\pi$ phase shifts across broadband channels.
\end{itemize}

Phase leakage scales with total phase difference $\delta\phi_\md(\sigma,t)$ and source spectrum $S(\sigma)$ \cite{Serabyn00}:
\begin{equation}
  N_\phi(t) = \frac14\int{S(\sigma)\Delta\phi_\md(\sigma,t)^2\dd\sigma}.
\label{eq-N-phi}
\end{equation}

Assuming zero mean phase bias, phase perturbations separate into mean wavelength path delay $\Delta\phi_\mmoy(t)$ and the phase dispersion over the wavelength range $\Delta\phi_\sigma(t)$:
\begin{equation}
  \Delta\phi_\md(\sigma,t) = \Delta\phi_\mmoy(t)+\Delta\phi_\sigma(t).
\end{equation}

$N_\phi$ can be decomposed into two independent terms, $N_\delta$ and $N_\sigma$. The optical path delay contribution $N_\delta$ is:
\begin{equation}
  N_\delta = \frac14\Delta\phi_\mmoy(t)^2 = (\pi\sigma_\mmoy\delta)^2,
  \label{eq-N-delta}
\end{equation}
where $\sigma_\mmoy$ is mean wavenumber.

The chromatic phase dispersion term $N_\sigma$ is:
\begin{equation}
  N_\sigma = \frac14\Delta\phi_\sigma(t)^2.
  \label{eq-N-sig}
\end{equation}

%-------------------------------------------------------------------------------
\subsubsection{Photometric Imbalance}
\label{sec-deseq-phot}

Intensity equality between arms is required to achieve deep destructive interference. Defining relative intensity imbalance $\varepsilon$ between arm fluxes $I_1$ and $I_2$ as:
\begin{equation}
  \varepsilon = \frac{I_1-I_2}{I_1+I_2},
  \label{eq-varepsilon}
\end{equation}
the photometric null depth contribution is \cite{Serabyn00}:
\begin{equation}
  N_\varepsilon = \frac14\varepsilon^2.
  \label{eq-N-eps}
\end{equation}

This term similarly decomposes into mean photometric imbalance $\varepsilon_\mmoy(t)$ and spectral power dispersion $\varepsilon_\sigma(t)$.

%-------------------------------------------------------------------------------
\subsubsection{Polarization Mismatches}
\label{sec-effets-polarisation}

Polarization errors comprise two effects:
\begin{itemize}
\item Differential phase shifts $\Delta\phi_\msp$ between transverse electric ($\mss$) and transverse magnetic ($\mpp$) polarization states:
  \begin{equation}
	N_\msp = \frac{1}{16}\Delta\phi_\msp^2;
    \label{eq-N-sp}
  \end{equation}
\item Polarization field orientation misalignment angle $\alpha_\mrot$ between arms:
  \begin{equation}
    N_\mrot = \frac14\alpha_\mrot^2.
    \label{eq-N-rot}
  \end{equation}
\end{itemize}

\bigskip

All terms must be strictly controlled to achieve ultra-deep nulling.

%¤¤¤¤¤¤¤¤¤¤¤¤¤¤¤¤¤¤¤¤¤¤¤¤¤¤¤¤¤¤¤¤¤¤¤¤¤¤¤¤¤¤¤¤¤¤¤¤¤¤¤¤¤¤¤¤¤¤¤¤¤¤¤¤¤¤¤¤¤¤¤¤¤¤¤¤¤¤¤
\subsection{Exoplanet Nulling Projects}
\label{sec-projets-detection}

%-------------------------------------------------------------------------------
\subsubsection{Ground-Based Nulling Projects}
\label{sec-projets-interferometres}

Ground-based nulling instruments were developed to test operational concepts.

At the Keck Observatory, KIN\footnote{Keck Interferometer Nuller} (installed 2004) combined the twin 10-meter telescopes, achieving $N \approx 10^{-3}$ in N-band ($10$~\mum) \cite{Serabyn06}. It surveyed debris disks, bright exozodiacal clouds, and young Pegasids. Keck interferometric modes were decommissioned in mid-2012 due to funding constraints.

At the LBTI, NIC\footnote{Nulling Infrared Camera} performs dual-band nulling (8–13~\mum via NOMIC\footnote{Nulling Optimized Mid-Infrared Camera} and 3–5~\mum imaging), targeting null depths of $10^{-4}$.

ESA's proposed GENIE\footnote{Ground-based European Nulling Interferometer Experiment} for the VLTI was canceled due to complex optical architecture and high atmospheric OPD jitter ($\sim 200$~nm RMS).

ALADDIN\footnote{Antarctic L-band Astrophysics Discovery Demonstrator for Interferometric Nulling} was proposed by V. Coudé du Foresto \cite{Coude06} for Dome C, Antarctica. Replicating GENIE's science goals outside the VLTI infrastructure, ALADDIN consisted of two 1-meter telescopes mounted on a rotating circular track raised 30~m above ground atmospheric boundary layer turbulence (Figure~\ref{fig-aladdin}).

\begin{figure} \centering
  \FIG{0.9}{false}{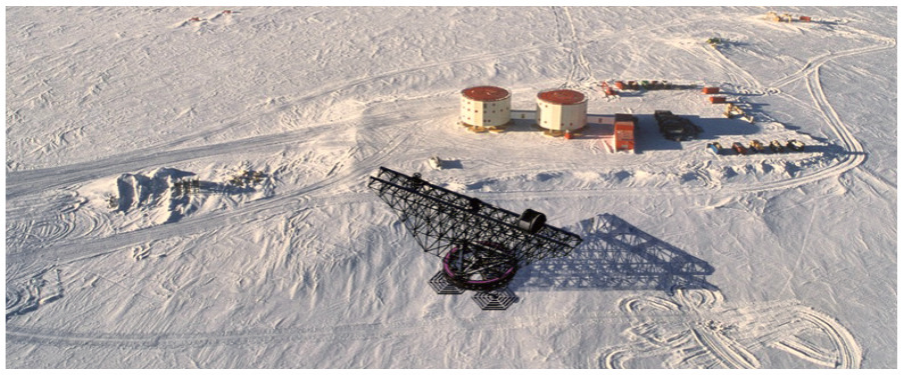}
  \caption{ALADDIN Antarctic nulling interferometer project.}
  \label{fig-aladdin}
\end{figure}

Antarctic environmental treaties requiring complete structure removal post-decommissioning stalled project approval.

%-------------------------------------------------------------------------------
\subsubsection{Darwin/TPF-I: Ambitious Missions for Earth-like Planet Characterization}
\label{sec-darwintpf-i}

Flagship space concepts—ESA's Darwin \cite{Leger96,Leger07} and NASA's TPF-I\footnote{Terrestrial Planet Finder Interferometer} \cite{Beichman99,Lawson00}—were designed to search for habitable Earth-like planets and analyze atmospheric biosignatures (H$_2$O, CO$_2$, O$_3$). Operating in the thermal infrared (6–20~\mum), planet-to-star contrast relaxes to $\sim 10^{-5}$ compared to $\sim 10^{-10}$ in the visible reflected light regime. While visible coronagraphy requires extreme optical precision, thermal nulling interferometers like Darwin/TPF-I trade contrast constraints ($N \le 10^{-5}$) against formation-flying baseline requirements.

NASA/ESA joint study groups evaluated multi-spacecraft array architectures combining 4 to 6 telescopes (Figure~\ref{fig-Darwin-tpfi}).

\begin{figure} \centering
  \subfloat[Alternative array configuration.]{\label{fig-tpfi}
    \FIGH{4.9}{false}{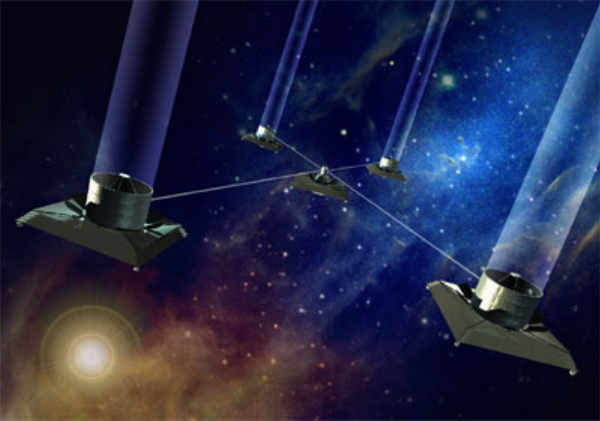}}\hfill
  \subfloat[Final baseline configuration.]{\label{fig-Darwin}
    \FIGH{4.9}{false}{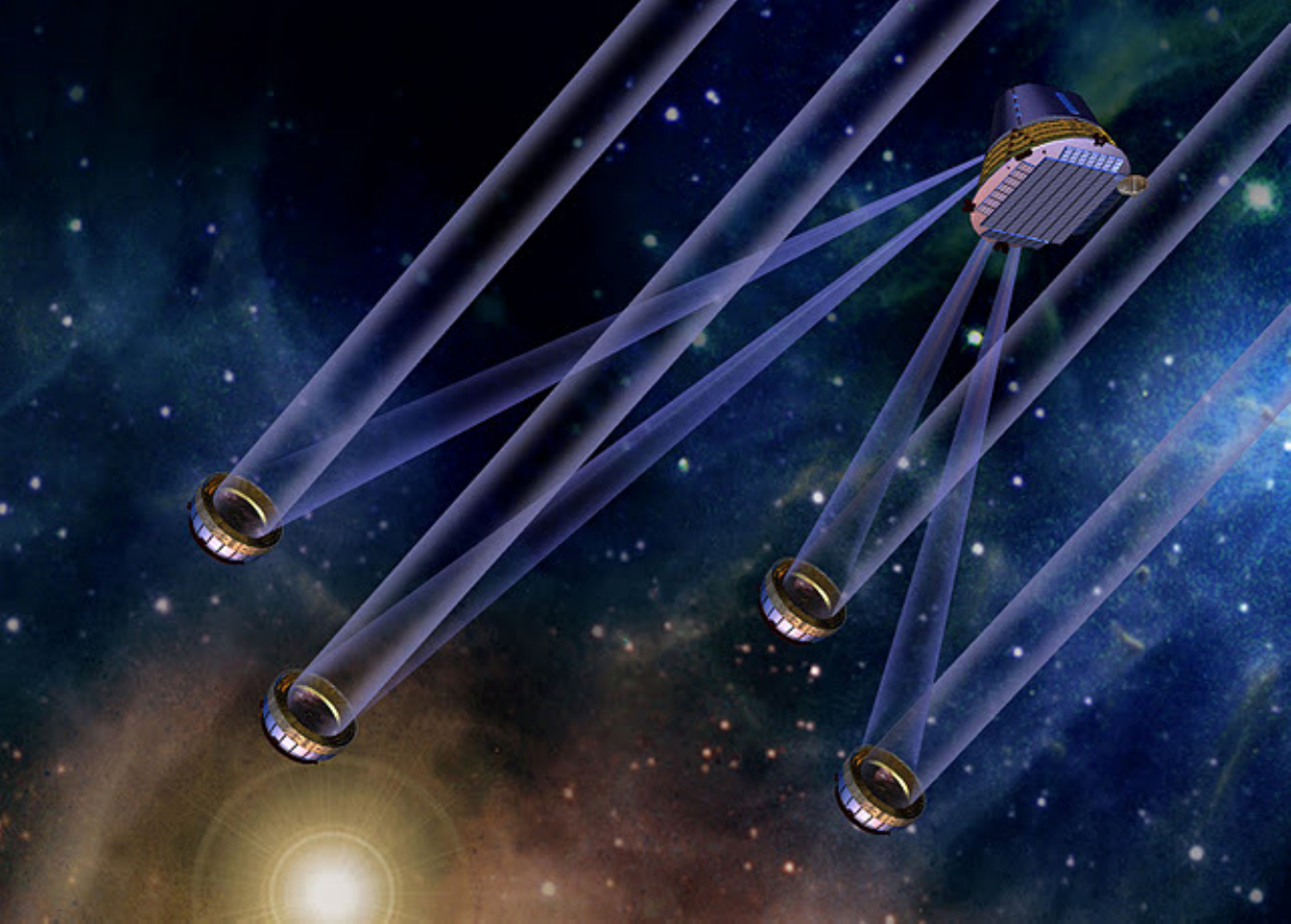}}
  \caption{Two concept configurations for the Darwin/TPF-I mission.}
  \label{fig-Darwin-tpfi}
\end{figure}

The consensus baseline converged on a 5-spacecraft array (Figure~\ref{fig-Darwin}) \cite{Mennesson97,Mennesson05}: four 2-meter off-axis collector free-flyers recombining signals at a central hub located at the Sun-Earth L2 Lagrangian point. Baselines scaled between 10 and 500~m to achieve targeted null depths of $10^{-5}$ stabilized to $10^{-9}$. Due to unproven formation-flying technology and high costs, these flagship missions were postponed.

%-------------------------------------------------------------------------------
\subsubsection{\peg: A Pathfinder for Pegasid Observation}
\label{sec-pegase-precurseur}

\peg was proposed to CNES (2004) and ESA (2007) \cite{Leduigou06a,Leduigou06b,Ollivier07} as a pathfinder mission to observe hot Jupiters, brown dwarfs, and protoplanetary disks using formation flying (Figure~\ref{fig-pegase}).

\begin{figure} \centering
  \FIG{0.7}{false}{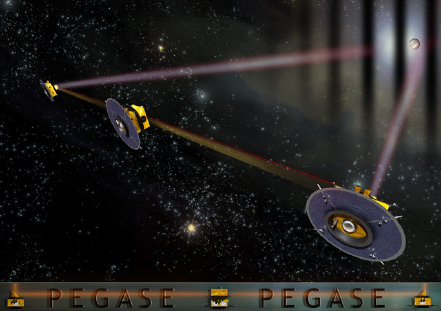}
  \caption{Artist's view of the \peg mission concept proposed to CNES and ESA.}
  \label{fig-pegase}
\end{figure}

Operating between 2.5 and 5~\mum, \peg aimed for a $10^{-4}$ null depth stabilized to $10^{-5}$ across 10–500~m baselines. Two 40~cm siderostat free-flyers directed light to a central beam combination hub. Achieving these specs required OPD stability below 2~nm RMS. Despite strong CNES support, ESA rejected the mission due to technology readiness gaps and cost considerations.

%-------------------------------------------------------------------------------
\subsubsection{Other Space Concepts}
\label{sec-autres-projets}

NASA/Goddard studied FKSI\footnote{Fourier-Kelvin Stellar Interferometer} \cite{Danchi06,Danchi08}, a non-formation-flying alternative featuring a rigid 12–20~m boom carrying two 50~cm siderostats cooled to 60~K, observing in the 3–8~\mum band (Figure~\ref{fig-fksi}).

\begin{figure} \centering
  \FIG{0.7}{false}{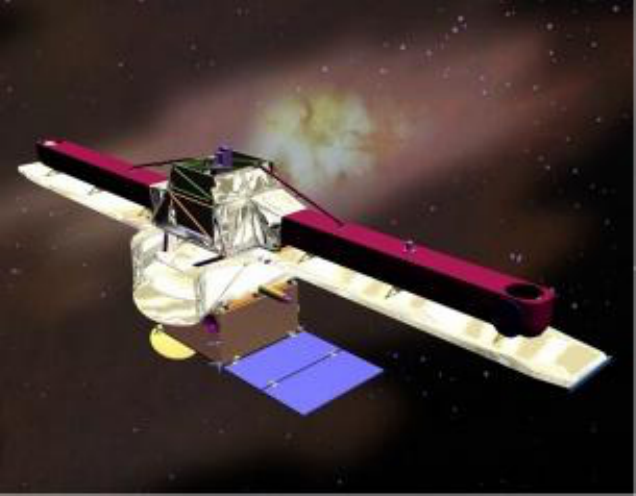}
  \caption{NASA/Goddard FKSI mission concept.}
  \label{fig-fksi}
\end{figure}

JPL proposed DAViNCI\footnote{Dilute Aperture Visible-Nulling-Coronagraph Imager} in 2009—a hybrid visible focal-plane nuller combining four 1.1~m mirrors on a non-formation-flying 2~m square frame ($N = 10^{-6}$ in visible/NIR bands; Figure~\ref{fig-davinci}).

\begin{figure} \centering
  \FIG{0.4}{false}{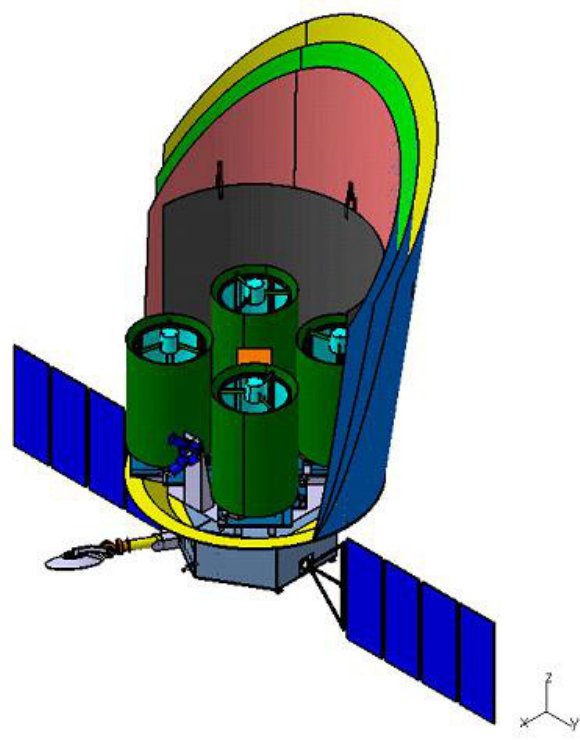}
  \caption{DAViNCI hybrid focal-plane nulling concept.}
  \label{fig-davinci}
\end{figure}

%¤¤¤¤¤¤¤¤¤¤¤¤¤¤¤¤¤¤¤¤¤¤¤¤¤¤¤¤¤¤¤¤¤¤¤¤¤¤¤¤¤¤¤¤¤¤¤¤¤¤¤¤¤¤¤¤¤¤¤¤¤¤¤¤¤¤¤¤¤¤¤¤¤¤¤¤¤¤¤
\subsection{State of the Art in Nulling Testbeds}
\label{sec--etat}

Testbeds were developed globally to demonstrate nulling performance. Figure~\ref{fig-etat-art} summarizes achieved null depths versus fractional bandwidth ($\Delta\lambda/\lambda$).

\begin{figure} \centering
  \FIG{0.774}{false}{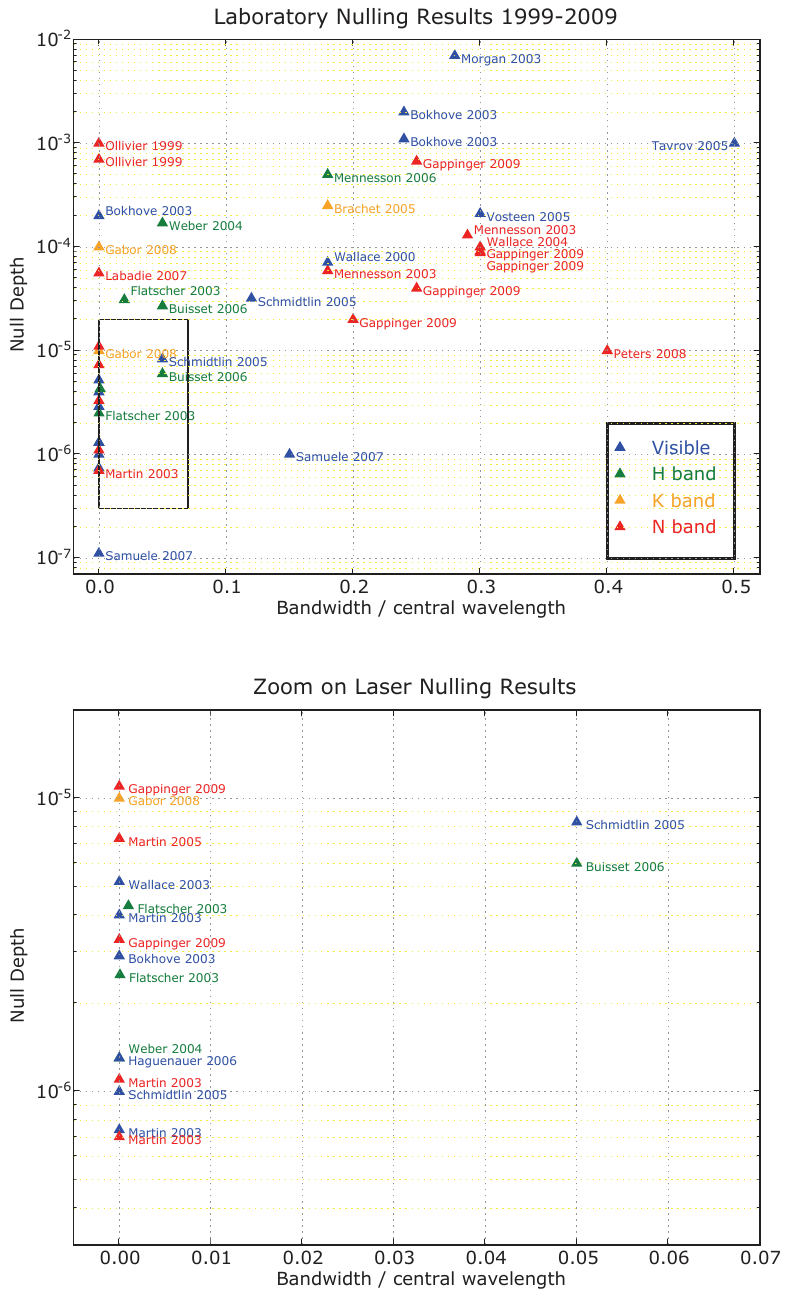}
  \caption[State of the art in nulling testbeds.]{State of the art in experimental null depth results across testbeds (1999–2009) \cre{P. Gabor}.}
  \label{fig-etat-art}
\end{figure}

Monochromatic tests achieved null depths below $10^{-5}$ up to $10^{-7}$ \cite{Samuele07}. However, broad bandwidths introduce chromatic dispersion errors that hinder performance. Only a few testbeds demonstrated $10^{-4}$ nulls over bandwidths exceeding 30\% \cite{Gappinger09,Peters08}.

JPL's Adaptive Nuller testbed \cite{Peters08} achieved a stable $10^{-5}$ null over a 34\% bandwidth centered at 10~\mum using a deformable mirror to correct spectral phase and intensity dispersion (Figure~\ref{fig-adapt-nuller}).

\begin{figure} \centering
  \subfloat[Adaptive nuller concept.]{\label{fig-adapt-nuller1}
    \FIGH{4.1}{false}{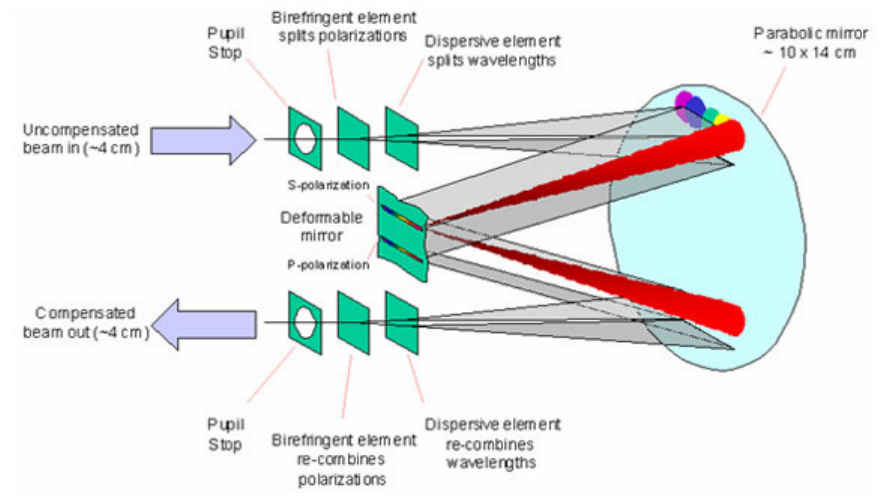}}\hfill
  \subfloat[Experimental testbed.]{\label{fig-adapt-nuller2}
    \FIGH{4.1}{false}{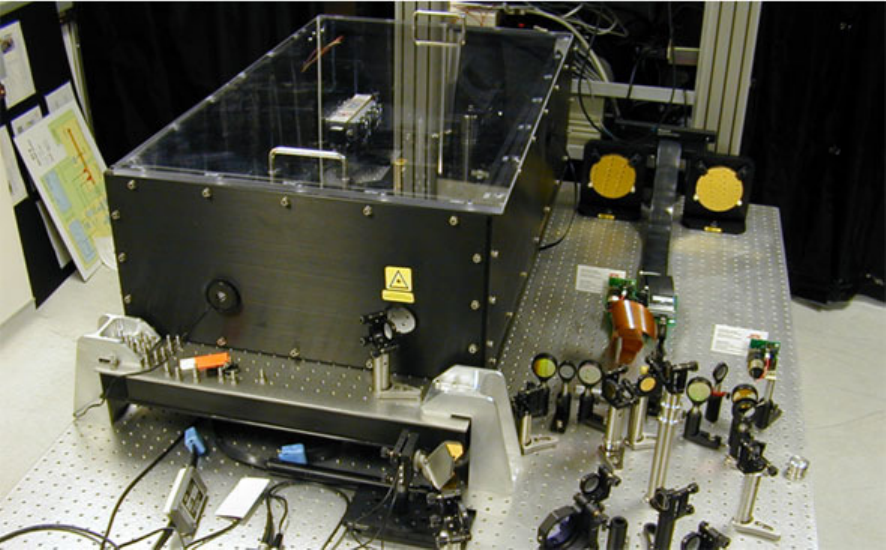}}
  \caption[The Adaptive Nuller testbed.]{Principle and experimental layout of JPL's Mid-Infrared Adaptive Nuller testbed.}
  \label{fig-adapt-nuller}
\end{figure}

%§§§§§§§§§§§§§§§§§§§§§§§§§§§§§§§§§§§§§§§§§§§§§§§§§§§§§§§§§§§§§§§§§§§§§§§§§§§§§§§
\section{Conclusion}
\label{sec-conclusion-1}

Direct exoplanet observation remains a formidably difficult task due to small angular separations, extreme flux contrasts, and potential exozodiacal background interference. Building on the legacy of historical astronomers like Cassini, modern optical instruments combine extreme adaptive optics and coronagraphy to reach diffraction-limited performance. Long-baseline interferometry and nulling techniques unlock even higher spatial resolutions and stellar suppression factors. Demonstrating these technologies under realistic disturbance environments on dedicated laboratory testbeds represents a critical stepping stone toward future space missions. This constitutes the main objective of my PhD work on the \pe testbed, presented in the following chapter.

%§§§§§§§§§§§§§§§§§§§§§§§§§§§§§§§§§§§§§§§§§§§§§§§§§§§§§§§§§§§§§§§§§§§§§§§§§§§§§§§
\chapter{\pe, Model of a Space-Based Nulling Interferometer}
\label{chap-persee}

\begin{flushright}
 \begin{minipage}{12cm}
   {\small \textit{One can judge from experiment, or one can blindly accept
       authority. To the scientific mind, experimental proof is all
       important and theory is merely a convenience in description, to be
       junked when it no longer fits.}}

   \raggedleft{{\small Robert A. Heinlein, Life-Line (1939)}}
 \end{minipage}
\end{flushright}

\minitoc

\bigskip

%§§§§§§§§§§§§§§§§§§§§§§§§§§§§§§§§§§§§§§§§§§§§§§§§§§§§§§§§§§§§§§§§§§§§§§§§§§§§§§§
\section{The \pe Testbed}
\label{sec-banc-persee}

%¤¤¤¤¤¤¤¤¤¤¤¤¤¤¤¤¤¤¤¤¤¤¤¤¤¤¤¤¤¤¤¤¤¤¤¤¤¤¤¤¤¤¤¤¤¤¤¤¤¤¤¤¤¤¤¤¤¤¤¤¤¤¤¤¤¤¤¤¤¤¤¤¤¤¤¤¤¤¤
\subsection{Project Context}
\label{sec-objectifs}

As seen in Section~\ref{sec--etat}, extremely deep null depths have been achieved on various testbeds worldwide. However, these testbeds focus primarily on cophasing and do not incorporate the typical disturbances encountered by a space interferometer. Indeed, whether for a project like FKSI or \peg, the optical path difference (OPD) is perturbed by structural flexures of the beam for the former, by the drift of formation-flying satellites for the latter, by thermo-elastic deformations, and by mechanical vibrations generated by internal components such as reaction wheels or cryogenic pumps.

It was precisely with the goal of comprehensively simulating a space mission like \peg that the \pe (\peg Experiment for Research and Stabilization of Extreme Extinction) testbed was conceived in 2006. Funded by CNES Research and Development and the Île-de-France region, it was developed by a consortium comprising CNES, the Institut d'Astrophysique Spatiale (IAS), LESIA at the Paris Observatory–Meudon, the DOTA/HRA\footnote{Département Optique Théorique et Appliquée / High Angular Resolution unit} unit of ONERA \tir{the Observatory and ONERA both being members of the GIS PHASE\footnote{Groupement d'Intérêt Scientifique Partenariat Haute-résolution Angulaire Sol-Espace}}, the FIZEAU laboratory at the Observatoire de la Côte d'Azur (OCA), and Thales Alenia Space (TAS).

\pe aims to combine a broadband mid-infrared nulling interferometer with an OPD and tip/tilt cophasing system, as well as a simulator for Guidance, Navigation, and Control (GNC) of formation-flying satellites that introduces realistic flight-like disturbances.

Although its design is based on the \peg project, the \pe testbed can also be adapted to other two-telescope concepts such as FKSI (Section~\ref{sec-autres-projets}) or ALADDIN (Section~\ref{sec-projets-interferometres}), as their optical architectures are very similar. It therefore aims to validate the integrated system on the ground by analyzing the coupling between the payload and the GNC, potentially relaxing certain GNC requirements.

%¤¤¤¤¤¤¤¤¤¤¤¤¤¤¤¤¤¤¤¤¤¤¤¤¤¤¤¤¤¤¤¤¤¤¤¤¤¤¤¤¤¤¤¤¤¤¤¤¤¤¤¤¤¤¤¤¤¤¤¤¤¤¤¤¤¤¤¤¤¤¤¤¤¤¤¤¤¤¤
\subsection{Objectives of \pe}
\label{sec-spec-exig}

The \pe laboratory demonstrator must validate the expected performance of the nulling interferometer proposed for the \peg mission. Quantitatively, its goals are to:
\begin{itemize}
\item Achieve a mean null depth of $\e{-4}$ with a stability of $\e{-5}$ over several hours across various spectral bands between 1.65 and 3.3~\mum;
\item Validate fringe acquisition under tracking rates up to 250~\mum.s$^{-1}$;
\item Characterize the maximum noise and drift levels tolerable by the two feedback control loops governing tip/tilt and OPD;
\item Study the interactions between pointing loops, OPD loops, and flux stability;
\item Guarantee differential path stability between the scientific measurement channel and the OPD control channel;
\item Implement calibration procedures taking cophasing loop measurements into account to optimize nulling depth performance;
\item Validate testbed operations under realistic external disturbances;
\item Study polarization effects, particularly those induced by siderostat rotation.
\end{itemize}

As seen in Section~\ref{sec--etat}, a null depth of $\e{-4}$ has been demonstrated multiple times globally \cite{Samuele07,Buisset06,Schmidtlin06,Gappinger09,Peters08}, though often in polarized light and never over such a broad spectral bandwidth. The key originality of the \pe testbed lies in reaching this nulling level despite the simultaneous injection of realistic formation-flying disturbances.

Such a demonstrator is indispensable for proving the practical feasibility of space-based nulling interferometry missions such as \peg or Darwin/TPF-I.

%¤¤¤¤¤¤¤¤¤¤¤¤¤¤¤¤¤¤¤¤¤¤¤¤¤¤¤¤¤¤¤¤¤¤¤¤¤¤¤¤¤¤¤¤¤¤¤¤¤¤¤¤¤¤¤¤¤¤¤¤¤¤¤¤¤¤¤¤¤¤¤¤¤¤¤¤¤¤¤
\subsection{Breakdown of the Nulling Budget}
\label{sec-allocations}

The testbed's ability to stabilize the null depth must be demonstrated across two timescales: 100~s and 10~h. The 100~s duration corresponds to an acquisition phase during which the two collector spacecraft drift under solar radiation pressure under reaction wheel control alone. Indeed, for the \peg project, cold gas thrusters were selected for satellite positioning and reaction wheels for attitude control; thus, satellite relative positions are corrected by gas thruster firings only once every 100~s. The 10~h duration represents the total integration time needed to accumulate the weak photon flux from an exoplanet.

However, the stability of nulling measurements depends strongly on detector integration time; longer integration times average out non-drift fluctuations. On \pe, a typical integration timeframe of 1~s is adopted. Several metrics are evaluated:
\begin{itemize}
\item The mean null depth over $T=10$~h, denoted $\moy[\Td]{N}$;
\item Its short-term stability over $T=100$~s for an integration time $\tau=1$~s, denoted $\sigma_{N,\tauu}$;
\item Its long-term stability over $T=10$~h, corresponding to the stability of 100~s mean null depths $\moy[\Tc]{N}$, denoted $\sigma_{N,\tauc}$.
\end{itemize}

Specifications were assigned to these parameters based on \peg performance requirements:
\begin{equation}
  \left\{
    \begin{array}{r@{\hs}l}
      \moy[\Td]{N} &= \e{-4} \\
      \sigma_{N,\tauu} &= 1.5\E{-5} \\
      \sigma_{N,\tauc} &= \e{-5}
    \end{array}
  \right..
\end{equation}

Starting from the null depth formulation presented in Section~\ref{sec-taux-extinction}, tolerances were allocated to each contributor in Equation~\eqref{eq-systeme-nulling} and translated into subsystem-level requirements (OPD stability, flux balance, polarization). Table~\ref{tab-allocnulling} outlines this nulling error budget.

\begin{table} \centering
  \caption{Allocation of nulling contributions.}
  \medskip
  \begin{tabular}{cccc}
    \hline \hline
    Term & $\moy[\Td]{N}$ & $\sigma_{N,\tauu}$ & $\sigma_{N,\tauc}$ \\
    \hline
    Flux balance & $2\E{-5}$ & $3\E{-6}$ & $\e{-6}$ \\
    Optical path difference & $3.5\E{-5}$ & $1.5\E{-5}$ & $7\E{-6}$ \\
    Chromatism & $3.5\E{-5}$ & --- & $\e{-6}$ \\
    Polarization & $\e{-5}$ & --- & $\e{-6}$ \\
    \hline
    \textbf{TOTAL} & $\boldsymbol{\e{-4}}$ & $\boldsymbol{1.5\E{-5}}$ &
    $\boldsymbol{\e{-5}}$ \\
    \hline \hline
  \end{tabular}
  \label{tab-allocnulling}  
\end{table}

Chromatism and polarization variations within each spectral channel are assumed to be static over 100~s timescales. Summing average contributions linearly represents a worst-case scenario.

This places stringent demands on subsystem performance that remain compatible with state-of-the-art testbed demonstrations. For example, OPD stability is set to 2~nm RMS, guaranteeing a null contribution under $1.5\E{-5}$ at the shortest wavelength (1.65~\mum). This level of control was achieved by Gappinger on JPL's \emph{Achromatic Nulling Testbed} \cite{Gappinger09} and by Peters on the \emph{Adaptive Nuller} testbed \cite{Peters09}. Photometric imbalance is budgeted at 0.7\%, matching results obtained by Gabor on the SYNAPSE testbed at IAS \cite{Gabor08a}. These targets are achievable relative to state-of-the-art benchmarks; the central challenge lies in maintaining this performance while actively rejecting injected flight-like disturbances.

%¤¤¤¤¤¤¤¤¤¤¤¤¤¤¤¤¤¤¤¤¤¤¤¤¤¤¤¤¤¤¤¤¤¤¤¤¤¤¤¤¤¤¤¤¤¤¤¤¤¤¤¤¤¤¤¤¤¤¤¤¤¤¤¤¤¤¤¤¤¤¤¤¤¤¤¤¤¤¤
\subsection[Paper presented at SPIE Astron. Telescopes and Instrumentation]{Paper presented at SPIE Astronomical Telescopes and Instrumentation, Marseille (2008)}
\label{sec-description-banc}

This paper by F. Cassaing \cite{Cassaing08}, presented at the end of \pe's definition phase, describes the testbed architecture in detail. Subsequent sections expand on key subsystems. For a comprehensive treatment of the cophasing system and beam combiner, see \cite{Houairi08b} and \cite{Jacquinod08}, respectively.

Modifications made to this baseline layout during my thesis are detailed in Chapter~\ref{sec-integr-devel}.

\newpage

\newcommand{\SPIEFC}[1]{\centerline{\FIG{1.}{false,viewport=55 135 540
      770,clip}{SPIE_2008_FC_#1}}}

\SPIEFC{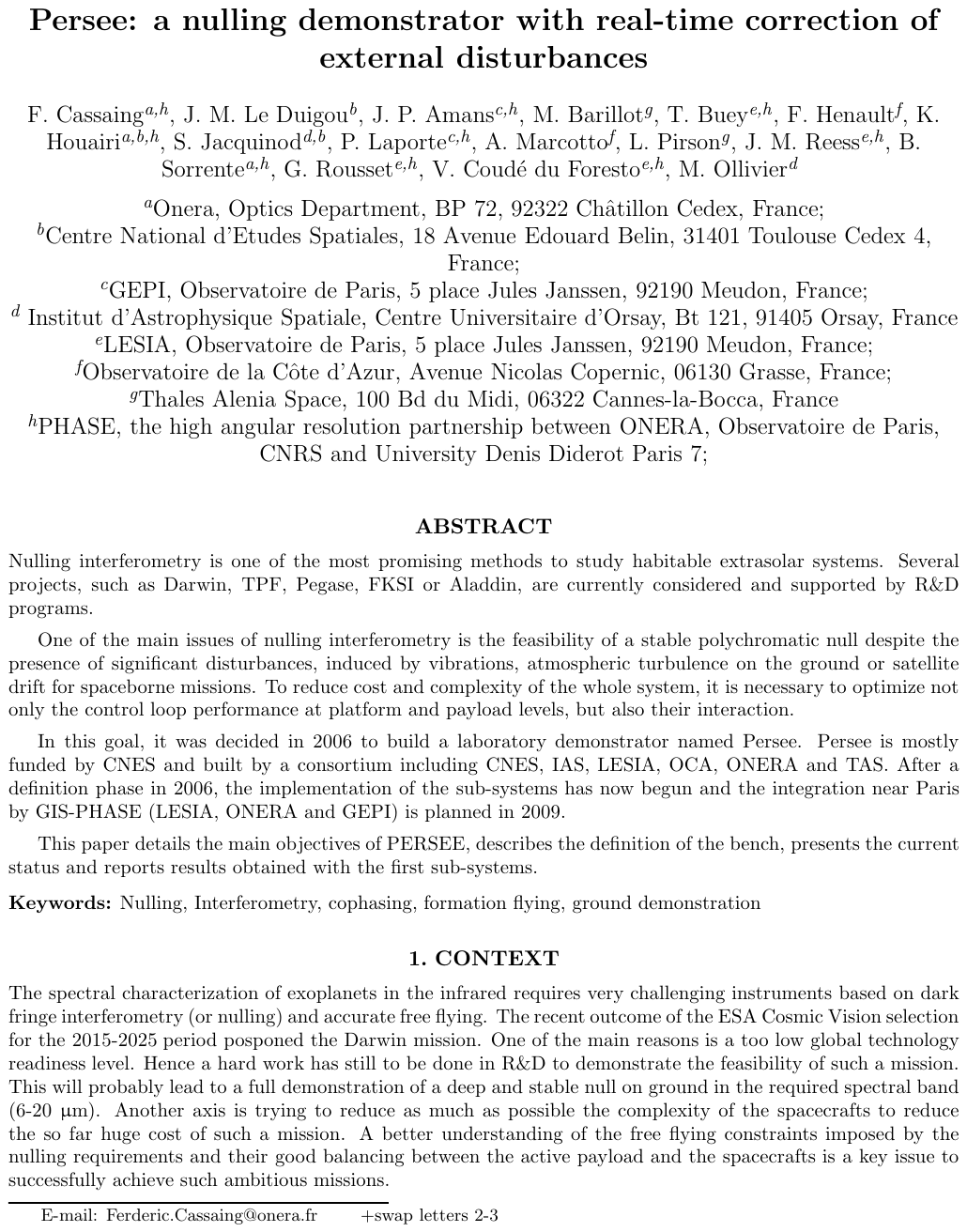}
\SPIEFC{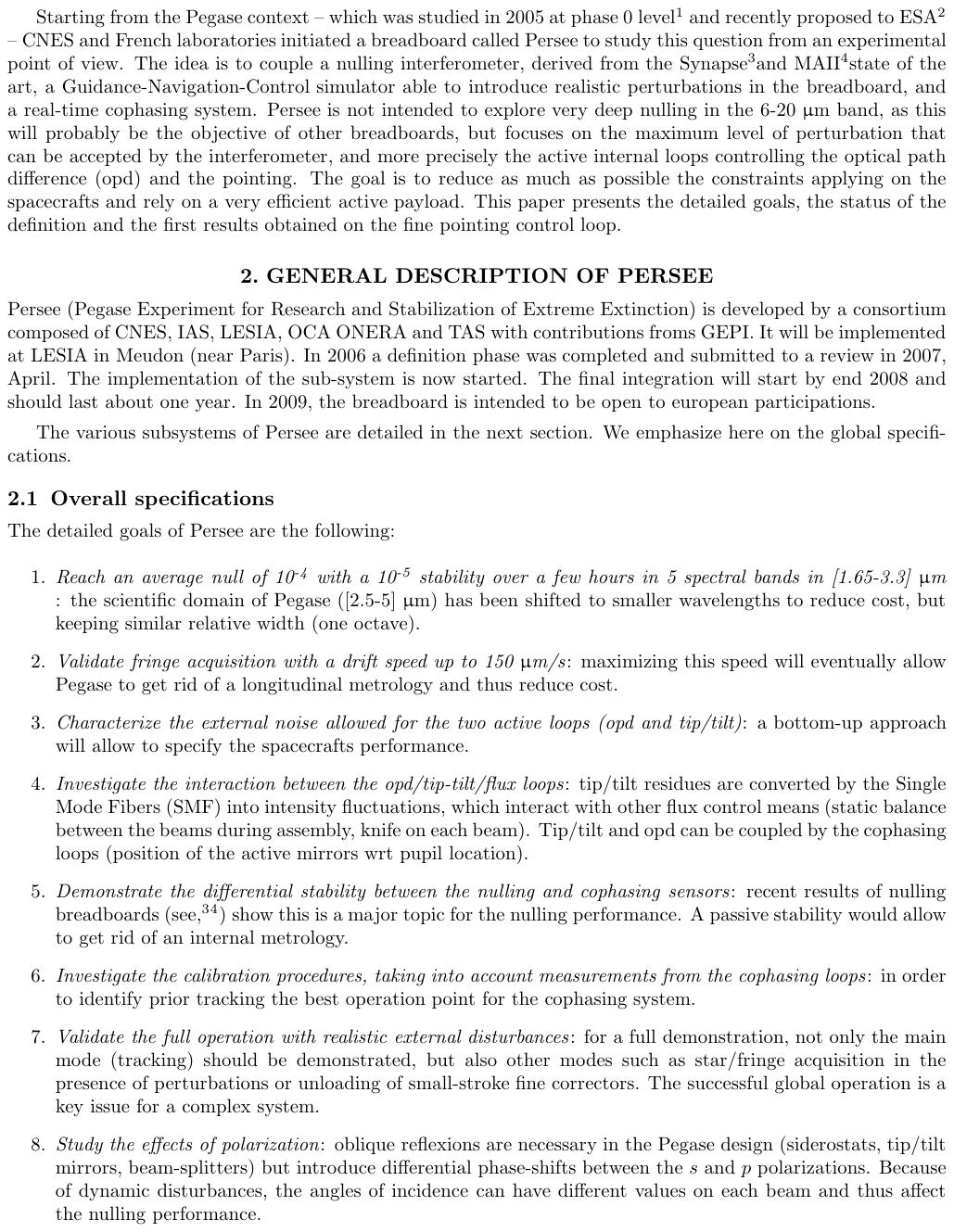}
\SPIEFC{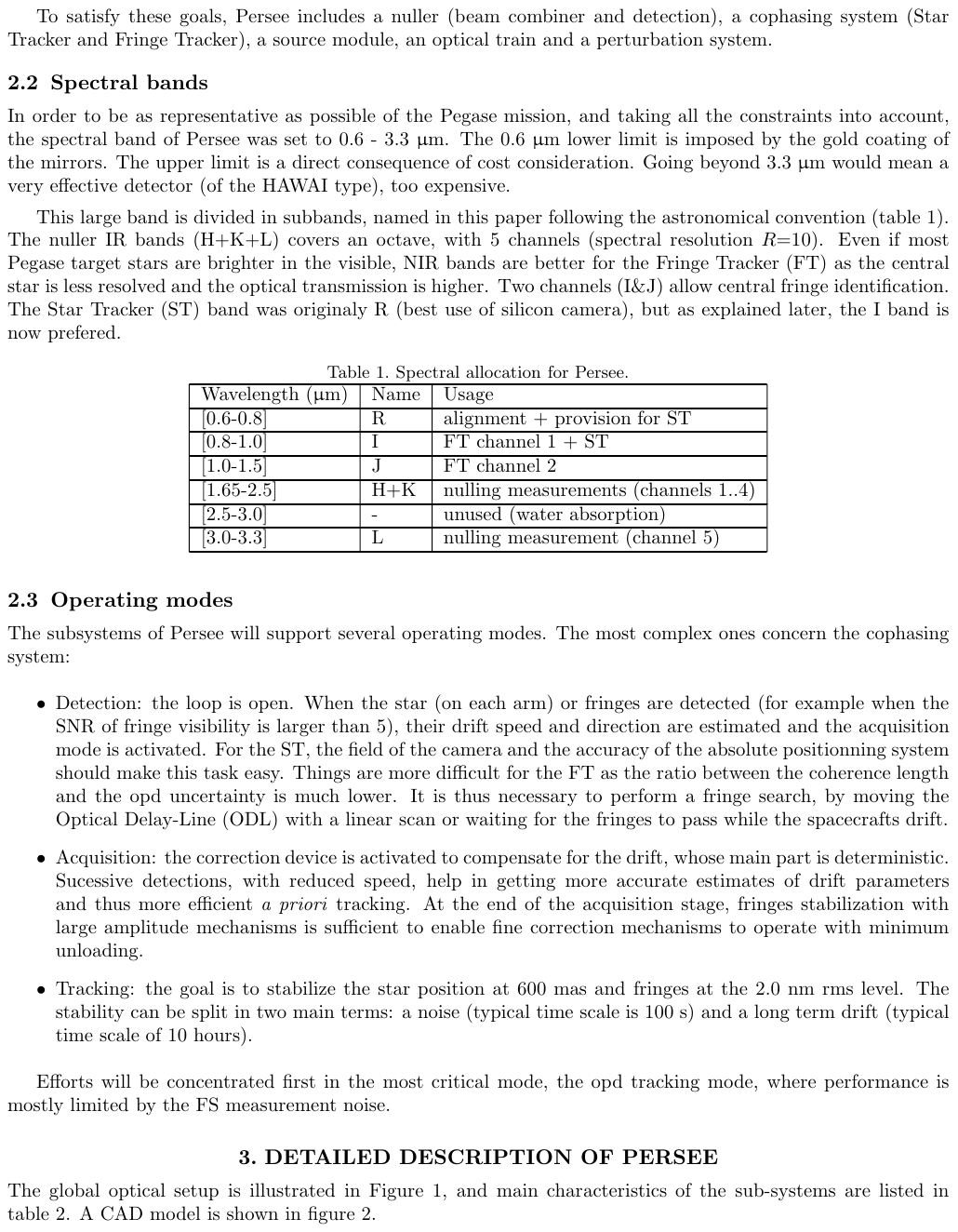}
\SPIEFC{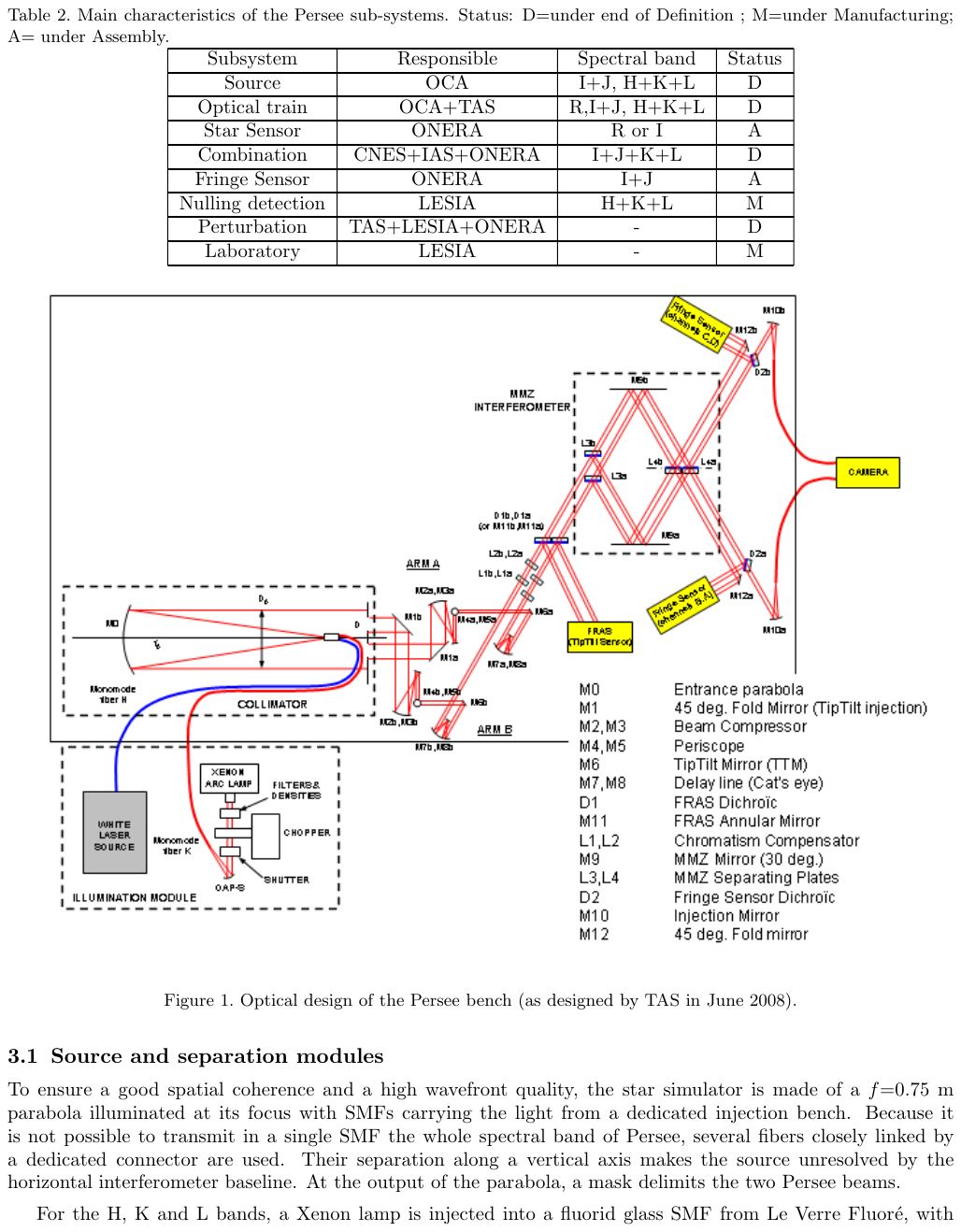}
\SPIEFC{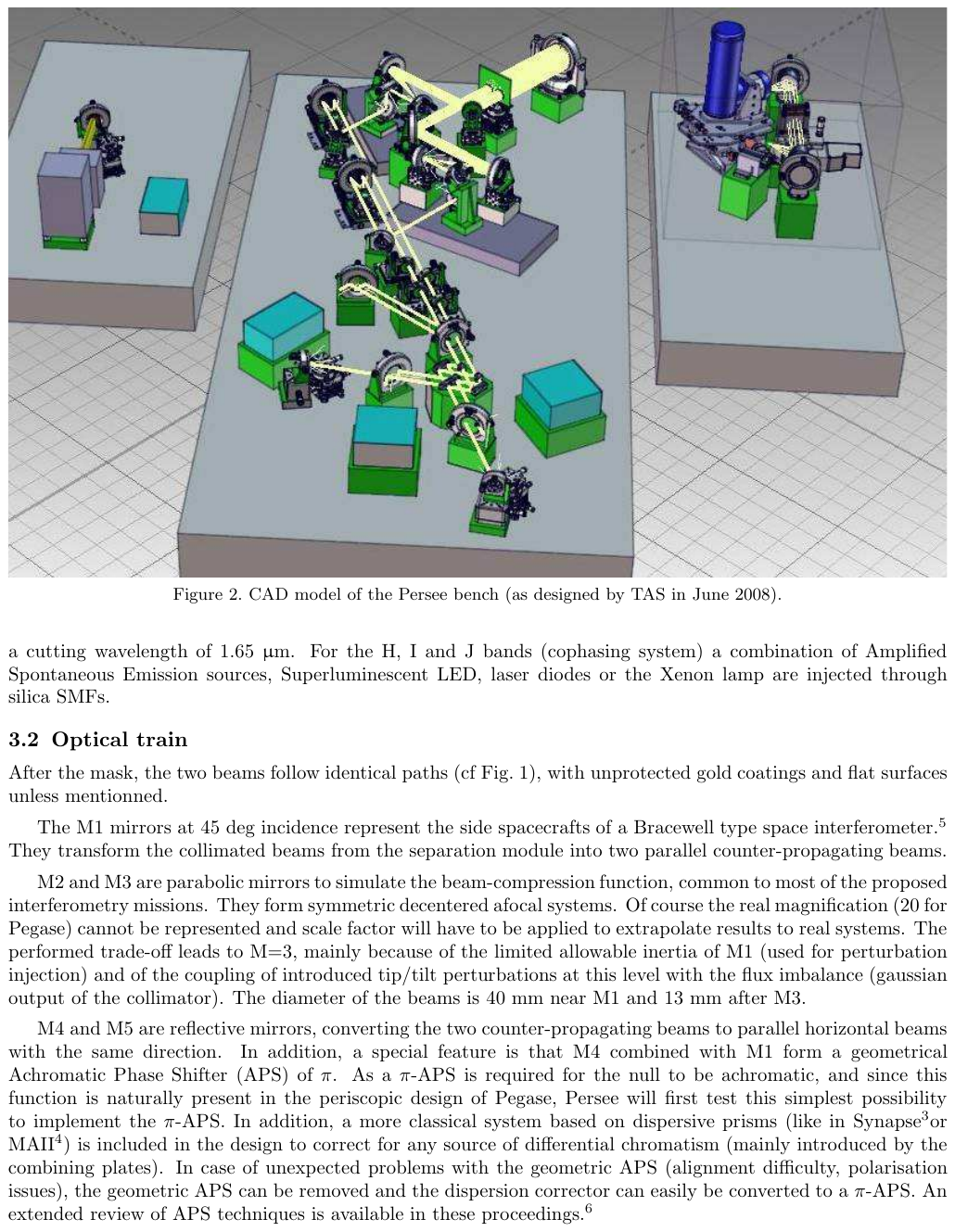}
\SPIEFC{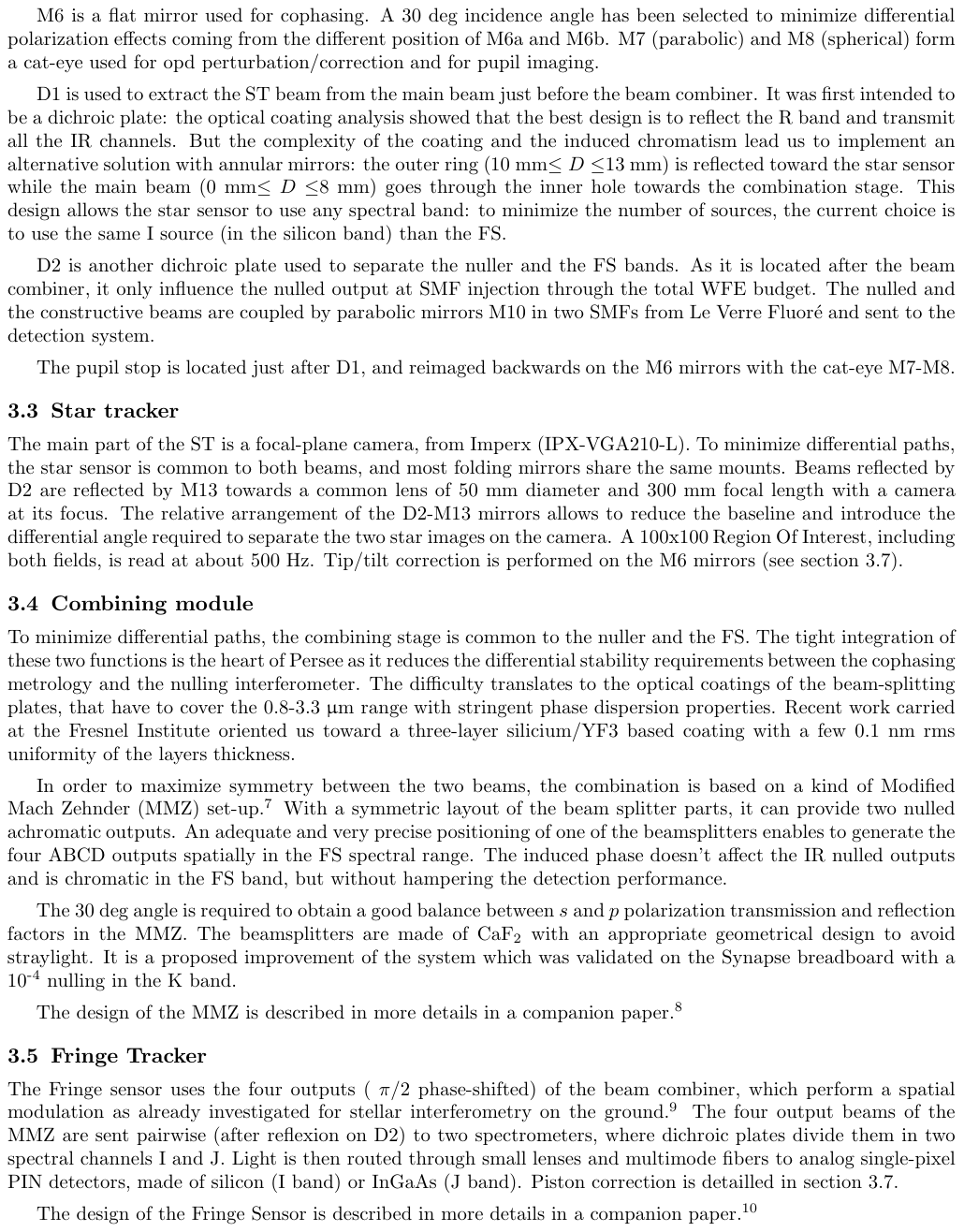}
\SPIEFC{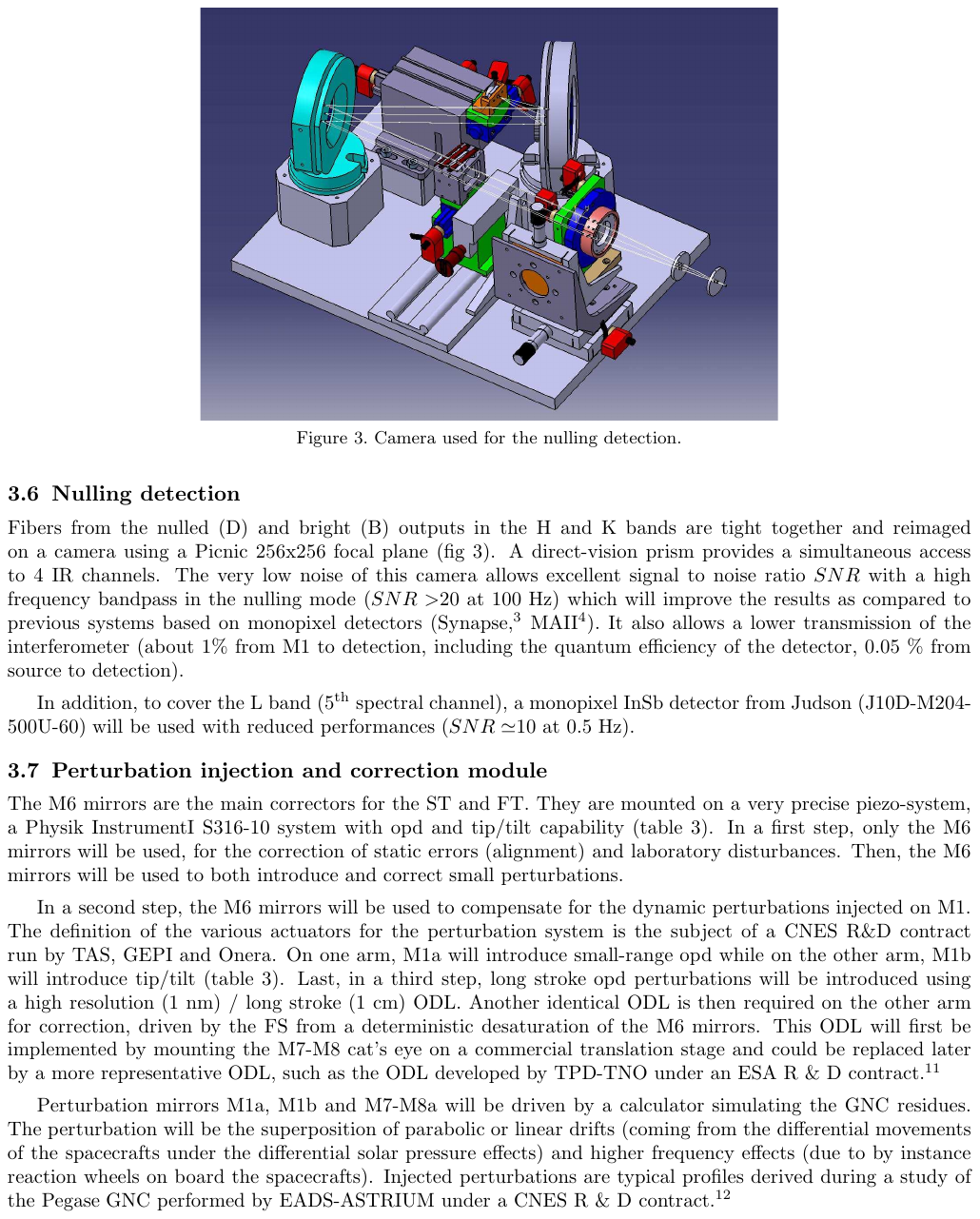}
\SPIEFC{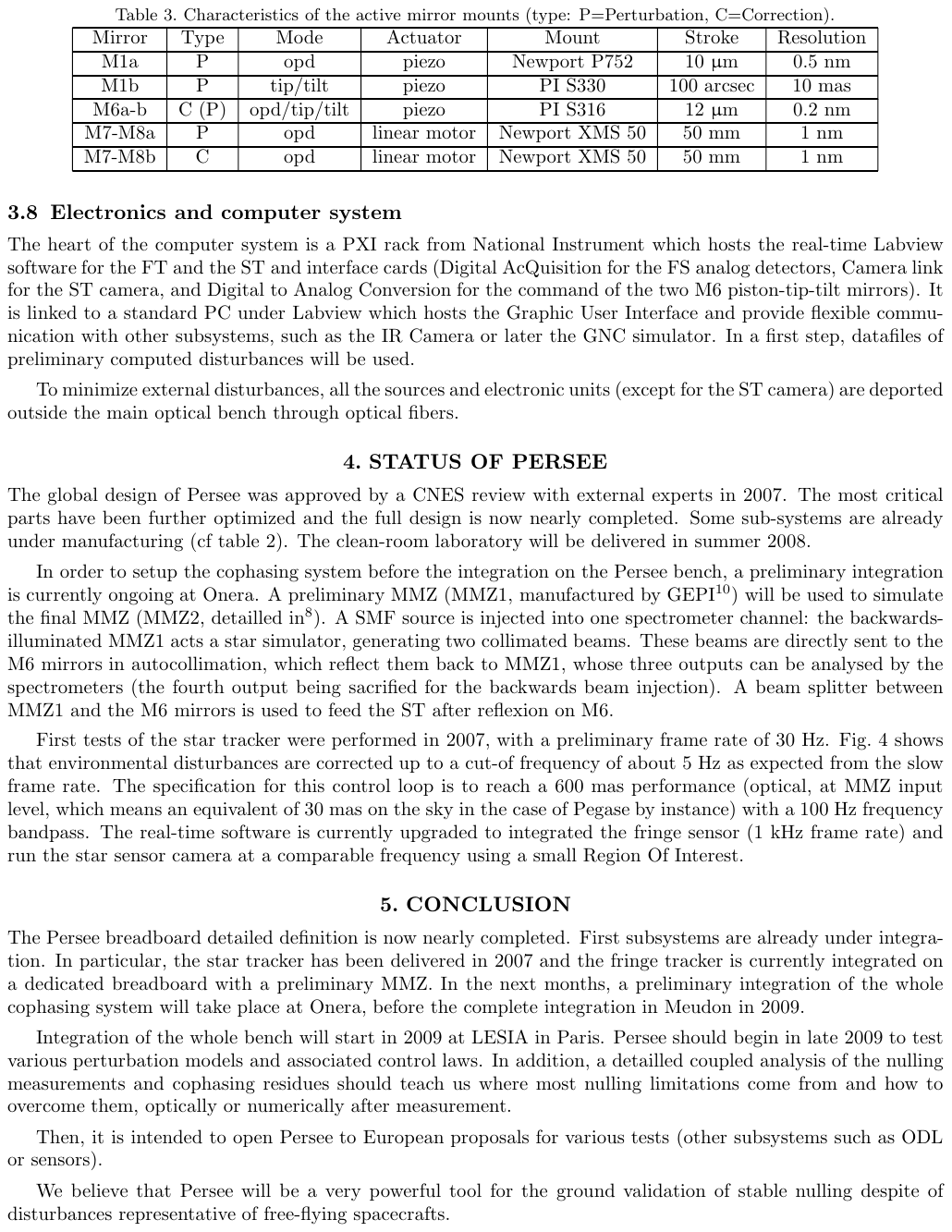}
\SPIEFC{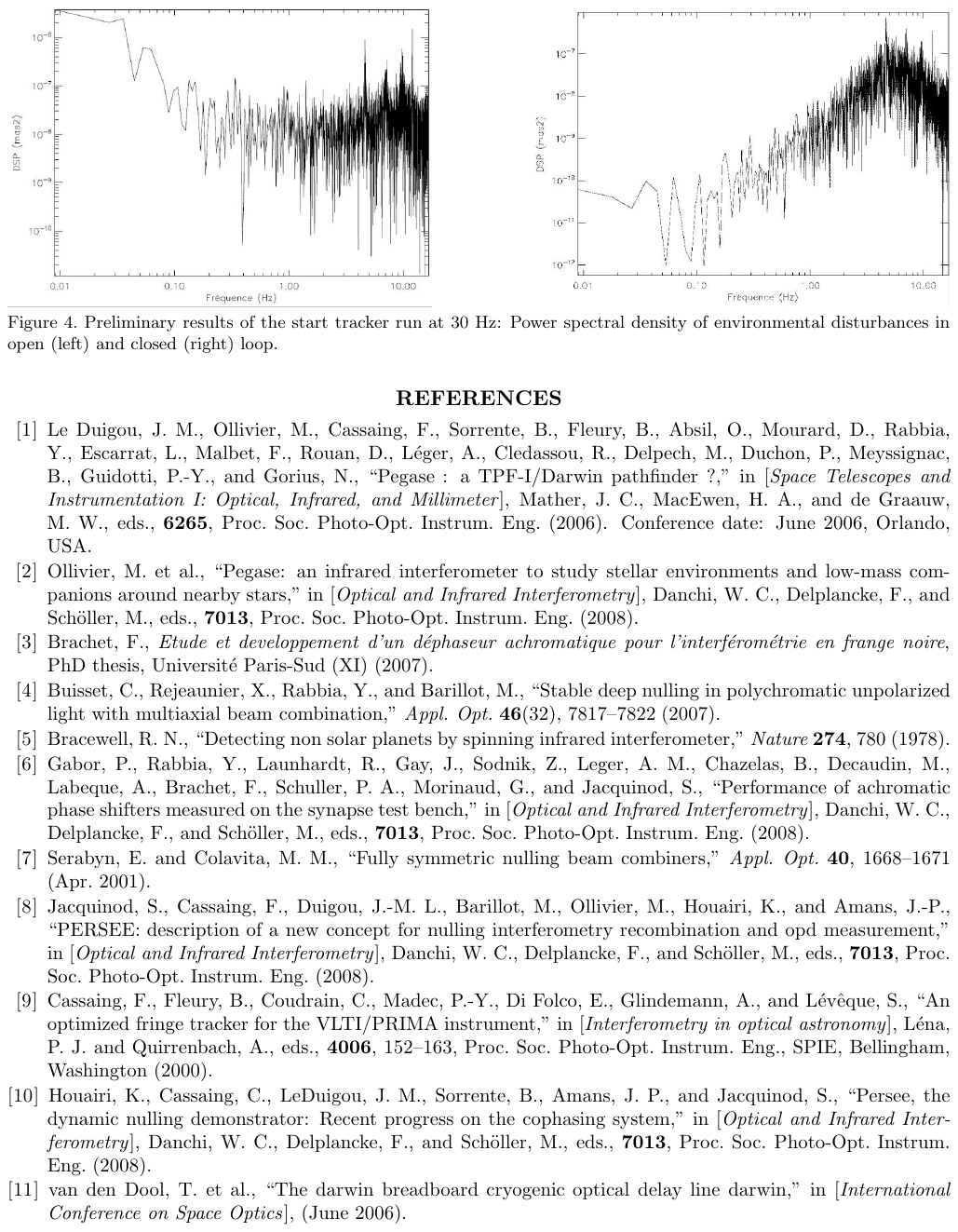}
\centerline{\FIG{1.}{false,viewport=55 745 540
      770,clip}{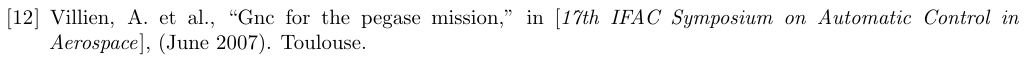}}

%§§§§§§§§§§§§§§§§§§§§§§§§§§§§§§§§§§§§§§§§§§§§§§§§§§§§§§§§§§§§§§§§§§§§§§§§§§§§§§§
\section{OPD Measurement for Beam Cophasing}
\label{sec-cophasage-telescopes}

Metrology is critical to nulling interferometry: beams must be cophased to nanometric precision. This requires measuring OPD continuously over a sufficient dynamic range. The cophasing methodologies described below formed the core of Kamel Houairi's PhD research on \pe \cite{Houairi09b}.

%¤¤¤¤¤¤¤¤¤¤¤¤¤¤¤¤¤¤¤¤¤¤¤¤¤¤¤¤¤¤¤¤¤¤¤¤¤¤¤¤¤¤¤¤¤¤¤¤¤¤¤¤¤¤¤¤¤¤¤¤¤¤¤¤¤¤¤¤¤¤¤¤¤¤¤¤¤¤¤
\subsection{Principle of ABCD Modulation}
\label{sec-desc-abcd}

An interferometer generates an interferogram that varies sinusoidally with OPD, with a period equal to the metrology wavelength $\lambda$. Measuring the phase of the interferogram yields the relative OPD between the beams; however, because phase measurements are $2\pi$-periodic, raw OPD estimates are $\lambda$-periodic.

Various techniques sample the interference pattern across at least two phase points to measure phase \cite{Cassaing01}. The standard technique in optical interferometry is Wyant's ABCD phase-stepping method \cite{Wyant75b,Shao88}, which samples 4 phase points in quadrature ($\pi/2$ phase separation), as illustrated in Figure~\ref{fig-abcd}.

\begin{figure} \centering
  \FIG{.35}{false}{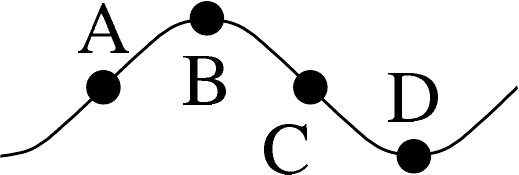}
  \caption{The four principal points on an interferogram.}
  \label{fig-abcd}
\end{figure}

The 4 points shown in Figure~\ref{fig-abcd} correspond to a specific alignment where point D sits at destructive interference:
\begin{itemize}
\item D (Destructive/Dark): Minimum intensity corresponding to the dark fringe—the operating state of a nuller;
\item B (Bright): Maximum intensity corresponding to the white fringe peak;
\item A (dAwn): Inflection point on the rising slope;
\item C (dusK=C): Inflection point on the falling slope.
\end{itemize}

When an additional OPD is introduced, these sample points shift, modulating their measured fluxes along the fringe profile (Figure~\ref{fig-abcd-mod}). Demodulating these intensity variations recovers the fringe phase shift and the differential OPD.

\begin{figure} \centering
  \FIG{.7}{false}{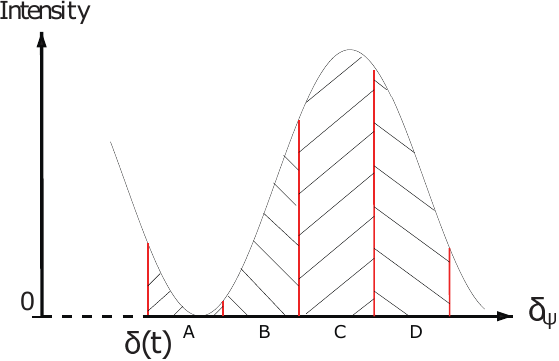}
  \caption[Principle of ABCD modulation.]{Principle of ABCD modulation \cre{K. Houairi}.}
  \label{fig-abcd-mod}
\end{figure}

For small OPD deviations, intensity changes are most pronounced at quadrature points A and C along the steep slopes. For nulling, maintaining point D at absolute minimum intensity is paramount.

Quadrature phase sampling is implemented using two main approaches:
\begin{itemize}
\item Temporal modulation: Phase points are acquired sequentially by stepping a delay line across four positions separated by $\lambda/4$ ($\pi/2$ in phase). Measurement bandwidth is constrained by mechanical actuator dynamics.

\item Spatial modulation: The beam combiner produces 4 simultaneous outputs phase-shifted by $\pi/2$. The combiner is calibrated to account for non-ideal phase quadrature shifts. Simultaneous sampling eliminates moving parts and supports high control loop bandwidths, benefiting spaceflight reliability.
\end{itemize}

%¤¤¤¤¤¤¤¤¤¤¤¤¤¤¤¤¤¤¤¤¤¤¤¤¤¤¤¤¤¤¤¤¤¤¤¤¤¤¤¤¤¤¤¤¤¤¤¤¤¤¤¤¤¤¤¤¤¤¤¤¤¤¤¤¤¤¤¤¤¤¤¤¤¤¤¤¤¤¤
\subsection{The Modified Mach-Zehnder: Core of the \pe Testbed}
\label{sec-desc-mmz}

Beam combination is performed by a Modified Mach-Zehnder (MMZ) interferometer \cite{Serabyn00}, developed during Sophie Jacquinod's PhD research \cite{Jacquinod10}. To minimize differential optical paths, the MMZ handles both science recombination and OPD metrology sensing. Figure~\ref{fig-principe-mmz} illustrates its optical schematic.

\begin{figure}  \centering
  \FIG{.7}{false}{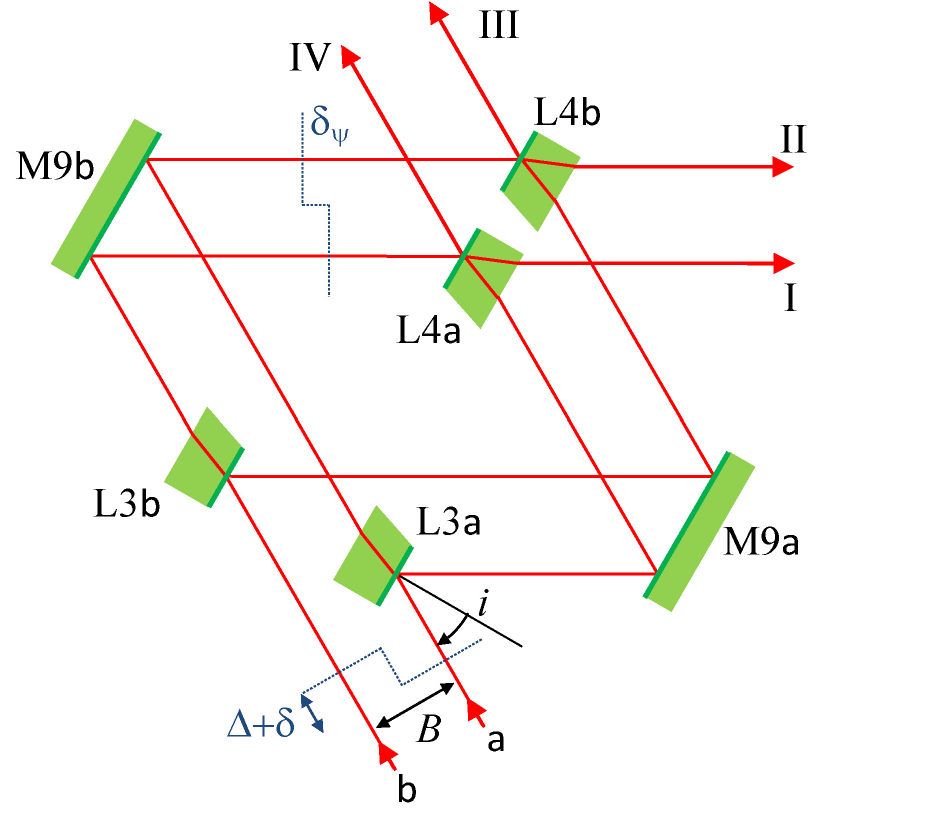}
  \caption{Principle of the \pe MMZ interferometer.}  \label{fig-principe-mmz}
\end{figure}

The MMZ operates as follows \cite{Jacquinod08}:
\begin{itemize}
\item Input beams $\ma$ and $\mb$ are split into pairs by CaF$_2$ beamsplitters L3$\ma$ and L3$\mb$;
\item Flat mirrors M9$\ma$ and M9$\mb$ fold the beams, keeping pairs parallel;
\item Recombining optics L4$\ma$ and L4$\mb$ (identical to L3$\ma$ and L3$\mb$) combine the beams into 4 outputs: I, II, III, and IV. The arrangement swaps beam pairs to interfere arms $\ma$ and $\mb$;
\item When all plates and mirrors are parallel, all four outputs exhibit uniform intensity regardless of input incidence angle $i$;
\item Under symmetric conditions, internal OPD $\delta_\psi$ is zero, making outputs III and IV identical. If input fluxes are equal, outputs I and II are also balanced. Output OPD is zero when the input wavefront features a step height $\Delta = B / \sin(i)$, where $B$ is beam separation and $i$ is incidence angle;
\item External OPD variations $\delta$ modulate intensity across output pairs;
\item Symmetric component layout around the central plane cancels global thermal expansion effects;
\item Beamsplitter substrates are thick enough to spatially separate internal secondary reflections. Trapezoidal substrate geometry directs ghost reflections out of the primary beam path.
\end{itemize}

The primary advantage of the MMZ lies in its symmetric outputs III and IV. Assuming beamsplitter intensity reflection $R$ and transmission $T$, both beams undergo identical $RT$ attenuation, ensuring balance in flux and phase. Symmetric recombination is essential for deep destructive interference. On asymmetric outputs I and II ($R^2$ and $T^2$), fluxes are unbalanced because achieving $R=T$ across broad bandwidths is unfeasible.

An ideal MMZ at zero OPD produces an achromatic constructive white fringe on symmetric outputs III and IV and a complementary destructive dark fringe on outputs I and II. Adding an Achromatic Phase Shifter (APS) providing a $\pi$ phase shift upstream converts symmetric outputs III and IV into an achromatic dark fringe, while outputs I and II become a chromatic bright fringe.

The Fringe Sensor (FS) measures external OPD $\delta$ to drive stabilization feedback loops. {ONERA}'s approach embeds the FS within the MMZ to eliminate differential path errors \cite{Houairi08b}. An internal phase delay $\delta_\psi$ is introduced into outputs I and IV without altering outputs II and III. Outputs I and IV measure phase (near points A and C) while outputs II and III record states B and D. This matches the ABCD demodulation algorithm described in Section~\ref{sec-desc-abcd}, allowing the FS to measure residual OPD $\delta$.

Because science extraction utilizes outputs B and D, half the MMZ output power is allocated to metrology, reducing science SNR by $\sqrt{2}$. Similarly, metrology sensing uses outputs A and C, reducing sensing SNR by $\sqrt{2}$. However, this integrated approach eliminates differential path drift while supplying real-time flux and contrast metrics for fringe tracking.

Dichroic beamsplitters downstream of the 4 MMZ outputs split science and metrology wavelengths at a 1.65~\mum cutoff. Science bands ($\BH$ and $\K$) are transmitted; metrology bands ($\I$ and $\J$) are reflected. A lower cutoff at 0.8~\mum bounds metrology light to 0.8–1.65~\mum.

Metrology light is split into $\I$ and $\J$ bands by 1~\mum cutoff dichroics to perform dual-wavelength fringe sensing. Dual-band phase sensing enables phase disambiguation algorithms that extend the dynamic measurement range (Section~\ref{sec-coherencage}).

%¤¤¤¤¤¤¤¤¤¤¤¤¤¤¤¤¤¤¤¤¤¤¤¤¤¤¤¤¤¤¤¤¤¤¤¤¤¤¤¤¤¤¤¤¤¤¤¤¤¤¤¤¤¤¤¤¤¤¤¤¤¤¤¤¤¤¤¤¤¤¤¤¤¤¤¤¤¤¤
\subsection{MMZ Output Intensity Equations}
\label{sec-mise-en}

In practice, beamsplitters exhibit slight absorption and phase delays. The quadrature phase shift across the 4 outputs is produced by translating beamsplitter L3$\ma$ perpendicular to its face, introducing a chromatic phase shift $\delta_\psi$. Consequently, phase shifts in the $\I$ and $\J$ bands differ from ideal quadrature. We derive MMZ output intensities as a function of these phase shifts and path differences.

%-------------------------------------------------------------------------------
\subsubsection{Diopter Reflection, Transmission, and Absorption Coefficients}
\label{sec-facteurs-reflexion}

We express output intensities using beamsplitter reflection and transmission coefficients:
\begin{itemize}
\item $r$: amplitude reflection coefficient in air;
\item $t$: amplitude transmission coefficient from air to CaF$_2$;
\item $r'$: amplitude reflection coefficient inside CaF$_2$;
\item $t'$: amplitude transmission coefficient from CaF$_2$ to air.
\end{itemize}

Denoting air and glass refractive indices as $n_\maa$ and $n_\mv$, intensity reflection and transmission factors are defined by:
\begin{equation}
   \left\{
    \begin{array}{r@{\hs}l}
      R &= rr^\ast \\[+6pt]
      T &= \dfrac{n_\mv}{n_\maa}tt^\ast
    \end{array}
  \right. \qquad \text{and} \qquad \left\{
    \begin{array}{r@{\hs}l}
      R' &= r'r'^\ast \\[+6pt]
      T' &= \dfrac{n_\maa}{n_\mv}t't'^\ast
    \end{array}
  \right.,
\end{equation}
where $z^\ast$ represents the complex conjugate of $z$.

Complex amplitude factors are expressed as:
\begin{equation}
  r = \GN{r} \times e^{i\phi_r} \qquad \text{and} \qquad t = \GN{t} \times
  e^{i\phi_t},
\end{equation}
satisfying $T = T'$ and $\phi_t = \phi_{t'}$.

Substrate absorption introduces a phase shift $\phi_\ml$ defined by:
\begin{equation}
  \phi_\ml = \phi_r+\phi_{r'}-2\phi_t.
\end{equation}

For an ideal non-absorbing diopter, energy conservation requires $R = R'$ and $\phi_\ml = \pi$.

MMZ beamsplitter plates feature a multi-layer semi-reflective coating on the active front surface, while the back surface (Ar) remains uncoated. Trapezoidal substrate geometry directs secondary back-surface reflections away from the primary beam path.

%-------------------------------------------------------------------------------
\subsubsection{Measured Intensity Formulations}
\label{sec-expression-intensites}

For an extended source, spatial extent reduces fringe visibility by the complex degree of spatial coherence $\mu$ (Section~\ref{sec-res-interf}). Denoting input arm intensities as $I_\ma$ and $I_\mb$ and back-surface transmission as $T_\mar$, output intensities at the FS are given by \cite{Houairi09b}:
\begin{equation}
  \left\{
    \begin{array}{r@{\hs}l}
      I_\mni &= RR'T^2_\mar I_\ma + T^2T^2_\mar I_\mb + 2 \mu\sqrt{RR'}TT^2_\mar 
      \sqrt{I_\ma I_\mb}\cos\GP{2\pi\sigma\GC{\delta+\delta_\psi}+\phi_\ml}
      \\[+6pt]
      I_\mnii &= T^2T^2_\mar I_\ma + RR'T^2_\mar I_\mb + 2 \mu\sqrt{RR'}TT^2_\mar 
      \sqrt{I_\ma I_\mb}\cos\GP{2\pi\sigma\delta-\phi_\ml} \\[+6pt]
      I_\mniii &= RTT_\mar I_\ma + RTT_\mar I_\mb + 2 \mu RTT_\mar 
      \sqrt{I_\ma I_\mb}\cos\GP{2\pi\sigma\delta} \\[+6pt]
      I_\mniv &=  RTT_\mar I_\ma + RTT_\mar I_\mb + 2 \mu RTT_\mar 
      \sqrt{I_\ma I_\mb}\cos\GP{2\pi\sigma\GC{\delta+\delta_\psi}}
    \end{array}
  \right..
\end{equation}

Here, substrates are assumed lossless. $I_\ma$, $I_\mb$, $\mu$, and $\delta$ are the four unknowns in this system of four equations.

Accounting for broadband chromatic effects requires integrating over the temporal degree of coherence $\gamma(\delta)$ (Section~\ref{sec-interferences-deux}). Assuming spectral differences across outputs are negligible, and incorporating substrate absorption, the full expressions become \cite{Houairi09b}:
\begin{equation}
  \left\{
    \begin{array}{r@{\hs}l}
      I_\mni &= \moy[\sigma]{RR'T^2_\mar}I_\ma+\moy[\sigma]{T^2T^2_\mar}I_\mb \\
      &{}+2\mu\moy[\sigma]{\sqrt{RR'}TT^2}\gamma(\delta+\delta_\psi)
      \sqrt{I_\ma I_\mb}\cos\GP{2\pi\sigma\GC{\delta+\delta_\psi}+\phi_\ml}
      \\[+6pt]
      I_\mnii &= \moy[\sigma]{T^2T^2_\mar}I_\ma+\moy[\sigma]{RR'T^2_\mar}I_\mb
      +2\mu\moy[\sigma]{\sqrt{RR'}TT^2}\gamma(\delta)\sqrt{I_\ma I_\mb}
      \cos\GP{2\pi\sigma\delta-\phi_\ml} \\[+6pt]
      I_\mniii &= \moy[\sigma]{RTT_\mar}I_\ma+\moy[\sigma]{RTT_\mar}I_\mb
      +2\mu\moy[\sigma]{RTT_\mar}\gamma(\delta)\sqrt{I_\ma I_\mb}
      \cos\GP{2\pi\sigma\delta} \\[+6pt]
      I_\mniv &= \moy[\sigma]{RTT_\mar}I_\ma+\moy[\sigma]{RTT_\mar}I_\mb
      +2\mu\moy[\sigma]{RTT_\mar}\gamma(\delta+\delta_\psi)\sqrt{I_\ma I_\mb}
      \cos\GP{2\pi\sigma\GC{\delta+\delta_\psi}}
    \end{array}
  \right..
\end{equation}

This represents a quasi-ABCD modulation scheme rather than ideal quadrature sampling.

Assuming the source coherence length exceeds internal OPD offset $\delta_\psi$, we approximate $\gamma(\delta) \simeq \gamma(\delta+\delta_\psi)$. Expanding trigonometric terms yields the linear matrix formulation:
\begin{equation}
  \underbrace{\begin{pmatrix}
      I_\mni \\
      I_\mnii \\
      I_\mniii \\  
      I_\mniv
    \end{pmatrix}}_{\V{Y}} = \underbrace{\begin{pmatrix}
      a_\mni & b_\mni & c_\mni & d_\mni \\
      a_\mnii & b_\mnii & c_\mnii & d_\mnii \\
      a_\mniii & b_\mniii & c_\mniii & d_\mniii \\
      a_\mniv & b_\mniv & c_\mniv & d_\mniv
    \end{pmatrix}}_{\V{M}} \cdot \underbrace{\begin{pmatrix}
      I_\ma \\
      I_\mb \\
      2\gamma(\delta)\sqrt{I_\ma I_\mb}\cos\GP{2\pi\sigma\delta} \\
      2\gamma(\delta)\sqrt{I_\ma I_\mb}\sin\GP{2\pi\sigma\delta}
    \end{pmatrix}}_{\V{X}}.
\label{eq-algo-direct}
\end{equation}

Vector $\V{Y}$ contains detected intensities, vector $\V{X}$ contains the four target parameters, and modulation matrix $\V{M}$ incorporates beamsplitter coefficients, spatial coherence, and instrumental phase shifts $2\pi\sigma\delta_\psi$ and $\phi_\ml$.

When $\V{M}$ is non-singular ($\det \V{M} \neq 0$), demodulation matrix $\V{D} = \V{M}^{-1}$ recovers state vector $\V{X}$:
\begin{equation}
\V{X} = \V{D} \cdot \V{Y}.
\end{equation}

%¤¤¤¤¤¤¤¤¤¤¤¤¤¤¤¤¤¤¤¤¤¤¤¤¤¤¤¤¤¤¤¤¤¤¤¤¤¤¤¤¤¤¤¤¤¤¤¤¤¤¤¤¤¤¤¤¤¤¤¤¤¤¤¤¤¤¤¤¤¤¤¤¤¤¤¤¤¤¤
\subsection{Calibration of the Demodulation Matrix}
\label{sec-etalonnage-abcd}

Vector $\V{X}$ yields residual OPD $\delta$, individual beam intensities, and temporal coherence:
\begin{equation}
  \left\{
    \begin{array}{r@{\hs}l}
      I_\ma &= X_1 \\[+6pt]
      I_\mb &= X_2 \\[+6pt]
      \gamma &= \dfrac{X_3^2+X_4^2}{4X_1X_2}\\[+6pt]
      \delta &= \dfrac{\lambda}{2\pi}\arctan\GP{\dfrac{X_3}{X_4}}
    \end{array}
  \right. \qquad \text{with} \qquad \V{X} = 
  \begin{pmatrix}
    X_1 \\
    X_2 \\
    X_3 \\
    X_4 \\
  \end{pmatrix}.
\end{equation}

OPD estimates drive beam cophasing, while individual beam intensities and visibility metrics provide real-time diagnostics. On ground-based interferometers, monitoring individual beam coupling into single-mode fibers helps validate OPD measurements under atmospheric fluctuations.

Deriving $\V{X}$ from raw output intensities $\V{Y}$ requires determining demodulation matrix $\V{D}$ and subtracting dark current biases.

The calibration procedure developed by K. Houairi \cite{Houairi09b} follows four steps:
\begin{itemize}
\item Both arms $\ma$ and $\mb$ are blocked to measure detector dark currents;
\item Arm $\mb$ is blocked to measure background intensities from arm $\ma$, yielding coefficients $a_i$ ($i \in \{I, II, III, IV\}$) of matrix $\V{M}$;
\item Arm $\ma$ is blocked to measure background intensities from arm $\mb$, yielding coefficients $b_i$ ($i \in \{I, II, III, IV\}$);
\item A calibrated piston modulation is applied with both arms open. Subtracting DC offsets derived from $a_i$ and $b_i$ yields the real and imaginary components of the interference terms, supplying coefficients $c_i$ and $d_i$.
\end{itemize}

Inverting calibrated matrix $\V{M}$ yields demodulation matrix $\V{D}$.

This calibration also extracts internal phase shift $\phi_\ml$ and internal offset $\delta_\psi$. Performing calibrations across metrology bands determines effective wavelength ratios (critical for phase unwrapping algorithms, Section~\ref{sec-estim-elargi}) and evaluates wavelength-dependent absorption dispersion $\phi_\ml$.

%§§§§§§§§§§§§§§§§§§§§§§§§§§§§§§§§§§§§§§§§§§§§§§§§§§§§§§§§§§§§§§§§§§§§§§§§§§§§§§§
\section{Fringe Disambiguation for Dynamic Range Extension}
\label{sec-coherencage}

%¤¤¤¤¤¤¤¤¤¤¤¤¤¤¤¤¤¤¤¤¤¤¤¤¤¤¤¤¤¤¤¤¤¤¤¤¤¤¤¤¤¤¤¤¤¤¤¤¤¤¤¤¤¤¤¤¤¤¤¤¤¤¤¤¤¤¤¤¤¤¤¤¤¤¤¤¤¤¤
\subsection{Simple Estimator}
\label{sec-estim-simple}

Single-wavelength phase measurements are limited by $2\pi$ phase ambiguity to a range of $\pm\lambda/2$ around zero OPD, as illustrated in Figure~\ref{fig-multi-lambda} \cite{Houairi09a}.

\begin{figure} \centering
  \FIG{.7}{false}{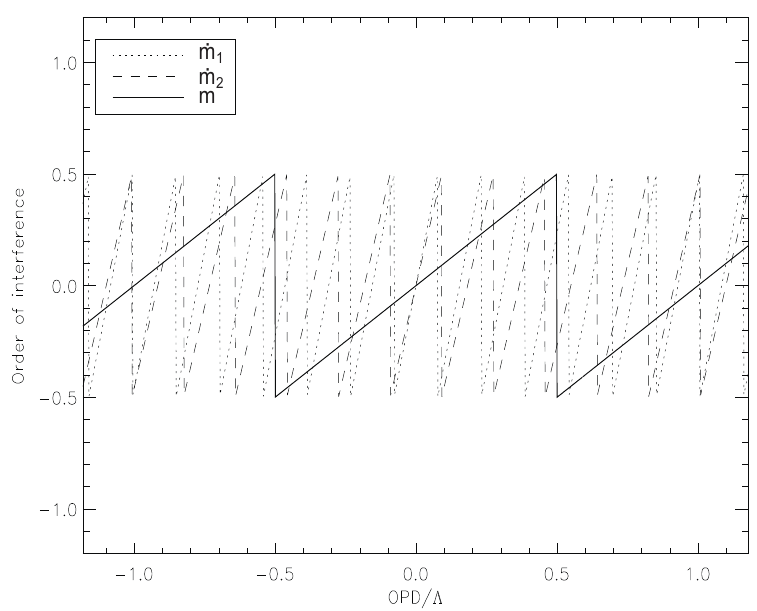}
  \caption[Principle of dual-wavelength fringe disambiguation.]{Principle of dual-wavelength fringe disambiguation \cre{K. Houairi}.}
  \label{fig-multi-lambda}
\end{figure}

Measuring phase at two distinct wavelengths ($\lambda_1=1.31$~\mum and $\lambda_2=1.55$~\mum) shifts phase-wrapping boundaries across wavelengths.

Multi-wavelength algorithms use these phase wrap differences to expand the unambiguous measurement range.

Defining fractional interference order $m_i$ at wavelength $\lambda_i$:
\begin{equation}
  m_i=\frac{\delta}{\lambda_i}.
\end{equation}

Decomposing $m_i$ into integer order $\overline m_i$ and fractional remainder $\dot m_i$ yields:
\begin{equation}
  \delta = \GP{\overline m_i+\dot m_i}\lambda_i \qquad \text{with} \qquad
  \left\{
    \begin{array}{rcl}
      \overline m_i &\in& \mathbb{Z} \\
      \dot m_i &\in& \left]-0.5;0.5\right]\\
    \end{array} \right..
\end{equation}

A single wavelength measures fractional remainder $\dot m_i$ but cannot resolve integer order $\overline m_i$. A simple dual-wavelength estimator \cite{Wyant71,Polhemus73} computes:
\begin{equation}
  \widehat \delta = \widehat m\Lambda \qquad \text{with} \qquad \widehat m
  = \wrap{\widehat{\dot m}_1-\widehat{\dot m}_2},
\end{equation}

where $\widehat \delta$ is estimated OPD and $\widehat{\dot m}_i$ is estimated fractional remainder in channel $i$. The wrap function wraps values into the interval $\left]-0.5;0.5\right]$.

Synthetic beat wavelength $\Lambda$ defines the extended measurement range:
\begin{equation}
  \Lambda = \frac{\lambda_1\lambda_2}{\lambda_2-\lambda_1}.
\end{equation}

For $\lambda_1=1.31$~\mum and $\lambda_2=1.55$~\mum, $\Lambda=8.46$~\mum, expanding dynamic range sixfold.

%¤¤¤¤¤¤¤¤¤¤¤¤¤¤¤¤¤¤¤¤¤¤¤¤¤¤¤¤¤¤¤¤¤¤¤¤¤¤¤¤¤¤¤¤¤¤¤¤¤¤¤¤¤¤¤¤¤¤¤¤¤¤¤¤¤¤¤¤¤¤¤¤¤¤¤¤¤¤¤
\subsection{Extended Estimator}
\label{sec-estim-elargi}

As seen in Figure~\ref{fig-multi-lambda}, the simple estimator can be sensitive to phase noise at points where fractional remainders coincide. K. Houairi developed an extended estimator using fractional approximations of the wavelength ratio $r_\lambda = \lambda_1/\lambda_2$ \cite{Houairi09a}.

Relating OPD expressions across channels:
\begin{equation}
  \delta = \lambda_1\GP{\overline m_1+\dot m_1} = \lambda_2\GP{\overline
    m_2+\dot m_2},
\end{equation}
yields:
\begin{equation}
  r_\lambda\overline m_1+r_\lambda\dot m_1 = \overline m_2+\dot m_2,
\end{equation}

Approximating wavelength ratio $r_\lambda$ as coprime fraction $p/q$ with estimation error $\varepsilon_f$:
\begin{equation}
  r_\lambda = \frac{\lambda_1}{\lambda_2} = \frac{p}{q}+\varepsilon_f.
\end{equation}

Bézout's identity states that for coprime integers $p$ and $q$, an integer $k$ exists such that:
\begin{equation}
  k \times p \equiv 1 \quad (\mod q).
\end{equation}

The extended OPD estimator $\widehat \delta_\oplus$ is given by:
\begin{equation}
  \widehat \delta_\oplus = \widehat m_\oplus\Lambda_\oplus \qquad \text{with}
  \qquad \widehat m_\oplus = q \times \wrap{\frac{k}{q} \times
    \overline{q\GP{\widehat{\dot m}_2-\widehat{r}_\lambda\widehat{\dot
          m}_1}}}+\widehat{\dot m}_1.
\end{equation}

Extended range $\Lambda_\oplus$ scales as an integer multiple of $\lambda_1$:
\begin{equation}
  \Lambda_\oplus = q\lambda_1.
\end{equation}

Full mathematical derivations are detailed in \cite{Houairi09a}. Selecting larger coprime integers $p$ and $q$ expands measurement range, though sensitivity to noise and wavelength calibration errors increases.

\medskip

I implemented this extended estimator on \pe to expand OPD dynamic range. During experimental validation with broadband light split by dichroics (Section~\ref{sec-desc-mmz}), I identified operational bounds not observed during initial monochromatic testing \cite{Houairi09b} (Section~\ref{sec-resultat-etalonnage}).

%§§§§§§§§§§§§§§§§§§§§§§§§§§§§§§§§§§§§§§§§§§§§§§§§§§§§§§§§§§§§§§§§§§§§§§§§§§§§§§§
\section{LQG Control for Vibration Rejection}
\label{sec-commande-lqg}

Mechanical vibrations pose significant challenges for exoplanet instruments. High-contrast coronagraphs and nullers require strict metrological stability; uncompensated vibrations degrade pointing (tip/tilt) or phase delay (piston). Passive damping reduces vibration amplitudes but cannot eliminate sharp spectral peaks.

An active vibration mitigation scheme deployed at the VLTI reduced OPD residuals from 470~nm RMS down to 360~nm RMS using accelerometers. A second-stage feedforward scheme, Vibration Tracking (VTK), reduced residuals further to 260~nm RMS \cite{DiLieto08}. However, VTK requires manual identification of disturbance frequencies and assumes undamped harmonic modes, making it sensitive to frequency shifts.

ONERA developed an active control framework \cite{Petit08,Meimon10} for VLT's SPHERE extreme AO system (Section~\ref{sec-optique-adaptative}). Applicable to tip/tilt and piston errors, it utilizes Linear Quadratic Gaussian (LQG) control to minimize residual phase variance. Optimal LQG control requires accurate disturbance models. S. Meimon formulated an identification algorithm \cite{Meimon10} that extracts disturbance models automatically from open-loop data (Section~\ref{sec-identification}).

Because vibrations directly impact \pe's cophasing performance under flight-like disturbances, I integrated LQG control and online identification into the real-time control loop to maintain nanometric OPD stability.

%¤¤¤¤¤¤¤¤¤¤¤¤¤¤¤¤¤¤¤¤¤¤¤¤¤¤¤¤¤¤¤¤¤¤¤¤¤¤¤¤¤¤¤¤¤¤¤¤¤¤¤¤¤¤¤¤¤¤¤¤¤¤¤¤¤¤¤¤¤¤¤¤¤¤¤¤¤¤¤
\subsection{Single-Mode Feedback Control Loop}
\label{sec-modelisation-temporelle}

%-------------------------------------------------------------------------------
\subsubsection{Time-Domain Model}
\label{sec-modele-temporel}

Assuming decoupled control modes (piston, tip, tilt), single-mode formulations apply independently. Figure~\ref{fig-bloc-diag} shows the block diagram of a standard AO/interferometry feedback control loop.

\begin{figure} \centering
  \FIG{.7}{false}{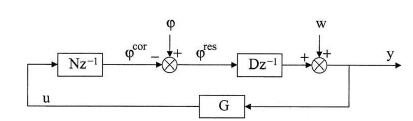}
  \caption{Block diagram of an AO or interferometry feedback control loop.}
  \label{fig-bloc-diag}
\end{figure}

The loop actuator converts control voltage $u$ into corrective phase $\phi^{cor}$. Corrective phase compensates disturbance phase $\phi=\phi^{per}$, leaving residual phase $\phi^{res}$. Residual phase is sampled by a sensor. Measurement $y$ includes additive zero-mean white Gaussian noise $w$ with variance $\sigma_w^2$. Controller $G$ processes measurement $y$ to generate new actuator command $u$.

Processing, readout, and actuation delays introduce latency into the loop (Figure~\ref{fig-chrono}).

\begin{figure} \centering
  \FIG{.7}{false}{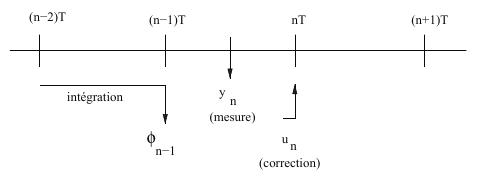}
  \caption{Timeline of feedback loop execution.}
  \label{fig-chrono}
\end{figure}

Loop frequency is set by sensor integration time.

Over interval $[(n-2)T, (n-1)T]$, the sensor integrates residual phase $\phi^{res}_{n-1} = \phi^{per}_{n-1} - \phi^{cor}_{n-1}$. Assuming linear instantaneous actuator response characterized by scalar gain $N$:
\begin{equation}
  \phi^{cor}_{n-1} = Nu_{n-2}.
  \label{eq-phicor}
\end{equation}

Actuator command $u_{n-2}$ computed during step $n-2$ applies corrective phase $\phi^{cor}_{n-1}$ during step $n-1$.

At time $(n-1)T$, sensor integration completes. Assuming linear sensor response with gain $D$, measurement $y_n$ integrated over period $T$ with noise $w_n$ is:
\begin{align}
  y_n &= D\phi^{res}_{n-1}+w_n \nonumber\\
  &= D\GP{\phi^{per}_{n-1}-Nu_{n-2}}+w_n.
  \label{eq-yn}
\end{align}

Controller $G$ processes $y_n$ to compute command $u_n$, applied over interval $[nT, (n+1)T]$. System latency equals two full frames (half-frame integration, full-frame computation, half-frame actuation). Linear controller $G$ acts on measurements via convolution \cite{Petit08}:
\begin{equation}
  u_n = (G \ast y)_n.
  \label{eq-un}
\end{equation}

%-------------------------------------------------------------------------------
\subsubsection{Closed-Loop Transfer Functions}
\label{sec-fonction-transfert}

Using Z-transforms ($\tilde{x}(z)$) simplifies discrete-time stability analysis. A delay of $n$ frames transforms to multiplication by $z^{-n}$.

Transforming Equation~\eqref{eq-yn} yields:
\begin{equation}
  \tilde{y}(z) = Dz^{-1}\tilde{\phi}^{res}(z)+\tilde{w}(z).
\end{equation}

Transforming Equation~\eqref{eq-un} ($\tilde{u}(z) = \tilde{G}(z)\tilde{y}(z)$) gives:
\begin{equation}
  \tilde{u}(z) =
  \tilde{G}(z)Dz^{-1}\tilde{\phi}^{res}(z)+\tilde{G}(z)\tilde{w}(z).
\end{equation}

Using $\tilde{\phi}^{cor}(z) = \tilde{\phi}^{per}(z)-\tilde{\phi}^{res}(z) = Nz^{-1}\tilde{u}(z)$ from Equation~\eqref{eq-phicor} yields:
\begin{equation}
  \GP{1+N\tilde{G}(z)Dz^{-2}}\tilde{\phi}^{res}(z) =
  \tilde{\phi}^{per}(z)-N\tilde{G}(z)z^{-1}\tilde{w}(z).
\end{equation}

This highlights two main transfer functions:
\begin{itemize}
\item Rejection transfer function $E(z)$ (for $\tilde{w}(z) = 0$):
  \begin{equation}
    E(z) = \frac{\tilde{\phi}^{res}(z)}{\tilde{\phi}^{per}(z)} = \GP{1+N\tilde{G}(z)Dz^{-2}}^{-1}.
  \end{equation}
  This matches standard unity-feedback formulations, the Closed-Loop Tranfer Function $\text{CLTF} = (1+\text{OLTF})^{-1}$, with Open-Loop Transfer Function $\text{OLTF} = N\tilde{G}(z)Dz^{-2}$).

\item Noise transfer function $H_w(z)$ (for $\tilde{\phi}^{per}(z) = 0$):
  \begin{equation}
    H_w(z) = \frac{\tilde{\phi}^{res}(z)}{\tilde{w}(z)} = -\GP{N\tilde{G}(z)z^{-1}}\GP{1+N\tilde{G}(z)Dz^{-2}}^{-1}.
  \end{equation}
\end{itemize}

These functions define frequency-domain disturbance rejection and noise propagation.

%-------------------------------------------------------------------------------
\subsubsection{Integrator Control Baseline}
\label{sec-rappel-integrateur}

For a standard integrator controller with gain $g$:
\begin{equation}
  u_n = u_{n-1}+g\,y_n,
\end{equation}
yielding Z-transform:
\begin{equation}
  \tilde{G}(z) = \frac{g}{1-z^{-1}}.
\end{equation}

Figure~\ref{fig-transfert} plots rejection and noise transfer functions across integrator gains.

\begin{figure} \centering
  \subfloat[Rejection transfer function.]{\label{fig-trans-rej}
    \FIG{0.49}{false}{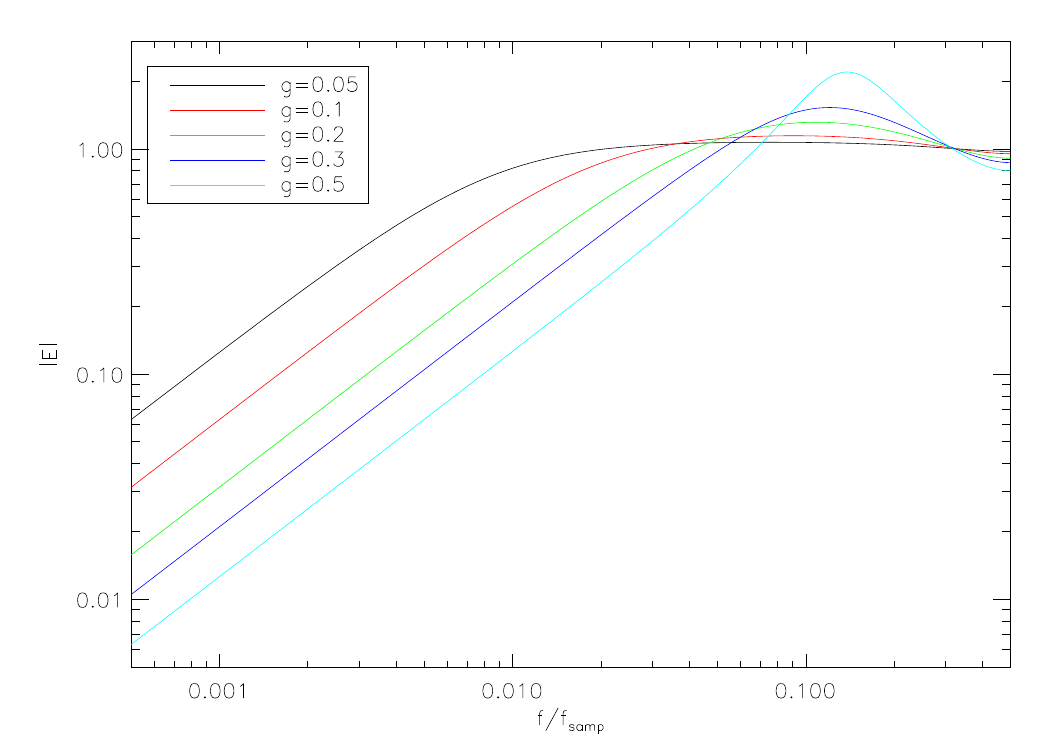}}\hfill
  \subfloat[Noise transfer function.]{\label{fig-trans-bruit}
    \FIG{0.49}{false}{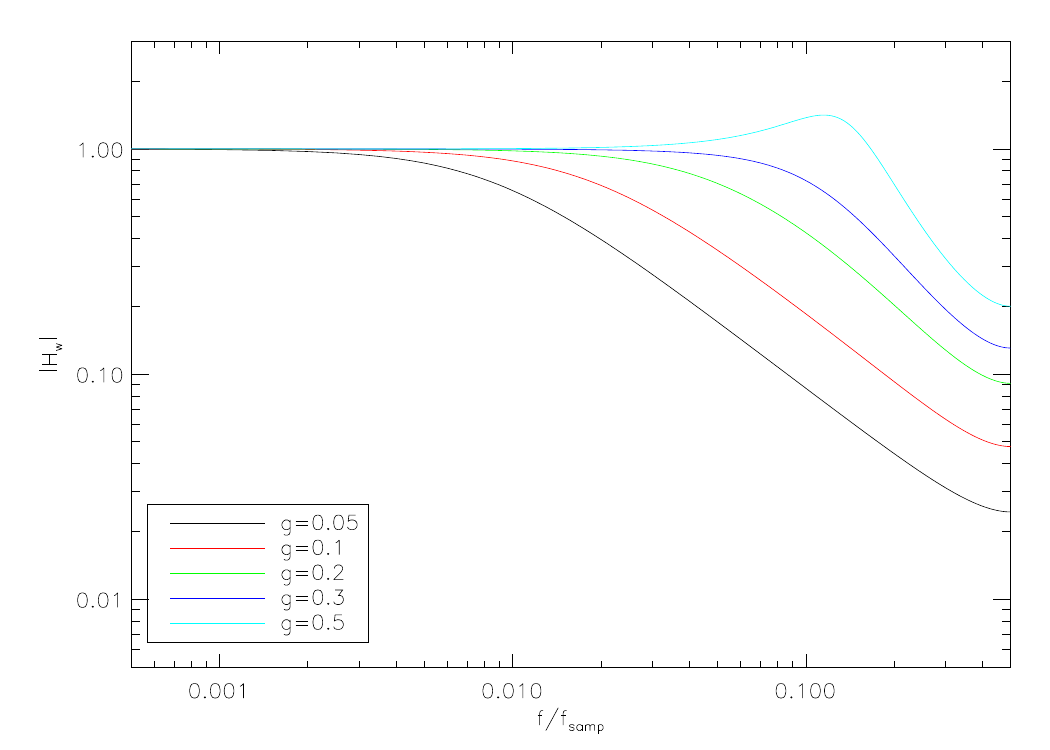}}
  \caption[Integrator transfer functions versus gain.]{Integrator rejection and noise transfer functions versus gain.}
  \label{fig-transfert}
\end{figure}

Figure~\ref{fig-trans-rej} exhibits characteristic $-20$~dB/decade low-frequency rejection. Beyond loop cutoff frequency, latency induces overshoot and amplifies noise (Figure~\ref{fig-trans-bruit}).

%¤¤¤¤¤¤¤¤¤¤¤¤¤¤¤¤¤¤¤¤¤¤¤¤¤¤¤¤¤¤¤¤¤¤¤¤¤¤¤¤¤¤¤¤¤¤¤¤¤¤¤¤¤¤¤¤¤¤¤¤¤¤¤¤¤¤¤¤¤¤¤¤¤¤¤¤¤¤¤
\subsection{Disturbance Modeling}
\label{sec-modelisation-temporelle-1}

At frame $n$, disturbance $\phi^{per}_n$ is modeled as a sum of independent processes:
\begin{itemize}
\item Low-frequency (LF) component $\phi^{LF}_n$ (atmospheric turbulence ground-side, or satellite drift space-side);
\item $N_{vib}$ narrow-band vibration modes $\phi^{vib,i}_n$. Characterized by narrow spectral peaks, vibration modes are predictable and can be actively rejected.
\end{itemize}

Summing independent contributions yields:
\begin{equation}
  \phi^{per}_n = \phi^{LF}_n+\sum_{i=1}^{N_{vib}}{\phi^{vib,i}_n}.
  \label{eq-phiper}
\end{equation}

Each vibration $\phi^{vib,i}_n$ discretizes a continuous mechanical mode specified by frequency $f^{vib,i}$ and damping factor $k^{vib,i}$.

Vibration dynamics are approximated by a 2nd-order Auto-Regressive process (AR2) \cite{Petit08}. For linear parameters $a^{vib,i}_1$ and $a^{vib,i}_2$ driven by zero-mean Gaussian white noise $\upsilon^{vib, i}_n$ with variance $(\sigma_{\upsilon}^{vib,i})^2$:
\begin{equation}
  \phi^{vib,i}_{n+1} =
  a^{vib,i}_1\phi^{vib,i}_n+a^{vib,i}_2\phi^{vib,i}_{n-1}+\upsilon^{vib,i}_n.
  \label{eq-phivib}
\end{equation}

AR2 coefficients relate to frequency $f^{vib,i}$, damping $k^{vib,i}$, and sampling period $T$ \cite{Meimon10}:
\begin{align}
  a^{vib,i}_1 &= 2\exp\GP{-2\pi
    k^{vib,i}f^{vib,i}T}\cos\GP{2\pi\sqrt{1-\GP{k^{vib,i}}^2} f^{vib,i}T} \\
  a^{vib,i}_2 &= -\exp\GP{-4\pi k^{vib,i}f^{vib,i}T}.
\end{align}

Varying $k^{vib,i}$ adjusts peak width in Power Spectral Density (PSD) plots (Figure~\ref{fig-ar2}). For $k^{vib,i} > 0.7$, the AR2 PSD decreases monotonically, modeling low-frequency noise.

\begin{figure} \centering
  \FIG{.5}{false}{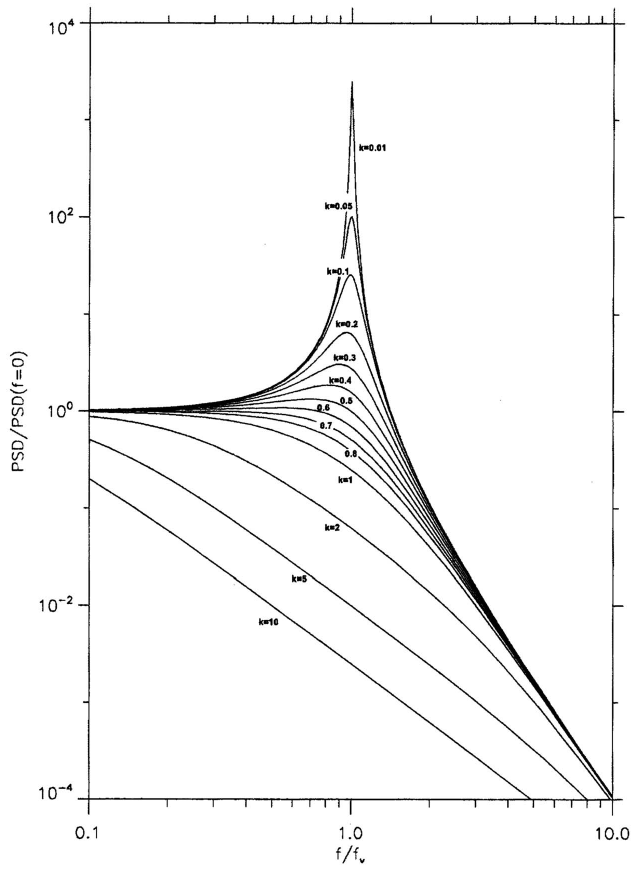}
  \caption[Relative PSDs of an AR2 model for various damping factors $k^{vib,i}$.]{Relative PSDs of an AR2 model for various damping factors $k^{vib,i}$ \cre{S. Meimon}.}
  \label{fig-ar2}
\end{figure}

Low-frequency disturbance $\phi^{BF}$ is similarly structured as an AR2 model:
\begin{equation}
  \phi^{BF}_{n+1} = a^{BF}_1\phi^{BF}_n+a^{BF}_2\phi^{BF}_{n-1}+\upsilon^{BF}_n,
  \label{eq-phibf}
\end{equation}
with parameters $a^{BF}_1$, $a^{BF}_2$ and white noise driving variance $(\sigma_{\upsilon}^{BF})^2$.

%¤¤¤¤¤¤¤¤¤¤¤¤¤¤¤¤¤¤¤¤¤¤¤¤¤¤¤¤¤¤¤¤¤¤¤¤¤¤¤¤¤¤¤¤¤¤¤¤¤¤¤¤¤¤¤¤¤¤¤¤¤¤¤¤¤¤¤¤¤¤¤¤¤¤¤¤¤¤¤
\subsection[Principle of LQG Control for Single-Mode Disturbances]{Principle of Linear Quadratic Gaussian (LQG) Control for Single-Mode Disturbances}
\label{sec-principe-commande}

%-------------------------------------------------------------------------------
\subsubsection{State-Space Representation}
\label{sec-modelisation-espace}

LQG controllers use state-space formulations \cite{Kulcsar06}. Combining Equations~\eqref{eq-phiper}, \eqref{eq-phivib}, and \eqref{eq-phibf} yields:
\begin{equation}
  \underbrace{\begin{pmatrix}
      \begin{array}{@{}c@{}}
        \phi^{BF}_{n+1}\\
        \phi^{BF}_{n}\\
        \hline
        \phi^{vib,1}_{n+1}\\
        \phi^{vib,1}_{n}\\
        \hline
        \phi^{vib,2}_{n+1}\\
        \phi^{vib,2}_{n}\\
        \hline
        \vdots
      \end{array}
    \end{pmatrix}}_{\V{x}_{n+1}}=\underbrace{\begin{pmatrix}
      \begin{array}{@{}c@{\,}c@{\,}|@{\,}c@{\,}c@{\,}|@{\,}c@{\,}c@{\,}|c@{}}
        a^{BF}_1 & a^{BF}_2 & 0 & 0 & 0 & 0 & \cdots\\
        1 & 0 & 0 & 0 & 0 & 0 & \cdots\\
        \hline
        0 & 0 & a^{vib,1}_1 & a^{vib,1}_2 & 0 & 0 & \cdots\\
        0 & 0 & 1 & 0 & 0 & 0 & \cdots\\
        \hline
        0 & 0 & 0 & 0 & a^{vib,2}_1 & a^{vib,2}_2 & \cdots\\
        0 & 0 & 0 & 0 & 1 & 0 & \cdots\\
        \hline
        \vdots & \vdots & \vdots & \vdots & \vdots & \vdots & \ddots
      \end{array}
    \end{pmatrix}}_{\V{A}} \underbrace{\begin{pmatrix}
      \begin{array}{@{}c@{}}
        \phi^{BF}_{n}\\
        \phi^{BF}_{n-1}\\
        \hline
        \phi^{vib,1}_{n}\\
        \phi^{vib,1}_{n-1}\\
        \hline
        \phi^{vib,2}_{n}\\
        \phi^{vib,2}_{n-1}\\
        \hline
        \vdots
      \end{array}
    \end{pmatrix}}_{\V{x}_n} +\underbrace{\begin{pmatrix}
      \begin{array}{@{}c@{}}
        \upsilon^{BF}_{n}\\
        0\\
        \hline
        \upsilon^{vib,1}_{n}\\
        0\\
        \hline
        \upsilon^{vib,2}_{n}\\
        0\\
        \hline
        \vdots
      \end{array}
    \end{pmatrix}}_{\V{\upsilon}_n},
  \label{eq-phi-mat}
\end{equation}

where $\V{x}_n$ is the state vector of phase components, $\V{A}$ is the state transition matrix containing AR2 parameters, and $\V{\upsilon}_n$ is process noise.

Equation~\eqref{eq-yn} converts to matrix form:
\begin{equation}
  y_n = \underbrace{\begin{pmatrix}
      \begin{array}{cc|cc|cc|c}
        0 & D & 0 & D & 0 & D & \cdots \\
      \end{array}
    \end{pmatrix}}_{\V{C}} \V{x}_n-DNu_{n-2}+w_n.
  \label{eq-yn2}
\end{equation}

This establishes state-space equations:
\begin{align}
  \V{x}_{n+1} &= \V{A}\V{x}_n+\upsilon_n \label{eq-xn1}\\
  y_n &= \V{C} \V{x}_n-DNu_{n-2}+w_n.
  \label{eq-yn1}
\end{align}

%-------------------------------------------------------------------------------
\subsubsection{LQG Control via Kalman Filtering}
\label{sec-etapes-controleur}

The LQG framework pairs a Kalman filter estimator with state feedback gains.

We define state estimation metrics:
\begin{itemize}
\item $\hat{\V{x}}_{n/n'}$: estimate of $\V{x}_n$ at step $n$ given measurements up to $n'$;
\item $\hat{y}_{n/n'}$: estimate of measurement $y_n$ given data up to $n'$;
\item $\tilde{\V{x}}_{n/n'} = \V{x}_n-\hat{\V{x}}_{n/n'}$: estimation error with covariance matrix $\V{\Sigma}_{n/n'}$;
\item $\tilde{y}_{n/n'} = y_n - \hat{y}_{n/n'}$: innovation term.
\end{itemize}

Predicted measurement $\hat{y}_{n/n-1}$ is:
\begin{equation}
  \hat{y}_{n/n-1} = \V{C}\hat{\V{x}}_{n/n-1}-DNu_{n-2}.
\end{equation}

Updated state estimate $\hat{\V{x}}_{n/n}$ follows \cite{Petit08}:
\begin{equation}
  \hat{\V{x}}_{n/n} = \hat{\V{x}}_{n/n-1}+\V{H}_n\tilde{y}_{n/n-1},
  \label{eq-xnn}
\end{equation}
where Kalman gain $\V{H}_n$ incorporates measurement noise variance $\sigma_w^2$ via the Riccati equation:
\begin{equation}
  \V{H}_n =
  \V{\Sigma}_{n/n-1}\V{C}^T\GP{\V{C}\V{\Sigma}_{n/n-1}\V{C}^T+\sigma_w^2}^{-1}.
\end{equation}

State error covariance matrix $\V{\Sigma}_{n/n}$ updates as:
\begin{equation}
  \V{\Sigma}_{n/n} = \V{\Sigma}_{n/n-1}-\V{H}_n\V{C}\V{\Sigma}_{n/n-1}.
\end{equation}

Assuming asymptotic steady-state covariance $\V{\Sigma}_\infty$ simplifies real-time computation:
\begin{equation}
  \V{\Sigma}_\infty = \V{\Sigma}_\infty-\V{H}_\infty\V{C}\V{\Sigma}_\infty,
\end{equation}
with steady-state Kalman gain $\V{H}_\infty$:
\begin{equation} \label{eq-H-infty}
  \V{H}_\infty =
  \V{\Sigma}_\infty\V{C}^T\GP{\V{C}\V{\Sigma}_\infty\V{C}^T+\sigma_w^2}^{-1}.
\end{equation}

The state update simplifies to:
\begin{equation}
  \hat{\V{x}}_{n/n} = \hat{\V{x}}_{n/n-1}+\V{H}_\infty\tilde{y}_{n/n-1}.
  \label{eq-maj}
\end{equation}

Predicting state $\hat{\V{x}}_{n+1/n}$ using model $\V{A}$ yields:
\begin{equation}
  \hat{\V{x}}_{n+1/n} = \V{A}\hat{\V{x}}_{n/n}.
  \label{eq-prediction}
\end{equation}

Control command $u_n$ is derived via feedback matrix $\V{K}$:
\begin{equation}
  u_n = -\V{K}\hat{\V{x}}_{n+1/n},
  \label{eq-commande}
\end{equation}
with:
\begin{equation}
  \V{K} = -N^{-1}\begin{pmatrix}
	\begin{array}{cc|cc|cc|c}
      0 & 1 & 0 & 1 & 0 & 1 & \cdots \\
    \end{array}
  \end{pmatrix}.
\end{equation}

Reconstructing open-loop equivalents yields pseudo open-loop measurement $\hat{y}_n^{BO}$:
\begin{equation}
  \hat{y}_n^{BO} = y_n+DNu_{n-2}.
  \label{eq-pseudobo}
\end{equation}

%-------------------------------------------------------------------------------
\subsubsection{LQG Transfer Function}
\label{sec-fonction-transfert-1}

Taking Z-transforms of updates (Equation~\eqref{eq-maj}) and predictions (Equation~\eqref{eq-prediction}) yields:
\begin{equation}
  \tilde{\hat{\V{x}}}(z) =
  \V{A}z^{-1}\tilde{\hat{\V{x}}}(z)+\V{A}\V{H}_\infty
  \GP{\tilde{y}(z)-\V{C}z^{-1}\tilde{\hat{\V{x}}}(z)+DNz^{-2}\tilde{u}(z)}.
\end{equation}

Transforming control equation~\eqref{eq-commande} ($\tilde{u}(z) = -\V{K}\tilde{\hat{\V{x}}}(z)$) yields:
\begin{equation}
  \GC{\V{Id}-\V{A}\GP{\V{Id}-\V{H}_\infty\V{C}}z^{-1}+
    \V{A}\V{H}_\infty\V{K}DNz^{-2}}\tilde{\hat{\V{x}}}(z) =
  \V{A}\V{H}_\infty\tilde{y}(z).
\end{equation}

The LQG transfer function $\tilde{G}(z)$ is:
\begin{equation}
  \tilde{G}(z) = -\V{K}\GC{\V{Id}-\V{A}\GP{\V{Id}-\V{H}_\infty\V{C}}z^{-1}+
    \V{A}\V{H}_\infty\V{K}DNz^{-2}}^{-1}\V{A}\V{H}_\infty.
  \label{eq-trans-lqg}
\end{equation}

Unlike fixed integrator transfer functions, $\tilde{G}(z)$ adapts dynamically to estimated disturbance parameters.

%¤¤¤¤¤¤¤¤¤¤¤¤¤¤¤¤¤¤¤¤¤¤¤¤¤¤¤¤¤¤¤¤¤¤¤¤¤¤¤¤¤¤¤¤¤¤¤¤¤¤¤¤¤¤¤¤¤¤¤¤¤¤¤¤¤¤¤¤¤¤¤¤¤¤¤¤¤¤¤
\subsection{Disturbance Identification Method}
\label{sec-identification}

LQG performance depends on disturbance model fidelity. When disturbances evolve, parameter identification must run online at operational update rates.

Identification operates on pseudo open-loop spectra reconstructed from closed-loop data using Equation~\eqref{eq-pseudobo}.

Disturbance PSDs combine:
\begin{itemize}
\item Low-frequency drift;
\item Discrete vibration peaks;
\item Measurement noise.
\end{itemize}

Sine we only work with limited number of data points, we are not actually looking at the PSD, but at an estimation called periodogram. Identification models raw periodograms using a finite number of data points. Unsmoothed periodograms exhibit unity SNR across frequencies up to Nyquist limit $f_\mech/2$.

High-frequency periodogram bins above threshold $f_w$ estimate white measurement noise levels. Threshold $f_w$ is set above known vibration frequencies to avoid overestimating noise levels.

Low-frequency AR2 parameters are fitted across interval $[f_1, f_2]$. As shown in Figure~\ref{fig-clip}, periodograms are approximated by power-law fits (green dashed line). Spectral peaks exceeding twice this baseline fit are clipped (green regions). Fitting low-frequency AR2 parameters to clipped periodograms minimizes residual power.

\begin{figure} \centering
  \FIG{.7}{false}{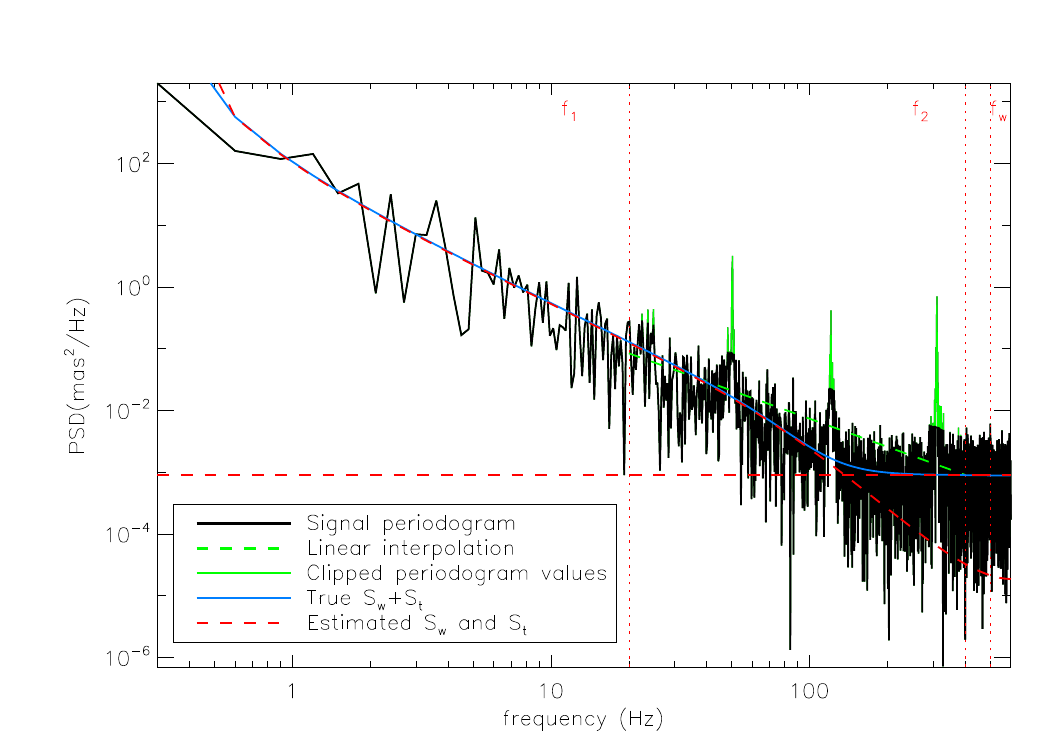}
  \caption[Low-frequency component and measurement noise identification.]{Identification of low-frequency component and measurement noise \cre{S. Meimon}.}
  \label{fig-clip}
\end{figure}

Subtractions isolate vibration peaks, identified iteratively by power ranking. Peak frequency, amplitude, and damping parameters are fitted by residual power minimization and subtracted sequentially.

Ref. \cite{Meimon10} provides a detailed mathematical description, including performance validations on simulated SPHERE datasets and on-sky VLT/NAOS data.

%§§§§§§§§§§§§§§§§§§§§§§§§§§§§§§§§§§§§§§§§§§§§§§§§§§§§§§§§§§§§§§§§§§§§§§§§§§§§§§§
\section{Thesis Context and Contributions}
\label{sec-ma-these}

The \pe testbed was established following Phase-0 studies for the \peg mission. Preliminary definition involved a French multi-partner consortium led by CNES, dividing subsystem responsibilities by institutional expertise.

Two PhD theses initiated the project:
\begin{itemize}
\item Kamel Houairi \cite{Houairi09b}: Cophasing definition and testing;
\item Sophie Jacquinod \cite{Jacquinod10}: MMZ beam combiner design and realization.
\end{itemize}

A prototype combiner (MMZ1, 0.8–1.5~\mum) developed by ONERA and GEPI enabled initial cophasing trials prior to completing \pe's broadband MMZ2 (0.8–3.3~\mum).

When my PhD began in November 2008, subsystem procurement was underway. Algorithms (ABCD modulation, fringe unwrapping, LQG control) were undergoing preliminary testing or conceptual design.

Working within a 6-partner consortium (Figure~\ref{fig-equipe}), project activities represented 20–25 full-time equivalent years, with hardware costs around €700k (€3M overall including personnel).

\begin{figure} \centering
  \FIG{.9}{false}{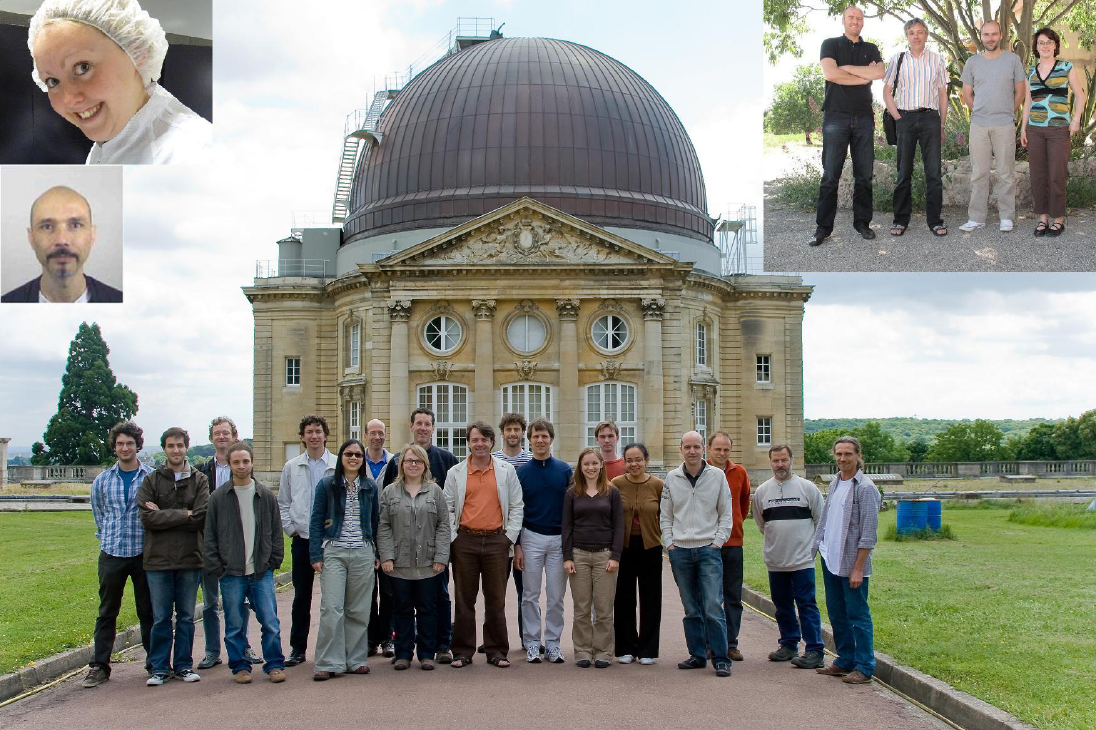}
  \caption{Group photograph of the \pe team.}
  \label{fig-equipe}
\end{figure}

Cleanroom installation at Paris Observatory–Meudon began in March 2009. Initial work involved upgrading control software for ONERA's MMZ1, expanding execution rates from <200~Hz to 1~kHz. I developed real-time control software and diagnostic packages. Testbed integration proceeded in stages to validate subsystems sequentially (Chapter~\ref{sec-integr-devel}). I participated in optical alignments alongside dedicated LESIA engineers Laurie Pham (joined late 2008) and \'Emilie Lhomé (joined late 2009). Over 75\% of my time between February 2009 and manuscript drafting was spent on site at Meudon. Figure~\ref{fig-calendrier} outlines the project timeline, institutional responsibilities, and associated PhD theses.

\begin{figure}  \centering
  \FIGA{90}{0.84}{false}{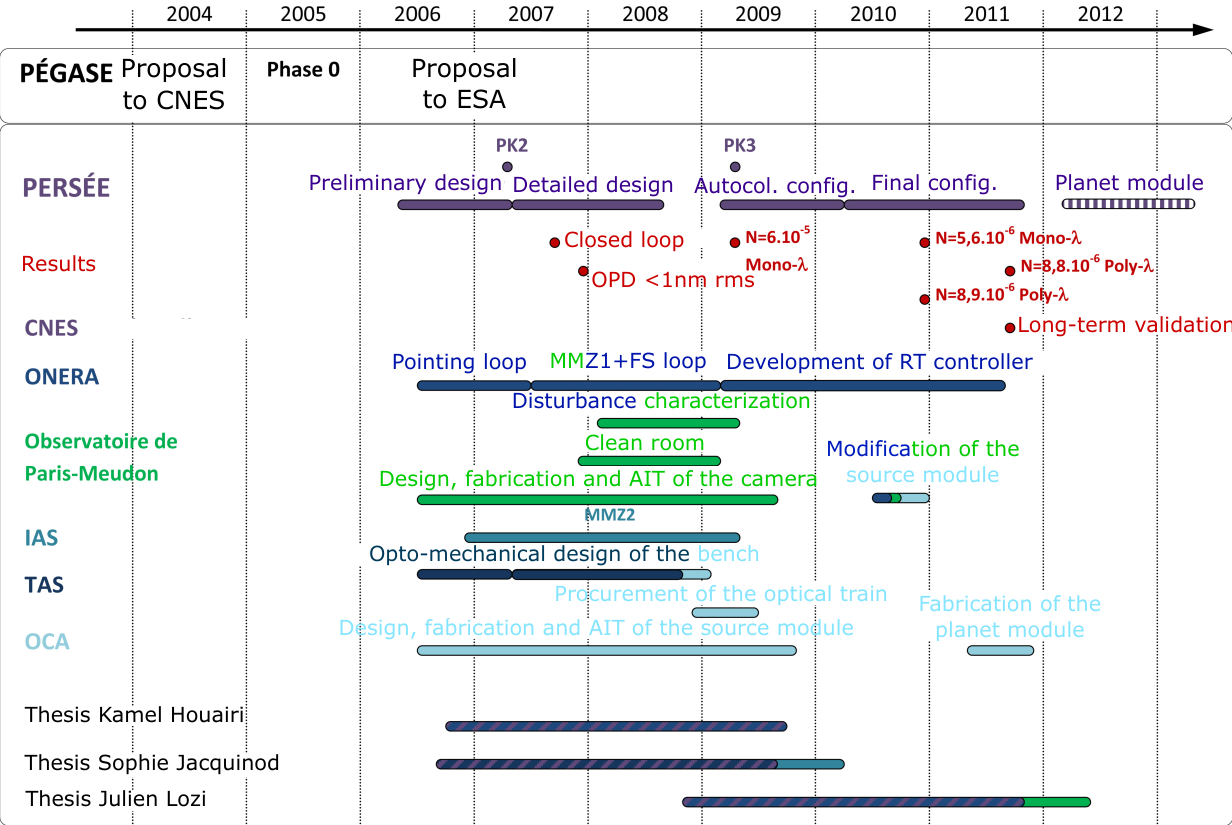}
  \caption{\pe project schedule, institutional partners, and associated PhD theses.}
  \label{fig-calendrier}
\end{figure}

My thesis focuses on integrating the \pe testbed, evaluating subsystem performance, and analyzing inter-subsystem couplings. I aim to validate system performance specifications by optimizing cophasing control loops critical to nulling stability. Experimental results will be extrapolated to space mission scale (\peg) to identify critical design parameters and evaluate potential relaxation of formation-flying control requirements, directly impacting future mission feasibility and cost.

%§§§§§§§§§§§§§§§§§§§§§§§§§§§§§§§§§§§§§§§§§§§§§§§§§§§§§§§§§§§§§§§§§§§§§§§§§§§§§§§
\chapter{Integration and Development of the \pe Testbed}
\label{sec-integr-devel}

\begin{flushright}
 \begin{minipage}{12cm}
   {\small \textit{The truth may be puzzling. It may take some work to
       grapple with. It may be counterintuitive. It may contradict deeply
       held prejudices. It may not be consonant with what we desperately
       want to be true. But our preferences do not determine what is true.
       We have a method, and that method helps us to reach not absolute
       truth, only asymptotic approaches to the truth. never there, just
       closer and closer, always finding vast new oceans of undiscovered
       possibilities. Cleverly designed experiments are the key.}}

   \raggedleft{{\small Carl Sagan, "Wonder and Skepticism", Skeptical
       Enquirer Vol.~19, Issue~1 (1995)}}
 \end{minipage}
\end{flushright}

\minitoc

\bigskip

%§§§§§§§§§§§§§§§§§§§§§§§§§§§§§§§§§§§§§§§§§§§§§§§§§§§§§§§§§§§§§§§§§§§§§§§§§§§§§§§
\section{Introduction}
\label{sec-introduction}

As seen in the previous chapter, the \pe testbed simulates a space-based nulling interferometer, specifically modeling the concept of the \peg mission. To this end, it includes various dedicated modules: an achromatic phase shifter combined with a symmetric interferometric combiner to produce an achromatic dark fringe that extinguishes the host star, a cophasing system to stabilize this dark fringe, and a system that simulates typical \peg disturbances. In this chapter, I describe the integration of the testbed, and more specifically the successive steps implemented to evaluate the individual subsystems. Indeed, these components were developed across different laboratories \tir{the MMZ at IAS, the cophasing system at ONERA, the optomechanics at OCA, and the cleanroom environment as well as the science camera at LESIA}. The primary objective of my PhD work was to bring these subsystems together into an integrated operational system and to characterize their key performance parameters.

%§§§§§§§§§§§§§§§§§§§§§§§§§§§§§§§§§§§§§§§§§§§§§§§§§§§§§§§§§§§§§§§§§§§§§§§§§§§§§§§
\section{Intermediate Autocollimation Setup}
\label{sec-conf-autocol}

%¤¤¤¤¤¤¤¤¤¤¤¤¤¤¤¤¤¤¤¤¤¤¤¤¤¤¤¤¤¤¤¤¤¤¤¤¤¤¤¤¤¤¤¤¤¤¤¤¤¤¤¤¤¤¤¤¤¤¤¤¤¤¤¤¤¤¤¤¤¤¤¤¤¤¤¤¤¤¤
\subsection{Testbed Configuration at ONERA}
\label{sec-configuration-banc}

At the beginning of my thesis in late 2008, the testbed was not yet integrated at the Meudon Observatory. The Modified Mach-Zehnder (MMZ) was undergoing assembly at IAS\footnote{Institut d'Astrophysique Spatiale}, while other hardware components were being delivered to the Fizeau Laboratory at the Observatoire de la Côte d'Azur and to ONERA. The cophasing subsystem was first tested at ONERA using an initial prototype combiner (MMZ No. 1) fabricated at the GEPI\footnote{Galaxies, Étoiles, Physique et Instrumentation} laboratory. The experimental setup is shown in Figure~\ref{fig-mmz1}.

\begin{figure} \centering
  \FIG{0.5}{false}{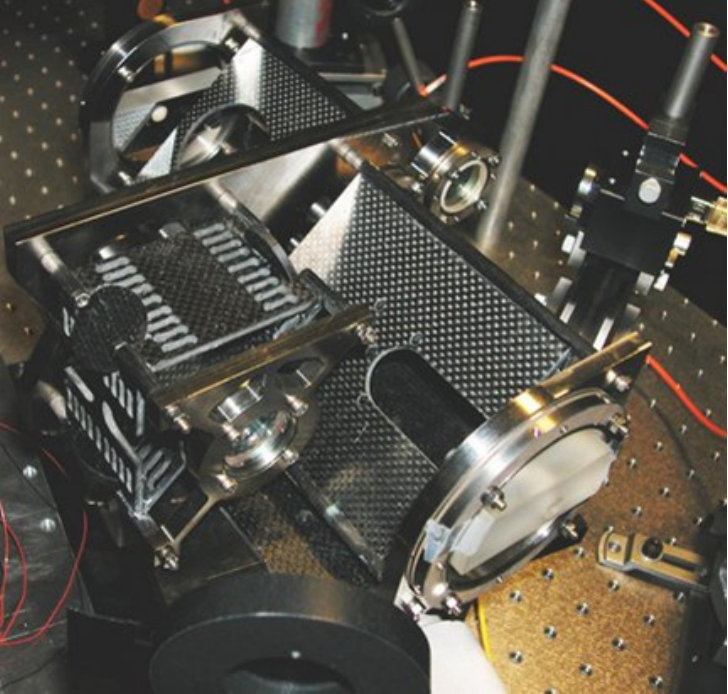}
  \caption{MMZ No. 1, used to test the cophasing subsystem at ONERA.}
  \label{fig-mmz1}
\end{figure}

To evaluate the cophasing subsystem, an \ac configuration was set up. Light was injected backward through one of the MMZ outputs; the input beam was then divided by amplitude splitting into two beams at the semi-reflective plates, forming the two interferometer arms. Two flat mirrors mounted on piston-tip/tilt actuators were positioned perpendicularly at the end of each arm to reflect light back along the incident path. Light propagated a second time through the interferometer in the forward direction. The optical layout is shown in Figure~\ref{fig-banc-onera}.

\begin{figure} \centering
  \FIG{1.}{false}{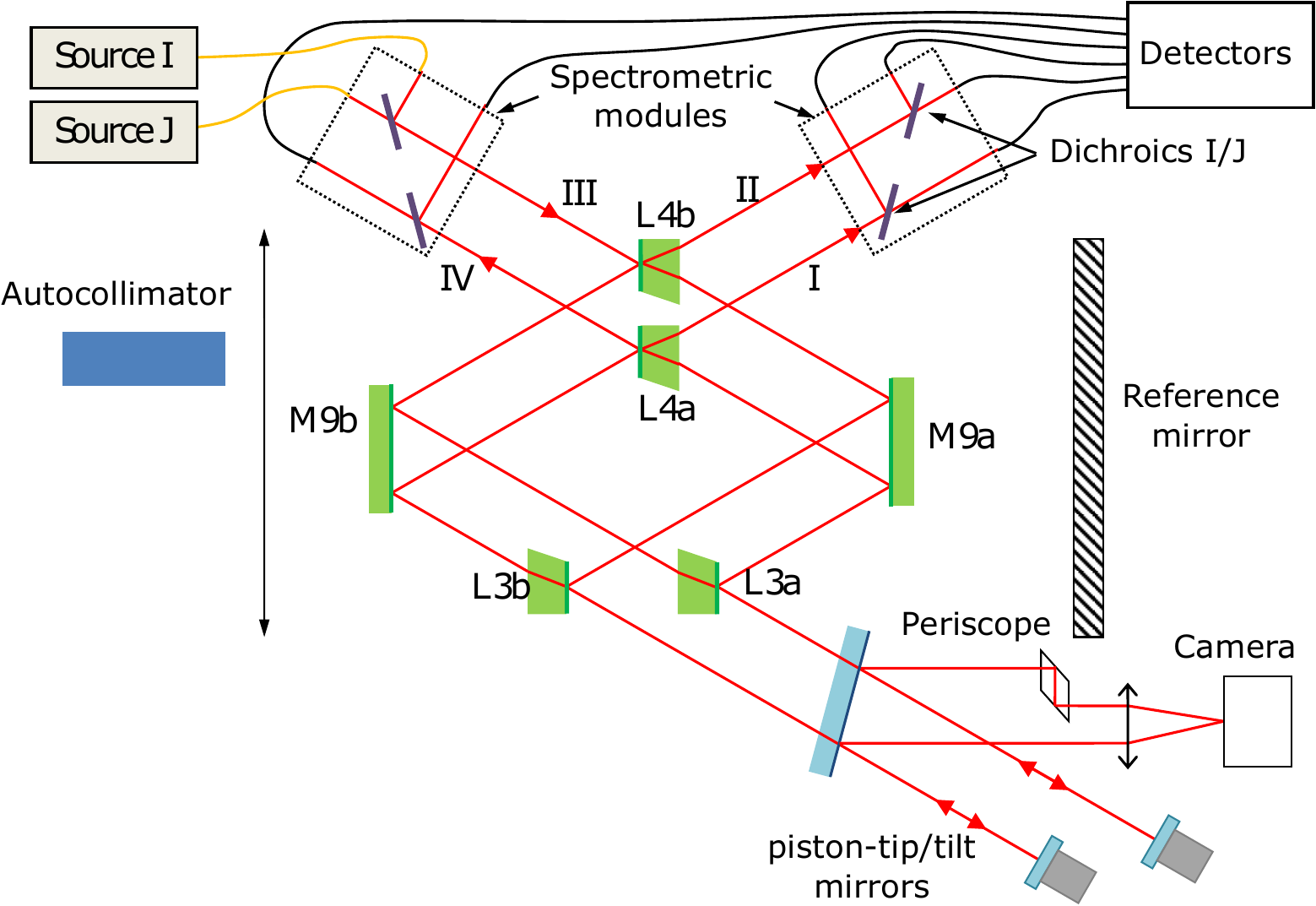}
  \caption{Optical schematic of the autocollimation setup constructed at ONERA.}
  \label{fig-banc-onera}
\end{figure}

The MMZ (comprising 2 flat mirrors and 4 semi-reflective plates) sits at the center of Figure~\ref{fig-banc-onera}. Above it are the spectrometer modules, where output beams are split into $\I$ and $\J$ bands by dichroic plates and focused into multimode optical fibers using achromatic doublets. Two light sources ($\I$ and $\J$) are injected into one of these outputs—an 830~nm laser diode for the $\I$ band, and a 1.32~\mum Superluminescent Light Emitting Diode (SLED) for the $\J$ band. The two active retro-mirrors sit at the input of the MMZ (bottom of the diagram). Upstream, a beamsplitter directs a fraction of the reflected light toward the FRAS\footnote{Field Relative Angle Sensor} subsystem for tip/tilt sensing. Because the MMZ optical surfaces must remain strictly parallel, they were aligned relative to a master flat mirror covering all optics using a translation-mounted autocollimating telescope.

This setup served exclusively to evaluate cophasing performance for the upcoming \pe testbed. Beamsplitter coatings were optimized for the $\I$ and $\J$ bands below 1.65~\mum and did not cover the mid-infrared science channel. Furthermore, nulling depth measurements could not be performed in this setup because MMZ output III (the balanced dark-fringe output) was occupied by the injection sources. Rather than quasi-ABCD sampling, the phase was demodulated via quasi-ABC sampling across 3 output phase points. While photometric demodulation could no longer separate $I_\ma$ and $I_\mb$ individually (yielding total intensity $I_\ma+I_\mb$), phase extraction remained operational.

This initial setup confirmed the core performance of the cophasing architecture, achieving residual optical path difference (OPD) stability of 0.45~nm RMS at a real-time control loop rate near 200~Hz. Further details regarding this testbed are provided in the PhD thesis of K. Houairi \cite{Houairi09b}.

%¤¤¤¤¤¤¤¤¤¤¤¤¤¤¤¤¤¤¤¤¤¤¤¤¤¤¤¤¤¤¤¤¤¤¤¤¤¤¤¤¤¤¤¤¤¤¤¤¤¤¤¤¤¤¤¤¤¤¤¤¤¤¤¤¤¤¤¤¤¤¤¤¤¤¤¤¤¤¤
\subsection{Initial Testbed Setup at Meudon}
\label{sec-premiere-configuration}

When MMZ No. 2 was installed in February 2009 at the Meudon Observatory cleanroom, we replicated ONERA's autocollimation setup with key modifications to validate the new broadband MMZ and the upgraded real-time control software. The revised layout is shown in Figure~\ref{fig-banc-autocol}.

\begin{figure} \centering
  \FIG{0.9}{false}{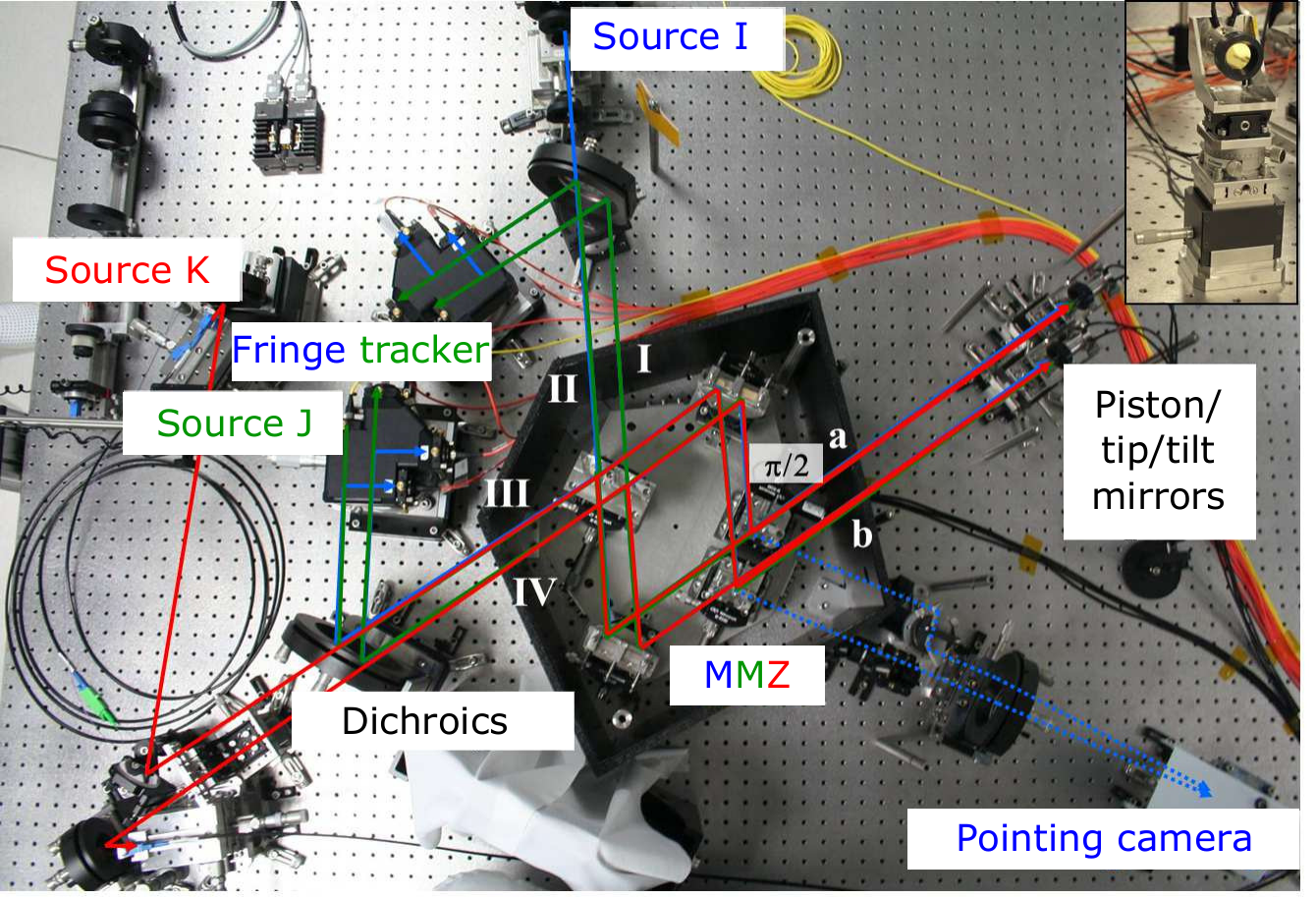}
  \caption{Autocollimation setup implemented on the \pe testbed.}
  \label{fig-banc-autocol}
\end{figure}

The principal modifications were as follows:
\begin{itemize}
\item The $\I$-band source was moved out of the spectrometer unit and injected directly into output II using a standalone beamsplitter and injection stage. Consequently, phase demodulation operated as quasi-ABC in the $\J$ band and as quasi-ABCD in the $\I$ band. Significant photometric imbalance was present in the $\I$ band because output II is asymmetric in its number of internal reflections and transmissions. Nevertheless, phase measurement remained functional: interference fringes were formed, and the matrix formalisms derived in Section~\ref{sec-expression-intensites} remained valid provided the calibrated interaction matrix was non-singular.
\item A monochromatic source in the science band ($\K$), consisting of a 2.32~\mum laser diode, was injected into MMZ output III. This guaranteed photometric balance across arms following beam splitting. Monochromatic nulling measurements were performed at output IV, which is also photometrically symmetric. Following cophasing validation, this enabled direct measurement of monochromatic nulling depth on MMZ No. 2.
\item Unlike the ONERA testbed, which lacked a dedicated science channel and used a beamsplitter to feed the FRAS, this setup avoided introducing differential phase shifts via auxiliary beamsplitters. Tip/tilt sensing was performed using internal ghost reflections escaping from the MMZ substrate edges. Figure~\ref{fig-lame} illustrates ray tracing through a trapezoidal MMZ plate.

  \begin{figure} \centering
    \FIG{0.7}{false}{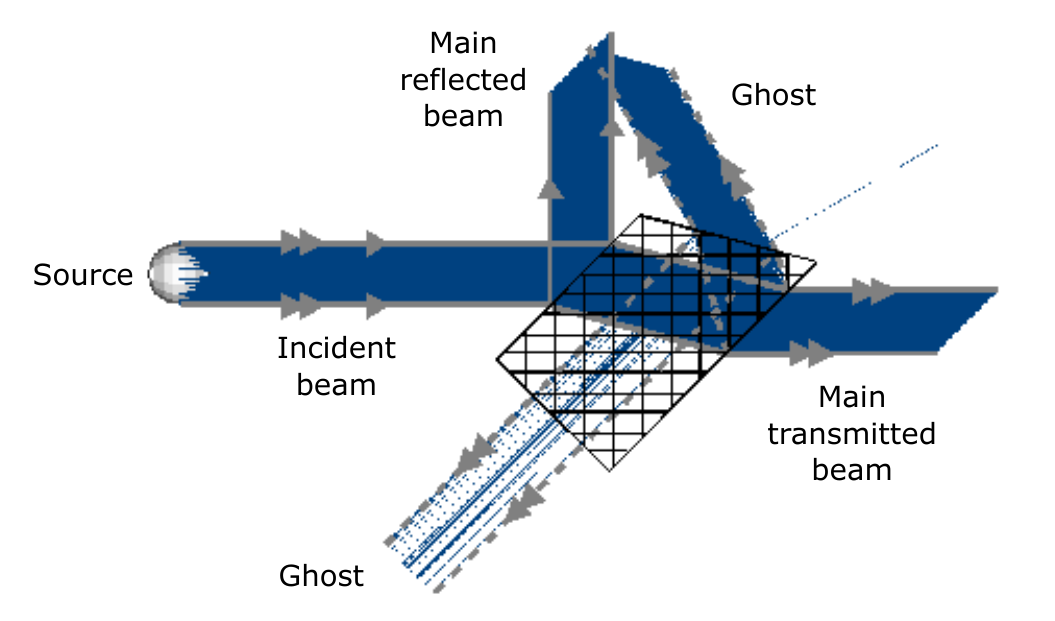}
    \caption{Secondary ghost reflections in a trapezoidal MMZ substrate.}
    \label{fig-lame}
  \end{figure}

  As shown in Figure~\ref{fig-lame}, secondary beams exit parallel to the active optical faces owing to the substrate's trapezoidal geometry and thickness. These ghost reflections exiting plates L3$\ma$ and L3$\mb$ were imaged onto a CCD camera to extract tip/tilt errors.
\end{itemize}

Using this configuration, beam cophasing was validated (Section~\ref{sec-perf-piston}), and deep monochromatic null depth measurements were achieved (Section~\ref{sec-premieres-mesures}). These tests confirmed MMZ No. 2 manufacturing quality and allowed us to isolate intrinsic combiner aberrations from upstream setup errors prior to final system integration.

%¤¤¤¤¤¤¤¤¤¤¤¤¤¤¤¤¤¤¤¤¤¤¤¤¤¤¤¤¤¤¤¤¤¤¤¤¤¤¤¤¤¤¤¤¤¤¤¤¤¤¤¤¤¤¤¤¤¤¤¤¤¤¤¤¤¤¤¤¤¤¤¤¤¤¤¤¤¤¤
\subsection{Second Intermediate Setup at Meudon}
\label{sec-deuxieme-configuration}

A second intermediate integration step involved installing the long-stroke Delay Lines (DL) between the MMZ and the active actuators in the autocollimation path to evaluate their residual mechanical noise levels. Figure~\ref{fig-lar} shows the delay lines integrated on the bench.

\begin{figure} \centering
  \FIG{0.7}{false}{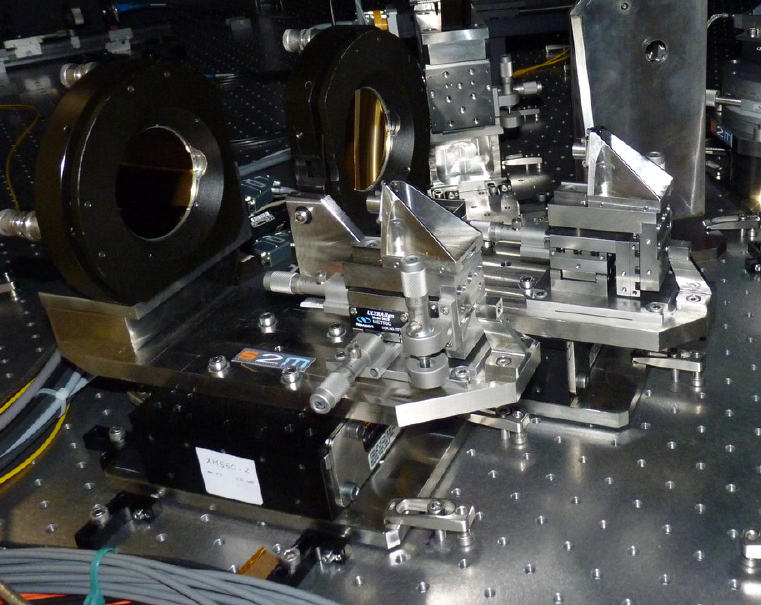}
  \caption{The two long-stroke delay lines of the \pe testbed.}
  \label{fig-lar}
\end{figure}

These delay lines consist of cat's-eye retroreflectors \tir{a parabolic mirror M7 and a focal spherical mirror M8} mounted on motorized linear translation stages. The mechanical stroke spans 50~mm, providing a maximum optical path delay range of 200~mm. They simulate large-amplitude OPD drifts representative of parabolic trajectory offsets in formation-flying satellites. One stage injects the dynamic path disturbance while the second tracks and compensates it. In closed-loop tracking, both stages translate synchronously.

When inactive, the stages are locked together with a rigid bracket (visible in Figure~\ref{fig-lar}) to eliminate mechanical jitter and drift during baseline testing.

This intermediate configuration allowed us to measure the impact of the delay lines on cophasing stability and to quantify their vibration spectrum (Section~\ref{sec-perf-piston}).

%§§§§§§§§§§§§§§§§§§§§§§§§§§§§§§§§§§§§§§§§§§§§§§§§§§§§§§§§§§§§§§§§§§§§§§§§§§§§§§§
\section{Final Testbed Architecture}
\label{sec-conf-finale}

%¤¤¤¤¤¤¤¤¤¤¤¤¤¤¤¤¤¤¤¤¤¤¤¤¤¤¤¤¤¤¤¤¤¤¤¤¤¤¤¤¤¤¤¤¤¤¤¤¤¤¤¤¤¤¤¤¤¤¤¤¤¤¤¤¤¤¤¤¤¤¤¤¤¤¤¤¤¤¤
\subsection{Description}
\label{sec-description}

\begin{figure} \centering
  \FIGA{90}{0.9}{false}{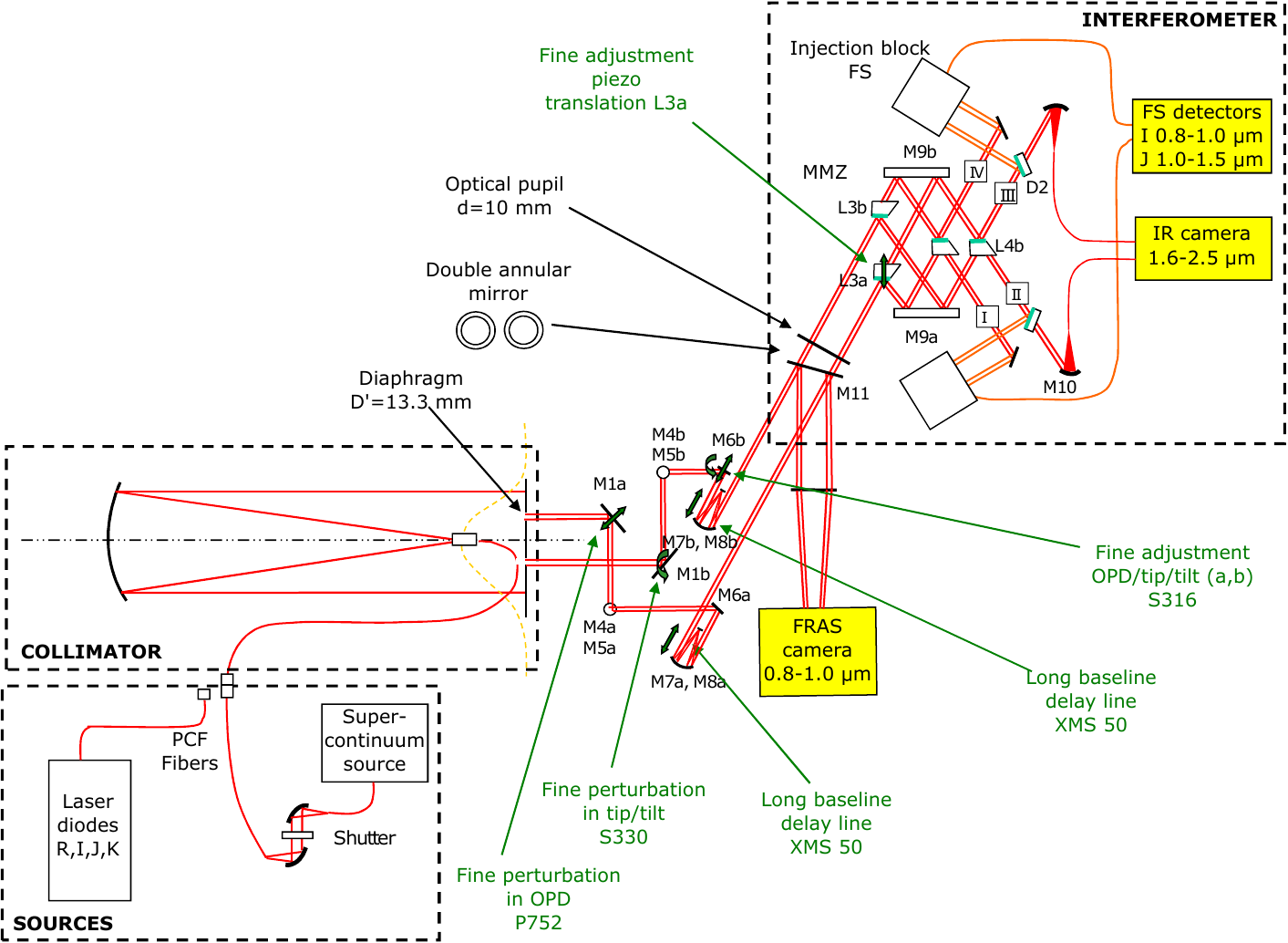}
  \caption{Final optical layout of the \pe testbed.}
  \label{fig-config-finale}
\end{figure}

The final configuration used for broadband nulling measurements (Figure~\ref{fig-config-finale}) matches the baseline design detailed in Section~\ref{sec-description-banc}, incorporating minor adjustments detailed in Section~\ref{sec-modifications-apportees}.

Following the optical propagation path, the main bench includes the following elements:
\begin{itemize}
\item \textbf{Collimator M0} (Figure~\ref{fig-M0}), an unprotected gold-coated off-axis parabolic mirror.

  \begin{figure} \centering
    \FIG{0.7}{false}{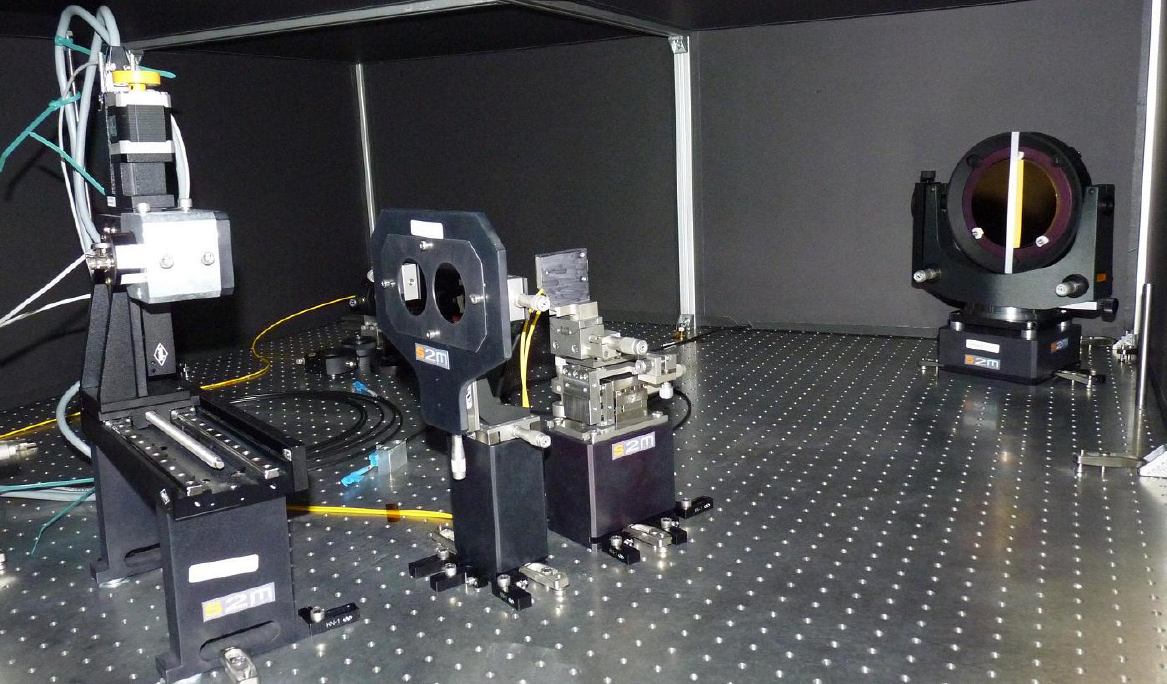}
    \caption{Collimator M0 with beam masks.}
    \label{fig-M0}
  \end{figure}

  An optical fiber output delivering source light is placed at the focus of M0. The collimated beam is divided into two sub-pupils by an aperture mask (Figure~\ref{fig-M0}). In the baseline architecture, sub-pupils spanned 40~mm in diameter across a $b=50$~mm baseline and were compressed threefold by off-axis afocal telescopes to simulate satellite collector optics. In the final layout, these afocal compressors were omitted to minimize optical alignment complexity and reduce thermal/vibration sources. Sub-pupil diameters were scaled down to 13~mm across the same 50~mm baseline.

\item \textbf{Siderostats M1$\boldsymbol{\ma}$ and M1$\boldsymbol{\mb}$} (Figure~\ref{fig-M1}), simulating the steerable collector mirrors of the \peg mission.

  \begin{figure} \centering
    \subfloat[Mirror M1$\ma$ mounted on a piezo piston stage.]{
      \label{fig-M1a} \FIGH{8}{false}{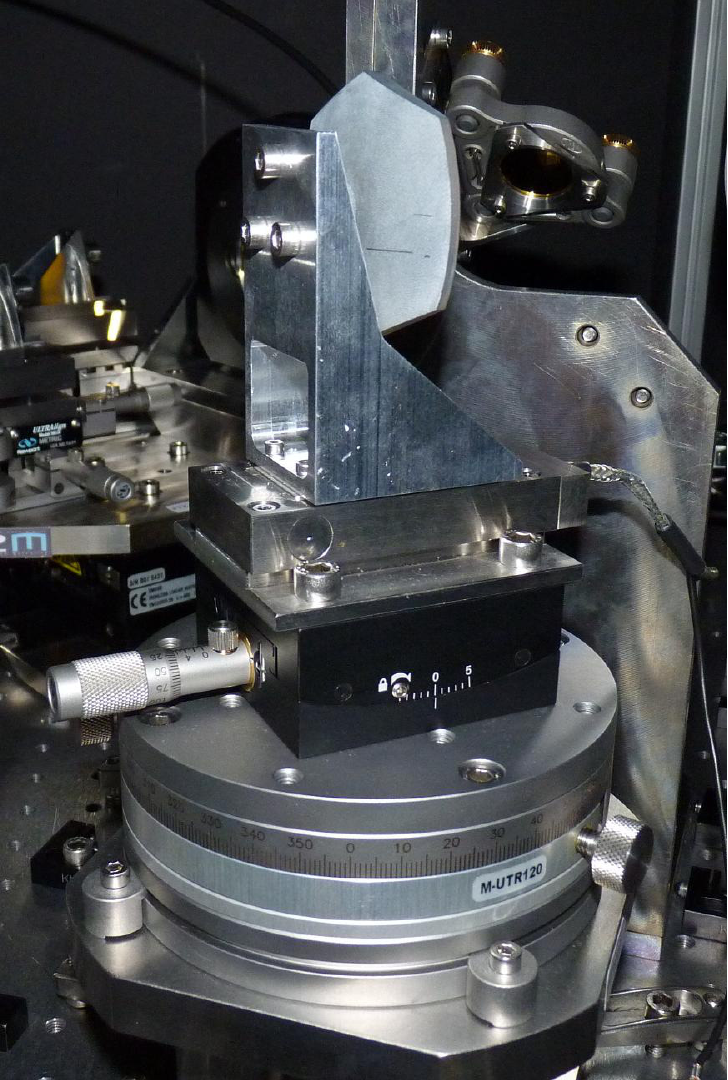}}
    \hspace{10pt}\subfloat[Mirror M1$\mb$ mounted on a piezo tip/tilt stage.]{\label{fig-M1b} \FIGH{8}{false}{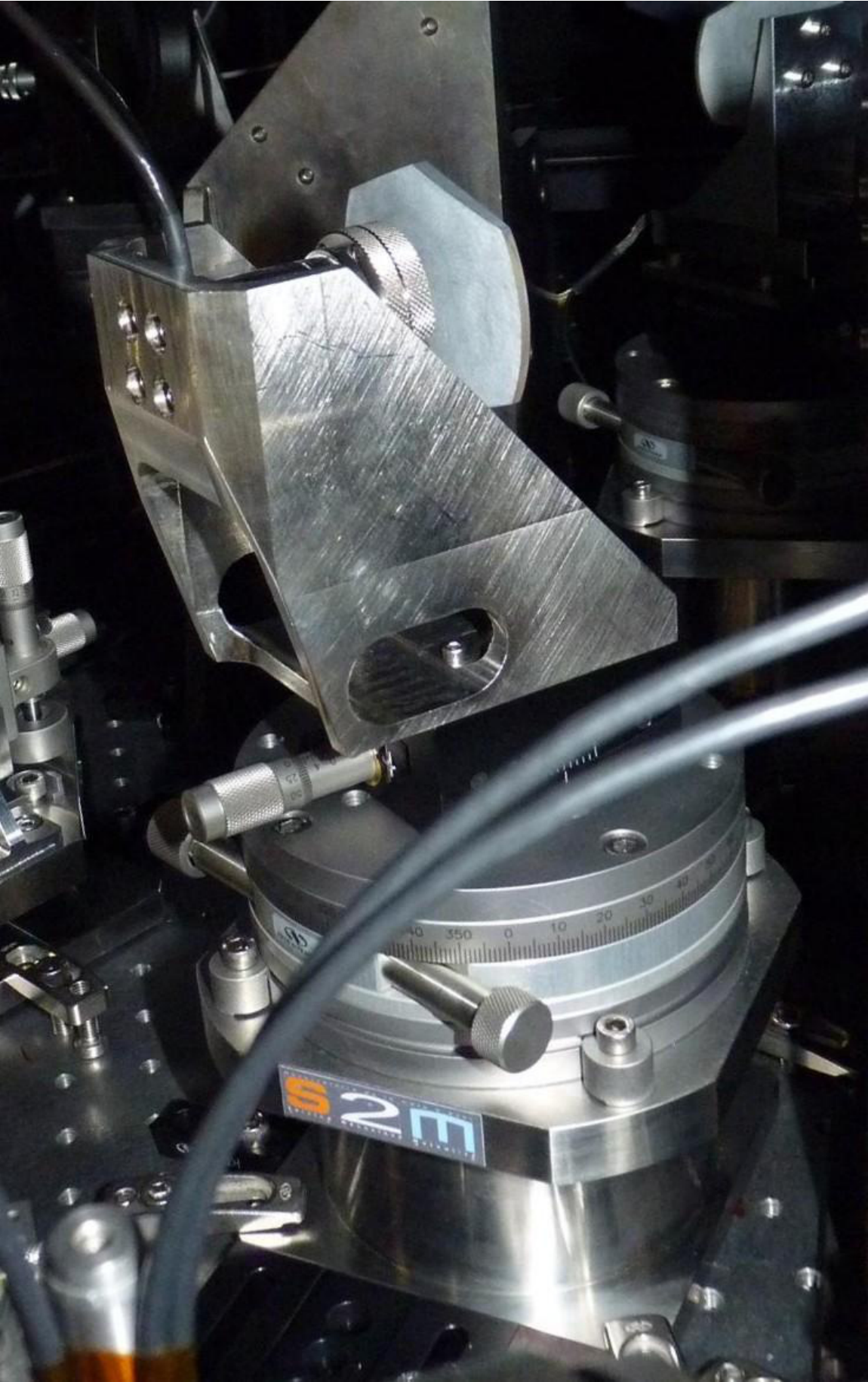}}
    \caption{Siderostat mirrors M1$\ma$ and M1$\mb$ integrated on the bench.}
    \label{fig-M1}
  \end{figure}
	
  These mirrors are mounted on piezoelectric stages \tir{piston for M1$\ma$ and tip/tilt for M1$\mb$}. Calibrated disturbances are driven through these stages to simulate siderostat jitter during flight.
	
\item \textbf{Periscopes M4$\boldsymbol{\ma}$--M5$\boldsymbol{\ma}$ and M4$\boldsymbol{\mb}$--M5$\boldsymbol{\mb}$} (Figure~\ref{fig-APS}). Mirrors M4 pair with mirrors M1 to form the geometric $\pi$ Achromatic Phase Shifter (APS). Mirrors M5 fold the beams back parallel to the optical table plane.
	
  \begin{figure} \centering
    \FIG{0.7}{false}{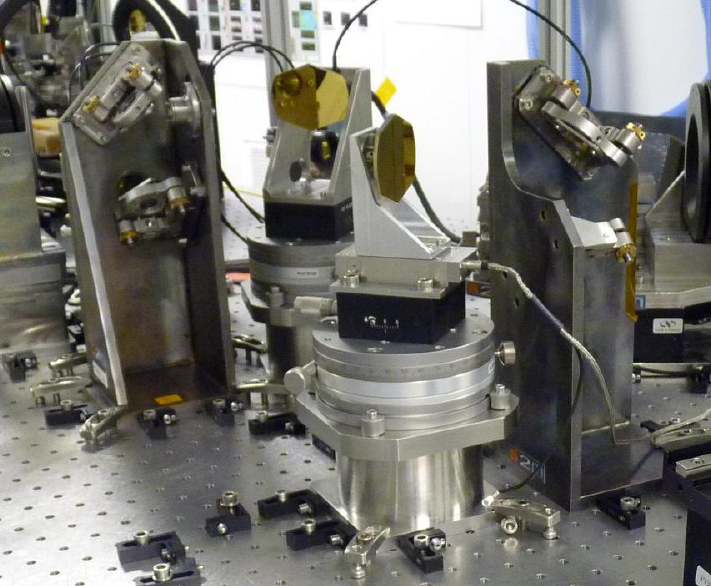}
    \caption[Geometric $\pi$ achromatic phase shifter.]{Geometric $\pi$ achromatic phase shifter formed by the M1 mirrors and periscopes.}
    \label{fig-APS}
  \end{figure}

  Downstream of the periscopes sit active piston-tip/tilt mirrors M6, oriented at a 30\degree incidence angle relative to the optical path. The long-stroke delay lines follow, handling dynamic path disturbance injection and tracking.

\item \textbf{Annular Mirror M11} (Figure~\ref{fig-M11}), reflecting outer beam annuli toward the tip/tilt sensor.
	
  \begin{figure} \centering
    \FIG{0.7}{false}{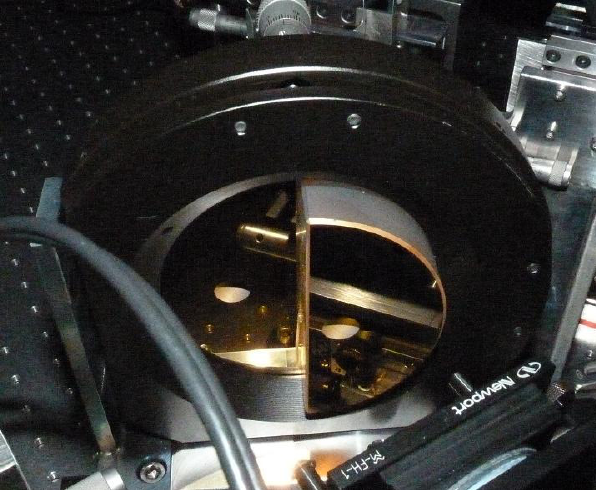}
    \caption{Annular pick-off mirror M11 for the tip/tilt sensor.}
    \label{fig-M11}
  \end{figure}

  Mirror M11 features a stepped geometry to narrow the beam centerline distance, allowing a single focusing lens to image both beams onto a CCD camera. The mirror contains two 10~mm clear apertures that transmit central beam cores to the combiner while picking off outer beam perimeters for tip/tilt sensing.

\end{itemize}

Downstream of mirror M11, the optics match the autocollimation configuration: beams recombine within the MMZ, phase is sampled across the 4 outputs by spectrometer modules, and the science signal is coupled into a single-mode fiber feeding the infrared detector.

A dispersion bi-prism positioned upstream of the science camera spreads incoming light, enabling null depth measurements across 9 spectral channels spanning 1.65~\mum to 2.45~\mum. The science detection channel is described in Section~\ref{sec-description-voie-scientifique}.

%¤¤¤¤¤¤¤¤¤¤¤¤¤¤¤¤¤¤¤¤¤¤¤¤¤¤¤¤¤¤¤¤¤¤¤¤¤¤¤¤¤¤¤¤¤¤¤¤¤¤¤¤¤¤¤¤¤¤¤¤¤¤¤¤¤¤¤¤¤¤¤¤¤¤¤¤¤¤¤
\subsection{Alignment Procedure}
\label{sec-methode-alignement}

%-------------------------------------------------------------------------------
\subsubsection{Tolerances}
\label{sec-tolerances}

A collimated light beam possesses 6 degrees of freedom adjustable via reflection: piston (tuned at the end of alignment), beam rotation (governing polarization field alignment, Section~\ref{sec-effets-polarisation}), two pointing angles (tip and tilt), and two transverse positions ($x$ and $y$, with $z$ along the propagation axis). Each planar reflection requires aligning beam pointing angle and centration. Collimation quality must also be maintained across parabolic surfaces.

Alignment tolerances were derived from Zemax ray-tracing models of \pe to meet null depth targets. Polarization constraints drove these tolerances, as off-axis reflections alter polarization states with incidence angle. Because the two interferometer arms are symmetric, differential alignment errors between arms $\ma$ and $\mb$ dominate null depth degradation. Table~\ref{tab-tolerences} summarizes the tolerances derived from optical modeling for arm alignment angles. Here, $g$ denotes the local gravity vector, $\perp$M$_i$ is the normal to mirror M$_i$, and $u\ \forall\ v$ represents the bisector between vectors $u$ and $v$.

\begin{table} \centering
  \caption{Alignment tolerances established by Zemax simulations of the testbed.}
  \medskip
  \renewcommand{\arraystretch}{1.25}
  \begin{tabular}[table]{cccc}
    \hline \hline Element & Reference & Type & Tolerance \\
    \hline
    Fiber at M0 focus & M0 focus & $x$ and $y$ & $\pm 36$~\mum \\
    $\perp$M0 & $\perp$M$_{\mref,2}$ & tip/tilt & $\pm 10$~arcsec \\
    $\perp$M1$\ma$ and $\perp$M1$\mb$ & $\perp$M$_{\mref,1}\ \forall\ \perp$M$_{\mref,2}$ & tip/tilt & $\pm 10$~arcsec \\
    $\perp$M4$\ma$ and $\perp$M4$\mb$ & $\perp$M$_{\mref,1}\ \forall\ g$ & tip/tilt & $\pm 20$~arcsec \\
    $\perp$M5$\ma$ and $\perp$M5$\mb$ & $\perp$M$_{\mref,2}\ \forall\ g$ & tip/tilt & $\pm 20$~arcsec \\
    DL $\ma$ and DL $\mb$ & MMZ input axes & tip/tilt & $\pm 1$~arcmin \\
    M8$\ma$ and M8$\mb$ & M7$\ma$ and M7$\mb$ focus & $x$, $y$, and $z$ & $\pm 0.1$~mm \\
    MMZ optics & M$_{\mref,1}$ & tip/tilt & $\pm 10$~arcsec \\
    \hline \hline
  \end{tabular}
  \label{tab-tolerences}
\end{table}

Reference mirrors M$_{\mref,1}$ and M$_{\mref,2}$ in Table~\ref{tab-tolerences} are flat optical references used during assembly. As shown in Figure~\ref{fig-miroirs-ref}, M$_{\mref,1}$ is aligned parallel to the MMZ plates to guide combiner alignment. Mirror M$_{\mref,2}$ sits perpendicular to M$_{\mref,1}$, facing the collimator and aligned perpendicular to its optical axis and to the achromatic phase shifter outputs.

\begin{figure} \centering
  \FIG{0.9}{false}{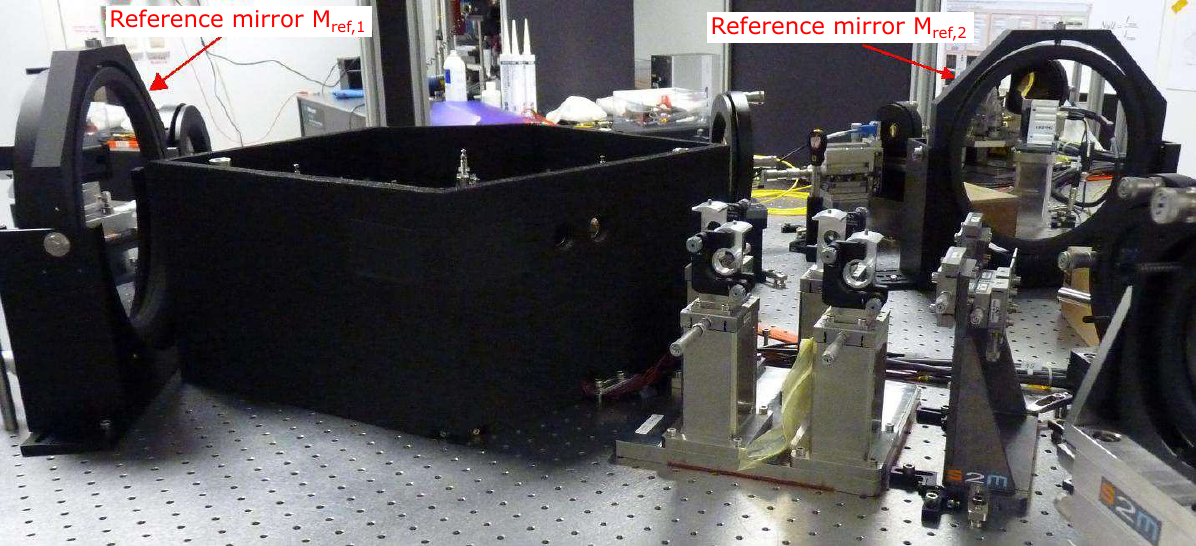}
  \caption{Alignment reference mirrors.}
  \label{fig-miroirs-ref}
\end{figure}

In addition to angular tolerances, several spatial constraints apply:
\begin{itemize}
\item Beams must be centered on piston-tip/tilt correction mirrors M6 to prevent beam walk during angular corrections and to minimize coupling between piston and tip/tilt modes.
\item Beams must be centered on the apertures of annular mirror M11, as aperture spacing is fixed. Similarly, the spacing between the collimator mask apertures is fixed, bounding lateral alignment margins.
\item Beams must avoid vignetting along internal mounts, particularly inside the MMZ, leaving a transverse positioning margin of $\sim 1$~mm in $x$ and $y$.
\end{itemize}

%-------------------------------------------------------------------------------
\subsubsection{Alignment Equipment}
\label{sec-materiel}

Meeting the requirements in Table~\ref{tab-tolerences} requires high-precision optical alignment tools. Alignment is established relative to reference flats and parallel axes using an industrial theodolite (Figure~\ref{fig-theo}).

\begin{figure} \centering
  \FIG{0.4}{false}{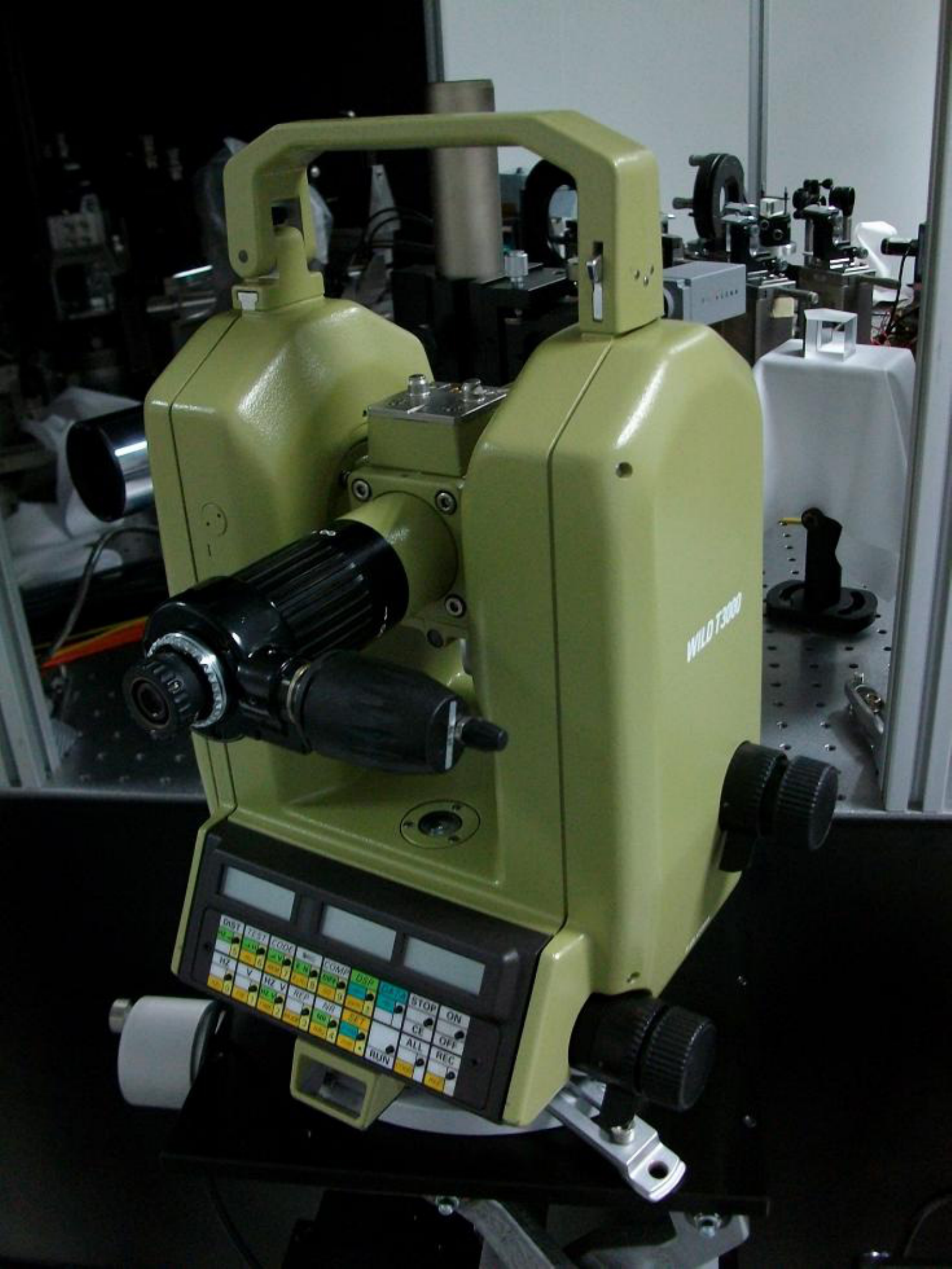}
  \caption{Theodolite used during optical alignment.}
  \label{fig-theo}
\end{figure}

The theodolite combines an autocollimating telescope with precision azimuth and elevation encoders calibrated to local gravity, enabling angular measurements to $\sim 1$~arcsec precision. Local gravity serves as the primary vertical reference rather than the optical table surface.

To align separated components or retro-reflected paths, precision reference prisms are used (Figure~\ref{fig-optiques-align}).

\begin{figure} \centering
  \subfloat[Penta-prism.]{\label{fig-optiques-align1}
    \FIGH{5}{false}{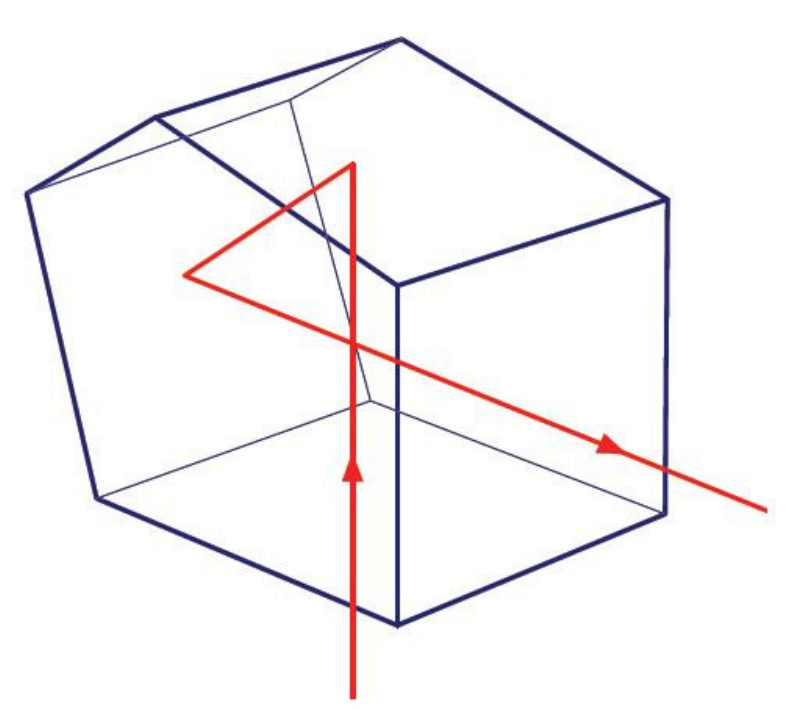}}
  \hspace{10pt}\subfloat[Hollow retroreflector.]{\label{fig-optiques-align2}
    \FIGH{5}{false}{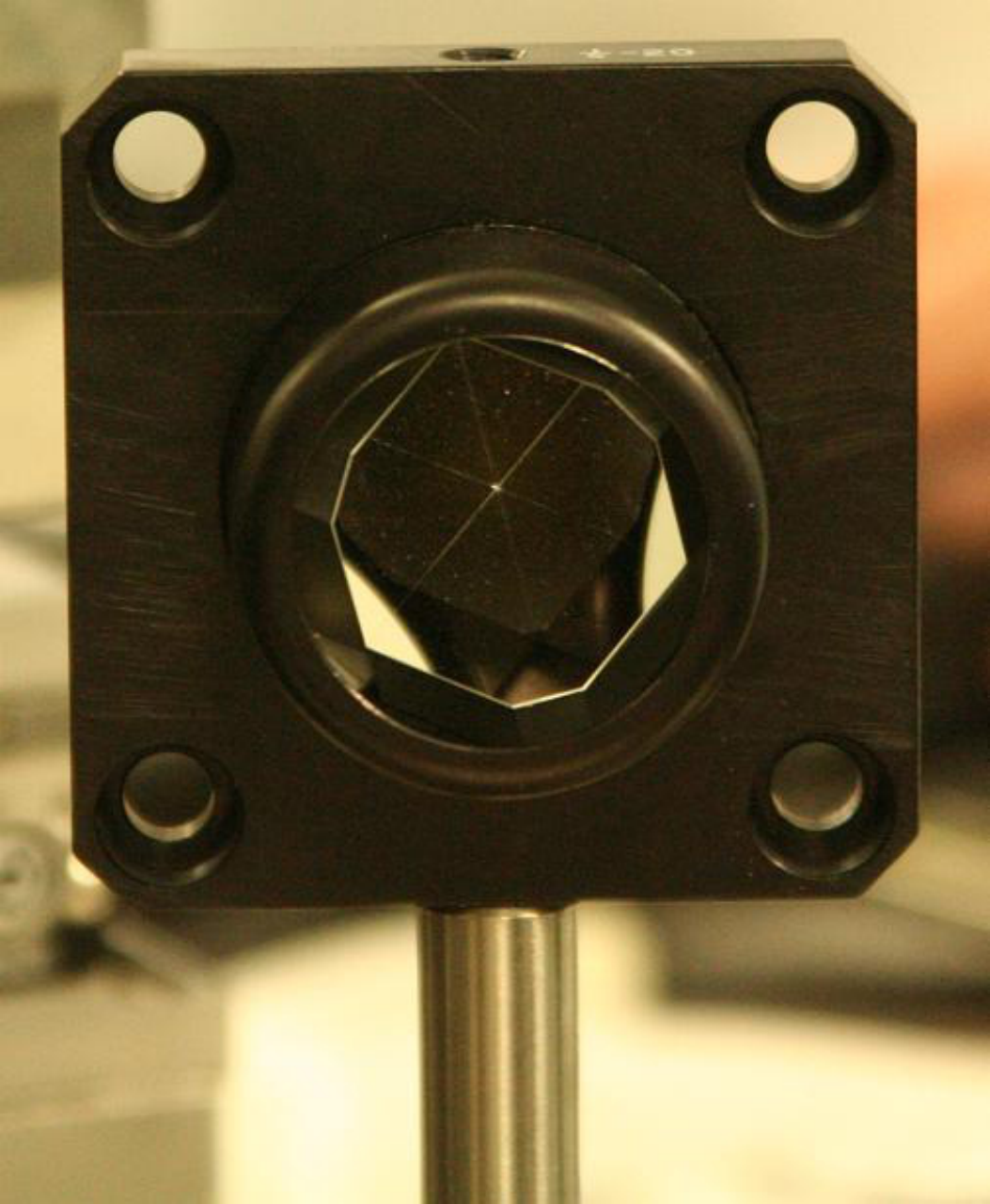}}
  \caption{Optical references used during alignment.}
  \label{fig-optiques-align}
\end{figure}

A penta-prism deviates incoming beams by 90\degree regardless of orientation angle (Figure~\ref{fig-optiques-align1}). The calibrated penta-prism maintains a $90\degree \pm 1$~arcsec deviation. Hollow retroreflectors (Figure~\ref{fig-optiques-align2}) return beams parallel to their incident paths to within 1~arcsec.

%-------------------------------------------------------------------------------
\subsubsection{Alignment Sequence}
\label{sec-etapes}

The complete optical alignment procedure is documented in a dedicated 70-page manual; the primary steps are summarized below (refer to Figure~\ref{fig-config-finale}).

Local gravity defines the master vertical reference, establishing an optical plane perpendicular to gravity.

Master reference mirror M$_{\mref,1}$ was placed on the optical bench according to CAD coordinates and aligned parallel to gravity. Reference mirror M$_{\mref,2}$ was aligned parallel to gravity and perpendicular to M$_{\mref,1}$.

First, all optical surfaces inside the MMZ were aligned parallel to one another using reference mirror M$_{\mref,1}$.

Injecting light into MMZ output III generates two parallel reverse-propagating beams (arms $\ma$ and $\mb$). Source injection was aligned so these output beams lie in a plane perpendicular to gravity, oriented at 30\degree relative to the normal of reference mirror M$_{\mref,1}$. Theodolite azimuth angles defined this baseline orientation, establishing the tracking axis for the delay lines (Figure~\ref{fig-lar}).

Mirrors M6 (set to mid-stroke) were adjusted so reverse beams hit reference mirror M$_{\mref,2}$ at normal incidence. Periscope mirrors M5 were adjusted to align reflected beams parallel to gravity using the penta-prism to redirect vertical paths into the horizontal plane for theodolite measurement. Periscope mirrors M4 were then adjusted until both reflected beams lay in the horizontal plane perpendicular to M$_{\mref,2}$.

A calibration source was placed at the focus of collimator M0 and adjusted until the collimated beam output was perpendicular to M$_{\mref,2}$.

Finally, siderostat mirrors M1 were adjusted until collimated forward beams and reverse beams overlapped. Beam centration across mounts was verified against CAD models throughout the process. During alignment, active M1 actuators were powered down in their mechanical rest positions.

This CAD-guided alignment established initial zero OPD to within a few millimeters. To reduce zero-OPD offset within the stroke range of the active M6 mirrors (<100~\mum), one delay line stage was manually translated while monitoring interference fringes. Coarse path equalization was performed using a long-coherence source ($\I$-band laser diode, then $\J$-band SLED), followed by fine equalization using broadband white light. Residual path difference was reduced manually to under 1~\mum.

Using the theodolite and precision prisms, angular alignment errors were constrained below 2~arcsec—well within the 10~arcsec tolerance requirement. This procedure does not directly measure differential polarization rotation between arms, however. Residual tip and tilt errors compensated downstream by M6 mirrors can rotate polarization orientation along propagation axes, adding a polarization leakage term to the null depth (Section~\ref{sec--etude}).

%-------------------------------------------------------------------------------
\subsubsection{Spectrometer Alignment Verification}
\label{sec-align-bloc}

The spectrometer modules housing the fringe sensor detectors were aligned in $x, y, z$ translation and tip/tilt to maximize fiber coupling efficiency.

Coupling efficiency was verified by scanning focal spots across fiber cores using steering mirrors M6 (detailed in Section~\ref{sec-procedure-recherche}). The resulting coupling map represents the spatial coupling profile of the multimode fiber convolved with the local beam Point Spread Function (PSF). Alignment maps are shown in Figure~\ref{fig-fibres-al}.

\begin{figure} \centering
  \subfloat[Sub-optimal alignment.]{\label{fig-fibres-al1}
    \FIG{0.7}{false}{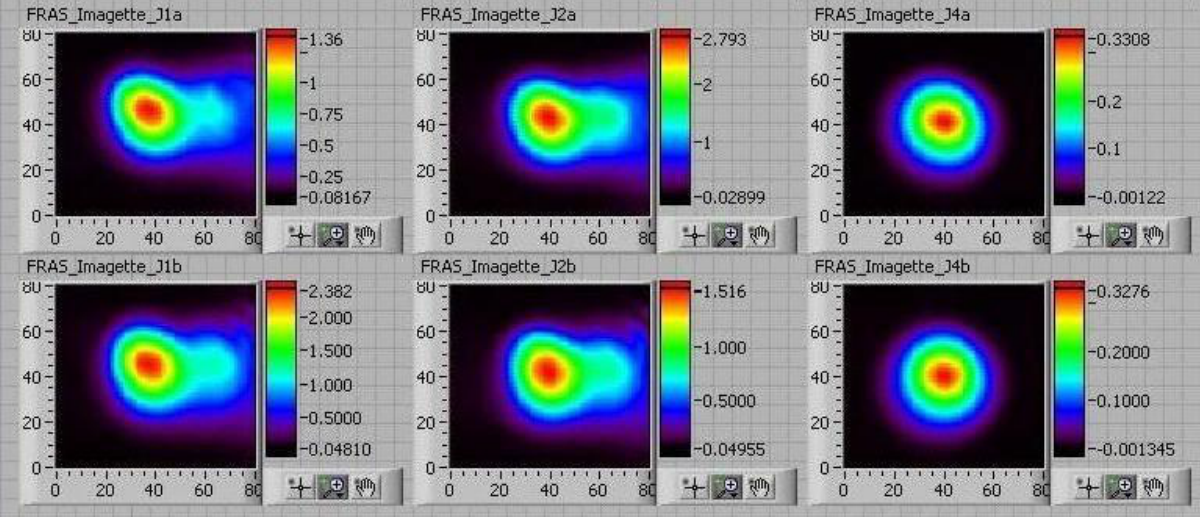}} \\
  \subfloat[Corrected alignment.]{\label{fig-fibres-al2}
    \FIG{0.7}{false}{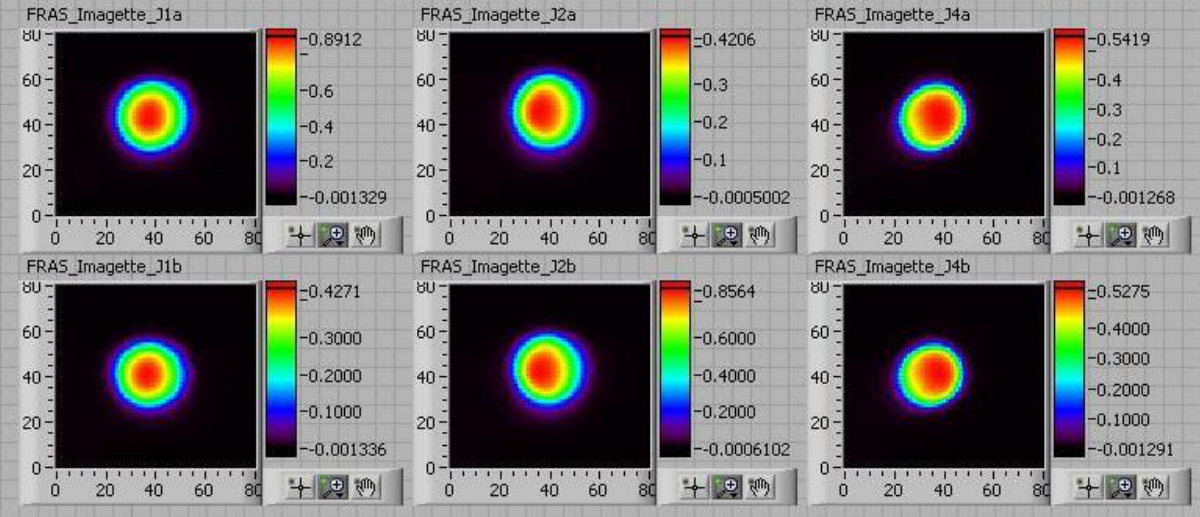}}
  \caption[Optimization of spectrometer module alignment.]{Optimization of spectrometer module alignment across outputs I, II, and IV ($\J$ band) for arms $\ma$ and $\mb$.}
  \label{fig-fibres-al}
\end{figure}

Figure~\ref{fig-fibres-al} shows coupled flux maps across outputs I, II, and IV ($\J$ band) for both arms under two alignment states. In Figure~\ref{fig-fibres-al1}, coupling profiles on outputs I and II show asymmetric distortion compared to output IV due to beam vignetting.

After adjusting spectrometer positioning and lens focus, coupling maps improved (Figure~\ref{fig-fibres-al2}). Multimode fiber coupling profiles approach top-hat functions convolved with the optical PSF; well-focused, un-vignetted PSFs yield symmetric, uniform coupling maps.

Section~\ref{sec-fourier-FS} details the impact of spectrometer alignment on fringe sensor accuracy and presents an automated alignment optimization method.

%§§§§§§§§§§§§§§§§§§§§§§§§§§§§§§§§§§§§§§§§§§§§§§§§§§§§§§§§§§§§§§§§§§§§§§§§§§§§§§§
\section{Baseline Modifications}
\label{sec-modifications-apportees}

%¤¤¤¤¤¤¤¤¤¤¤¤¤¤¤¤¤¤¤¤¤¤¤¤¤¤¤¤¤¤¤¤¤¤¤¤¤¤¤¤¤¤¤¤¤¤¤¤¤¤¤¤¤¤¤¤¤¤¤¤¤¤¤¤¤¤¤¤¤¤¤¤¤¤¤¤¤¤¤
\subsection{Source Module Redesign}
\label{sec-modification-module}

Transitioning to broadband operations revealed a limitation in the original source module design (Section~\ref{sec-identification-probleme}). We implemented a replacement source and custom injection stage (Section~\ref{sec-realisation-nouveau}) while building a modular injection bench to maintain compatibility with monochromatic testing (Section~\ref{sec-nouveau-banc}).

%-------------------------------------------------------------------------------
\subsubsection{Issue Identification}
\label{sec-identification-probleme}

In the initial design, cophasing was driven by two monochromatic sources ($\I$ and $\J$), while science measurements used a 2.32~\mum laser diode or a broadband blackbody source ([1.65–2.5]~\mum). These sources were delivered via individual optical fibers grouped together at the focus of collimator M0. Fibers were mounted side-by-side with touching cladding (125~\mum center-to-center spacing), placing the science fiber between the $\I$ and $\J$ metrology fibers (Figure~\ref{fig-F0}).

\begin{figure} \centering
  \subfloat[Bench mounting.]{\label{fig-F01}
    \FIGH{6.5}{false}{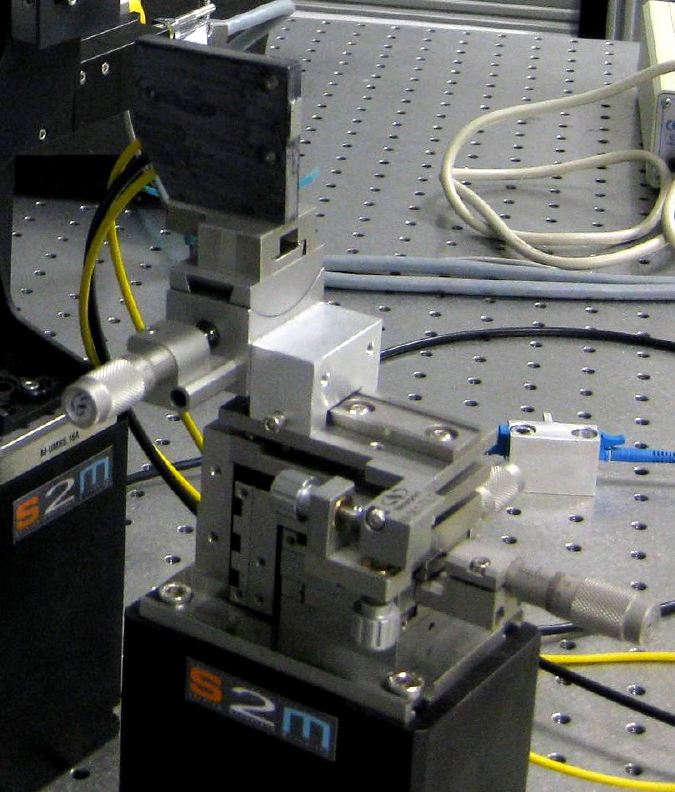}}
  \hspace{10pt}\subfloat[Microscopic view of the three fiber cores, with red light injected into $\I$.]{\label{fig-F02}
    \FIGH{6.5}{false}{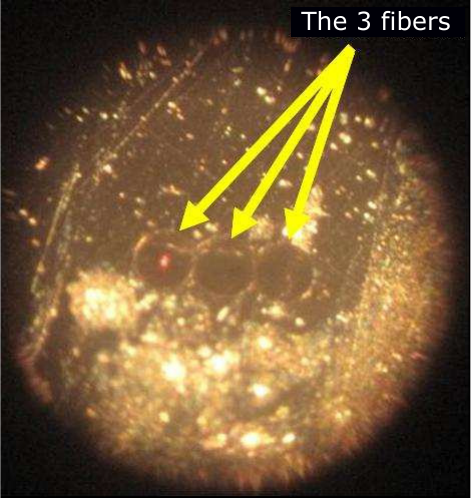}}
  \caption{Fiber holder positioning three source fibers at the M0 collimator focus.}
  \label{fig-F0}
\end{figure}

This arrangement placed three displaced point sources at the focus of parabolic collimator M0. Off-axis sources passing through the perioscopic APS generate anti-symmetric beam tilts in the pupil, introducing tilt fringes that degrade fringe visibility in the metrology channels. Figure~\ref{fig-frangesIJK} shows interference fringes recorded in the metrology channels during OPD ramps.

\begin{figure} \centering
  \subfloat[Fringes in channel $\I$.]{\label{fig-frangesIK}
    \FIG{0.49}{false}{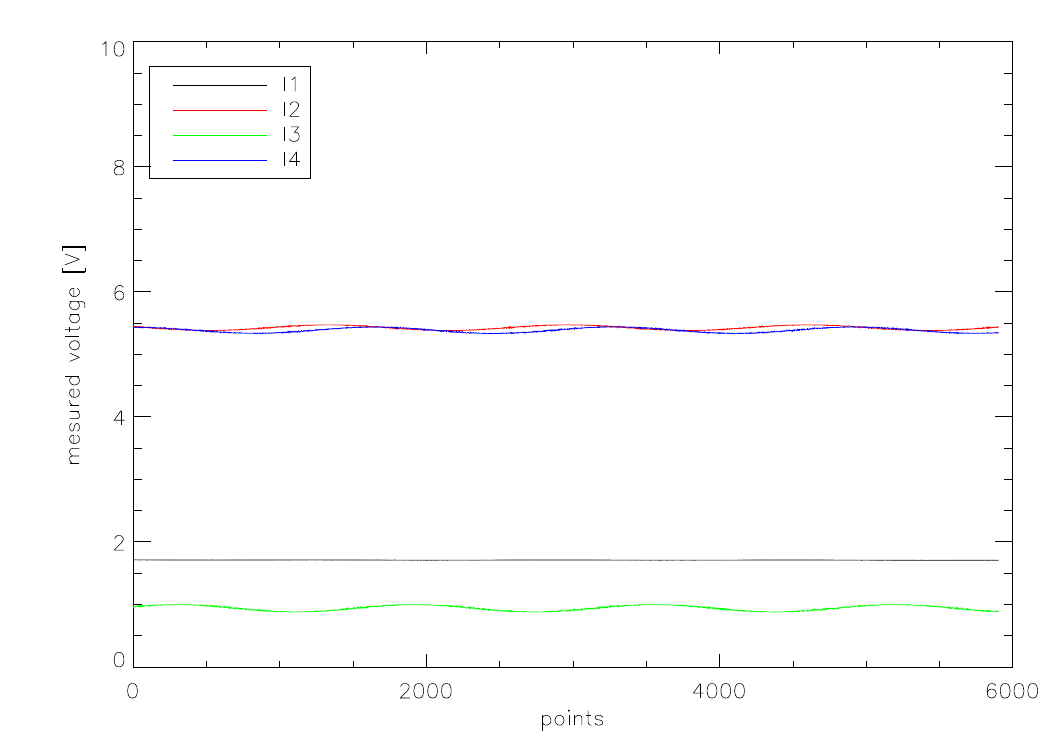}}
  \hfill\subfloat[Fringes in channel $\J$.]{\label{fig-frangesJK}
    \FIG{0.49}{false}{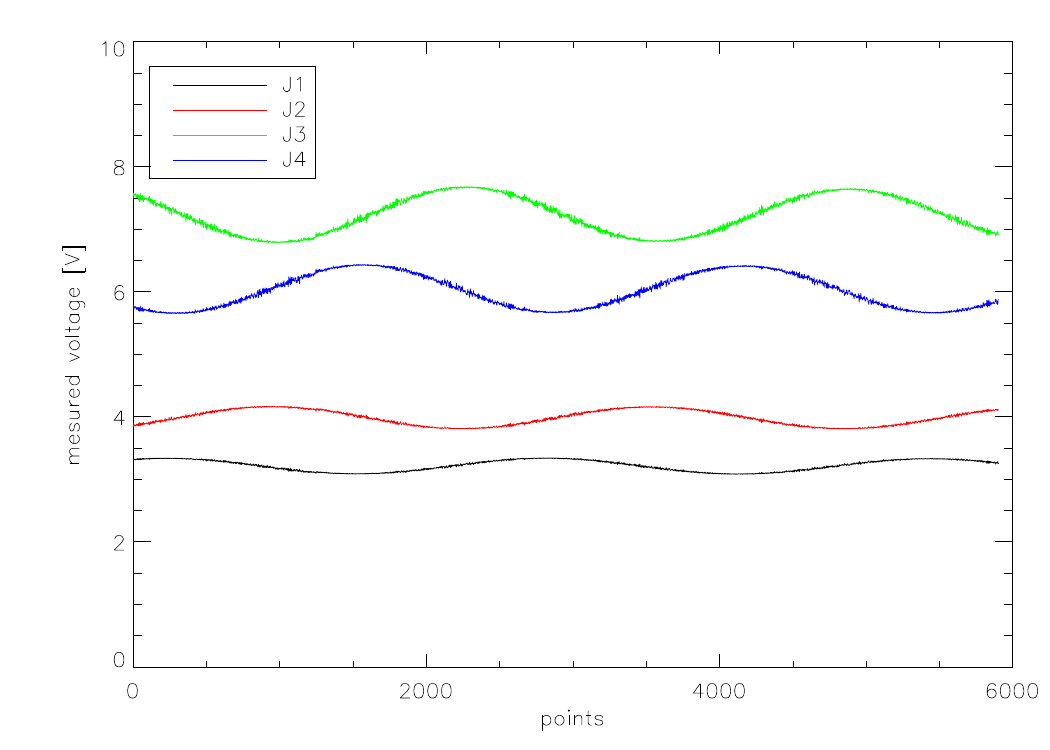}}
  \caption{Metrology fringe signals obtained with the science fiber positioned at the M0 focus.}
  \label{fig-frangesIJK}
\end{figure}

Figure~\ref{fig-frangesIK} exhibits near-zero fringe visibility in the $\I$ band, while visibility in the $\J$ band remains low (Figure~\ref{fig-frangesJK}). Figure~\ref{fig-contraste} plots modeled fringe contrast versus off-axis fiber displacement at M0.

\begin{figure} \centering
  \FIG{0.7}{false}{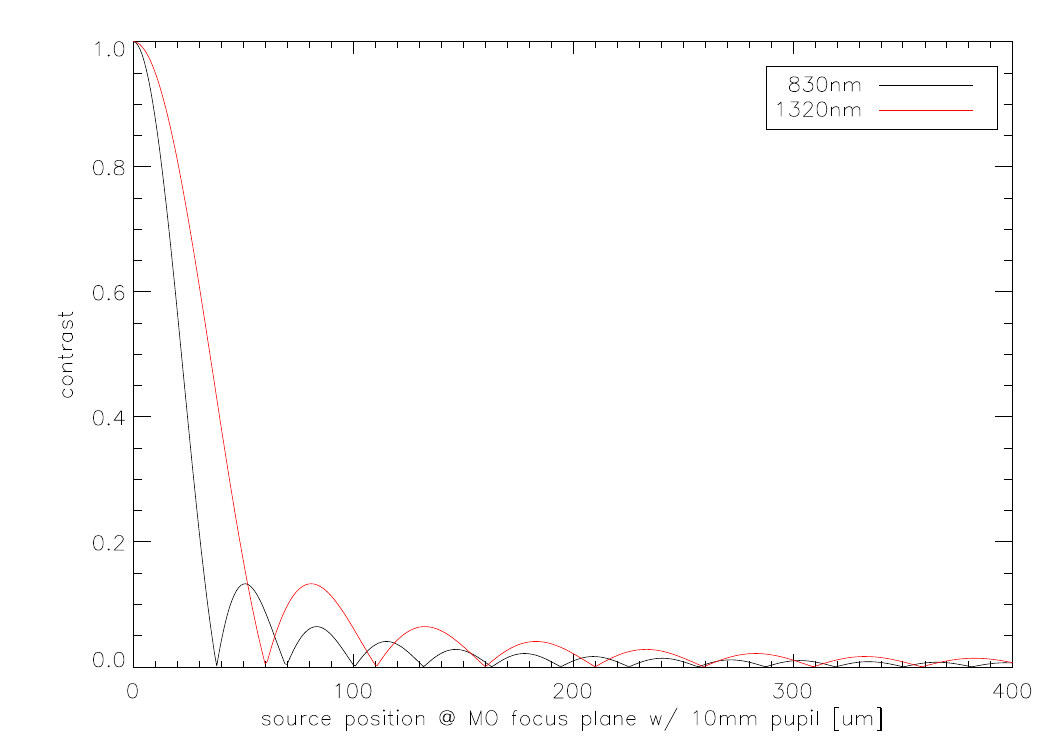}
  \caption{Modeled fringe contrast versus off-axis fiber displacement at the M0 focus.}
  \label{fig-contraste}
\end{figure}

At 125~\mum offset, $\I$-band contrast sits near a zero-crossing, while $\J$-band contrast reaches a local peak of $\sim 0.05$, matching experimental data in Figure~\ref{fig-frangesIJK}. With the 3$\times$ beam compressors installed, angular offsets scale threefold (equivalent to a 375~\mum source offset), reducing contrast to near zero across both metrology bands. Maintaining adequate fringe contrast requires fiber core separations below 30~\mum, which cannot be achieved by bundling individual fibers.

This necessitated a solution where science light and cophasing light emerge from a single shared fiber core. Because the broadband science source emits light across $\I$, $\J$, and $\K$ bands, a single source can feed both metrology and science channels—matching the \peg flight concept of using target star photons for cophasing. Fringe signals recorded using a broad-spectrum Xenon lamp are shown in Figure~\ref{fig-frangesXeIJ}.

\begin{figure} \centering
  \subfloat[Channel $\I$ fringes.]{\label{fig-frangesXeI}
    \FIG{0.49}{false}{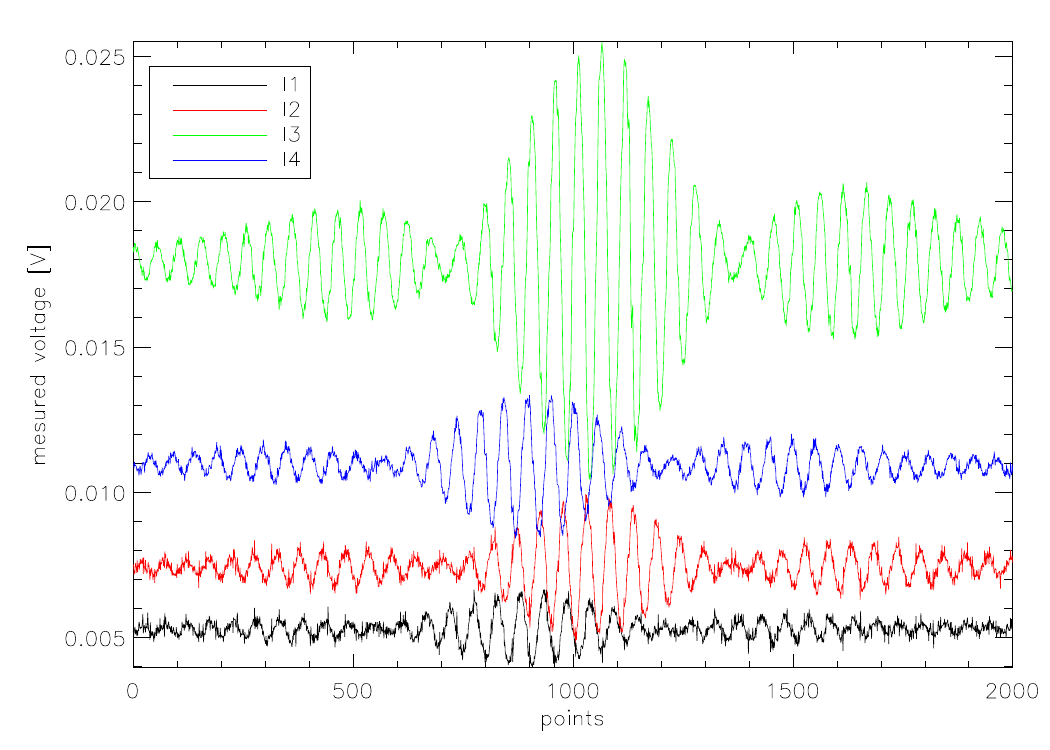}}
  \hfill\subfloat[Channel $\J$ fringes.]{\label{fig-frangesXeJ}
    \FIG{0.49}{false}{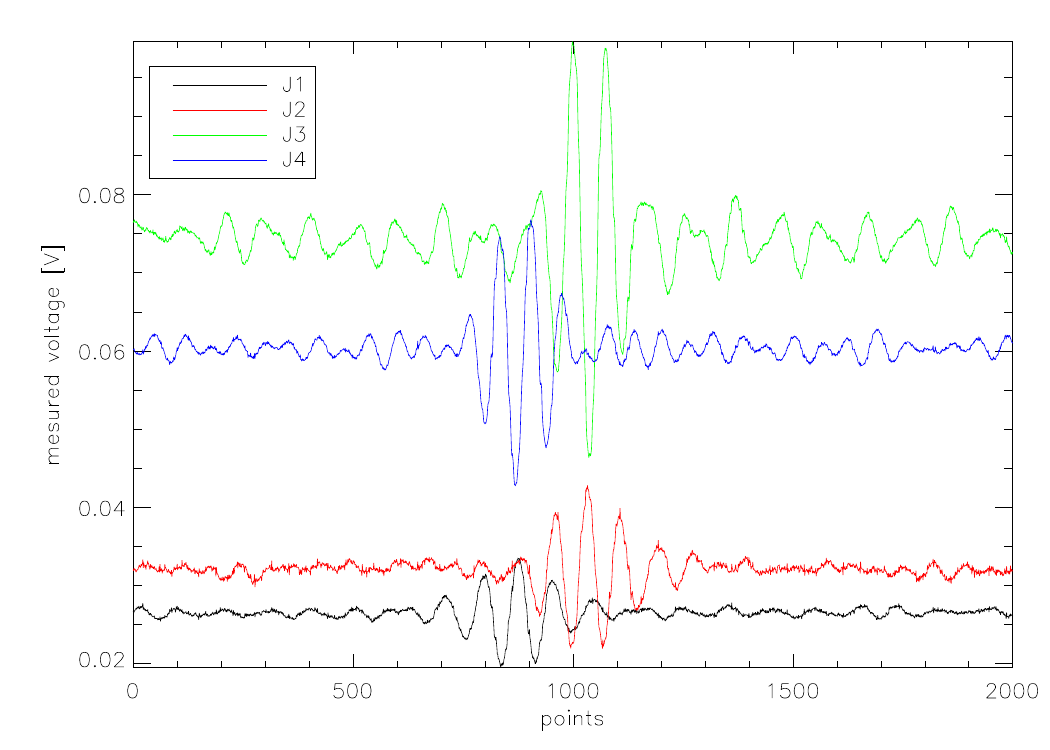}}
  \caption{Fringe signals obtained using a Xenon arc lamp for cophasing.}
  \label{fig-frangesXeIJ}
\end{figure}

While fringe contrast improved, the Signal-to-Noise Ratio (SNR) was degraded. Peak signal levels remained far below the 10~V detector saturation limit despite maximum amplifier gain. Geometrical etendue mismatch between the Xenon arc lamp and the single-mode fiber core resulted in poor coupling efficiency (<0.1\% transmitted). A higher-brightness broadband source covering 0.6~\mum to 2.5~\mum was required.

%-------------------------------------------------------------------------------
\subsubsection{A Supercontinuum Source for Stellar Simulation}
\label{sec-realisation-nouveau}

We selected a supercontinuum laser source ("white laser") driven by a pulsed 1064~nm Nd:YAG pump laser coupled into a highly non-linear Photonic Crystal Fiber (PCF). Non-linear spectral broadening produces a continuous spectrum from 400~nm to 2.5~\mum delivered through a single fiber core.

While a supercontinuum source was evaluated at ONERA, its low pulse repetition rate (8~kHz) alias-coupled with the 250~kHz fringe sensor readout rate. We procured a high-repetition-rate source (20~MHz, Fianium) providing continuous output across the 400~nm– 2.5~\mum range.

To prevent optical damage from high energy density, the output end of the Fianium source fiber uses an unconnectorized aluminum housing sealed with an AR window. This output could not be directly connected to standard fiber patch cords. Mounting the source directly on the bench was unfeasible as it blocked red alignment lasers and prevented monochromatic testing.

We designed an intermediate injection stage to couple light from the Fianium source into a single master fiber mounted at the M0 collimator focus (Figure~\ref{fig-banc-fianium}).

\begin{figure} \centering
  \FIG{0.7}{false}{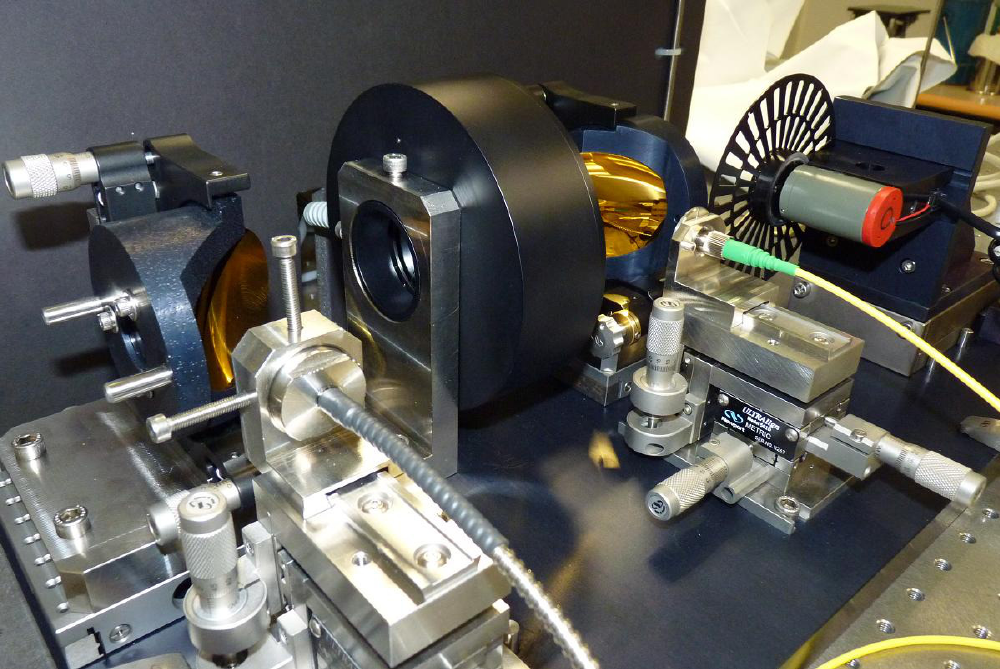}
  \caption{Injection bench for the Fianium supercontinuum source.}
  \label{fig-banc-fianium}
\end{figure}

The injection stage uses two off-axis parabolic mirrors. An electro-mechanical shutter mounted between the mirrors provides computer-controlled beam blanking. An optical chopper wheel is included for lock-in detection during single-pixel power measurements.

The delivery fiber must maintain single-mode propagation across 0.8~\mum to 2.5~\mum. We replaced the original Fluoride glass fiber (Le Verre Fluor\'e, single-mode only above 1.65~\mum) with an endless single-mode pure silica PCF fiber that maintains single-mode guidance across the full silica transmission window.

High optical power focused at the PCF fiber tip induced localized air heating and thermal turbulence, causing flux coupling fluctuations (Figure~\ref{fig-fluct-fi}).

\begin{figure} \centering
  \FIG{0.7}{false}{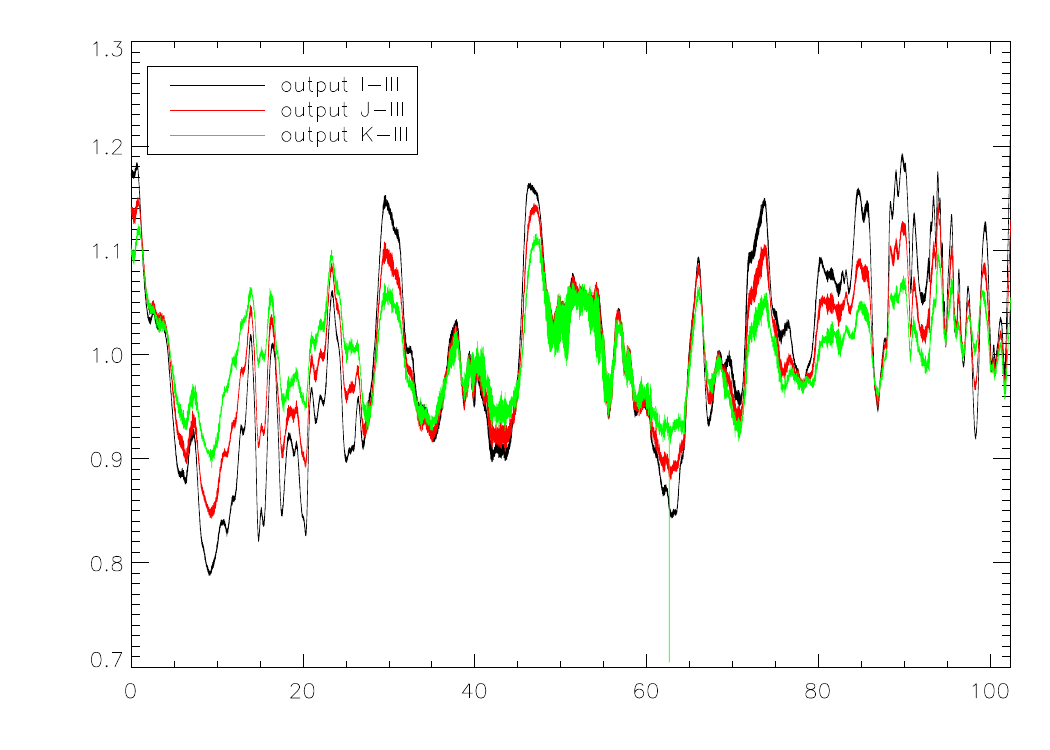}
  \caption{Relative intensity fluctuations across the three spectral bands.}
  \label{fig-fluct-fi}
\end{figure}

Figure~\ref{fig-fluct-fi} plots relative intensity fluctuations (normalized to mean power) over time across spectral bands. Fluctuation amplitudes exhibit chromatic dependency, increasing at shorter wavelengths—a behavior characteristic of fiber coupling jitter. Power Spectral Density (PSD) analysis of these fluctuations is shown in Figure~\ref{fig-PSD-flux-max}.

\begin{figure} \centering
  \subfloat[PSD of intensity in $\I$ and $\J$ bands.]{\label{fig-PSD-flux-max1}
    \FIG{0.49}{false}{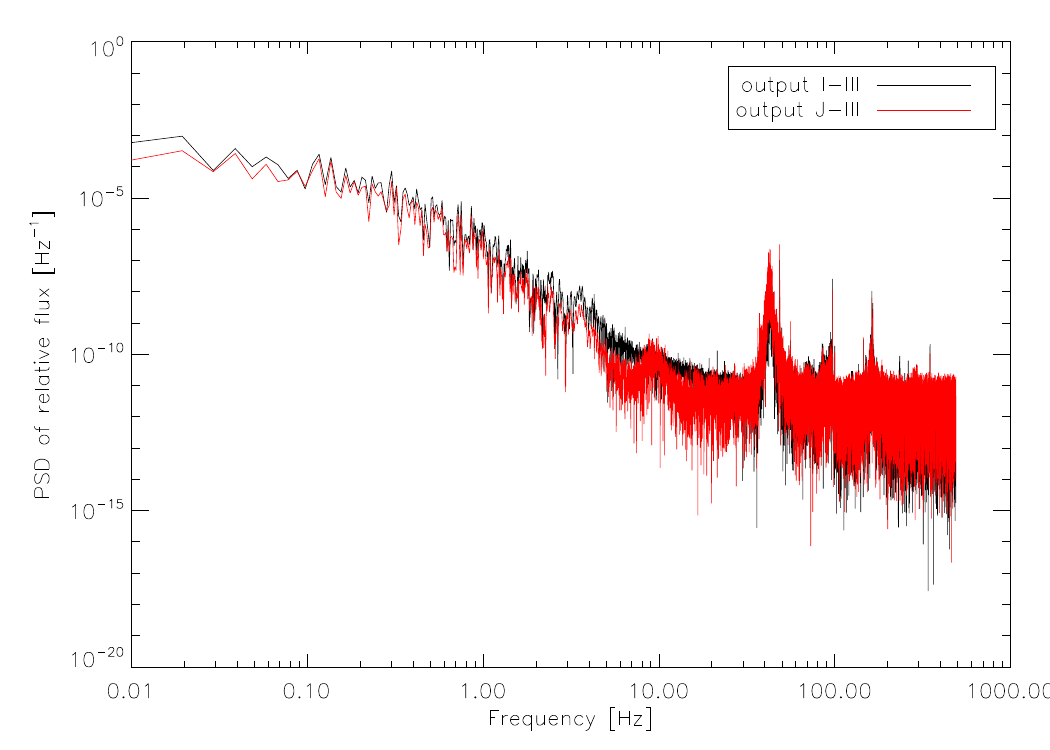}}
  \hfill\subfloat[PSD of intensity in $\K$ band.]{\label{fig-PSD-flux-max2}
    \FIG{0.49}{false}{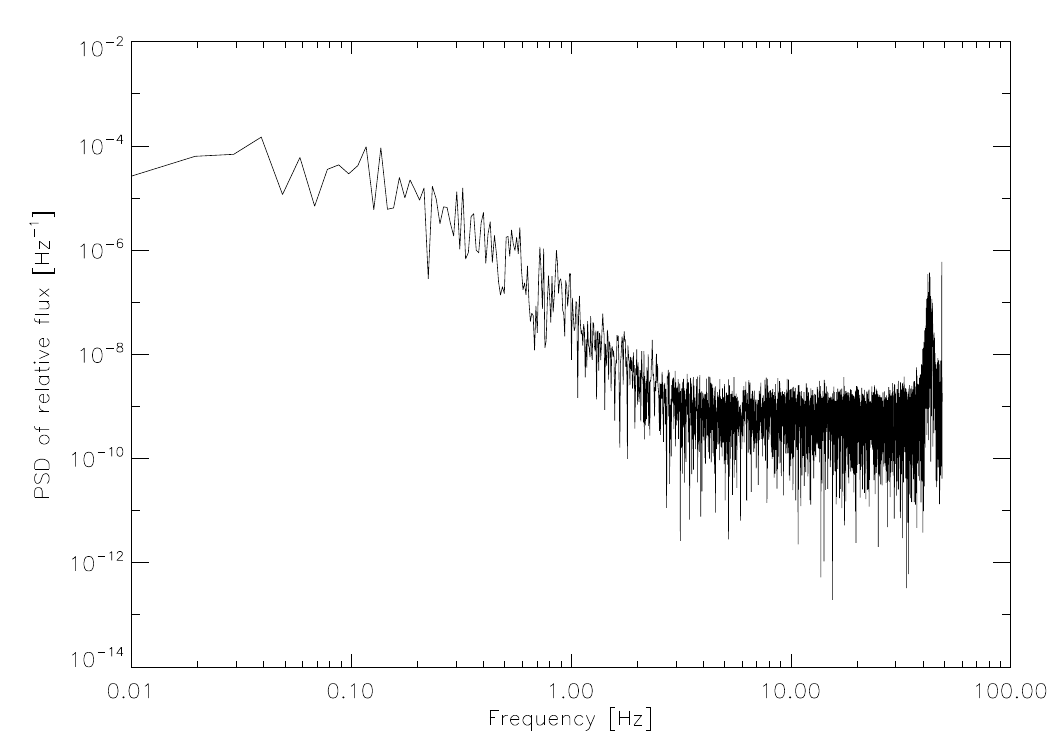}}
  \caption{Intensity PSDs in $\I$, $\J$, and $\K$ bands under maximum input power.}
  \label{fig-PSD-flux-max}
\end{figure}

At low frequencies (<0.3~Hz), noise spectra match convective air turbulence profiles. Discrete vibration peaks are also present:
\begin{itemize}
\item A strong 40~Hz mechanical vibration originating from the source cooling fan or injection bench, visible across metrology and science detectors;
\item A 50~Hz line power frequency peak across all PSDs;
\item Minor peaks at 100~Hz and 170~Hz detected by the high-speed fringe sensor.
\end{itemize}

Under maximum output power, light levels exceeded detector saturation limits. We introduced a slight defocus at the fiber coupling stage by shifting the fiber tip forward toward the parabolic mirror. Defocussing lowered total throughput while suppressing relative coupling fluctuations (Figure~\ref{fig-PSD-flux-min}).

\begin{figure} \centering
  \subfloat[PSD of intensity in $\I$ and $\J$ bands.]{\label{fig-PSD-flux-min1}
    \FIG{0.49}{false}{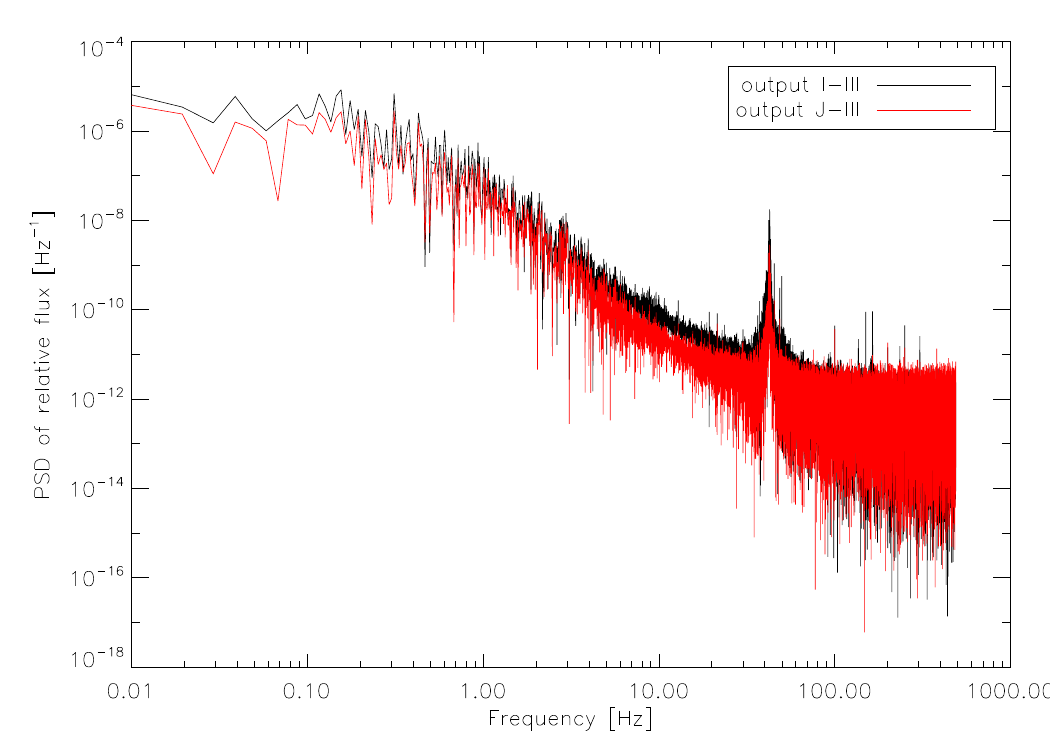}}
  \hfill\subfloat[PSD of intensity in $\K$ band.]{\label{fig-PSD-flux-min2}
    \FIG{0.49}{false}{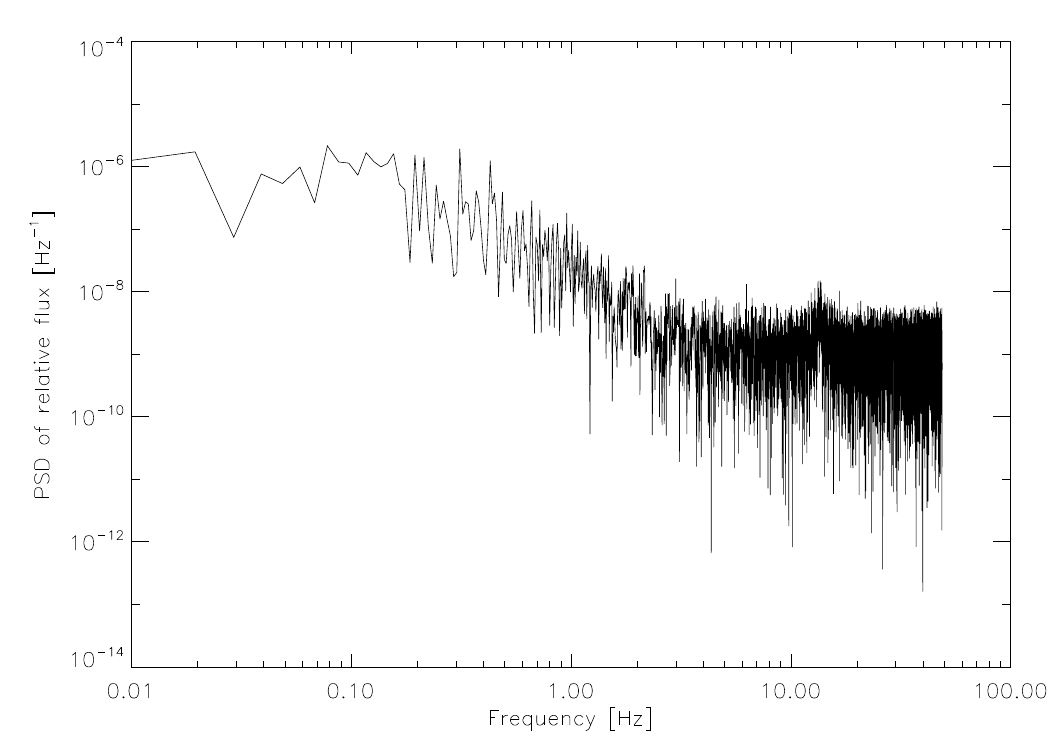}}
  \caption{Intensity PSDs in $\I$, $\J$, and $\K$ bands under reduced fiber coupling.}
  \label{fig-PSD-flux-min}
\end{figure}

Slightly detuning fiber coupling reduced low-frequency noise power by two orders of magnitude (10$\times$ reduction in fluctuation amplitude). Vibration peaks were similarly attenuated.

%-------------------------------------------------------------------------------
\subsubsection{Noise Characterization Versus Optical Power}
\label{sec--etude-1}

We evaluated signal variance against mean intensity in the fringe sensor after filtering low-frequency convective drift and mechanical lines. Results for the $\J$ band are shown in Figure~\ref{fig-var-flux-fi}.

\begin{figure} \centering
  \FIG{0.7}{false}{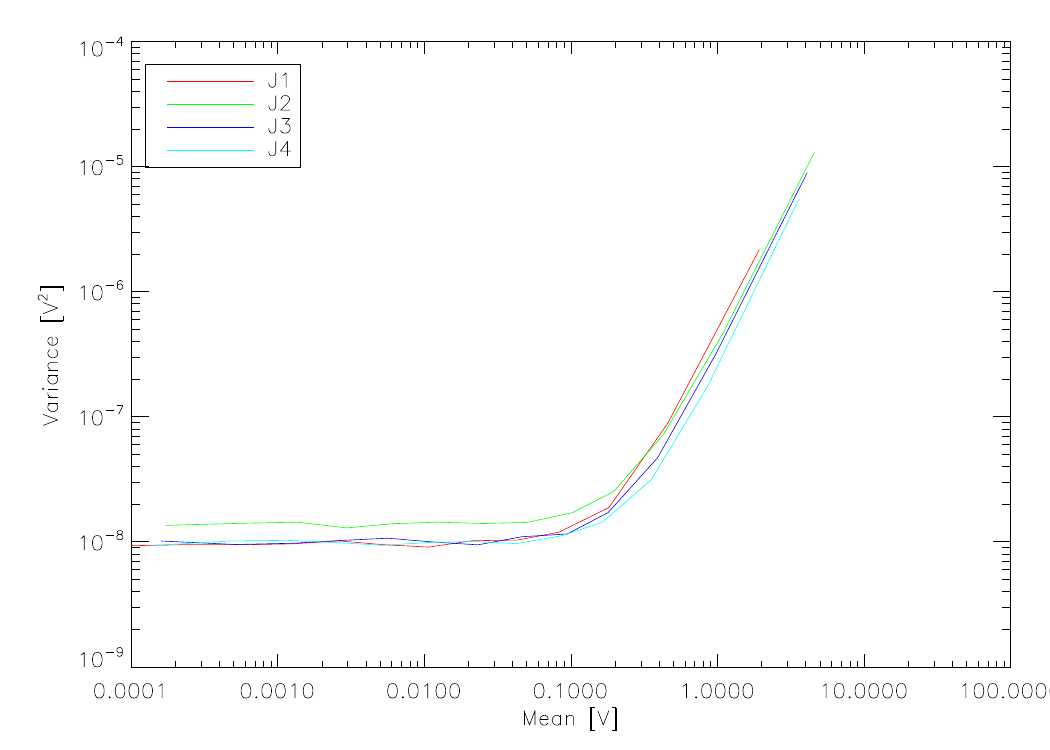}
  \caption{Signal variance versus mean intensity in the fringe sensor.}
  \label{fig-var-flux-fi}
\end{figure}

At low signal levels (<0.1~V), noise is dominated by detector read noise (constant variance). Above 0.1~V, signal variance scales quadratically with mean intensity ($\text{variance} \propto I^2$), indicating that noise scales linearly with signal amplitude. In photon shot-noise limited regimes, variance scales linearly with signal ($\text{variance} \propto I$). Quadratic scaling indicates that performance at high flux is limited by source intensity jitter or amplifier electronics rather than photon statistics, yielding a constant SNR across 0.1~V to 10~V signal levels.

A similar noise assessment was performed on the science camera. Figure~\ref{fig-var-flux-cam} plots pixel variance against mean photon counts across dark fringe pixels.

\begin{figure} \centering
  \FIG{0.7}{false}{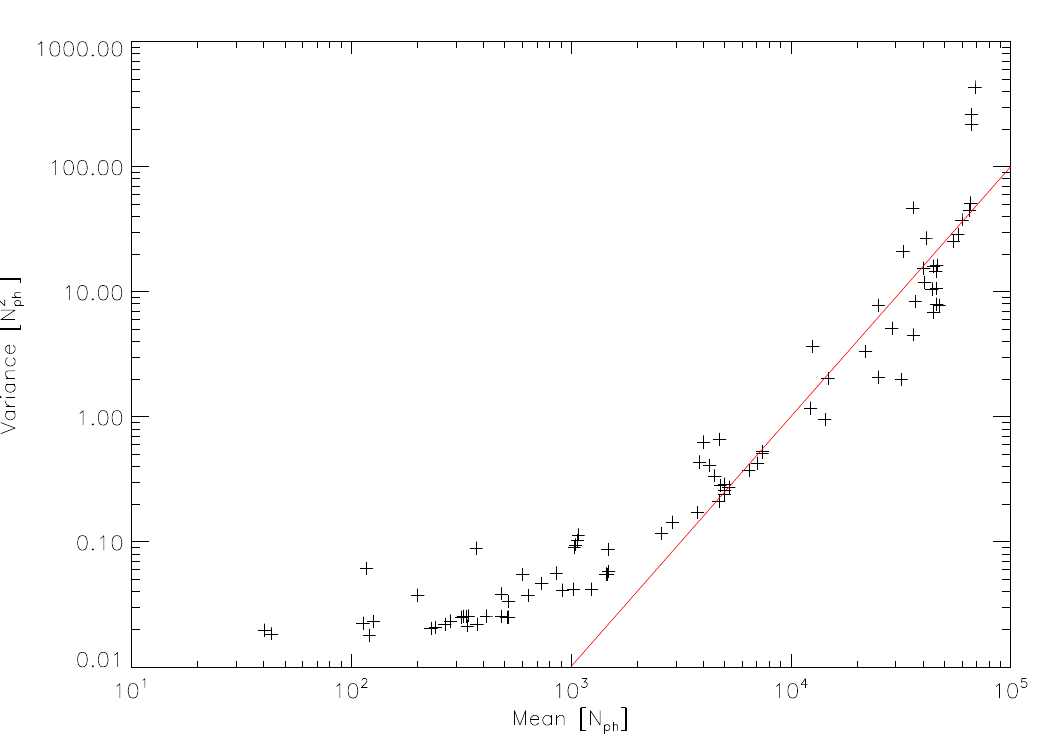}
  \caption{Signal variance versus mean photon counts on the science camera.}
  \label{fig-var-flux-cam}
\end{figure}

The science camera exhibits identical quadratic variance scaling at high flux levels and flat read-noise limits at low flux, confirming that source intensity fluctuations dominate at high signal levels.

%-------------------------------------------------------------------------------
\subsubsection{Modular Source Bench}
\label{sec-nouveau-banc}

To enable monochromatic testing, a modular source bench was constructed for the laser diodes and SLED. Light is delivered through the master PCF fiber mounted at the M0 focus, avoiding optical realignment when switching between monochromatic and supercontinuum sources. Source configurations are illustrated in Figure~\ref{fig-banc-source}.

\begin{figure} \centering
  \FIG{0.9}{false}{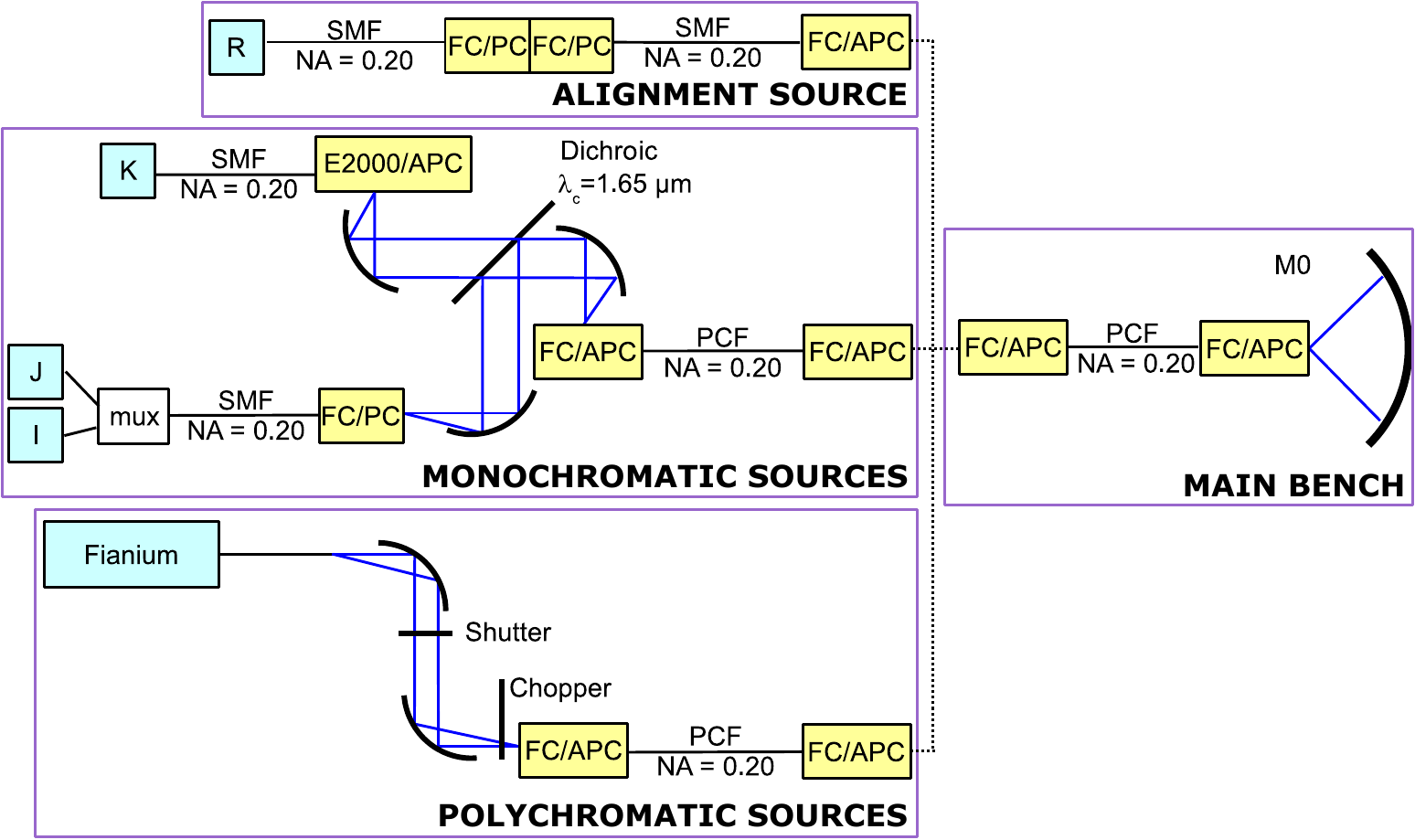}
  \caption{Modular injection setup for switching light sources.}
  \label{fig-banc-source}
\end{figure}

A fiber multiplexer combines the two metrology wavelengths into a single fiber output. This output and the 2.32~\mum laser diode are combined using a 1.65~\mum cutoff dichroic beamsplitter and focused into the master PCF fiber (Figure~\ref{fig-banc-IJK}).

\begin{figure} \centering
  \FIG{0.7}{false}{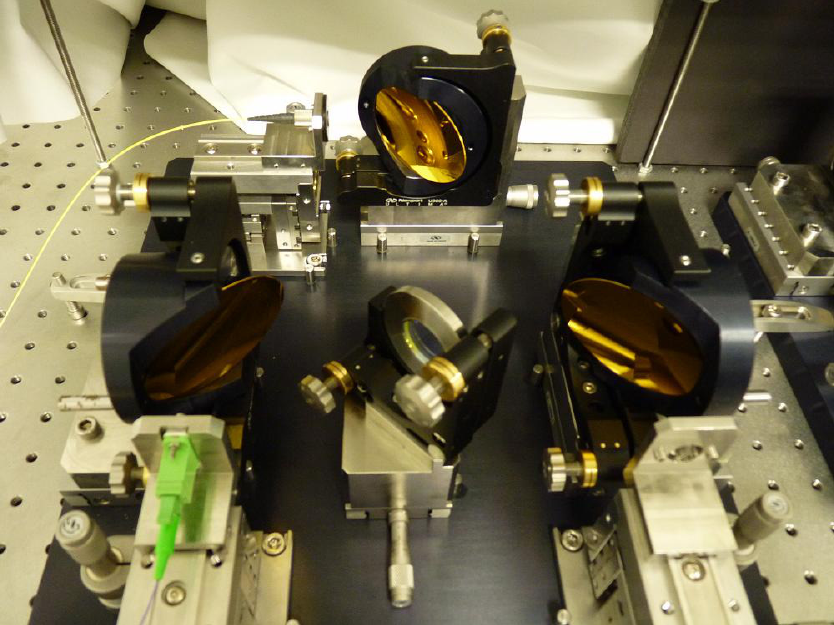}
  \caption{Injection bench for monochromatic sources.}
  \label{fig-banc-IJK}
\end{figure}

Red laser diodes (635~nm or 675~nm) can be connected to perform visual optical alignment. Single-mode fibers (SMF) connect auxiliary sources to the injection stage.

Auxiliary sources are mounted in parallel to maintain fiber alignment across parabolic injection optics without disturbing FC/APC fiber connectors.

%¤¤¤¤¤¤¤¤¤¤¤¤¤¤¤¤¤¤¤¤¤¤¤¤¤¤¤¤¤¤¤¤¤¤¤¤¤¤¤¤¤¤¤¤¤¤¤¤¤¤¤¤¤¤¤¤¤¤¤¤¤¤¤¤¤¤¤¤¤¤¤¤¤¤¤¤¤¤¤
\subsection{Photometric Reference Channel Redesign}
\label{sec-modification-voie}

Calculating null depth $N$ requires normalizing destructive output intensity $I_\mmin$ by a constructive reference flux $I_\mref$. While constructive peak flux can be measured sequentially by scanning OPD, real-time source fluctuations require simultaneous sampling of a reference flux $I_\mref$. Section~\ref{sec-objectif-creation} details how null depth is calculated from synchronized $I_\mmin$ and $I_\mref$ measurements.

In the initial design, reference flux $I_\mref$ was sampled from MMZ output II. Output II delivers complementary quasi-constructive interference, but presents two operational drawbacks:
\begin{itemize}
\item Fiber positioning on output II must be aligned relative to output III, making simultaneous coupling optimization difficult;
\item Output II light does not represent a pure constructive interference state. Because beamsplitter absorption breaks symmetry on output II, its baseline phase shift deviates from zero ($\sim -20\degree$ in our setup). Sampling along the slope of the fringe function rather than at the constructive peak leaves reference measurements sensitive to residual OPD jitter.
\end{itemize}

To improve measurement stability, I developed a reference channel that samples light upstream of the MMZ combiner. Light reflected off collimator M0 is only partially transmitted through the dual-aperture mask (Figure~\ref{fig-M0}); we pick off a portion of the unused reflected beam. The reference pick-off setup is shown in Figure~\ref{fig-I-ref}.

\begin{figure} \centering
  \FIG{0.7}{false}{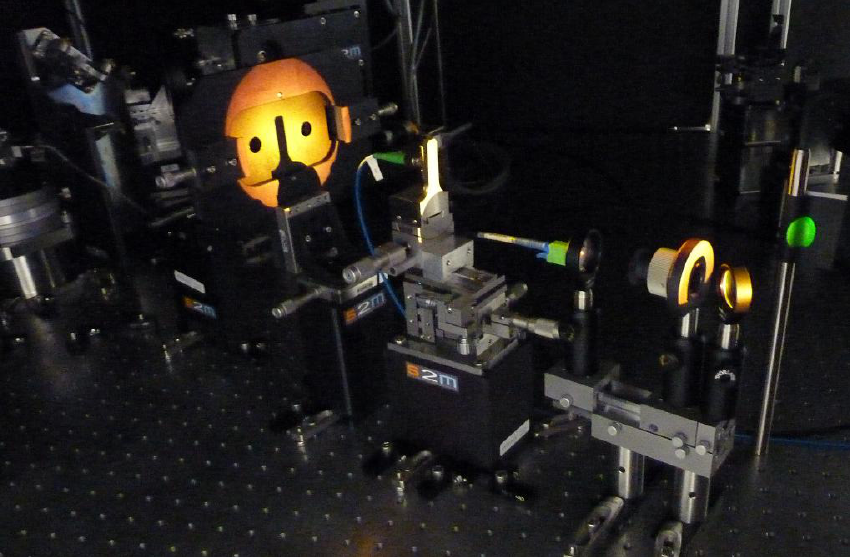}
  \caption{Photometric reference pick-off located near the M0 focus.}
  \label{fig-I-ref}
\end{figure}

Figure~\ref{fig-I-ref} shows the dual-aperture mask and the beam footprint reflected off M0. Omitting the afocal beam compressors reduced the mask clear apertures to 13~mm, leaving unused light around the perimeter. A pick-off lens placed near the source fiber holder collects a portion of this outer beam and focuses it into a Fluoride glass fiber (LVF) connected to the science camera. A 45\degree dichroic plate ($\lambda_\mc = 1.65$~\mum) mirrors the spectral response of the science channel. The fiber tip is intentionally positioned out of focus to attenuate intensity and prevent detector saturation.

Positioning the fiber out of focus renders reference intensity measurements insensitive to local mechanical vibration while providing an interference-free reference signal.

%§§§§§§§§§§§§§§§§§§§§§§§§§§§§§§§§§§§§§§§§§§§§§§§§§§§§§§§§§§§§§§§§§§§§§§§§§§§§§§§
\section{Control Software and User Interface}
\label{sec-developpement-interface}

The \pe testbed is controlled in real time by LabVIEW RT software that synchronizes detector readouts and executes cophasing control loops. Data streams are logged for post-processing analysis.

Initial cophasing tests used a first-generation software package focused on low-level real-time tasks:
\begin{itemize}
\item Hardware interfacing with loop sensors and actuators;
\item Execution of real-time control algorithms;
\item Data logging;
\item Injection of calibrated path disturbances.
\end{itemize}

This early package supported closed-loop operations up to a loop rate of $\sim 100$~Hz—well below the 1~kHz specification.

While this software was used to validate opto-mechanical hardware and refine calibration routines, I developed a redesigned real-time software system. Thread execution and deterministic timing were restructured using LabVIEW RT primitives to achieve 1~kHz closed-loop execution. Control loop execution dynamics are detailed in Section~\ref{sec-description-architecture}.

A remote Graphical User Interface (GUI) was developed to interface with the real-time target. The GUI was updated to accommodate the expanded calibration routines and real-time control modes. It displays live diagnostic data and transmits control commands to real-time loops. Figure~\ref{fig-ihm-marche} shows the main control interface during closed-loop operation.

\begin{figure} \centering
  \FIG{1.}{false}{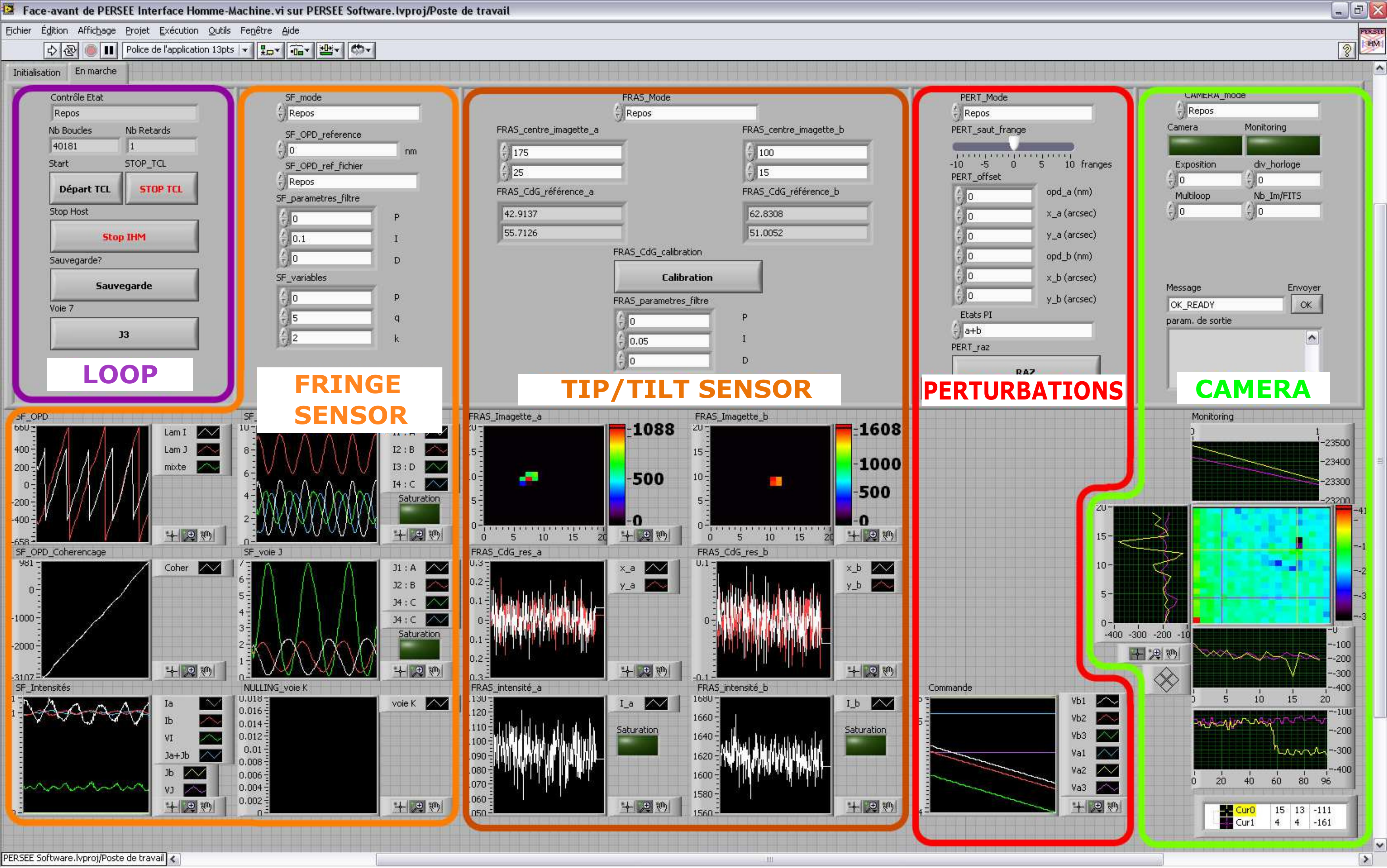}
  \caption{GUI control panel during active closed-loop operations.}
  \label{fig-ihm-marche}
\end{figure}

The GUI is structured as a subsystem matrix, with columns representing functional modules. The upper section groups operational controls into two categories:
\begin{itemize}
\item \textbf{Initialization Parameters:} System settings loaded at startup to minimize real-time network traffic;
\item \textbf{Runtime Parameters:} Operational variables adjustable during execution, updated on the real-time target at 2~Hz.
\end{itemize}

The lower section displays graphical diagnostic indicators, including raw sensor arrays, computed state metrics, and actuator command signals. Real-time data arrays are buffered and pushed to the GUI in 16-sample blocks at 2~Hz (yielding a 32~Hz display refresh rate).

The GUI columns in Figure~\ref{fig-ihm-marche} correspond to key subsystems:
\begin{itemize}
\item \textbf{Real-Time Loop:} Global controls governing loop execution, state toggles, and logging controls;
\item \textbf{Piston Loop:} Parameters for OPD cophasing, displaying raw detector levels and estimated phase across both metrology channels;
\item \textbf{Tip/Tilt Loop:} Controls for pointing tracking, reference setpoints, CCD sub-array displays, and calculated beam centroids;
\item \textbf{Disturbance Generator:} Controls for injecting calibrated path disturbances into active M6 mirrors via pre-recorded files or direct manual offsets, alongside real-time drive voltage displays;
\item \textbf{Science Camera:} Controls for exposure times, frame counts, and real-time array readouts to monitor dark fringe saturation levels.
\end{itemize}

Automated execution scripts were implemented to coordinate multi-step measurement sequences during extended testing (Section~\ref{sec-etude-preliminaire}).

%§§§§§§§§§§§§§§§§§§§§§§§§§§§§§§§§§§§§§§§§§§§§§§§§§§§§§§§§§§§§§§§§§§§§§§§§§§§§§§§
\section{Fourier Transform Spectral Calibration}
\label{sec-etalonnage-FTS}

%¤¤¤¤¤¤¤¤¤¤¤¤¤¤¤¤¤¤¤¤¤¤¤¤¤¤¤¤¤¤¤¤¤¤¤¤¤¤¤¤¤¤¤¤¤¤¤¤¤¤¤¤¤¤¤¤¤¤¤¤¤¤¤¤¤¤¤¤¤¤¤¤¤¤¤¤¤¤¤
\subsection{Measurement Principle}
\label{sec-principe-mesure}

Fourier Transform Spectroscopy (FTS) derives source spectra from interferograms recorded across scanned optical path delays. While FTS is typically implemented on Michelson interferometers using a scanning mirror stage, the \pe testbed's MMZ combiner and integrated delay lines can be operated as an FTS spectrometer.

For wavenumber $\sigma$, output intensity $I_\sigma$ as a function of OPD $\delta$ is given by:
\begin{equation}
  I_\sigma(\delta,\sigma) =
  \tilde{I}(\sigma)\GC{1+V(\sigma)\cos\GP{2\pi\sigma\delta+\phi(\sigma)}},
  \label{eq-fts1}
\end{equation}
where $\tilde{I}(\sigma) = I_\ma(\sigma)+I_\mb(\sigma)$ represents input power, $I_\ma(\sigma)$ and $I_\mb(\sigma)$ are individual arm intensities, and $V(\sigma)$ accounts for spatial/temporal fringe visibility factors (source size, spectral bandwidth, arm transmissions, wavefront errors).

The goal is to determine spectral distribution $\tilde{I}(\sigma)$. Total output intensity is:
\begin{equation}
  I(\delta) = \int_0^{\infty}I_\sigma(\delta,\sigma)\dd\sigma.
\end{equation}

The Fourier Transform of $I(\delta)$, denoted $\widehat{I}(\sigma)$, is:
\begin{align}
  \widehat{I}(\sigma) = &
  \int_{-\infty}^{\infty}I(\delta)e^{-i2\pi\sigma\delta}\dd\delta \nonumber\\
  = & \int_{-\infty}^{\infty}\int_0^{\infty}\tilde{I}(\sigma_0)[1+V(\sigma_0)
  \cos(2\pi\sigma_0\delta+\phi(\sigma_0))]e^{-i2\pi\sigma\delta}\dd\sigma_0
  \dd\delta \nonumber\\
  = & \int_0^{\infty}\tilde{I}(\sigma_0)\dd\sigma_0\int_{-\infty}^{\infty}
  e^{-i2\pi\sigma\delta}\dd\delta \nonumber\\
  & {}+\frac12\int_{-\infty}^{\infty}\int_0^{\infty}\tilde{I}(\sigma_0)V(\sigma_0)
  \exp(-i2\pi(\sigma-\sigma_0)\delta+i\phi(\sigma_0))\dd\sigma_0 \dd\delta
  \nonumber\\
  & {}+\frac12\int_{-\infty}^{\infty}\int_0^{\infty}\tilde{I}(\sigma_0)V(\sigma_0)
  \exp(-i2\pi(\sigma+\sigma_0)\delta-i\phi(\sigma_0))\dd\sigma_0 \dd\delta.
\end{align}

Defining total integrated intensity $I_0=\int_0^{\infty}\tilde{I}(\sigma)\dd\sigma$ and Dirac delta function $\delta_0$ yields:
\begin{align}
  \widehat{I}(\sigma) = &\ I_0\delta_0(\sigma) +
  \frac12\int_0^{\infty}\tilde{I}(\sigma_0)V(\sigma_0)e^{i\phi(\sigma_0)}
  \int_{-\infty}^{\infty}\exp(-i2\pi(\sigma-\sigma_0)\delta)\dd\delta
  \dd\sigma_0 \nonumber\\
  & {}+\frac12\int_0^{\infty}\tilde{I}(\sigma_0)V(\sigma_0)e^{-i\phi(\sigma_0)}
  \int_{-\infty}^{\infty}\exp(-i2\pi(\sigma+\sigma_0)\delta)\dd\delta
  \dd\sigma_0 \nonumber\\
  = &\ I_0\delta_0(\sigma) +
  \frac12\int_0^{\infty}\tilde{I}(\sigma_0)V(\sigma_0)
  e^{i\phi(\sigma_0)}\delta_0(\sigma-\sigma_0)\dd\sigma_0 \nonumber\\
  & {}+\frac12\int_0^{\infty}\tilde{I}(\sigma_0)V(\sigma_0)
  e^{-i\phi(\sigma_0)}\delta_0(\sigma+\sigma_0)\dd\sigma_0.
\end{align}

Because integration is restricted to positive wavenumbers ($\sigma_0 \ge 0$), the final term vanishes, leaving:
\begin{equation}
  \widehat{I}(\sigma) = I_0\delta_0(\sigma)+
  \frac{1}{2}e^{i\phi(\sigma)}\tilde{I}(\sigma)V(\sigma).
  \label{eq-spectre-mmz}
\end{equation}

For an ideal Michelson with unit contrast ($V=1$) and zero phase shift ($\phi=0$), this reduces to:
\begin{equation}
  \widehat{I}(\sigma)=I_0\delta_0(\sigma)+\frac{1}{2}\tilde{I}(\sigma),
\end{equation}
yielding a DC component at zero frequency and the source spectrum scaled by 0.5 at non-zero frequencies.

Equation~\eqref{eq-spectre-mmz} shows that if visibility $V(\sigma)$ is slowly varying, the measured transform recovers the source spectrum scaled by $V/2$ and phase-shifted by $\phi(\sigma)$.

Practically, OPD is scanned linearly at constant velocity $v_0$ ($\delta = v_0 t$).

One long-stroke delay line stage is driven at $v_0 = 10$~\mum.s$^{-1}$ over duration $T = 100$~s while keeping the second stage fixed. Readout rates are $f_\mfs = 1$~kHz for metrology detector arrays ($\I$ and $\J$ bands) and $f_\mcam = 100$~Hz for the science camera ($\K$ band). The total scanned optical path stroke is $\Delta = v_0 T = 1$~mm, with sampling steps $\delta_\mfs = v_0/f_\mfs = 10$~nm for the fringe sensor and $\delta_\mcam = v_0/f_\mcam = 100$~nm for the science camera.

Spectral resolution depends on scan length $\Delta$ and windowing functions \cite{Thompson01}:
\begin{equation}
  \Delta\sigma = \frac{\eta_\mf}{2\Delta},
  \label{eq-res-fts-sig}
\end{equation}
where $\eta_\mf$ is a windowing factor ($\eta_\mf = 1.2$ for unwindowed boxcar sampling; $\eta_\mf = 2$ for Hann windowing). Unwindowed resolution is $\Delta\sigma = 6\E{-4}$~\imum.

Resolving power $R = \lambda/\Delta\lambda$ scales inversely with wavelength:
\begin{equation}
  R = \frac{\lambda}{\Delta\lambda} = \frac{2\Delta}{\eta_\mf\lambda}.
  \label{eq-res-fts}
\end{equation}

Evaluated at mean band wavelengths, resolving power reaches $R_{\mfs,\I} \approx 2000$, $R_{\mfs,\J} \approx 1300$, and $R_{\mcam} \approx 800$.

%¤¤¤¤¤¤¤¤¤¤¤¤¤¤¤¤¤¤¤¤¤¤¤¤¤¤¤¤¤¤¤¤¤¤¤¤¤¤¤¤¤¤¤¤¤¤¤¤¤¤¤¤¤¤¤¤¤¤¤¤¤¤¤¤¤¤¤¤¤¤¤¤¤¤¤¤¤¤¤
\subsection{Stage Non-Linearity Correction}
\label{sec-influence-correction}

Assuming perfectly linear translation yields the uncorrected raw spectra shown on the left side of Figure~\ref{fig-fts}.

\begin{figure} \centering
  \subfloat[Band $\I$, uncorrected spectrum.]{\label{fig-fts-I-n}
    \FIG{0.475}{false}{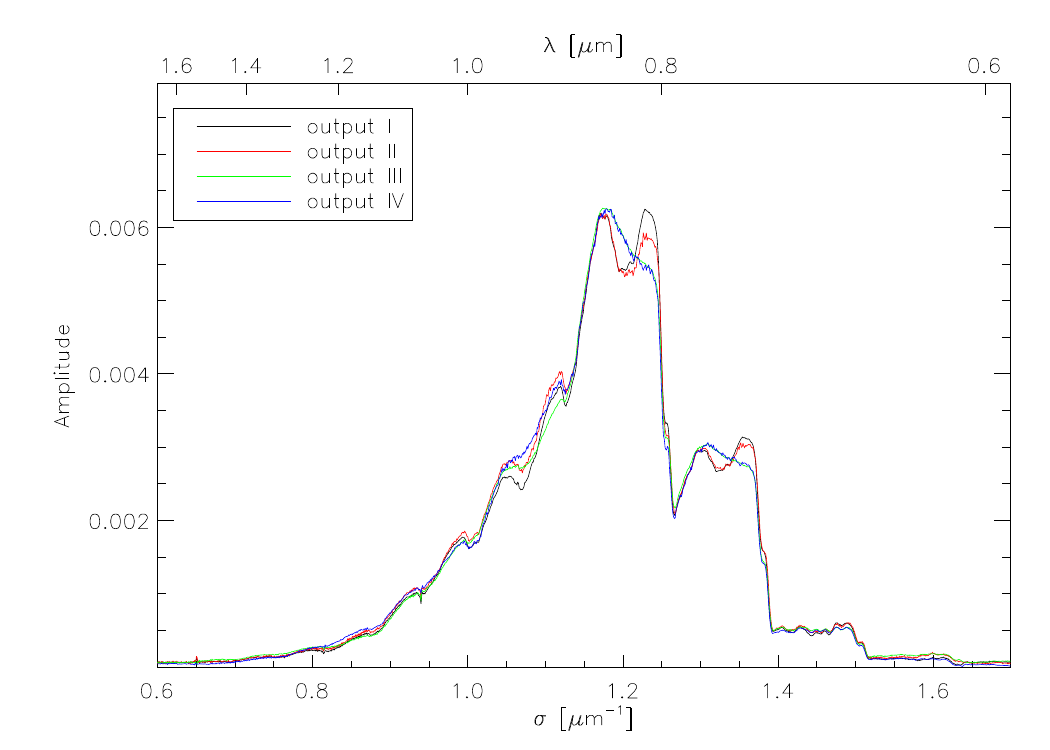}}
  \hfill\subfloat[Band $\I$, corrected spectrum.]{\label{fig-fts-I}
    \FIG{0.475}{false}{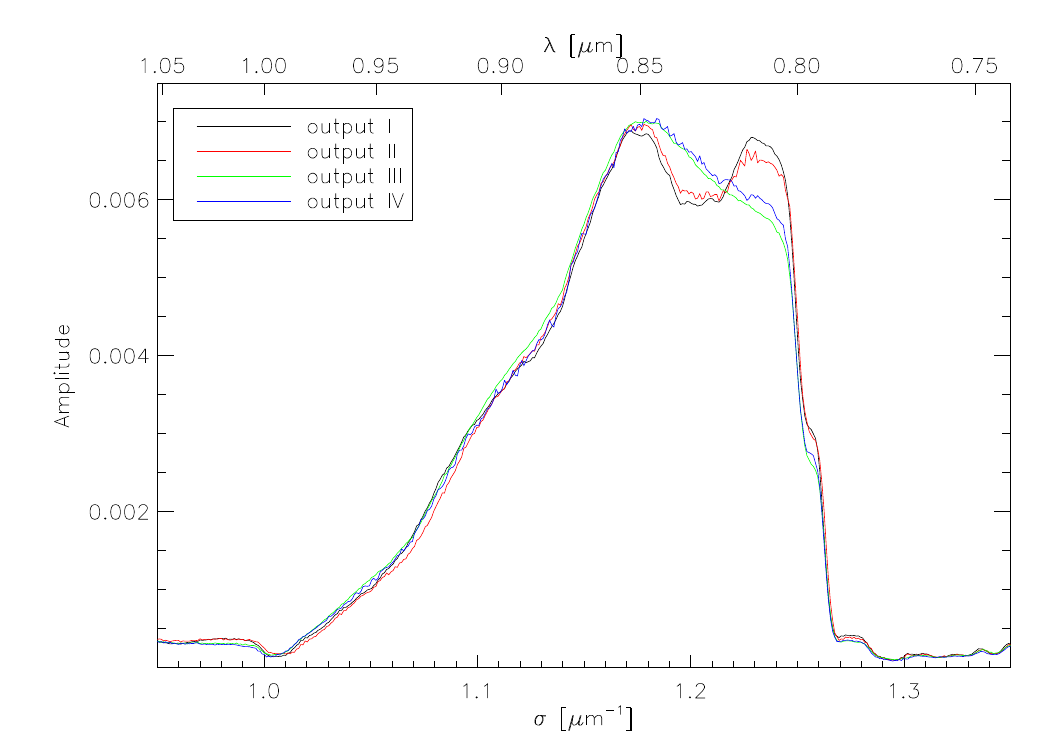}}\\
  \subfloat[Band $\J$, uncorrected spectrum.]{\label{fig-fts-J-n}
    \FIG{0.475}{false}{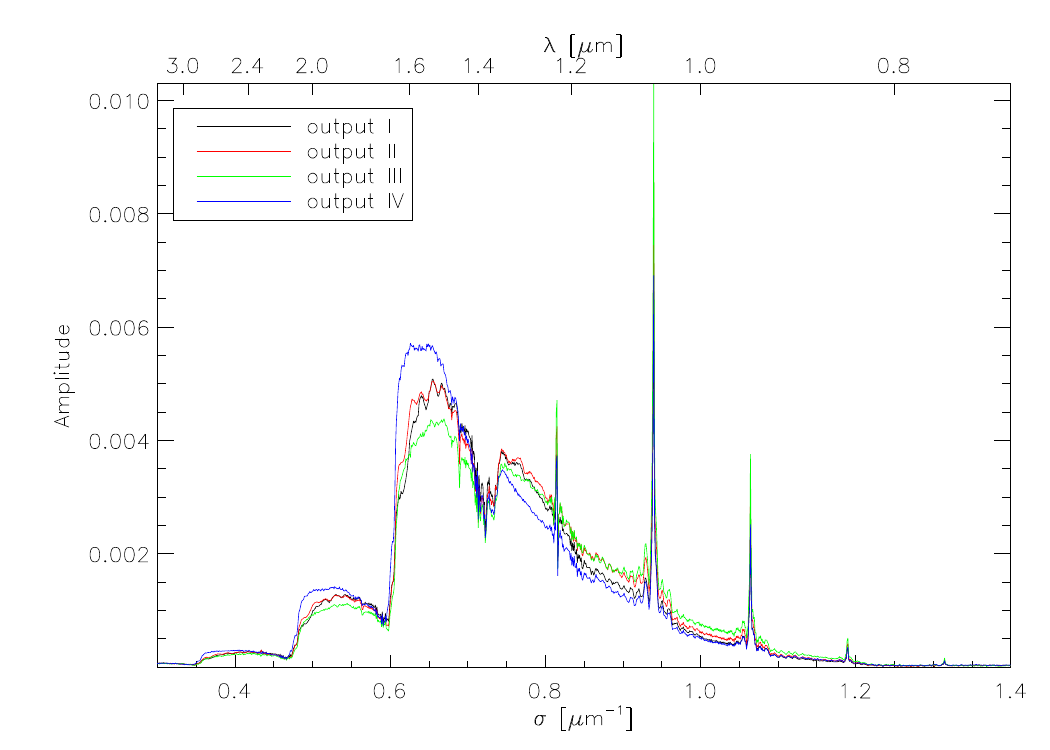}}
  \hfill\subfloat[Band $\J$, corrected spectrum.]{\label{fig-fts-J}
    \FIG{0.475}{false}{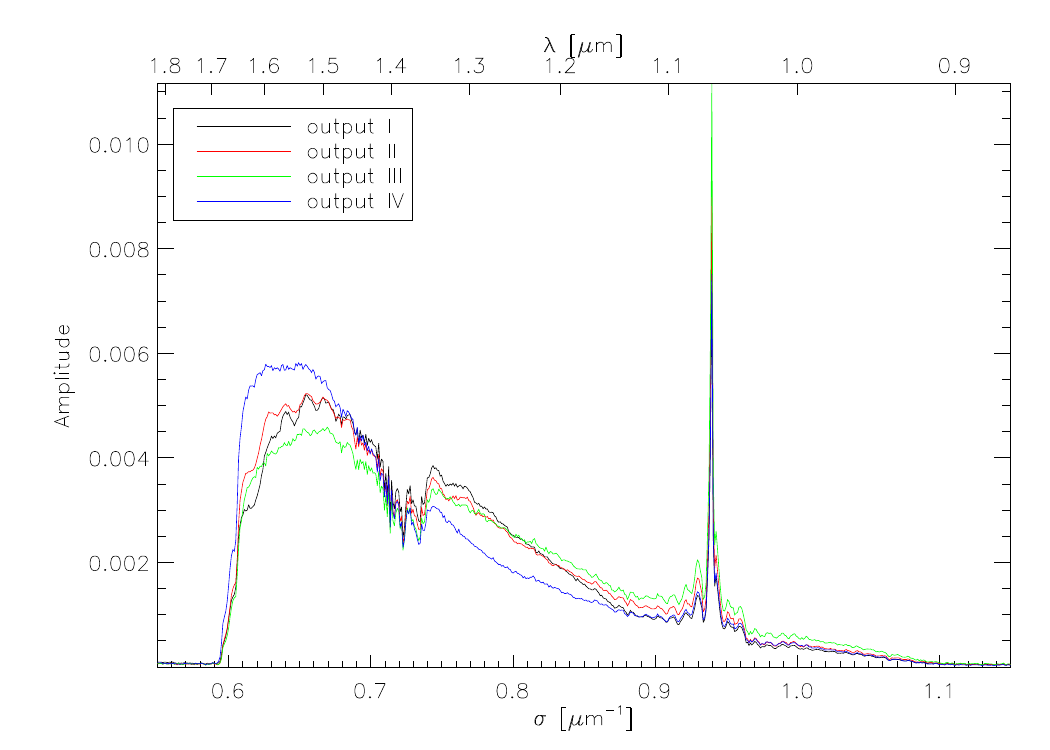}}\\
  \subfloat[Bands $\BH$--$\K$, uncorrected spectrum.]{\label{fig-fts-K-n}
    \FIG{0.475}{false}{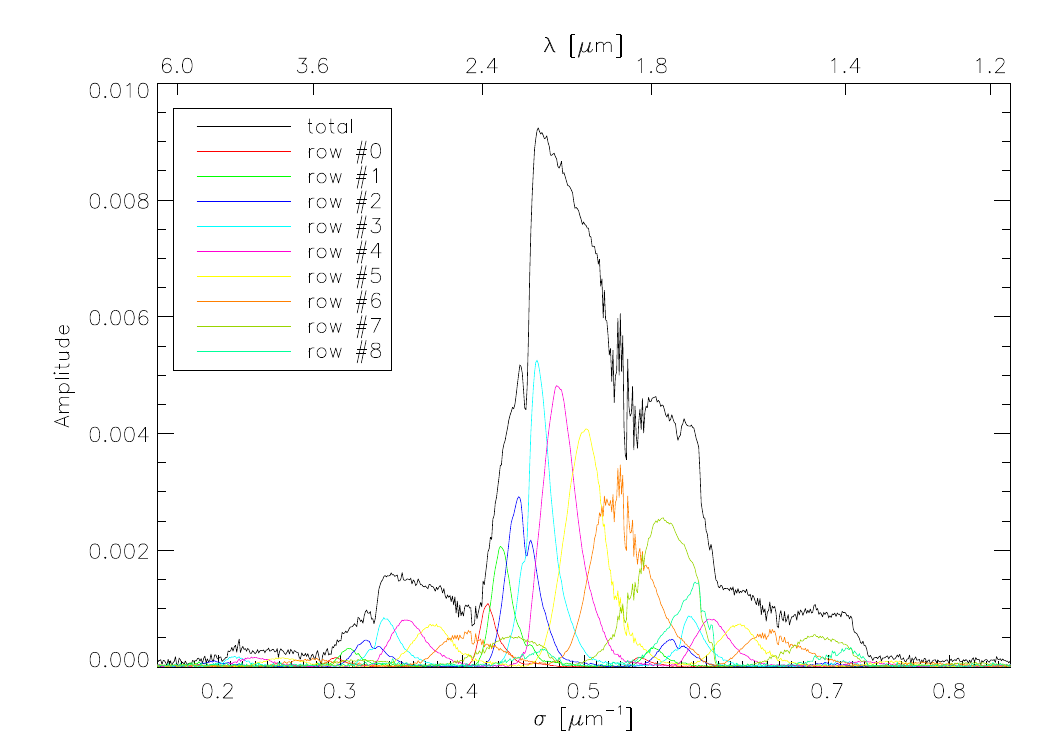}}
  \hfill\subfloat[Bands $\BH$--$\K$, corrected spectrum.]{\label{fig-fts-K}
    \FIG{0.475}{false}{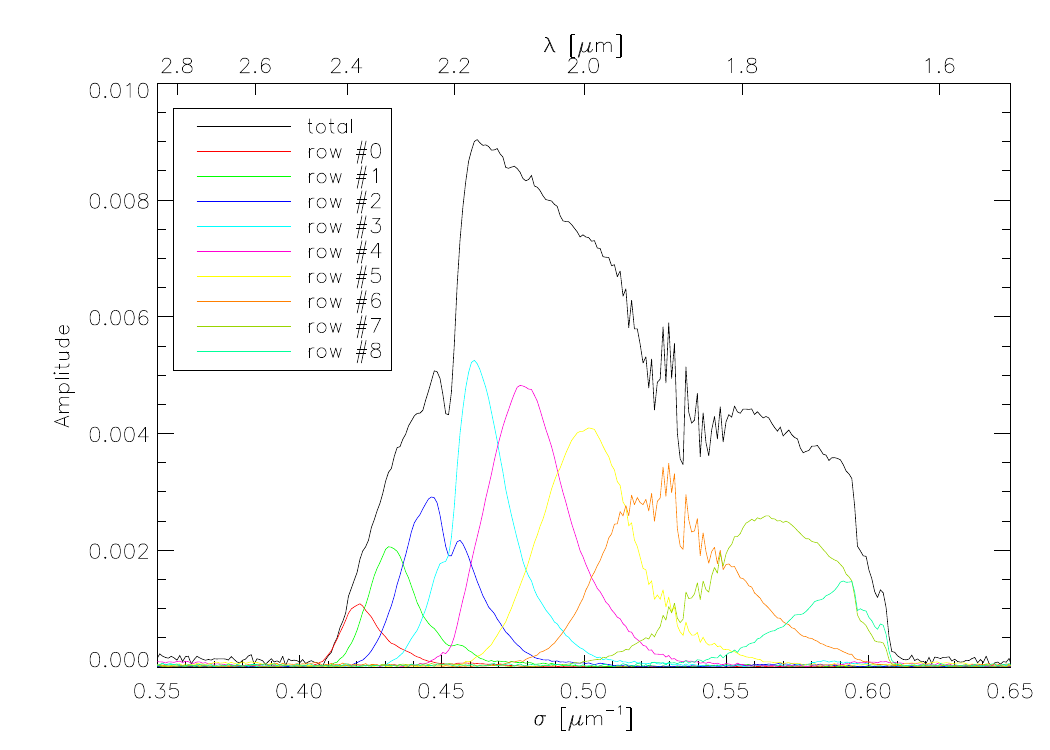}}\\
  \caption[Spectra across spectral channels.]{Estimated source spectra across metrology bands ($\I$ and $\J$) and science channels ($\BH$ and $\K$). Left panels show raw spectra; right panels show non-linearity corrected spectra.}
  \label{fig-fts}
\end{figure}

Symmetric ghost side-lobes appear flanking the primary spectral peaks in the uncorrected spectra. These side-lobes are artifacts that fall outside physical detector passbands. Side-lobe spacing remains constant across channels ($\Delta\sigma = 0.125$~\imum). Envelope analysis reveals that these artifacts stem from periodic velocity ripple in the translation stage with a spatial period of $1/\Delta\sigma = 8$~\mum. Scanned OPD follows:
\begin{equation}
  \delta_\mper(t)=\delta+A_\mper\cos\GP{2\pi\Delta\sigma\delta+\phi_\mper}.
\end{equation}

Equation~\eqref{eq-fts1} becomes:
\begin{equation}
  I_\sigma(\delta,\sigma) =
  \tilde{I}(\sigma)\GC{1+V(\sigma)\cos\GP{2\pi\sigma\GC{\delta+
        A_\mper\cos\GP{2\pi\Delta\sigma\delta+\phi_\mper}}+\phi(\sigma)}}.
  \label{eq-interf-fts}
\end{equation}

Expanding phase modulation using Bessel functions:
\begin{equation}
  \cos(\alpha+\beta\sin\gamma)=\sum^{+\infty}_{k=-\infty}J_k(\beta)
  \cos(\alpha+k\gamma),
\end{equation}
yields:
\begin{eqnarray}
  \lefteqn{I_\sigma(\delta,\sigma) = } \\
  & & \tilde{I}(\sigma)\GC{1+V(\sigma)\sum^{+\infty}_{k=-\infty}
    J_k\GP{2\pi\sigma A_\mper}\cos\GP{2\pi(\sigma+k\Delta\sigma)\delta+
      \phi(\sigma)+k\phi_\mper+k\frac{\pi}{2}}}. \nonumber
\end{eqnarray}

Defining $\sigma_k = \sigma-k\Delta\sigma$, the Fourier Transform becomes:
\begin{equation}
  \widehat{I}(\sigma) =  
  I_0\delta_0(\sigma)+\frac{1}{2}\sum^{+\infty}_{k=-\infty} 
  J_k(2\pi\sigma_kA_\mper) 
  e^{i\GP{\phi(\sigma_k)+k\phi_\mper+k\frac{\pi}{2}}}
  I(\sigma_k)V(\sigma_k). 
\end{equation}

This confirms side-lobe generation at multiples of $\Delta\sigma$, with first-order side-lobe amplitudes governed by $J_1(2\pi(\sigma\pm\Delta\sigma)A_\mper)$.

Resampling time-series data onto a corrected, uniform OPD grid eliminates these side-lobes. Fitting side-lobe rejection across science channels yielded ripple parameters $A_\mper=110$~nm and $\phi_\mper=0$.

%¤¤¤¤¤¤¤¤¤¤¤¤¤¤¤¤¤¤¤¤¤¤¤¤¤¤¤¤¤¤¤¤¤¤¤¤¤¤¤¤¤¤¤¤¤¤¤¤¤¤¤¤¤¤¤¤¤¤¤¤¤¤¤¤¤¤¤¤¤¤¤¤¤¤¤¤¤¤¤
\subsection{Spectral Analysis Results}
\label{sec-resultats-analyse}

Corrected spectra are shown on the right side of Figure~\ref{fig-fts}. First-order side-lobes are suppressed across science channels. Residual second-order side-lobes fall outside active passbands and do not affect characterization.

In the $\J$-band spectrum, a strong residual peak from the 1064~nm supercontinuum pump laser is present. Atmospheric water vapor absorption features are visible near 1.4~\mum (edge of $\J$ band) and 1.8~\mum (between $\BH$ and $\K$ bands).

Metrology band $\I$ cuts off below 790~nm, matching the dichroic beamsplitter short-wavelength reflection boundary. The long-wavelength boundary near 1~\mum is set by the $\I/\J$ dichroic splitter. The gradual response roll-off between 0.85~\mum and 1~\mum reflects silicon detector quantum efficiency roll-off near its indirect bandgap edge.

In the $\J$ band, short-wavelength response extends down to 900~nm. For InGaAs detectors, this extended blue response is caused by multimode fiber coupling offsets induced by lens longitudinal chromatic aberration. The sharp long-wavelength cutoff at 1.65~\mum is set by the science/metrology dichroic splitter.

For the science camera, short-wavelength response starts at 1.64~\mum (matching the $\J$-band cutoff) and extends to 2.44~\mum, bounded by supercontinuum source emission limits and atmospheric absorption.

The 1064~nm pump laser line introduces non-linearities in fringe sensor phase estimates. Near zero OPD, effective mean wavelength is 1370~nm; far from zero OPD, residual 1064~nm pump laser fringes dominate. Installing a narrow-band notch filter at 1064~nm would eliminate this offset.

Furthermore, subtle throughput differences across the 4 MMZ outputs shift effective mean wavelengths between channels. Adjusting spectrometer focusing lenses alters output channel passbands, tuning mean channel wavelengths. Section~\ref{sec-fourier-FS} evaluates the impact of these channel mismatches.

%¤¤¤¤¤¤¤¤¤¤¤¤¤¤¤¤¤¤¤¤¤¤¤¤¤¤¤¤¤¤¤¤¤¤¤¤¤¤¤¤¤¤¤¤¤¤¤¤¤¤¤¤¤¤¤¤¤¤¤¤¤¤¤¤¤¤¤¤¤¤¤¤¤¤¤¤¤¤¤
\subsection{Fractional Bandwidth Calculation}
\label{sec-calcul-largeur}

Comparing \pe null depth results (Chapter~\ref{sec-performance-nulling}) with other nulling testbeds requires adopting a consistent fractional bandwidth metric, defined per channel $i$ as $(\Delta\lambda/\lambda)_i = (\Delta\sigma/\sigma)_i$. This requires defining mean wavenumber $\sigma_i$ and spectral bandwidth $\Delta\sigma_i$ for each channel.

Passbands on the science camera deviate from ideal Gaussian profiles. Treating the full spectrum as a top-hat passband from 0.42 to 0.6~\imum yields a fractional bandwidth of $\sim 35\%$. Testbeds like JPL's \emph{Achromatic Nuller} \cite{Peters08} evaluate Gaussian-like profiles using Full Width at Half Maximum (FWHM). For non-Gaussian passbands, FWHM is ill-defined.

We evaluate bandwidth using an equivalent energy fraction matching Gaussian passbands. For a Gaussian distribution with standard deviation $\sigma$, integrated power $\eta_a$ within interval $a$ centered on the mean is given by the error function:
\begin{equation}
  \eta_a = \erf{\frac{a}{2\sqrt{2}\sigma}}.
\end{equation}

Because FWHM equals $2\sqrt{2\ln2}\sigma$, the integrated power within the FWHM spans $\eta_\mfwhm = \erf{\sqrt{\ln2}} = 76\%$. We define effective passband width $\Delta\sigma_i$ as the spectral interval enclosing 76\% of total channel power.

Applying this 76\% power criterion to the total integrated spectrum (Figure~\ref{fig-fts-K}, black line) yields an integrated fractional bandwidth of $(\Delta\sigma/\sigma)_\mint = 23\%$. This metric applies if output flux is summed onto a single detector.

Because \pe disperses light across $n=9$ science channels, nulling depth is evaluated per channel with equal weighting. Channel passbands derived from Figure~\ref{fig-fts-K} are plotted in Figure~\ref{fig-canal}.

\begin{figure} \centering
  \subfloat[Mean wavenumber per science channel with spectral passband width.]{\label{fig-sigma-vs-canal}
    \FIG{0.49}{false}{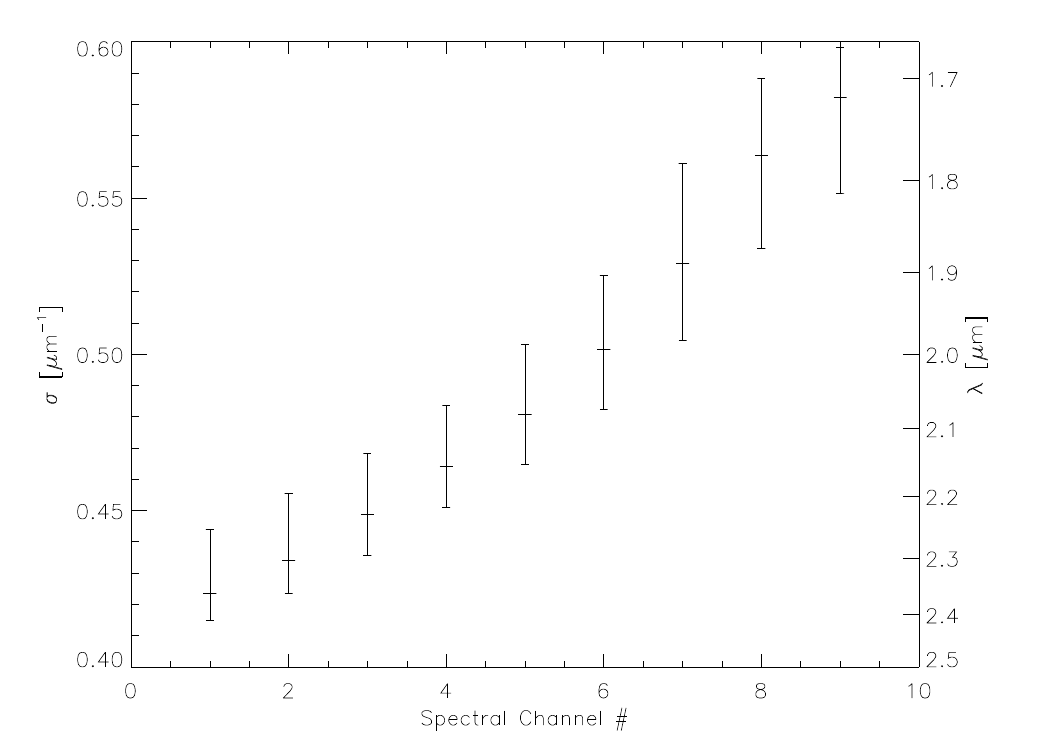}}
  \hfill\subfloat[Fractional bandwidth per science channel.]{\label{fig-bandwidth-vs-canal}
    \FIG{0.49}{false}{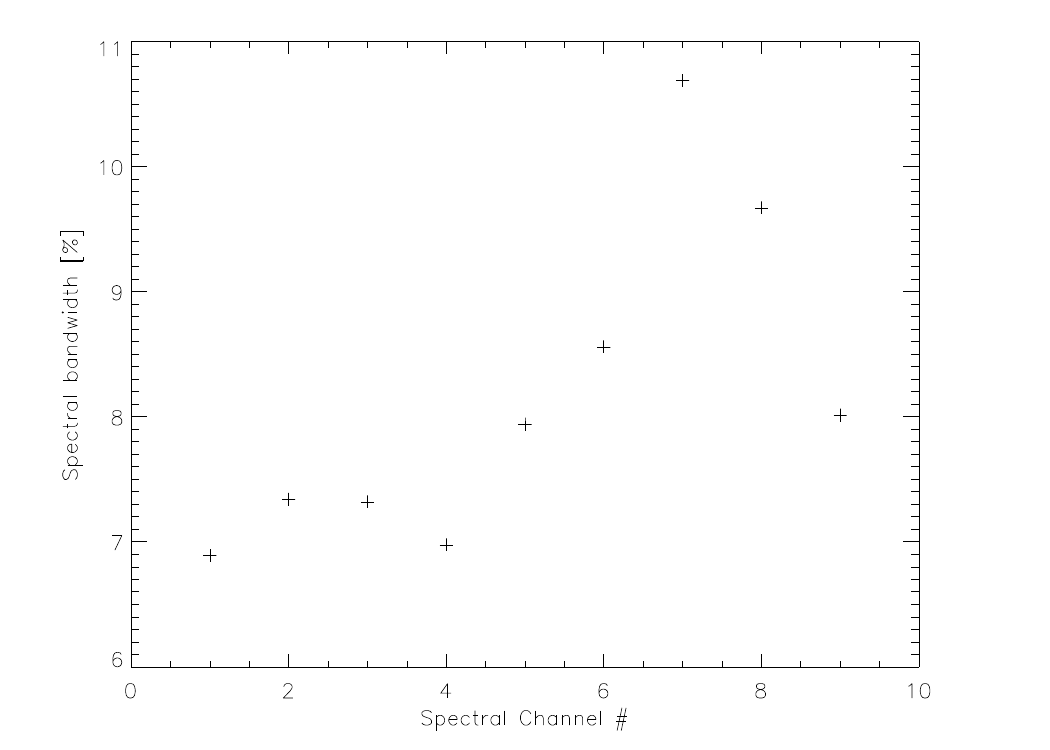}}
  \caption{Spectral characterization of the 9 science channels.}
  \label{fig-canal}
\end{figure}

Figure~\ref{fig-sigma-vs-canal} displays mean wavenumbers and passband widths per channel. Individual channel fractional bandwidths hover near 10\% (Figure~\ref{fig-bandwidth-vs-canal}).

Total effective bandwidth spanning the outer channel boundaries is given by:
\begin{equation}
  \Delta\sigma_\mmoy =
  \GP{\sigma_n+\frac{\Delta\sigma_n}{2}}-\GP{\sigma_1-\frac{\Delta\sigma_1}{2}}.
\end{equation}

This defines a total coverage fractional bandwidth of $\boldsymbol{(\Delta\sigma/\sigma)_\mmoy = 37\%}$ centered at mean wavenumber $\boldsymbol{\sigma_\mmoy = 0.5}$~\bimum (mean wavelength $\boldsymbol{\lambda_\mmoy = 2.0}$~\bmum).

%§§§§§§§§§§§§§§§§§§§§§§§§§§§§§§§§§§§§§§§§§§§§§§§§§§§§§§§§§§§§§§§§§§§§§§§§§§§§§§§
\section{Conclusion}
\label{sec-conclusion}

The \pe testbed was integrated through staged configurations of increasing complexity to isolate individual subsystem errors. Integration required key baseline design modifications. The source injection module was redesigned to deliver broadband light through a single optical fiber core, and the photometric reference channel was relocated upstream of the MMZ combiner to eliminate phase-dependent reference noise. Alignment and calibration procedures were refined, and a real-time control interface was developed to manage tracking loops, disturbance injection, and automated testing sequences. An FTS calibration method was implemented to characterize passbands across metrology and science detectors. With optical alignment precision exceeding baseline specifications, the testbed is fully operational for high-precision cophasing and deep nulling performance evaluation.

%§§§§§§§§§§§§§§§§§§§§§§§§§§§§§§§§§§§§§§§§§§§§§§§§§§§§§§§§§§§§§§§§§§§§§§§§§§§§§§§
\chapter{Optimization of Disturbance Correction}
\label{sec-optimisation-boucles}

\begin{flushright}
 \begin{minipage}{12cm} {\small \textit{There is in all things a pattern that
       is part of our universe. It has symmetry, elegance, and grace
       ---~those qualities you find always in that the true artist captures.
       You can find it in the turning of the seasons, the way sand trails
       along a ridge, in the branch clusters of the creosote bush of the
       pattern of its leaves. We try to copy these patterns in our lives and
       in our society, seeking the rhythms, the dances, the forms that
       comfort. Yet, it is possible to see peril in the finding of ultimate
       perfection. It is clear that the ultimate pattern contains its own
       fixity. In such perfection, all things move towards death.}}

\raggedleft{{\small Frank Herbert, Dune (1965)}}
\end{minipage}
\end{flushright}

\minitoc

\bigskip

%§§§§§§§§§§§§§§§§§§§§§§§§§§§§§§§§§§§§§§§§§§§§§§§§§§§§§§§§§§§§§§§§§§§§§§§§§§§§§§§
\section{Introduction}
\label{sec-introduction-1}

Controlling the phase of an optical system—whether a coronagraph or a nulling interferometer—is essential for high-contrast imaging. In this chapter, I primarily describe the cophasing control system of the \pe testbed, independently of scientific null depth measurements. I describe in details both feedback loops (piston and tip/tilt) and analyze their residual tracking errors. Because \pe aims to validate scientific nulling performance under realistic formation-flying disturbances, I analyze disturbance rejection using two control laws: a baseline integrator and an Linear Quadratic Gaussian (LQG) controller (introduced in Section~\ref{sec-commande-lqg}), optimized to reject narrow-band mechanical vibrations. Finally, I investigate the application of the LQG controller to the \sce coronagraph, focusing specifically on tip/tilt vibration rejection.

%§§§§§§§§§§§§§§§§§§§§§§§§§§§§§§§§§§§§§§§§§§§§§§§§§§§§§§§§§§§§§§§§§§§§§§§§§§§§§§§
\section{Description of \pe's Cophasing Control}
\label{sec-description-controle}

%¤¤¤¤¤¤¤¤¤¤¤¤¤¤¤¤¤¤¤¤¤¤¤¤¤¤¤¤¤¤¤¤¤¤¤¤¤¤¤¤¤¤¤¤¤¤¤¤¤¤¤¤¤¤¤¤¤¤¤¤¤¤¤¤¤¤¤¤¤¤¤¤¤¤¤¤¤¤¤
\subsection{Detailed Description of the Control Loops}
\label{sec-description-detaillee}

%-------------------------------------------------------------------------------
\subsubsection{Characteristics of \pe's Loops}
\label{sec-caracteristiques-boucles}

As discussed in Section~\ref{sec-taux-extinction}, nulling interferometers rely on single-mode optical fibers to filter out higher-order spatial modes that degrade null depth. However, low-order modes subject to dynamic jitter—specifically differential path delay (piston) and individual arm pointing (tip/tilt)—must be actively controlled to maintain deep, stable extinction over time. Furthermore, as highlighted in Section~\ref{sec-objectifs}, \pe is designed to simulate the operational environment of a formation-flying mission. Dynamic disturbances representative of such missions are injected in both piston and tip/tilt modes. Without adequate active control, these high-amplitude perturbations prevent reaching the optical stability required for exoplanet detection.

The piston loop uses a Modified Mach-Zehnder Fringe Sensor (FS) implementing spatial quasi-ABCD modulation (Section~\ref{sec-desc-abcd}) to extract differential path delay. As outlined in Section~\ref{sec-desc-mmz}, the fringe sensor acquires eight signals: four quadrature phase outputs across two spectral metrology channels, supporting fringe unwrapping over an extended dynamic range (Section~\ref{sec-coherencage}). Raw signals are multiplexed at 250~kHz, yielding an effective sampling rate per signal of 250~kHz$/8 = 31.25$~kHz. To match real-time processing constraints, 32 consecutive samples are averaged per control cycle, yielding an operational frame rate $f_\mfs = 31250\text{~Hz}/32 \simeq 1$~kHz. This averaging reduces sensor measurement noise by a factor of $\sqrt{32}=5.6$.

The tip/tilt loop utilizes a Field Relative Angle Sensor (FRAS) comprising a CCD camera imaging both beams through a focusing doublet. An annular pick-off mirror reflects the outer annulus of each beam at a 7~arcmin relative angle. The doublet forms two distinct focal spots on the CCD detector (one per arm) separated by ~70~pixels, allowing independent centroid extraction within dedicated sub-aperture windows. Relative spot motions directly yield tip and tilt tracking errors for each beam. The FRAS loop frame rate is bounded by the maximum camera readout speed and set to a sub-multiple of the piston loop rate to simplify real-time scheduling. The loop rate is fixed at $f_\mfras = f_\mfs/5 \simeq 200$~Hz. Figure~\ref{fig-fs-fras} shows both sensor subsystems: the FS in Figure~\ref{fig-fs} and the FRAS in Figure~\ref{fig-fras}.

\begin{figure} \centering \subfloat[The FS, comprising the Modified Mach-Zehnder (black box, center) and spectrometer units (foreground) with dichroics and multimode fiber couplers (orange fibers).]{\label{fig-fs}\FIG{0.49}{false}{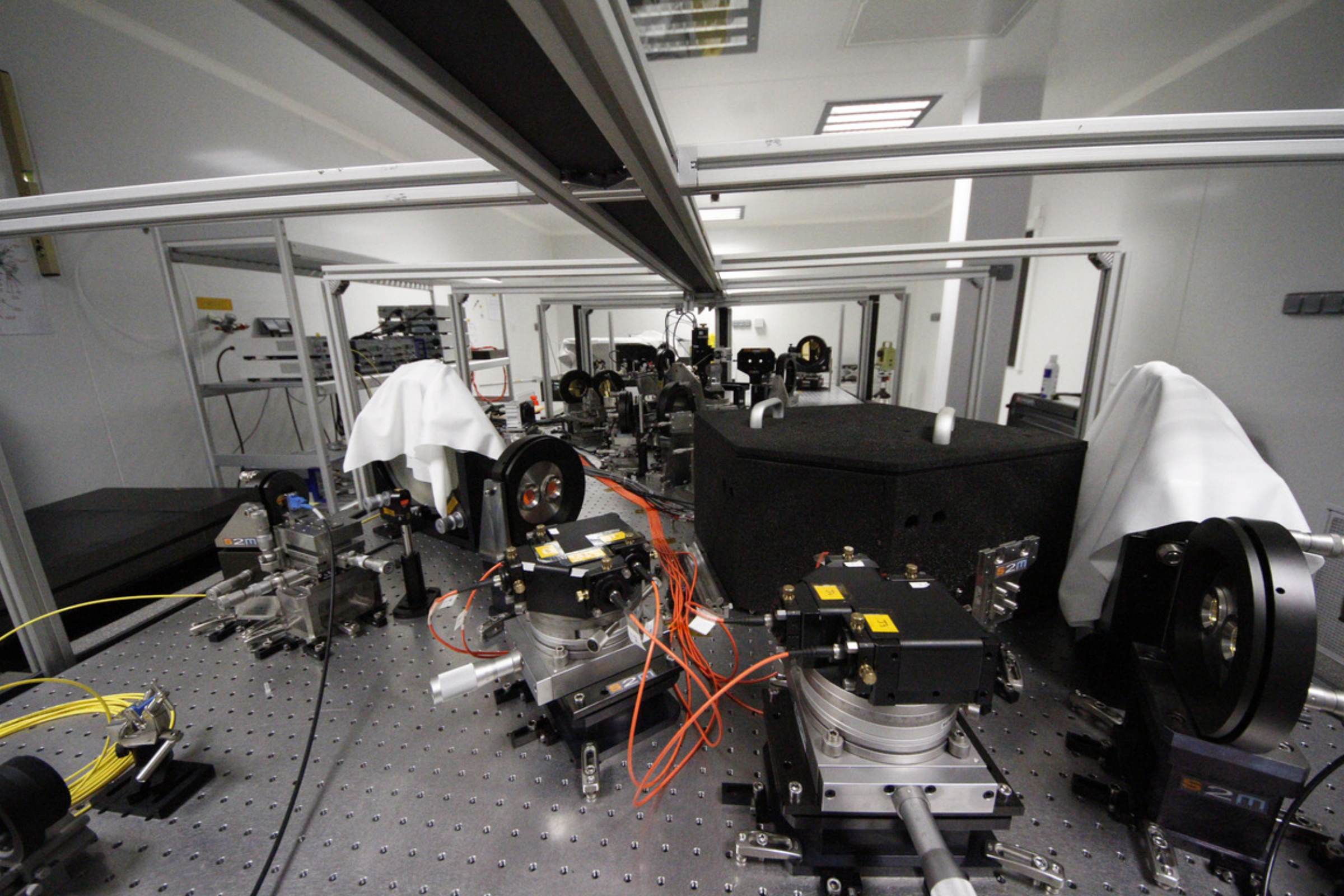}} \hfill\subfloat[The FRAS, comprising annular mirror M11, a focusing lens, and a CCD camera.]{\label{fig-fras}\FIG{0.49}{false}{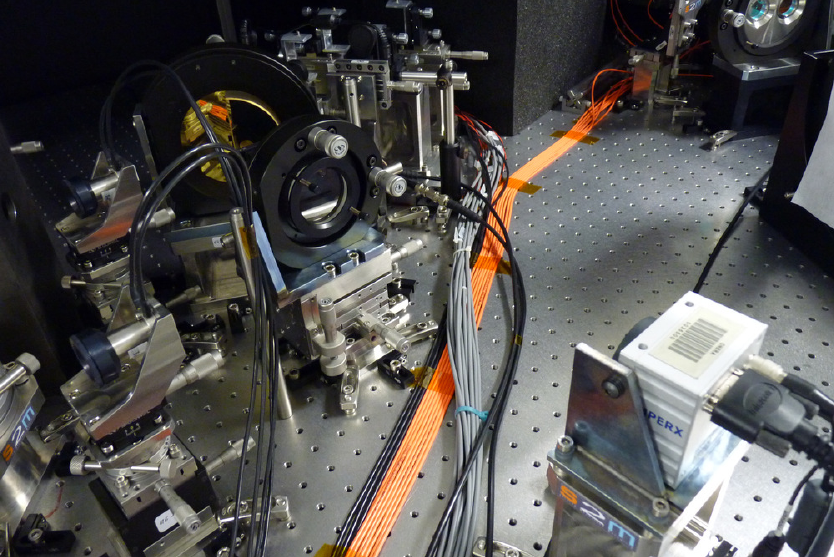}}
  \caption{Piston (FS) and tip/tilt (FRAS) sensors integrated on the \pe testbed.}
  \label{fig-fs-fras}
\end{figure}

Active correction of tip/tilt and differential piston is performed by two flat M6 mirrors mounted on piezoelectric piston-tip/tilt stages (Physik Instrumente S316, one per arm), shown in Figure~\ref{fig-M6}. Each stage incorporates three piezoelectric transducers (PZT) arranged at $120\degree$ intervals. Linear combinations of the three PZT voltages drive piston, tip, and tilt adjustments. Because both control loops share these physical actuators, they are partially coupled. Piston corrections can be driven via a single stage or distributed across both arms.

\begin{figure}\centering
  \FIG{0.5}{false}{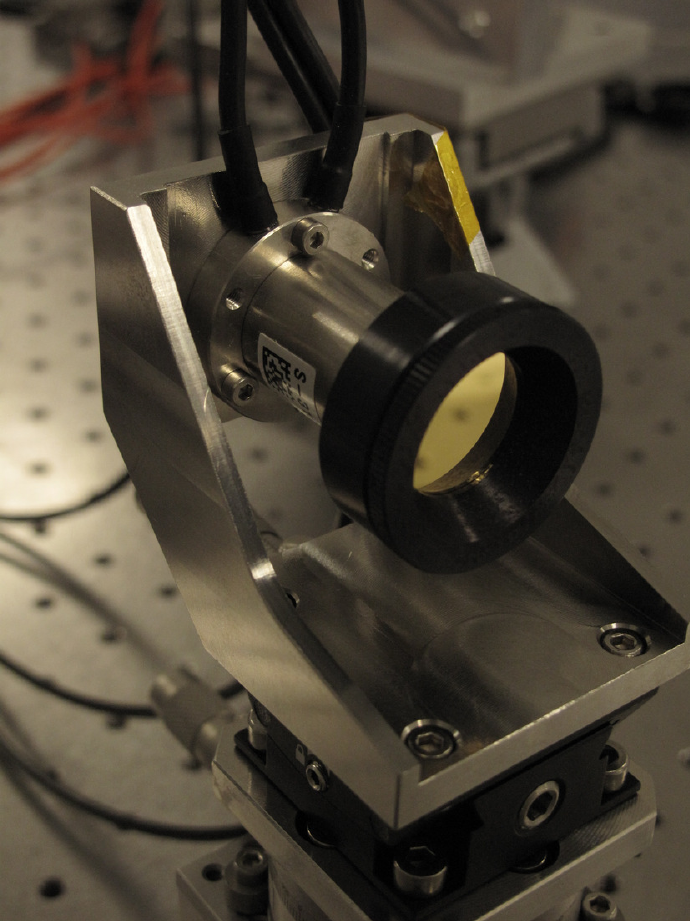}
  \caption{Piezoelectric M6 actuators used on \pe (one per arm).}
  \label{fig-M6}
\end{figure}

%-------------------------------------------------------------------------------
\subsubsection{Real-Time Software Architecture}
\label{sec-description-architecture}

The real-time control system is structured across prioritized execution threads. In deterministic real-time computing, performance relies on execution predictability and low timing jitter rather than pure processing speed. Critical tasks are triggered synchronously by a master hardware clock, while non-deterministic routines execute in the remaining frame margin.

\begin{figure} \centering
  \FIG{0.8}{false}{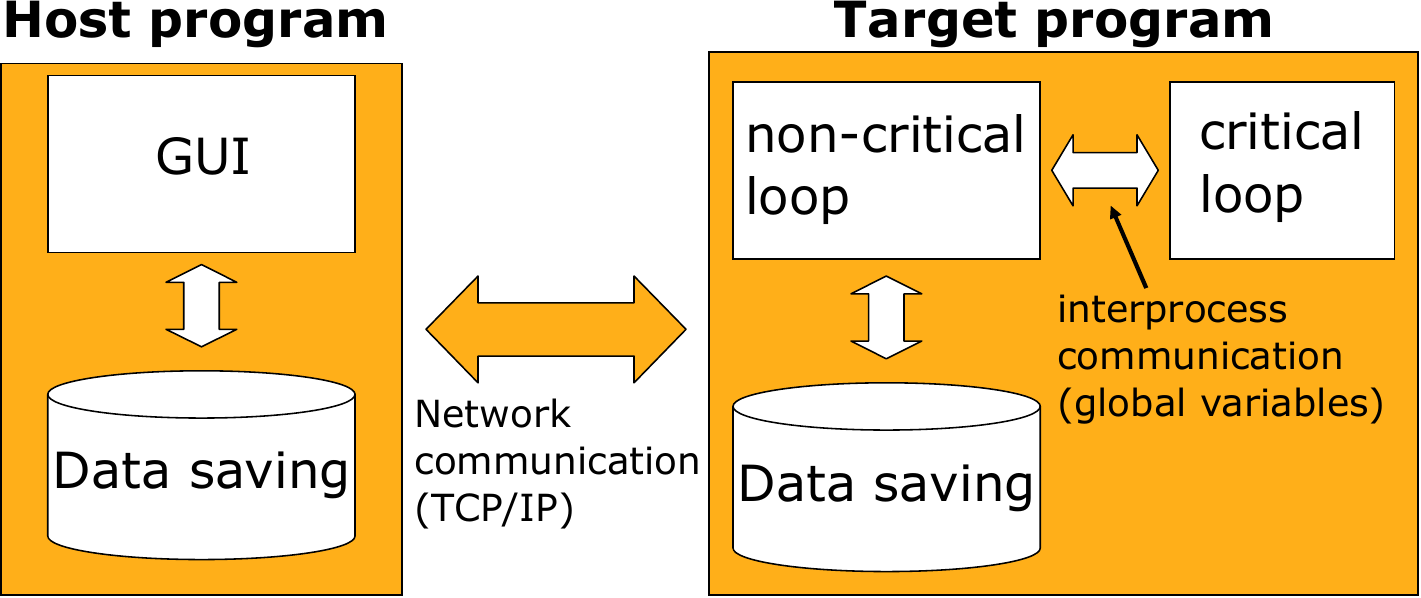}
  \caption{Real-time control architecture of the \pe testbed.}
  \label{fig-archi-rt}
\end{figure}

The software architecture (Figure~\ref{fig-archi-rt}) is split into three functional layers:
\begin{itemize}
\item \textbf{Time Critical Loop (TCL):} Handles time-sensitive operations, including sensor data acquisition, control law calculation, and Digital-to-Analog (DAC) voltage output. The TCL is driven by a hardware clock divided down from a 20~MHz system clock in the real-time chassis to minimize timing jitter;
\item \textbf{Non-Critical Loop (NCL):} Interprocesses with the TCL via shared memory structures to execute lower-priority tasks, such as host communication and disk logging;
\item \textbf{Graphical User Interface (GUI):} Runs on an external host PC, communicating asynchronously with the NCL over TCP/IP. The GUI manages loop states (start/stop), logs data, and receives downsampled diagnostic data at 15~Hz for real-time visualization.
\end{itemize}

The TCL and NCL threads run deterministically on the dedicated real-time target, while the GUI runs on a standard desktop PC.

\begin{figure} \centering
  \FIG{1.}{false}{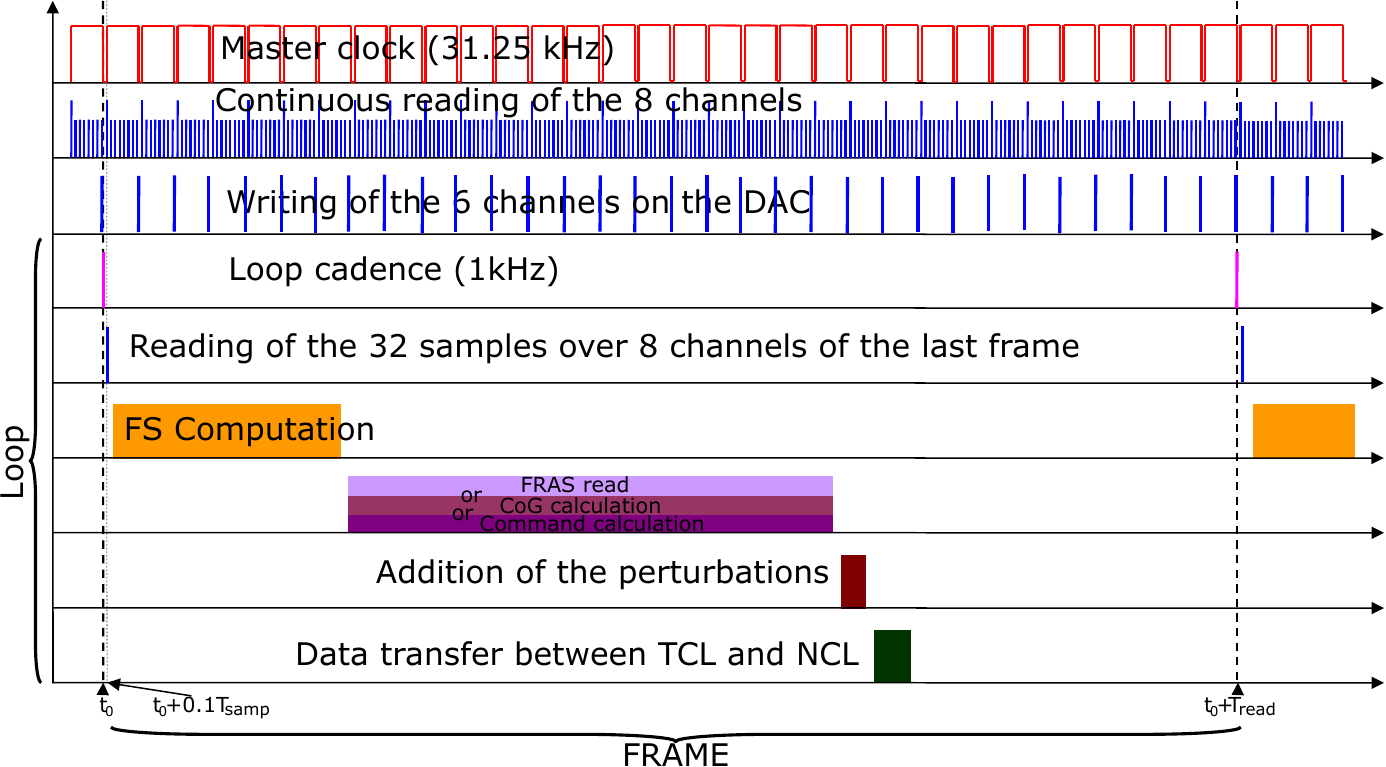}
  \caption[Real-time execution timeline during a single piston loop frame.]{Execution timeline of real-time control tasks within a single piston frame. Tip/tilt processing is multiplexed across 5 consecutive piston frames.}
  \label{fig-chronogrammes}
\end{figure}

Figure~\ref{fig-chronogrammes} details the task execution timeline over a single piston frame. Tasks for the tip/tilt loop (image readout, centroiding, command computation) are multiplexed across 5 consecutive piston frames. Unallocated frame margins are reserved for NCL background processing.

%-------------------------------------------------------------------------------
\subsubsection{Calibration Sequence}
\label{sec-chronologie-etalonnage}

System operation relies on sequential calibration routines that characterize sensor responses and actuator transfer matrices. The overall calibration workflow follows specific steps outlined below:

The first step (Section~\ref{sec-etalonnage-M6}) calibrates the M6 actuators (twin S316 piezo stages). This converts physical piston (in nm) and tip/tilt commands (in arcsec) into 6 drive voltages (V) applied to the PZT elements. This calibration is static unless optomechanical geometry changes.

Next, nominal mirror positions maximizing single-mode fiber coupling into the science channel are identified. A 2D tip/tilt raster scan (Section~\ref{sec-procedure-recherche}) maps focal spot coupling across the fiber core face, extracting optimal setpoints for the tip/tilt tracking loop. Between operational runs, a streamlined routine (Section~\ref{sec-procedure-correction}) tracks and corrects local thermal drifts. While this routine optimizes per-arm coupling, input power balance is adjusted at the M0 collimator focus by tilting the primary source delivery fiber.

The Fringe Sensor is then calibrated (Sections~\ref{sec-etalonnage-abcd} and \ref{sec-etalonnage-FS}). The calibration yields a transformation matrix for each metrology channel converting the four MMZ output intensities (V) into total power per arm, visibility, and differential path delay. This step also extracts central metrology wavelengths required for fringe unwrapping.

When running LQG control, disturbance models (combining lab environments and injected jitter) are identified (Sections~\ref{sec-identification} and \ref{sec-adapt-ident}).

This workflow builds upon calibration methods established by K. Houairi \cite{Houairi09b} during initial testbed studies at ONERA, adapted here for \pe.

%-------------------------------------------------------------------------------
\subsection{Calibration of Piston-Tip/Tilt Actuators}
\label{sec-etalonnage-M6}

The movable M6 mirrors are driven by 6 control voltages $u_i$ (3 per stage), computed from FRAS and FS error signals. To decouple control modes and account for geometric sensitivities, the S316 stages are driven in modal piston-tip/tilt space derived from measurements $y_i$ rather than directly in voltage space. Assuming identical PZT responses, uniform voltage shifts produce pure piston, while differential voltages generate tip/tilt. While this simplified formulation was used during early testing at ONERA, operational tests on \pe revealed residual cross-coupling between piston and tip/tilt. To eliminate mode cross-talk, an explicit calibration method was implemented.

%¤¤¤¤¤¤¤¤¤¤¤¤¤¤¤¤¤¤¤¤¤¤¤¤¤¤¤¤¤¤¤¤¤¤¤¤¤¤¤¤¤¤¤¤¤¤¤¤¤¤¤¤¤¤¤¤¤¤¤¤¤¤¤¤¤¤¤¤¤¤¤¤¤¤¤¤¤¤¤
\subsubsection{Interaction Matrix and Control Matrix}
\label{sec-interaction-commande} 

Actuator calibration relies on establishing the linear relationship between applied drive voltages $\V{u}$ and resulting modal motions $\V{y}$. This relationship is inverted to yield control matrix $\V{M}_\mc$:
\begin{equation}
  \V{u}=\V{M}_\mc\V{y}.
\end{equation}

To decouple both arms, $\V{M}_\mc$ is structured as a block-diagonal matrix:
\begin{equation}
  \V{M}_\mc = \begin{pmatrix}
      \begin{array}{c|c}
        \V{M}_\mc^\ma & \V{0} \\
        \hline
        \V{0} & \V{M}_\mc^\mb \\
      \end{array}
    \end{pmatrix}.
\end{equation}

The setup features 6 control voltages feeding 6 PZT elements, but only 5 physical measurements: differential piston from the FS, and tip/tilt per arm from the FRAS. To map 5 sensor inputs onto 6 actuator outputs, an adjustable piston distribution factor $\alpha_\mcor$ is introduced: a fraction $\alpha_\mcor$ of the total piston command is assigned to arm $\ma$, while $(1-\alpha_\mcor)$ is applied to arm $\mb$. Setting $\alpha_\mcor=0.5$ balances corrections symmetrically across both arms, whereas $\alpha_\mcor=0$ or $1$ drives piston correction exclusively through a single stage.

Control matrix $\V{M}_\mc$ is obtained by inverting interaction matrix $\V{M}_\mi$:
\begin{equation}
  \V{y}=\V{M}_\mi\V{u}.
\end{equation}

Interaction matrix $\V{M}_\mi$ is measured experimentally by recording sensor displacements $\V{\Delta y}$ produced by voltage steps $\V{\Delta u}$:
\begin{equation}
  \V{\Delta y}=\V{M}_\mi\V{\Delta u}.
\end{equation}

Decoupled arms yield a block-diagonal interaction matrix:
\begin{equation}
  \V{M}_\mi = \begin{pmatrix}
      \begin{array}{c|c}
        \V{M}_\mi^\ma & \V{0} \\
        \hline
        \V{0} & \V{M}_\mi^\mb \\
      \end{array}
    \end{pmatrix}.
\end{equation}

%¤¤¤¤¤¤¤¤¤¤¤¤¤¤¤¤¤¤¤¤¤¤¤¤¤¤¤¤¤¤¤¤¤¤¤¤¤¤¤¤¤¤¤¤¤¤¤¤¤¤¤¤¤¤¤¤¤¤¤¤¤¤¤¤¤¤¤¤¤¤¤¤¤¤¤¤¤¤¤
\subsubsection{Interaction Matrix Derivation}
\label{sec--etablissement}

Measuring the full interaction matrix $\V{M}_\mi^j$ per arm $j \in \{\ma,\mb\}$ requires applying voltage steps $\V{\Delta V}^j$ ($\pm 0.5$~V around nominal operating points). Resulting centroid shifts $(\Delta x^j, \Delta y^j)$ are recorded on the FRAS. Applying symmetric steps across arms maintains parallel alignment without altering fringe contrast. However, because large tip/tilt steps shift inter-arm optical phases, raw FS phase measurements during tip/tilt scans cannot be used directly to extract piston coefficients.

Instead, a truncated interaction matrix $\check{\V{M}}_\mi^j$ (dimension $2 \times 3$, units in arcsec.V$^{-1}$) is derived per arm from FRAS measurements alone:
\begin{equation}
  \begin{pmatrix} \Delta x^j\\ \Delta y^j \end{pmatrix}=\check{\V{M}}_\mi^j
  \begin{pmatrix}
    \Delta V_1^j \\
    \Delta V_2^j \\
    \Delta V_3^j
  \end{pmatrix}.
  \label{eq-mat-inter}
\end{equation}

Piston sensitivities omitted from $\check{\V{M}}_\mi^j$ are reconstructed using system geometry and SVD decomposition.

%¤¤¤¤¤¤¤¤¤¤¤¤¤¤¤¤¤¤¤¤¤¤¤¤¤¤¤¤¤¤¤¤¤¤¤¤¤¤¤¤¤¤¤¤¤¤¤¤¤¤¤¤¤¤¤¤¤¤¤¤¤¤¤¤¤¤¤¤¤¤¤¤¤¤¤¤¤¤¤
\subsubsection{Singular Value Decomposition (SVD) of the Interaction Matrix}
\label{sec-calcul-svd}

Deriving control matrix $\V{M}_\mc$ requires pseudo-inverting non-square ($2 \times 3$) matrix $\check{\V{M}}_\mi^j$. We perform Singular Value Decomposition (SVD) on $\check{\V{M}}_\mi^j$:
\begin{equation}
  \check{\V{M}}_\mi^j =
  \GP{\V{M}_\my^j}^T.\V{M}_\ms^j.\V{M}_\muu^j,
  \label{svd}
\end{equation}
where:
\begin{itemize}
\item $\V{M}_\ms^j$ is a diagonal matrix of singular values;
\item $\V{M}_\my^j$ is an orthonormal matrix defining measurement modes;
\item $\V{M}_\muu^j$ is an orthonormal matrix defining actuator voltage modes.
\end{itemize}

SVD decomposes matrix $\check{\V{M}}_\mi^j$ into singular values $s_0$, $s_1$, and $s_2$ corresponding to piston, tip, and tilt modes:
\begin{itemize}
\item Singular values $s_1$ and $s_2$ capture angular sensitivities along orthogonal tip and tilt axes.
\item Singular value $s_0 \approx 0$ corresponds to pure piston, which produces no spot motion on the FRAS.
\end{itemize}

%¤¤¤¤¤¤¤¤¤¤¤¤¤¤¤¤¤¤¤¤¤¤¤¤¤¤¤¤¤¤¤¤¤¤¤¤¤¤¤¤¤¤¤¤¤¤¤¤¤¤¤¤¤¤¤¤¤¤¤¤¤¤¤¤¤¤¤¤¤¤¤¤¤¤¤¤¤¤¤
\subsubsection{Control Matrix Calculation}
\label{sec-calc-mat-com}

Control matrix $\V{M}_\mc$ is calculated via pseudo-inversion:
\begin{equation}
  \V{M}_\mc^j =
  \GP{\V{M}_\muu^j}^T\GP{\V{M}_\ms^j}^\dagger\V{M}_\my^j,
  \label{matcom_inter}
\end{equation}
where $^\dagger$ represents pseudo-inversion of singular matrix $\V{M}_\ms^j$.

Three inversion strategies can be considered:
\begin{enumerate}
\item \textbf{Filtered inversion:} The unobserved piston mode is zeroed out in singular space:
  \begin{equation}
    \GP{\V{M}_\ms^j}^\dagger = \begin{pmatrix}
      0 & 0 & 0 \\
      0 & s_1^{-1} & 0 \\
      0 & 0 & s_2^{-1} \\
    \end{pmatrix}.
  \end{equation}
  This yields a $3 \times 2$ control matrix that ignores piston errors. This is unsuitable for \pe because active piston control is required.
\item \textbf{Heuristic extension:} A third column is appended using manufacturer-specified nominal PZT sensitivities $k$ (in \mum.V$^{-1}$) assumed equal across all actuators.
\item \textbf{Full SVD inversion:} The first row of voltage matrix $\V{M}_\muu^j$ defines the linear combination driving pure piston. Because $\V{M}_\muu^j$ is orthonormal, absolute scaling is fixed by setting singular value $\V{M}_\ms[0,0]$ using geometric modeling or wavelength-calibrated phase steps.
\end{enumerate}

Option 2 was used during initial testbed setup. However, assuming nominal manufacturer sensitivities required manual gain adjustments when changing mirror incidence angles (e.g., autocollimination vs. 30\degree incidence).

Option 3 (full SVD inversion) was subsequently adopted, exploiting measured mode shapes while calibrating absolute piston scaling $\V{M}_\ms[0,0]$ from optical path steps.

%¤¤¤¤¤¤¤¤¤¤¤¤¤¤¤¤¤¤¤¤¤¤¤¤¤¤¤¤¤¤¤¤¤¤¤¤¤¤¤¤¤¤¤¤¤¤¤¤¤¤¤¤¤¤¤¤¤¤¤¤¤¤¤¤¤¤¤¤¤¤¤¤¤¤¤¤¤¤¤
\subsubsection{Sampling Resolution Effects in FRAS Calibration}
\label{sec-influence-resolution}

The FRAS measures angular offsets by calculating focal spot centroids on a CCD camera. The focusing lens focal length ($F_\mfras = 260$~mm) was sized to sample the diffraction spot over ~2 pixels ($p_{pix} = 7.5$~\mum) for a beam diameter $D_\mfras = 13.33$~mm over $0.6$–$0.8$~\mum wavelengths. At $\lambda=0.7$~\mum, the diffraction spot size is:
\begin{equation*}
  \frac{\lambda F_\mfras}{D_\mfras} = 13.7\text{~\mum},
\end{equation*}
corresponding to $1.8$~pixels per resolution element. However, replacing the dichroic pick-offs with annular pick-off mirrors directed the full source bandwidth (0.4–1.0~\mum, peaking near 0.5~\mum) onto the CCD. At shorter wavelengths, diffraction spot size shrunk to $9.8$~\mum ($1.3$~pixels), violating Nyquist-Shannon spatial sampling criteria (>2 pixels). This spatial under-sampling produced periodic centroiding errors, shown in Figures~\ref{fig-non-lin-fras-x} and \ref{fig-non-lin-fras-y}.

\begin{figure} \centering
  \subfloat[Measurement along the $x$-axis.]{\label{fig-non-lin-fras-xx}
    \FIG{0.49}{false}{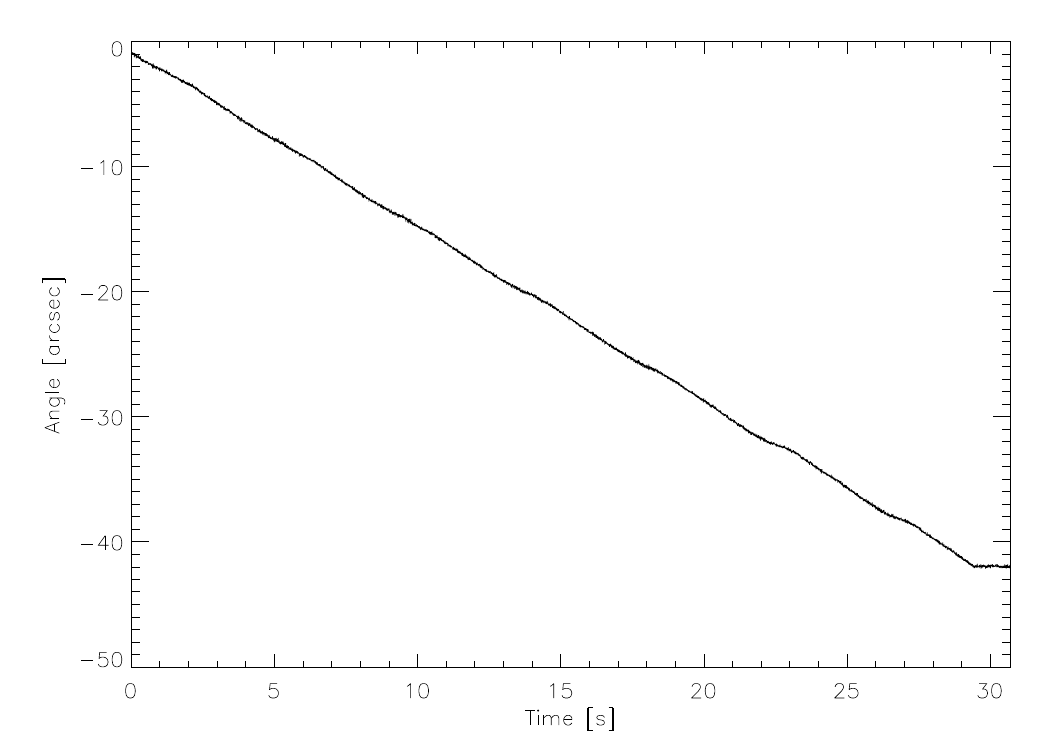}}
  \hfill\subfloat[Measurement along the $y$-axis.]{\label{fig-non-lin-fras-xy}
    \FIG{0.49}{false}{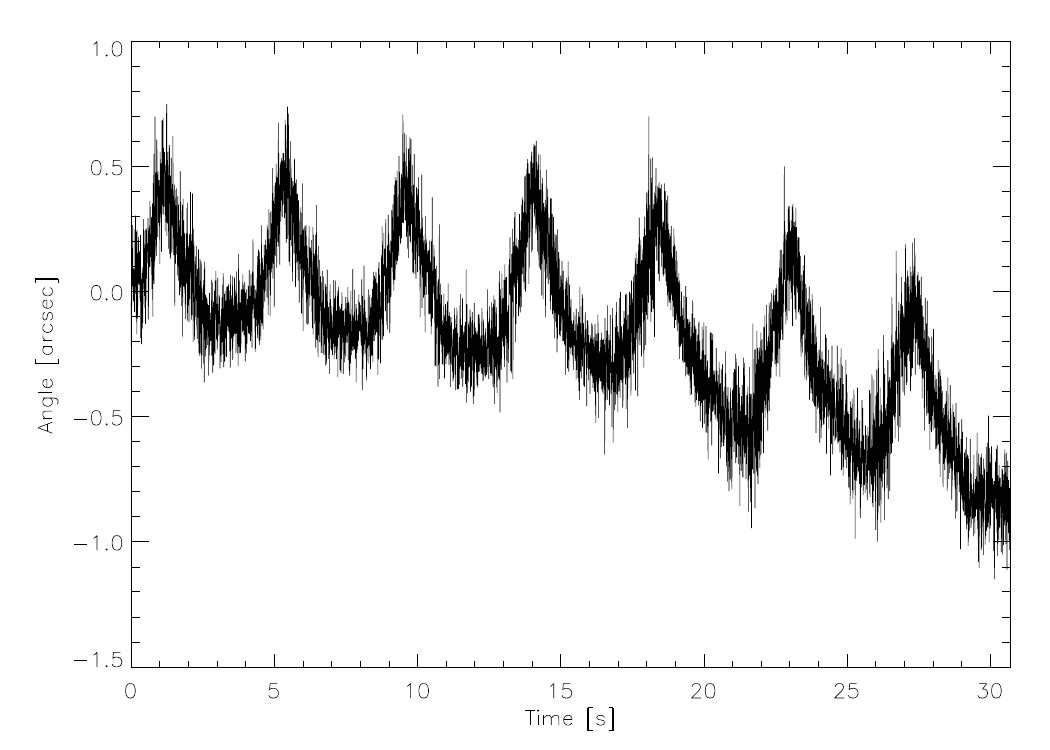}}
  \caption[FRAS non-linearities under $x$-axis tilt ramps.]{FRAS response along $x$ and $y$ axes during an $x$-axis tilt ramp driven by the S330 piezo stage.}
  \label{fig-non-lin-fras-x}
\end{figure}

\begin{figure} \centering
  \subfloat[Measurement along the $x$-axis.]{\label{fig-non-lin-fras-yx}
    \FIG{0.49}{false}{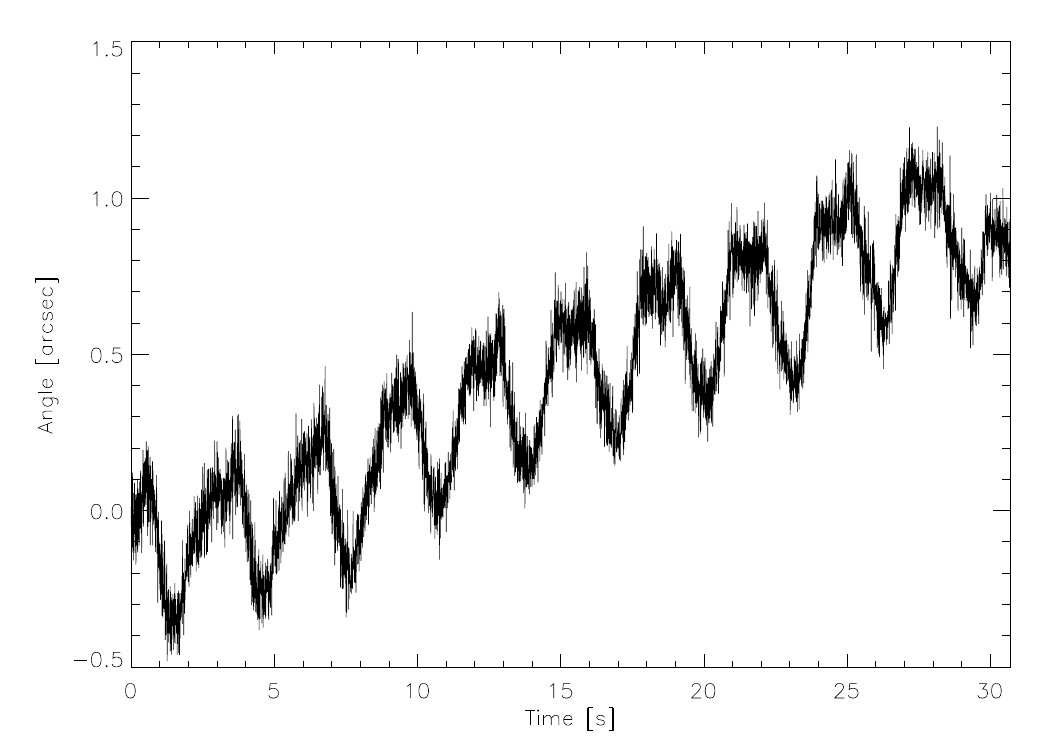}}
  \hfill\subfloat[Measurement along the $y$-axis.]{\label{fig-non-lin-fras-yy}
    \FIG{0.49}{false}{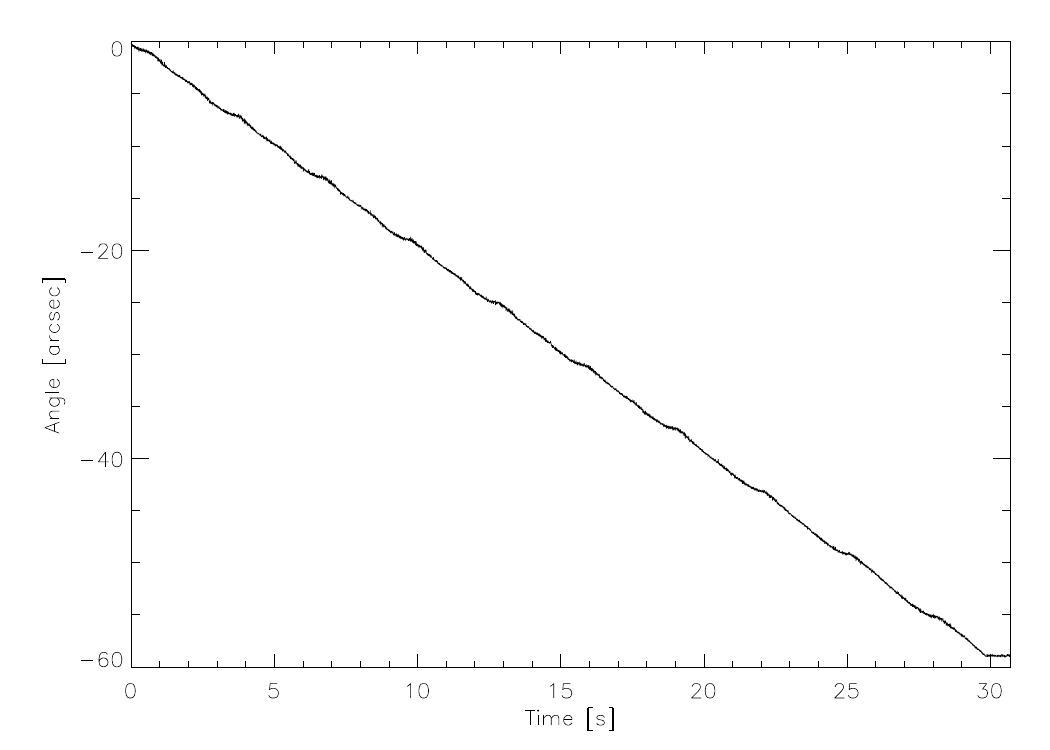}}
  \caption[FRAS non-linearities under $y$-axis tilt ramps.]{FRAS response along $x$ and $y$ axes during a $y$-axis tilt ramp driven by the S330 piezo stage.}
  \label{fig-non-lin-fras-y}
\end{figure}

Figures~\ref{fig-non-lin-fras-x} and \ref{fig-non-lin-fras-y} plot FRAS centroid measurements as a 60~arcsec ramp over 30~s is driven through steering mirror M1$\mb$ (S330 stage).

In Figure~\ref{fig-non-lin-fras-xx}, the measured $x$-axis angular excursion spans 42~arcsec because the 45\degree mirror geometry scales mechanical $x$-axis motion by $1/\sqrt{2}$ relative to $y$-axis motion.

Cross-axis coupling in Figures~\ref{fig-non-lin-fras-xy} and \ref{fig-non-lin-fras-yx} (1~arcsec $y$-motion during $x$-ramps) reveals a 2\degree angular misalignment between S330 actuator axes and the CCD camera frame.

Additionally, periodic non-linear ripples with a period of 6~arcsec (corresponding to 1 CCD pixel) affect the measurements. Spatial under-sampling causes focal spot energy to modulate as centroids cross pixel boundaries and thresholding masks. This introduced systematic centroiding errors exceeding $0.1$~arcsec.

%¤¤¤¤¤¤¤¤¤¤¤¤¤¤¤¤¤¤¤¤¤¤¤¤¤¤¤¤¤¤¤¤¤¤¤¤¤¤¤¤¤¤¤¤¤¤¤¤¤¤¤¤¤¤¤¤¤¤¤¤¤¤¤¤¤¤¤¤¤¤¤¤¤¤¤¤¤¤¤
\subsubsection{Ramp-Based Calibration Method}
\label{sec-procedure-rampes}

To mitigate spatial sampling artifacts, voltage steps were replaced by continuous bidirectional linear ramps driven at constant velocity. Interaction matrix $\check{\V{M}}_\mi^j$ was extracted from measured velocity slopes:
\begin{equation}
  \begin{pmatrix}
	\der[t]{x} \\[+6pt]
	\der[t]{y}
  \end{pmatrix} = \V{M}_\mi
  \begin{pmatrix}
	\der[t]{V_1} \\[+6pt]
	\der[t]{V_2} \\[+6pt]
	\der[t]{V_3}
  \end{pmatrix}.
\end{equation}

Averaging across multiple pixel crossings smoothed out local sub-pixel non-linearities, yielding a more robust calibration.

Additionally, a high-pass optical filter (cutoff wavelength $0.8$~\mum) was installed upstream of the FRAS focusing lens. Truncating shorter wavelengths enlarged the diffraction spot, restoring Nyquist sampling.

\begin{figure} \centering
  \FIG{0.9}{false}{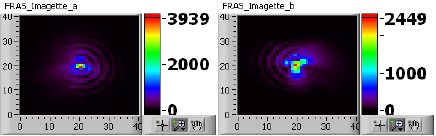}
  \caption{Focal spots from both arms imaged on the FRAS camera.}
  \label{fig-image-fras}
\end{figure}

Figure~\ref{fig-image-fras} displays sample FRAS sub-aperture images. Sub-frame windows span $20 \times 20$~pixels ($120 \times 120$~arcsec). Outer diffraction rings are prominent due to the 75\% central linear obscuration of the annular pick-off mirror. Low-order comatic aberration is visible, particularly in arm $\mb$. Centroid tracking across these spot profiles provides the error signals driving tip/tilt calibration.

\begin{figure} \centering
  \subfloat[Time-series centroid tracks.]{\label{fig-calib-fras1}
    \FIGH{5.5}{false}{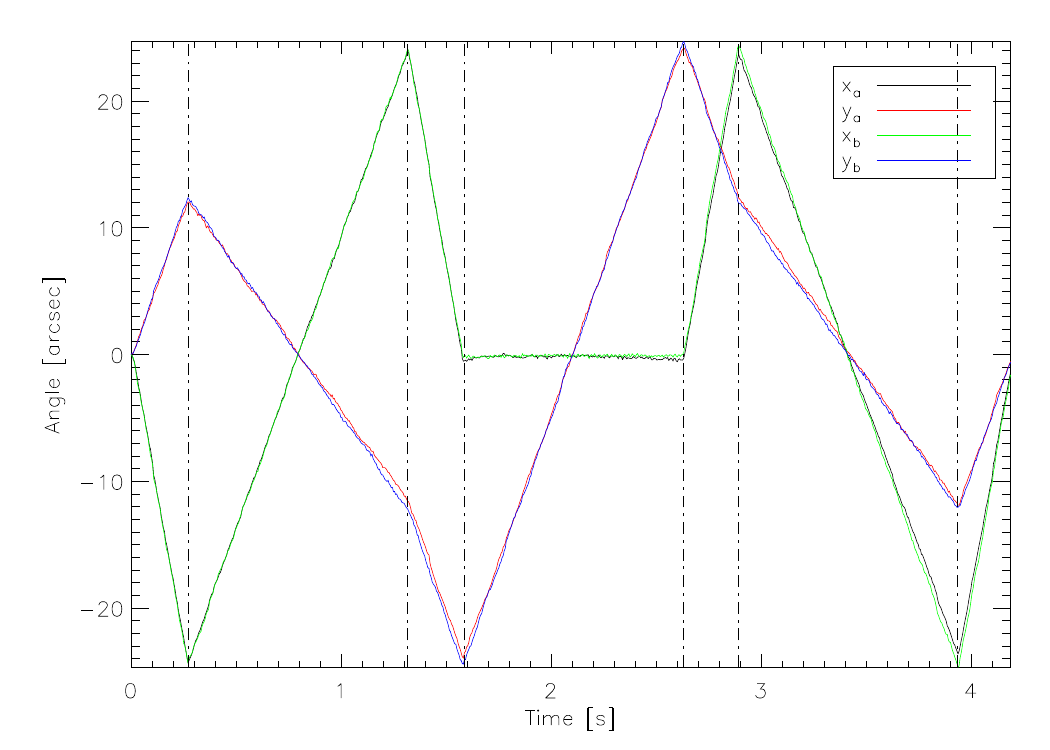}}
  \hfill\subfloat[2D $x$-$y$ position plots.]{\label{fig-calib-fras2}
    \FIGH{5.5}{false}{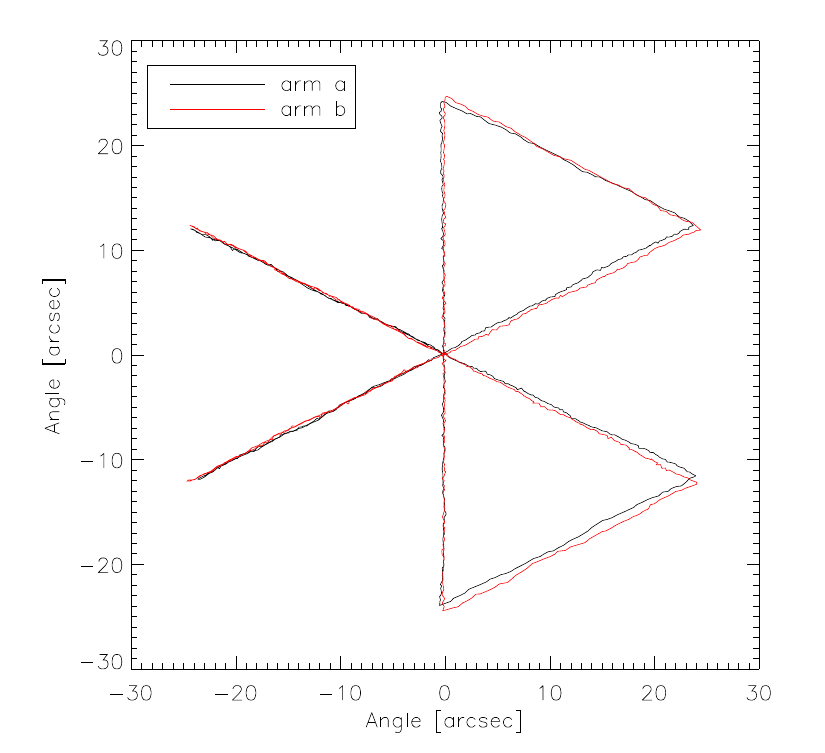}}
  \caption[FRAS calibration dataset.]{Centroid tracks and 2D position plots recorded during FRAS actuator calibration.}
  \label{fig-calib-fras}
\end{figure}

Figure~\ref{fig-calib-fras} shows centroid tracks recorded during actuator calibration in autocollimination. The 2D $x$-$y$ plot (Figure~\ref{fig-calib-fras2}) shows the regular hexagonal pattern generated by driving the three 120\degree PZT channels. In the final bench layout (30\degree incidence on M6), the pattern compresses vertically by $\cos(30\degree)$. Combining ramp-based fitting with optical bandpass filtering successfully eliminated sub-pixel ripples.

%-------------------------------------------------------------------------------
\subsection{Actuator Characterization}
\label{sec-analyse-actionneurs}

%¤¤¤¤¤¤¤¤¤¤¤¤¤¤¤¤¤¤¤¤¤¤¤¤¤¤¤¤¤¤¤¤¤¤¤¤¤¤¤¤¤¤¤¤¤¤¤¤¤¤¤¤¤¤¤¤¤¤¤¤¤¤¤¤¤¤¤¤¤¤¤¤¤¤¤¤¤¤¤
\subsubsection{Linearity and Hysteresis}
\label{sec-linearite}

Optimal feedback control requires linear actuator responses with minimal hysteresis. Piston linearity is particularly critical for deep nulling. We evaluated M6 stage piston response using the fringe sensor by driving a 16~\mum bidirectional displacement ramp. Results are plotted in Figure~\ref{fig-hyst}.

\begin{figure} \centering
\vspace{-10pt}
  \subfloat[Closed-loop strain-gauge feedback ON.]{\label{fig-hyst-avjau}
    \FIG{.46}{false}{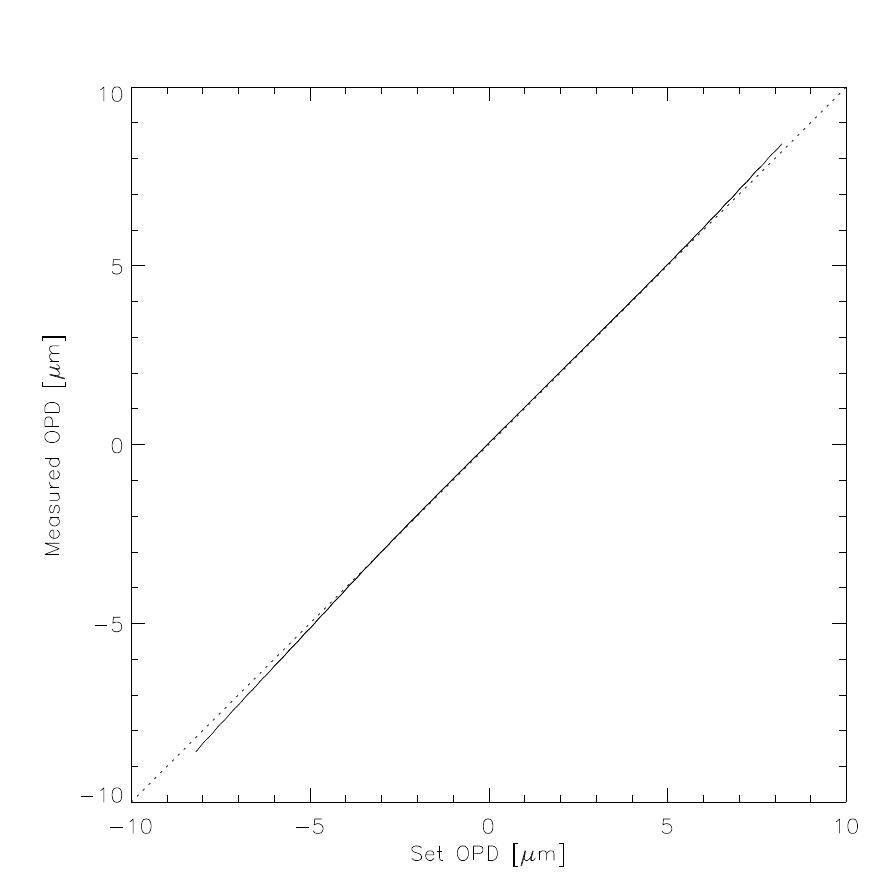}}
  \hfill\subfloat[Open-loop strain-gauge feedback OFF.]{\label{fig-hyst-ssjau}
    \FIG{.46}{false}{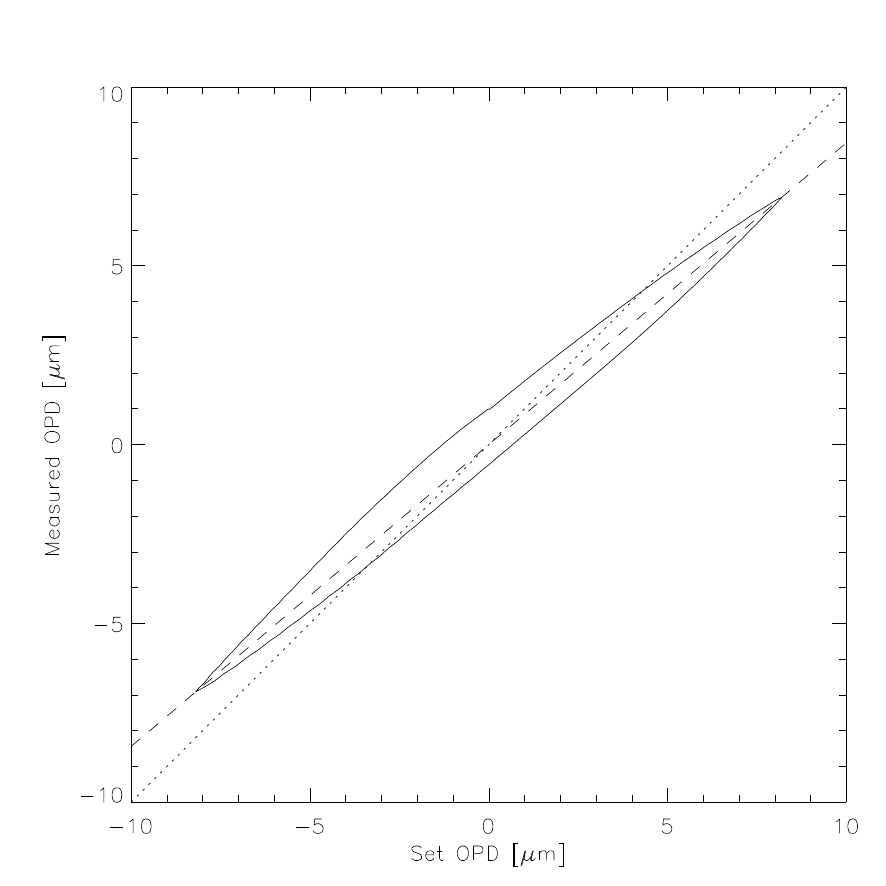}}
  \caption{Linearity and hysteresis curves for an M6 piezo actuator.}
  \label{fig-hyst}
\end{figure}

Figure~\ref{fig-hyst-avjau} shows stage response with internal strain-gauge feedback active. Response is highly linear without measurable hysteresis. Minor slope deviations reflect slight metrology wavelength calibration offsets. Internal strain-gauge sensors eliminate inherent PZT hysteresis. When strain-gauge feedback is disabled (Figure~\ref{fig-hyst-ssjau}), pronounced hysteresis (>2~\mum loop width) and a 16\% gain reduction appear (total displacement drops from 16~\mum to $13.4$~\mum).

Disabling strain gauges eliminates sensor electronics noise, reducing actuator position noise twofold according to manufacturer specifications. However, the resulting hysteresis must be evaluated against performance trade-offs.

%¤¤¤¤¤¤¤¤¤¤¤¤¤¤¤¤¤¤¤¤¤¤¤¤¤¤¤¤¤¤¤¤¤¤¤¤¤¤¤¤¤¤¤¤¤¤¤¤¤¤¤¤¤¤¤¤¤¤¤¤¤¤¤¤¤¤¤¤¤¤¤¤¤¤¤¤¤¤¤
\subsubsection{Step Response}
\label{sec-rep-temp}

Actuator dynamic response was evaluated by applying 300~nm piston steps and measuring path displacement with the fringe sensor. Both S316 stages were tested with strain gauges enabled and disabled (Figure~\ref{fig-step-m6}).

\begin{figure} \centering
  \subfloat[S316 stage driving mirror M6$\ma$.]{\label{fig-step-m6a}
    \FIG{0.49}{false}{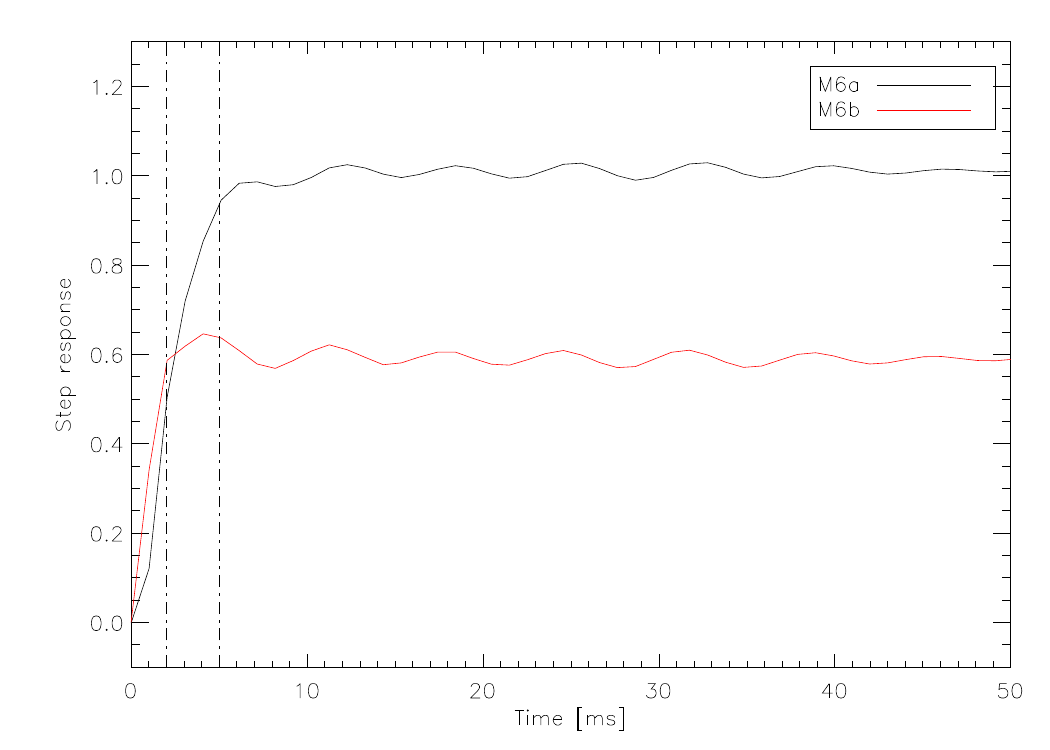}}
  \hfill\subfloat[S316 stage driving mirror M6$\mb$.]{\label{fig-step-m6b}
    \FIG{0.49}{false}{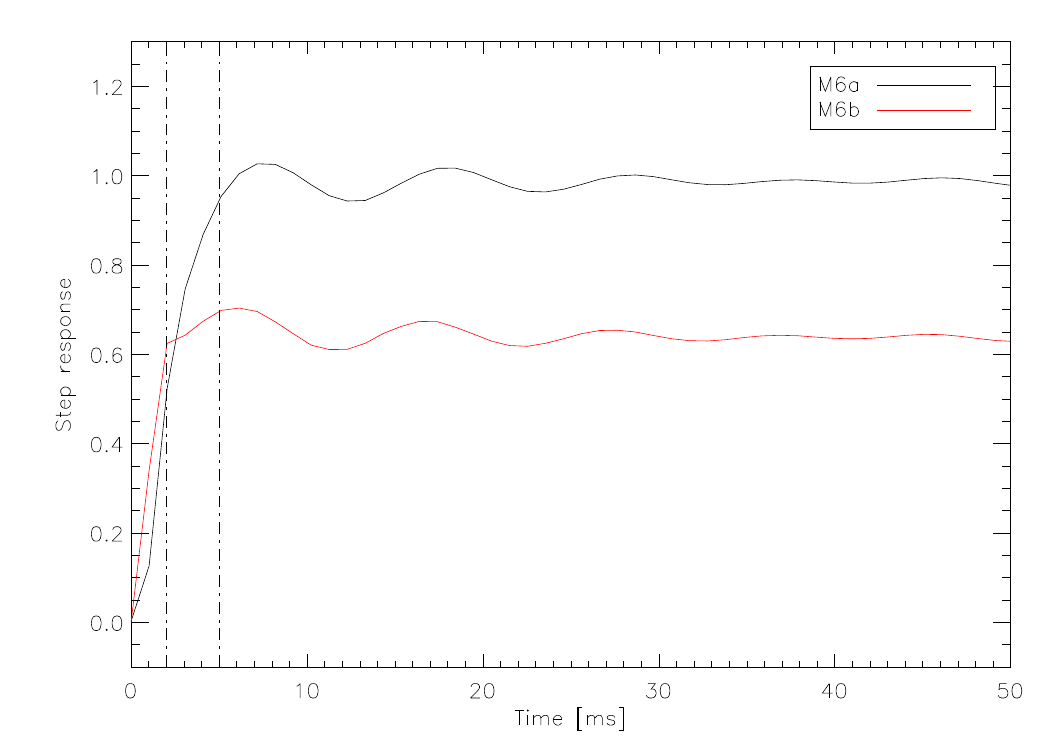}}
  \caption[Step response of the twin S316 piezo stages.]{Piston step responses for both S316 stages with strain gauges active (black) and disabled (red).}
  \label{fig-step-m6}
\end{figure}

Step responses to 10~nm steps yielded identical normalized settling curves, confirming linear small-signal dynamics. With strain gauges enabled, settling times span ~5 fringe sensor frames (5~ms). Disabling strain gauges speeds up rise times to 2~ms, but introduces a 40\% steady-state gain offset.

Figure~\ref{fig-step-m6a} exhibits a 140~Hz mechanical ringing mode (7~ms period) on stage M6$\ma$, while stage M6$\mb$ (Figure~\ref{fig-step-m6b}) rings at 125~Hz (8~ms period). Accelerometer testing confirmed these structural resonance frequencies. Frequency differences between nominally identical mounts stem from differential mounting torque, cabling tension, or spring preloads.

%§§§§§§§§§§§§§§§§§§§§§§§§§§§§§§§§§§§§§§§§§§§§§§§§§§§§§§§§§§§§§§§§§§§§§§§§§§§§§§§
\section{Characterization of \pe's Tip/Tilt Control}
\label{sec-optimisation-FRAS}

%-------------------------------------------------------------------------------
\subsection{Tip/Tilt Sensor Calibration}
\label{sec-ref-tt}

By optical design, 1 pixel on the FRAS camera corresponds to 6~arcsec on the testbed. Sky-equivalent angles are obtained by dividing by the magnification factor of the afocal beam compressors (20$\times$ for \peg; $1\times$ on \pe without compressors). Operational tracking requires identifying the reference pixel coordinates that maximize single-mode fiber coupling into the science channel.

%¤¤¤¤¤¤¤¤¤¤¤¤¤¤¤¤¤¤¤¤¤¤¤¤¤¤¤¤¤¤¤¤¤¤¤¤¤¤¤¤¤¤¤¤¤¤¤¤¤¤¤¤¤¤¤¤¤¤¤¤¤¤¤¤¤¤¤¤¤¤¤¤¤¤¤¤¤¤¤
\subsubsection{Identification of Optimal Setpoints}
\label{sec-procedure-recherche}

Optimal setpoints are mapped by scanning each arm sequentially while blocking the opposite arm. Motorized beam shutters (Figure~\ref{fig-shutter}) were installed on each arm to replace manual piezo offsets, preventing mechanical stress and eliminating parasitic stray light during single-arm scans.

\begin{figure} \centering
  \FIG{.7}{false}{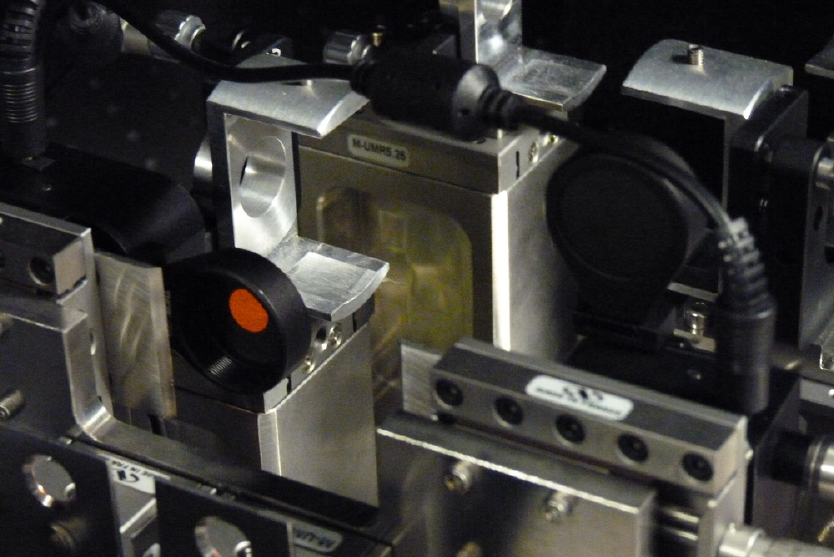}
  \caption[Motorized beam shutters.]{Motorized shutters for single-arm beam isolation (shutter $\mb$ closed, shutter $\ma$ open).}
  \label{fig-shutter}
\end{figure}

The calibration routine executes a 2D tip/tilt raster scan while logging coupled intensity on the science camera. Figure~\ref{fig-balayage1} illustrates the scan pattern; Figure~\ref{fig-balayage2} shows the recorded intensity sequence (arm $\ma$ scan followed by arm $\mb$ scan).

\begin{figure} \centering
  \subfloat[Raster scan pattern.]{\label{fig-balayage1}
    \FIG{0.42}{false}{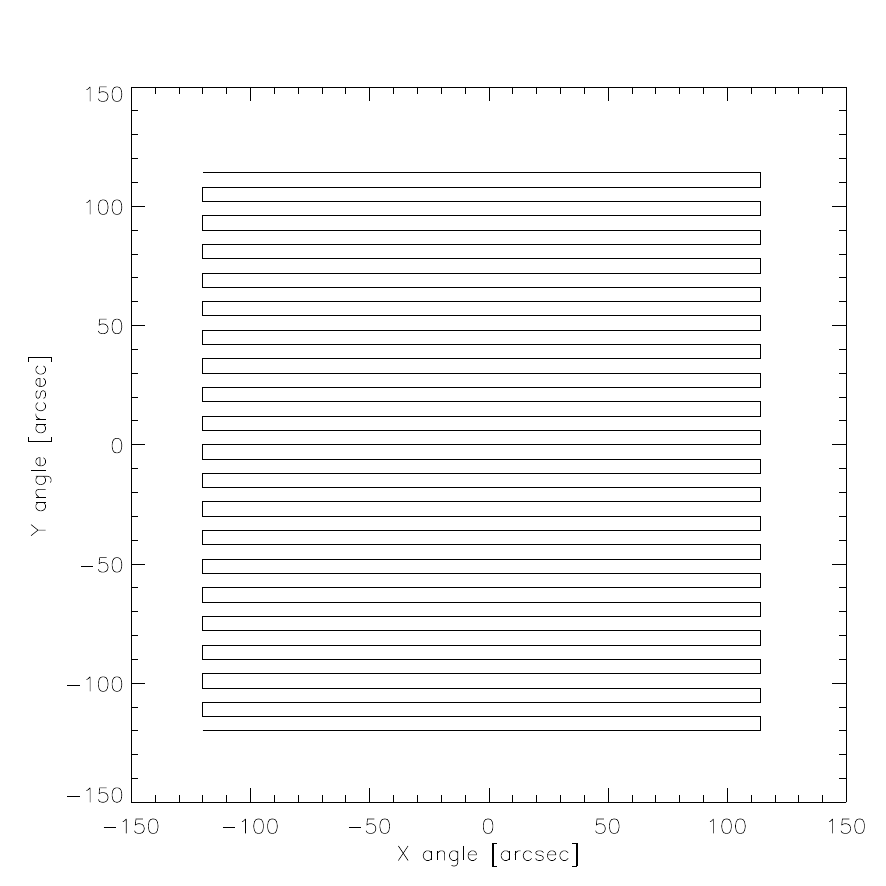}}
  \hfill\subfloat[Time-series intensity.]{\label{fig-balayage2}
    \FIG{0.56}{false}{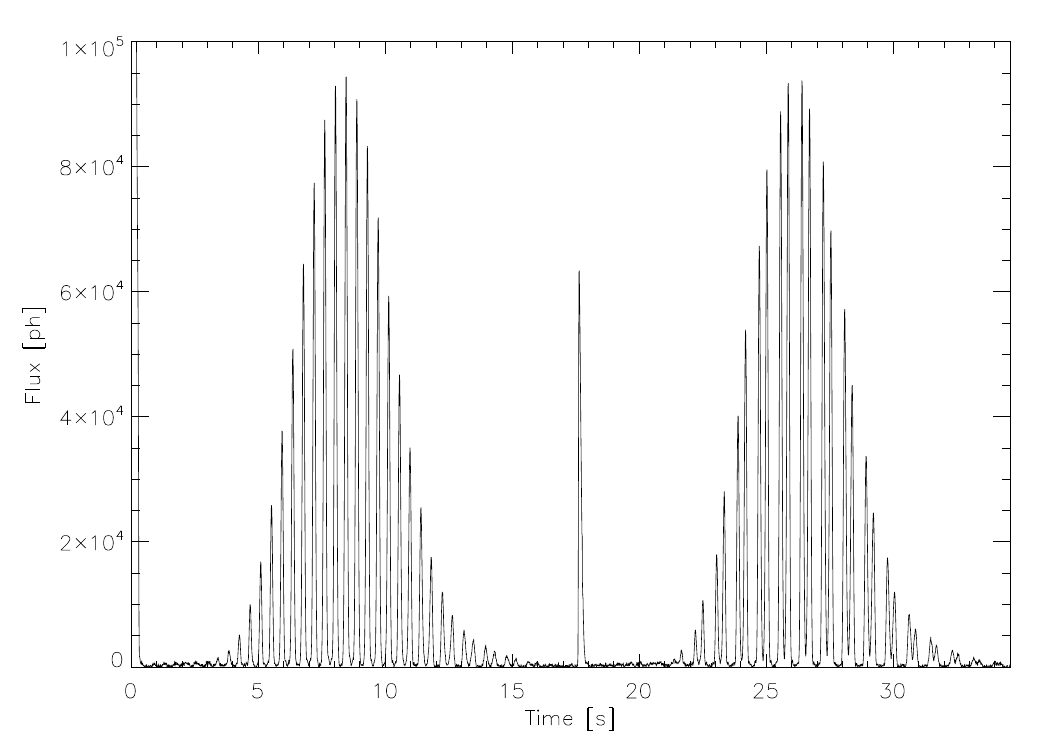}}
  \caption[Fiber coupling calibration dataset.]{Tip/tilt raster scan pattern and total integrated science camera intensity.}
  \label{fig-balayage}
\end{figure}

2D coupling maps are reconstructed by folding time-series data using known scan coordinates. Processing delays, actuator rise times, and camera readout latency introduce time offsets between command issuance and data acquisition. Exploiting scan symmetry along $x$-axis sweeps, these latency offsets are measured and interpolated out. Reconstructed coupling maps per spectral channel are shown in Figure~\ref{fig-recons}.

\begin{figure} \centering
  \FIG{1.}{false}{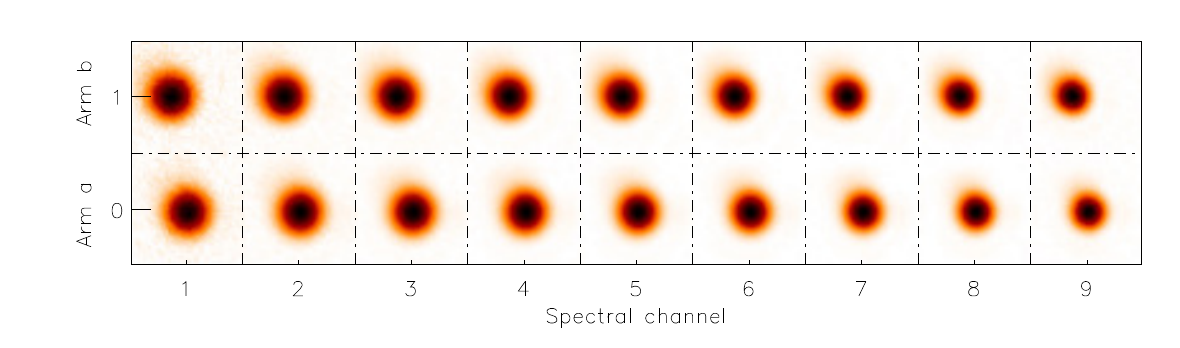}
  \caption[Single-mode fiber coupling efficiency maps.]{Single-mode fiber coupling efficiency maps across science camera spectral channels (wavelengths decrease from left to right).}
  \label{fig-recons}
\end{figure}

Figure~\ref{fig-recons} maps fiber coupling efficiency—representing the spatial overlap integral between the optical PSF and the fiber fundamental mode.

According to \cite{Ruilier01}, single-mode fiber coupling efficiency $I$ is approximated by:
\begin{equation}
  I \simeq I_{\minj,\mmax}\exp\GC{-\frac45\GP{a_2^2+a_3^2}},
  \label{eq-inj}
\end{equation}
where $a_2$ and $a_3$ are Zernike tip and tilt coefficients. These coefficients relate to physical pointing angles $\alpha$ and $\beta$ via:
\begin{equation}
  \left\{
    \begin{array}{r@{\hs}c@{\hs}l}
      \alpha-\alpha_\mref &=& \dfrac{2\lambda a_2}{\pi D} \\[+6pt]
      \beta-\beta_\mref &=& \dfrac{2\lambda a_3}{\pi D} \\
    \end{array}
  \right.,
  \label{eq-inj-coeff}
\end{equation}
where $D$ is beam diameter ($D=9$~mm), $\alpha$ and $\beta$ are tip/tilt angles, and $\alpha_\mref, \beta_\mref$ are setpoints for peak coupling.

The Full Width at Half Maximum (FWHM) coupling diameter $\theta_\minj$ is:
\begin{equation}
  \theta_\minj = \frac{2\sqrt{5\ln2}}{\pi}\frac{\lambda}{D}.
\end{equation}

Coupling spot size scales linearly with wavelength $\lambda$. On \pe, $\theta_\minj$ ranges from 45 to 67~arcsec across science channels. In Figure~\ref{fig-recons} (120~arcsec field per box), spot sizes shrink visually at shorter wavelengths (left to right).

Centroids calculated across broadband-integrated maps establish setpoint coordinates $(\alpha_\mref, \beta_\mref)$ for both arms. Evaluating coupling maps across individual spectral channels confirmed that peak setpoints are achromatic, demonstrating that fiber coupling optimization is independent of wavelength.

Reference spot coordinates on the FRAS are saved as operational setpoints for closed-loop tracking.

Because full 2D raster scans are time-consuming (Figure~\ref{fig-balayage2}), a fast setpoint optimization routine was developed to track local thermal drifts between operational runs.

%¤¤¤¤¤¤¤¤¤¤¤¤¤¤¤¤¤¤¤¤¤¤¤¤¤¤¤¤¤¤¤¤¤¤¤¤¤¤¤¤¤¤¤¤¤¤¤¤¤¤¤¤¤¤¤¤¤¤¤¤¤¤¤¤¤¤¤¤¤¤¤¤¤¤¤¤¤¤¤
\subsubsection{Fine Setpoint Optimization Routine}
\label{sec-procedure-correction}

The fine optimization routine operates in closed loop around current setpoints.

For small angular offsets, expanding Equations~\eqref{eq-inj} and \eqref{eq-inj-coeff} yields:
\begin{equation}
  I(\alpha,\beta) \simeq I_{\minj,\mmax}\GP{1-\frac15\GP{\pi\sigma D}^2
  \GC{\GP{\alpha-\alpha_\mref}^2+\GP{\beta-\beta_\mref}^2}}.
\label{eq-flux-angle}
\end{equation}

Approximation error remains below 0.06\% at 6~arcsec offsets (1\% at 12~arcsec), validating parabolic fits near coupling peaks.

Modulation offsets $\pm\theta_\varepsilon = \pm 6$~arcsec (10\% of diffraction spot size) are applied sequentially along orthogonal axes. Coupled intensity values are recorded:
\begin{equation}
  \left\{
    \begin{array}{r@{\hs}c@{\hs}l}
      I_0 &=& I(\alpha_0,\beta_0) \\
      I_{\alpha_-} &=& I(\alpha_0-\theta_\varepsilon,\beta_0) \\
      I_{\alpha_+} &=& I(\alpha_0+\theta_\varepsilon,\beta_0) \\
      I_{\beta_-} &=& I(\alpha_0,\beta_0-\theta_\varepsilon) \\
      I_{\beta_+} &=& I(\alpha_0,\beta_0+\theta_\varepsilon)
    \end{array}
  \right..
\end{equation}

Assuming source power $I_{\minj,\mmax}$ is stable (or normalized using the reference channel, Section~\ref{sec-modification-voie}), parabolic fitting along $\alpha$ gives:
\begin{equation}
  \left\{
    \begin{array}{r@{\hs}c@{\hs}l}
      I_0 &=& I_{\minj,\mmax}\GP{1-\dfrac15\GP{\pi\sigma D}^2
        \GC{\GP{\alpha_0-\alpha_\mref}^2+\GP{\beta_0-\beta_\mref}^2}} \\
      I_{\alpha_+} &=& I_{\minj,\mmax}\GP{1-\dfrac15\GP{\pi\sigma D}^2
        \GC{\GP{\alpha_0+\theta_\varepsilon-\alpha_\mref}^2+
          \GP{\beta_0-\beta_\mref}^2}} \\
      I_{\alpha_-} &=& I_{\minj,\mmax}\GP{1-\dfrac15\GP{\pi\sigma D}^2
        \GC{\GP{\alpha_0-\theta_\varepsilon-\alpha_\mref}^2+
          \GP{\beta_0-\beta_\mref}^2}} \\
    \end{array}
  \right.
\end{equation}

Evaluating intensity differences yields:
\begin{equation}
  I_{\alpha_+}-I_{\alpha_-} = \dfrac45I_{\minj,\mmax}\GP{\pi\sigma
    D}^2\GP{\alpha_0-\alpha_\mref}\theta_\varepsilon,
\label{eq-Ip-m}
\end{equation}
and
\begin{align}
  I_{\alpha_+}+I_{\alpha_-} &= 2I_{\minj,\mmax}\GP{1-\dfrac15\GP{\pi\sigma D}^2
    \GC{\GP{\alpha_0-\alpha_\mref}^2+\theta_\varepsilon^2
      \GP{\beta_0-\beta_\mref}^2}} \nonumber\\
  &= 2I_0+\dfrac25I_{\minj,\mmax}\GP{\pi\sigma D}^2\theta_\varepsilon^2.
\label{eq-Ip+m}
\end{align}

Solving Equations~\eqref{eq-Ip-m} and \eqref{eq-Ip+m} updates setpoint coordinates:
\begin{equation}
  \left\{
    \begin{array}{r@{\hs}c@{\hs}l}
      \alpha_\mref &=& \alpha_0-\dfrac{I_{\alpha_+}-I_{\alpha_-}} 
      {2\GP{I_{\alpha_+}+I_{\alpha_-}-2I_0}}\cdot\theta_\varepsilon \\[+12pt]
      \beta_\mref &=& \beta_0-\dfrac{I_{\beta_+}-I_{\beta_-}} 
      {2\GP{I_{\beta_+}+I_{\beta_-}-2I_0}}\cdot\theta_\varepsilon \\
    \end{array}
  \right..
\end{equation}

Figure~\ref{fig-calib-injfin} shows intense sampling during fine optimization. A 6~arcsec offset reduces throughput by ~2\% at 2~\mum. Corrective updates are typically under 0.5~arcsec (0.9\% of spot size).

\begin{figure} \centering
  \FIG{.7}{false}{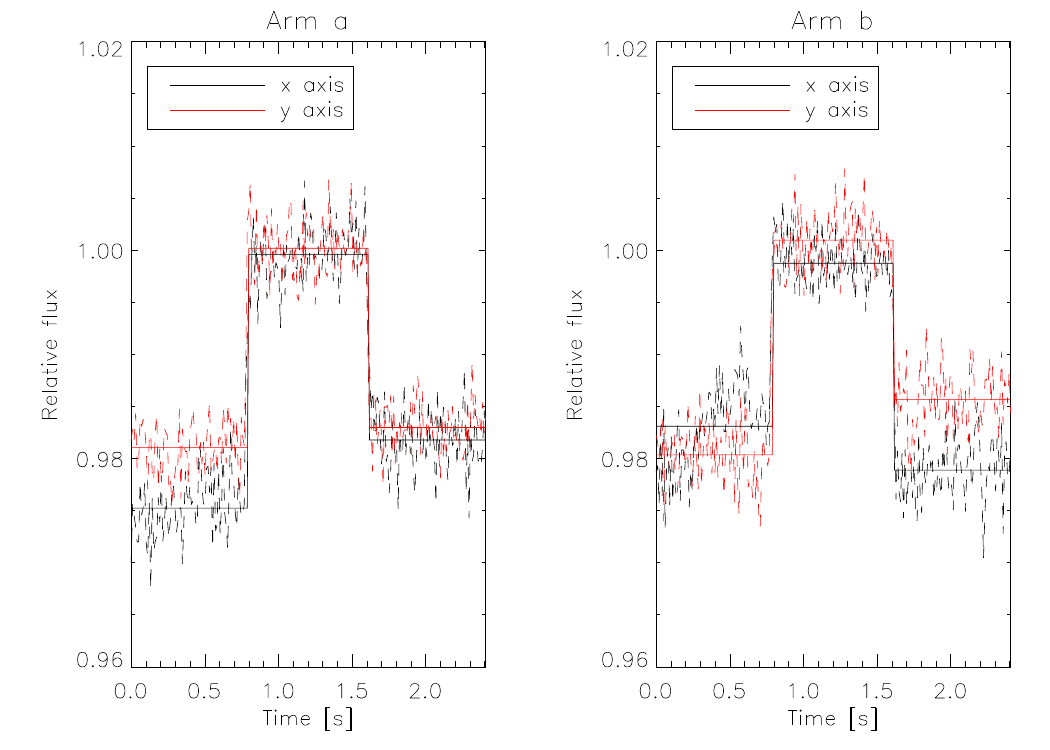}
  \caption{Intensity signal recorded during fine setpoint optimization.}
  \label{fig-calib-injfin}
\end{figure}

%-------------------------------------------------------------------------------
\subsection{Analysis of Tip/Tilt Control Performance}
\label{sec-analyse-tt-int}

%¤¤¤¤¤¤¤¤¤¤¤¤¤¤¤¤¤¤¤¤¤¤¤¤¤¤¤¤¤¤¤¤¤¤¤¤¤¤¤¤¤¤¤¤¤¤¤¤¤¤¤¤¤¤¤¤¤¤¤¤¤¤¤¤¤¤¤¤¤¤¤¤¤¤¤¤¤¤¤
\subsubsection{Rejection Transfer Function}
\label{sec-fonction-transfert-2}

The tip/tilt loop utilizes an integrator controller (Section~\ref{sec-rappel-integrateur}) running at 200~Hz ($f_\mfs/5$). Staggering execution tasks across piston frames (Section~\ref{sec-description-architecture}) reduces loop latency from 2.0 frames down to 1.8 frames. Figure~\ref{fig-FRASrej} plots rejection transfer functions across integrator gains.

\begin{figure} \centering
  \FIG{.7}{false}{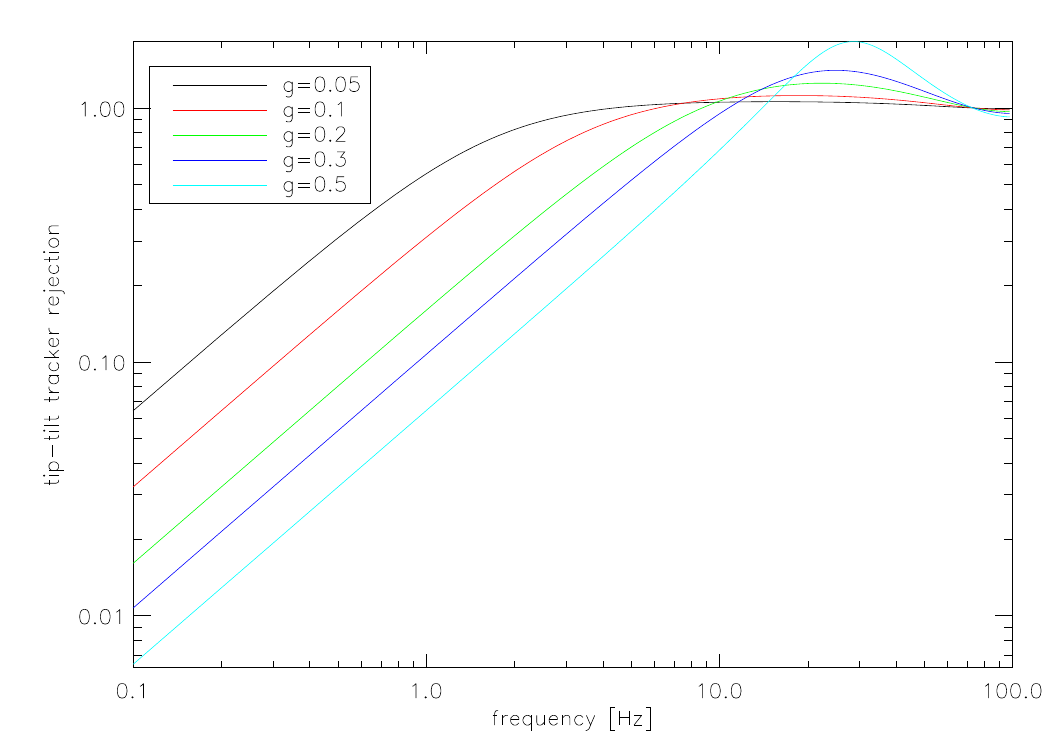}
  \caption[Tip/tilt loop rejection transfer function.]{Rejection transfer function of the tip/tilt loop across integrator gains.}
  \label{fig-FRASrej}
\end{figure}

Reduced latency lowers high-frequency overshoot compared to standard 2-frame delays (Figure~\ref{fig-trans-rej}). Peak overshoot drops from 2.2 to 1.8 for gain $g=0.5$. Loop $-3$~dB rejection cutoff frequency scales from 1.5~Hz to 10~Hz as gain increases from 0.05 to 0.5. Rejection is limited to disturbance frequencies below $f_\mfras/20$.

%¤¤¤¤¤¤¤¤¤¤¤¤¤¤¤¤¤¤¤¤¤¤¤¤¤¤¤¤¤¤¤¤¤¤¤¤¤¤¤¤¤¤¤¤¤¤¤¤¤¤¤¤¤¤¤¤¤¤¤¤¤¤¤¤¤¤¤¤¤¤¤¤¤¤¤¤¤¤¤
\subsubsection{Tip/Tilt Tracking Performance}
\label{sec-performances-boucle}

Figure~\ref{fig-FRASpsd_BOssj} displays open-loop tip/tilt PSDs recorded on the FRAS.

\begin{figure} \centering
  \FIG{.9}{false}{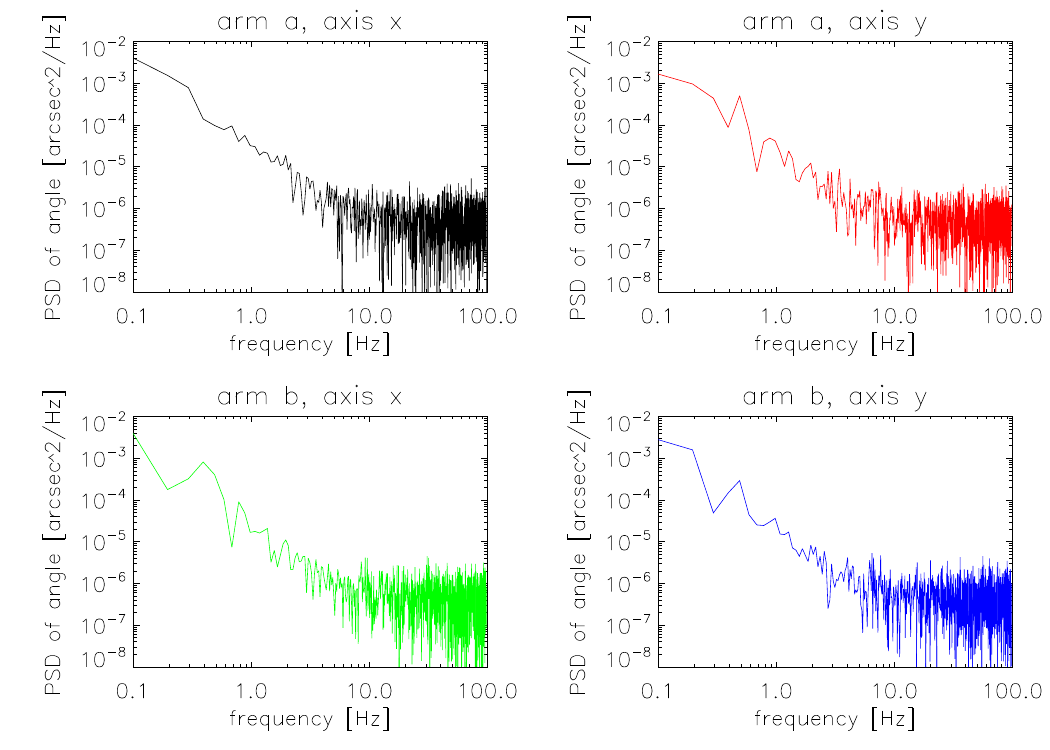}
  \caption{Open-loop tip/tilt PSDs recorded across both arms.}
  \label{fig-FRASpsd_BOssj}
\end{figure}

Uncorrected pointing jitter measures about 500~mas RMS. Open-loop spectra (Figure~\ref{fig-FRASpsd_BOssj}) are dominated by low-frequency air turbulence, flattening into white sensor noise at high frequencies. Because open-loop power is concentrated below 5~Hz, moderate integrator gains suffice. Figure~\ref{fig-FRASpsd_BFssj} plots closed-loop PSDs for $g=0.05$.

\begin{figure} \centering
  \FIG{.9}{false}{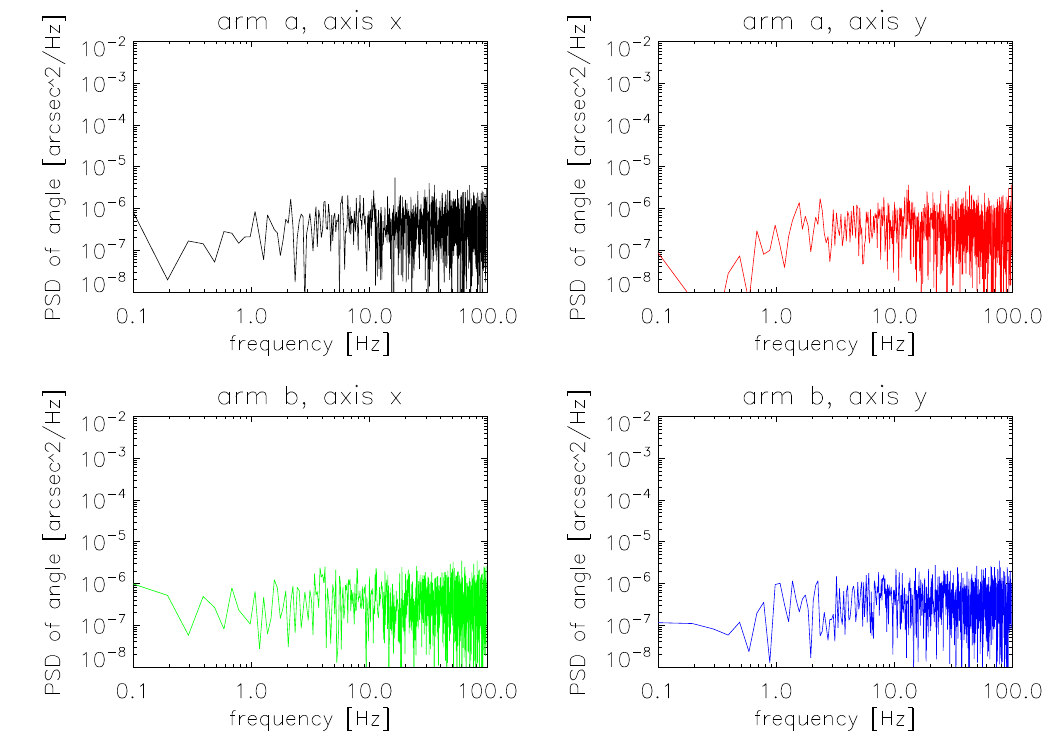}
  \caption{Closed-loop tip/tilt PSDs across both arms.}
    \label{fig-FRASpsd_BFssj}
\end{figure}

Closing the loop flattens the PSDs to the sensor noise floor (Figure~\ref{fig-FRASpsd_BFssj}). Increasing gain to $0.2$ provides no further rejection while increasing high-frequency noise amplification. Furthermore, high tip/tilt gains risk feeding residual cross-talk into the piston loop.

\begin{figure} \centering
  \FIG{0.7}{false}{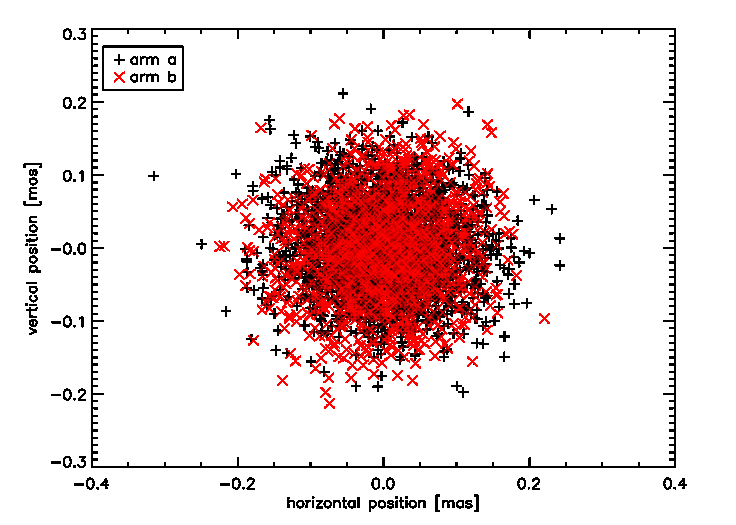}
  \caption{Closed-loop angular tracking errors over 10~s.}
  \label{fig-FRASxy_best}
\end{figure}

As shown in Figure~\ref{fig-FRASxy_best}, closed-loop tracking errors reach $\boldsymbol{\sigma_{\theta,\ma} \simeq \sigma_{\theta,\mb} \simeq \sigma_\theta = 56}$~\textbf{mas RMS} per axis—well below the 100~mas RMS specification. Symmetric performance is achieved across both arms despite differences in spot comatic aberration (Figure~\ref{fig-image-fras}).

%¤¤¤¤¤¤¤¤¤¤¤¤¤¤¤¤¤¤¤¤¤¤¤¤¤¤¤¤¤¤¤¤¤¤¤¤¤¤¤¤¤¤¤¤¤¤¤¤¤¤¤¤¤¤¤¤¤¤¤¤¤¤¤¤¤¤¤¤¤¤¤¤¤¤¤¤¤¤¤
\subsubsection{Flux Sensitivity}
\label{sec-sensibilite-au}

Tracking performance was evaluated against available input flux by defocussing the delivery fiber while holding optical density filters constant. Received flux was calculated in photon counts ($N_\mph$) per sub-aperture. Figure~\ref{fig-sigma-tt-vs-flux} plots residual tracking jitter against photon count.

\begin{figure} \centering
  \FIG{0.7}{false}{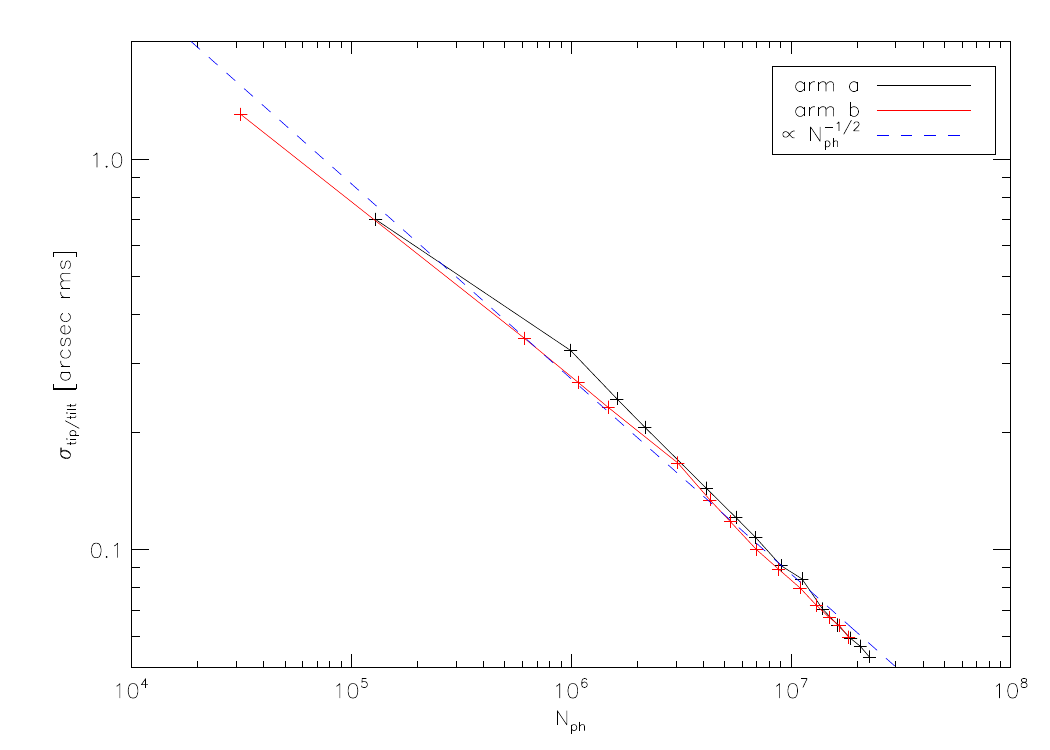}
  \caption{Residual tip/tilt tracking error versus input photon flux per arm.}
  \label{fig-sigma-tt-vs-flux}
\end{figure}

In log-log space, residual tracking error $\sigma_\theta$ follows a $-1/2$ power law against photon count $N_\mph$:
\begin{equation}
  \sigma_\theta \propto N_\mph^{-\frac12}.
\end{equation}

Tip/tilt tracking is then photon-noise limited. Maintaining tracking jitter below 600~mas RMS requires a minimum flux of $2\E{5}$~photons per arm per frame.

Although source intensity fluctuations affect total power (Section~\ref{sec--etude-1}), spatial centroiding is unaffected by common-mode power shifts. Shot noise across sub-aperture pixels remains the primary tracking noise source.

%§§§§§§§§§§§§§§§§§§§§§§§§§§§§§§§§§§§§§§§§§§§§§§§§§§§§§§§§§§§§§§§§§§§§§§§§§§§§§§§
\section{Characterization of \pe's Piston Control}
\label{sec-optimisation-FS}

%-------------------------------------------------------------------------------
\subsection{Fringe Sensor Calibration}
\label{sec-etalonnage-FS}

Extracting differential path delay (piston) requires calibrating the Modified Mach-Zehnder Fringe Sensor (Section~\ref{sec-desc-mmz}) using spatial quasi-ABCD modulation (Section~\ref{sec-desc-abcd}). The calibration methodology was established by K. Houairi \cite{Houairi09b} on ONERA's prototype combiner. Adapting this routine to broadband operation on \pe revealed limitations that required the modifications described below.

%¤¤¤¤¤¤¤¤¤¤¤¤¤¤¤¤¤¤¤¤¤¤¤¤¤¤¤¤¤¤¤¤¤¤¤¤¤¤¤¤¤¤¤¤¤¤¤¤¤¤¤¤¤¤¤¤¤¤¤¤¤¤¤¤¤¤¤¤¤¤¤¤¤¤¤¤¤¤¤
\subsubsection{Monochromatic Versus Broadband Calibration}
\label{sec-monochrom-vs-polychrom}

Initial autocollimination tests used narrow-band metrology sources: an 830~nm laser diode (linewidth <1~nm, coherence length >1~mm) and a 1320~nm SLED (bandwidth 40~nm, coherence length 40~\mum). Figure~\ref{fig-calib-mono} displays calibration scans recorded with these sources.

\begin{figure} \centering
  \subfloat[$\I$-band laser source.]{\label{fig-calib-mono-I}
    \FIG{0.49}{false}{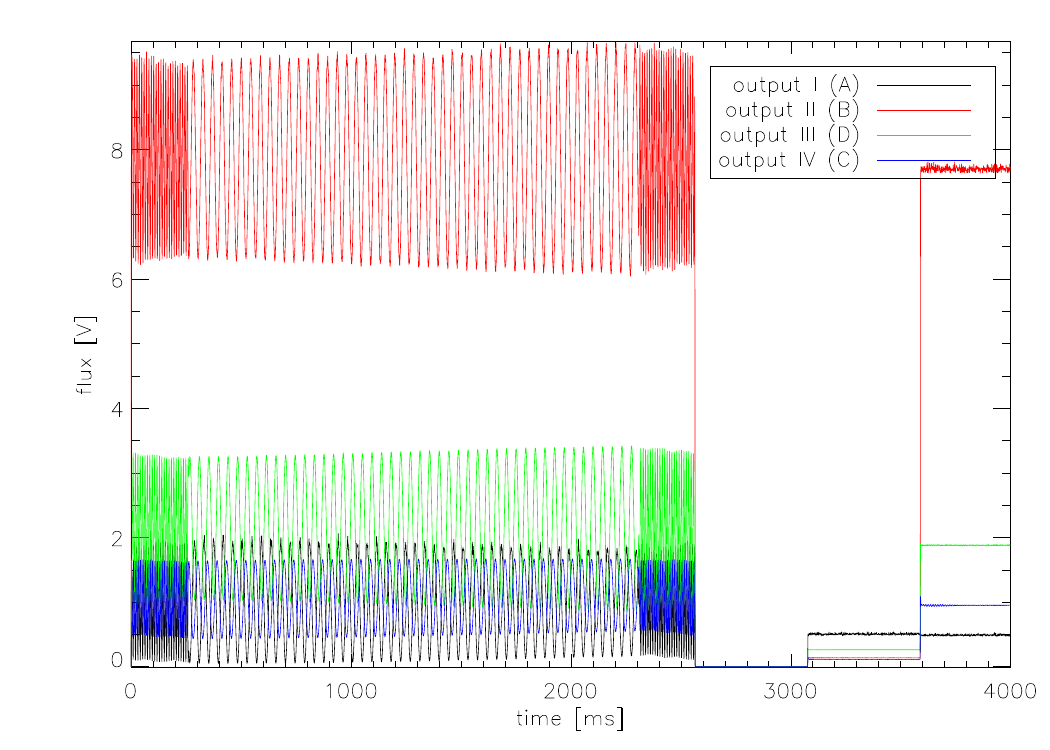}}
  \hfill\subfloat[$\J$-band SLED source.]{\label{fig-calib-mono-J}
    \FIG{0.49}{false}{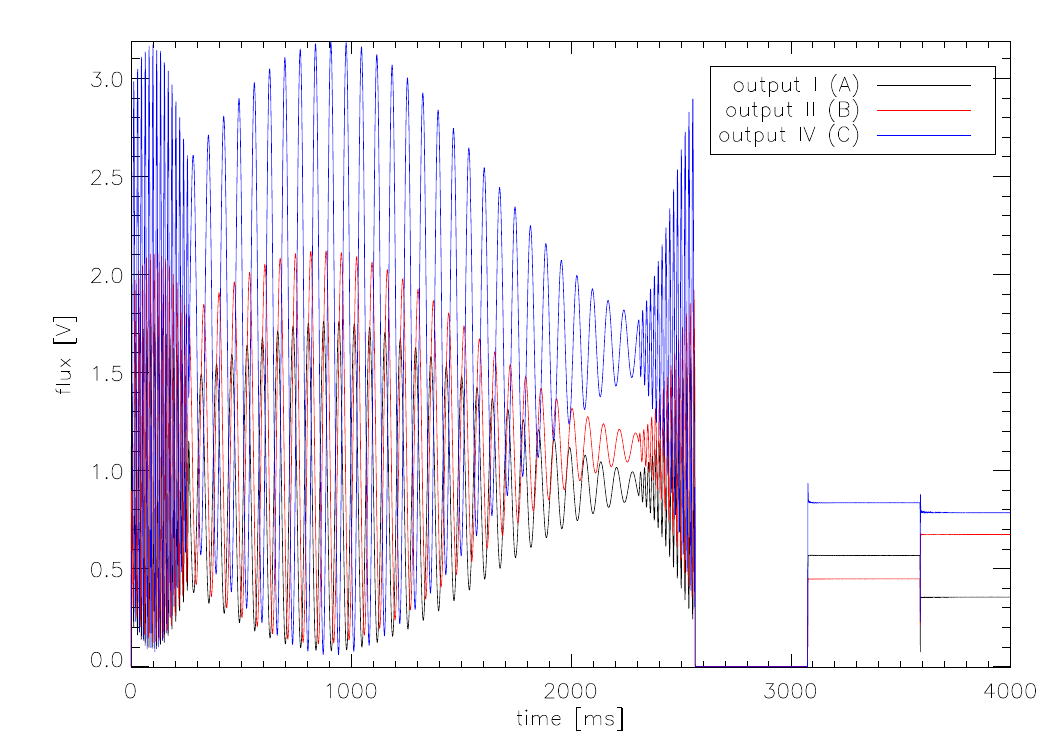}}
  \caption[Fringe sensor calibration datasets using narrow-band sources.]{Calibration scans in $\I$ and $\J$ bands using narrow-band metrology sources.}
  \label{fig-calib-mono}
\end{figure}

The calibration sequence comprises four steps:
\begin{enumerate}
\item A 38~\mum piston ramp bounded by fast centering sweeps;
\item Dark current measurement (beams misaligned);
\item Background intensity measurement in arm $\ma$ (arm $\mb$ blocked);
\item Background intensity measurement in arm $\mb$ (arm $\ma$ blocked).
\end{enumerate}

In Figure~\ref{fig-calib-mono-I}, envelope variations in the $\I$ band stem from minor tip/tilt cross-talk during piezo sweeps driven by uncalibrated S316 matrices (Section~\ref{sec-calc-mat-com}), which distorts spatial fringe contrast.

Figure~\ref{fig-calib-mono-J} exhibits a distinct coherence envelope from the SLED. While identifying zero OPD in $\I$-band laser fringes is impossible due to long coherence lengths, $\J$-band SLED fringes display a distinct visibility peak at zero OPD. Figure~\ref{fig-XMS50-parabole} plots unwrapped path delay measurements across large linear sweeps.

\begin{figure} \centering
  \FIG{0.7}{false}{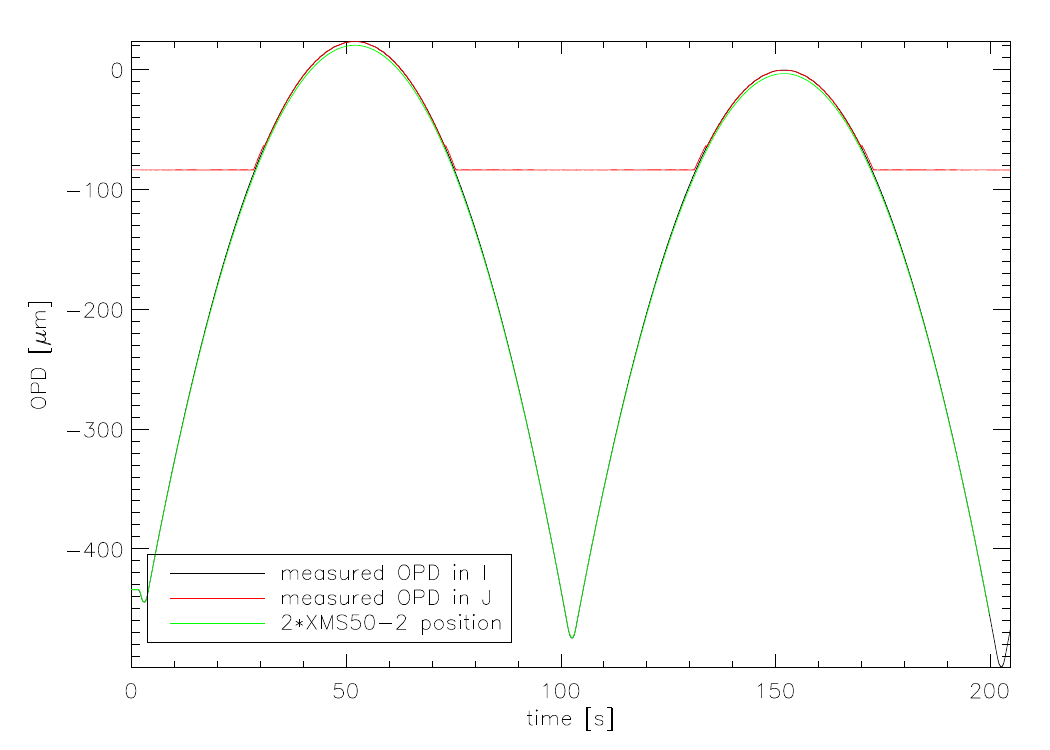}
  \caption{Unwrapped path delay measurements recorded across large parabolic sweeps.}
  \label{fig-XMS50-parabole}
\end{figure}

Parabolic path delays were driven using an XMS50 long-stroke delay stage. Unwrapped $\I$-band laser phase tracks stage displacement accurately over >10~mm ranges (0.4\% scale error). In the $\J$ band, phase unwrapping fails beyond $\delta = -65$~\mum due to coherence envelope attenuation, limiting $\J$-band tracking range to 170~\mum.

These early calibrations assumed static metrology wavelengths. However, thermal shifts in laser diodes alter operational wavelengths, degrading multi-wavelength fringe unwrapping (Section~\ref{sec-coherencage}).

For science testing, the narrow-band sources were replaced by the broadband Fianium supercontinuum source. Dichroic splitters divide the broadband light into $\I$ (0.8–1.0~\mum) and $\J$ (1.0–1.65~\mum) metrology channels.

Broadband metrology channels yield substantially narrower coherence envelopes (Figure~\ref{fig-calib-poly}).

\begin{figure} \centering
  \subfloat[Broadband $\I$ band.]{\label{fig-calib-poly-I}
    \FIG{0.49}{false}{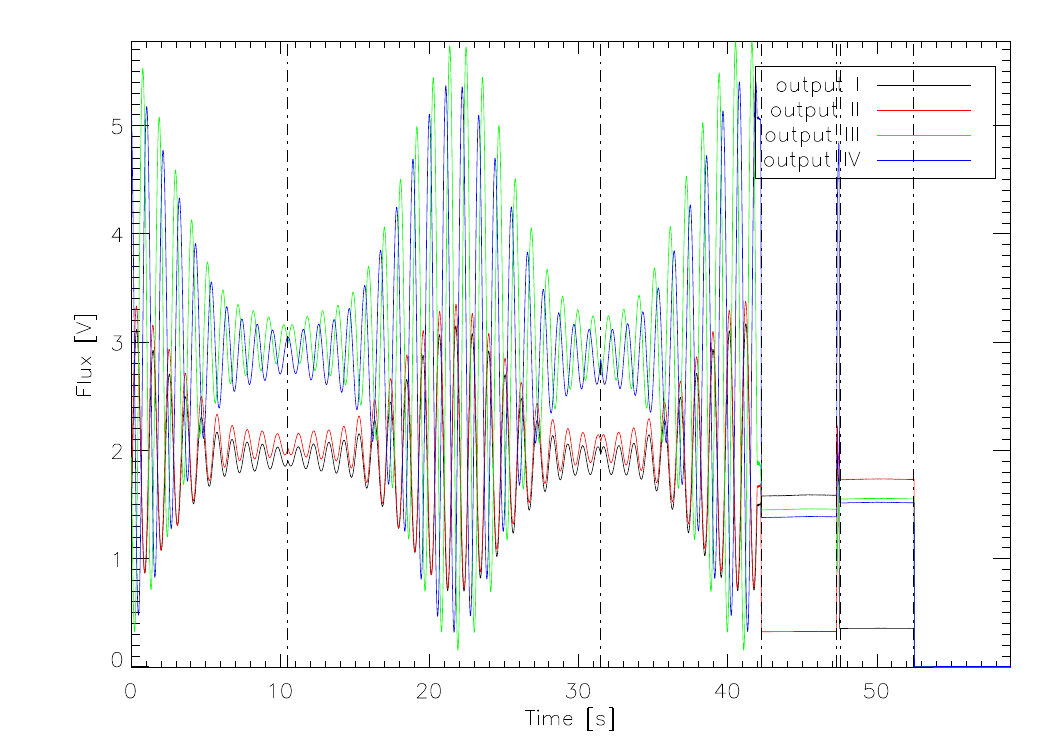}}
  \hfill\subfloat[Broadband $\J$ band.]{\label{fig-calib-poly-J}
    \FIG{0.49}{false}{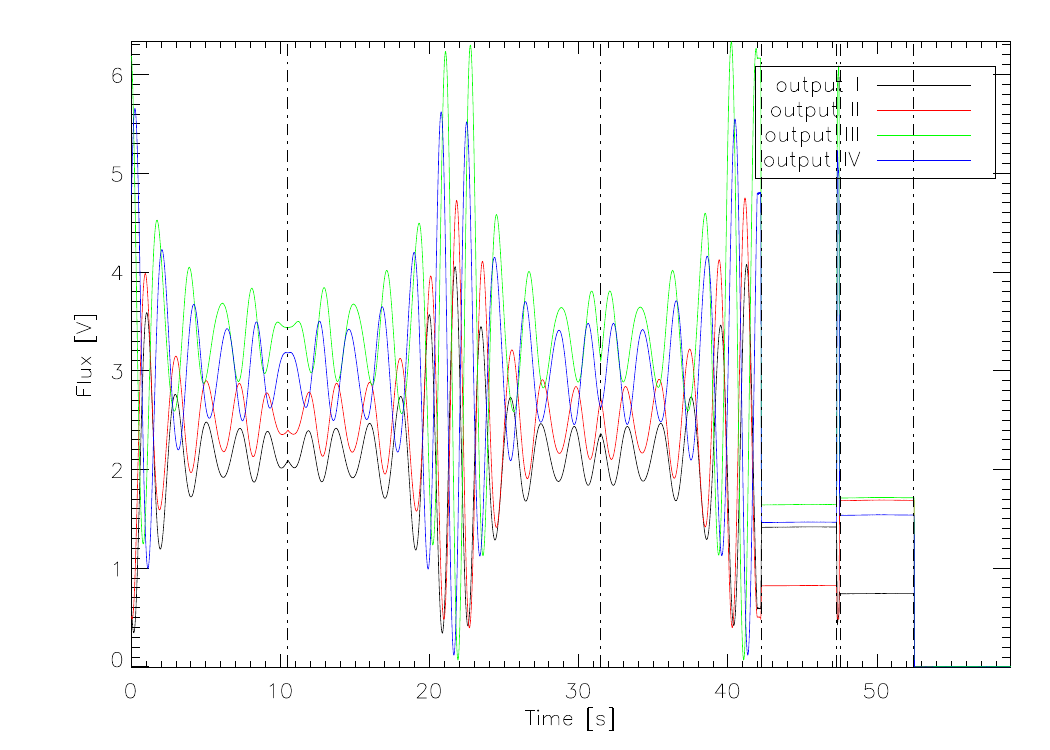}}
  \caption[Broadband fringe sensor calibration datasets.]{Calibration scans recorded in $\I$ and $\J$ metrology bands using the broadband source.}
  \label{fig-calib-poly}
\end{figure}

In Figure~\ref{fig-calib-poly}, the piston ramp spans 16~\mum with symmetric forward and reverse sweeps. Frame timing symmetry is used to extract processing latency (Section~\ref{sec-calcul-retard}).

K. Houairi established that matching coherence lengths between $\I$ and $\J$ channels optimizes fringe unwrapping. Setting the dichroic cutoff at 1.0~\mum for an operational passband of 0.8–1.5~\mum balanced both envelopes. Shifting the upper cutoff to 1.65~\mum narrowed the $\J$-band coherence envelope relative to the $\I$ band (Figure~\ref{fig-calib-poly}). $\I$-band fringes remain regular, whereas $\J$-band fringes display variable fringe spacing and output-dependent envelope shapes.

Accurate extended fringe unwrapping (Section~\ref{sec-estim-elargi}) requires precise knowledge of the effective wavelength ratio $r_\lambda = \lambda_1/\lambda_2$. This prompted an in-situ spectral calibration routine.

%¤¤¤¤¤¤¤¤¤¤¤¤¤¤¤¤¤¤¤¤¤¤¤¤¤¤¤¤¤¤¤¤¤¤¤¤¤¤¤¤¤¤¤¤¤¤¤¤¤¤¤¤¤¤¤¤¤¤¤¤¤¤¤¤¤¤¤¤¤¤¤¤¤¤¤¤¤¤¤
\subsubsection{Fourier-Transform Spectral Calibration}
\label{sec-fourier-FS}

Effective metrology wavelengths are extracted from calibration scans using Fourier Transform Spectroscopy (Section~\ref{sec-etalonnage-FTS}).

Calibration sweeps run at $0.8$~\mum.s$^{-1}$ over a $4.8$~\mum window around zero OPD at a 100~Hz loop rate. Applying Hann windowing yields a spectral resolution $\Delta\sigma = 0.2$~\imum ($R_{\mfs,\I} \approx 6$, $R_{\mfs,\J} \approx 4$). Low-resolution spectra suffice to extract central channel wavelengths. Figure~\ref{fig-fts-ij} displays extracted metrology passbands under two alignment states.

\begin{figure} \centering
  \subfloat[$\I$ and $\J$ passbands with misaligned $\J$-IV output lens.]{\label{fig-fts-ij-1}
    \FIG{0.49}{false}{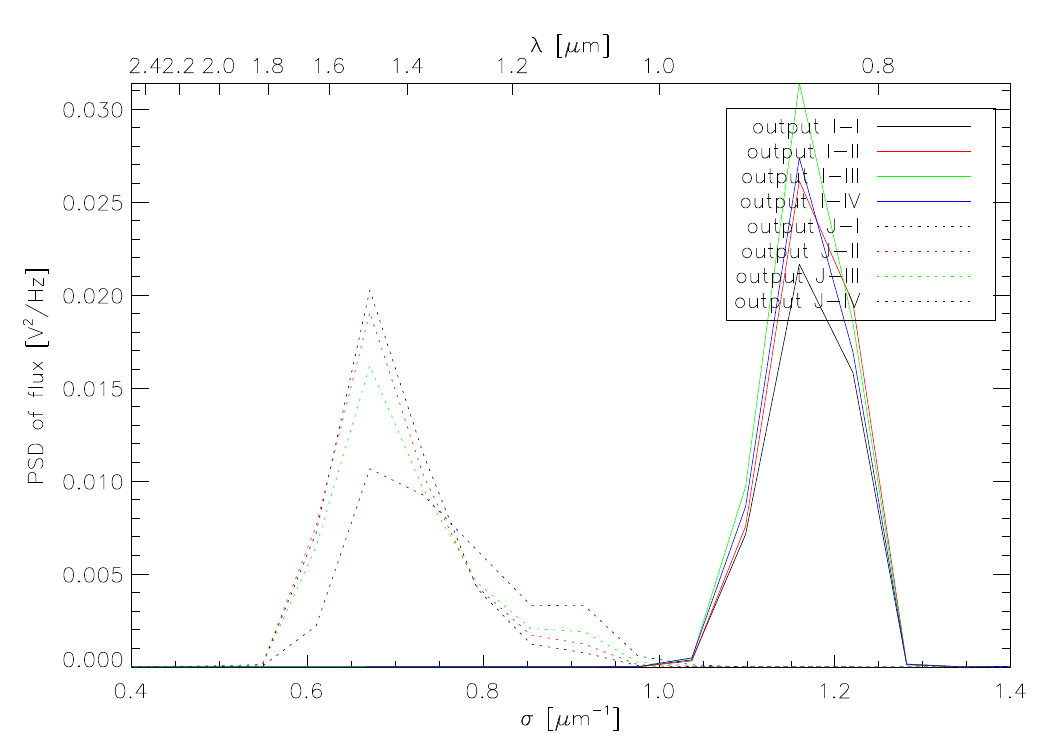}}
  \hfill\subfloat[$\I$ and $\J$ passbands with corrected $\J$-IV alignment.]{\label{fig-fts-ij-2}
    \FIG{0.49}{false}{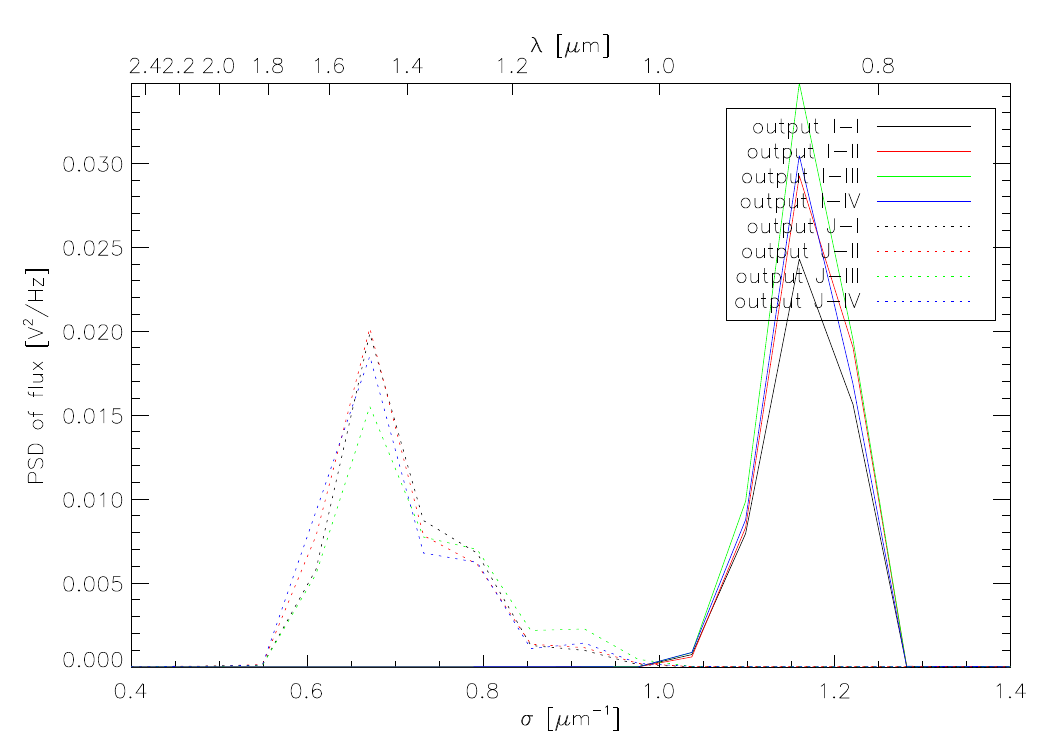}}
  \caption{Metrology passbands extracted via FTS under different alignment states.}
  \label{fig-fts-ij}
\end{figure}

Figure~\ref{fig-fts-ij-1} shows an unaligned state where output IV in the $\J$ band peaks at $1.25$~\mum, while outputs I–III peak near $1.35$~\mum. Disparate passbands produce phase mismatches that distort the quasi-ABCD interaction matrix (Equation~\ref{eq-algo-direct}). These passband shifts stem from defocusing in the multimode fiber coupling doublets. Adjusting fiber position aligns output passbands (Figure~\ref{fig-fts-ij-2}).

Inter-channel phase shifts are derived simultaneously from complex Fourier phases relative to output III (dark fringe output).

%¤¤¤¤¤¤¤¤¤¤¤¤¤¤¤¤¤¤¤¤¤¤¤¤¤¤¤¤¤¤¤¤¤¤¤¤¤¤¤¤¤¤¤¤¤¤¤¤¤¤¤¤¤¤¤¤¤¤¤¤¤¤¤¤¤¤¤¤¤¤¤¤¤¤¤¤¤¤¤
\subsubsection{Latency Extraction via Sweep Symmetry}
\label{sec-calcul-retard}

System latency introduces phase offsets between commanded actuator positions and recorded fringe signals.

Coarse zero OPD is set manually to within $\pm 5$~\mum using the long-stroke delay lines. Driving symmetric bidirectional sweeps through the M6 piezo stages yields symmetric fringe profiles.

Calculating the cross-correlation between forward and reverse fringe scans extracts the system latency offset.

This symmetry method requires hysteresis-free actuator sweeps (achieved when S316 strain gauges are enabled, Section~\ref{sec-linearite}).

Zero OPD in the science channel is aligned to the metrology setpoint using joint camera-FS calibration scans (Section~\ref{sec-proc-camera}).

%¤¤¤¤¤¤¤¤¤¤¤¤¤¤¤¤¤¤¤¤¤¤¤¤¤¤¤¤¤¤¤¤¤¤¤¤¤¤¤¤¤¤¤¤¤¤¤¤¤¤¤¤¤¤¤¤¤¤¤¤¤¤¤¤¤¤¤¤¤¤¤¤¤¤¤¤¤¤¤
\subsubsection{Demodulation Matrix Validation}
\label{sec-resultat-etalonnage}

Calibrated demodulation matrices $\V{D}^k$ ($k \in \{\I,\J\}$) reconstruct state vectors $\V{X}_I^k$ from output intensities $\V{Y}_I^k$:
\begin{equation}
  \V{X}_I^k = \V{D}^k\V{Y}_I^k,
\end{equation}
where
\begin{equation}
  \V{Y}_I^k = \begin{pmatrix}
    I_\mni^k \\
    I_\mnii^k \\
    I_\mniii^k \\
    I_\mniv^k
  \end{pmatrix},
\end{equation}

and $\V{X}_I^k$ contains the demodulated intensities:
\begin{equation}
  \V{X}_I^k = \begin{pmatrix}
    \breve{I}_\ma^k\\
    \breve{I}_\mb^k\\
    2\breve{\mu}^k\sqrt{\breve{I}_\ma^k\breve{I}_\mb^k}
    \cos\GP{2\pi\sigma_\mmoy^k\delta}\\
    2\breve{\mu}^k\sqrt{\breve{I}_\ma^k\breve{I}_\mb^k}
    \sin\GP{2\pi\sigma_\mmoy^k\delta}
  \end{pmatrix},
\end{equation}
where $\breve{I}_\ma^k$ and $\breve{I}_\mb^k$ are the relative fluxes in arms $\ma$ and $\mb$, $\breve{\mu}^k$ is the relative visibility, and $\sigma_\mmoy^k$
the average wave number.

Figure~\ref{fig-inter-ij} plots calibration fringe scans $\V{Y}_I^k$.

\begin{figure} \centering
  \subfloat[$\I$-band calibration fringes.]{\label{fig-inter-i}\FIG{0.49}{false}{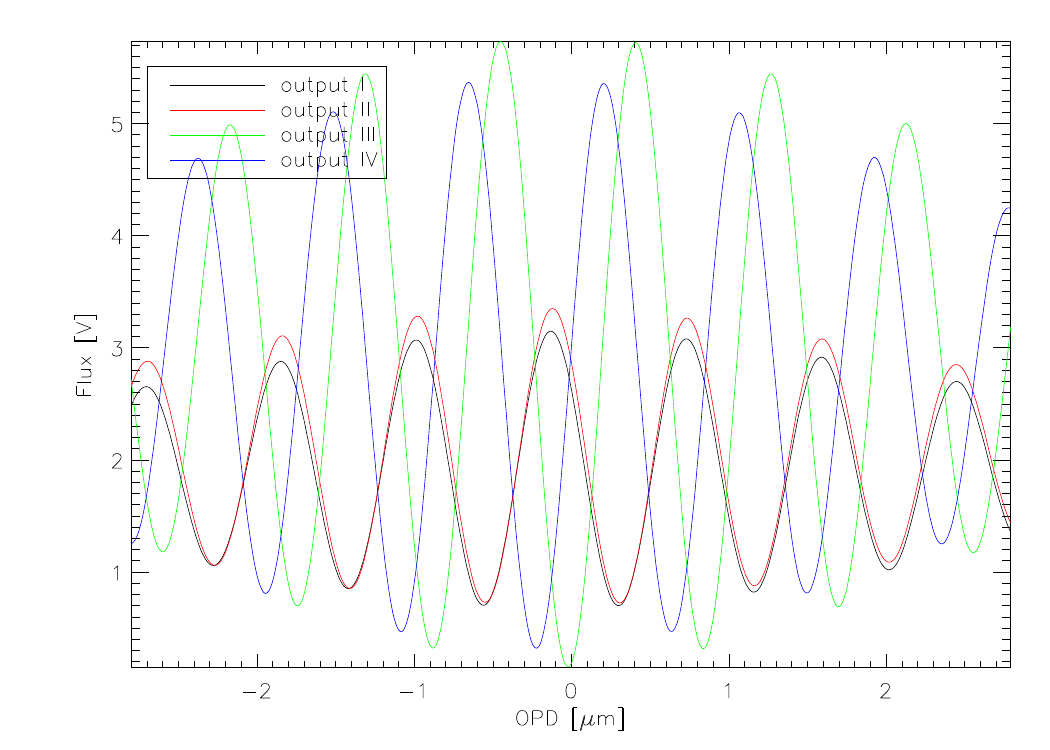}}
  \hfill\subfloat[$\J$-band calibration fringes.]{\label{fig-inter-j}\FIG{0.49}{false}{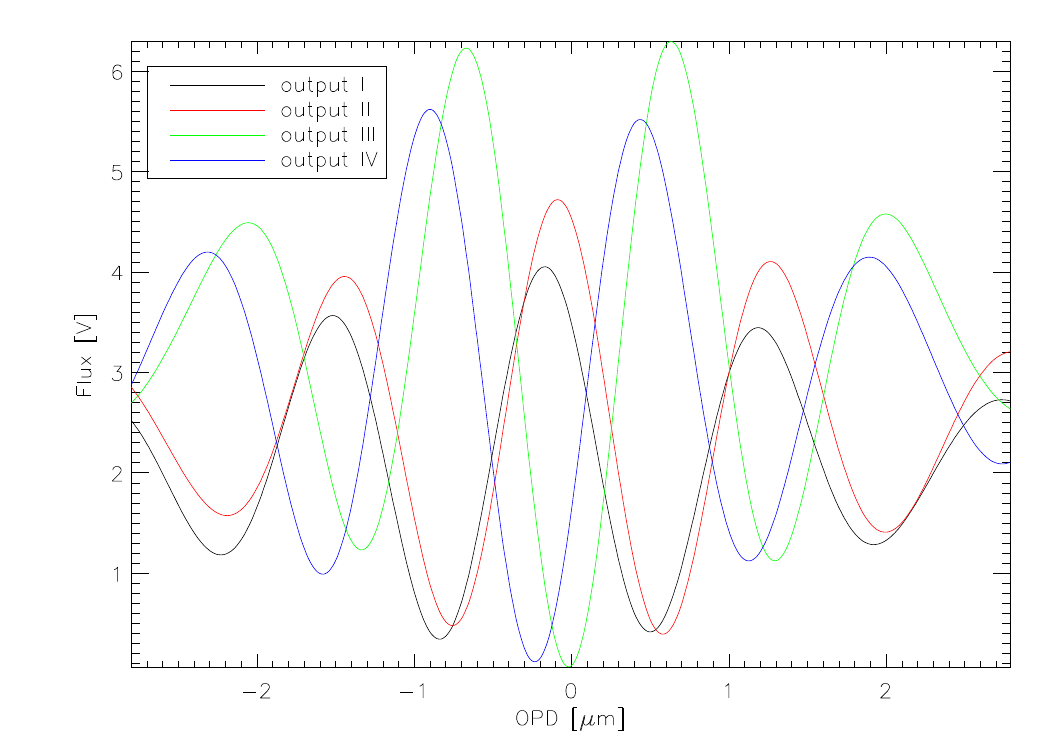}}
  \caption{Raw calibration fringe signals recorded across MMZ outputs.}
  \label{fig-inter-ij}
\end{figure}

In Figure~\ref{fig-inter-ij}, output III operates in anti-phase relative to output II, while output IV opposes output I. Non-ideal internal phase shifts deviate outputs I and II from exact quadrature. Fringe contrast is higher on symmetric outputs III and IV than on asymmetric outputs I and II. Substrate alignment residual errors ($\sim$5~arcsec) reduce contrast on output IV relative to output III. Multi-mode fiber coupling renders fringe sensor outputs sensitive to optical aberrations, yielding higher contrast in the $\J$ band than in the $\I$ band.

Multiplying output intensities $\V{Y}_I^k$ by demodulation matrix $\V{D}^k$ yields relative arm intensities $\breve{I}_\ma^k$ and $\breve{I}_\mb^k$ (Figure~\ref{fig-flux}).

\begin{figure} \centering
  \FIG{0.7}{false}{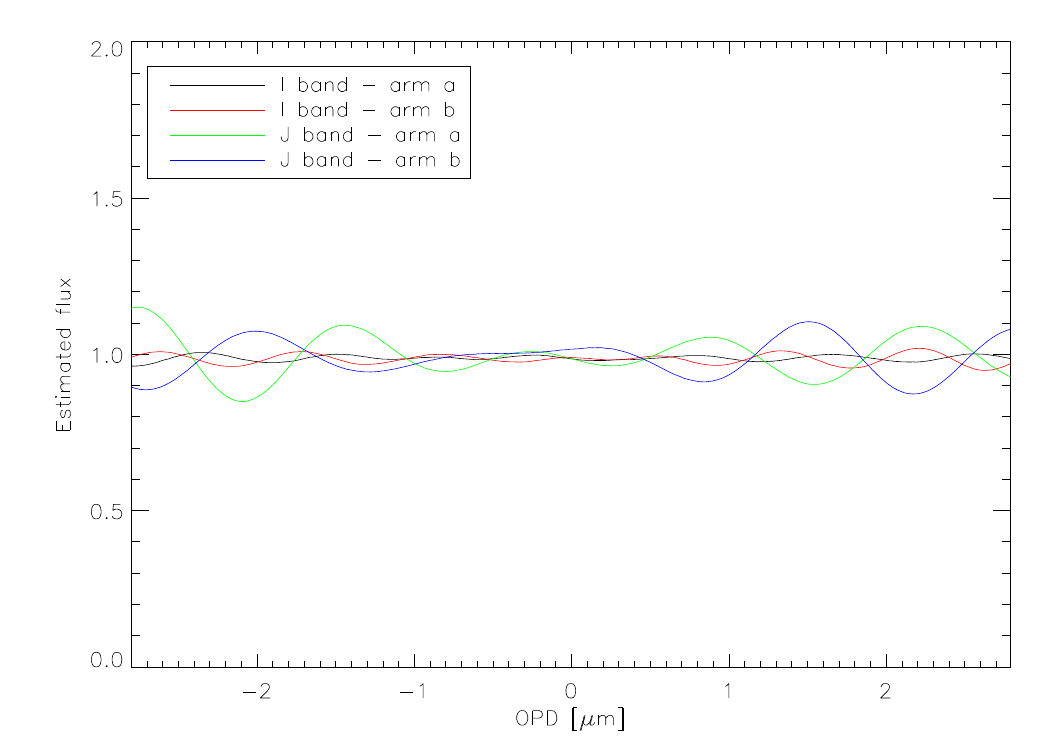}
  \caption{Reconstructed relative arm intensities $\breve{I}_\ma$ and $\breve{I}_\mb$ in $\I$ and $\J$ bands.}
  \label{fig-flux}
\end{figure}

Reconstructed relative fluxes hover near 1.0. Flux estimation error near zero OPD is ~1\% in $\I$ band (5\% at 2~\mum OPD offset). In the $\J$ band, passband dispersion increases flux error to 12\% at 2~\mum OPD. Residual ripples reflect fringe phase cross-talk.

Fringe visibility $\breve{\mu}^k$ is extracted from quadrature components (Figure~\ref{fig-visi}).

\begin{figure} \centering
  \FIG{0.7}{false}{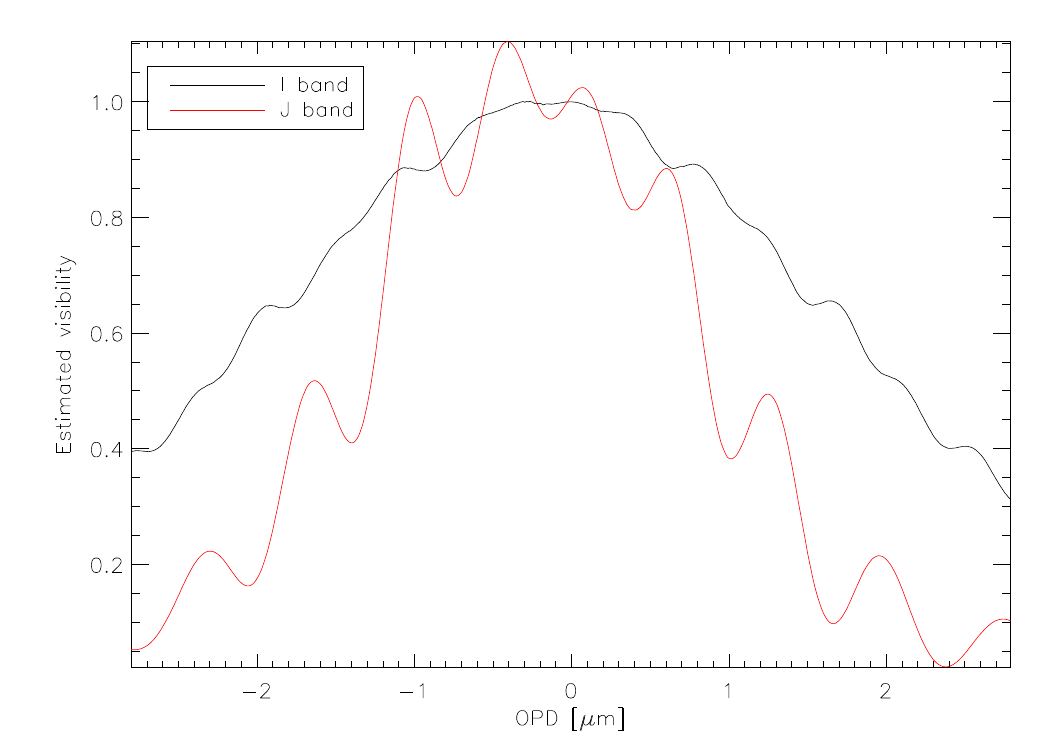}
  \caption{Reconstructed relative fringe visibility in $\I$ and $\J$ bands.}
  \label{fig-visi}
\end{figure}

Figure~\ref{fig-visi} displays visibility estimates. Residual ripples at $\lambda/2$ periods (430~nm in $\I$; 680~nm in $\J$) stem from phase quadrature deviations. Measured FWHM coherence envelope widths are $\Delta_\mfwhm^\I \approx 4$~\mum and $\Delta_\mfwhm^\J \approx 2.5$~\mum.

Phase angle calculations extract unwrapped OPD estimates (Figure~\ref{fig-opd-ij}).

\begin{figure} \centering
  \FIG{0.7}{false}{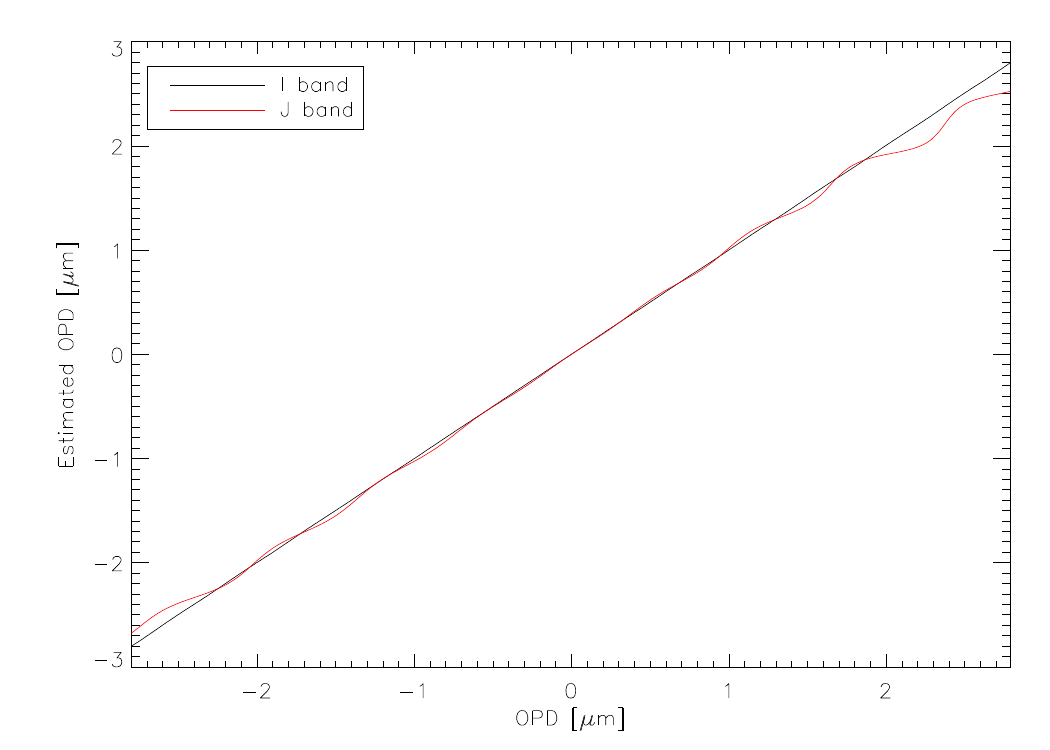}
  \caption[Unwrapped OPD estimates.]{Unwrapped OPD estimates in $\I$ and $\J$ bands versus commanded path delay.}
  \label{fig-opd-ij}
\end{figure}

Figure~\ref{fig-opd-err-ij} plots absolute estimation errors ($\delta_\text{est} - \delta_\text{true}$).

\begin{figure} \centering
  \subfloat[$\I$-band estimation error.]{\label{fig-opd-err-i}
    \FIG{0.49}{false}{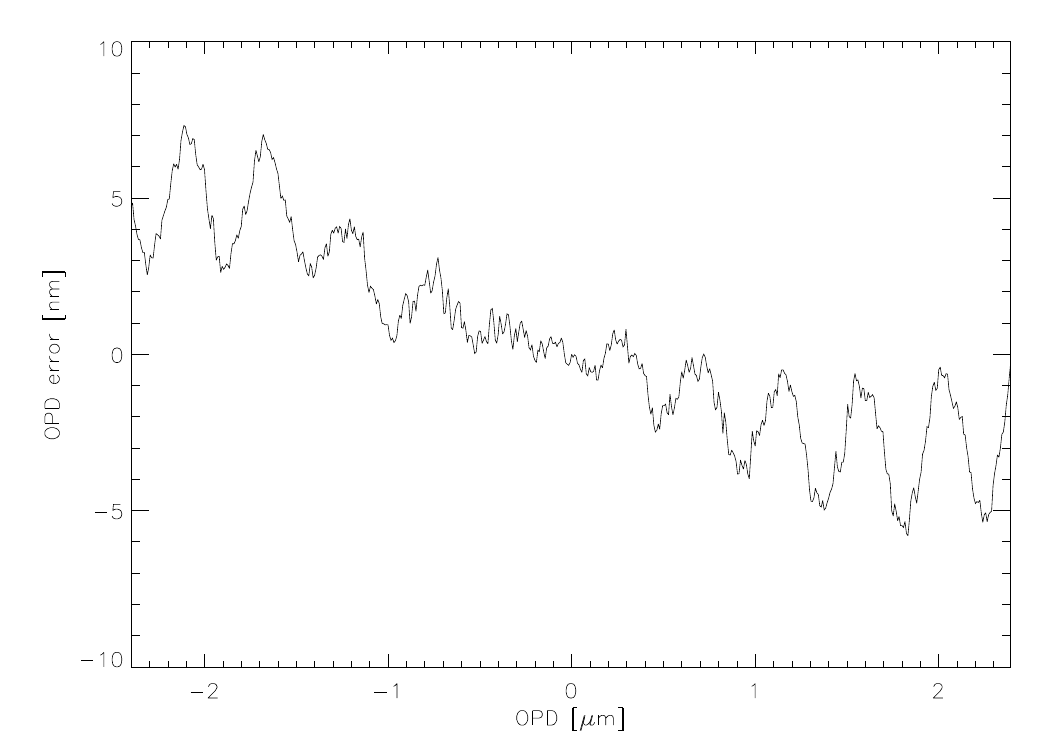}}
  \hfill\subfloat[$\J$-band estimation error.]{\label{fig-opd-err-j}
    \FIG{0.49}{false}{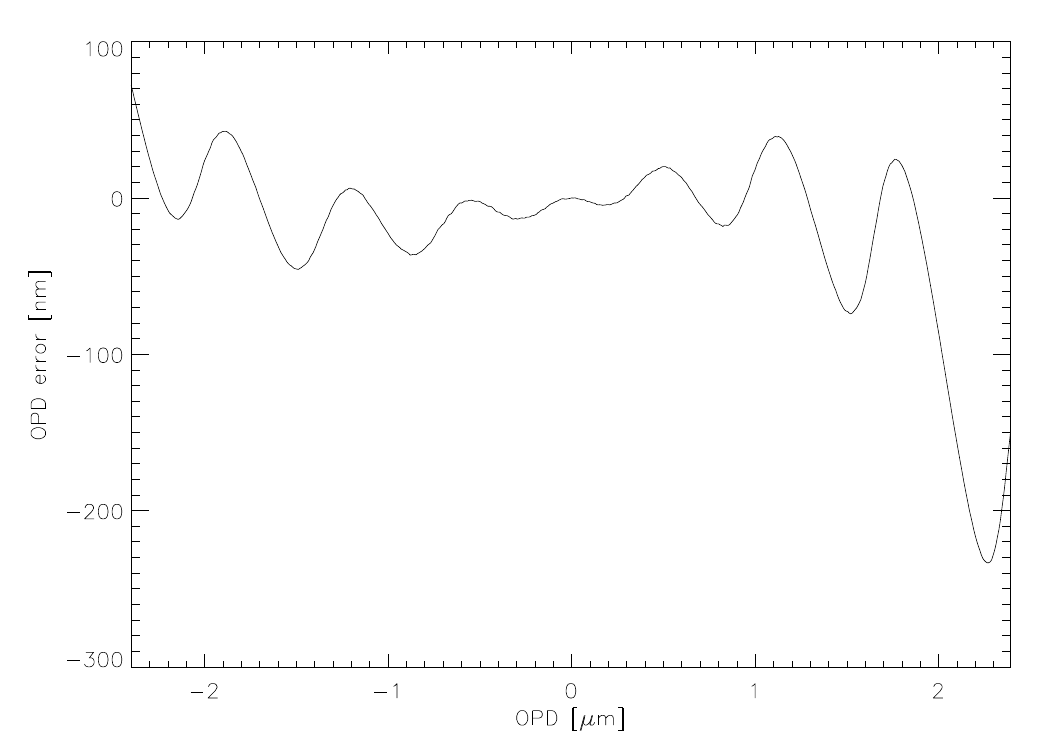}}
  \caption{Absolute path delay estimation errors across metrology channels.}
  \label{fig-opd-err-ij}
\end{figure}

In Figure~\ref{fig-opd-err-i}, $\I$-band path delay errors remain bounded within $-8$ to $+5$~nm across a 16~\mum range (2~nm RMS near zero OPD). A residual linear slope reflects a $0.17$\% ($1.4$~nm) wavelength calibration offset. Subtracting this linear trend leaves residual cyclic ripples of 5~nm peak-to-peak.

In the $\J$ band (Figure~\ref{fig-opd-err-j}), estimation errors reach 20~nm near zero OPD and increase at larger path delays. Disparate channel performance complicates fringe unwrapping (Figure~\ref{fig-coher}).

\begin{figure} \centering
  \FIG{0.7}{false}{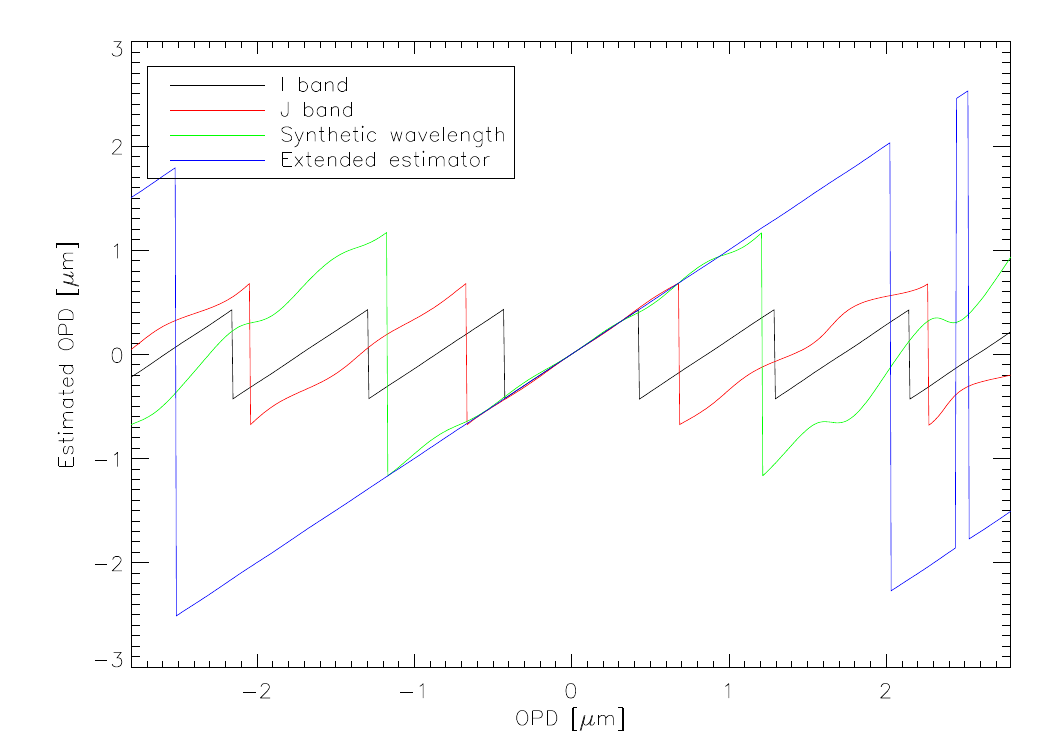}
  \caption{Dual-wavelength unwrapped path delay estimates.}
  \label{fig-coher}
\end{figure}

Single synthetic wavelength unwrapping provides a linear range of $2.5$~\mum. The extended estimator (Section~\ref{sec-estim-elargi}) with $q=8, k=5$ expands the linear range to $4.5$~\mum (Figure~\ref{fig-coher}).

%-------------------------------------------------------------------------------
\subsection{Piston Control Analysis with an Integrator}
\label{sec-analyse-piston-int}

%¤¤¤¤¤¤¤¤¤¤¤¤¤¤¤¤¤¤¤¤¤¤¤¤¤¤¤¤¤¤¤¤¤¤¤¤¤¤¤¤¤¤¤¤¤¤¤¤¤¤¤¤¤¤¤¤¤¤¤¤¤¤¤¤¤¤¤¤¤¤¤¤¤¤¤¤¤¤¤
\subsubsection{Rejection Transfer Function}
\label{sec-fonction-transfert-3}

The piston loop operates at 1~kHz ($31250/32$~Hz) with a 2-frame processing latency. The integrator rejection transfer function matches Figure~\ref{fig-trans-rej}, plotted in Figure~\ref{fig-FSrej} for various gains.

\begin{figure} \centering
  \FIG{.7}{false}{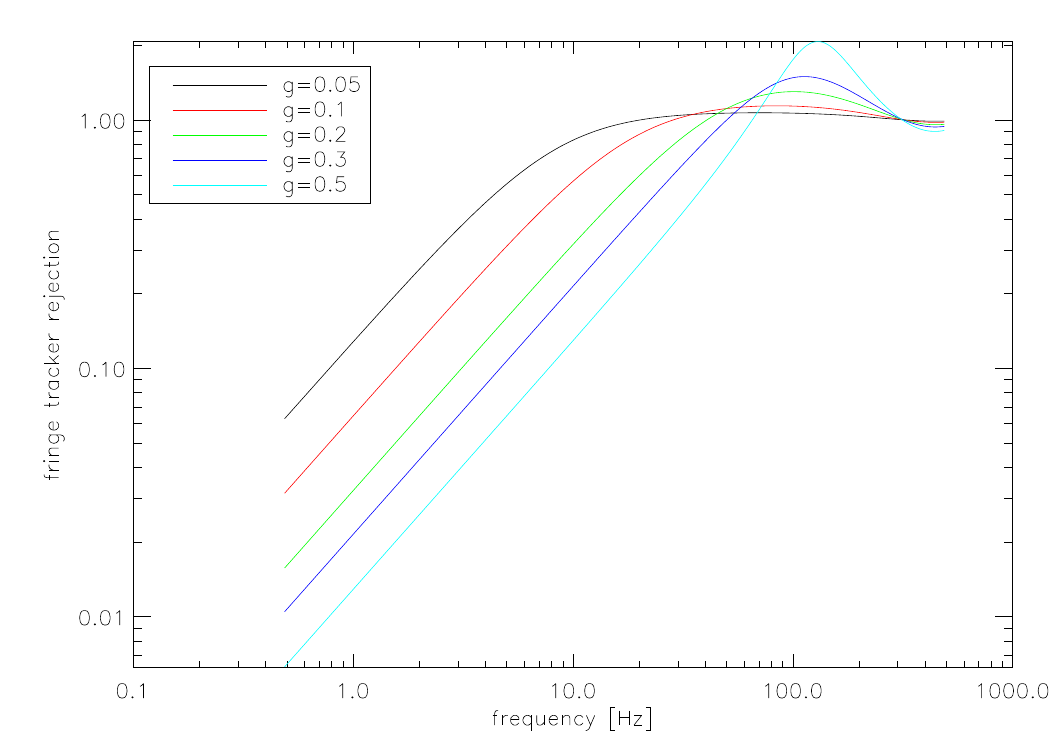}
  \caption[Piston loop rejection transfer function.]{Rejection transfer function of the piston loop across integrator gains.}
  \label{fig-FSrej}
\end{figure}

Latency introduces high-frequency amplification between 60~Hz and 300~Hz. To avoid exciting mechanical resonance modes within this band, integrator gain is set to $g=0.1$, yielding a 25~Hz closed-loop bandwidth.

%¤¤¤¤¤¤¤¤¤¤¤¤¤¤¤¤¤¤¤¤¤¤¤¤¤¤¤¤¤¤¤¤¤¤¤¤¤¤¤¤¤¤¤¤¤¤¤¤¤¤¤¤¤¤¤¤¤¤¤¤¤¤¤¤¤¤¤¤¤¤¤¤¤¤¤¤¤¤¤
\subsubsection{Optical Path Disturbance Sources}
\label{sec-elements-pert}

Optical path delay is perturbed by internal and external environmental noise. Pneumatic isolation legs under the optical table filter ground vibrations. Environmental noise sources include:
\begin{itemize}
\item Nearby rail traffic (RER C tunnel located 80~m below the building, Figure~\ref{fig-plan});
\item Acoustic noise from aircraft (Villacoublay Air Base 107 and Paris Heliport) exciting structural resonances (Figure~\ref{fig-pert-avion}).
\end{itemize}

\begin{figure} \centering
  \FIG{.7}{false}{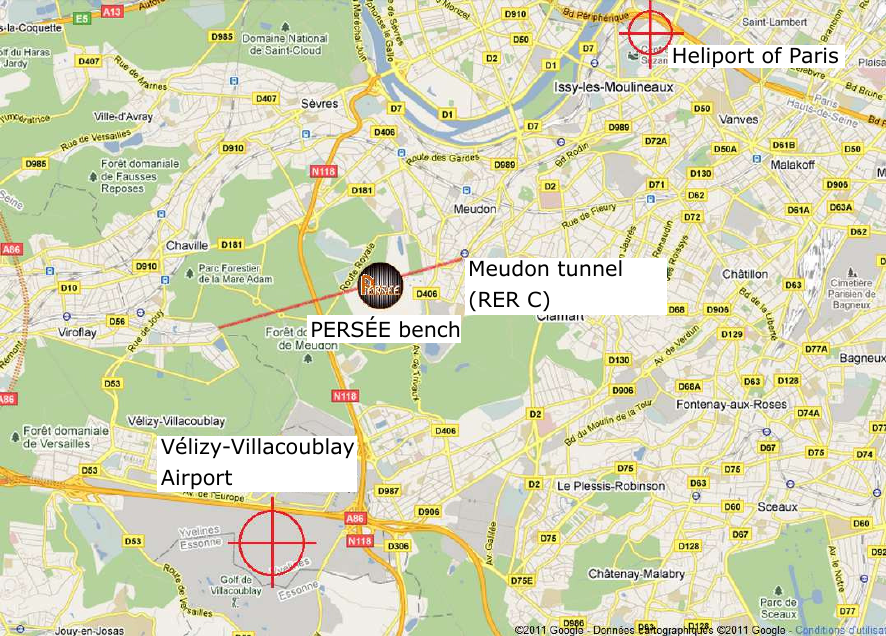}
  \caption{Map of the Meudon Observatory surroundings.}
  \label{fig-plan}
\end{figure}

\begin{figure} \centering
  \FIG{.7}{false}{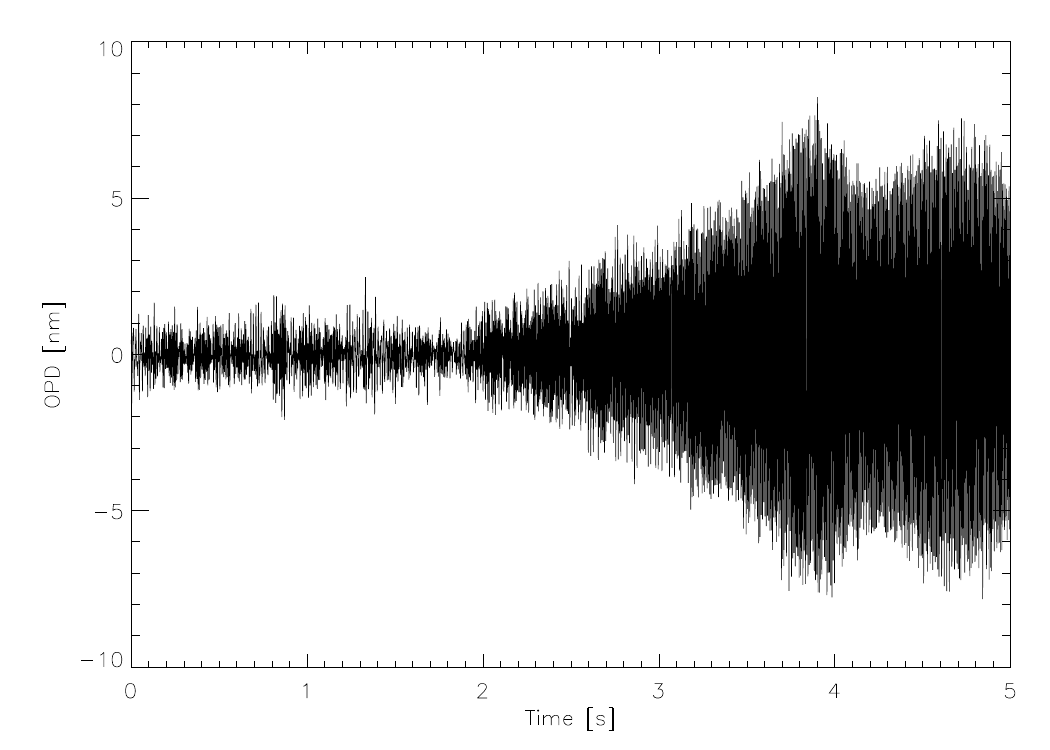}
  \caption{Closed-loop OPD jitter induced by acoustic aircraft noise.}
  \label{fig-pert-avion}
\end{figure}

Acoustic noise expands path delay jitter to >5~nm RMS (Figure~\ref{fig-pert-avion}) as Doppler shifts sweep engine frequencies across structural resonance modes. Enclosing the bench in a 5~cm thermo-acoustic foam enclosure mitigated acoustic excitation and thermal turbulence. Long-term stability tests were conducted at night.

Continuous acoustic noise sources include:
\begin{itemize}
\item Air conditioning systems ($5.9$~nm RMS OPD jitter);
\item Cleanroom overpressure ventilation ($6.5$~nm RMS OPD jitter);
\item Electronics cooling fans.
\end{itemize}

HVAC systems were powered down during precision testing. Electronics racks were relocated to an adjoining room.

Line power ripple (50~Hz and odd harmonics at 150~Hz, 250~Hz) introduced by S316 piezo drivers was eliminated using isolation transformers.

Disabling S316 strain gauges eliminates driver sensor noise, though internal strain gauges remained active during long-term nulling runs to maintain stage linearity.

%¤¤¤¤¤¤¤¤¤¤¤¤¤¤¤¤¤¤¤¤¤¤¤¤¤¤¤¤¤¤¤¤¤¤¤¤¤¤¤¤¤¤¤¤¤¤¤¤¤¤¤¤¤¤¤¤¤¤¤¤¤¤¤¤¤¤¤¤¤¤¤¤¤¤¤¤¤¤¤
\subsubsection{Piston Loop Performance}
\label{sec-perf-piston}

Piston rejection performance was evaluated across three integration setups:
\begin{enumerate}
\item Autocollimination setup (MMZ and M6 mirrors only);
\item Autocollimination setup with long-stroke delay lines (DL locked or active);
\item Final setup (M0 collimator, M1 siderostats, periscopes, M6 mirrors, DL, MMZ).
\end{enumerate}

Tests were performed with strain gauges disabled and optimal light levels. Figure~\ref{fig-FSpsdLAR} compares closed-loop path delay PSDs for setups 1 and 2 (DL locked).

\begin{figure} \centering
  \FIG{.7}{false}{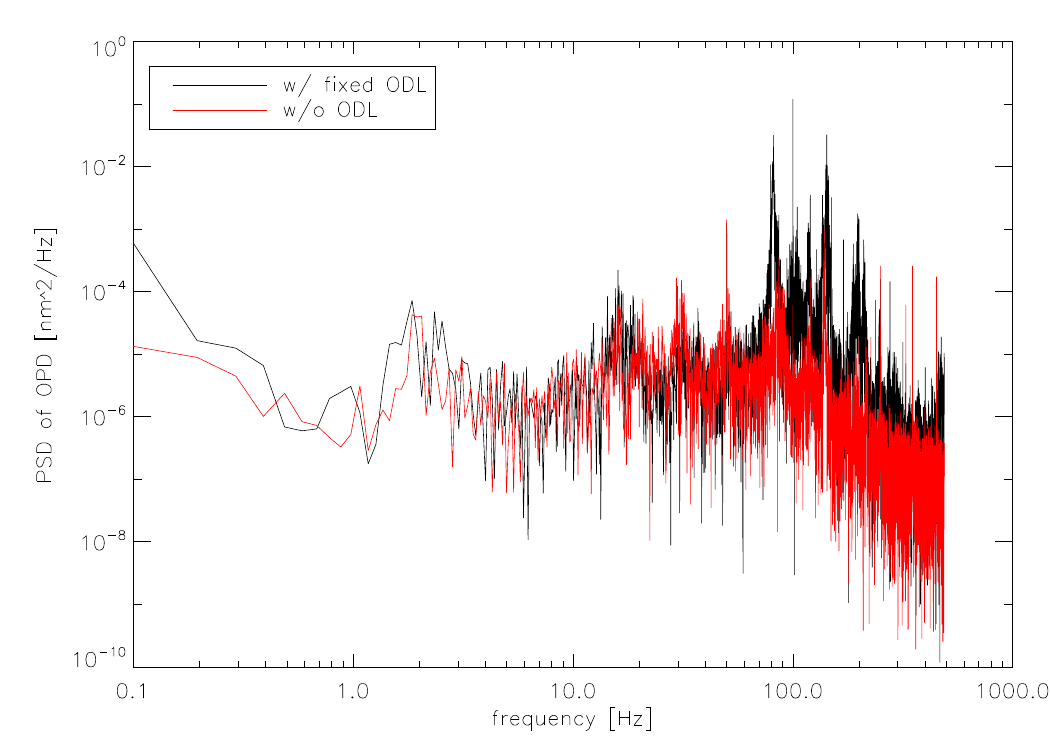}
  \caption[Impact of delay lines on closed-loop path delay PSDs.]{Closed-loop path delay PSDs in autocollimination with and without locked delay lines.}
  \label{fig-FSpsdLAR}
\end{figure}

In setup 1, residual path delay jitter reached $0.22$~nm RMS—well below the 1~nm RMS target. Spectral peaks correspond to:
\begin{itemize}
\item 2~Hz: Pneumatic table isolation mode;
\item 15, 25~Hz: Table flexure modes;
\item 50~Hz: Line power residual;
\item 80, 110~Hz: Flexure modes of M6$\mb$ and M6$\ma$ mounts;
\item 125, 140~Hz: Internal S316 piezo resonance modes (Section~\ref{sec-rep-temp});
\item 138~Hz: Dominant MMZ structure resonance mode (M9a/M9b mirror mounts);
\item 150, 250, 350, 450~Hz: Line power odd harmonics;
\item 165, 177, 188, 218~Hz: MMZ plate and L3 beamsplitter structural modes.
\end{itemize}

Locking the delay lines (setup 2) introduced an 80~Hz structural mode ($>1$~nm RMS). Loosening locking brackets attenuated this mode.

Figure~\ref{fig-FSpsdLARon} evaluates path delay jitter when delay line tracking loops are active.

\begin{figure} \centering
  \FIG{.7}{false}{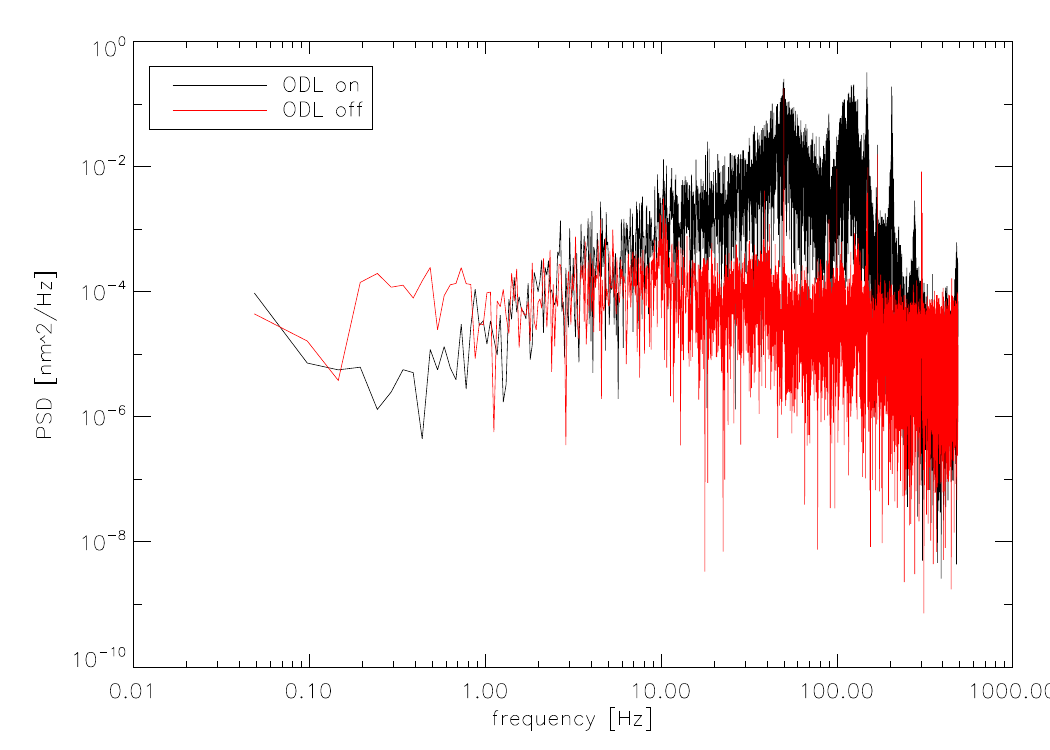}
  \caption[Impact of active delay lines on path delay PSDs.]{Closed-loop path delay PSDs with delay lines locked (red) and actively tracking (black).}
  \label{fig-FSpsdLARon}
\end{figure}

Active delay line tracking increased path delay jitter to 10~nm RMS (peaking at 50~Hz). Delay line tracking is used exclusively during coarse fringe search sequences.

Figure~\ref{fig-effet-jauges} shows the impact of enabling S316 strain-gauge feedback.

\begin{figure} \centering
  \FIG{.7}{false}{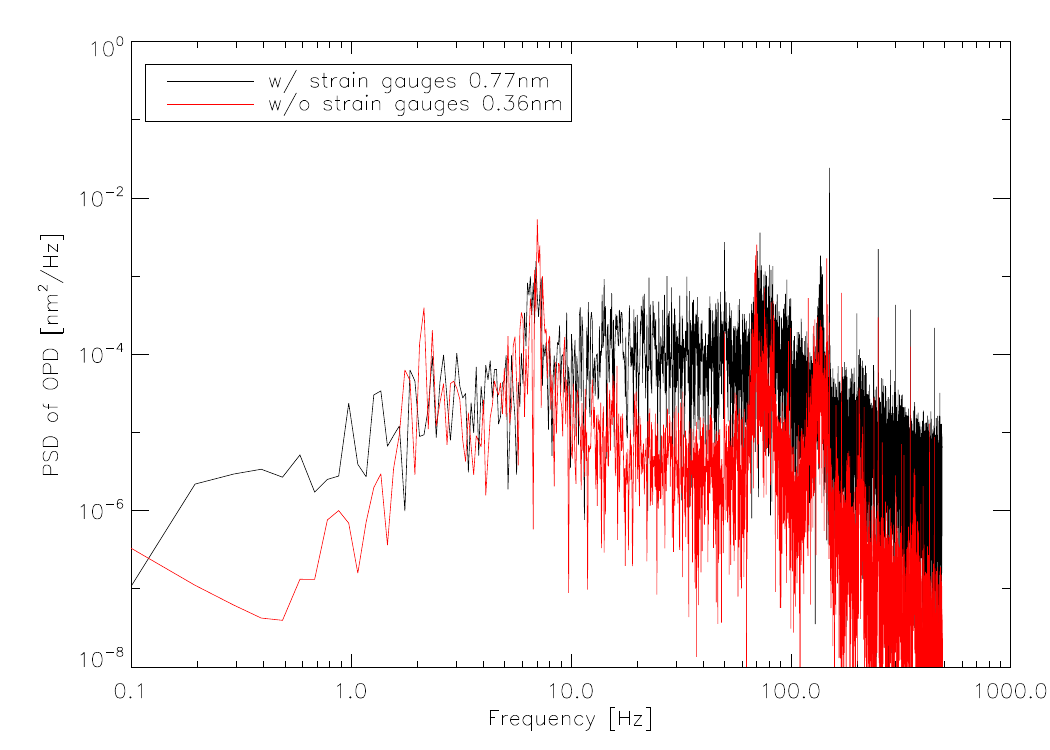}
  \caption[Impact of strain-gauge feedback on closed-loop path delay PSDs.]{Closed-loop path delay PSDs with strain gauges enabled (black) and disabled (red).}
  \label{fig-effet-jauges}
\end{figure}

Enabling strain-gauge feedback amplifies 50~Hz power harmonics and increases baseline sensor noise. Adjusting pneumatic isolation legs eliminated a 7~Hz table mode.

Under optimal lab conditions, closed-loop path delay jitter reached $\boldsymbol{0.3}$~\textbf{nm RMS} over 100~s ($0.4$~nm RMS over extended runs). Enabling strain gauges increased jitter slightly to $0.7$~nm RMS—still below the 1~nm RMS requirement.

Injecting dynamic formation-flying disturbances requires control strategies beyond simple integrators, as evaluated in the following section.

%§§§§§§§§§§§§§§§§§§§§§§§§§§§§§§§§§§§§§§§§§§§§§§§§§§§§§§§§§§§§§§§§§§§§§§§§§§§§§§§
\section[LQG Piston Control Optimization]{LQG Piston Control Optimization}
\label{sec-lqg-persee}

Applying LQG control and online identification (Section~\ref{sec-commande-lqg}) to \pe's piston loop yielded results compiled for peer-reviewed publication \cite{Lozi12a}.

%-------------------------------------------------------------------------------
\subsection{Adapting Disturbance Identification to \pe}
\label{sec-adapt-ident}

%¤¤¤¤¤¤¤¤¤¤¤¤¤¤¤¤¤¤¤¤¤¤¤¤¤¤¤¤¤¤¤¤¤¤¤¤¤¤¤¤¤¤¤¤¤¤¤¤¤¤¤¤¤¤¤¤¤¤¤¤¤¤¤¤¤¤¤¤¤¤¤¤¤¤¤¤¤¤¤
\subsubsection{Mean Path Delay Handling}
\label{sec-gestion-moyenne}

Nulling interferometry requires locking path delay to zero mean offset. Standard integrator controllers zero mean path errors naturally because their rejection transfer function vanishes at DC ($f=0$) (Figure~\ref{fig-FSrej}). Standard LQG controllers assume zero-mean disturbances, leaving uncorrected DC biases when applied to path delay tracking (Figure~\ref{fig-int-kal}).

\begin{figure} \centering
  \FIG{.7}{false}{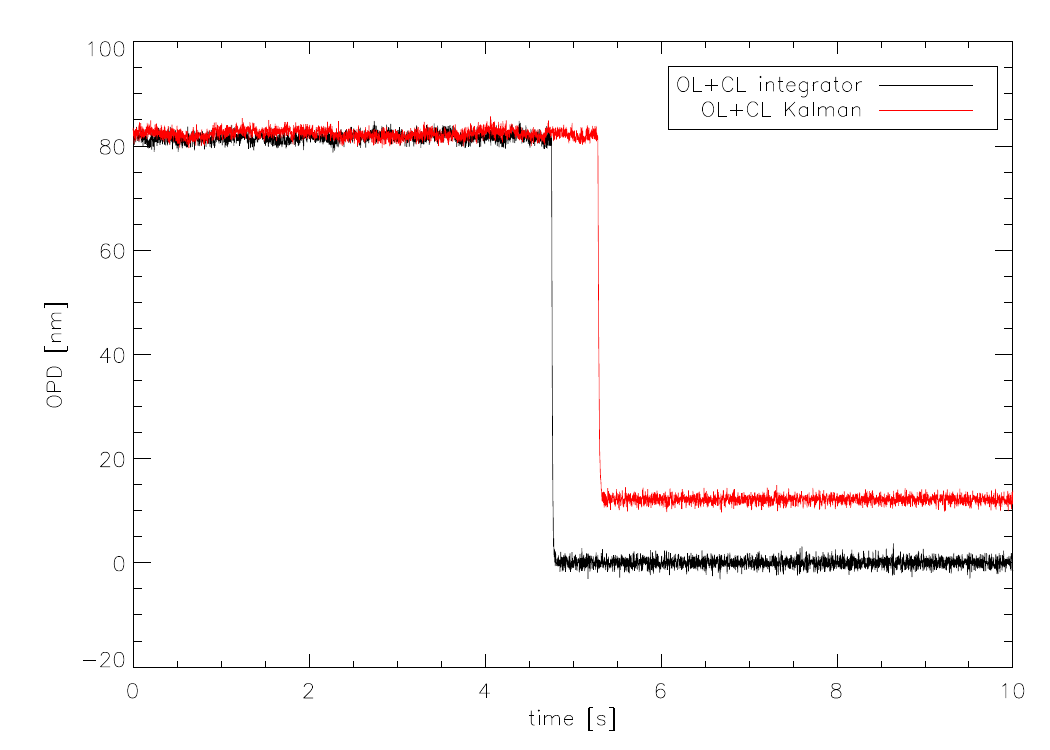}
  \caption[Closed-loop step responses.]{Closed-loop time series comparing integrator and standard LQG responses under a static path offset.}
  \label{fig-int-kal}
\end{figure}

In Figure~\ref{fig-int-kal}, an 82~nm open-loop path offset is zeroed by the integrator, whereas the standard LQG leaves a 12~nm residual DC bias.

Four methods address mean path delay correction:
\begin{itemize}
\item Outer-loop offset integration;
\item Constraining the low-frequency AR2 disturbance model (Equation~\ref{eq-phibf}) to satisfy $a^\mbf_1+a^\mbf_2 = 1$, forcing zero rejection at DC at the expense of model optimality;
\item Adding a static state variable $\phi^\msta_n$ to state vector $\V{x}_n$ (Equation~\ref{eq-phi-mat}):
  \begin{equation}\label{eq-phi-mat2}
    \begin{pmatrix}
      \begin{array}{@{}c@{}}
      	\phi^\msta_{n+1}\\
      	\hline
        \phi^\mbf_{n+1}\\
        \phi^\mbf_{n}\\
        \hline
        \phi^{\mvib,1}_{n+1}\\
        \phi^{\mvib,1}_{n}\\
        \hline
        \phi^{\mvib,2}_{n+1}\\
        \phi^{\mvib,2}_{n}\\
        \hline
        \vdots
      \end{array}
    \end{pmatrix}=\begin{pmatrix}
      \begin{array}{@{}c@{\,}|@{\,}c@{\,}c@{\,}|@{\,}c@{\,}c@{\,}|@{\,}c@{\,}c@{\,}|c@{}}
      	1 & 0 & 0 & 0 & 0 & 0 & 0 & \cdots\\
      	\hline
        0 & a^\mbf_1 & a^\mbf_2 & 0 & 0 & 0 & 0 & \cdots\\
        0 & 1 & 0 & 0 & 0 & 0 & 0 & \cdots\\
        \hline
        0 & 0 & 0 & a^{\mvib,1}_1 & a^{\mvib,1}_2 & 0 & 0 & \cdots\\
        0 & 0 & 0 & 1 & 0 & 0 & 0 & \cdots\\
        \hline
        0 & 0 & 0 & 0 & 0 & a^{\mvib,2}_1 & a^{\mvib,2}_2 & \cdots\\
        0 & 0 & 0 & 0 & 0 & 1 & 0 & \cdots\\
        \hline
        \vdots & \vdots & \vdots & \vdots & \vdots & \vdots & \vdots & \ddots
      \end{array}
    \end{pmatrix} \begin{pmatrix}
      \begin{array}{@{}c@{}}
      	\phi^\msta_n\\
      	\hline
        \phi^\mbf_n\\
        \phi^\mbf_{n-1}\\
        \hline
        \phi^{\mvib,1}_n\\
        \phi^{\mvib,1}_{n-1}\\
        \hline
        \phi^{\mvib,2}_n\\
        \phi^{\mvib,2}_{n-1}\\
        \hline
        \vdots
      \end{array}
    \end{pmatrix} +\begin{pmatrix}
      \begin{array}{@{}c@{}}
      	0\\
      	\hline
        \upsilon^\mbf_n\\
        0\\
        \hline
        \upsilon^{\mvib,1}_n\\
        0\\
        \hline
        \upsilon^{\mvib,2}_n\\
        0\\
        \hline
        \vdots
      \end{array}
    \end{pmatrix}.
  \end{equation}
  This approach can complicate distinguishing static offsets from low-frequency drift \cite{Petit06};
\item Adding an explicit sensor bias term $y^\msta_n$ ($y^\msta_{n+1} = y^\msta_n$) to measurement equation~\eqref{eq-yn2}:
  \begin{equation}
    y_n = \V{C}\V{x}_n-DNu_{n-2}+y^\msta_n+w_n.
    \label{eq-yn3} 
  \end{equation}
  This decouples sensor offsets from phase dynamics \cite{Petit06}.
\end{itemize}

Method 2 was implemented on \pe for zero-mean path delay tracking.

%¤¤¤¤¤¤¤¤¤¤¤¤¤¤¤¤¤¤¤¤¤¤¤¤¤¤¤¤¤¤¤¤¤¤¤¤¤¤¤¤¤¤¤¤¤¤¤¤¤¤¤¤¤¤¤¤¤¤¤¤¤¤¤¤¤¤¤¤¤¤¤¤¤¤¤¤¤¤¤
\subsubsection{Excluded Frequency Bands}
\label{sec-vibrations-inter}

Actively driving control commands at structural resonance frequencies risks triggering mechanical instability. The identification algorithm was modified to exclude known mechanical resonance bands (S316 piezo modes, MMZ mount modes) from vibration fitting.

%¤¤¤¤¤¤¤¤¤¤¤¤¤¤¤¤¤¤¤¤¤¤¤¤¤¤¤¤¤¤¤¤¤¤¤¤¤¤¤¤¤¤¤¤¤¤¤¤¤¤¤¤¤¤¤¤¤¤¤¤¤¤¤¤¤¤¤¤¤¤¤¤¤¤¤¤¤¤¤
\subsubsection{Measurement Noise Overestimation}
\label{sec-surestim-bruit}

Uncertainties in identified disturbance parameters can degrade closed-loop stability. Robustness is improved by scaling the estimated measurement noise variance $\hat{w}_n$ by a factor $k_w$ when solving steady-state Kalman gains $H_\infty$ (Equation~\ref{eq-H-infty}) \cite{Petit06}:
\begin{equation}
  \hat{w}'_n = k_w \hat{w}_n.
\end{equation}

Scaling factors $k_w = 50$ to $100$ provided optimal control robustness (Section~\ref{sec-etude-surestim}).

%-------------------------------------------------------------------------------
\subsection{Performance Evaluation Without Injected Disturbances}
\label{sec-resultats}

%¤¤¤¤¤¤¤¤¤¤¤¤¤¤¤¤¤¤¤¤¤¤¤¤¤¤¤¤¤¤¤¤¤¤¤¤¤¤¤¤¤¤¤¤¤¤¤¤¤¤¤¤¤¤¤¤¤¤¤¤¤¤¤¤¤¤¤¤¤¤¤¤¤¤¤¤¤¤¤
\subsubsection{Test Protocol}
\label{sec-protocole-mesure}

To decouple disturbance injection from active correction, perturbations were driven through stage M6$\ma$, while corrections were applied via stage M6$\mb$.

Identification ran on 20~s open-loop datasets (20,000 samples) over frequency range $f_1 = 5$~Hz to $f_2 = 225$~Hz, with measurement noise estimated across $f_w = 300$~Hz to $f_\mfs/2 = 500$~Hz.

Tests ran with S316 strain gauges disabled. PZT gain factor $N=0.6$ (Equation~\ref{eq-phicor}) was updated in the LQG model. Baseline integrator gain was set to $g=0.4$.

Injected disturbances comprised low-frequency satellite positioning noise (flat PSD from 0 to 1~Hz, $-40$~dB/decade roll-off above 1~Hz, amplitude 10~nm RMS) and discrete vibration modes.

%¤¤¤¤¤¤¤¤¤¤¤¤¤¤¤¤¤¤¤¤¤¤¤¤¤¤¤¤¤¤¤¤¤¤¤¤¤¤¤¤¤¤¤¤¤¤¤¤¤¤¤¤¤¤¤¤¤¤¤¤¤¤¤¤¤¤¤¤¤¤¤¤¤¤¤¤¤¤¤
\subsubsection{Environmental Disturbance Characterization}
\label{sec-descr-envir}

Figure~\ref{fig-p-asjau} plots open-loop path delay PSDs without injected disturbances.

\begin{figure} \centering
  \FIG{.7}{false}{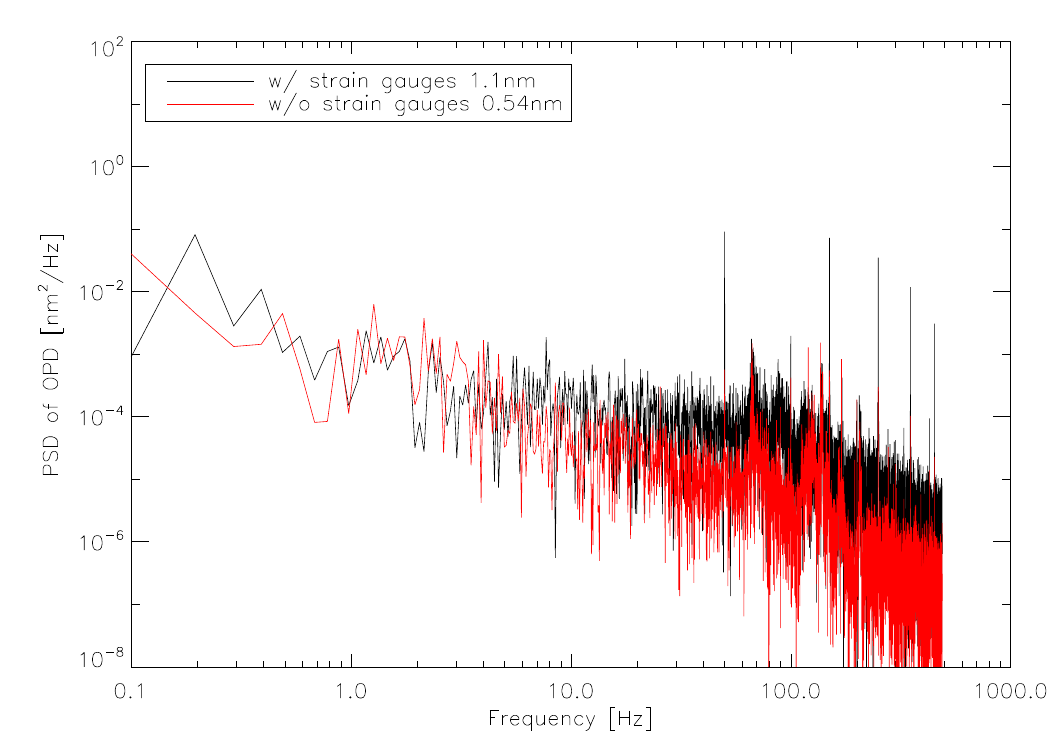}
  \caption{Open-loop path delay PSDs with strain gauges enabled and disabled.}
  \label{fig-p-asjau}
\end{figure}

Uncorrected lab path delay jitter measures $1.1$~nm RMS with strain gauges enabled and $0.5$~nm RMS with strain gauges disabled. Disabling strain gauges eliminates 50~Hz line power harmonics.

Integrated cumulative power spectra $S_\mi(f)$ are defined as:
\begin{equation}
  S_\mi(f) = \GP{2\int_0^{f}S(f')\dd f'-S(0)}^\frac12.
\end{equation}

Cumulative spectra $S_\mi(f)$ display discrete vibrations as steps and broadband noise as smooth curves, evaluating total RMS jitter at $f = f_\mfs$. Figure~\ref{fig-pi-oil-sspert} compares integrator and LQG performance under lab environments.

\begin{figure} \centering
  \subfloat[Path delay PSDs.]{\label{fig-p-oil-sspert}
    \FIG{0.49}{false}{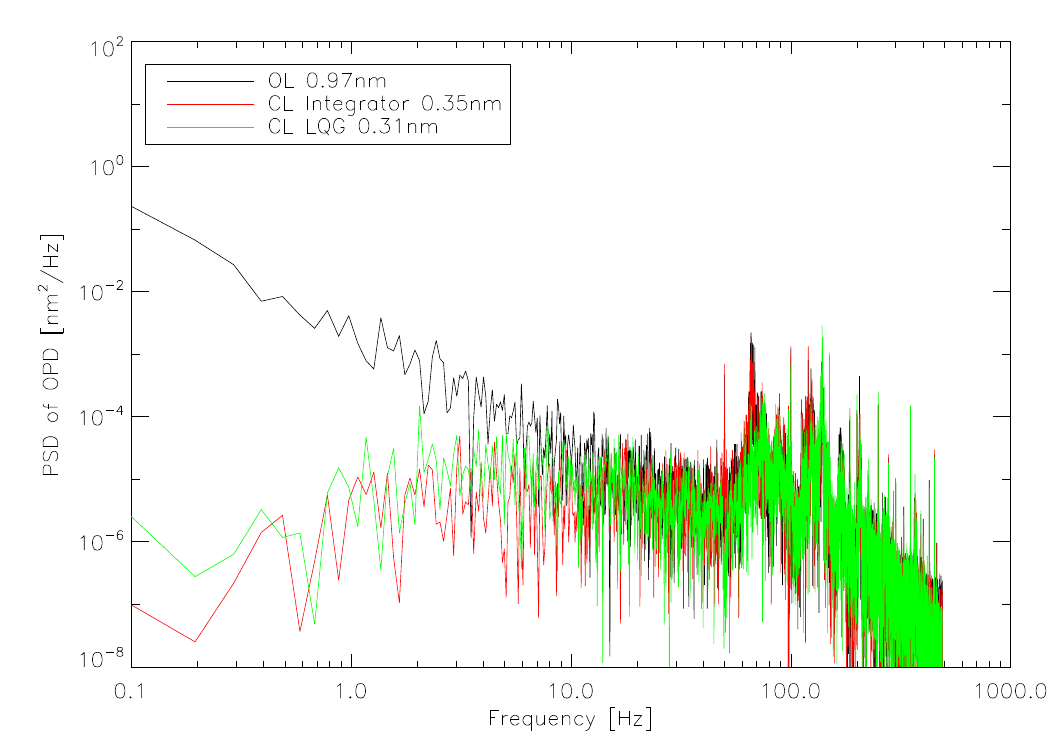}}
  \hfill\subfloat[Cumulative power spectra.]{\label{fig-i-oil-sspert}
    \FIG{0.49}{false}{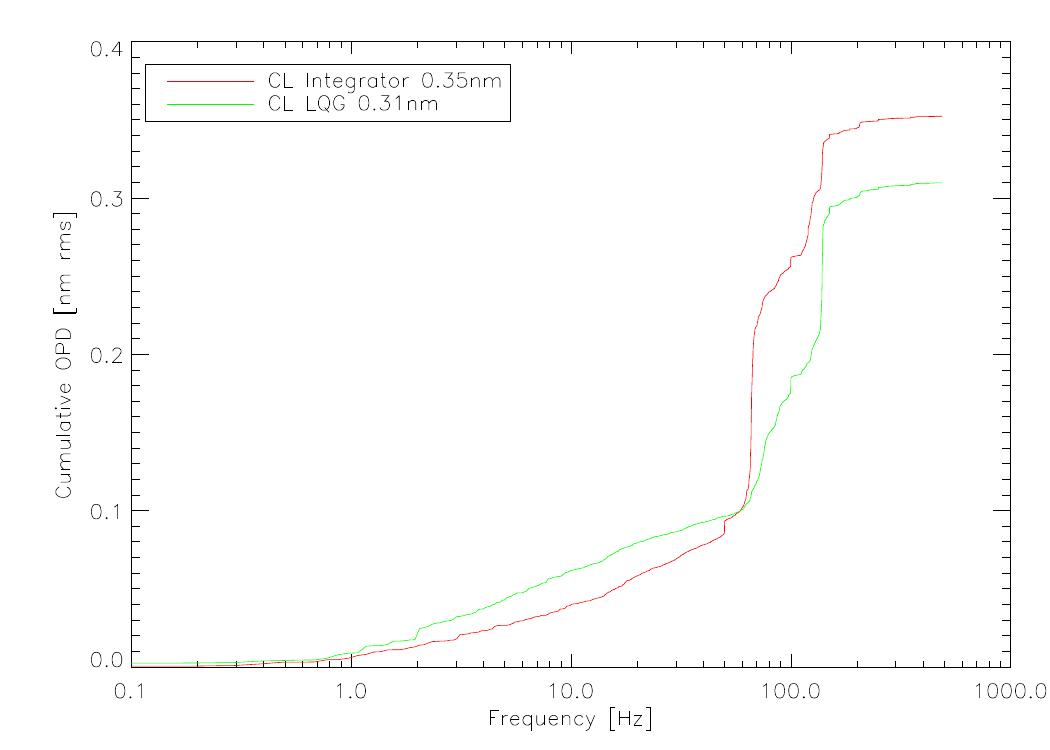}}
  \caption[Closed-loop performance under ambient lab noise.]{Path delay PSDs and cumulative power spectra under ambient lab noise.}
  \label{fig-pi-oil-sspert}
\end{figure}

The LQG filter identified lab modes at 67, 69, 120, 205~Hz, and 50~Hz line noise. Closed-loop RMS path delay jitter dropped from $0.97$~nm RMS (open loop) down to $0.35$~nm RMS with the integrator and $0.31$~nm RMS with LQG. An 11\% reduction in RMS path delay jitter corresponds to a 22\% reduction in quadratic null depth leakage.

%¤¤¤¤¤¤¤¤¤¤¤¤¤¤¤¤¤¤¤¤¤¤¤¤¤¤¤¤¤¤¤¤¤¤¤¤¤¤¤¤¤¤¤¤¤¤¤¤¤¤¤¤¤¤¤¤¤¤¤¤¤¤¤¤¤¤¤¤¤¤¤¤¤¤¤¤¤¤¤
\subsubsection{Noise Overestimation Factor Analysis}
\label{sec-etude-surestim}

Figure~\ref{fig-pi-l-kw} evaluates LQG performance across noise scaling factors $k_w$.

\begin{figure} \centering
  \subfloat[Path delay PSDs.]{\label{fig-p-l-kw}
    \FIG{0.49}{false}{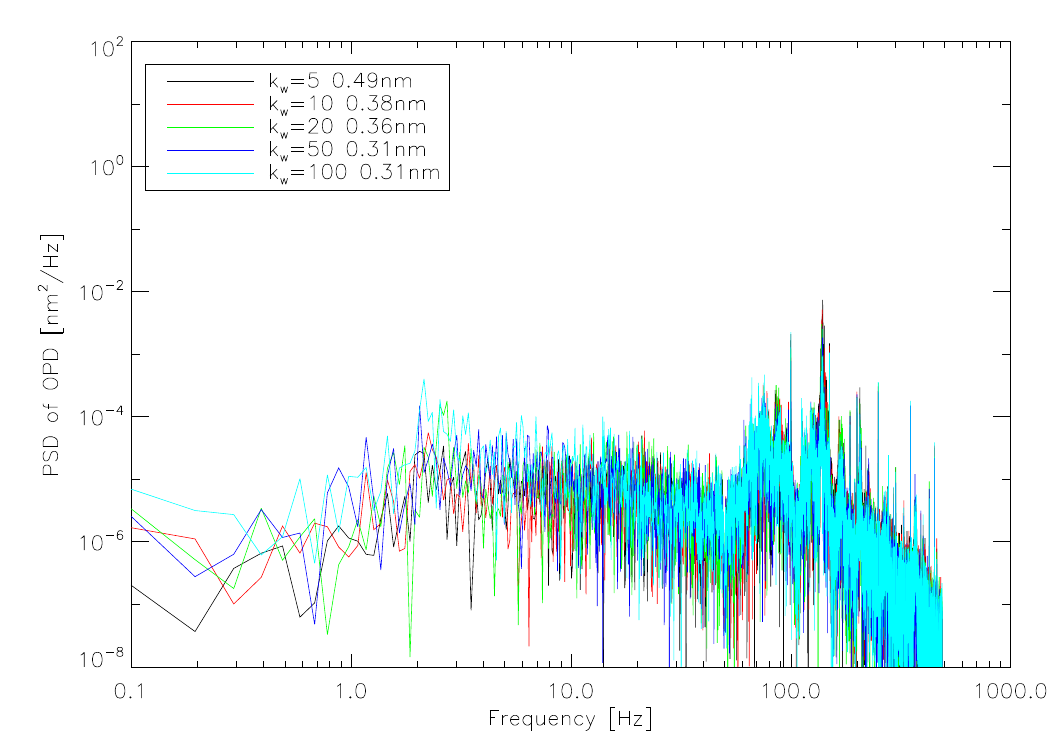}}
  \hfill\subfloat[Cumulative power spectra.]{\label{fig-i-l-kw}
    \FIG{0.49}{false}{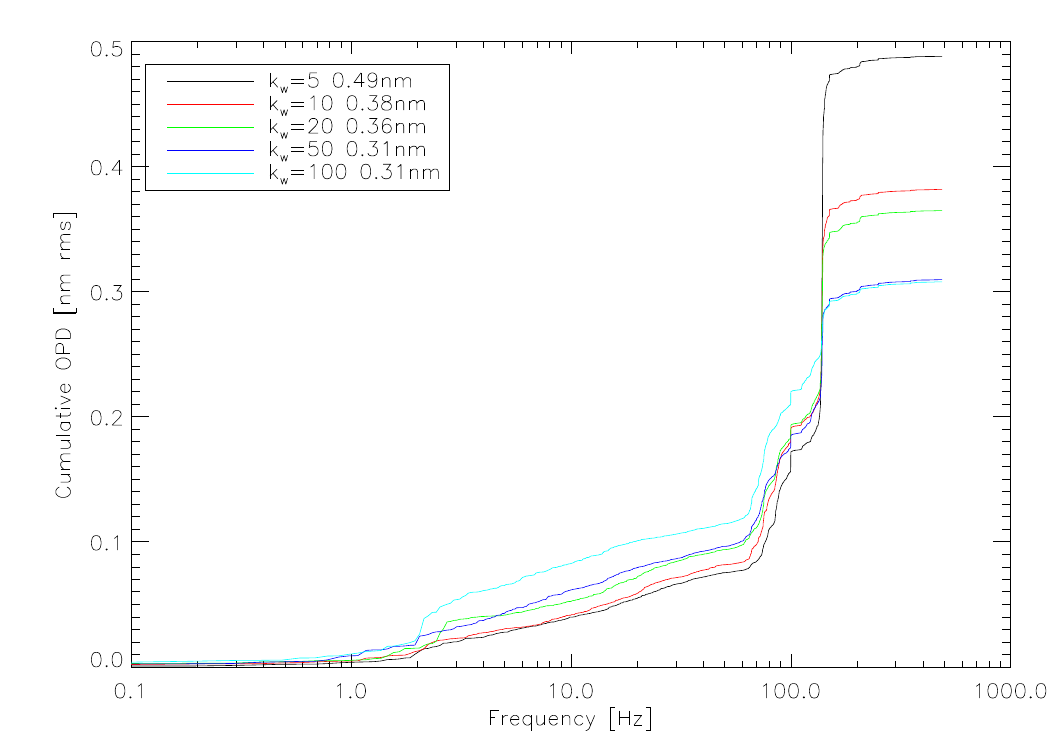}}
  \caption[Impact of noise overestimation factor.]{LQG path delay PSDs and cumulative power across noise scaling factors $k_w$.}
  \label{fig-pi-l-kw}
\end{figure}

Larger scaling factors $k_w$ reduce gain overshooting near piezo resonance modes. Optimal performance ($0.31$~nm RMS) was obtained for $k_w = 50$ to $100$.

%¤¤¤¤¤¤¤¤¤¤¤¤¤¤¤¤¤¤¤¤¤¤¤¤¤¤¤¤¤¤¤¤¤¤¤¤¤¤¤¤¤¤¤¤¤¤¤¤¤¤¤¤¤¤¤¤¤¤¤¤¤¤¤¤¤¤¤¤¤¤¤¤¤¤¤¤¤¤¤
\subsubsection{Flux Sensitivity}
\label{sec-sensib-flux}

Controller performance was evaluated across light levels at maximum detector gain ($10^8$~V/A). Figures~\ref{fig-pi-i-flux} and \ref{fig-pi-l-flux} plot PSDs for the integrator and LQG across flux levels.

\begin{figure} \centering
  \subfloat[Path delay PSDs.]{\label{fig-p-i-flux}
    \FIG{0.47}{false}{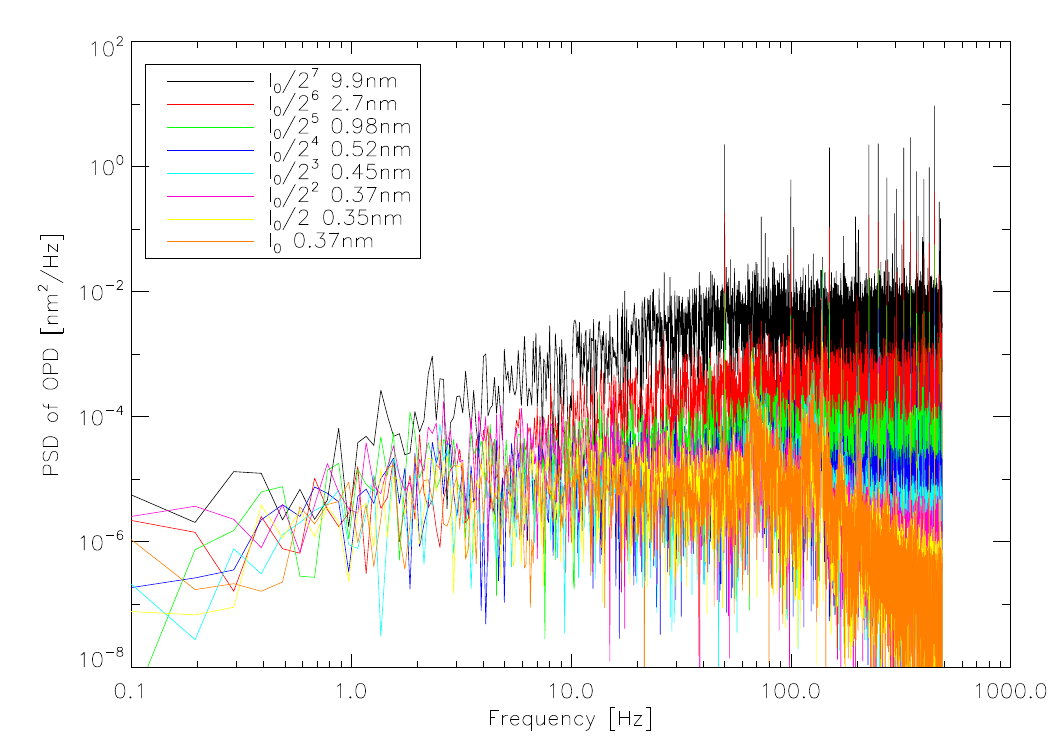}}
  \hfill\subfloat[Cumulative power spectra.]{\label{fig-i-i-flux}
    \FIG{0.47}{false}{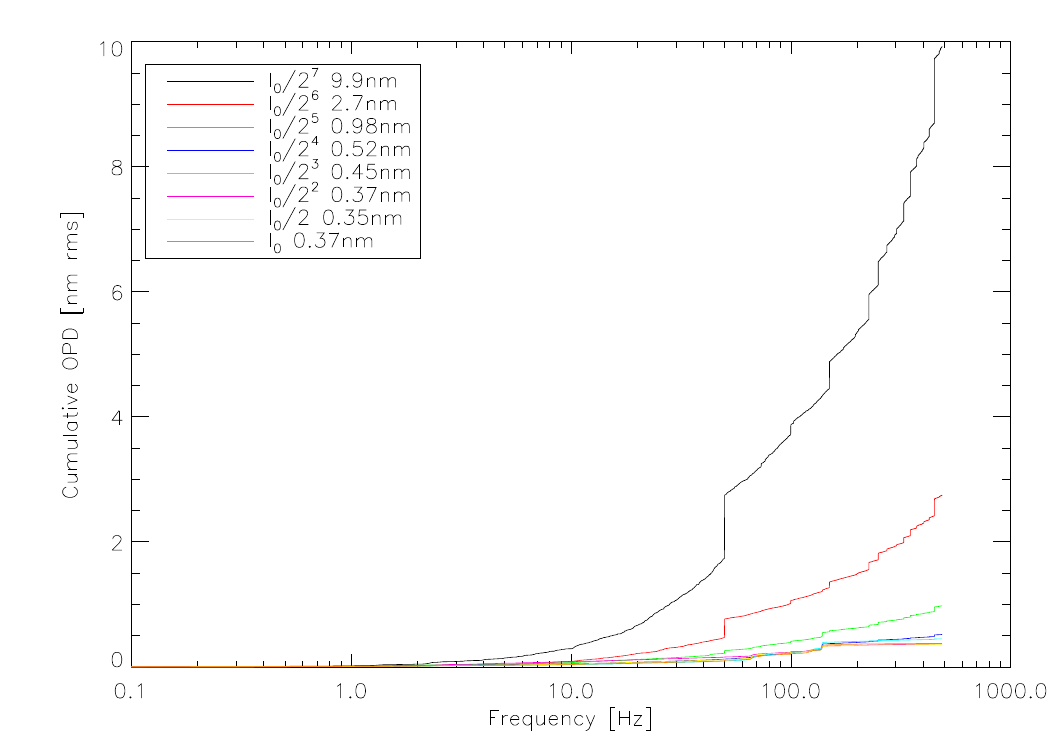}}
  \caption{Integrator path delay PSDs across flux levels.}
  \label{fig-pi-i-flux}
\end{figure}

\begin{figure} \centering
  \subfloat[Path delay PSDs.]{\label{fig-p-l-flux}
    \FIG{0.47}{false}{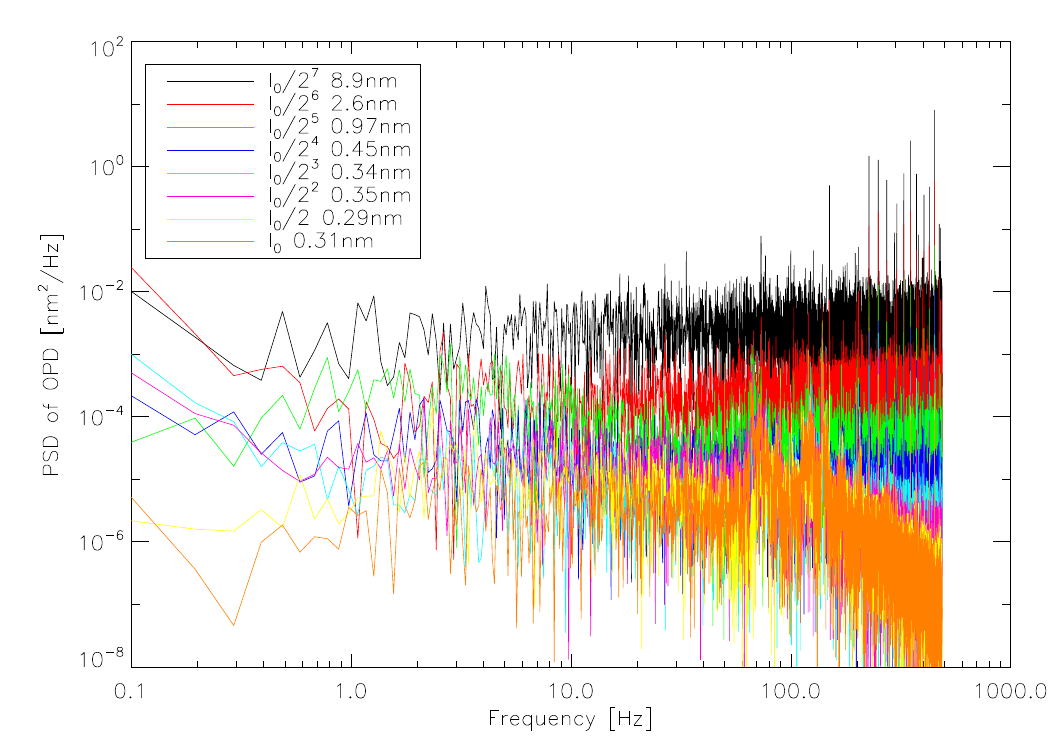}}
  \hfill\subfloat[Cumulative power spectra.]{\label{fig-i-l-flux}
    \FIG{0.47}{false}{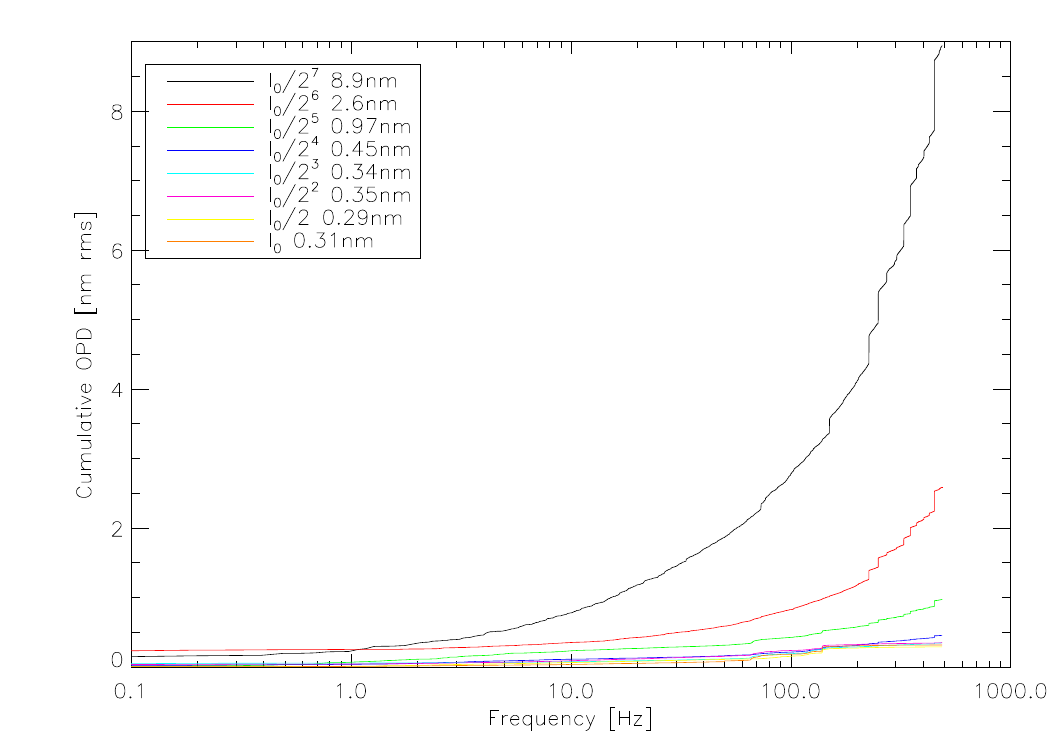}}
  \caption{LQG path delay PSDs across flux levels.}
  \label{fig-pi-l-flux}
\end{figure}

At high light levels, performance is limited by lab structural vibrations. Lowering light levels increases white sensor noise. The LQG controller maintains notch rejection at 50~Hz line noise across flux levels. Figure~\ref{fig-res-flux} plots residual path delay jitter against photon flux.

\begin{figure} \centering
  \FIG{.5}{false}{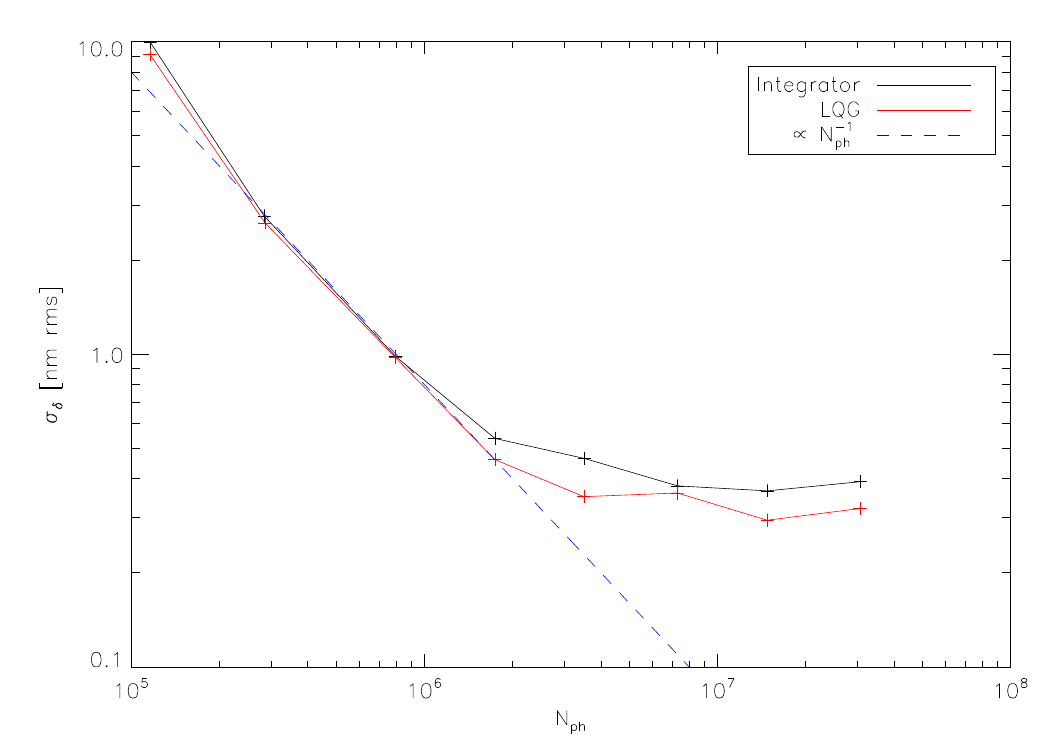}
  \caption{Residual path delay jitter versus input photon flux.}
  \label{fig-res-flux}
\end{figure}

Maintaining residual path delay jitter below requirements requires a minimum metrology flux of $3\E{5}$~photons in the $\I$ band ($\text{SNR} = 440$). Below this threshold, residual jitter scales as $N_\mph^{-1}$, reflecting source intensity noise limitations (Section~\ref{sec--etude-1}).

%-------------------------------------------------------------------------------
\subsection{Identification Performance Under Injected Disturbances}
\label{sec-analyse-identif}

%¤¤¤¤¤¤¤¤¤¤¤¤¤¤¤¤¤¤¤¤¤¤¤¤¤¤¤¤¤¤¤¤¤¤¤¤¤¤¤¤¤¤¤¤¤¤¤¤¤¤¤¤¤¤¤¤¤¤¤¤¤¤¤¤¤¤¤¤¤¤¤¤¤¤¤¤¤¤¤
\subsubsection{Low-Frequency Drift Identification}
\label{sec-identification-bf}

We evaluate low-frequency satellite positioning drift correction.

We define variance metrics:
\begin{itemize}
\item $\sigma_{\mbo,\mper}^2$: open-loop variance with low-frequency drift from satellite pointing and vibrations;
\item $\sigma_{c,\mper}^2$: closed-loop variance for controller $c \in \{\mint, \mlqg\}$;
\item $\sigma_{\mbo,\mps}^2$: open-loop variance with low-frequency drift from satellite pointing (SP) only;
\item $\sigma_{c,\mps}^2$: closed-loop variance with low-frequency drift only;
\item $\sigma_{\mbo,\mexp}^2, \sigma_{c,\mexp}^2$: baseline open-loop and closed-loop variances without injected disturbances;
\item $\sigma_{\minj,\mps}^2, \sigma_{\minj,\mvib}^2$: injected low-frequency drift and vibration variances.
\end{itemize}

Summing independent variance terms yields:
\begin{equation}
  \sigma_{\mbo,\mper}^2 =
  \sigma_{\mbo,\mexp}^2+\sigma_{\minj,\mps}^2+\sigma_{\minj,\mvib}^2.
  \label{eq-sig-per}
\end{equation}

For low-frequency drift alone:
\begin{equation}
  \sigma_{\mbo,\mps}^2 = \sigma_{\mbo,\mexp}^2+\sigma_{\minj,\mps}^2.
  \label{eq-sig-PS}
\end{equation}

Residual low-frequency power fraction $\rho_{c,\mps}$ for controller $c$ is defined as:
\begin{equation}
  \rho_{c,\mps} = \frac{\sigma_{c,\mps}^2-\sigma_{c,\mexp}^2}
  {\sigma_{\mbo,\mps}^2-\sigma_{c,\mexp}^2}.
  \label{eq-rho-PS-c}
\end{equation}

$\rho_{c,\mps} = 0\%$ indicates complete disturbance rejection. Figure~\ref{fig-pi-oil-pertrms} and Table~\ref{tab-cor-PS} evaluate low-frequency drift rejection.

\begin{figure} \centering
  \subfloat[Path delay PSDs.]{\label{fig-p-oil-pertrms}
    \FIG{0.49}{false}{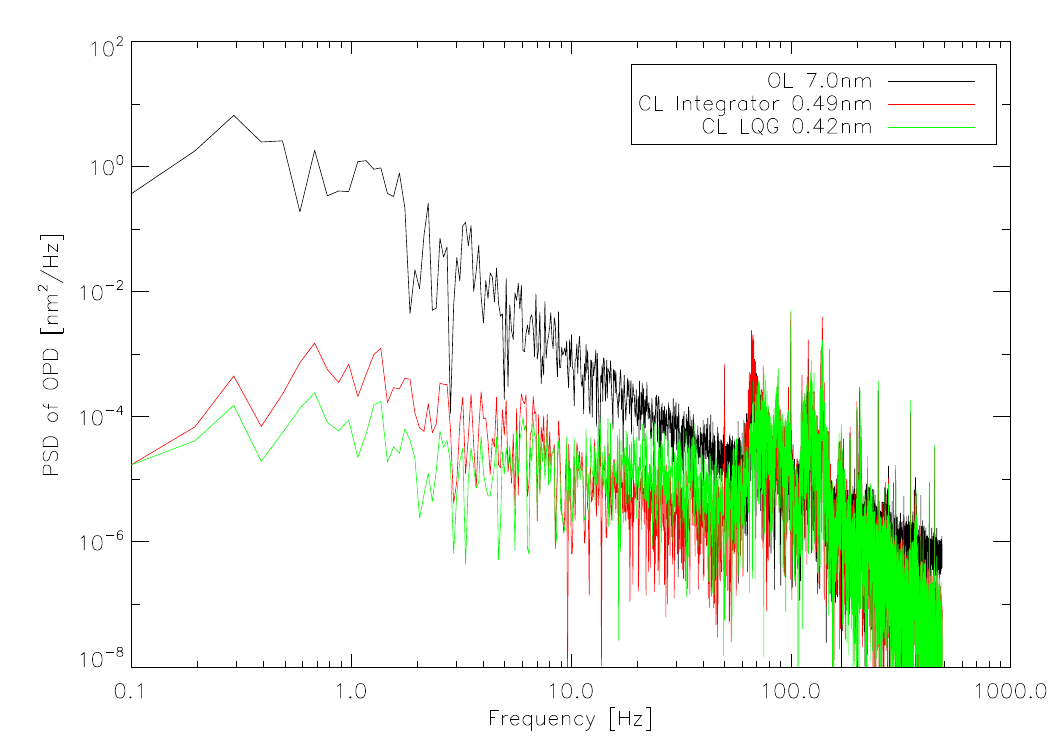}}
  \hfill\subfloat[Cumulative power spectra.]{\label{fig-i-oil-pertrms}
    \FIG{0.49}{false}{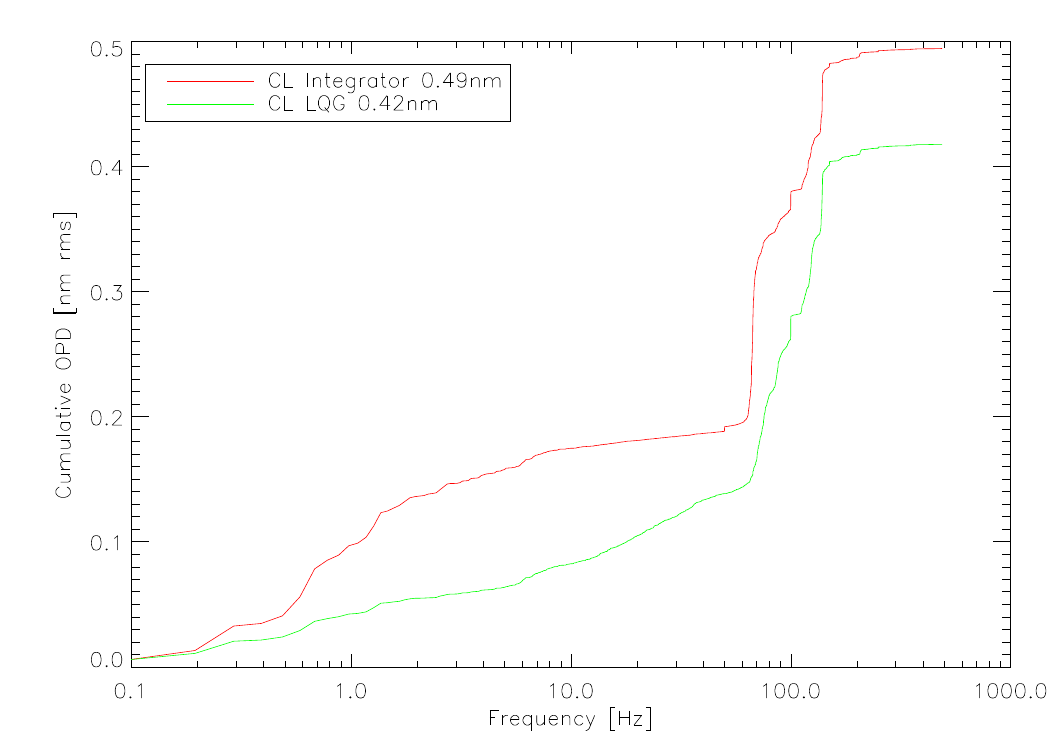}}
  \caption[Low-frequency positioning drift rejection.]{Low-frequency satellite positioning drift rejection under integrator and LQG control.}
\label{fig-pi-oil-pertrms}
\end{figure}

\begin{table} \centering
  \caption{Low-frequency positioning drift rejection under integrator and LQG control.}
  \medskip
  \begin{tabular}{ccc}
    \hline \hline
    Controller $c$ & $\sigma_{c,\mps}$ & $\rho_{c,\mps}$ \\
    \hline
    Integrator & $0.45$~nm~RMS & $0.17$\% \\
    LQG & $0.37$~nm~RMS & $0.09$\% \\
    \hline \hline
  \end{tabular}
  \label{tab-cor-PS}
\end{table}

Both controllers reject low-frequency positioning drift effectively ($\rho_{c,\mps} < 0.2\%$). The LQG controller achieves slightly lower residual jitter ($0.37$~nm RMS) by simultaneously suppressing ambient lab vibrations.

Residual vibration power fraction $\rho_c$ is evaluated as:
\begin{equation}
  \rho_c = \frac{\sigma_{c,\mper}^2-\sigma_{c,\mps}^2}
  {\sigma_{\mbo,\mper}^2-\sigma_{\mbo,\mps}^2-\sigma_{c,\mps}^2}.
  \label{eq-rho-c}
\end{equation}

%¤¤¤¤¤¤¤¤¤¤¤¤¤¤¤¤¤¤¤¤¤¤¤¤¤¤¤¤¤¤¤¤¤¤¤¤¤¤¤¤¤¤¤¤¤¤¤¤¤¤¤¤¤¤¤¤¤¤¤¤¤¤¤¤¤¤¤¤¤¤¤¤¤¤¤¤¤¤¤
\subsubsection{Vibration Frequency Sensitivity}
\label{sec-influence-frequence}

Single harmonic vibrations were injected across frequencies $f_\mvib$. Rejection performance is compiled in Table~\ref{tab-freq-vib}.

\begin{table} \centering
  \caption{Rejection performance across injected vibration frequencies.}
  \medskip
  \begin{tabular}{ccccc}
    \hline \hline
    $f_\mvib$ & $\sigma_{\mint,\mper}$ & $\rho_\mint$ & $\sigma_{\mlqg,\mper}$ &
    $\rho_\mlqg$ \\\relax
    [Hz] & [nm RMS] & [\%] & [nm RMS] & [\%] \\
    \hline
    $1.1$ & $0.49$ & $0.08$ & $0.38$ & $0.02$ \\
    $2.2$ & $0.58$ & $0.34$ & $0.40$ & $0.06$ \\
    $5.5$ & $0.87$ & $1.2$ & $0.47$ & $0.17$ \\
    11 & $1.5$ & $8.9$ & $0.38$ & $0.03$ \\
    22 & $3.0$ & 17 & $0.52$ & $0.27$ \\
    55 & $7.1$ & 92 & $0.65$ & $0.53$ \\
    110 & 11 & 220 & $0.4$ & $0.04$ \\
    220 & $4.4$ & 130 & $0.38$ & $0.04$ \\
    \hline \hline
  \end{tabular}
  \label{tab-freq-vib}
\end{table}

Integrator rejection degrades rapidly above 10~Hz. At 110~Hz, the integrator amplifies vibration power by 2.2$\times$ ($\rho_\mint = 220\%$) due to latency overshoot. Conversely, the LQG controller maintains flat rejection ($\rho_\mlqg \le 0.53\%$) across all tested frequencies.

For low-frequency vibrations ($1.1$ and $2.2$~Hz), modes were incorporated into the low-frequency AR2 model without performance loss.

These tests highlighted a feature of periodogram-based identification. Injected sinusoids are undamped pure harmonics. When discrete Fourier transforms are evaluated over finite time windows, non-integer cycle counts introduce spectral leakage, broadening the identified peak. The algorithm models this leakage as damping ($k^{\text{vib}} > 0$).

Figure~\ref{fig-p-sim} simulates this spectral leakage effect alongside experimental 22~Hz vibration data (Figure~\ref{fig-p-kvib}).

\begin{figure} \centering
  \subfloat[Simulated closed-loop PSDs under integer and non-integer sample windows.]{\label{fig-p-sim}
    \FIG{0.49}{false}{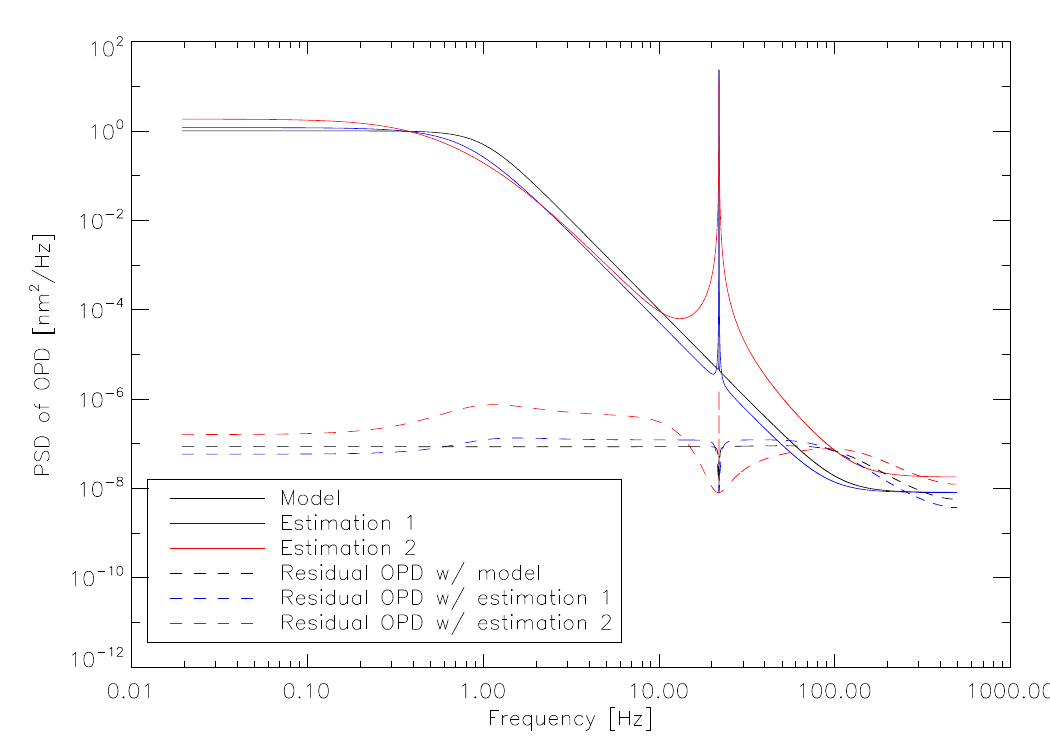}}
  \hfill\subfloat[Experimental closed-loop PSD under a 22~Hz injected vibration.]{\label{fig-p-kvib}
    \FIG{0.49}{false}{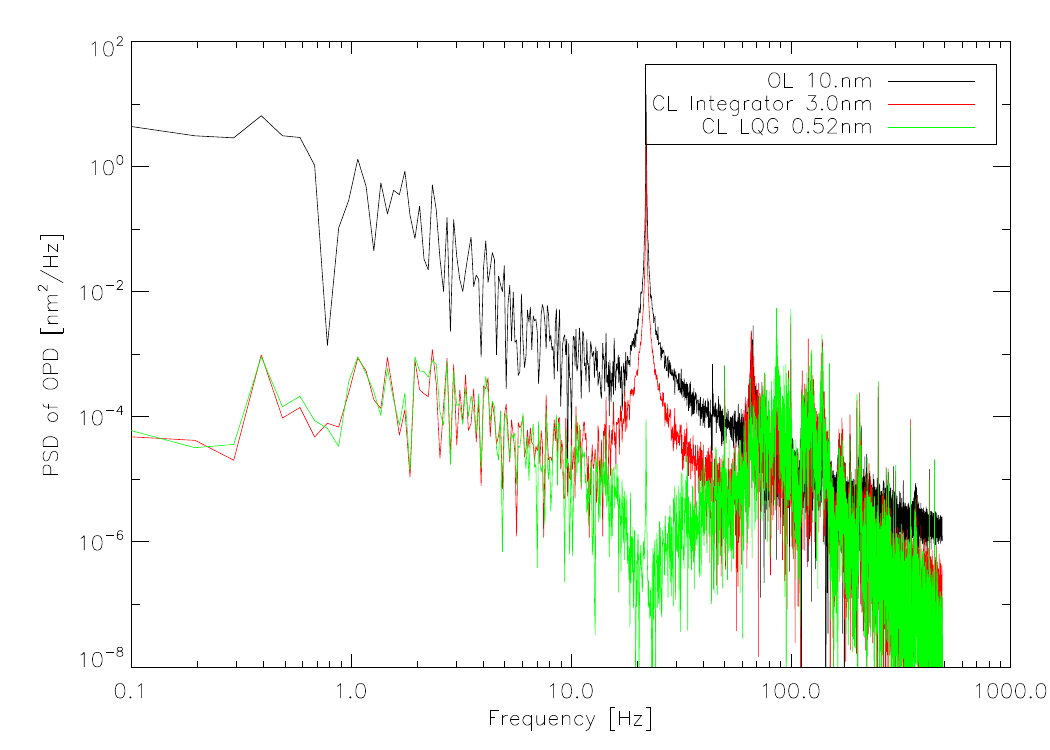}}
  \caption[Spectral leakage effect on LQG vibration notch filters.]{Impact of spectral leakage on identified damping and notch filter profiles.}
  \label{fig-p-sim-kvib}
\end{figure}

In Figure~\ref{fig-p-sim}, integer cycle windows (blue line) yield accurate low-damping models ($k = 10^{-5}$), generating deep notch filters. Non-integer windows (red line) broaden identified damping ($k = 5\cdot 10^{-4}$), over-correcting adjacent frequencies while leaving residual power at the peak. Experimental data (Figure~\ref{fig-p-kvib}) matches this predicted over-correction profile. Broadened damping models improve convergence speed and phase margin robustness against minor frequency shifts.

%¤¤¤¤¤¤¤¤¤¤¤¤¤¤¤¤¤¤¤¤¤¤¤¤¤¤¤¤¤¤¤¤¤¤¤¤¤¤¤¤¤¤¤¤¤¤¤¤¤¤¤¤¤¤¤¤¤¤¤¤¤¤¤¤¤¤¤¤¤¤¤¤¤¤¤¤¤¤¤
\subsubsection{Vibration Amplitude Sensitivity}
\label{sec-influence-amplitude}

LQG robustness against amplitude mismatch was evaluated by applying models identified under 2~nm (model I1) and 20~nm (model I2) 22~Hz vibrations to varying disturbance amplitudes $A_\mvib$. Figure~\ref{fig-pi-oil-amp} and Table~\ref{tab-amplitude-vib} summarize performance.

\begin{figure} \centering
  \subfloat[Path delay PSDs.]{\label{fig-p-oil-amp}
    \FIG{0.49}{false}{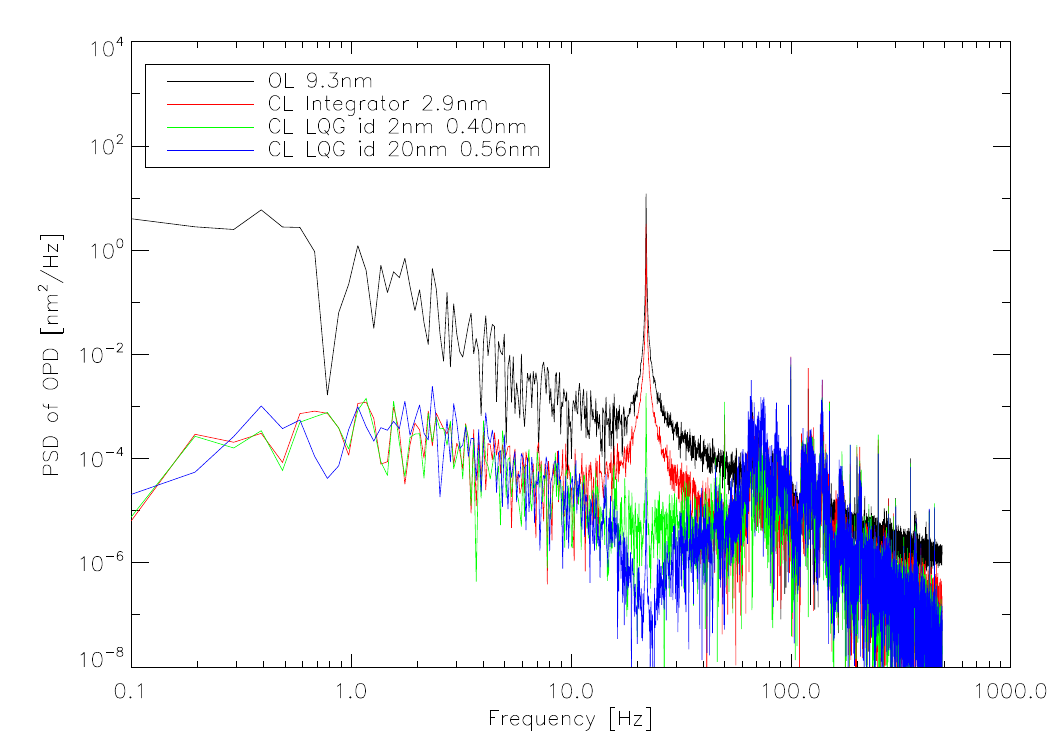}}
  \hfill\subfloat[Cumulative power spectra.]{\label{fig-i-oil-amp}
    \FIG{0.49}{false}{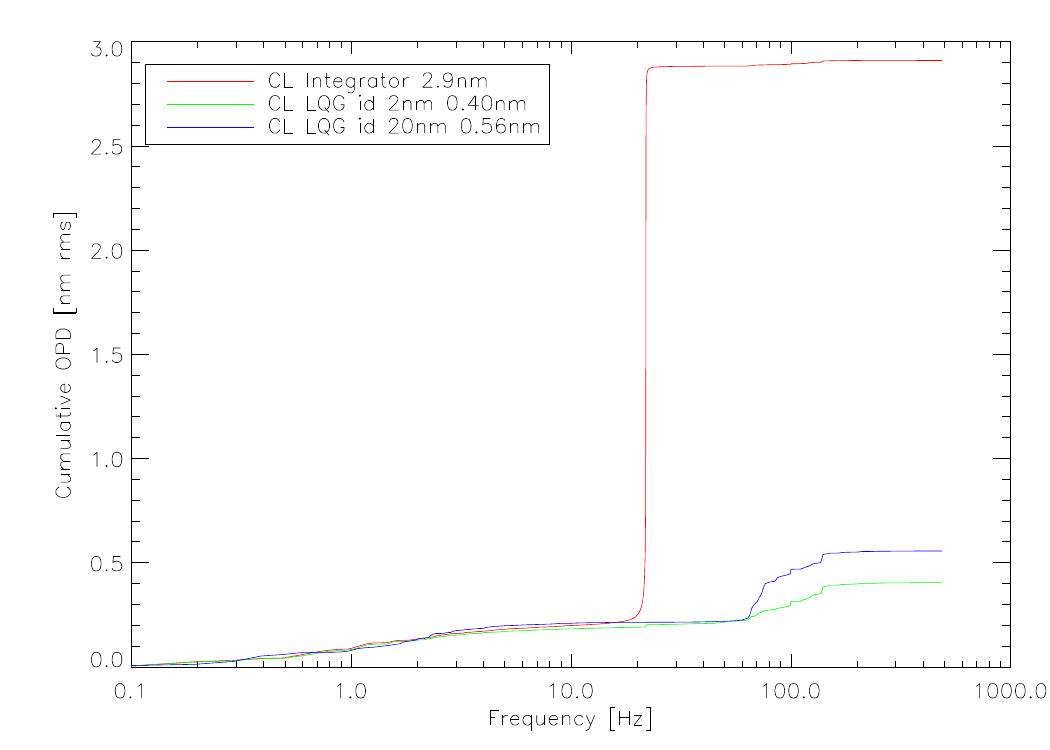}}
  \caption[Vibration amplitude sensitivity.]{Rejection of a 10~nm 22~Hz vibration using models identified at 2~nm (I1) and 20~nm (I2) amplitudes.}
  \label{fig-pi-oil-amp}
\end{figure}

\begin{table} \centering
  \caption[LQG performance across injected vibration amplitudes.]{LQG rejection performance across injected vibration amplitudes.}
  \medskip
  \begin{tabular}{ccccccc}
    \hline \hline
    & & & \multicolumn{2}{c}{I1} &
    \multicolumn{2}{c}{I2} \\
    \cline{4-5}
    \cline{6-7}
    $A_\mvib$ & $\sigma_{\mint,\mper}$ & $\rho_\mint$ & $\sigma_{\mlqg,\mper}$ &
    $\rho_\mlqg$ & $\sigma_{\mlqg,\mper}$ & $\rho_\mlqg$ \\\relax
    [nm] & [nm RMS] & [\%] & [nm RMS] & [\%] & [nm RMS] & [\%] \\
    \hline
    2 & $0.72$ & $4.7$ & $0.42$ & $0.60$ & $0.59$ & $3.1$ \\
    10 & $2.9$ & 22 & $0.40$ & $0.07$ & $0.56$ & $0.45$ \\
    20 & $5.8$ & 26 & $0.38$ & $0.01$ & $0.46$ & $0.06$ \\
    50 & 14 & 26 & $0.59$ & $0.03$ & $0.76$ & $0.06$ \\
    100 & 29 & 25 & $0.78$ & $0.01$ & $0.47$ & $0.00$ \\
    200 & 58 & 23 & $1.42$ & $0.01$ & $0.55$ & $0.00$ \\
    \hline \hline
  \end{tabular}
  \label{tab-amplitude-vib}
\end{table}

Integrator residual power fraction remains constant at $\sim$25\%. LQG control maintains residual path delay jitter below $0.8$~nm RMS even when disturbance amplitudes scale up 100-fold relative to the identification dataset. Identifying models under lower amplitudes (I1) avoids over-correcting adjacent frequencies, yielding better overall rejection.

%¤¤¤¤¤¤¤¤¤¤¤¤¤¤¤¤¤¤¤¤¤¤¤¤¤¤¤¤¤¤¤¤¤¤¤¤¤¤¤¤¤¤¤¤¤¤¤¤¤¤¤¤¤¤¤¤¤¤¤¤¤¤¤¤¤¤¤¤¤¤¤¤¤¤¤¤¤¤¤
\subsubsection{Closely Spaced Vibration Frequencies}
\label{sec-influence-distance}

We evaluated multi-peak identification by injecting two simultaneous vibrations: one fixed at 22~Hz and a second shifted by $\Delta f$. Results are compiled in Table~\ref{tab-dist-vib}.

\begin{table} \centering
  \caption[LQG performance under closely spaced vibration peaks.]{LQG performance under closely spaced vibration peaks.}
  \medskip
  \begin{tabular}{cccccc}
    \hline \hline
    $f_{\mvib,1}$ & $f_{\mvib,2}$ & $\sigma_{\mint,\mper}$ & $\rho_\mint$ &
    $\sigma_{\mlqg,\mper}$ & $\rho_\mlqg$ \\\relax
    [Hz] & [Hz] & [nm RMS] & [\%] & [nm RMS] & [\%] \\
    \hline
    22 & 27 & $4.6$ & 16 & $0.52$ & $0.10$ \\
    22 & 24 & $4.3$ & 16 & $0.58$ & $0.17$ \\
    22 & 23 & $4.2$ & 14 & $0.47$ & $0.06$ \\
    22 & $22.5$ & $4.1$ & 15 & $0.51$ & $0.11$ \\
    22 & $22.2$ & $4.1$ & 14 & $0.49$ & $0.09$ \\
    22 & $22.1$ & $4.1$ & 14 & $0.52$ & $0.11$ \\
    \hline \hline
  \end{tabular}
  \label{tab-dist-vib}
\end{table}

LQG rejection remains stable ($\rho_\mlqg \le 0.17\%$) down to frequency separations of $0.1$~Hz—below the $0.05$~Hz spectral resolution limit of a 20~s dataset. Figure~\ref{fig-p-oil-gap} plots PSDs for $\Delta f = 0.5$~Hz and $0.1$~Hz.

\begin{figure} \centering
  \subfloat[Vibration peaks at 22 and $22.5$~Hz.]{\label{fig-p-oil-gap1}\FIG{0.49}{false}{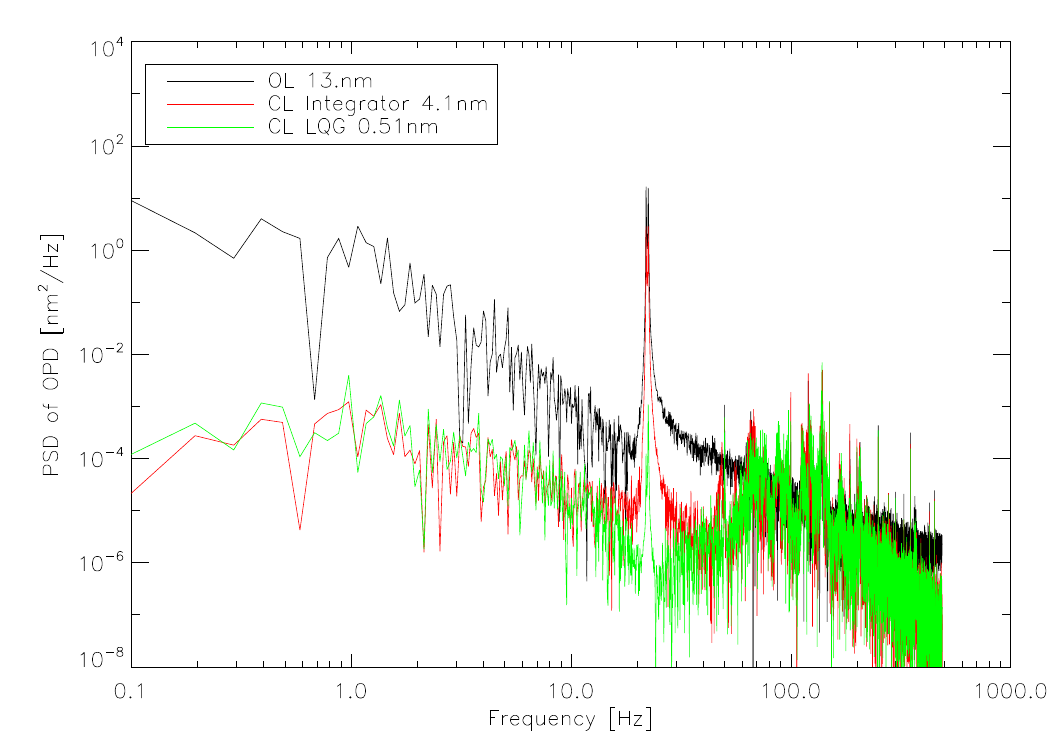}}
  \hfill\subfloat[Vibration peaks at 22 and $22.1$~Hz.]{\label{fig-p-oil-gap2}\FIG{0.49}{false}{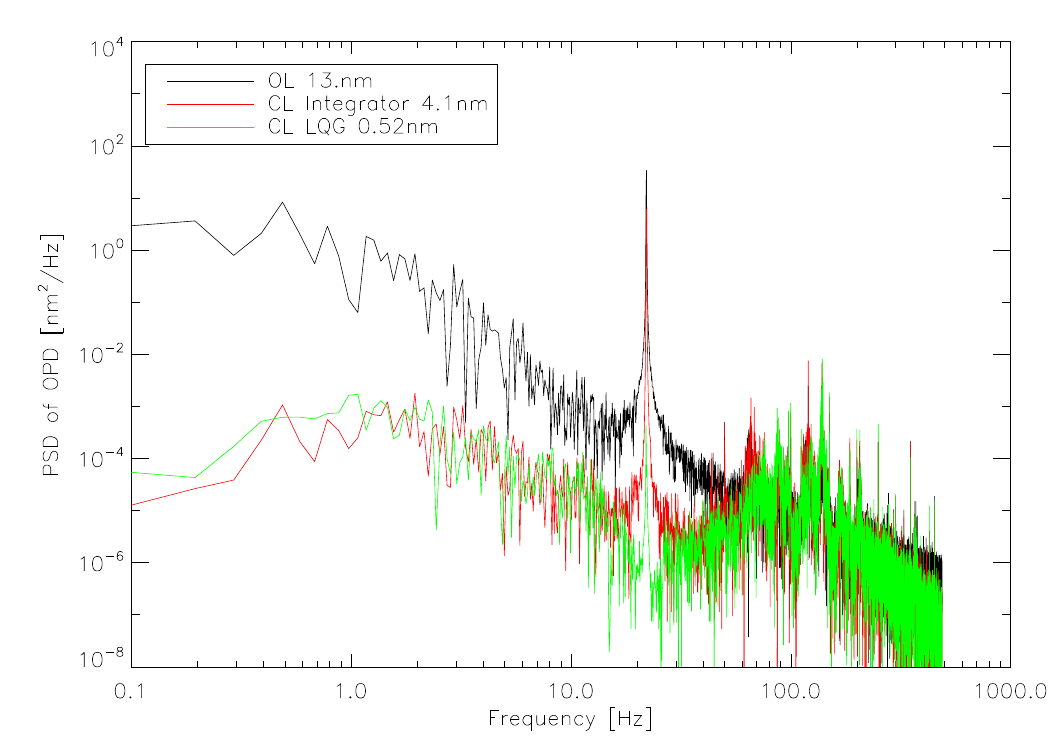}}
  \caption[Closed-loop PSDs for closely spaced vibration peaks.]{Closed-loop path delay PSDs under closely spaced vibration peaks ($\Delta f = 0.5$~Hz and $0.1$~Hz).}
  \label{fig-p-oil-gap}
\end{figure}

The identification routine resolves adjacent peaks, fitting a combined broadened state model when peaks overlap.

%¤¤¤¤¤¤¤¤¤¤¤¤¤¤¤¤¤¤¤¤¤¤¤¤¤¤¤¤¤¤¤¤¤¤¤¤¤¤¤¤¤¤¤¤¤¤¤¤¤¤¤¤¤¤¤¤¤¤¤¤¤¤¤¤¤¤¤¤¤¤¤¤¤¤¤¤¤¤¤
\subsubsection{Multi-Peak Vibration Rejection}
\label{sec-influence-nombre}

We tested multi-peak rejection by injecting up to $N_\mvib = 20$ simultaneous 10~nm sinusoids spaced evenly between 5 and 55~Hz with randomized phase offsets. Results are compiled in Table~\ref{tab-number-vib}.

\begin{table} \centering
  \caption{LQG rejection performance across number of injected vibration peaks.}
  \medskip
  \begin{tabular}{ccccc}
    \hline \hline
    $N_\mvib$ & $\sigma_{\mint,\mper}$ & $\rho_\mint$ &
    $\sigma_{\mlqg,\mper}$ & $\rho_\mlqg$ \\
    & [nm RMS] & [\%] & [nm RMS] & [\%] \\
    \hline
    2 & 7.2 & 41 & 0.57 & 0.15 \\
    5 & 10 & 37 & 0.56 & 0.06 \\
    10 & 14 & 41 & 0.63 & 0.05 \\
    20 & 20 & 39 & 0.91 & 0.07 \\
    \hline \hline
  \end{tabular}
  \label{tab-number-vib}
\end{table}

While integrator residual jitter increases proportionally to $\sqrt{N_\mvib}$ (39\% power leakage), LQG control maintains residual jitter below 1~nm RMS across 20 simultaneous modes ($\rho_\mlqg \le 0.15\%$). Processing load scales linearly with $N_\mvib$, remaining within 1~kHz real-time frame limits for 20 modes.

Figure~\ref{fig-pi-oil-numb} displays closed-loop performance under 10 simultaneous vibration modes.

\begin{figure} \centering
  \subfloat[Path delay PSDs.]{\label{fig-p-oil-numb}
    \FIG{0.49}{false}{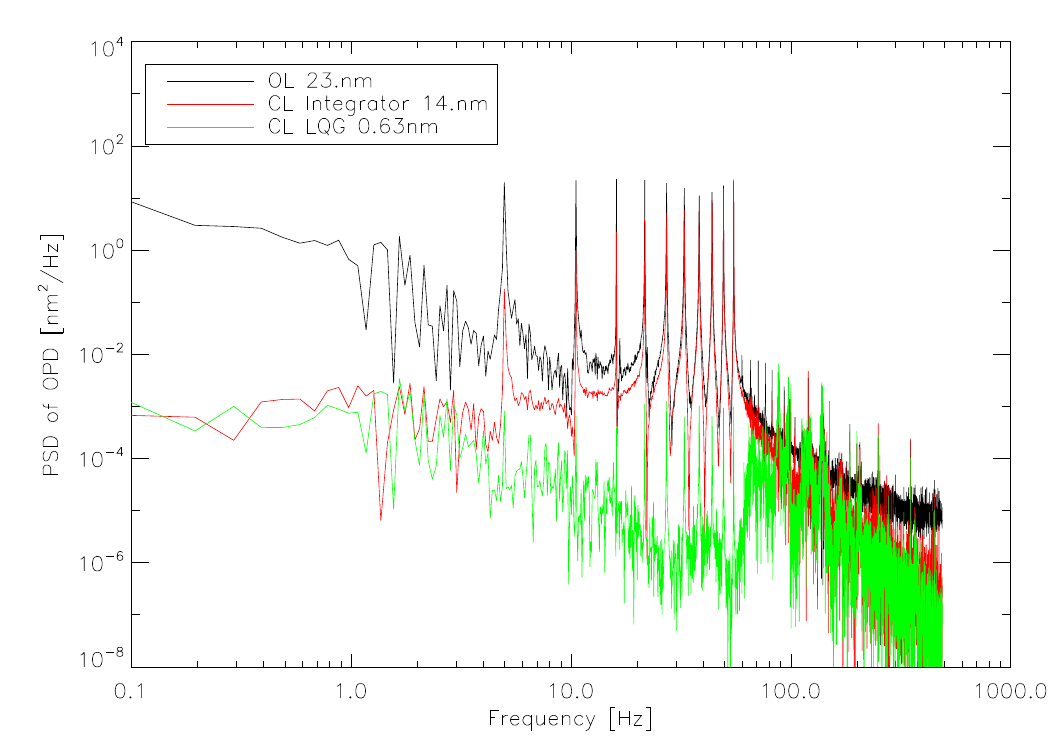}}
  \hfill\subfloat[Cumulative power spectra.]{\label{fig-i-oil-numb}
    \FIG{0.49}{false}{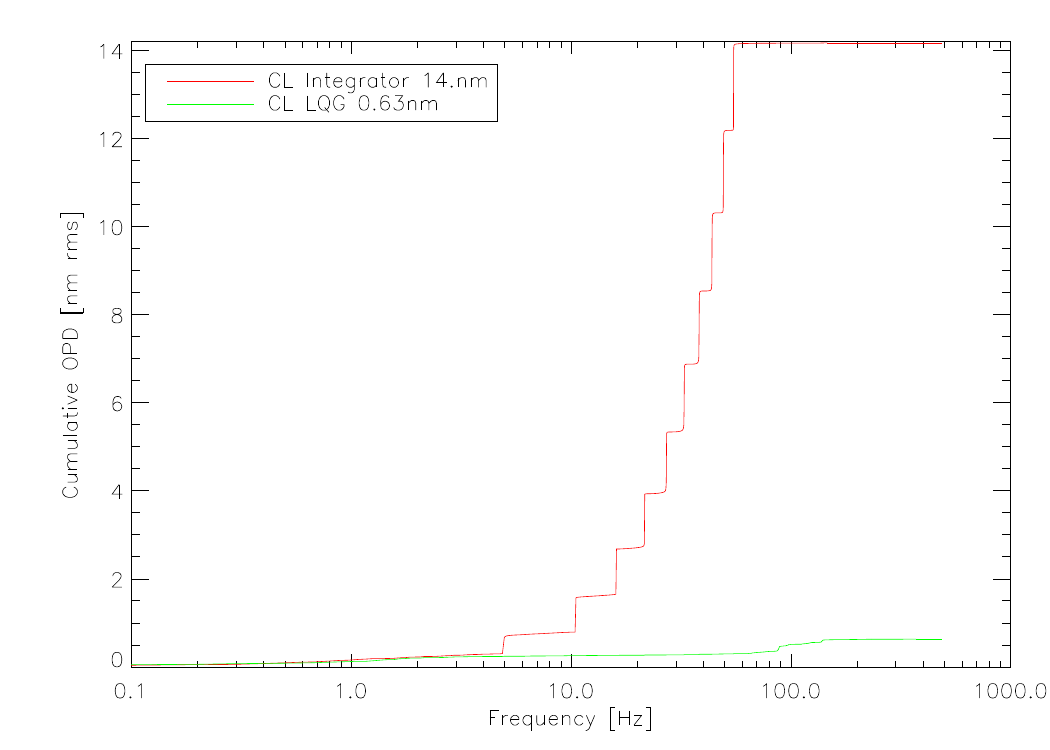}}
  \caption{Rejection of 10 simultaneous vibration modes spaced between 5 and 55~Hz.}
\label{fig-pi-oil-numb}
\end{figure}

The cumulative spectrum (Figure~\ref{fig-i-oil-numb}) highlights the contrast between integrator steps and LQG flat rejection.

%-------------------------------------------------------------------------------
\subsection{Rejection of Flight-Like Formation-Flying Disturbances}
\label{sec-analyse-typique}

Finally, we evaluated flight-like formation-flying disturbances combining satellite positioning drift ($\sigma_{\minj,\mps} = 3$~nm and $15$~nm RMS) with reaction wheel micro-vibrations. Six reaction wheels (2 per axis) were modeled operating near mean rotation frequencies $f_\mr = 1.5, 3.0,$ and $4.5$~Hz.

Wheel harmonics excite spacecraft structural modes:
\begin{itemize}
\item 15~Hz: Solar array flexible modes;
\item 55~Hz: Spacecraft main bus lateral mode;
\item 75, 85~Hz: Reaction wheel and optical bench cavity modes.
\end{itemize}

The top 40 harmonic lines were modeled. Figure~\ref{fig-p-simper} shows the simulated input spectrum for $f_\mr = 4.5$~Hz and $\sigma_{\minj,\mps} = 15$~nm RMS.

\begin{figure} \centering
  \FIG{0.7}{false}{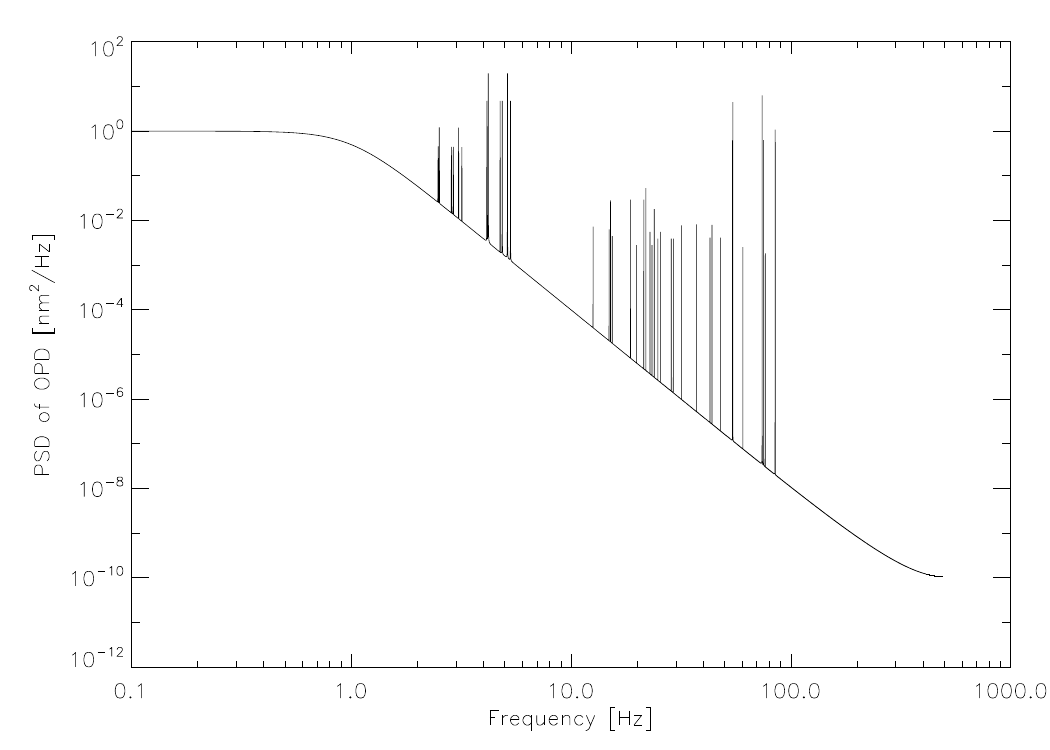}
  \caption[Simulated flight-like disturbance PSD.]{Simulated flight-like disturbance PSD ($f_\mr=4.5$~Hz, $\sigma_{\minj,\mps}=15$~nm RMS).}
  \label{fig-p-simper}
\end{figure}

Integrator gain was set to $g=0.5$; LQG identification targeted the 25 strongest lines. Results are compiled in Table~\ref{tab-real-dist} and Figures~\ref{fig-pi-oil-microvib-1} and \ref{fig-pi-oil-microvib-2}.

\begin{table} \centering
  \caption{Performance evaluation under flight-like formation-flying disturbances.}
  \medskip
  \begin{tabular}{cccccc}
    \hline \hline
    $f_\mr$ & $\sigma_{\minj,\mps}$ & $\sigma_{\mint,\mper}$ & $\rho_\mint$ &
    $\sigma_{\mlqg,\mper}$ & $\rho_\mlqg$ \\\relax
    [Hz] & [nm RMS] & [nm RMS] & [\%] & [nm RMS] & [\%] \\
    \hline
    $1.5$ & 3 & $0.62$ & $0.54$ & $0.41$ & $0.09$ \\
    $1.5$ & 15 & $0.58$ & $0.06$ & $0.44$ & $0.03$ \\
    $3.0$ & 3 & $2.2$ & 11 & $0.47$ & $0.20$ \\
    $3.0$ & 15 & $4.8$ & 13 & $0.51$ & $0.07$ \\
    $4.5$ & 3 & $4.3$ & 58 & $0.49$ & $0.34$ \\
    $4.5$ & 15 & $7.9$ & 22 & $0.77$ & $0.17$ \\
    \hline \hline
  \end{tabular}
  \label{tab-real-dist}
\end{table}

\begin{figure} \centering
  \subfloat[$f_\mr=1.5$~Hz: PSDs.]{\label{fig-p-oil-microvib-1a}
    \FIG{0.49}{false}{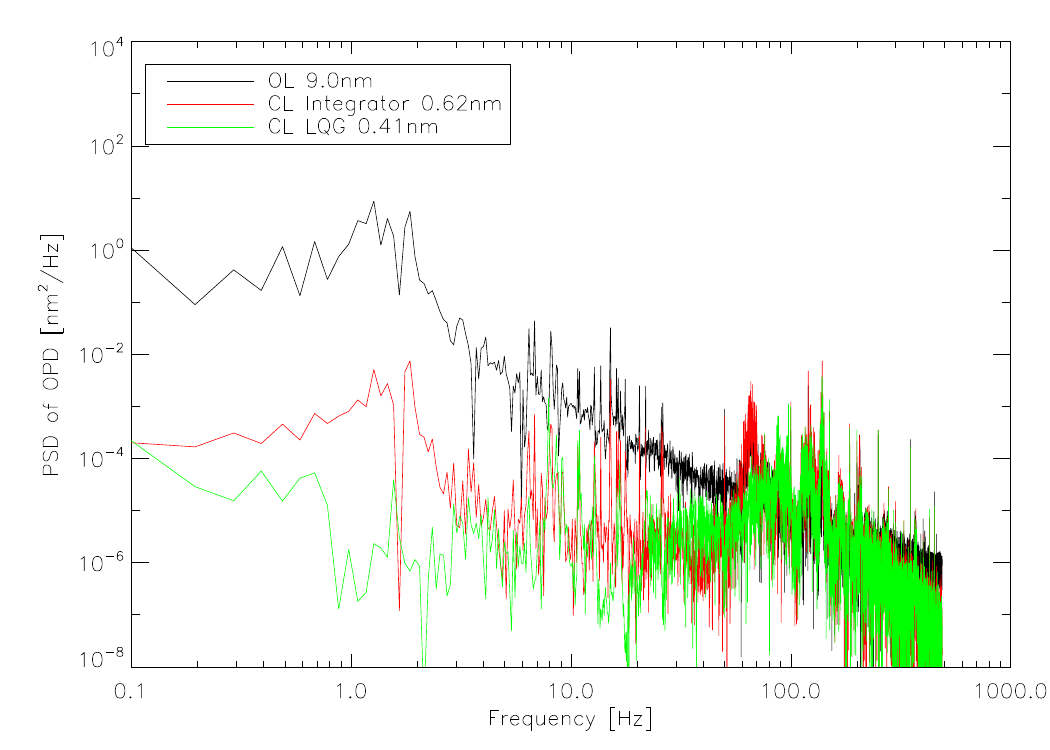}}
  \hfill\subfloat[$f_\mr=1.5$~Hz: Cumulative power.]{\label{fig-i-oil-microvib-1b}
    \FIG{0.49}{false}{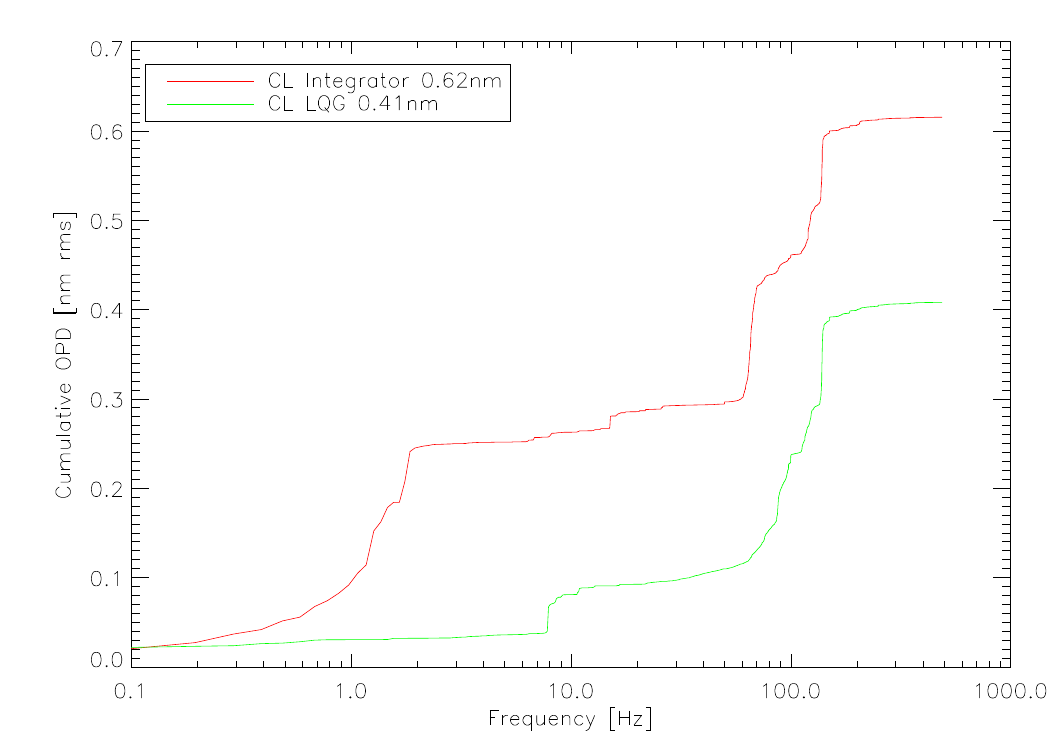}}\\
  \subfloat[$f_\mr=3$~Hz: PSDs.]{\label{fig-p-oil-microvib-1c}
    \FIG{0.49}{false}{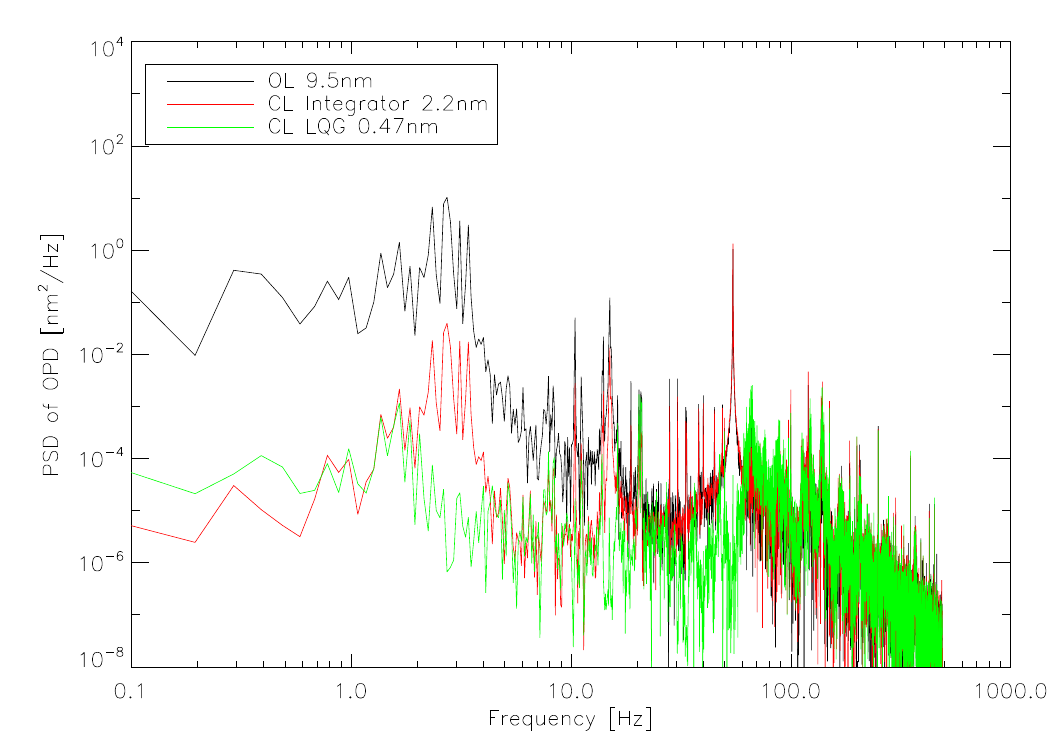}}
  \hfill\subfloat[$f_\mr=3$~Hz: Cumulative power.]{\label{fig-i-oil-microvib-1d}
    \FIG{0.49}{false}{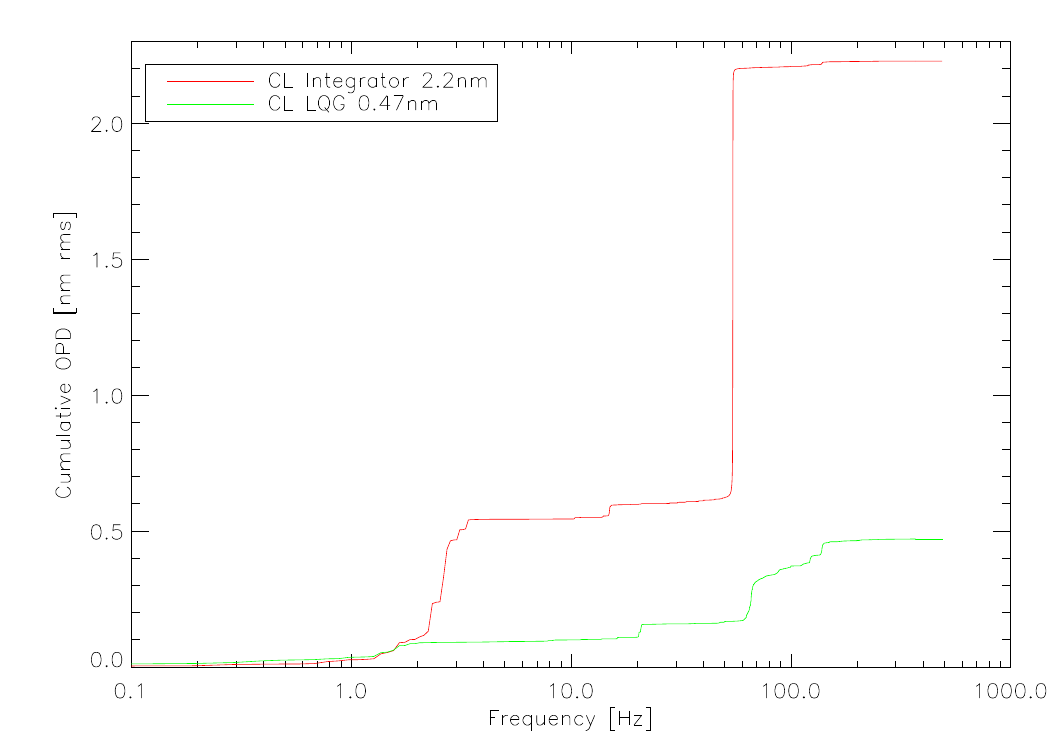}}\\
  \subfloat[$f_\mr=4.5$~Hz: PSDs.]{\label{fig-p-oil-microvib-1e}
    \FIG{0.49}{false}{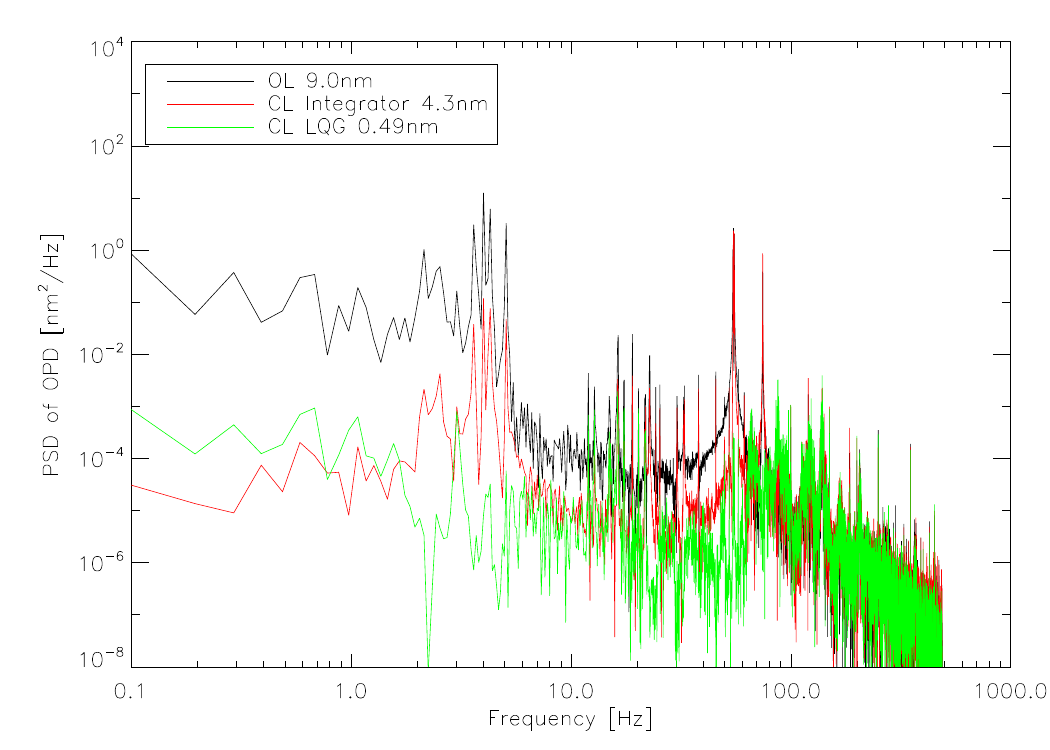}}
  \hfill\subfloat[$f_\mr=4.5$~Hz: Cumulative power.]{\label{fig-i-oil-microvib-1f}
    \FIG{0.49}{false}{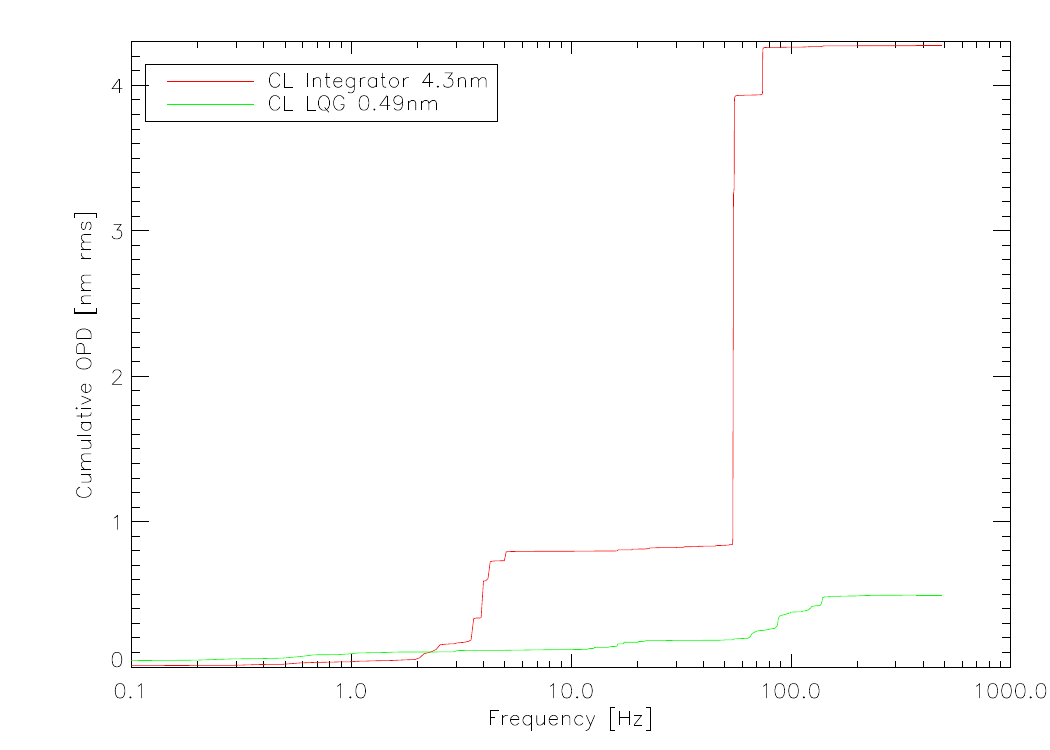}}\\
  \caption[Disturbance rejection for $\sigma_{\minj,\mps}=3$~nm RMS.]{Performance under flight-like disturbances for $\sigma_{\minj,\mps}=3$~nm RMS.}
  \label{fig-pi-oil-microvib-1}
\end{figure}

\begin{figure} \centering
  \subfloat[$f_\mr=1.5$~Hz: PSDs.]{\label{fig-p-oil-microvib-2a}
    \FIG{0.49}{false}{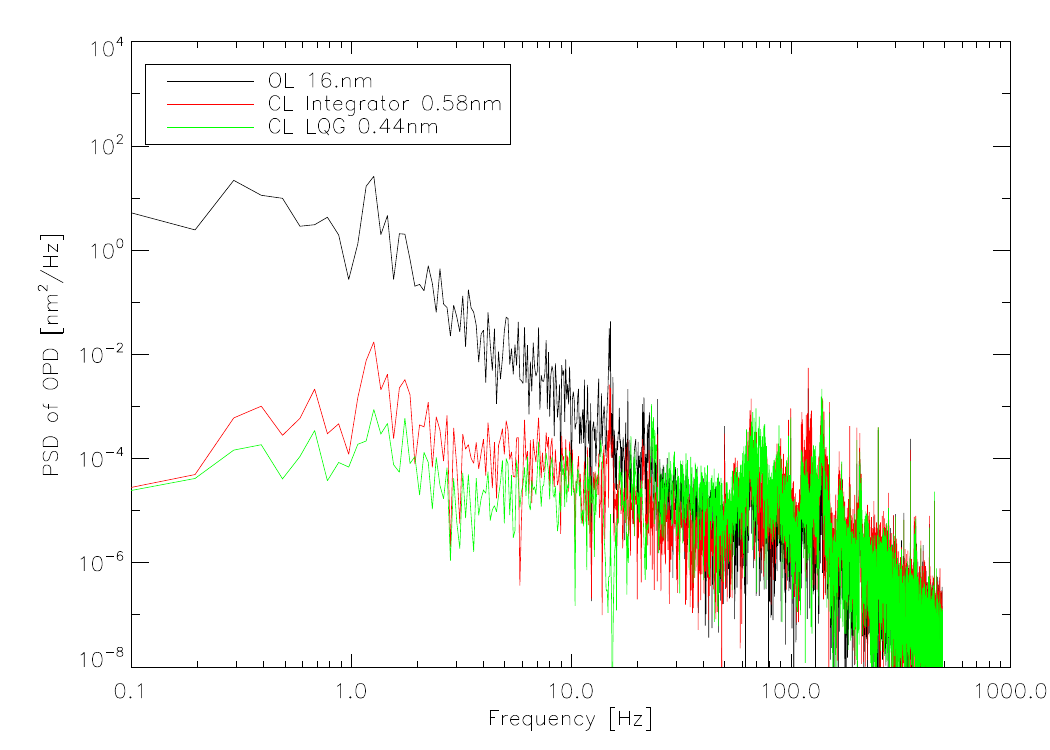}}
  \hfill\subfloat[$f_\mr=1.5$~Hz: Cumulative power.]{\label{fig-i-oil-microvib-2b}
    \FIG{0.49}{false}{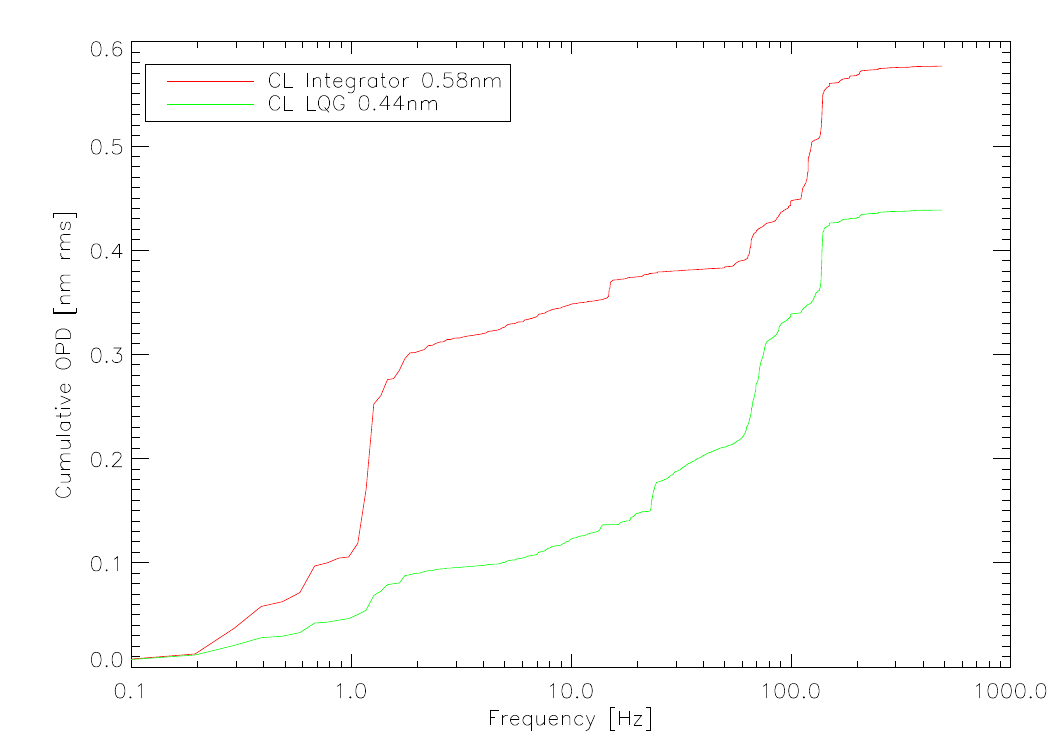}}\\
  \subfloat[$f_\mr=3$~Hz: PSDs.]{\label{fig-p-oil-microvib-2c}
    \FIG{0.49}{false}{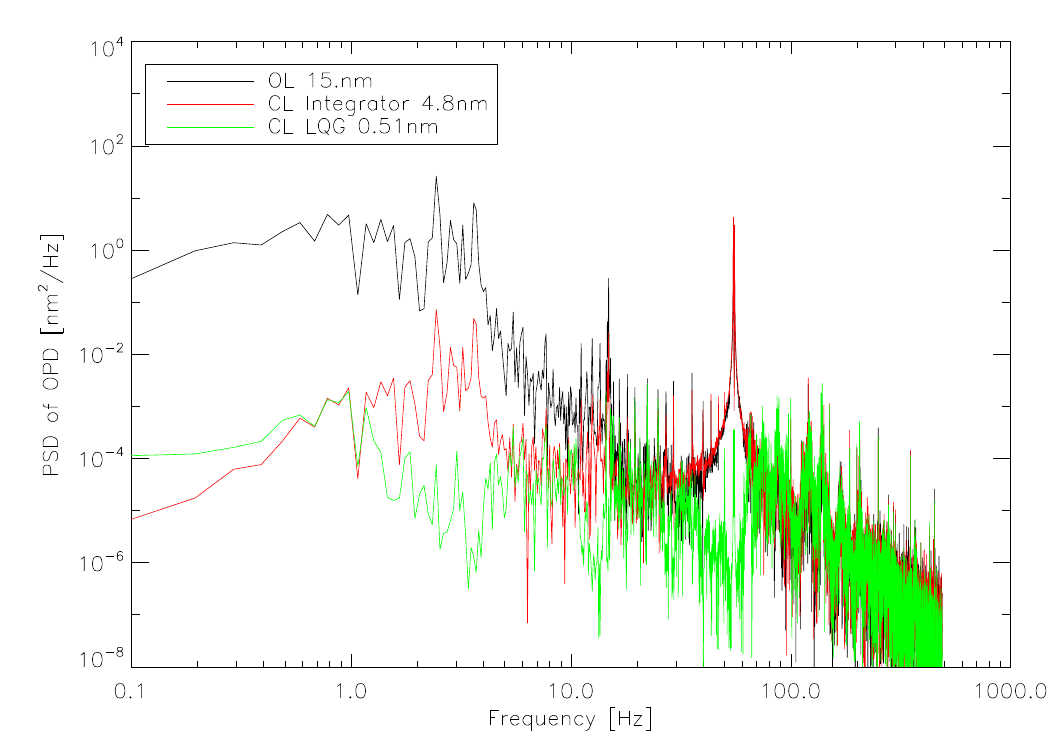}}
  \hfill\subfloat[$f_\mr=3$~Hz: Cumulative power.]{\label{fig-i-oil-microvib-2d}
    \FIG{0.49}{false}{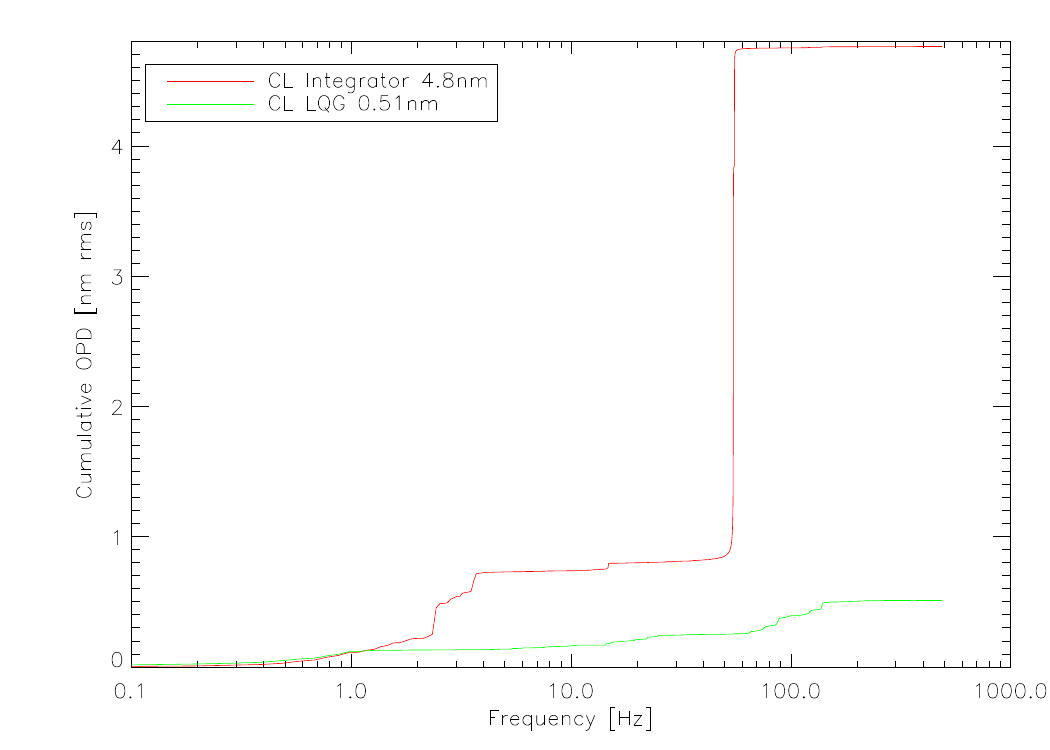}}\\
  \subfloat[$f_\mr=4.5$~Hz: PSDs.]{\label{fig-p-oil-microvib-2e}
    \FIG{0.49}{false}{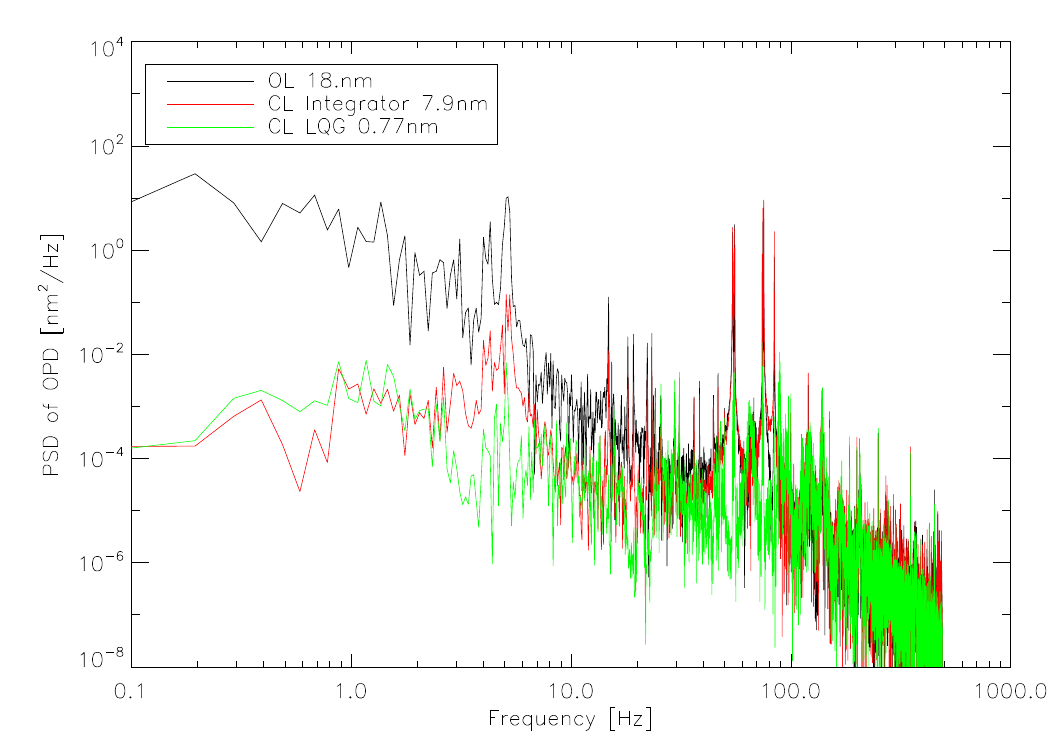}}
  \hfill\subfloat[$f_\mr=4.5$~Hz: Cumulative power.]{\label{fig-i-oil-microvib-2f}
    \FIG{0.49}{false}{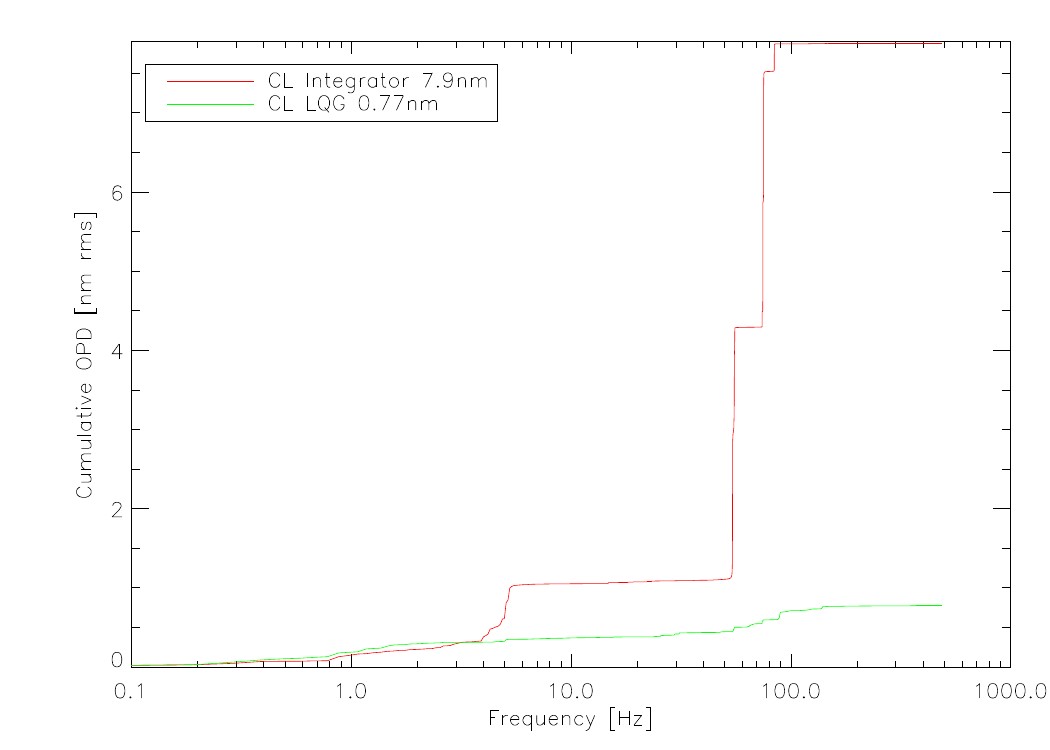}}\\
  \caption[Disturbance rejection for $\sigma_{\minj,\mps}=15$~nm RMS.]{Performance under flight-like disturbances for $\sigma_{\minj,\mps}=15$~nm RMS.}
  \label{fig-pi-oil-microvib-2}
\end{figure}

The integrator maintains path delay jitter below $0.6$~nm RMS only for $f_\mr = 1.5$~Hz, where harmonic lines fall below 20~Hz. At $f_\mr = 4.5$~Hz, integrator residual jitter expands to $7.9$~nm RMS.

In contrast, the LQG controller rejects wheel harmonics across all operational regimes, maintaining residual path delay jitter below $\boldsymbol{0.77}$~\textbf{nm RMS} under worst-case formation-flying disturbances—meeting the sub-nanometric stability requirement.

%¤¤¤¤¤¤¤¤¤¤¤¤¤¤¤¤¤¤¤¤¤¤¤¤¤¤¤¤¤¤¤¤¤¤¤¤¤¤¤¤¤¤¤¤¤¤¤¤¤¤¤¤¤¤¤¤¤¤¤¤¤¤¤¤¤¤¤¤¤¤¤¤¤¤¤¤¤¤¤
\subsection{Conclusion}

These experimental tests demonstrate that an LQG controller paired with online identification maintains sub-nanometric path delay stability under high-amplitude formation-flying disturbances (>20~nm RMS).

Adapting LQG control to space interferometry required solving static offset tracking and managing periodogram spectral leakage. Constraining the low-frequency AR2 model eliminated static path offsets, while periodogram over-estimation improved closed-loop robustness.

%§§§§§§§§§§§§§§§§§§§§§§§§§§§§§§§§§§§§§§§§§§§§§§§§§§§§§§§§§§§§§§§§§§§§§§§§§§§§§§§
\section{Vibration Analysis of the \sce Instrument}
\label{sec-scexao}

%-------------------------------------------------------------------------------
\subsection{Description of the Experiment}
\label{sec-description-experience}

As outlined in Sections~\ref{sec-optique-adaptative} and \ref{sec-coronographie}, \sce combines a PIAA coronagraph (Section~\ref{sec-coronographie}) with Classical Pupil Apodization (CPA) to achieve $10^{-6}$ contrast at $1\,\lambda/D$. Achieving this contrast requires minimizing wavefront phase and pointing jitter. Figure~\ref{fig-scexao-archi} presents the instrument block diagram.

\begin{figure} \centering
  \FIG{0.7}{false}{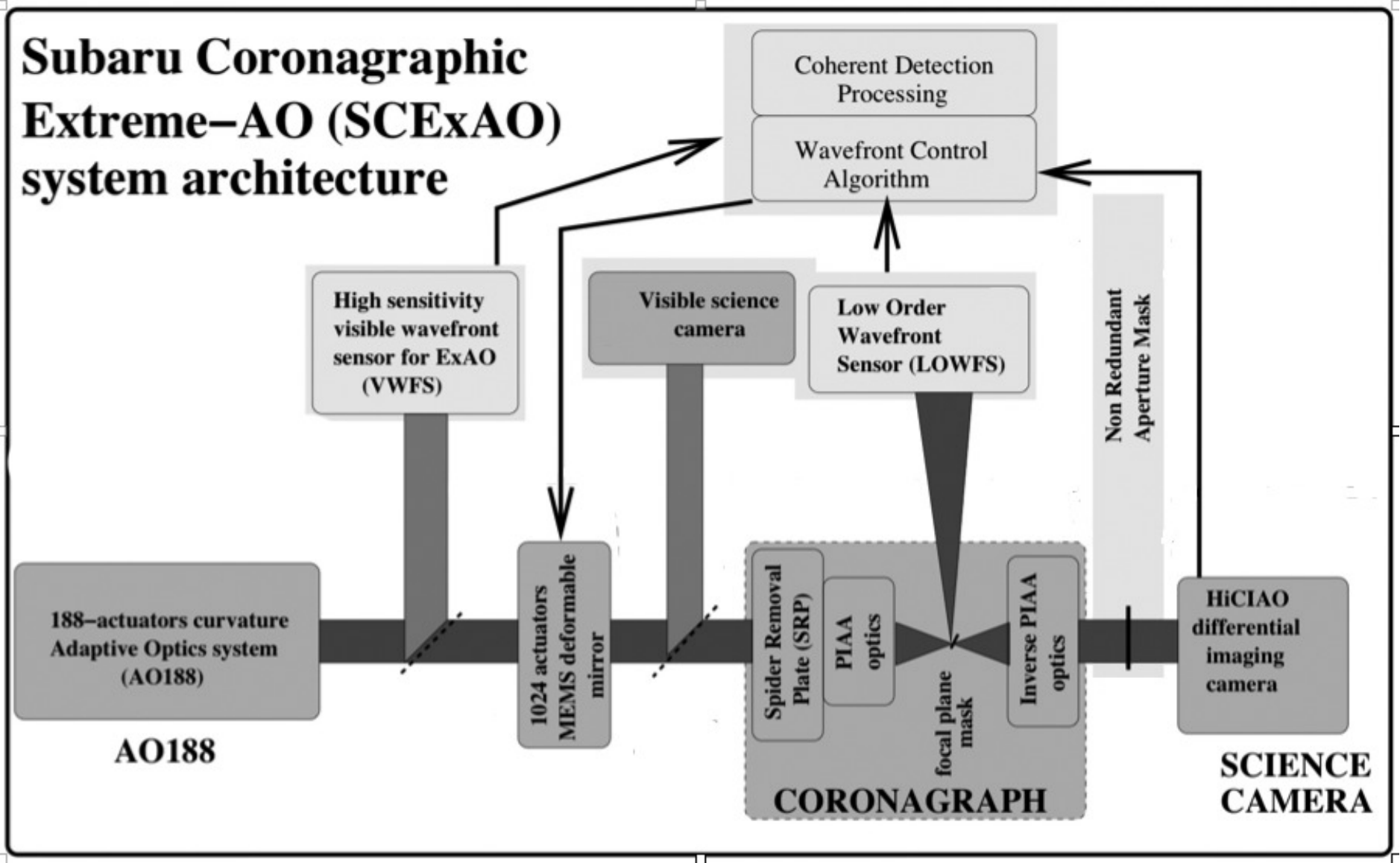}
  \caption{Block diagram of the \sce instrument.}
  \label{fig-scexao-archi}
\end{figure}

Upstream of the HiCIAO\footnote{High Contrast Instrument for the Subaru Next Generation Adaptive Optics} science camera, \sce incorporates multiple wavefront control stages:
\begin{itemize}
\item The Subaru AO188 visible AO system providing first-stage atmospheric wavefront correction;
\item An infrared Coronagraphic Low-Order Wavefront Sensor (CLOWFS) picking off focal mask reflections to drive tip/tilt and focus correction;
\item Phase-diversity calibration for quasi-static speckle suppression;
\item A visible pyramid wavefront sensor (CHEOPS) measuring dynamic speckles;
\item Speckle-nulling routines driving the deformable mirror;
\item A 1024-actuator ($32 \times 32$) MEMS deformable mirror.
\end{itemize}

During my undergraduate internship at the Institut d'Optique, I participated in \sce's initial optomechanical design and tested its coronagraphic optics. Results were published in \emph{Publications of the Astronomical Society of the Pacific} \cite{Lozi09}.

\newpage

\newcommand{\PASP}[1]{\centerline{\FIG{1.}{false,viewport=42 40 571 742,clip}{PASP_2009_#1}}}

 \PASP{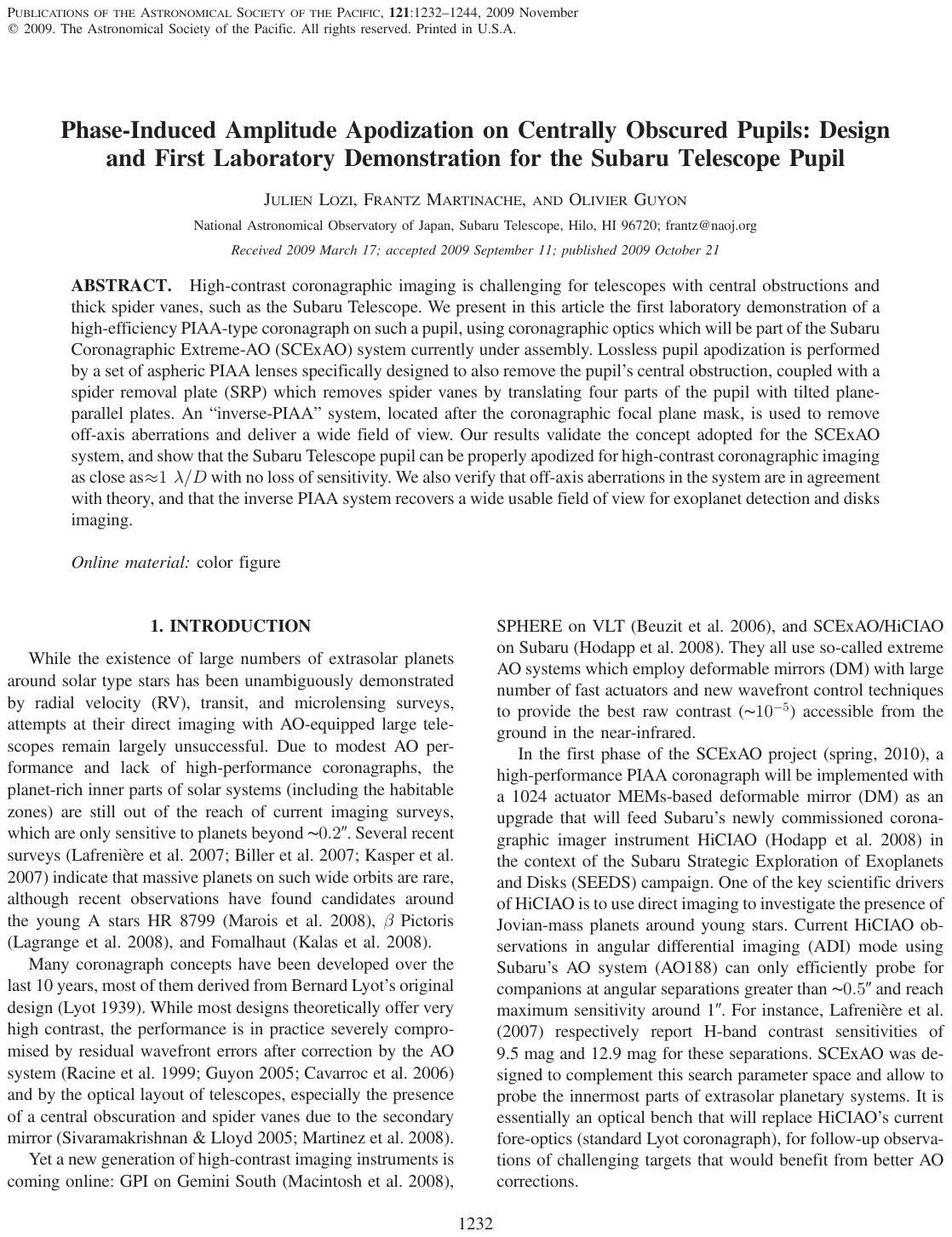}
 \PASP{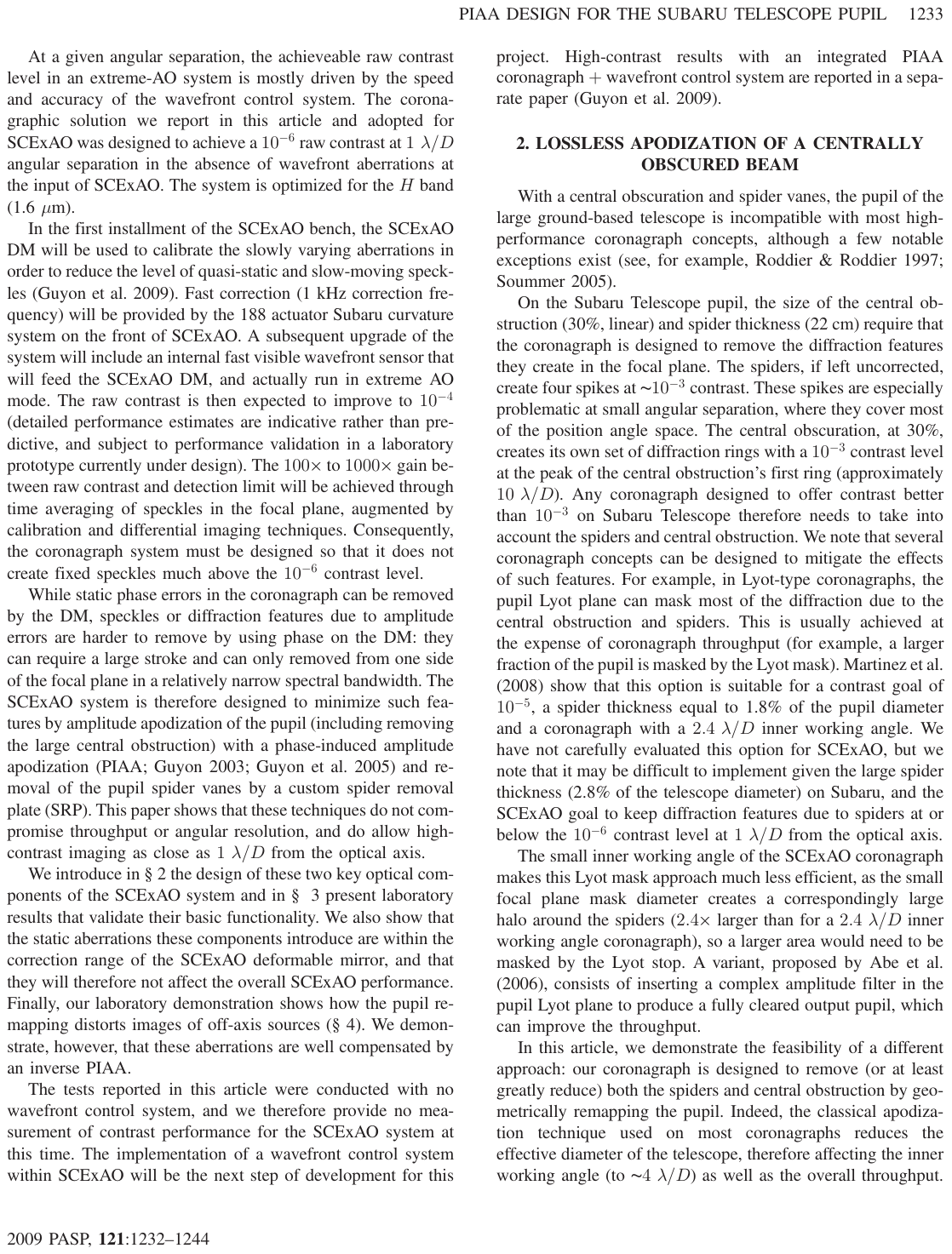}
 \PASP{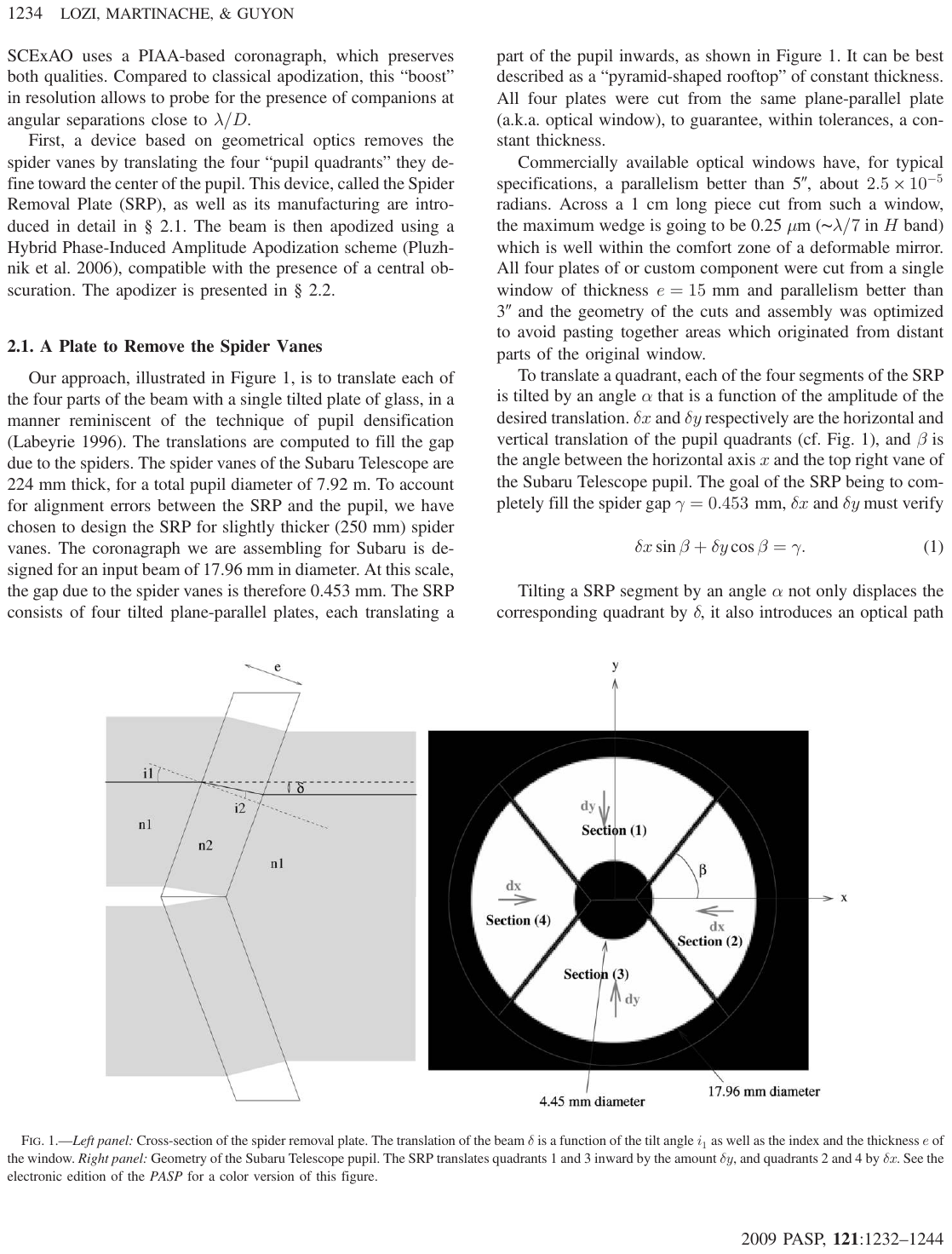}
 \PASP{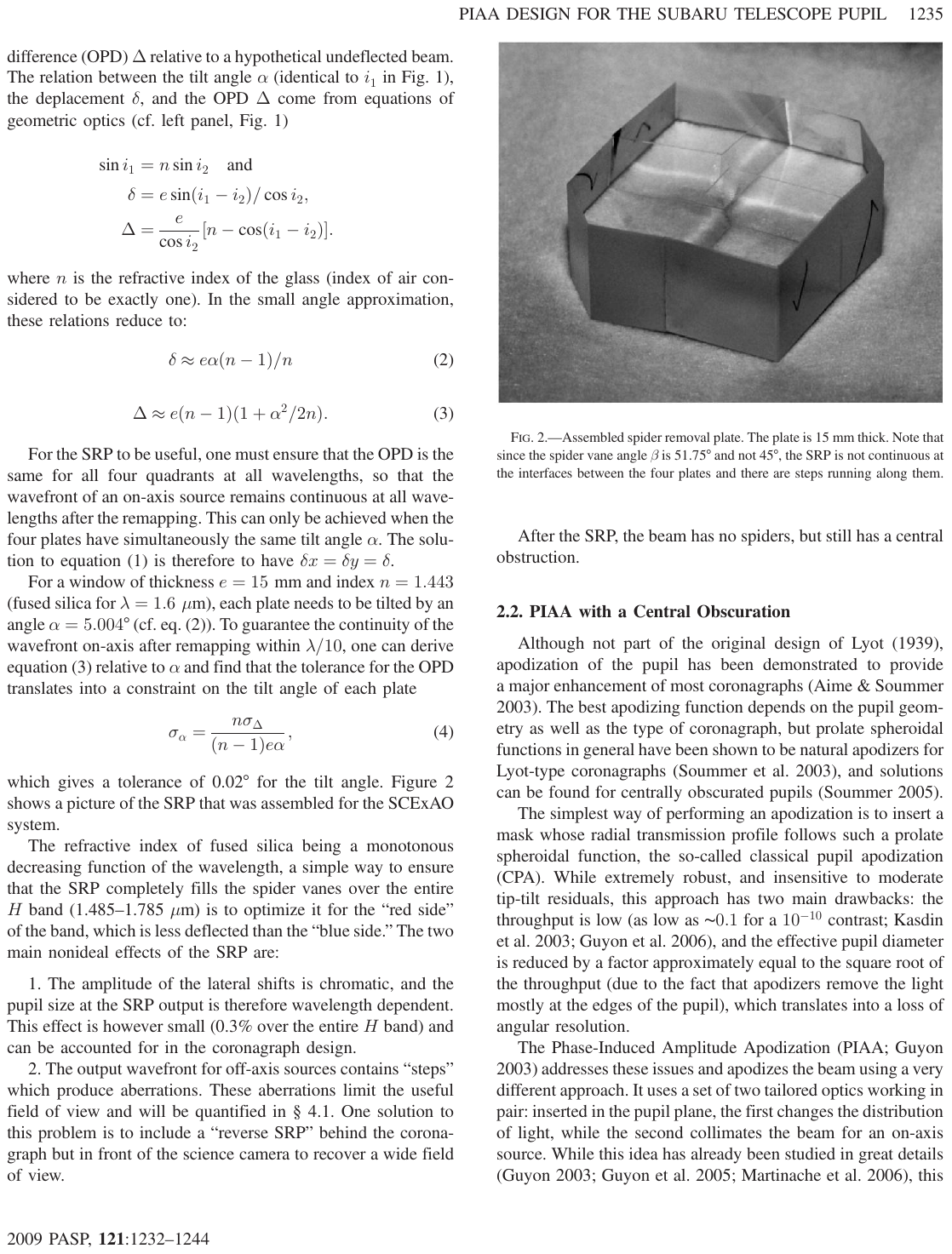}
 \PASP{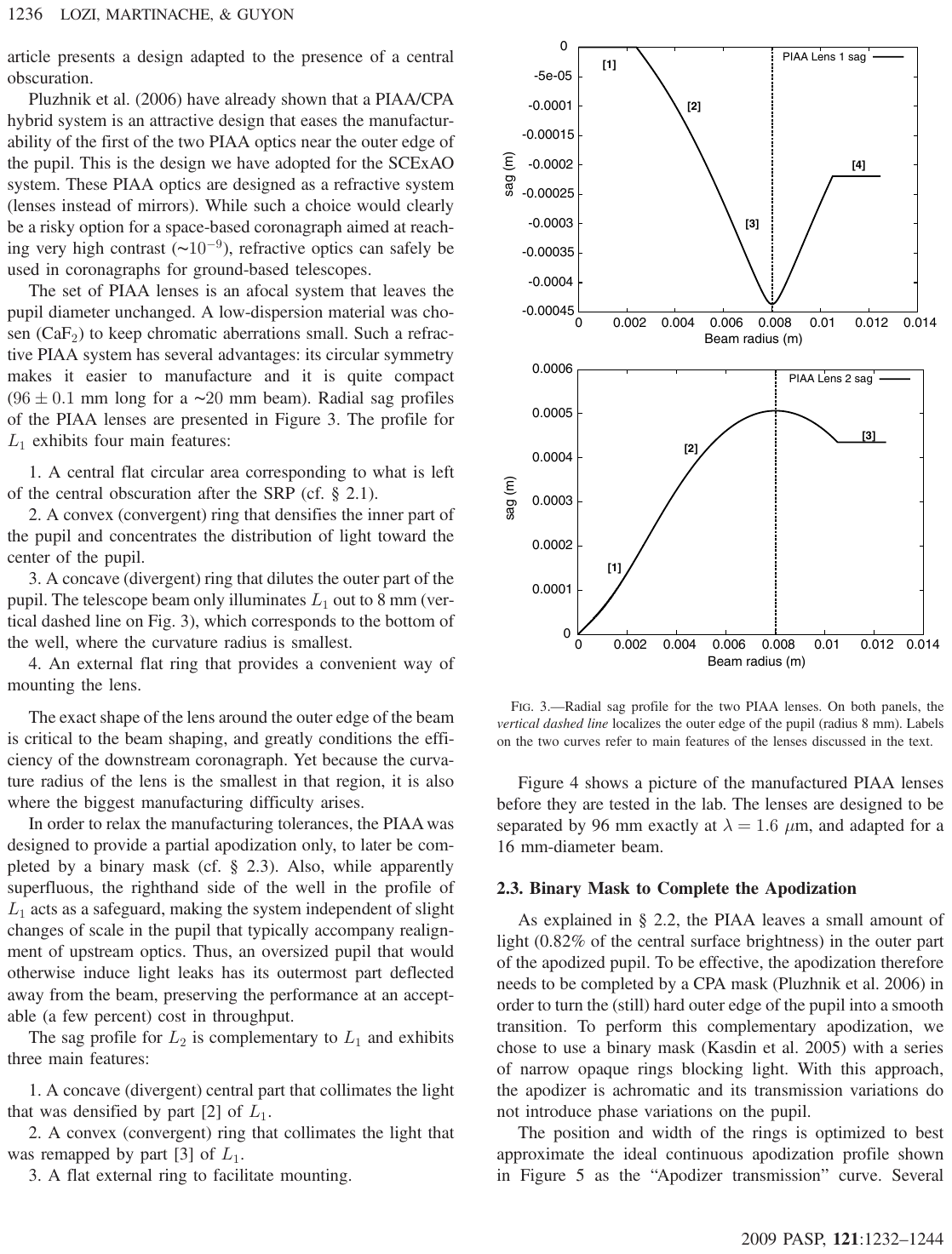}
 \PASP{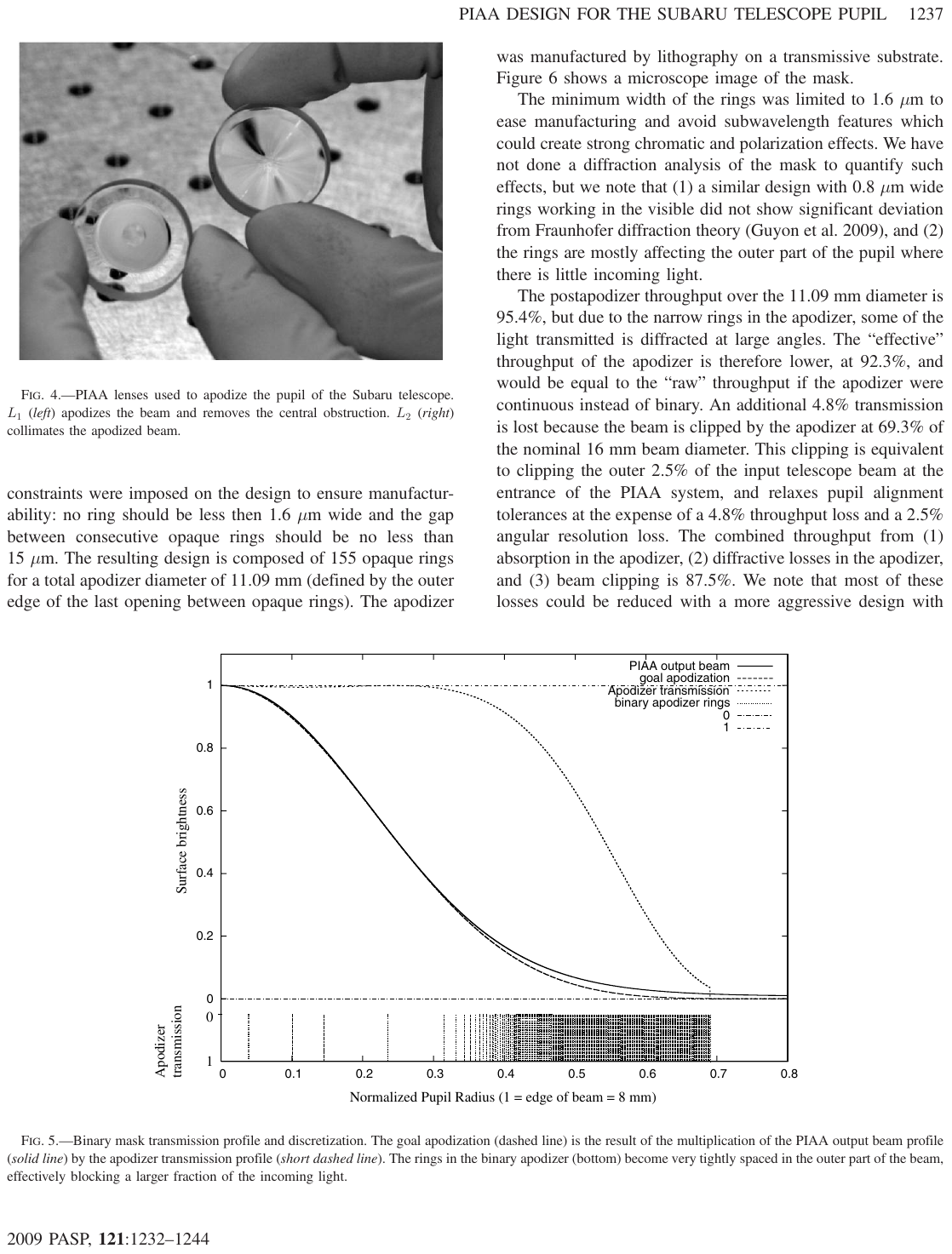}
 \PASP{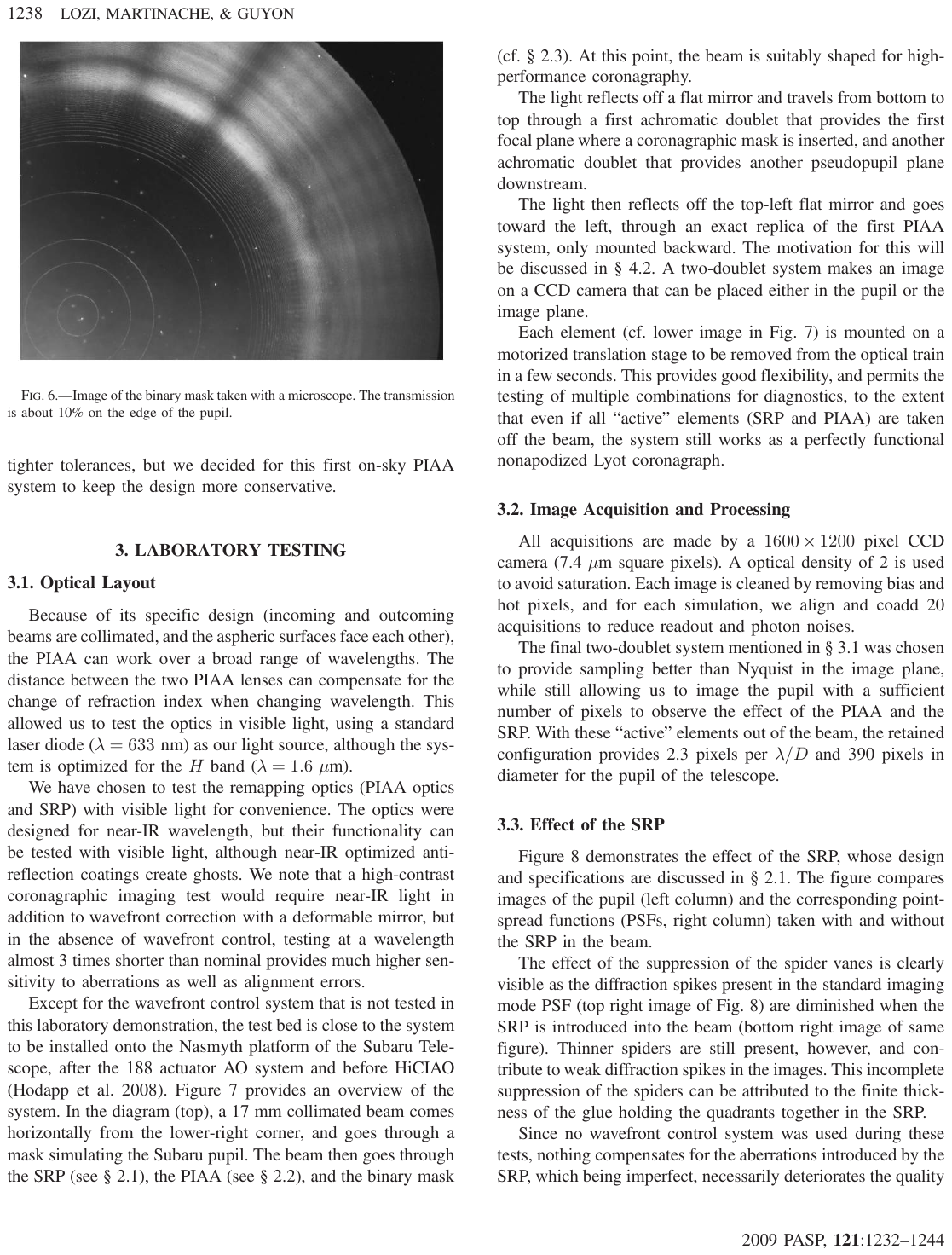}
 \PASP{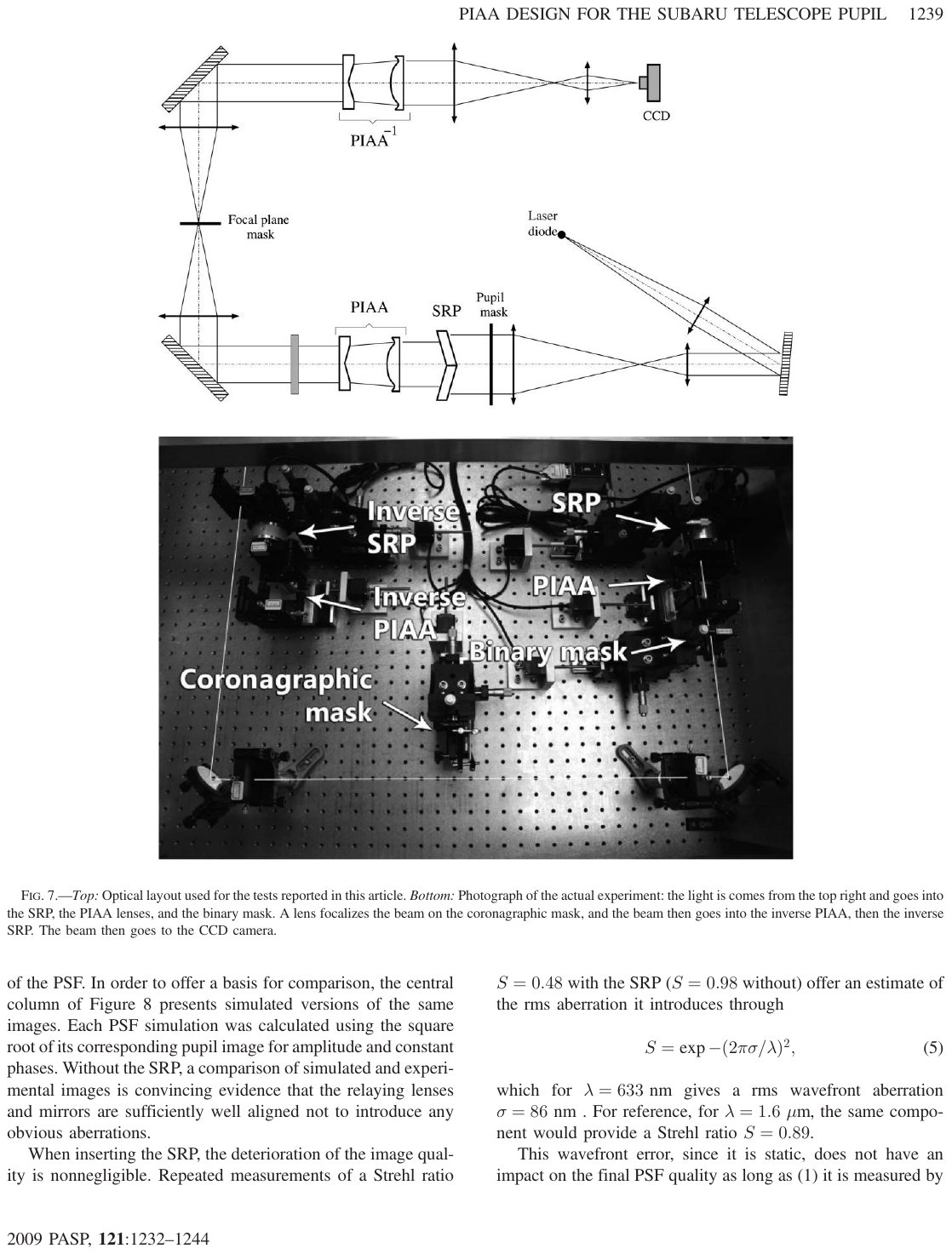}
 \PASP{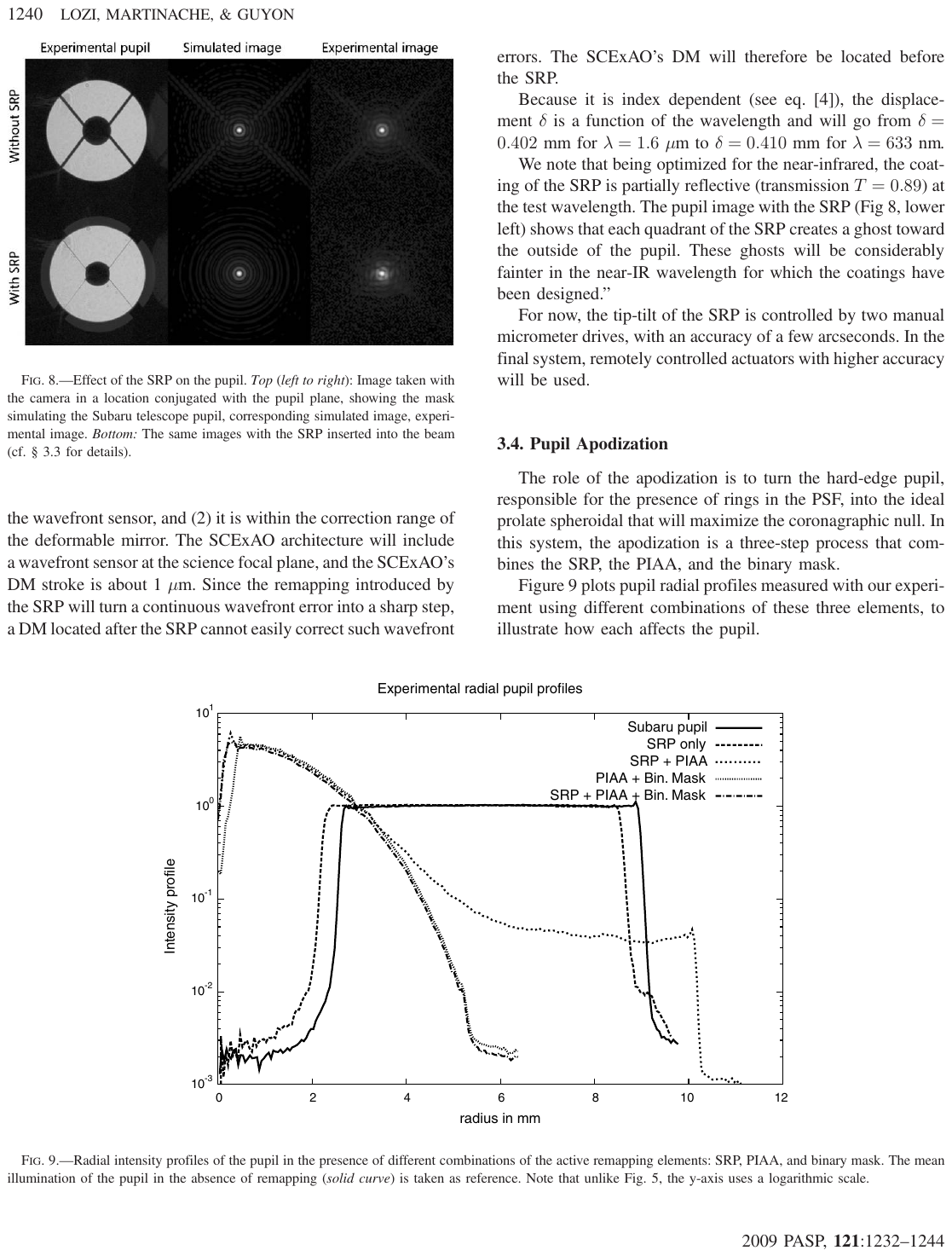}
 \PASP{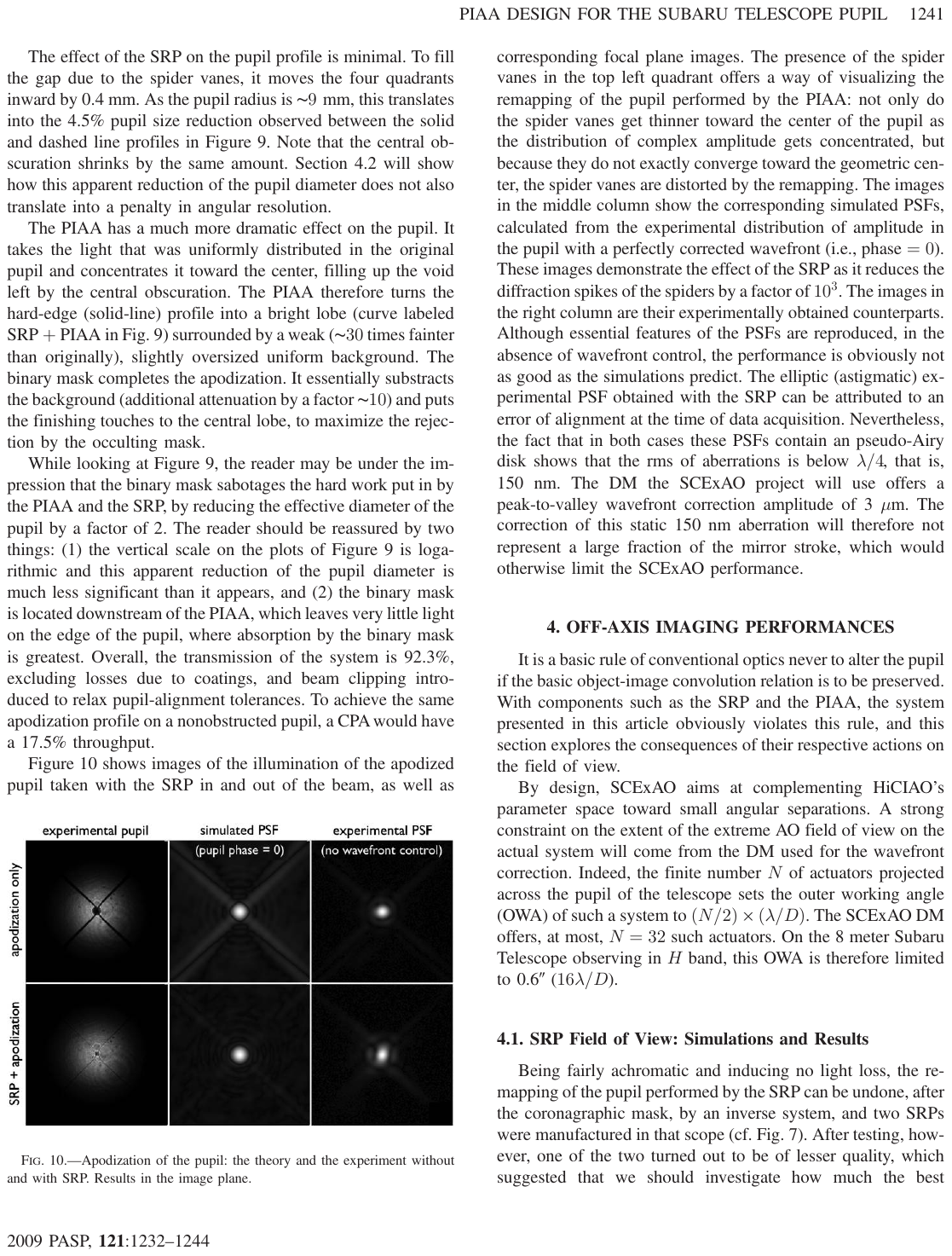}
 \PASP{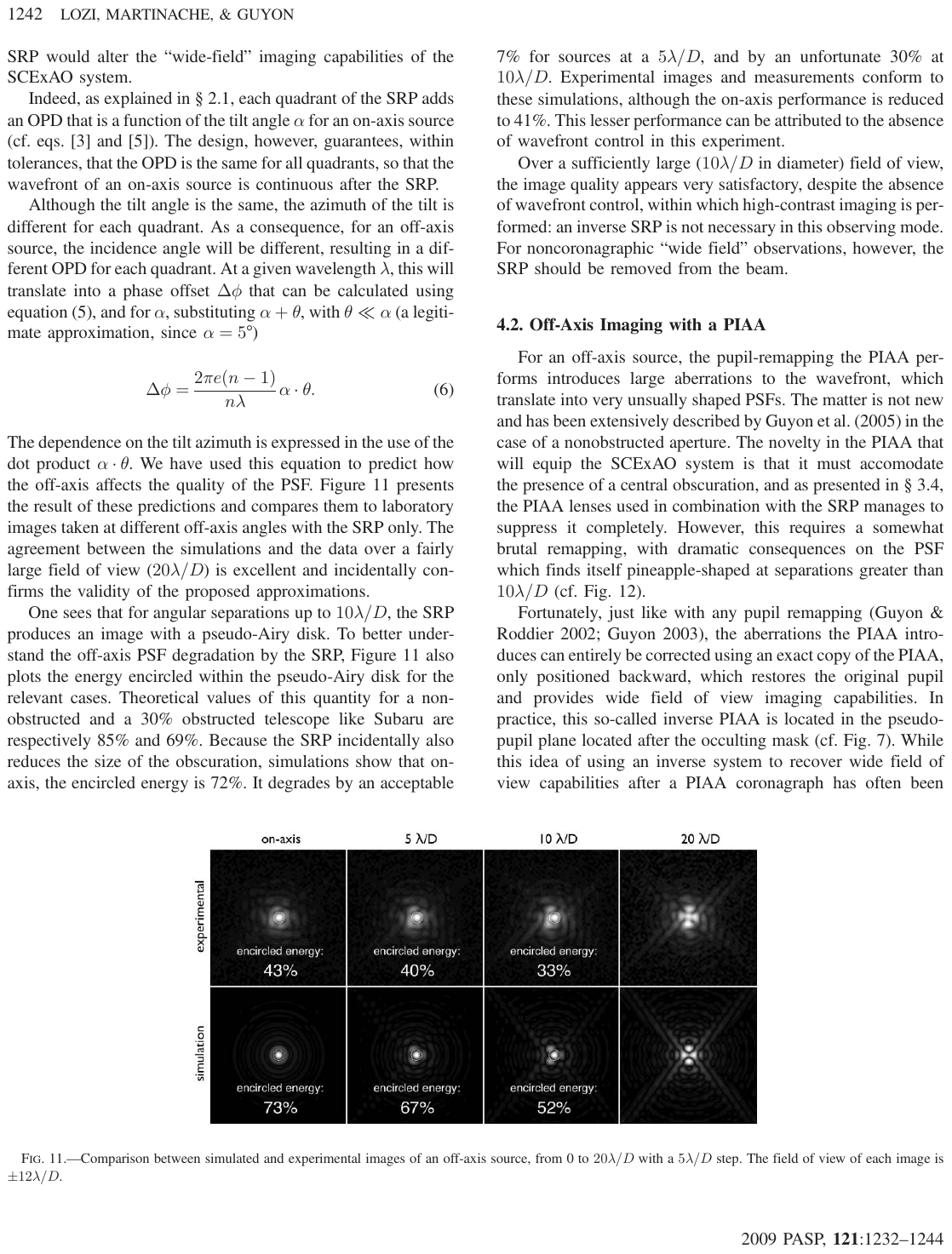}
 \PASP{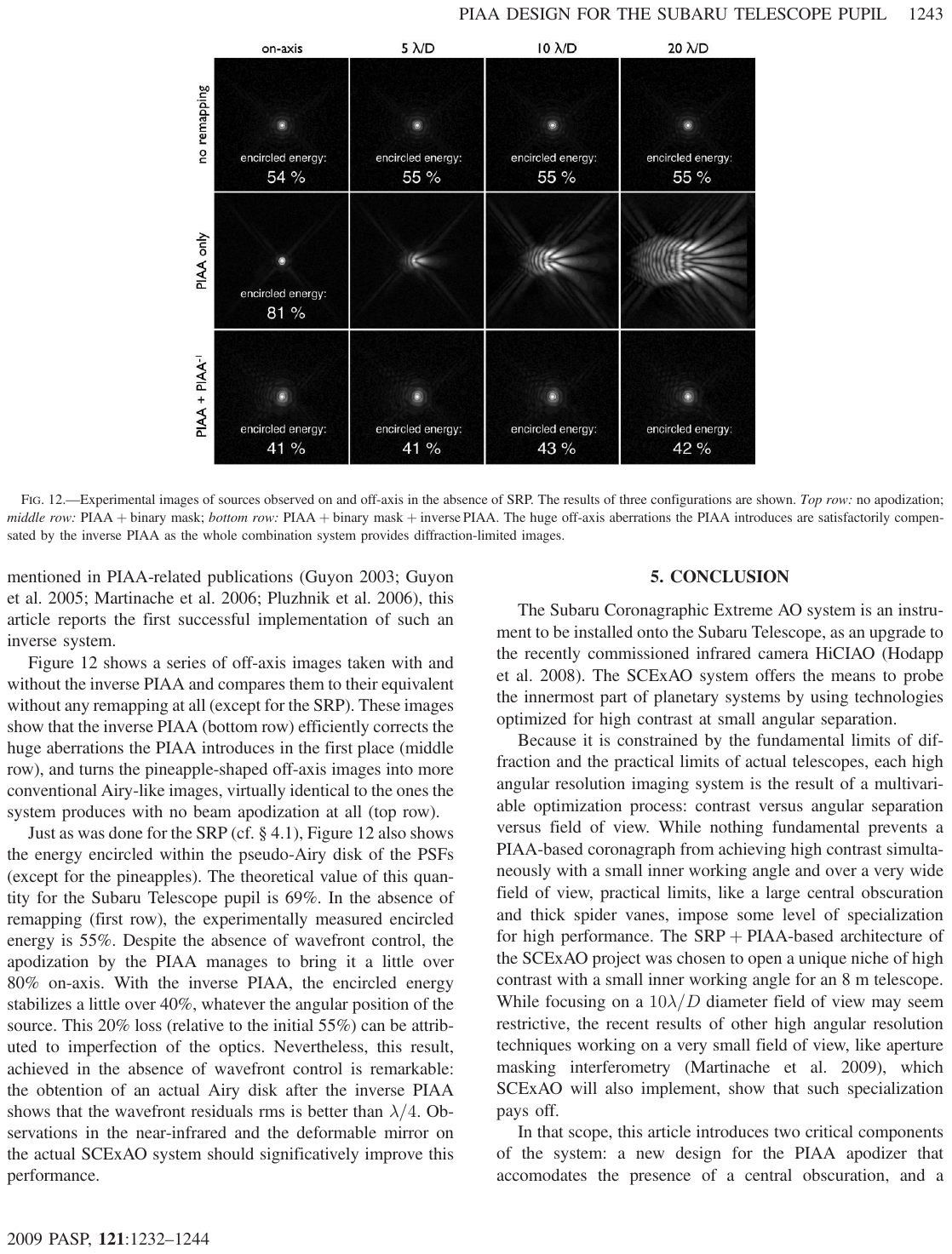}
 \PASP{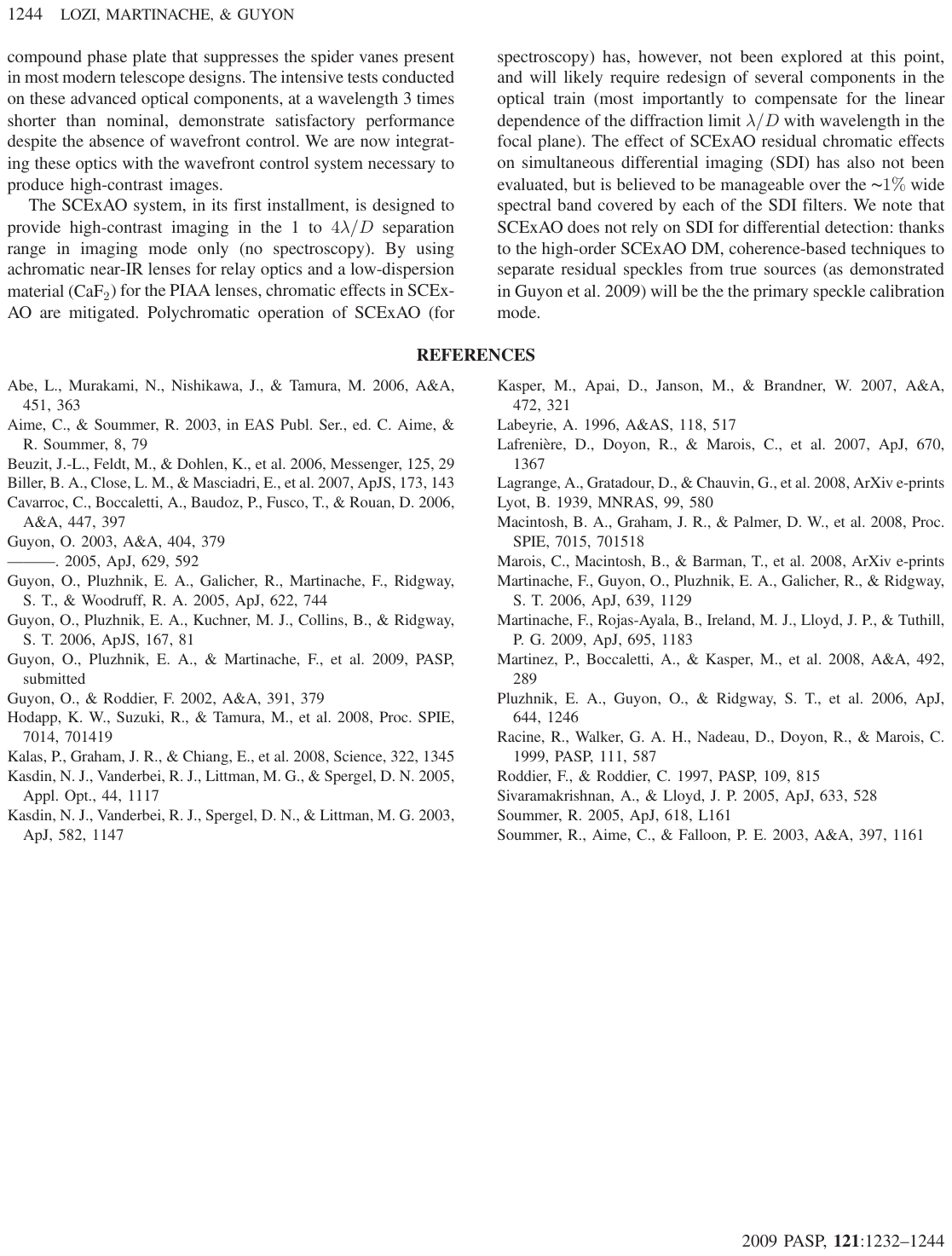}
\newpage

The testbed was subsequently upgraded and reconfigured for Nasmyth platform integration (Figure~\ref{fig-new-scexao}).

\begin{figure} \centering
  \FIGA{-90}{1.}{false}{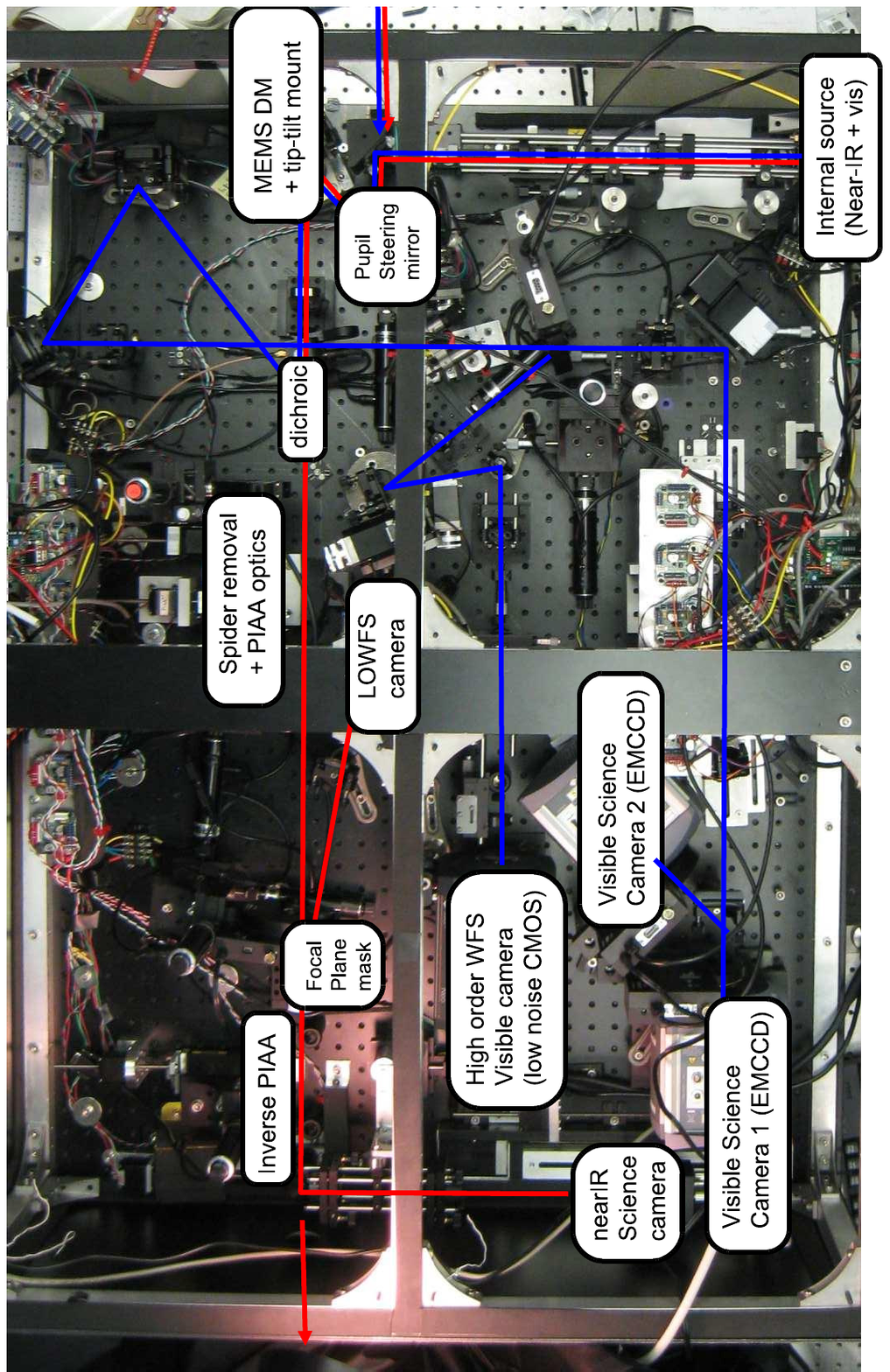}
  \caption{\sce testbed as installed on the Subaru Telescope.}
  \label{fig-new-scexao}
\end{figure}

Supported by ONERA funding, I participated in a commissioning run on \sce in February 2011, comprising daytime testing and two nighttime engineering runs. Closed-loop tip/tilt control was demonstrated, and images were acquired on visible and infrared science detectors.

Internal calibration sources allowed us to evaluate vibration signatures on the ground and at the telescope. Figure~\ref{fig-nasmyth} shows \sce installed on the Infrared Nasmyth platform of the Subaru Telescope between AO188 and HiCIAO.

\begin{figure} \centering
  \FIG{0.9}{false}{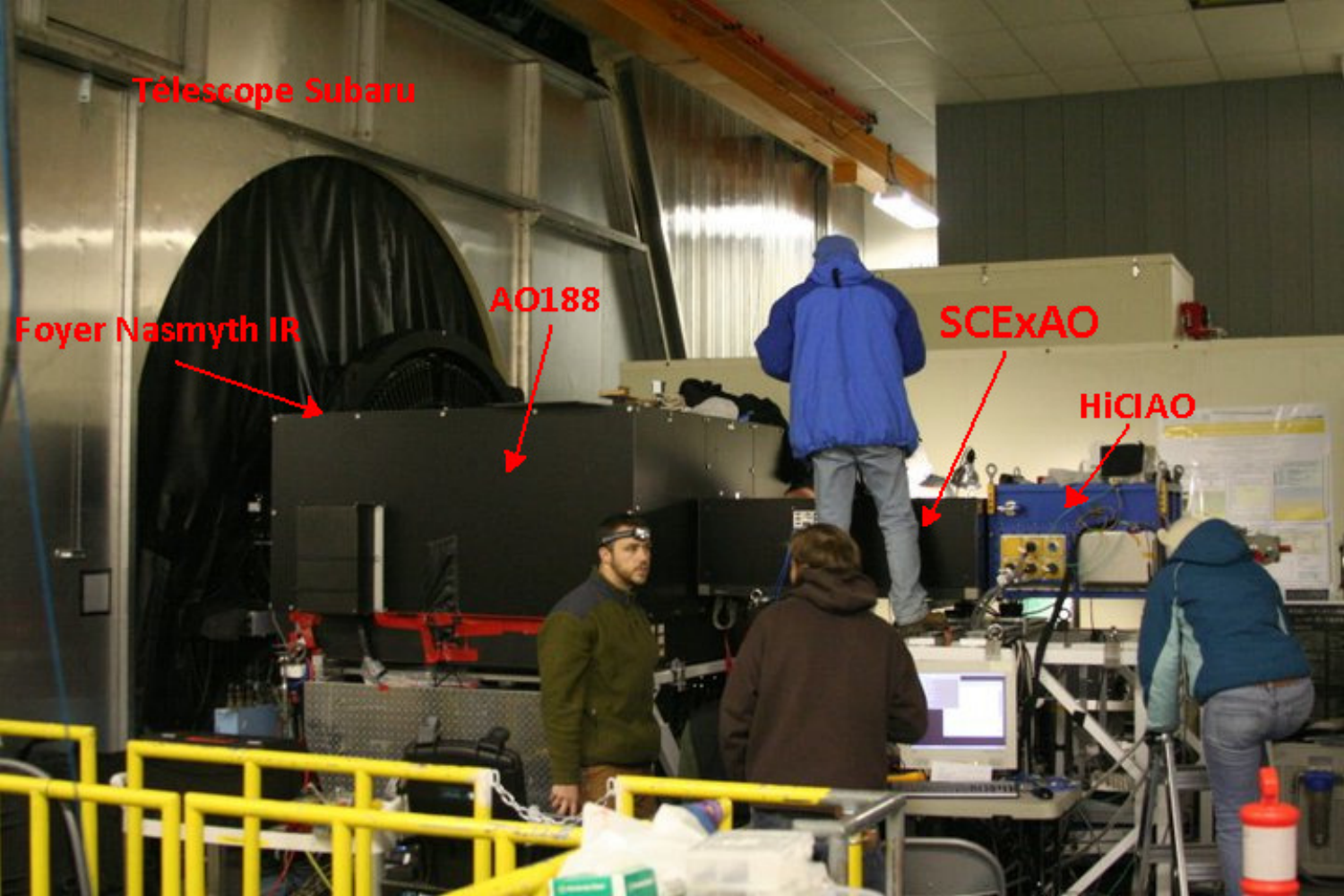}
  \caption[\sce installed on Subaru's IR Nasmyth platform.]{\sce installed on the IR Nasmyth platform of the Subaru Telescope between AO188 and HiCIAO.}
  \label{fig-nasmyth}
\end{figure}

%-------------------------------------------------------------------------------
\subsection{Vibration Analysis}
\label{sec-analyse-vibratoire}

%¤¤¤¤¤¤¤¤¤¤¤¤¤¤¤¤¤¤¤¤¤¤¤¤¤¤¤¤¤¤¤¤¤¤¤¤¤¤¤¤¤¤¤¤¤¤¤¤¤¤¤¤¤¤¤¤¤¤¤¤¤¤¤¤¤¤¤¤¤¤¤¤¤¤¤¤¤¤¤
\subsubsection{CLOWFS Jitter Analysis with Non-Deterministic Sampling}
\label{sec-analyse-jitter}

We analyze CLOWFS tip/tilt sensor data recorded during daytime testing using an internal light source.

Unlike \pe's deterministic real-time control, \sce's orginial tip/tilt loop ran asynchronously. Frame intervals varied depending on system task scheduling. Frame timestamps were recorded per acquisition.

Figure~\ref{fig-lowfs-fichier} plots time-series and 2D $x$-$y$ pointing offsets (detector horizontal $x$ and vertical $y$ axes). Driving voltages scale to sky angles at ~3~mas.V$_\text{piezo}^{-1}$.

\begin{figure} \centering
  \subfloat[Time-series pointing track.]{\label{fig-lowfs-temp}\FIGH{5.7}{false}{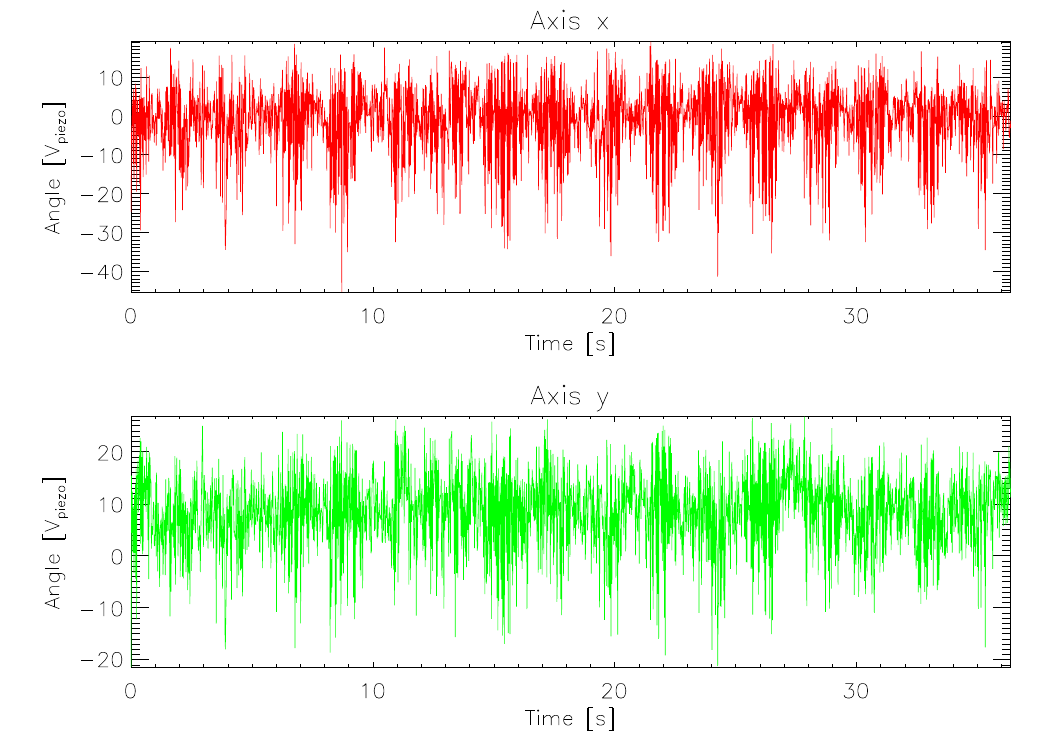}}
  \hfill\subfloat[2D $x$-$y$ position scatter plot.]{\label{fig-lowfs-tempxy}\FIGH{6.3}{false}{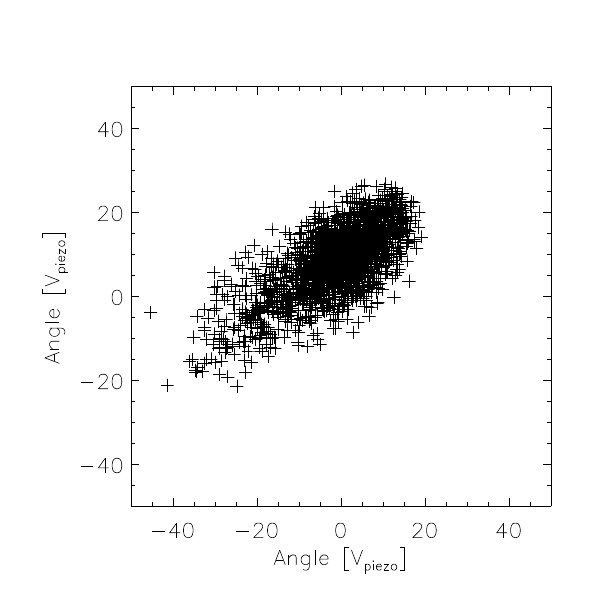}}
  \caption{CLOWFS tip/tilt tracking data.}
  \label{fig-lowfs-fichier}
\end{figure}

Pointing jitter in Figure~\ref{fig-lowfs-temp} is anisotropic, exhibiting a dominant axis oriented at ~45\degree. Time-series signals (Figure~\ref{fig-lowfs-tempxy}) display high-frequency beats indicative of two closely spaced mechanical vibration modes.

Evaluating PSDs requires addressing non-deterministic frame timing. Figure~\ref{fig-lowfs-tempo} analyzes frame delta-times and frequency distributions.

\begin{figure} \centering
  \subfloat[Frame interval time series.]{\label{fig-lowfs-time}\FIG{0.49}{false}{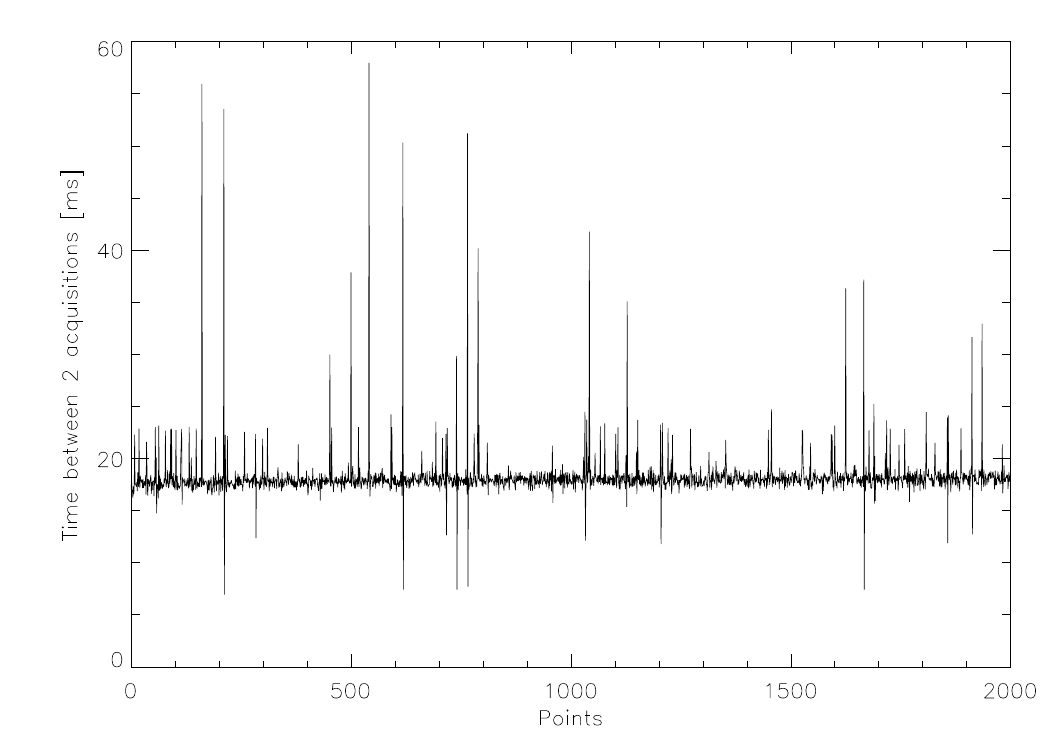}}
  \hfill\subfloat[Frame rate distribution histogram.]{\label{fig-lowfs-histo}\FIG{0.49}{false}{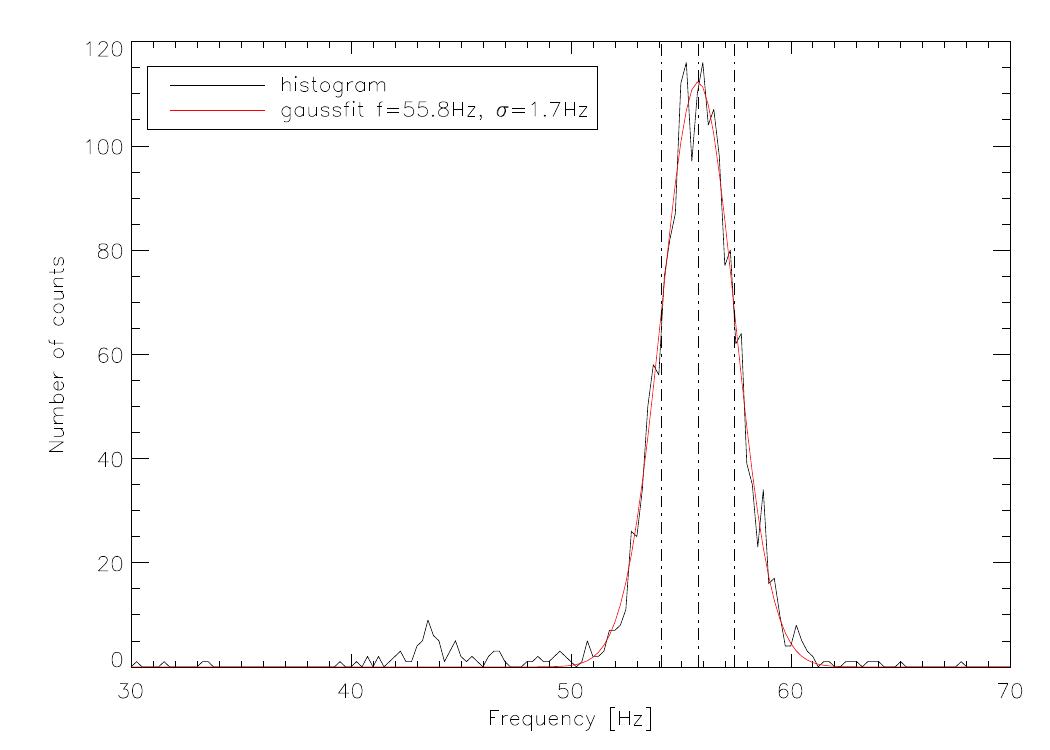}}
  \caption{CLOWFS frame rate timing analysis.}
  \label{fig-lowfs-tempo}
\end{figure}

Mean frame interval is 18~ms ($55.8$~Hz frame rate), but individual intervals range up to 40~ms due to task interrupts (Figure~\ref{fig-lowfs-time}). The central distribution (Figure~\ref{fig-lowfs-histo}) is Gaussian with standard deviation $1.7$~Hz.

PSDs were calculated using timestamp-interpolated resampling algorithms (Perrin \cite{Perrin96}). Data were rotated by 45\degree onto principal axes $x'$ and $y'$. Figure~\ref{fig-lowfs-pi} plots the resulting PSDs and cumulative power spectra.

\begin{figure} \centering
  \subfloat[Tip/tilt PSDs.]{\label{fig-lowfs-psd}\FIG{0.505}{false}{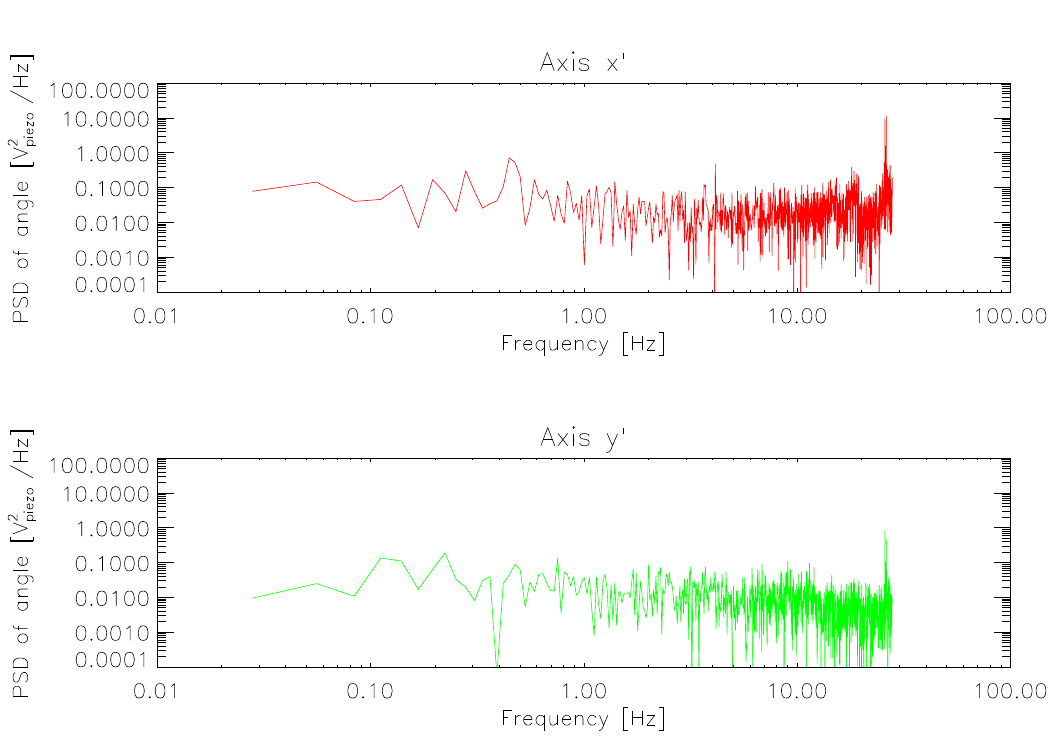}}
  \hfill\subfloat[Cumulative power spectra.]{\label{fig-lowfs-ic}\FIG{0.475}{false}{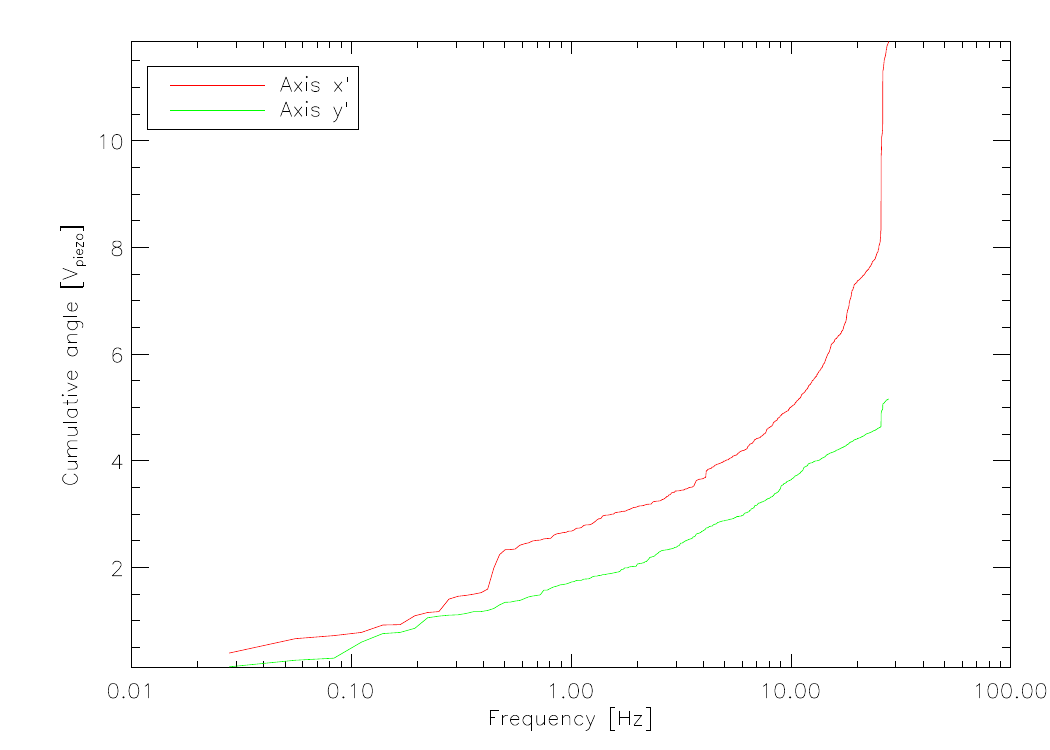}}
  \caption{CLOWFS tip/tilt PSDs and cumulative power along principal axes $x'$ and $y'$.}
  \label{fig-lowfs-pi}
\end{figure}

A sharp 26~Hz vibration mode dominates along principal axis $x'$ (Figure~\ref{fig-lowfs-psd}), exceeding $y'$-axis power by over an order of magnitude. Cumulative power (Figure~\ref{fig-lowfs-ic}) confirms a step at 26~Hz and low-frequency drift near $0.4$~Hz.

Figure~\ref{fig-lowfs-pzoom} compares high-frequency PSD profiles evaluated with and without timestamp interpolation.

\begin{figure} \centering
  \FIG{0.7}{false}{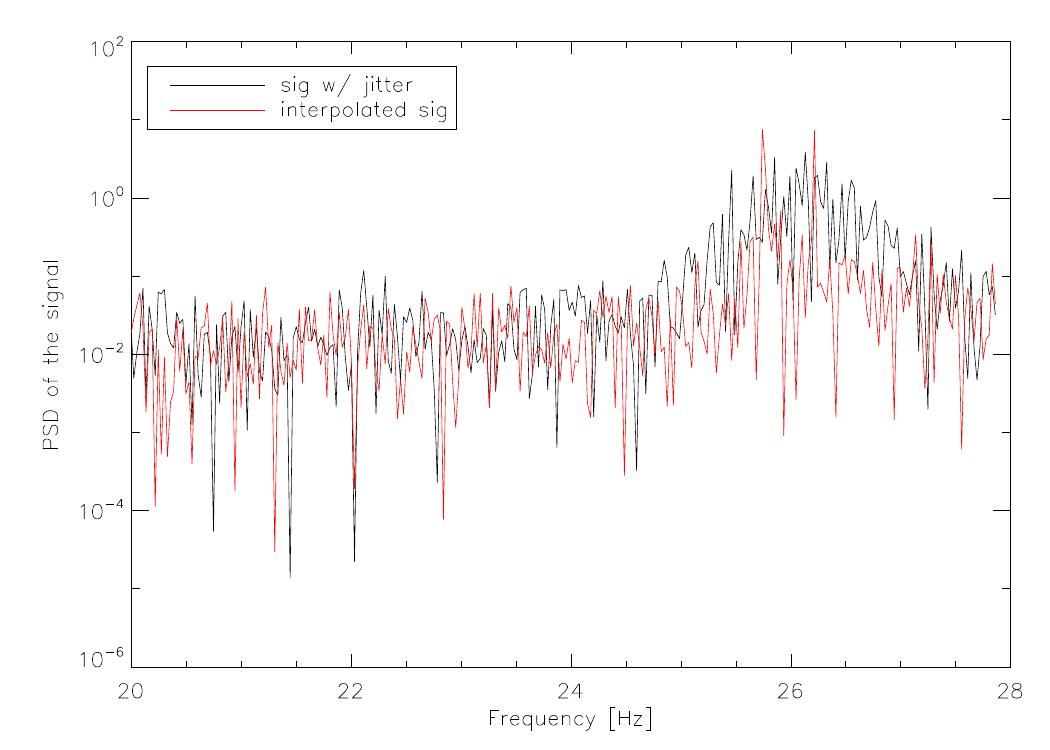}
  \caption{High-frequency PSD zoom with and without timestamp interpolation.}
  \label{fig-lowfs-pzoom}
\end{figure}

Interpolated processing resolves two distinct vibration peaks near 26~Hz separated by $0.5$~Hz.

Simulations incorporating Gaussian timing jitter (Figure~\ref{fig-lowfs-simzoom}) confirm that timestamp interpolation accurately recovers the twin 26~Hz modes without introducing spectral artifacts.

\begin{figure} \centering
  \subfloat[Simulated Gaussian timing jitter.]{\label{fig-lowfs-simz1}\FIG{0.49}{false}{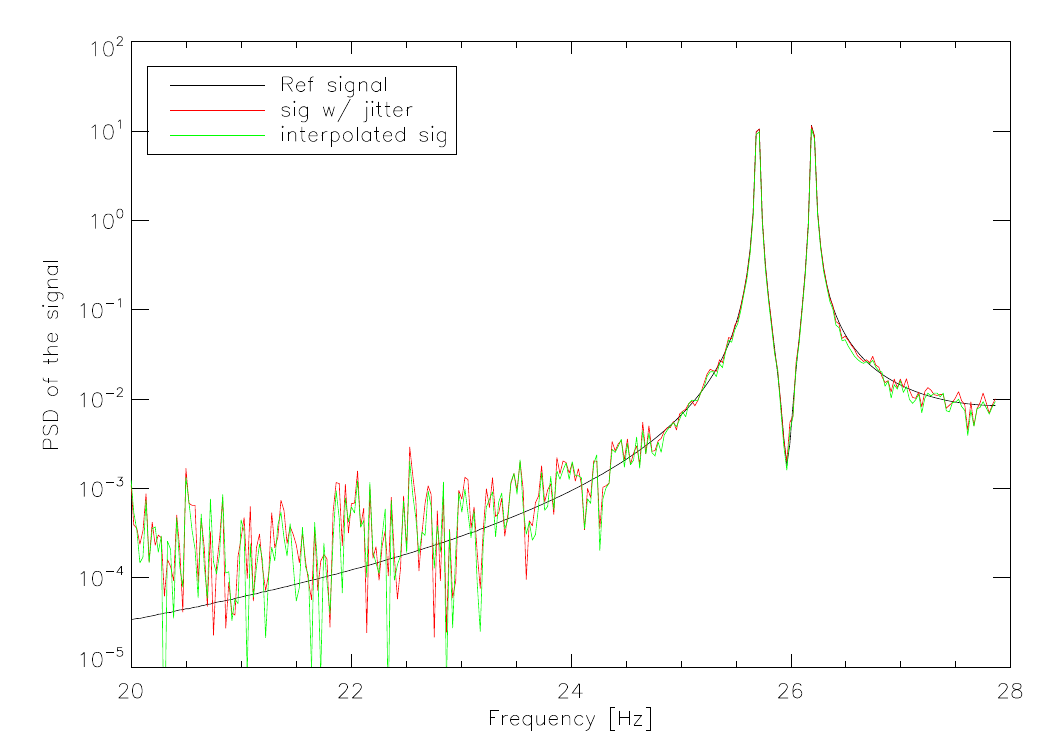}}
  \hfill\subfloat[Simulated measured timing jitter.]{\label{fig-lowfs-simz2}\FIG{0.49}{false}{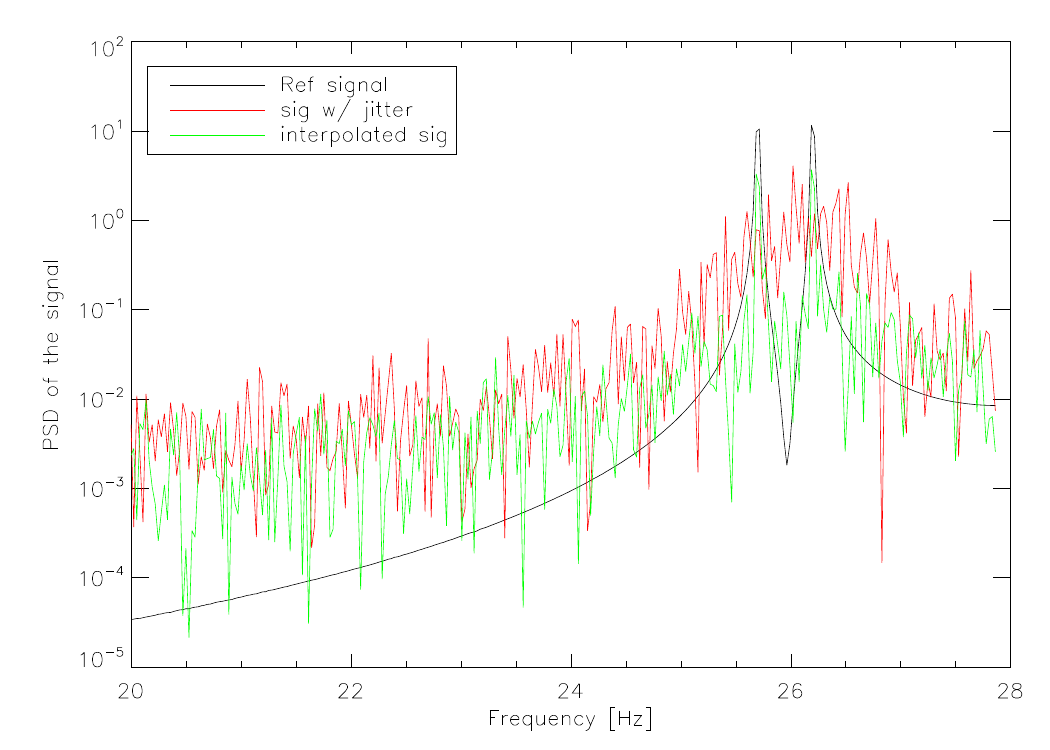}}
  \caption{High-frequency PSD simulations under timing jitter.}
  \label{fig-lowfs-simzoom}
\end{figure}

Applying LQG control requires identifying structural sources driving these 26~Hz modes, stabilizing frame timing, and raising sampling rates to avoid frequency aliasing near loop Nyquist limits.

%¤¤¤¤¤¤¤¤¤¤¤¤¤¤¤¤¤¤¤¤¤¤¤¤¤¤¤¤¤¤¤¤¤¤¤¤¤¤¤¤¤¤¤¤¤¤¤¤¤¤¤¤¤¤¤¤¤¤¤¤¤¤¤¤¤¤¤¤¤¤¤¤¤¤¤¤¤¤¤
\subsubsection{Infrared Science Camera Jitter}
\label{sec-cam-ir}

During nighttime testing, an infrared camera replacing HiCIAO recorded pointing jitter on Betelgeuse (plate scale 27~mas.pixel$^{-1}$; $1.6$~\mum diffraction spot sampled over 3 pixels). Figure~\ref{fig-ircam-fichier} plots pointing tracks.

\begin{figure} \centering
  \subfloat[Time-series pointing track.]{\label{fig-ircam-temp}\FIGH{5.7}{false}{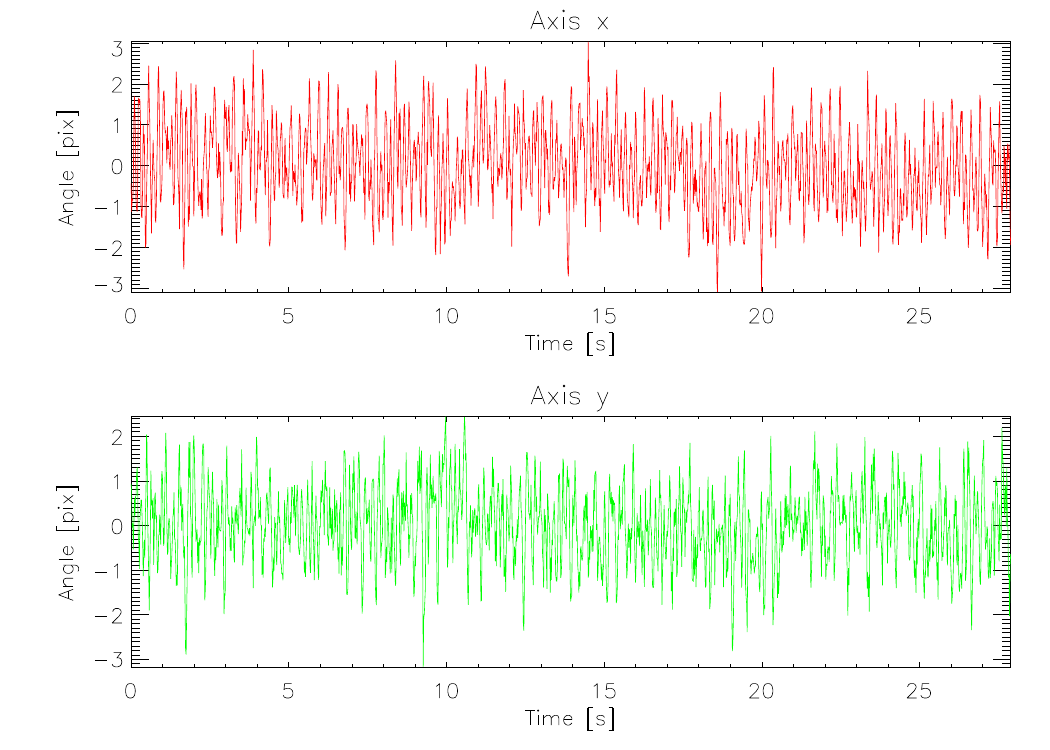}}
  \hfill\subfloat[2D $x$-$y$ position scatter plot.]{\label{fig-ircam-tempxy}\FIGH{6.3}{false}{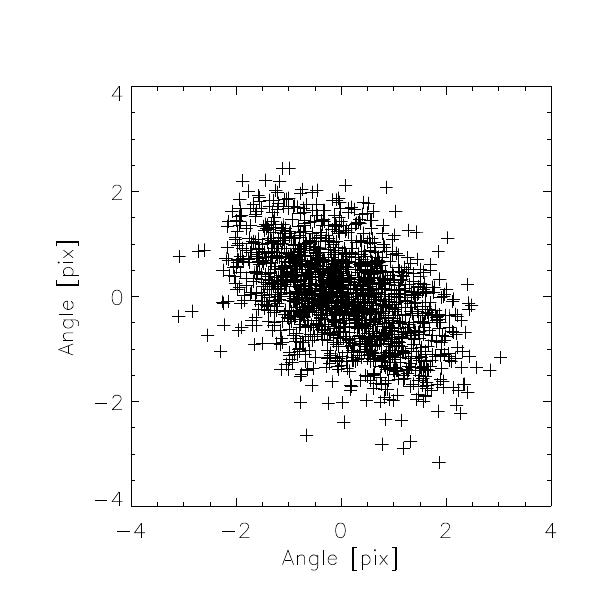}}
  \caption{Infrared camera pointing tracks recorded on Betelgeuse.}
  \label{fig-ircam-fichier}
\end{figure}

Pointing jitter in Figure~\ref{fig-ircam-temp} aligns along a $-45$\degree axis. Frame timing analysis (Figure~\ref{fig-ircam-tempo}) reveals periodic 1~s readout delays.

\begin{figure} \centering
  \subfloat[Frame interval time series.]{\label{fig-ircam-time}\FIG{0.49}{false}{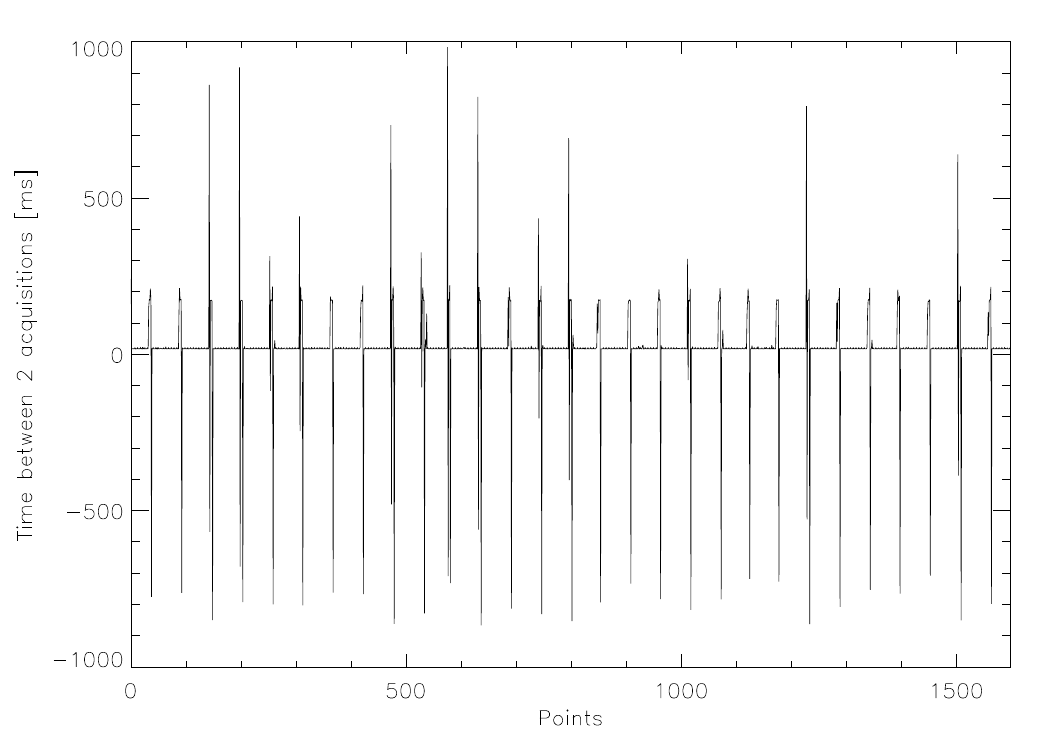}}
  \hfill\subfloat[Frame rate distribution histogram.]{\label{fig-ircam-histo}\FIG{0.49}{false}{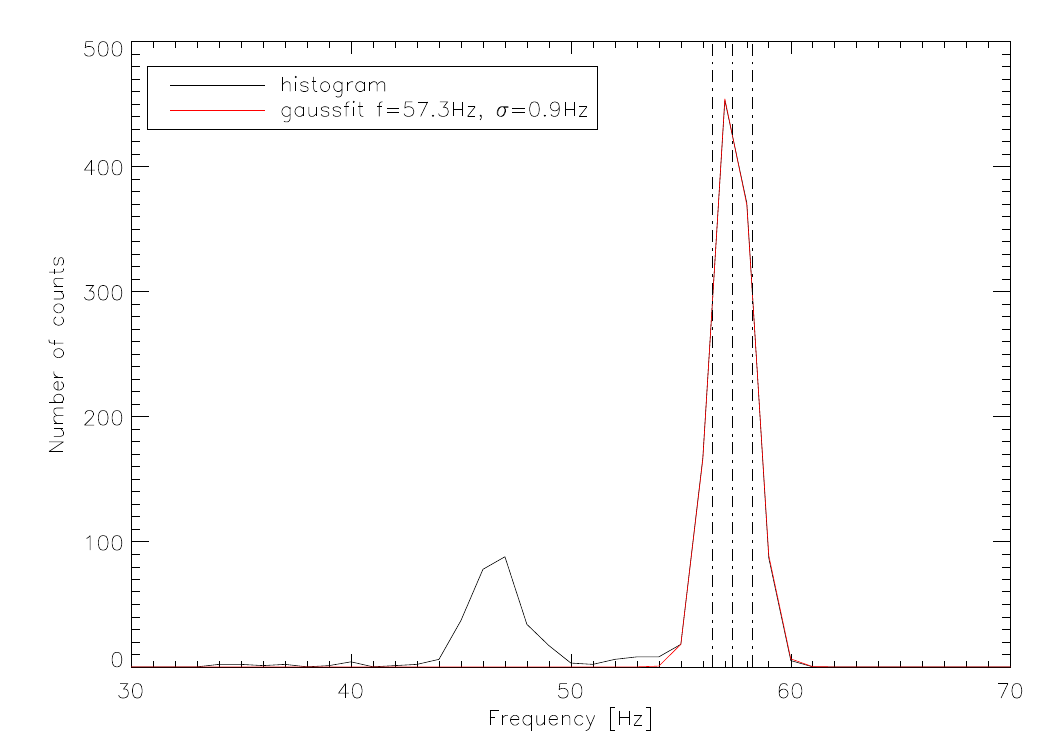}}
  \caption{Infrared camera frame timing analysis.}
  \label{fig-ircam-tempo}
\end{figure}

The primary frame rate peaks at $57.3$~Hz (Figure~\ref{fig-ircam-histo}). Timestamped PSDs rotated onto principal axes $x'$ and $y'$ are shown in Figure~\ref{fig-ircam-pi}.

\begin{figure} \centering
  \subfloat[Tip/tilt PSDs.]{\label{fig-ircam-psd}\FIG{0.505}{false}{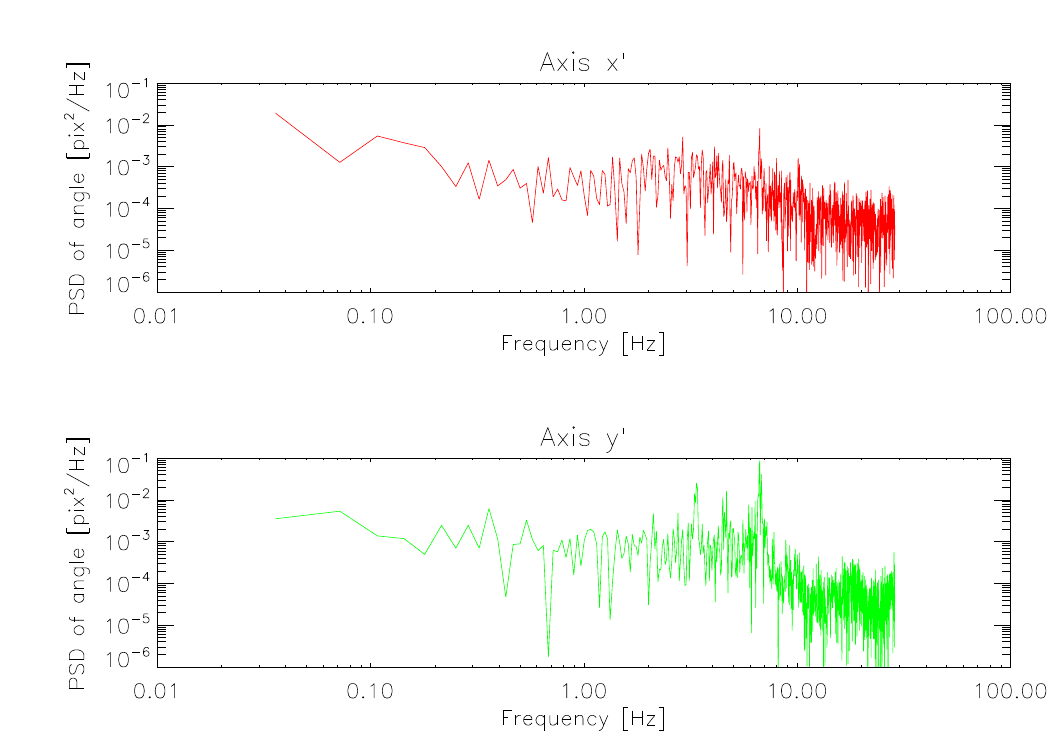}}
  \hfill\subfloat[Cumulative power spectra.]{\label{fig-ircam-ic}\FIG{0.475}{false}{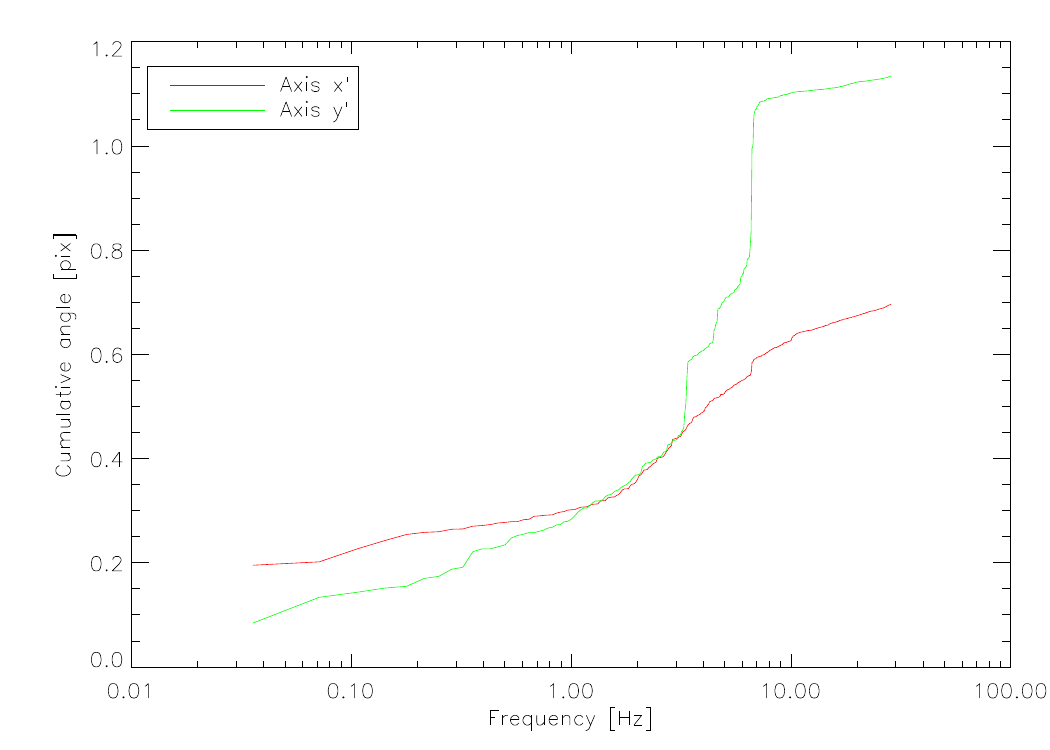}}
  \caption{Infrared camera tip/tilt PSDs and cumulative power along principal axes.}
  \label{fig-ircam-pi}
\end{figure}

The infrared camera spectra (Figure~\ref{fig-ircam-pi}) show low-frequency modes at 3, $4.5$, and 7~Hz along axis $y'$. The 26~Hz mode seen on the CLOWFS is absent, suggesting that 3–7~Hz modes may represent aliased higher-frequency vibrations or path-specific optical bench jitter.

%¤¤¤¤¤¤¤¤¤¤¤¤¤¤¤¤¤¤¤¤¤¤¤¤¤¤¤¤¤¤¤¤¤¤¤¤¤¤¤¤¤¤¤¤¤¤¤¤¤¤¤¤¤¤¤¤¤¤¤¤¤¤¤¤¤¤¤¤¤¤¤¤¤¤¤¤¤¤¤
\subsubsection{Visible EMCCD Camera Jitter}
\label{sec-cam-visible}

A high-speed visible EMCCD camera (1~kHz frame rate) operating downstream of the extreme AO stage was tested using daytime calibration light passing through AO188. The camera supports short-exposure lucky imaging (plate scale 6~mas.pixel$^{-1}$). The 1~kHz readout rate is locked to a deterministic internal clock.

Data were recorded across four operating states:
\begin{itemize}
\item AO188 loop OPEN or CLOSED;
\item HiCIAO cryo-cooler pump ON or OFF.
\end{itemize}

Figure~\ref{fig-viscam-xy} displays 2D $x$-$y$ pointing tracks across the four states.

\begin{figure} \centering
  \subfloat[HiCIAO OFF, AO188 OPEN.]{\label{fig-viscam-xy1}\FIG{0.35}{false}{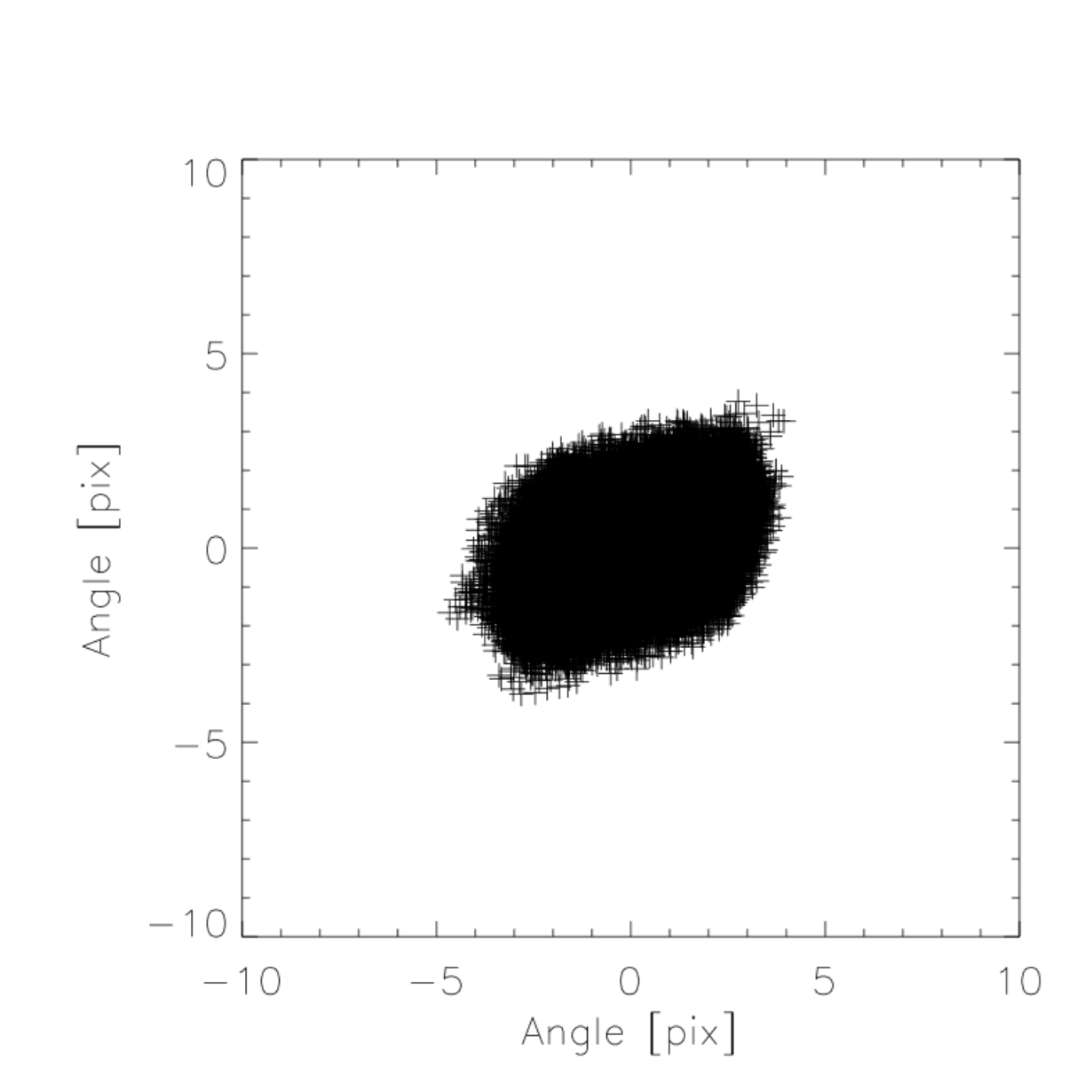}}
  \hspace{10pt}\subfloat[HiCIAO ON, AO188 OPEN.]{\label{fig-viscam-xy2}\FIG{0.35}{false}{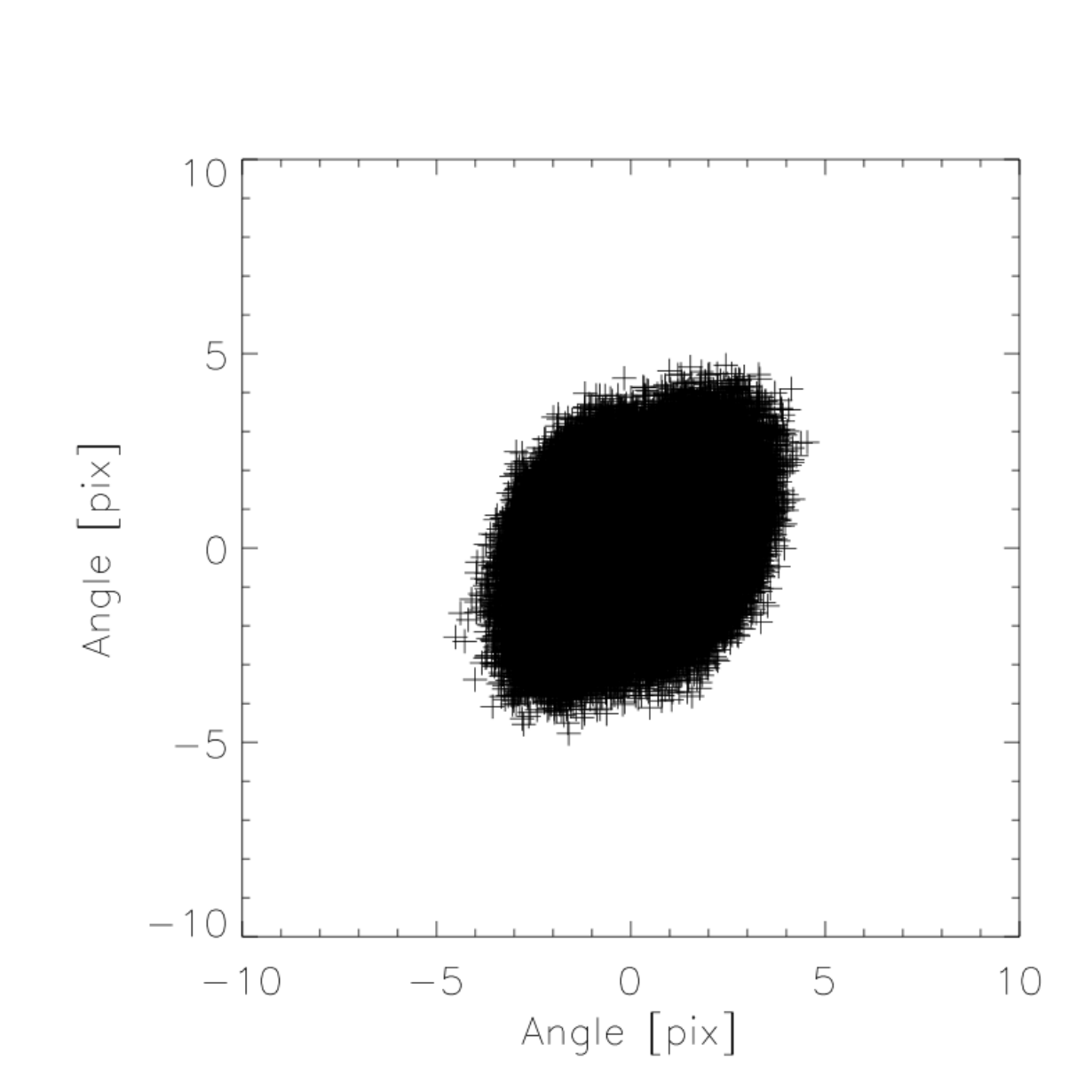}}\\
  \subfloat[HiCIAO OFF, AO188 CLOSED.]{\label{fig-viscam-xy3}\FIG{0.35}{false}{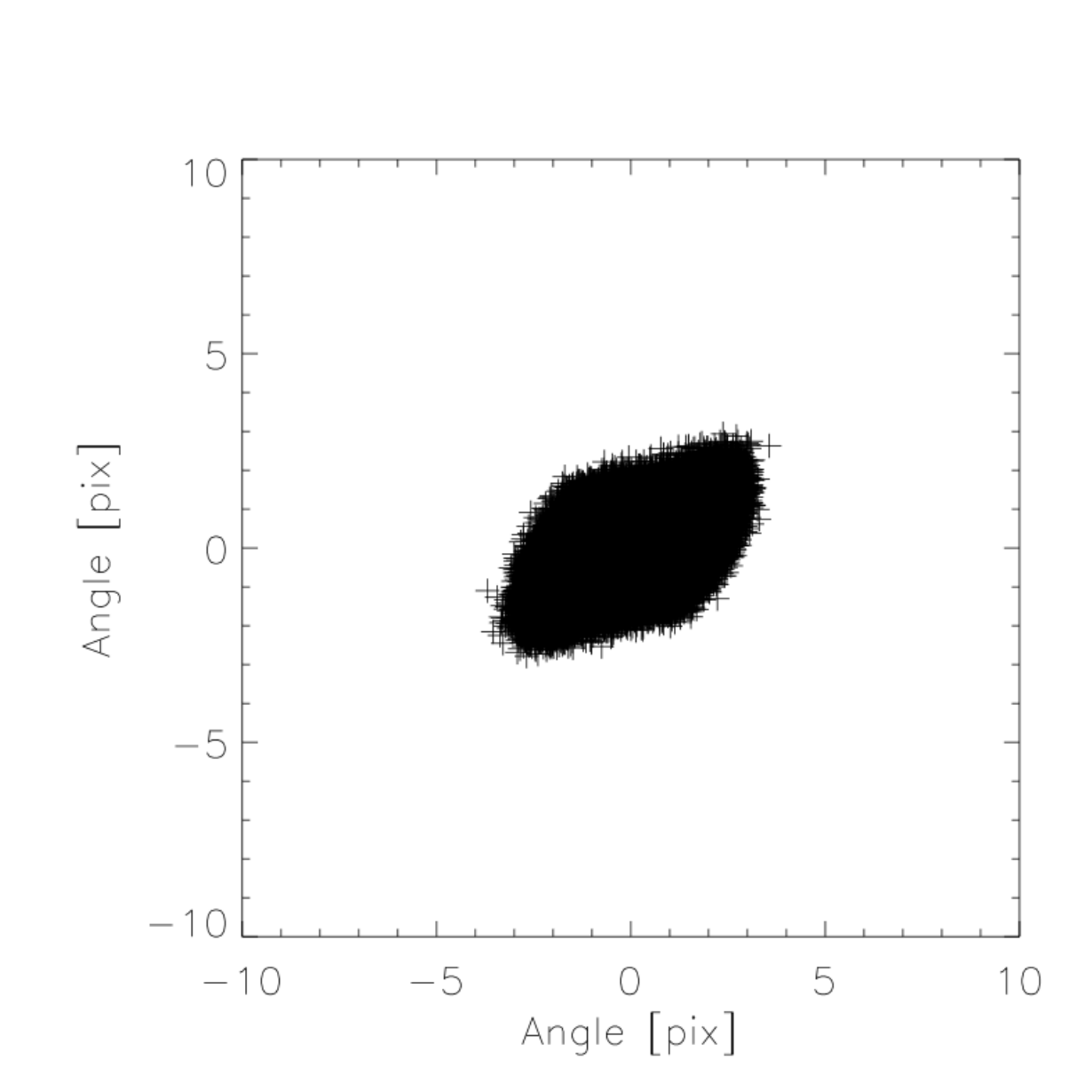}}
  \hspace{10pt}\subfloat[HiCIAO ON, AO188 CLOSED.]{\label{fig-viscam-xy4}\FIG{0.35}{false}{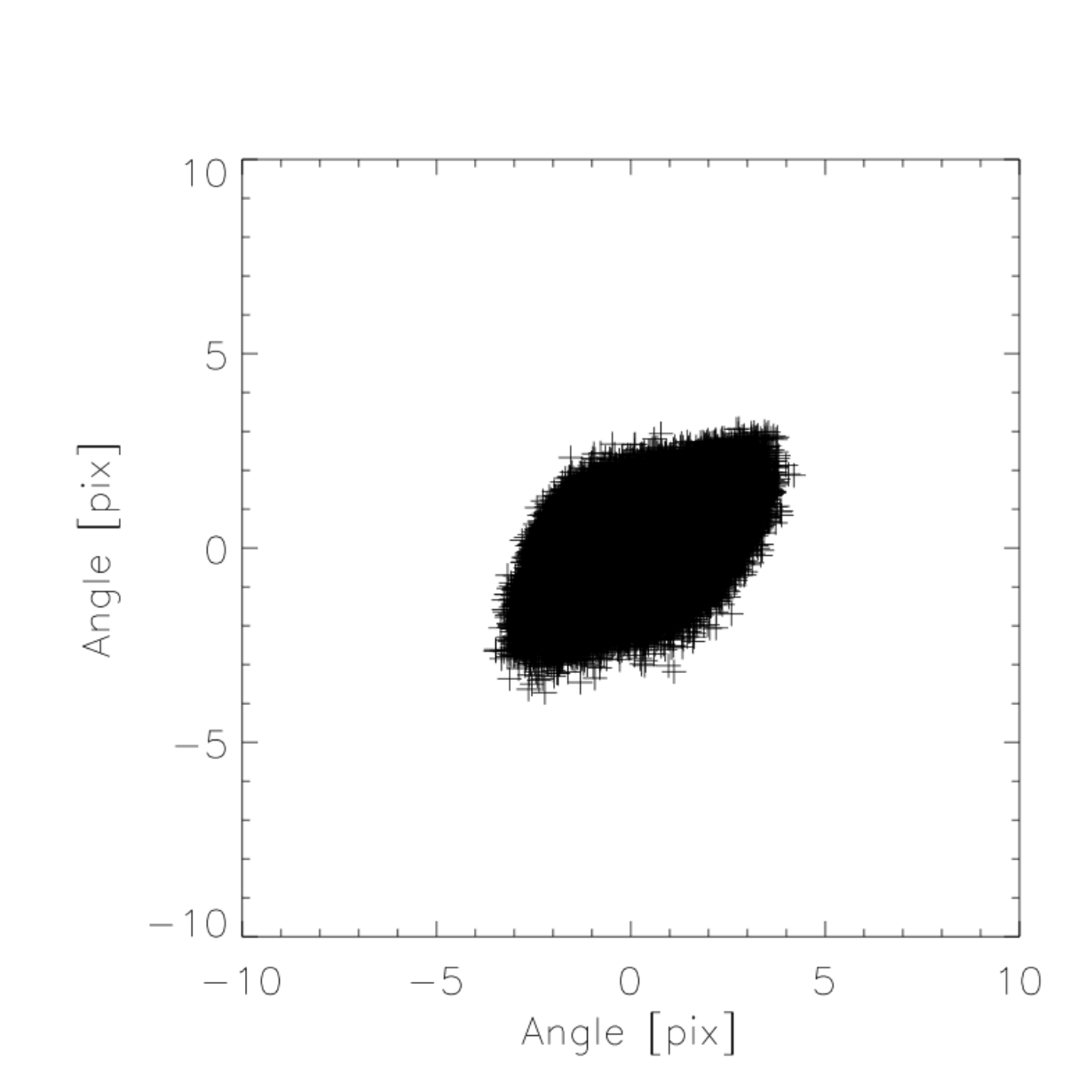}}
  \caption{Visible EMCCD tip/tilt tracking scatter plots across operating states.}
  \label{fig-viscam-xy}
\end{figure}

Closing the AO188 loop reduces pointing spread, while running the HiCIAO cryo-cooler expands jitter along a 45\degree axis.

Figure~\ref{fig-viscam-psd} plots tip/tilt PSDs rotated onto principal axes $x'$ and $y'$.

\begin{figure} \centering
  \subfloat[HiCIAO OFF, AO188 OPEN.]{\label{fig-viscam-psd1}\FIG{0.49}{false}{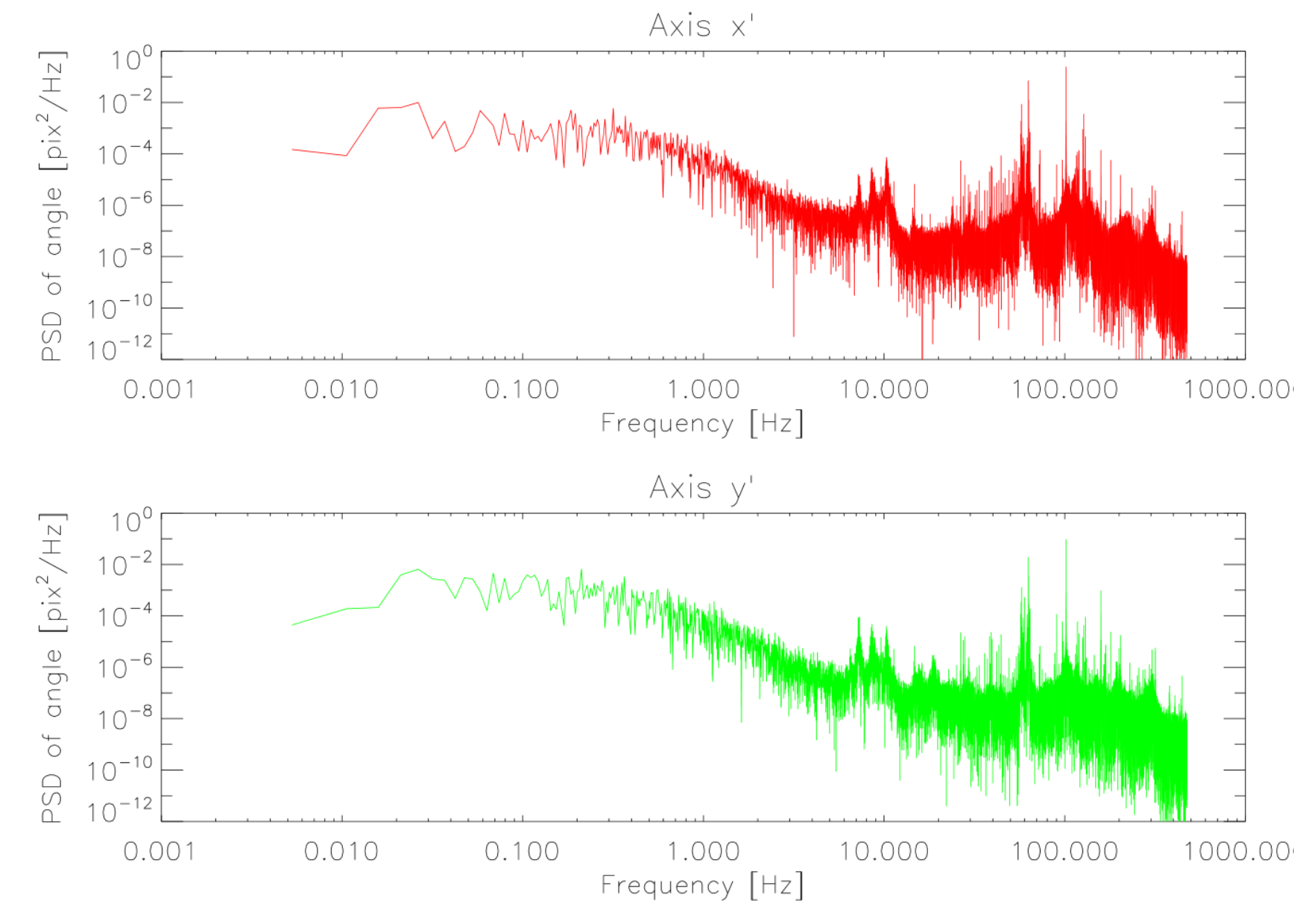}}
  \hfill\subfloat[HiCIAO ON, AO188 OPEN.]{\label{fig-viscam-psd2}\FIG{0.49}{false}{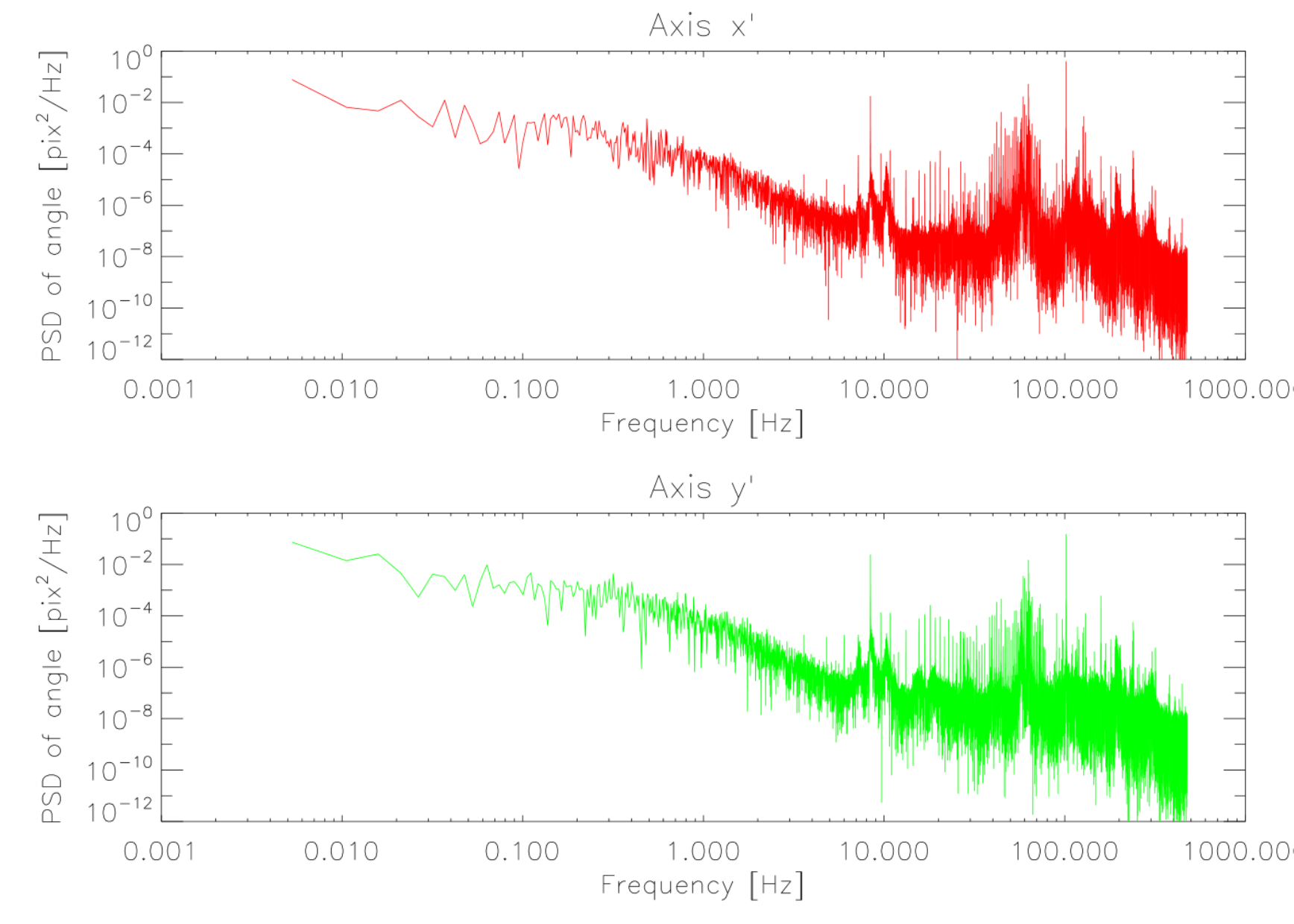}}\\
  \subfloat[HiCIAO OFF, AO188 CLOSED.]{\label{fig-viscam-psd3}\FIG{0.49}{false}{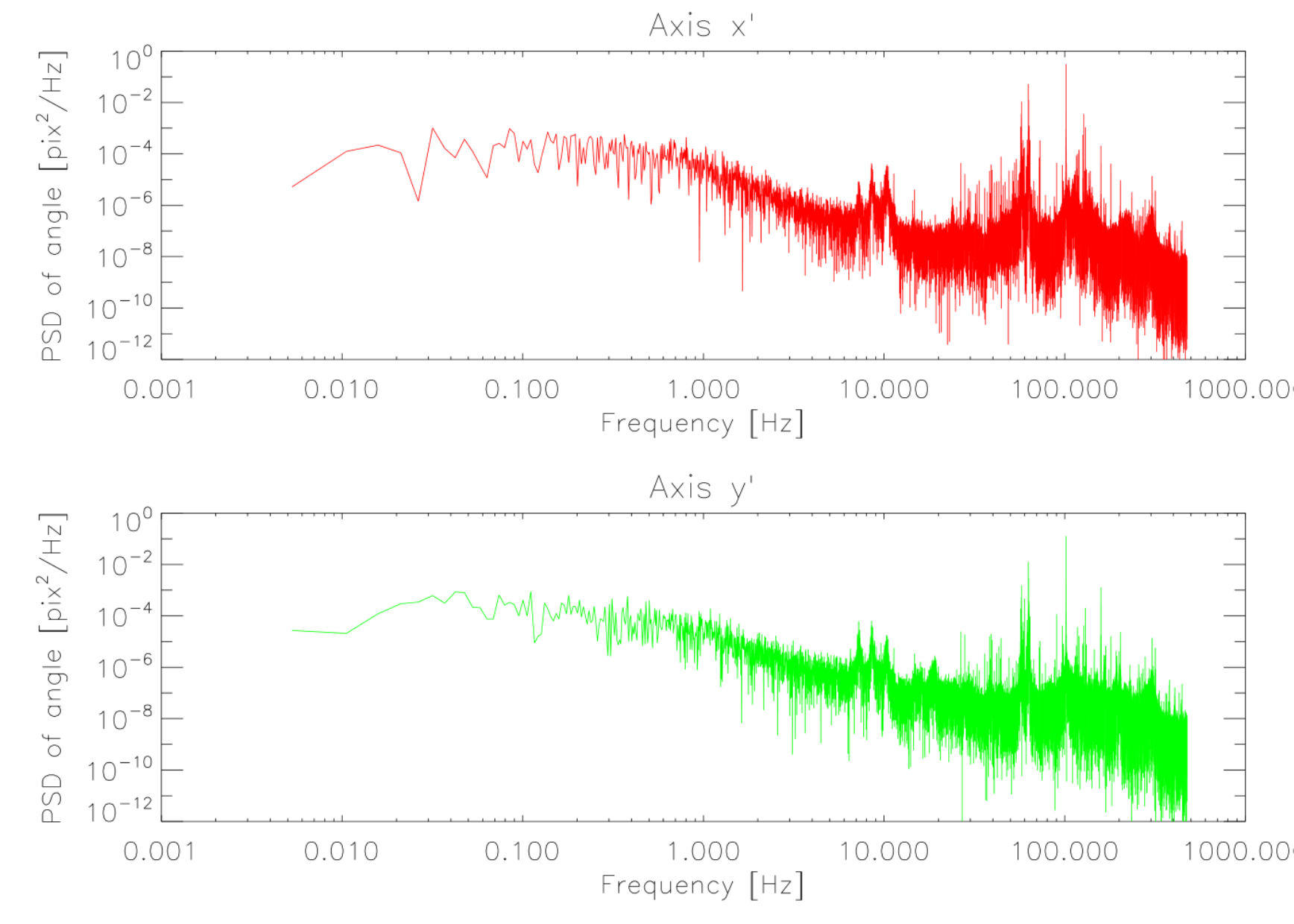}}
  \hfill\subfloat[HiCIAO ON, AO188 CLOSED.]{\label{fig-viscam-psd4}\FIG{0.49}{false}{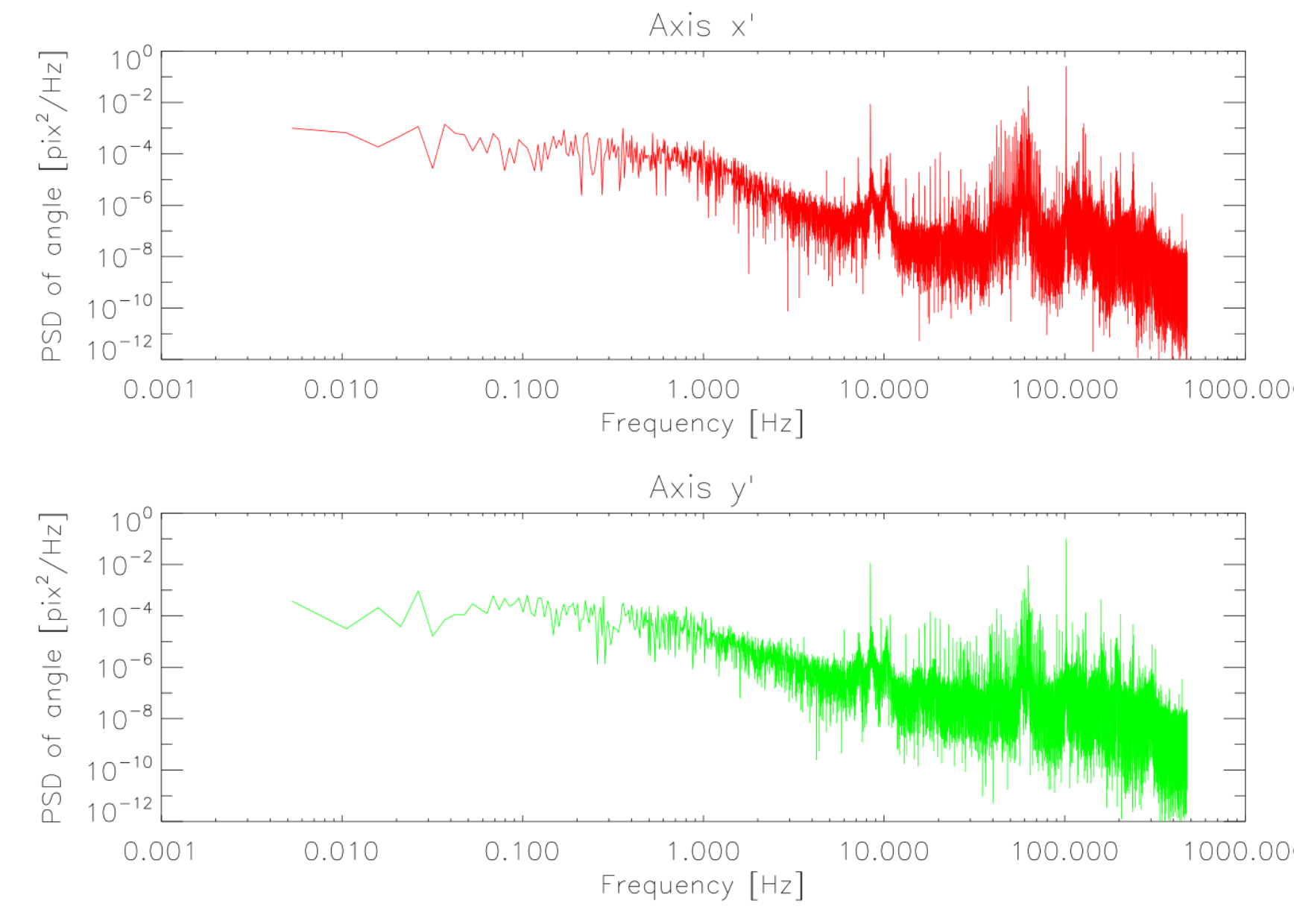}}
  \caption{Visible EMCCD tip/tilt PSDs across operating states.}
  \label{fig-viscam-psd}
\end{figure}

High-frequency vibration modes appear between 50~Hz and 200~Hz across all states. Closing AO188 suppresses drift below 1~Hz. Turning on HiCIAO introduces an 8~Hz vibration mode.

Figure~\ref{fig-viscam-ic} evaluates directional power distribution by computing cumulative power spectra across rotation angles.

\begin{figure} \centering
  \subfloat[HiCIAO OFF, AO188 OPEN.]{\label{fig-viscam-ic1}\FIG{0.49}{false}{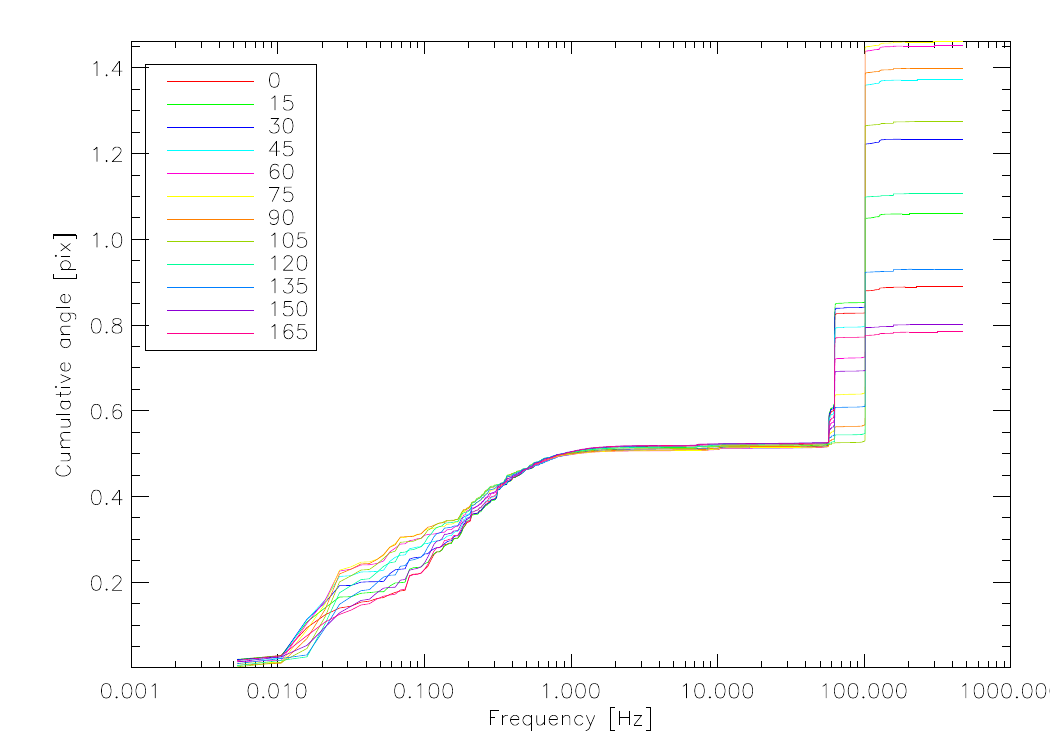}}
  \hfill\subfloat[HiCIAO ON, AO188 OPEN.]{\label{fig-viscam-ic2}\FIG{0.49}{false}{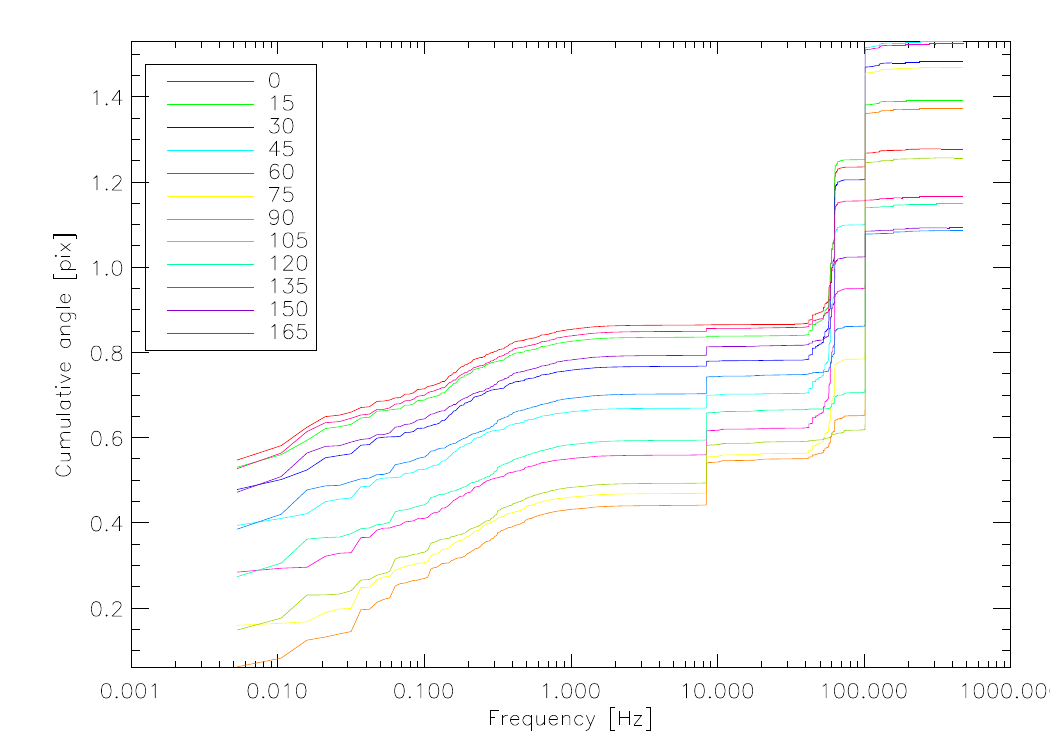}}\\
  \subfloat[HiCIAO OFF, AO188 CLOSED.]{\label{fig-viscam-ic3}\FIG{0.49}{false}{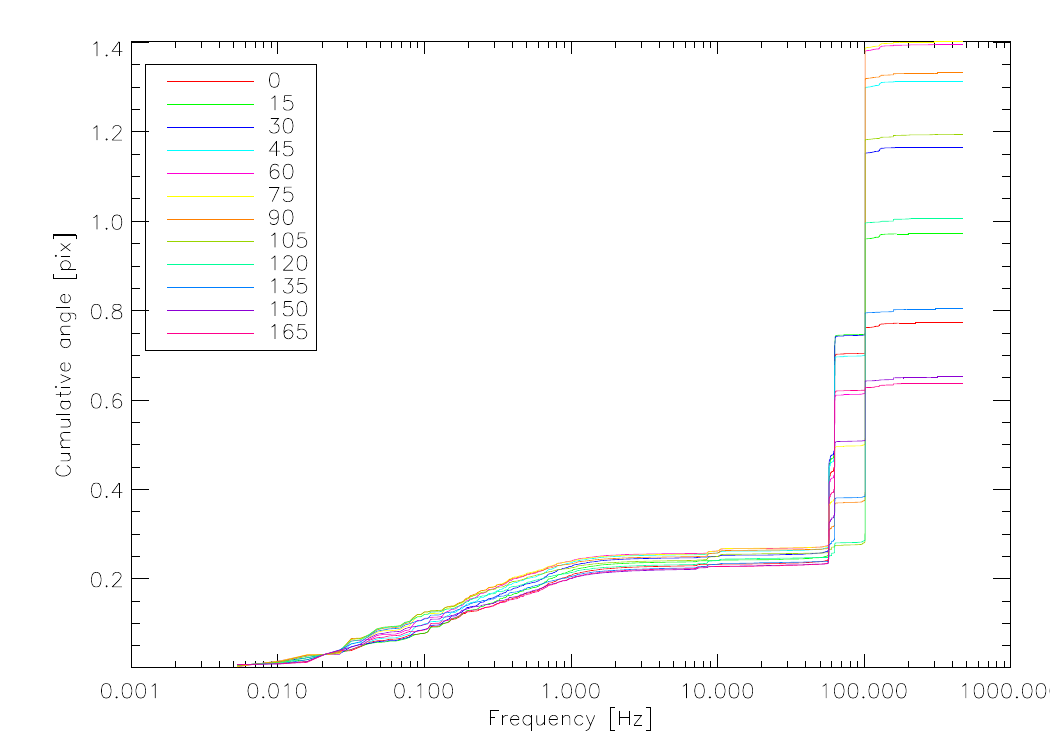}}
  \hfill\subfloat[HiCIAO ON, AO188 CLOSED.]{\label{fig-viscam-ic4}\FIG{0.49}{false}{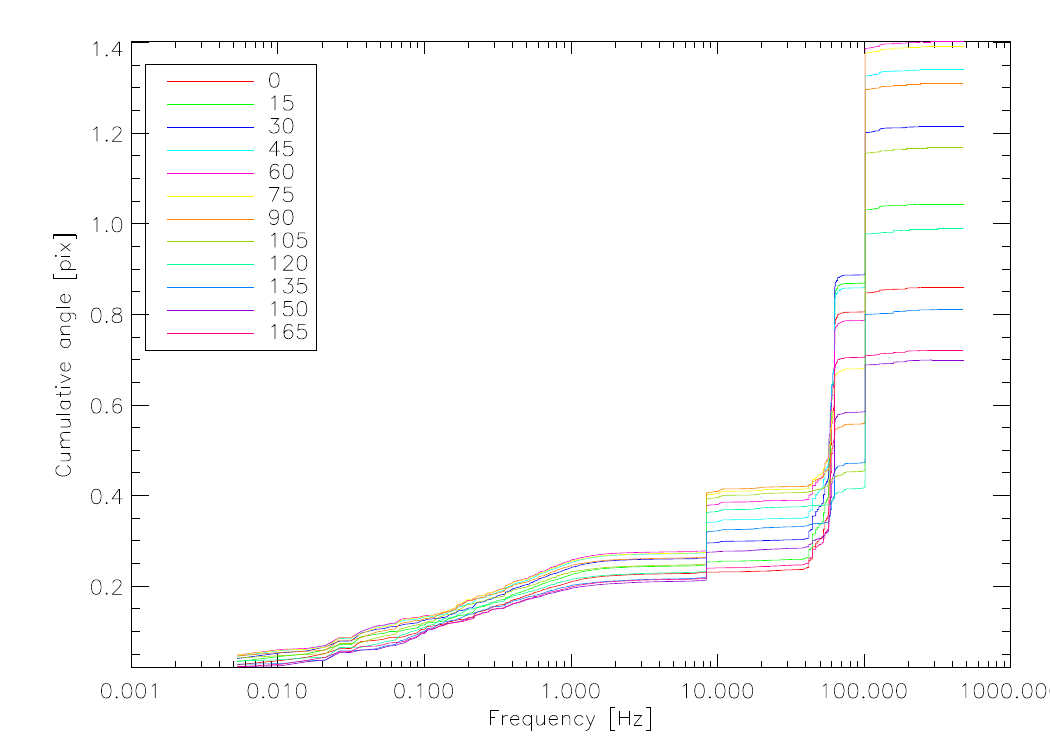}}
  \caption{Directional cumulative power spectra across rotation angles.}
  \label{fig-viscam-ic}
\end{figure}

Closing the AO188 loop reduces low-frequency power by a factor of 2. Three distinct mechanical modes emerge:
\begin{itemize}
\item 55~Hz mode: Directional peak at 30\degree (minimum at 120\degree);
\item 60~Hz mode: Line power harmonic peaking at 15\degree;
\item 100~Hz mode: Dominant structural resonance peaking at 75\degree.
\end{itemize}

Running the HiCIAO cryo-pump adds an 8~Hz horizontal mode. Deterministic 1~kHz sampling on the EMCCD avoids aliasing artifacts. Implementing LQG control on this channel could reduce tip/tilt jitter sixfold.

%-------------------------------------------------------------------------------
\subsection{Conclusions}
\label{sec-conclusions}

Daytime and nighttime engineering tests on \sce identified structural vibration sources. Non-deterministic frame timing on infrared detectors complicated spectral analysis and raised aliasing concerns. High-speed 1~kHz deterministic sampling on the visible EMCCD resolved structural modes at 55, 60, and 100~Hz, alongside an 8~Hz mode driven by the HiCIAO cryo-cooler pump.

Implementing LQG vibration control on \sce requires enforcing deterministic real-time hardware clocking, raising sensor readout rates, and matching metrology passbands between tracking sensors and science detectors.

%§§§§§§§§§§§§§§§§§§§§§§§§§§§§§§§§§§§§§§§§§§§§§§§§§§§§§§§§§§§§§§§§§§§§§§§§§§§§§§§
\chapter{Extinction Performance on \pe}
\label{sec-performance-nulling}

 \begin{flushright}
   \begin{minipage}{12cm} {\small \textit{Shadow is not the absence of light, merely the obstruction of the luminous rays by an opaque body. Shadow is of the nature of darkness. Light is of the nature of a luminous body; one conceals and the other reveals. They are always associated and inseparable from all objects. But shadow is a more powerful agent than light, for it can impede and entirely deprive bodies of their light, while light can never entirely expel shadow from a body, that is from an opaque body.}}

     \raggedleft{{\small The Notebooks of Leonardo da Vinci, III (1487-1508)}}
   \end{minipage}
 \end{flushright}

\minitoc

\bigskip

%§§§§§§§§§§§§§§§§§§§§§§§§§§§§§§§§§§§§§§§§§§§§§§§§§§§§§§§§§§§§§§§§§§§§§§§§§§§§§§§
\section{Introduction}
\label{sec-introd}

Having demonstrated the ability to stabilize the \pe testbed, we will study in this chapter the null depths obtained across different testbed configurations, under both monochromatic and polychromatic light. I will also present measurements recorded during the injection of disturbances, as well as long-term stability measurements. I will endeavor to describe as fully as possible the contributions that explain the results obtained, in order to best extrapolate these findings in Chapter~\ref{sec-extrapolation}. The results presented here will be the subject of a publication in a peer-reviewed journal \cite{Lozi12b}.

%§§§§§§§§§§§§§§§§§§§§§§§§§§§§§§§§§§§§§§§§§§§§§§§§§§§§§§§§§§§§§§§§§§§§§§§§§§§§§§§
\section{Null Depth in Monochromatic Light}
\label{sec-taux-extinction-1}

%¤¤¤¤¤¤¤¤¤¤¤¤¤¤¤¤¤¤¤¤¤¤¤¤¤¤¤¤¤¤¤¤¤¤¤¤¤¤¤¤¤¤¤¤¤¤¤¤¤¤¤¤¤¤¤¤¤¤¤¤¤¤¤¤¤¤¤¤¤¤¤¤¤¤¤¤¤¤¤
\subsection{Contributions to the Null Depth}
\label{sec-contrib-mono}

With a monochromatic source, certain terms in Equation~\eqref{eq-systeme-nulling} are simplified, notably all chromatic variation terms. The equation remains unchanged, as there is still a phase contribution, a flux imbalance contribution, and a polarization contribution, but these terms are described by:
\begin{itemize}
\item Phase contribution: the phase dispersion term $\Delta\phi_\sigma$ from Equation~\eqref{eq-N-sig} is zero, so the phase contribution is simply the contribution from the optical path difference,
  \begin{equation}
	N_\phi = N_\delta = \GP{\pi\sigma\delta}^2.
  \end{equation}
\item Photometric imbalance contribution: similarly, the flux imbalance dispersion term $\varepsilon_\sigma$ is zero, meaning that only the variation in flux imbalance contributes,
  \begin{equation}
	N_\varepsilon = \frac14\varepsilon^2.
  \end{equation}
\item Polarization contribution: for this contribution, operating in monochromatic light has no real effect. However, the source used is linearly polarized in $\mss$ or $\mpp$; thus, the $\Delta\phi_\msp$ term is zero, leaving only the rotation component in play, which yields
  \begin{equation}
	N_\mpol = N_\mrot = \frac14\alpha_\mrot^2.
  \end{equation}

  In reality, birefringence may appear in the interferometer plates if stress is too high in the assembly; this creates a modification of the incident linear polarization, so that the $\Delta\phi_\msp$ term may be non-zero, even with a linearly polarized source. The plates on \pe, being very thick and minimally constrained by their mount, should not generate significant birefringence. Reflection off a mirror also introduces a change in polarization state, represented by Jones matrices. It is to limit this effect that the MMZ has a 60\degree geometry instead of 90\degree. However, this means that even with perfectly linearly polarized light and perfectly aligned, stress-free optics, the light exiting the interferometer is not entirely linearly polarized in any case.

  Using a polarizer at the output of the interferometer would eliminate the $N_\mrot$ term by converting polarization angle issues into a photometric imbalance term. However, on the \pe testbed, I do not use a polarizer at the output of the MMZ.
\end{itemize}

%¤¤¤¤¤¤¤¤¤¤¤¤¤¤¤¤¤¤¤¤¤¤¤¤¤¤¤¤¤¤¤¤¤¤¤¤¤¤¤¤¤¤¤¤¤¤¤¤¤¤¤¤¤¤¤¤¤¤¤¤¤¤¤¤¤¤¤¤¤¤¤¤¤¤¤¤¤¤¤
\subsection{Discussion on Stellar Leakage}
\label{sec-discussion-sur}

The null depth term $N_\star$ associated with stellar leakage—i.e., with resolving the star—is, according to Equation~\eqref{eq-N-star}:
\begin{equation}
  N_\star = \frac{\pi^2}{16}(\sigma B\theta_\star)^2.
\end{equation}

On \pe, the angular diameter of the source is given by the ratio of the fiber head diameter $d$ to the focal length $f_0$ of parabola M0. The fiber head having a diameter of $4.3$~\mum (approximating the Gaussian by a top-hat) and the parabola having a focal length of 750~mm, considering baseline $B = 50$~mm and the wavenumber of the laser diode for the science measurement $\sigma = 0.43$~\imum \tir{i.e., a wavelength $\lambda = 2.32$~\mum}, we arrive at a stellar leakage term of $N_\star = 9\E{-3}$.

However, I obtained null depths far better than this, as will be seen below. In reality, this calculation is only valid for a spatially incoherent source, whereas in the case of \pe, the fiber being single-mode, the source is spatially coherent. For this type of source, there is no stellar leakage term, so we will consider
\begin{equation*}
  \boldsymbol{N_\star \simeq 0}.
\end{equation*}

%¤¤¤¤¤¤¤¤¤¤¤¤¤¤¤¤¤¤¤¤¤¤¤¤¤¤¤¤¤¤¤¤¤¤¤¤¤¤¤¤¤¤¤¤¤¤¤¤¤¤¤¤¤¤¤¤¤¤¤¤¤¤¤¤¤¤¤¤¤¤¤¤¤¤¤¤¤¤¤
\subsection{Initial Measurements in the Autocollimation Configuration}
\label{sec-premieres-mesures}

In the autocollimation configuration described in Section~\ref{sec-premiere-configuration}, it is impossible to place the periscopic achromatic phase shifter planned for the final configuration; indeed, a double pass through these periscopes creates a double phase shift of $\pi$, and it then has no net effect. The central achromatic fringe at OPD $\delta = 0$ is therefore bright. To measure a sufficiently low null depth, the source used for the science channel must necessarily be a laser source, ensuring a sufficiently long coherence length so that the envelope effect is negligible at $\delta = \pm\lambda/2$, the position of the dark fringe.

The science source used is a 2.32~\mum laser diode with a spectral width of less than 0.5~nm. Thus, the coherence length is greater than 10~mm, and the first dark fringe has a minimum null depth of $6\E{-8}$. Since the position of the dark fringe is not automatically determined in this configuration, it sometimes happens that we do not align with the first dark fringe, but rather with one of the subsequent fringes. Figure~\ref{fig-null-min-autocol} presents the minimum null depth for the first 10 dark fringes.

\begin{figure} \centering
  \FIG{.7}{false}{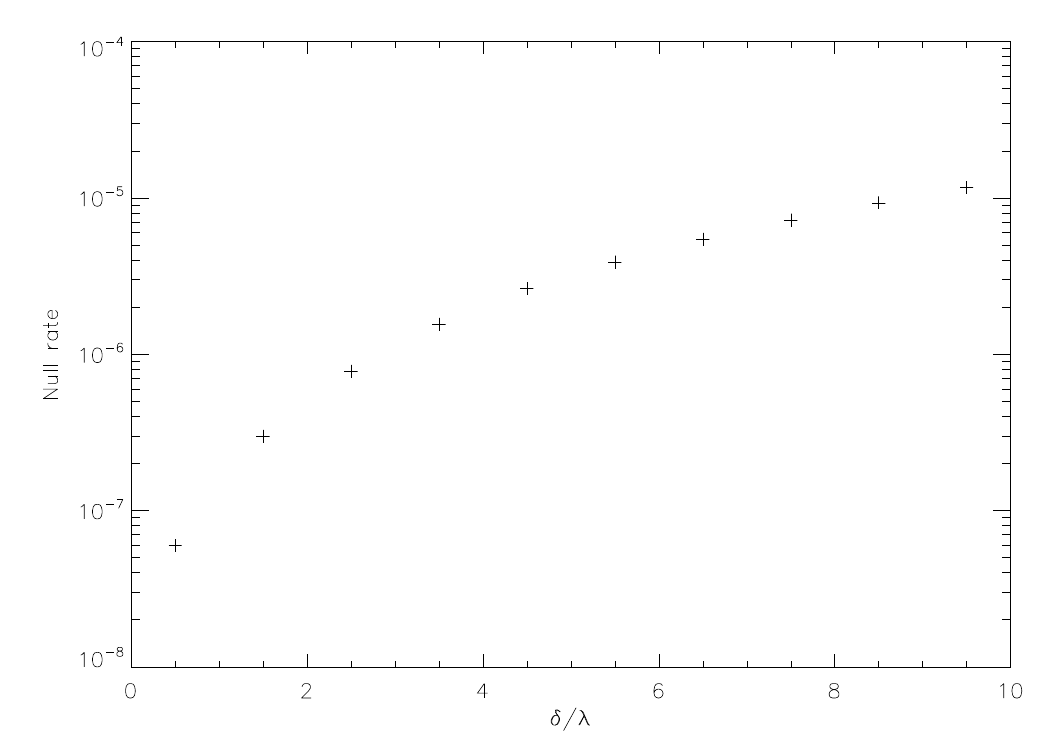}
  \caption{Minimum null depth due to the source spectrum envelope effect.}
  \label{fig-null-min-autocol}
\end{figure}

As seen in this figure, one must reach the 10th dark fringe to exceed an envelope contribution of $\e{-5}$. We therefore have a negligible envelope contribution to the null depth if we align within the first few dark fringes.

In the autocollimation configuration, since the central achromatic fringe is bright, the position of the dark fringe is difficult to estimate from the fringe sensor data. The central fringe being bright at output III of the MMZ, it is dark at output II. However, with the chromatic phase shift induced by the plates, the fringe minima in bands $\I$ and $\J$ are not located at the same OPDs. We arbitrarily choose the minimum of fringe $\J$ at output II as the OPD reference, but we must then manually adjust the reference position of the piston loop on the GUI to obtain the desired signal (dark or bright fringe).

Since the science camera was not yet usable for the autocollimation stage presented here, the science measurement was performed using a single-pixel detector measuring the flux at output III guided by a single-mode fiber. Background noise was filtered using synchronous detection, measuring the flux at a given frequency determined by a chopper placed at the $\K$ source injection. Having only a single science detection channel, it was impossible to measure the bright and dark fringes simultaneously. We therefore measured these two signals sequentially, adding an OPD of $\lambda/2$.

Figure~\ref{fig-null-autocol} presents a measurement of the bright and dark fringes, in one case with the piston feedback loop open (Figure~\ref{fig-noir-ns}), and in the other with the loop closed (Figure~\ref{fig-noir}).

\begin{figure} \centering
  \subfloat[Dark fringe signal with cophasing in open loop.]{
    \label{fig-noir-ns}\FIG{0.47}{false}{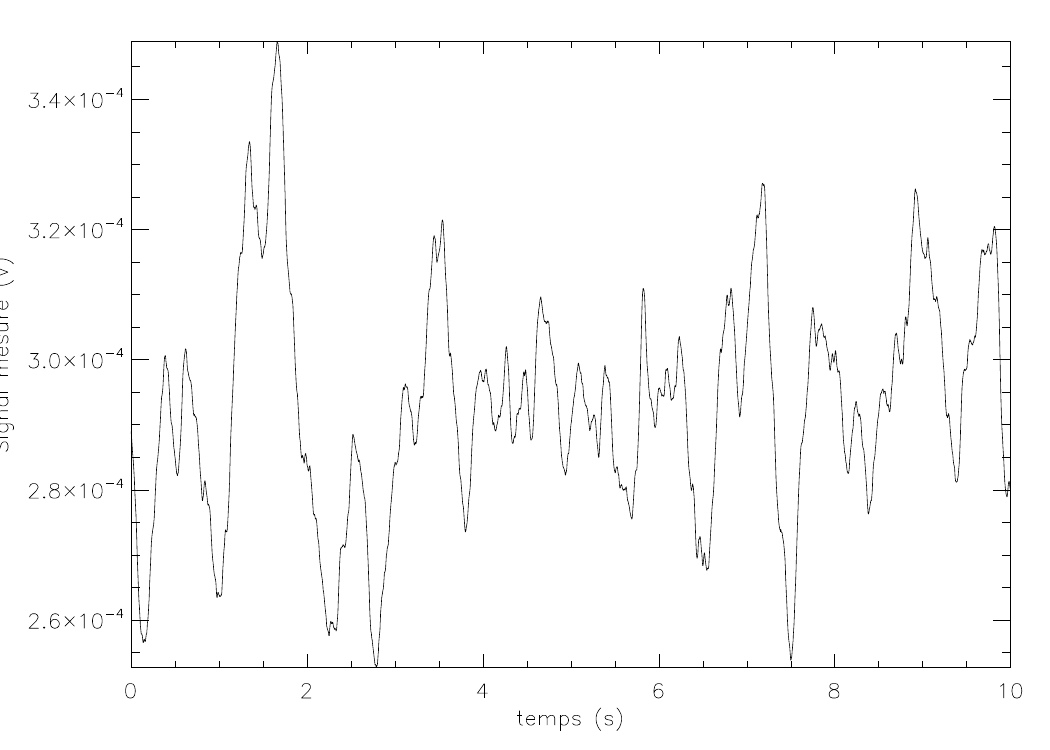}}
  \hfill\subfloat[Dark fringe signal with cophasing in closed loop.]{
    \label{fig-noir}\FIG{0.47}{false}{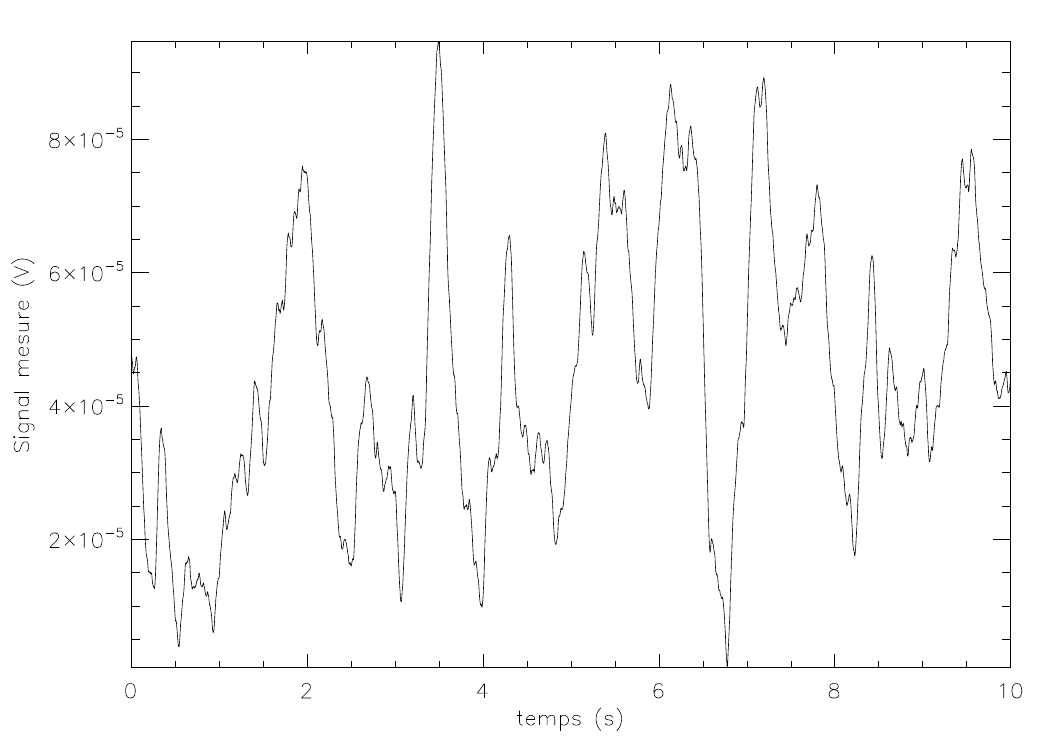}}\\
  \subfloat[Bright fringe signal.]{\label{fig-blanc}
    \FIG{0.47}{false}{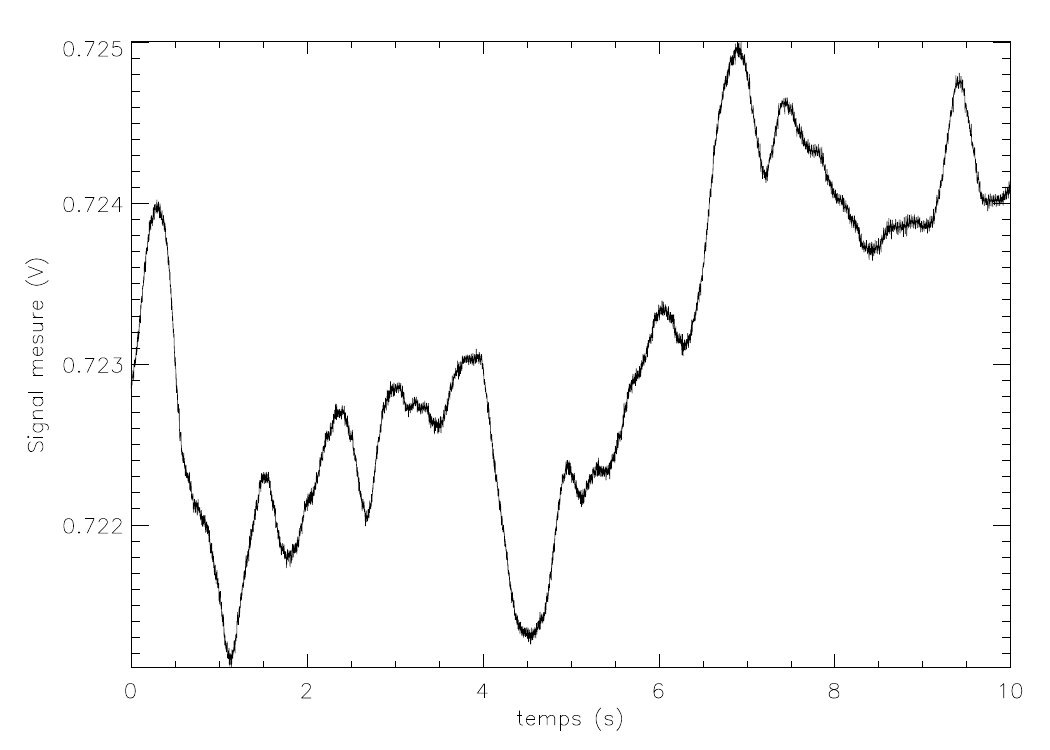}}
  \caption{Signals obtained during null depth measurements.}
  \label{fig-null-autocol}
\end{figure}

From these data, I obtained a null depth $N = 4\E{-4} \pm 3\E{-5}$ in open loop, and $N = 6\E{-5} \pm 3\E{-5}$ in closed loop.

The mean null depth is thus improved by a factor of 7, and yet the stability did not change. In both cases, the measured residual OPD is $\sigma_\delta = 0.7$~nm RMS, which gives an OPD contribution to the null depth of $9\E{-7}$; it is therefore not this residual that limits the null depth.

The observed variations are certainly due to temperature variations of the optics, which can severely degrade the null depth, as will be shown in the next section.

These measurements demonstrated first and foremost that with the \pe testbed, it is possible to achieve null depths below $\e{-4}$, even if for the moment this result was obtained using a monochromatic source.

%¤¤¤¤¤¤¤¤¤¤¤¤¤¤¤¤¤¤¤¤¤¤¤¤¤¤¤¤¤¤¤¤¤¤¤¤¤¤¤¤¤¤¤¤¤¤¤¤¤¤¤¤¤¤¤¤¤¤¤¤¤¤¤¤¤¤¤¤¤¤¤¤¤¤¤¤¤¤¤
\subsection{Thermal Stability}
\label{sec-stabilite-thermique}

In the quasi-ABCD algorithm presented in Section~\ref{sec-desc-abcd}, points A and C are the most sensitive to OPD variations because they are located roughly at the inflection points where the tangent is steepest, whereas points B and D, at the interference extrema, are insensitive to these variations. Thus, in \pe's MMZ, outputs I and IV contribute the most to the OPD measurement, whereas the science output is output III. A disturbance in the MMZ can therefore impact output III without affecting outputs I and IV, and vice versa. There are thus differential paths between the cophasing and the science channel that undergo drifts due to temperature variations in the MMZ.

To quantify these drifts, I measured the temperature of the MMZ at various points using sensors placed on the mounts, with a measurement accuracy of 10~mK. These sensors placed in the MMZ can be seen in Figure~\ref{fig-mmz-sondes}.

\begin{figure} \centering
  \FIG{.7}{false}{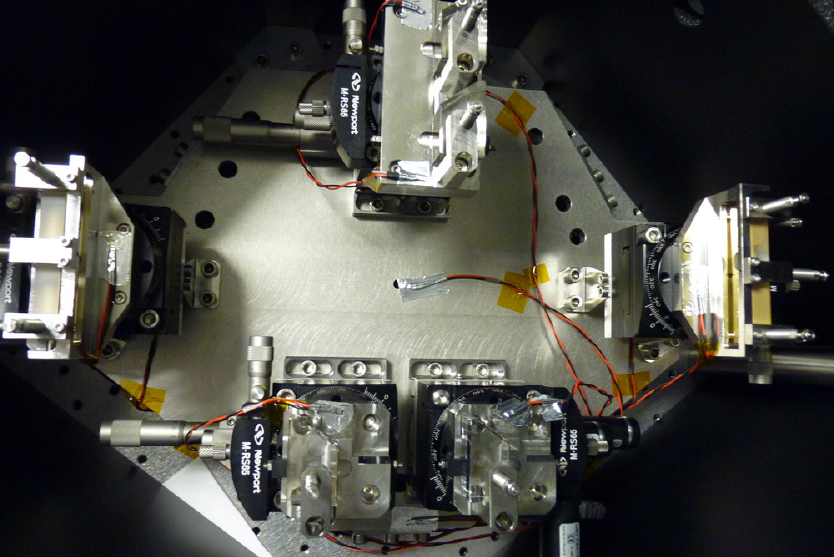}
  \caption[Temperature sensors placed in the MMZ.]{Temperature sensors placed in the MMZ (at the ends of the black and red cables).}
  \label{fig-mmz-sondes}
\end{figure}

I also measured the phase shift between outputs III and IV in the $\J$ band, which corresponds to a shift of the fringe packets, and is caused by the translation of plate L3$\ma$. I performed this measurement in the autocollimation configuration over three consecutive days, turning off the cleanroom air conditioning to allow the temperature to drift. The measurements are presented in Figure~\ref{fig-ddm-temp}.

\begin{figure} \centering
  \FIG{.7}{false}{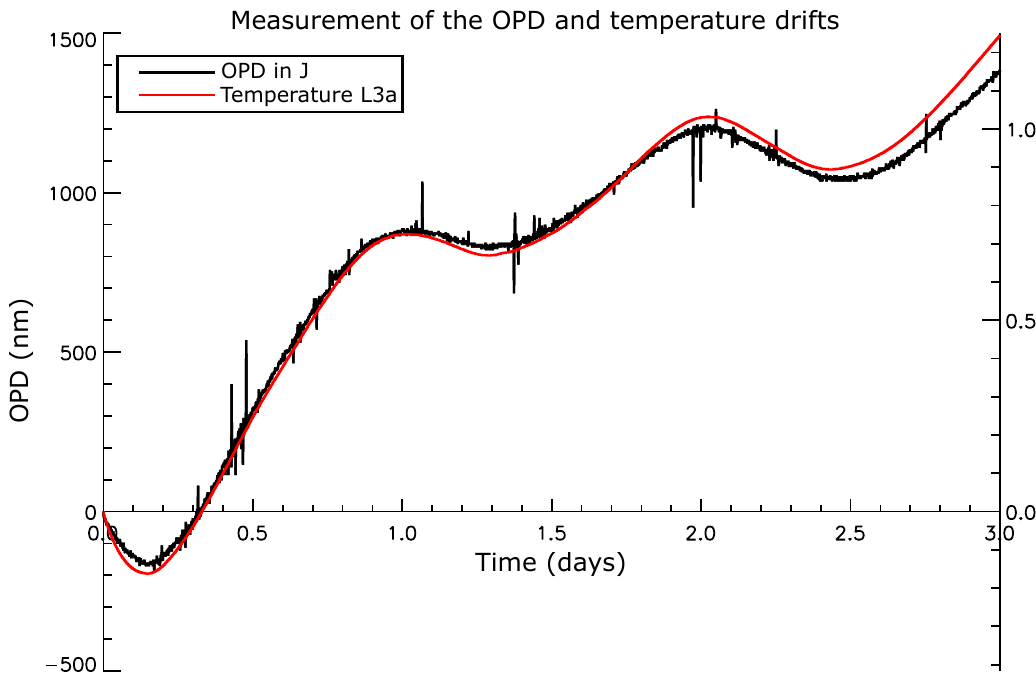}
  \caption{Correlation between OPD drift and temperature variation.}
  \label{fig-ddm-temp}
\end{figure}

The relationship between temperature variations and differential path drift is clearly established in this figure, with a coefficient of \textbf{1200~nm.K}$\boldsymbol{^{-1}}$. The temperature sensors are taped to the plate mounts, made of stainless steel, whose thermal expansion coefficient ($16.8\E{-6}$~K$^{-1}$) is almost identical to that of the CaF$_2$ plates ($16.7\E{-6}$~K$^{-1}$); they also have an equivalent thermal diffusivity (ability to transmit heat variations) at $4.2\E{-6}$~m$^2$.s$^{-1}$. The mount temperature thus gives a good indication of the plate temperature. On this curve, diurnal temperature variations are visible, with local maxima in the middle of the day.

With such a thermal drift coefficient, to achieve a null depth stability of $\e{-5}$ over 100~s as originally specified, OPD variations must remain below 1.75~nm at the lowest science wavelength, which requires a temperature drift of less than 1.5~mK over 100~s. This condition is quite stringent, especially in the autocollimation configuration; indeed, only the MMZ was enclosed, with a thin 1~cm foam layer that only partially dampens thermal variations. It is highly likely that this factor causes the variations in the null depth measured in Figure~\ref{fig-null-autocol}.

%¤¤¤¤¤¤¤¤¤¤¤¤¤¤¤¤¤¤¤¤¤¤¤¤¤¤¤¤¤¤¤¤¤¤¤¤¤¤¤¤¤¤¤¤¤¤¤¤¤¤¤¤¤¤¤¤¤¤¤¤¤¤¤¤¤¤¤¤¤¤¤¤¤¤¤¤¤¤¤
\subsection{Measurements with the Science Camera}
\label{sec-mesures-avec}

In the final configuration described in Section~\ref{sec-conf-finale}, I repeated the null depth measurement with the monochromatic source. Unlike the previous setup, only a single pass is made through the MMZ and through the other testbed optics, allowing the use of the geometric achromatic phase shifter. The dark fringe is therefore at zero OPD in channels $\I$, $\J$, and $\K$, excluding the chromatic effects of the plates; there is thus no longer a fringe envelope contribution as in the previous case.

In this setup, the single pass through the MMZ reduces differential path aberrations by a factor of 2; thus, the thermal drift coefficient is reduced to 600~nm.K$^{-1}$, meaning we can tolerate temperature variations of 3~mK over 100~s. In the meantime, we also installed the enclosure described in Section~\ref{sec-elements-pert}, which reduces thermal variations, as can be seen in Figure~\ref{fig-variations-temp}.

\begin{figure} \centering
  \FIG{.7}{false}{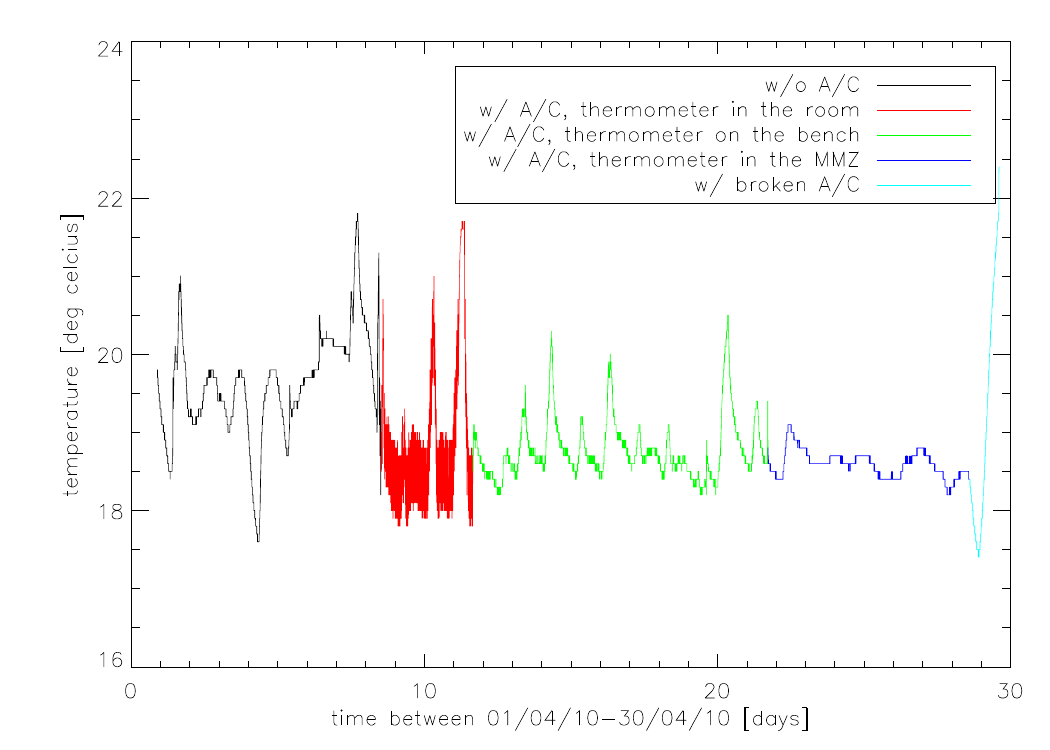}
  \caption{Temperature variations under different conditions.}
  \label{fig-variations-temp}
\end{figure}

At the time of the measurement, the air conditioning was not optimized at all, as seen in the red section of the curve. However, what is noteworthy in this plot is that the amplitude of temperature fluctuations is attenuated first by the testbed enclosure, and further by the MMZ enclosure.

In this configuration, I measure the null depth using the science camera instead of the single-pixel detector, but with the same methodology as before.

The result of the null depth measurement is presented in Figure~\ref{fig-null-mono}.

\begin{figure} \centering
  \FIG{.7}{false}{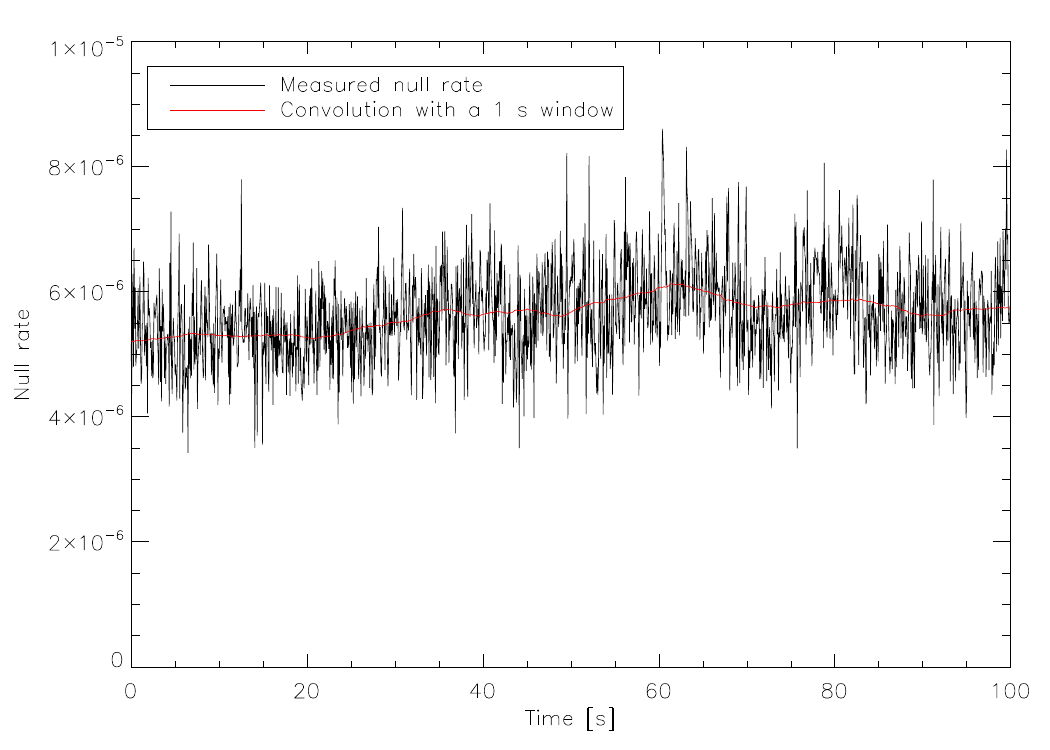}
  \caption{Monochromatic null depth measured with the science camera.}
  \label{fig-null-mono}
\end{figure}

For this acquisition, the mean value of the null depth is
\begin{equation*}
 \boldsymbol{\moy[\Tc]{N_\mmoy} = 5.6\E{-6}},
\end{equation*}
a value far superior to the testbed requirement of $\e{-4}$. This result is very promising for the transition to polychromatic light.

Since the null depth measurement is acquired simultaneously with the OPD measurement, I can simulate the null depth that would be obtained if OPD alone degraded it. This simulation is presented in Figure~\ref{fig-null-ddm-mono}.

\begin{figure} \centering
  \FIG{.7}{false}{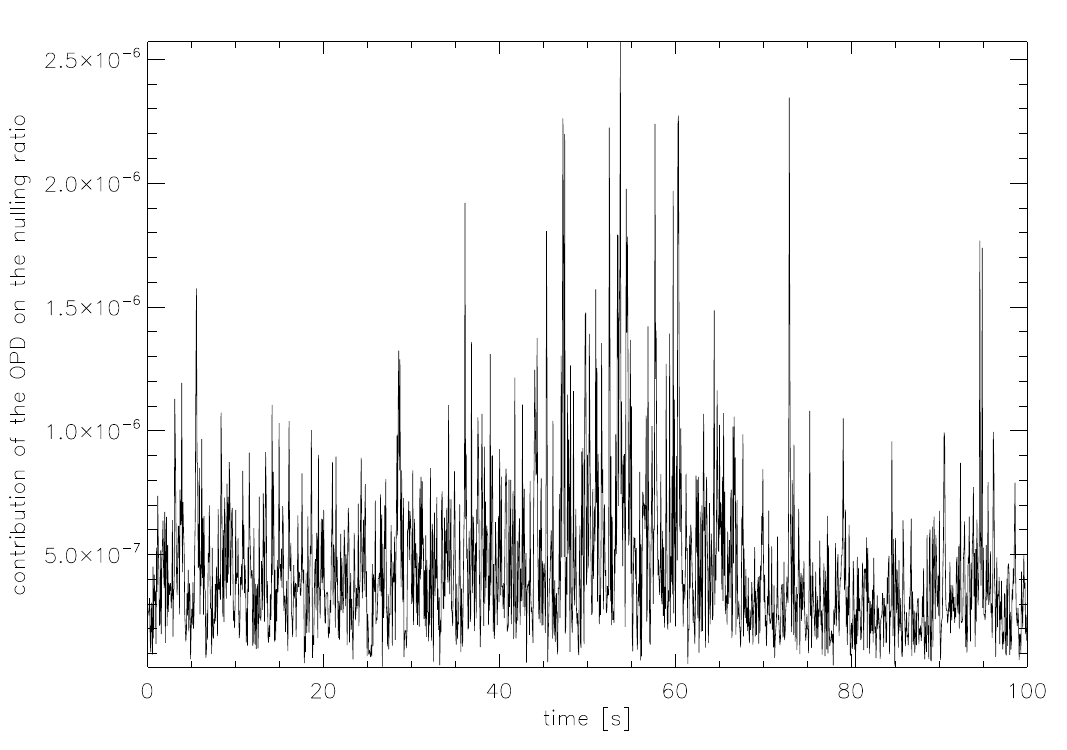}
  \caption{Simulation of the OPD contribution to the null depth.}
  \label{fig-null-ddm-mono}
\end{figure}

The mean value of this contribution is $3.3\E{-7}$, which is far below the measured null depth. The fluctuation amplitude is of the same order of magnitude, but I do not observe matching variations between the two signals. Other terms must therefore come into play in the measured null depth, both in mean value and in stability.

To analyze the null depth contributors more finely and optimize performance, a calibration procedure for the entire science channel was implemented, specifically for polychromatic nulling measurements.

%§§§§§§§§§§§§§§§§§§§§§§§§§§§§§§§§§§§§§§§§§§§§§§§§§§§§§§§§§§§§§§§§§§§§§§§§§§§§§§§
\section{Science Channel Calibration Procedure}
\label{sec-proc-camera}

%¤¤¤¤¤¤¤¤¤¤¤¤¤¤¤¤¤¤¤¤¤¤¤¤¤¤¤¤¤¤¤¤¤¤¤¤¤¤¤¤¤¤¤¤¤¤¤¤¤¤¤¤¤¤¤¤¤¤¤¤¤¤¤¤¤¤¤¤¤¤¤¤¤¤¤¤¤¤¤
\subsection{Description of the Science Channel} 
\label{sec-description-voie-scientifique}

In \pe's final configuration, we use a spectrometer to analyze the null depth across several wavelengths, verifying its uniformity across the spectrum.

The dark fringe output corresponds to output III of the MMZ. A dichroic plate separates the metrology $\I$ and $\J$ bands from the science band, transmitting wavelengths longer than 1.65~\mum. The beam is coupled into a single-mode fiber, which filters all beam aberrations except piston. This flux, along with a photometric reference flux, emerges onto an optical bench hosting the spectrometer. This setup is shown in Figure~\ref{fig-spectro}.

\begin{figure} \centering
  \subfloat[General view of the spectrometer.]{\label{fig-spectro1}
    \FIGH{6.5}{false}{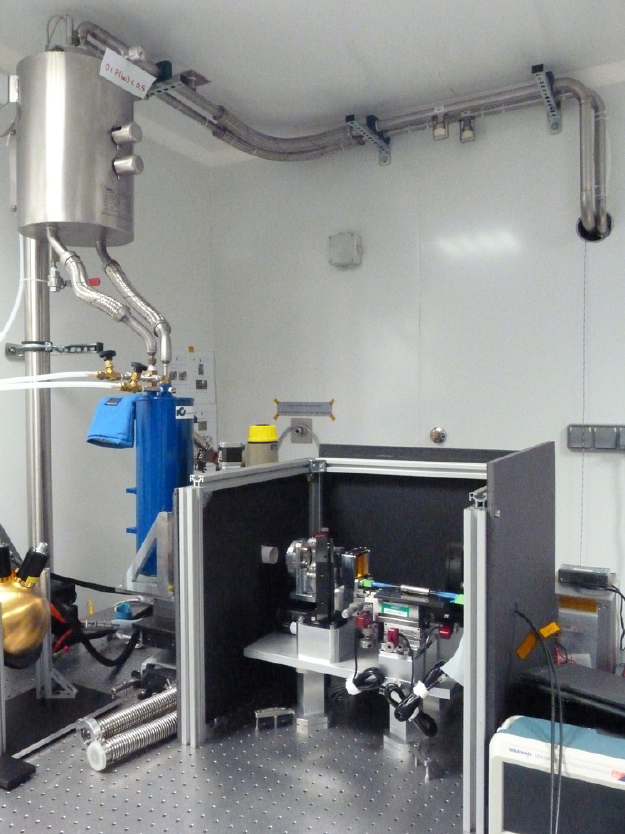}}
  \hfill\subfloat[Detail of the optical bench.]{\label{fig-spectro2}
    \FIGH{6.5}{false}{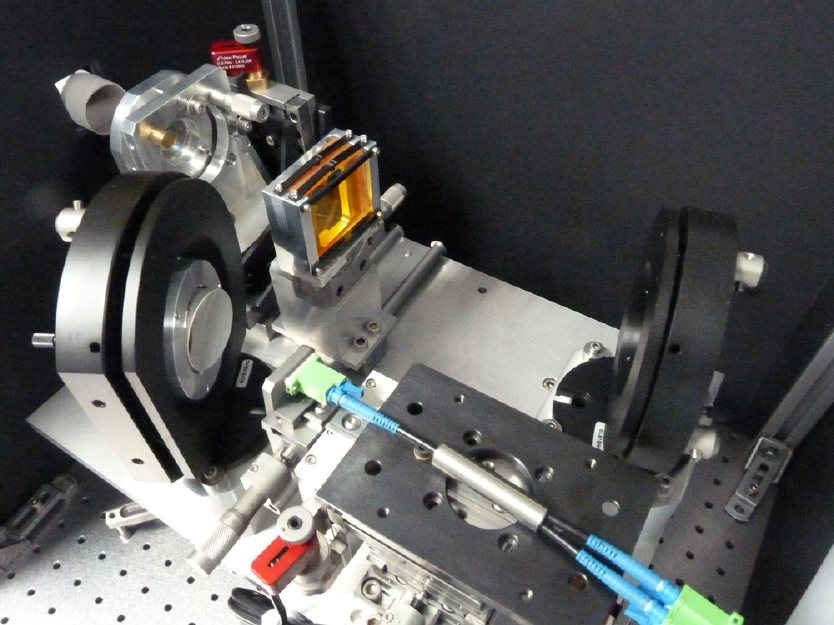}}
  \caption{Science channel spectrometer for null depth measurement.}
  \label{fig-spectro}
\end{figure}

The cores of the two optical fibers are positioned as close together as possible \tir{125~\mum apart} using a Y-connector, then placed at the focus of an off-axis parabolic mirror. The collimated beam is folded by a flat mirror and passes through a ZnSe and silica bi-prism that disperses the light according to wavelength while keeping the central beam axis on-axis. This dispersed light passes through a ZnSe/silica achromatic doublet and is focused onto the camera detector. The camera is a liquid-nitrogen-cooled $256 \times 256$ pixel PICNIC array operating at 77~K, with a read noise of 25 electrons. The two beams are dispersed across two lines of 9 pixels, yielding $n=9$ spectral channels measured over 2 pixels. The dark fringe and reference signals are separated by roughly ten pixels to prevent cross-talk.

Sub-frame images of $20 \times 20$ pixels are read out to maximize speed. Each frame has an integration time $\tau = 2.58$~ms and is preceded by a reset frame of identical duration after clearing pixel charges. Frame subtraction yields net integrated flux. Readouts are clock-synchronized with the cophasing system. The maximum frame rate is 120~Hz, but for simplicity, the camera rate is set to a sub-multiple of the fringe sensor loop rate: $f_\mcam = f_\mfs/10 = 31250/320$~Hz $\approx 100$~Hz.

Thus, I measure the null depth across several spectral channels, which requires defining a mean null depth to compare \pe's performance with state-of-the-art results presented in Section~\ref{sec--etat}.

%¤¤¤¤¤¤¤¤¤¤¤¤¤¤¤¤¤¤¤¤¤¤¤¤¤¤¤¤¤¤¤¤¤¤¤¤¤¤¤¤¤¤¤¤¤¤¤¤¤¤¤¤¤¤¤¤¤¤¤¤¤¤¤¤¤¤¤¤¤¤¤¤¤¤¤¤¤¤¤
\subsection{Discussion on Null Depth Calculation}
\label{sec-discussion-sur-1}

In the remainder of this chapter, I will present the best null depth results obtained on the testbed, detailing the methodology used to analyze these data.

First, it is necessary to clearly define the null depth to be calculated. For $n$ spectral channels, two values can be considered:
\begin{itemize}
\item By integrating over all camera channels, the measurement simulates a single-pixel detector, matching other nulling testbeds. In this case, the integrated null depth over the full band, $N_\mint$, is valid for a fractional bandwidth $(\Delta\sigma/\sigma)_\mint = 23\%$. This null is the flux-weighted sum ($F_i/F_\mtot$) of individual channel null depths $N_i$:
  \begin{equation}
    N_\mint = \sum_{i=1}^n\dfrac{F_i}{F_\mtot}N_i.
  \end{equation}
  
\item By averaging individual channel null depths, we obtain a mean null depth $N_\mmoy$ that is less dependent on the global spectral shape of the source through the instrument. This value applies to a fractional bandwidth $(\Delta\sigma/\sigma)_\mmoy = 37\%$:
  \begin{equation}
    N_\mmoy = \dfrac{1}{n}\sum_{i=1}^nN_i=\moy[i]{N_i}.
  \end{equation}
\end{itemize}

\bigskip

This second calculation generally yields a higher null depth value than the first; chromatic effects degrade performance near passband edges where relative flux drops. However, we favor the second metric because it corresponds to a wider bandwidth and is less sensitive to source spectral profiles.

\medskip

Having defined the mean null depth, we can analyze its characteristics from the individual channel null depths $N_i$.

Holding all other terms in Equation~\eqref{eq-systeme-nulling} constant, channel null depth $N_i$ varies with OPD $\delta$ according to:
\begin{equation}
  N_i(\delta)=N_{i,0}+\GP{\pi\sigma_i\GP{\delta-\delta_i}}^2,
\end{equation}
where $N_{i,0}$ is the minimum null depth in channel $i$, $\sigma_i$ is its mean wavenumber, and $\delta_i$ is the OPD that minimizes its null depth.

Thus, the mean null depth is given by:
\begin{align}
  N_\mmoy(\delta) & =
  \moy[i]{N_{i,0}}+\moy[i]{\GP{\pi\sigma_i\GP{\delta-\delta_i}}^2} \nonumber\\
  & = \moy[i]{N_{i,0}}+\moy[i]{\GP{\pi\sigma_i\delta_i}^2}-
  2\moy[i]{\GP{\pi\sigma_i}^2\delta_i}\delta+\moy[i]{\GP{\pi\sigma_i}^2}\delta^2.
  \label{eq-N_moy}
\end{align}

Writing $N(\delta) = \alpha+\beta\delta+\gamma\delta^2 = a+b(\delta-\delta_0)^2$, parameters satisfy:
\begin{equation}
  \left\{
    \begin{array}{r@{\hs}l}
      a &= \alpha-\dfrac{\beta^2}{4\gamma}\\
      b &= \gamma\\
      \delta_0 &= -\dfrac{\beta}{2\gamma}\\
    \end{array}
  \right..
\end{equation}

Equation~\eqref{eq-N_moy} rewrites as:
\begin{equation}
  N_\mmoy(\delta) =
  N_{\mmoy,0}+\GP{\pi\sigma_\mmoy\GP{\delta-\delta_\mmoy}}^2,
  \label{eq-N_moy2}
\end{equation}
with
\begin{equation}
  \left\{
    \begin{array}{r@{\hs}l}
      N_{\mmoy,0} &=
      \moy[i]{N_{i,0}}+\moy[i]{(\pi\sigma_i\delta_i)^2}
      -\dfrac{4\moy[i]{(\pi\sigma_i)^2\delta_i}^2}
      {4\moy[i]{(\pi\sigma_i)^2}}\\[+6pt]
      (\pi\sigma_\mmoy)^2 &= \moy[i]{(\pi\sigma_i)^2}\\[+6pt]
      \delta_\mmoy &=
      \dfrac{2\moy[i]{(\pi\sigma_i)^2\delta_i}}{2\moy[i]{(\pi\sigma_i)^2}}\\
    \end{array}
  \right..
\end{equation}

Which yields:
\begin{equation}
  \left\{
    \begin{array}{r@{\hs}l}
      N_{\mmoy,0} &= \moy[i]{N_{i,0}}+\pi^2\moy[i]{\GP{\sigma_i\delta_i}^2}-
      \GP{\pi\sigma_\mmoy\delta_\mmoy}^2\\[+6pt]
      \delta_\mmoy &= \frac{\displaystyle \moy[i]{\sigma_i^2\delta_i}} 
      {\displaystyle \sigma_\mmoy^2}\\
    \end{array}
  \right.,
  \label{eq-param_N_moy}
\end{equation}
where
\begin{equation}
  \sigma_\mmoy^2 = \moy[i]{\sigma_i^2}.
\end{equation}

Wavenumber $\sigma_\mmoy$ is the root-mean-square average of channel wavenumbers. We set the origin of the OPD scale such that $\delta_\mmoy=0$. Thus:
\begin{equation}
  N_\mmoy(\delta) = N_{\mmoy,0}+\GP{\pi\sigma_\mmoy\delta}^2,
  \qquad \text{with} \qquad \left\{
\begin{array}{r@{\hs}l}
  N_{\mmoy,0} &= \moy[i]{N_{i,0}}+\pi^2\moy[i]{(\sigma_i\delta_i)^2}\\[+6pt]
  \sigma_\mmoy &= \sqrt{\moy[i]{\sigma_i^2}}\\[+6pt]
  \moy[i]{\sigma_i^2\delta_i} &= 0\\
\end{array}
\right..
\end{equation}

The minimum achievable null depth is given by the average of channel minima degraded by chromatic dispersion $\pi^2\moy[i]{(\sigma_i\delta_i)^2}$.

Equation~\eqref{eq-param_N_moy} defines optimal OPD offset $\delta_\mmoy$ as the weighted average of channel offsets $\delta_i$ (weighted by $\sigma_i^2$). Minimizing overall null depth requires adjusting calibration offsets so $\delta_\mref = \delta_\mmoy = 0$. The calibration routine implementing this optimization is described in Section~\ref{sec-optimisation-position}. Despite optimization, thermal drift introduces residual offsets $\delta_\mref \neq 0$.

%¤¤¤¤¤¤¤¤¤¤¤¤¤¤¤¤¤¤¤¤¤¤¤¤¤¤¤¤¤¤¤¤¤¤¤¤¤¤¤¤¤¤¤¤¤¤¤¤¤¤¤¤¤¤¤¤¤¤¤¤¤¤¤¤¤¤¤¤¤¤¤¤¤¤¤¤¤¤¤
\subsection{Calibration Objectives} 
\label{sec-objectif-creation}

The science camera records destructive signal $I_\mmin$ (MMZ output III) and reference signal $I_\mref$.

Null depth $N$ is defined in Equation~\eqref{eq-taux-extinction} as:
\begin{equation}
  N = \frac{I_\mmin}{I_\mmax}.
\end{equation}

Signal $I_\mmin$ is measured directly on the camera. Evaluating $I_\mmax$ requires calibration. Under stable illumination, constructive peak flux $I_\mmax$ can be measured by introducing a $\lambda/2$ phase shift with neutral density filters inserted to prevent camera saturation. This method was used for monochromatic runs.

In broadband light, $I_\mmax$ can be defined either as the bright fringe adjacent to the dark fringe ($\lambda/2$ offset) or as the theoretical zero-OPD constructive peak without the $\pi$ phase shift. Because the former definition includes spectral coherence envelope attenuation, we adopt the latter. $I_\mmax$ is defined as the zero-OPD constructive peak intensity.

Because supercontinuum coupling fluctuates (Section~\ref{sec-modification-module}), $I_\mmax$ must be scaled dynamically using reference signal $I_\mref$ (Section~\ref{sec-modification-voie}). Reference signal $I_\mref$ is proportional to source intensity and $I_\mmax$. Calibration determines this scaling factor.

Neutral density filters extend the dynamic range of the camera beyond its native $10^4$ limit. Calibrated filters are inserted during $I_\mmax$ measurement and removed during dark fringe acquisition.

%¤¤¤¤¤¤¤¤¤¤¤¤¤¤¤¤¤¤¤¤¤¤¤¤¤¤¤¤¤¤¤¤¤¤¤¤¤¤¤¤¤¤¤¤¤¤¤¤¤¤¤¤¤¤¤¤¤¤¤¤¤¤¤¤¤¤¤¤¤¤¤¤¤¤¤¤¤¤¤
\subsection{Neutral Density Filter Calibration}
\label{sec-mesure-densites}

Three custom neutral density filters (nominal optical densities $D = 3, 4, 5$ across 1.65–2.45~\mum) were measured in-house.

Transmissions were calibrated by measuring intensity ratios across combinations ($D=3$ vs. $4$, $D=4$ vs. $5$, and $D=3+4$ vs. $5$) within the camera's linear range.

Adjusting source power yielded normalized intensity ratios $I'_{j,k} = I_{j,k}/I_{j,k,\mref}$ across filter setups $j$ and pick-off positions $k$:
\begin{equation}
  \left\{
    \begin{array}{r@{\hs}c@{\hs}l}
      \rho_1 =& \dfrac{T_5}{T_3} &= \dfrac{I'_{5,1}}{I'_{3,1}} \\[+6pt]
      \rho_2 =& \dfrac{T_5}{T_4} &= \dfrac{I'_{5,1}}{I'_{4,1}} \\[+6pt]
      \rho_3 =& \dfrac{T_3T_4}{T_5} &= \dfrac{I'_{3+4,2}}{I'_{5,2}}\\
    \end{array}
  \right..
\end{equation}

Transmissions $T_j$ were calculated via:
\begin{equation}
\left\{
    \begin{array}{r@{\hs}l}
      T_3 &= \rho_2\rho_3 \\[+6pt]
      T_4 &= \rho_1\rho_3 \\[+6pt]
      T_5 &= \rho_1\rho_2\rho_3 \\
    \end{array}
  \right..
\end{equation}

Spectral optical densities $D_j = -\log T_j$ are plotted in Figure~\ref{fig-densite}.

\begin{figure} \centering
  \FIG{.7}{false}{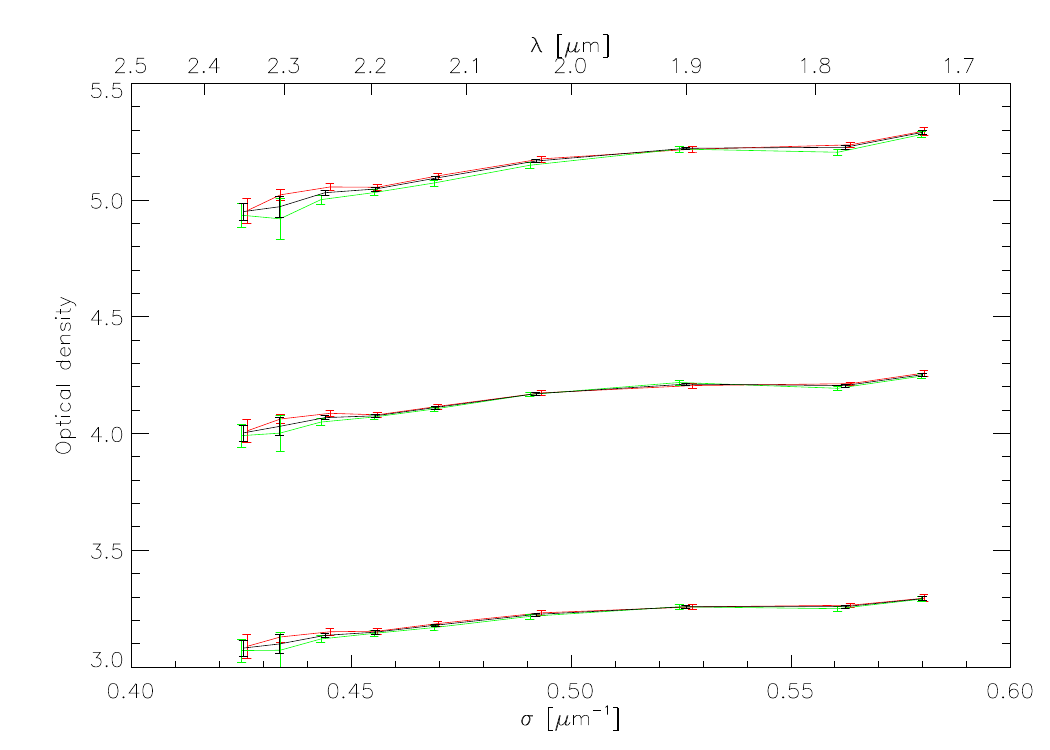}
  \caption{Calibrated optical density spectra for the three filters.}
  \label{fig-densite}
\end{figure}

The spectra display flat profiles with minor slopes toward shorter wavelengths.

Filter $D=4$ was selected for subsequent nulling calibration sequences.

%¤¤¤¤¤¤¤¤¤¤¤¤¤¤¤¤¤¤¤¤¤¤¤¤¤¤¤¤¤¤¤¤¤¤¤¤¤¤¤¤¤¤¤¤¤¤¤¤¤¤¤¤¤¤¤¤¤¤¤¤¤¤¤¤¤¤¤¤¤¤¤¤¤¤¤¤¤¤¤
\subsection{Science Camera Calibration}
\label{sec-etalonnage-camera}

Science channel calibration uses data recorded on the camera during fringe sensor sweeps (Section~\ref{sec-etalonnage-FS}). Figure~\ref{fig-data-k} shows sample calibration data.

\begin{figure} \centering
  \FIG{.7}{false}{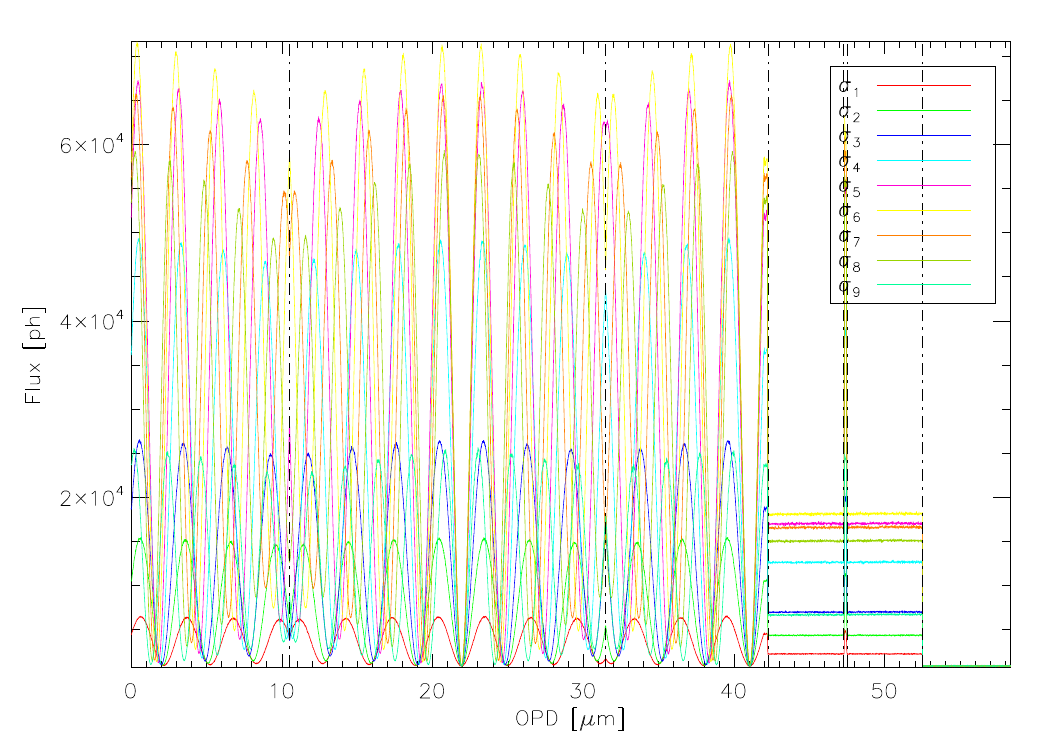}
  \caption{Fringe signals recorded on the science camera during calibration sweeps.}
  \label{fig-data-k}
\end{figure}

The calibration sequence records fringe sweeps, individual arm fluxes, and dark current. These data yield normalization factors for $I_\mmax$ scaling.

%-------------------------------------------------------------------------------
\subsubsection{Source Intensity Fluctuation Correction}
\label{sec-correction-variations}

Convective thermal air currents at the source coupling stage introduce low-frequency drift and high-frequency jitter (40~Hz mechanical mode, 50~Hz line noise, quadratic noise floor; Section~\ref{sec-modification-module}).

Slight readout timing offsets between $I_\mmin$ and $I_\mref$ prevent direct point-by-point division. Reference signals are smoothed using a moving average window to extract low-frequency drift $\tilde{I}_\mref$ (Figure~\ref{fig-fluct-source}).

\begin{figure} \centering
  \FIG{.7}{false}{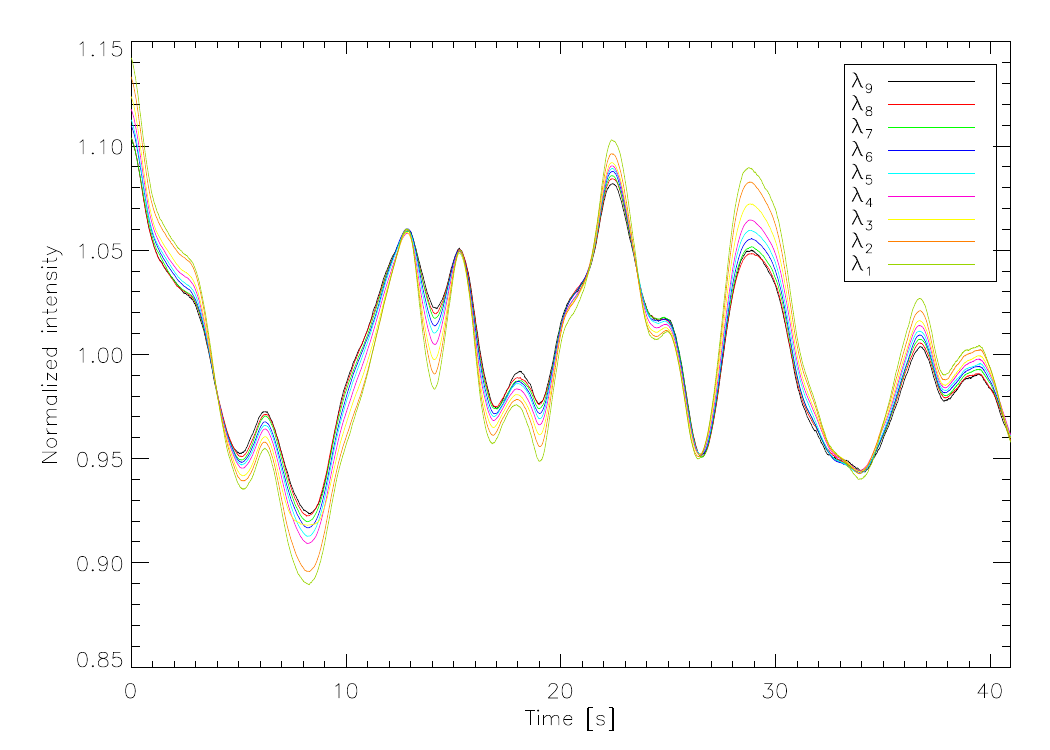}
  \caption[Source intensity fluctuations.]{Normalized source intensity drift across the 9 science channels ($\lambda_1 \approx 1.7$~\mum to $\lambda_9 \approx 2.4$~\mum).}
  \label{fig-fluct-source}
\end{figure}

Drift amplitude increases at shorter wavelengths (Figure~\ref{fig-fluct-source}).

Defocussing fiber coupling reduces drift amplitudes tenfold during operational runs.

%-------------------------------------------------------------------------------
\subsubsection{Null Depth Scaling Factor}
\label{sec-etalonnage-calcul}

Two-beam interference without an achromatic phase shift follows:
\begin{equation}
  I = I_\ma+I_\mb+2\sqrt{I_\ma I_\mb}\mu|\gamma(\delta)|\cos\GP{2\pi\sigma\delta},
\end{equation}
where $\gamma$ is temporal coherence ($|\gamma(0)|=1$) and spatial coherence $\mu \approx 1$. Peak constructive flux $I_\mmax$ is:
\begin{equation}
  I_\mmax = I_\ma+I_\mb+2\sqrt{I_\ma I_\mb}.
\end{equation}

Because the $\pi$ phase shifter cannot be removed during testing, $I_\mmax$ is calculated from single-arm intensities $I_\ma$ and $I_\mb$ measured through neutral density filter $T_j$. A calibration factor $M$ is defined per channel $i$:
\begin{equation}
  M = \frac{T_j\tilde{I}_\mref}{I_\ma+I_\mb+2\sqrt{I_\ma I_\mb}}.
\end{equation}

During nulling runs, channel null depths are computed via:
\begin{equation}
  N = \frac{I_\mmin}{I_\mmax} = \frac{I_\mmin}{I_\ma+I_\mb+2\sqrt{I_\ma I_\mb}}
  = \frac{M}{T_j}\frac{I_\mmin}{\tilde{I}_\mref}.
\end{equation}

%-------------------------------------------------------------------------------
\subsubsection{Photometric Imbalance Evaluation}
\label{sec-calcul-desequilibre}

Single-arm fluxes $I_\ma$ and $I_\mb$ yield relative photometric imbalance $\varepsilon$ (Section~\ref{sec-deseq-phot}):
\begin{equation}
  \varepsilon = \frac{\moy[t]{I_\mb}-\moy[t]{I_\ma}}
  {\moy[t]{I_\mb}+\moy[t]{I_\ma}}.
\end{equation}

Figure~\ref{fig-varepsilon} displays measured photometric imbalance across science channels.

\begin{figure} \centering
  \FIG{.7}{false}{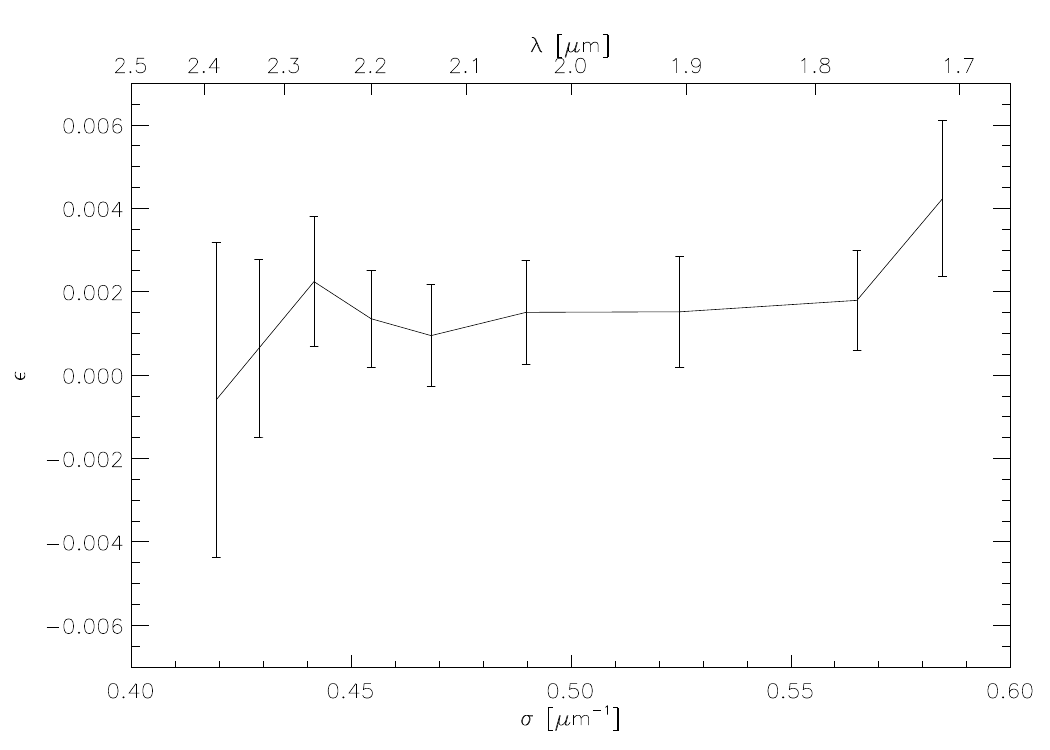}
  \caption{Measured photometric imbalance across science channels.}
  \label{fig-varepsilon}
\end{figure}

Mean photometric imbalance measures $\boldsymbol{0.16\%}$ with a spectral dispersion of $\boldsymbol{0.14\%}$—four times lower than the $0.7\%$ requirement (Section~\ref{sec-allocations}).

Pupil-plane intensity spatial trimmers built into the testbed were unneeded due to this intrinsic balance.

%-------------------------------------------------------------------------------
\subsubsection{Interference Parameter Extraction}
\label{sec-determ-caract}

Fourier transforms of calibration fringe sweeps (Figure~\ref{fig-franges-cam}) extract central channel wavelengths and differential chromatic phase shifts.

\begin{figure} \centering
  \FIG{.7}{false}{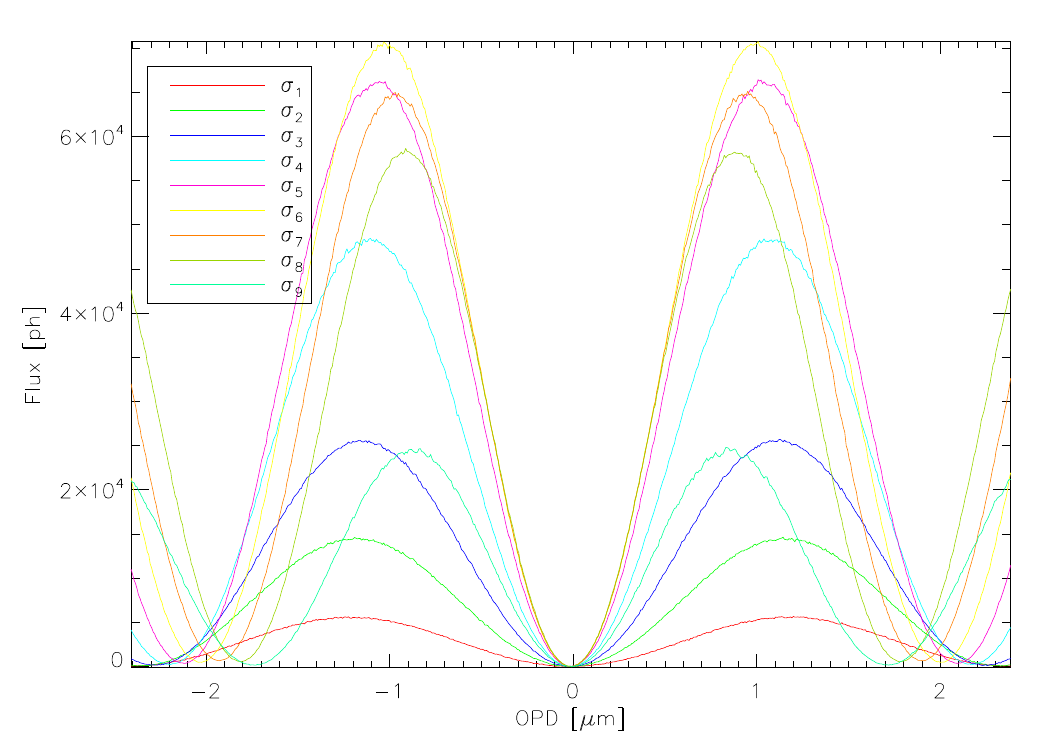}
  \caption{Science camera fringe signals across channels.}
  \label{fig-franges-cam}
\end{figure}

Phase angle $\phi_i$ evaluated at channel wavenumber $\sigma_i$ converts to relative OPD offset $\delta_i$ relative to optimal setpoint $\delta_\mmoy$. Derived parameters are plotted in Figure~\ref{fig-caract-cam}.

\begin{figure} \centering
  \subfloat[Central channel wavelengths.]{\label{fig-lambda-cam}
    \FIG{0.49}{false}{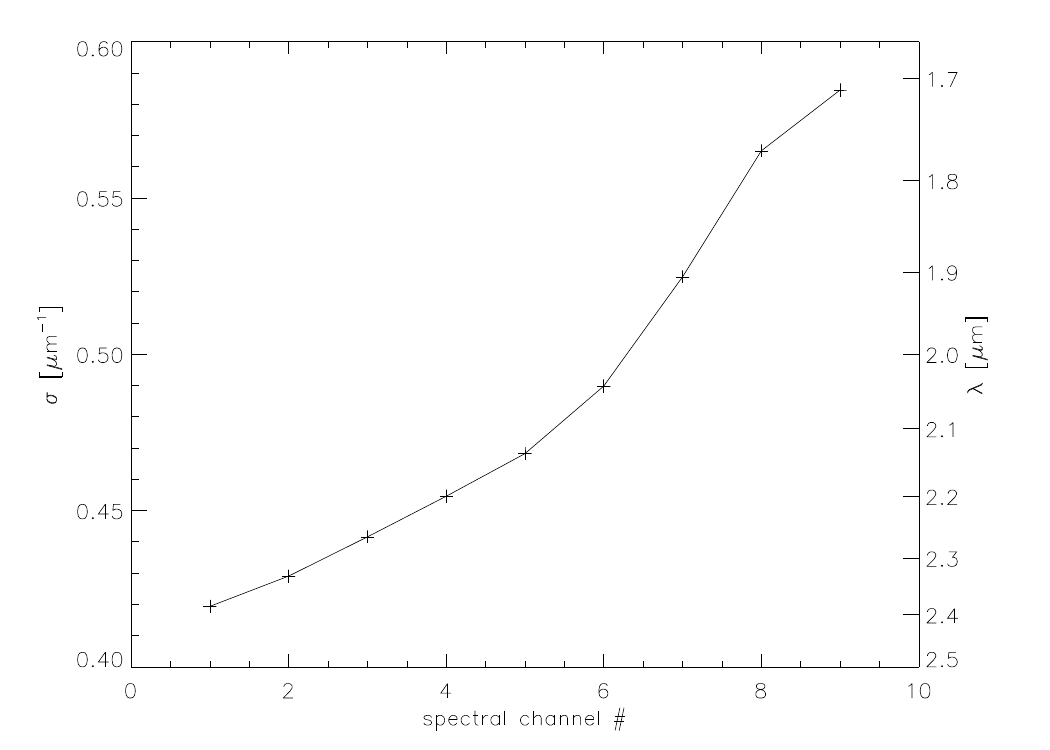}}
  \hfill\subfloat[Differential chromatic dispersion.]{\label{fig-chrom}
    \FIG{0.49}{false}{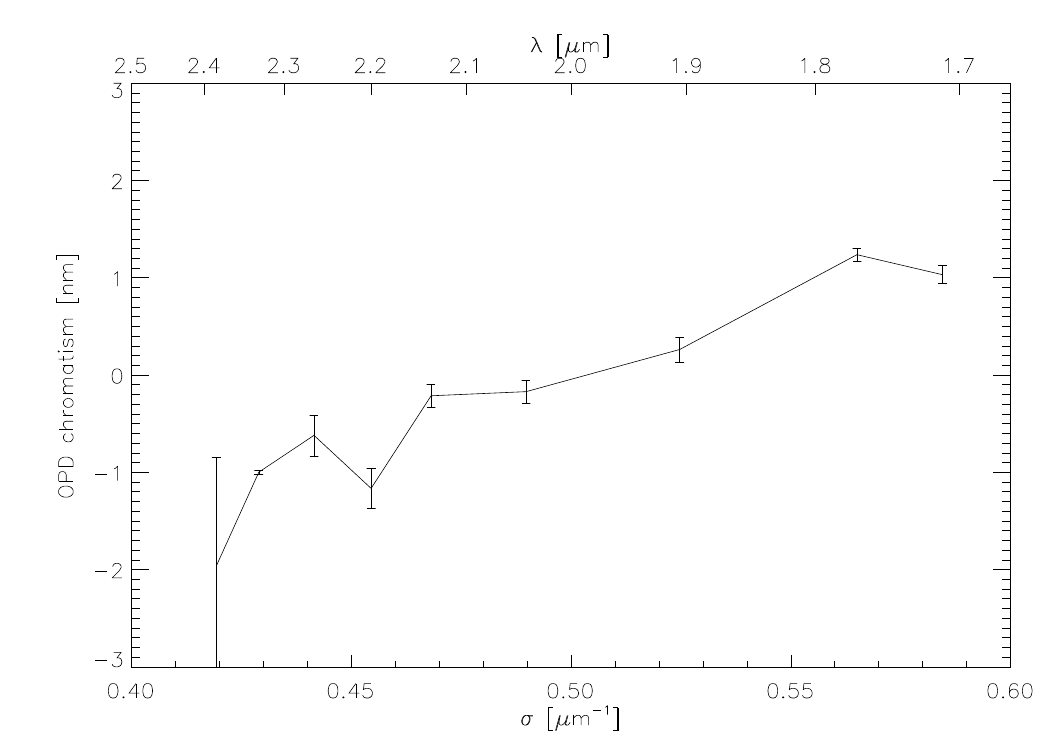}}
  \caption{Derived science channel wavelengths and chromatic dispersion offsets.}
  \label{fig-caract-cam}
\end{figure}

Tracking $\sigma_i$ monitors horizontal bi-prism dispersion alignment on the detector array.

Chromatic path offsets $\delta_i$ are measured to 0.1~nm precision. Figure~\ref{fig-chrom} shows low differential chromatic dispersion ($\sim \mathbf{3\text{~\bf nm}}$ peak-to-peak).

%-------------------------------------------------------------------------------
\subsubsection{Fringe Sensor Setpoint Optimization}
\label{sec-optimisation-position}

Fringe sensor reference setpoints can drift slightly relative to the optimal science dark fringe position.

We optimize setpoints using a 3-point dithering routine based on the NULLTIMATE protocol \cite{Gabor08a,Gabor08b}. Quadratic parabolic fitting across 3 OPD setpoints locates the global nulling minimum.

Setpoints are modulated by $\pm\delta_\varepsilon = \pm 2$~nm around nominal position under active closed-loop cophasing. A 2~nm offset increases null depth by $N_{\delta_\varepsilon} = 10^{-5}$ at 2~\mum. With filter $D=4$, detector saturation occurs above $N > 3\E{-4}$ (corresponding to a 9~nm path offset). Accounting for residual jitter ($3\sigma_\delta \approx 2.4$~nm) and a 3~nm initial offset, peak path delay remains under 7.4~nm ($N \approx 1.8\E{-4}$), preserving linear detector response.

Mean null depths $N_0, N_{+\varepsilon}, N_{-\varepsilon}$ are recorded at offsets $\delta_0, \delta_0+\delta_\varepsilon, \delta_0-\delta_\varepsilon$.

Equation~\eqref{eq-N_moy2} gives:
\begin{equation}
  N_\mmoy(\delta) = N_{\mmoy,0}+\GP{\pi\sigma_\mmoy\GP{\delta+\delta_\mref}}^2+w_N,
  \label{eq-N-moy}
\end{equation}
where $\delta_\mref$ is the fringe sensor setpoint offset and $w_N$ is zero-mean measurement noise.

Assuming static underlying null depth $N_{\mmoy,0}$, measured nulls follow:
\begin{equation}
  \left\{
    \begin{array}{r@{\hs}l}
      N_0 &= N_{\mmoy,0}+\GP{\pi\sigma_\mmoy\GP{\delta_0+\delta_\mref}}^2 \\[+6pt]
      N_{+\varepsilon} &=
      N_{\mmoy,0}+\GP{\pi\sigma_\mmoy
        \GP{\delta_0+\delta_\varepsilon+\delta_\mref}}^2 \\[+6pt]
      N_{-\varepsilon} &= N_{\mmoy,0}+
      \GP{\pi\sigma_\mmoy\GP{\delta_0-\delta_\varepsilon+\delta_\mref}}^2
      \\[+6pt]
    \end{array}
  \right..
\end{equation}

Evaluating differences yields:
\begin{equation}
  \left\{
    \begin{array}{r@{\hs}l}
      N_{+\varepsilon}-N_{-\varepsilon} &=
      4\GP{\pi\sigma_\mmoy}^2\GP{\delta_0+\delta_\mref}\delta_\varepsilon \\[+6pt]
      N_{+\varepsilon}+N_{-\varepsilon} &=
      2(N_{\mmoy,0}+\GP{\pi\sigma_\mmoy}^2\GP{\GP{\delta_0+\delta_\mref}^2+
        \delta_{\varepsilon}^2} =
      2\GP{N_0+\GP{\pi\sigma_\mmoy}^2\delta_{\varepsilon}^2} \\
    \end{array}
  \right..
\end{equation}

Solving for setpoint offset $\delta_\mref$ gives:
\begin{equation}
  \delta_\mref = -\delta_0+\frac{N_{+\varepsilon}-N_{-\varepsilon}}
  {2\GP{N_{+\varepsilon}+N_{-\varepsilon}-2N_0}}\cdot\delta_{\varepsilon}.
\end{equation}

Figure~\ref{fig-dithering} illustrates a dithering calibration sequence.

\begin{figure} \centering
  \subfloat[Time-series sequence across the 3 path steps.]{\label{fig-dithering1} \FIG{0.49}{false}{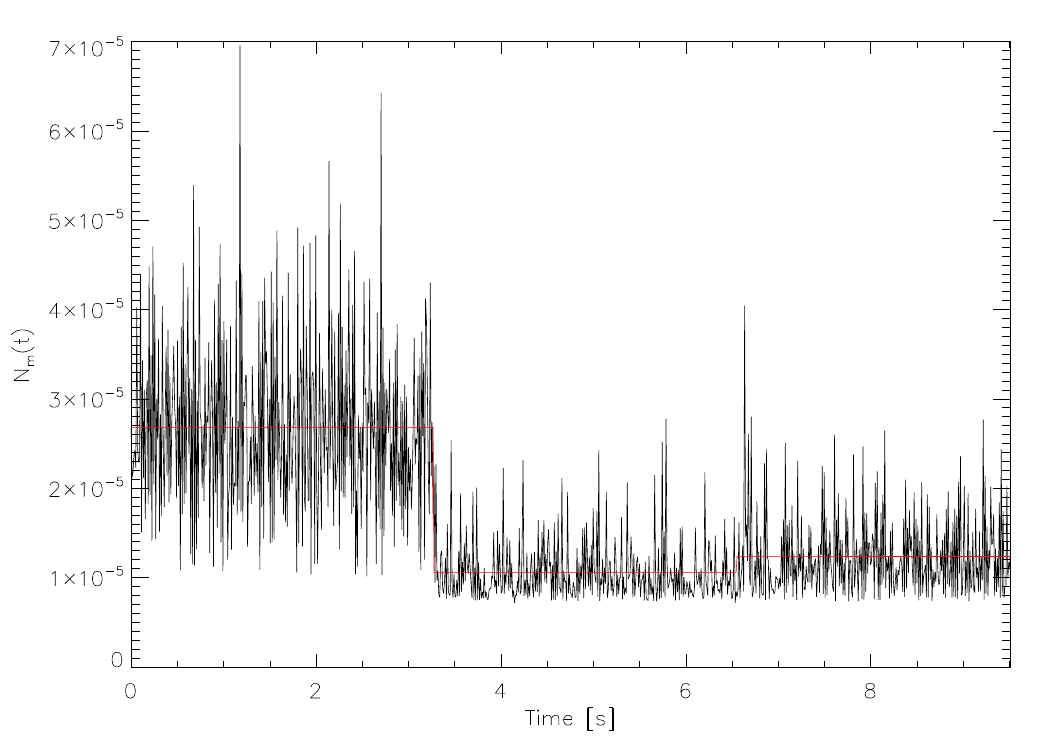}}
  \hfill\subfloat[Spectral null profiles across the 3 path steps.]{\label{fig-dithering2}\FIG{0.49}{false}{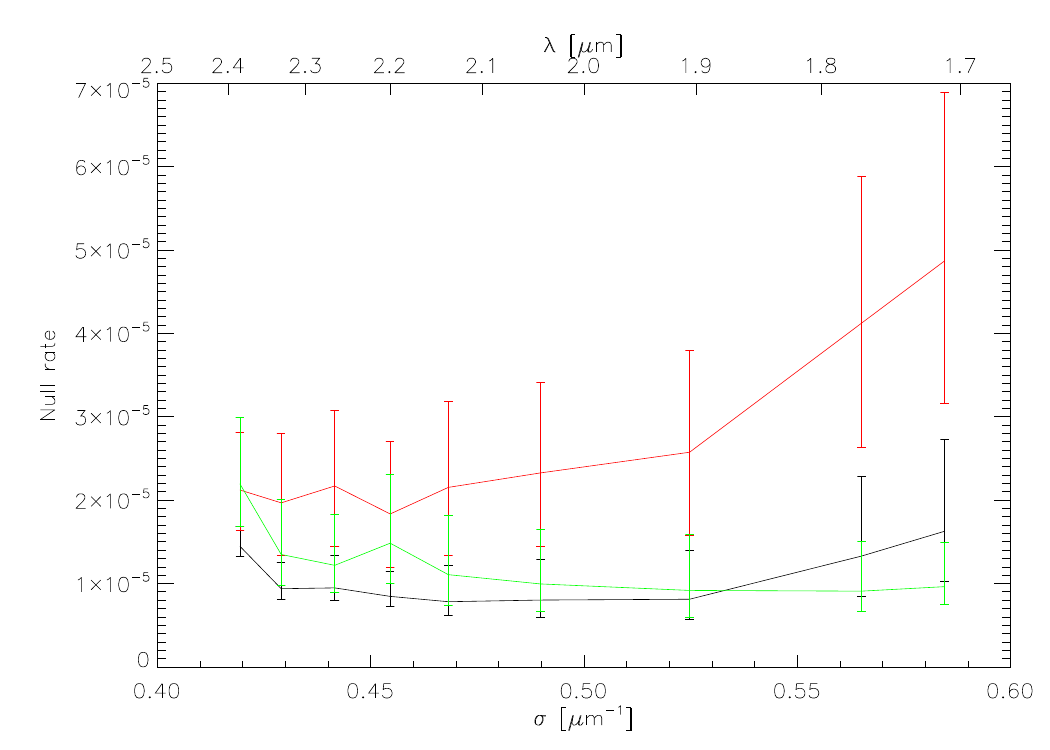}}
  \caption{Fringe sensor setpoint optimization via 3-point OPD dithering.}
  \label{fig-dithering}
\end{figure}

In Figure~\ref{fig-dithering1}, setpoint shifts alter mean null depth and noise variance without affecting baseline floor limits. In this example, the optimal setpoint correction was $\delta_\mref = -0.8$~nm. Figure~\ref{fig-dithering2} highlights chromatic dispersion: shorter wavelengths favor a $-2$~nm offset, while longer wavelengths favor 0~nm, matching the chromatic dispersion curve in Figure~\ref{fig-chrom}. Evaluating parabolic fits per channel measures chromatic dispersion with high precision.

Because thermal drift shifts $\delta_\mref$ over time, science runs were executed immediately following setpoint optimization.

%§§§§§§§§§§§§§§§§§§§§§§§§§§§§§§§§§§§§§§§§§§§§§§§§§§§§§§§§§§§§§§§§§§§§§§§§§§§§§§§
\section{100-Second Null Depth Performance}
\label{sec-analyse-te}

%¤¤¤¤¤¤¤¤¤¤¤¤¤¤¤¤¤¤¤¤¤¤¤¤¤¤¤¤¤¤¤¤¤¤¤¤¤¤¤¤¤¤¤¤¤¤¤¤¤¤¤¤¤¤¤¤¤¤¤¤¤¤¤¤¤¤¤¤¤¤¤¤¤¤¤¤¤¤¤
\subsection{Experimental Results}
\label{sec-presentation-resultats}

Following calibration, null depth was recorded over a nominal 100~s duration ($T=100$~s), matching \peg inter-thruster acquisition phases. The camera frame rate was $100$~Hz ($\tau = 2.58$~ms). The best 100~s performance dataset is shown in Figure~\ref{fig-best-null}.

\begin{figure} \centering
  \subfloat[Time series of mean null depth $N_\mmoy$.]{\label{fig-best-null1}
    \FIG{0.49}{false}{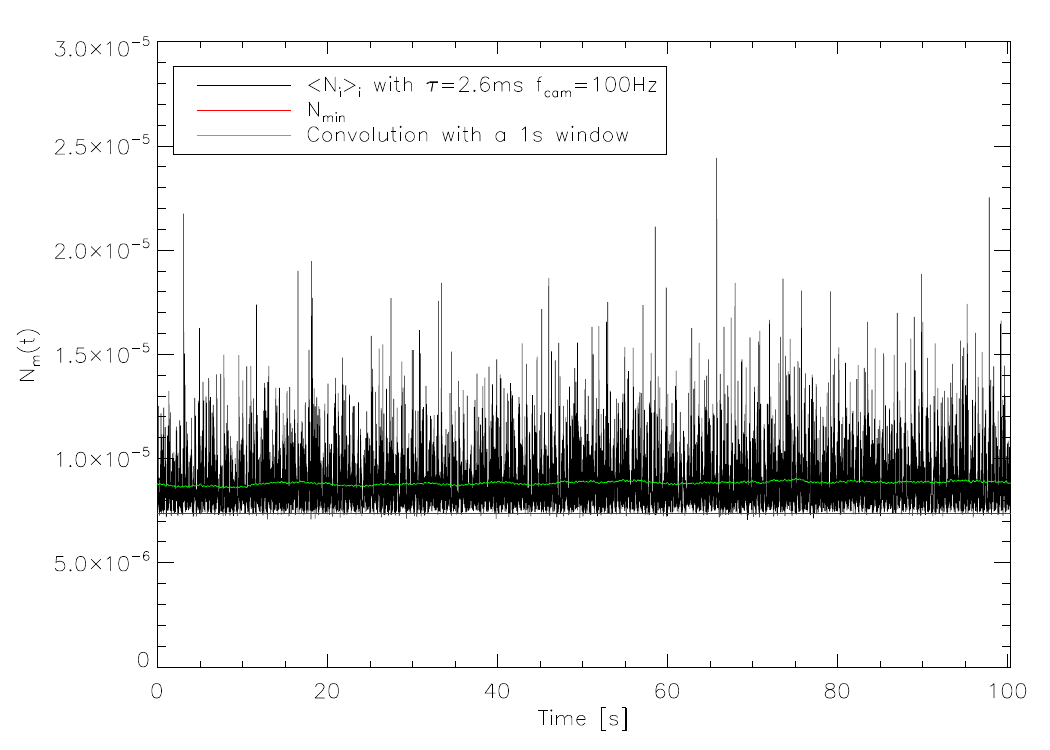}}
  \hfill\subfloat[Mean null depth per science channel.]{\label{fig-best-null2}
    \FIG{0.49}{false}{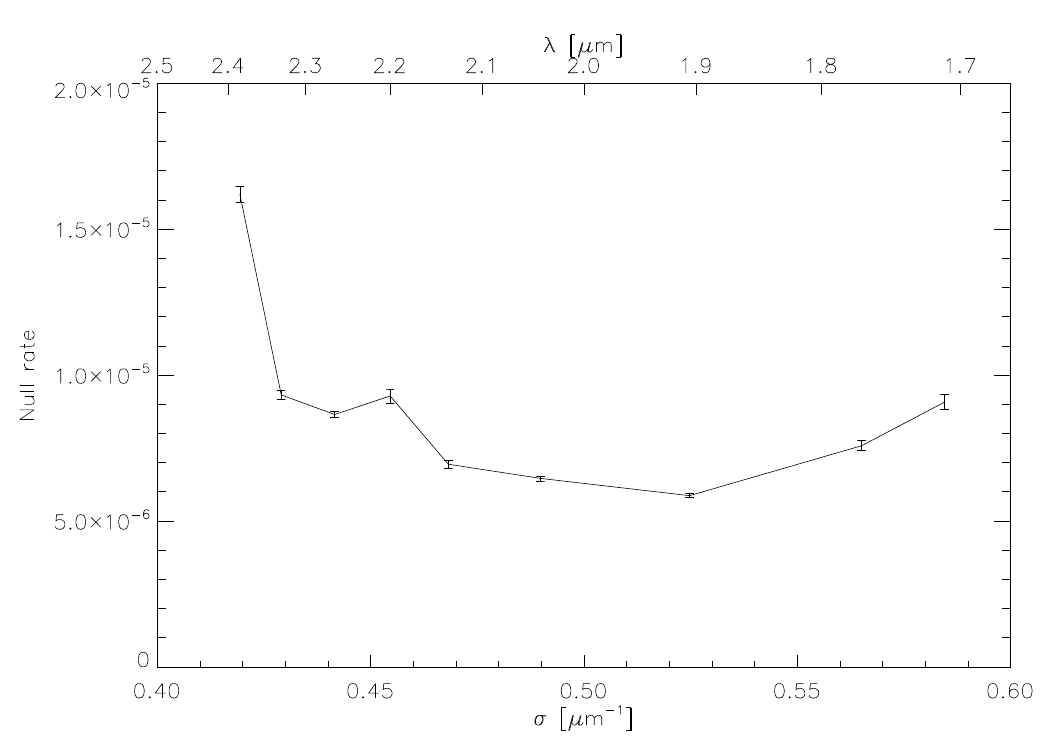}}
  \caption{Best 100~s null depth performance dataset.}
  \label{fig-best-null}
\end{figure}

The mean null depth integrated across a fractional bandwidth $(\Delta\sigma/\sigma)_\mmoy = 37\%$ reached:
\begin{equation*}
  \boldsymbol{\moy[\Tc]{N_\mmoy} = 8.8\E{-6}}.
\end{equation*}

In Figure~\ref{fig-best-null1}, $N_\mmin$ equals underlying static null $N_{\mmoy,0}$. Null stability $\sigma_{N,\tau}$ scales with camera integration time $\tau$. Binning frame data to simulate a 1~s integration time ($\tau=1$~s) yields short-term stability $\sigma_{N,\taud} = 1.5\E{-6}$ and:
\begin{equation*}
  \boldsymbol{\sigma_{N,\tauu} = 9\E{-8}}.
\end{equation*}

Figure~\ref{fig-best-null2} plots mean null depth per channel. Null depths range from $5.9\E{-6}$ to $1.62\E{-5}$ across the spectrum, with 8 of 9 channels achieving nulls deeper than $10^{-5}$. Excluding the edge channel at 2.4~\mum yields a mean null depth of $\boldsymbol{\moy[\Tc]{N_\mmoy} = 7.9\E{-6}}$ over a $(\Delta\sigma/\sigma)_\mmoy = 35\%$ bandwidth.

%¤¤¤¤¤¤¤¤¤¤¤¤¤¤¤¤¤¤¤¤¤¤¤¤¤¤¤¤¤¤¤¤¤¤¤¤¤¤¤¤¤¤¤¤¤¤¤¤¤¤¤¤¤¤¤¤¤¤¤¤¤¤¤¤¤¤¤¤¤¤¤¤¤¤¤¤¤¤¤
\subsection{Phase Perturbations}
\label{sec-perturbations-phase-1}

%-------------------------------------------------------------------------------
\subsubsection{Chromatism and OPD Coupling}
\label{sec-couplage-entre}

Serabyn's error budget assumes zero-mean path delay jitter ($\delta_w$), decoupling path jitter from broadband dispersion \cite{Serabyn01}.

On \pe, null depths are evaluated independently across 9 science channels. Global setpoints leave residual chromatic offsets $\delta_i$ in individual channels.

Intra-channel chromatic dispersion $\Delta\delta_{\sigma,i}(\sigma)$ varies around mean channel wavenumber $\sigma_i$:
\begin{equation}
  \delta_{\sigma,i}(\sigma) = \delta_i+\Delta\delta_{\sigma,i}(\sigma).
\end{equation}

Assuming flat channel passbands $S_i(\sigma)$:
\begin{equation}
  \int S_i(\sigma)\sigma^2\Delta\delta_{\sigma,i}(\sigma) \dd\sigma = 0.
  \label{eq-prop-chrom}
\end{equation}

At step $t$, phase-induced null depth leakage in channel $i$ is:
\begin{equation}
  N_{\phi,i}(t) = \int \pi^2S_i(\sigma)\sigma^2\GP{\delta_i+
    \Delta\delta_{\sigma,i}(\sigma)+\delta_\mref(t)+\delta_w(t)}^2 \dd\sigma.
\end{equation}

Expanding terms gives:
\begin{equation}
  N_{\phi,i}(t) = \int \pi^2S_i(\sigma)\sigma^2\GC{
    \delta_i^2+\Delta\delta_{\sigma,i}(\sigma)^2+\delta_\mref(t)^2
    +\delta_w(t)^2} \dd\sigma + \text{cross terms}.
\end{equation}

Approximating $\sigma_i^2 \simeq \int S_i(\sigma)\sigma^2\dd\sigma$ yields:
\begin{align}
  N_{\phi,i}(t) = &\ \pi^2\sigma_i^2\GC{\delta_i^2+\delta_\mref(t)^2
    +\delta_w(t)^2}+\pi^2\int S_i(\sigma)\sigma^2
  \Delta\delta_{\sigma,i}(\sigma)^2 \dd\sigma
  +2\pi^2\sigma_i^2\delta_i\delta_\mref(t) \\
  & {}+2\pi^2\sigma_i^2\GC{\delta_i+\delta_\mref(t)}\delta_w(t)
  +2\pi^2\GC{\delta_i+\delta_\mref(t)+\delta_w(t)}\int
  S_i(\sigma)\sigma^2\Delta\delta_{\sigma,i}(\sigma) \dd\sigma. \nonumber
  \label{eq-N-phii}
\end{align}

Applying Equation~\eqref{eq-prop-chrom} simplifies $N_{\phi,i}(t)$ to:
\begin{align}
  N_{\phi,i}(t) = &\ \pi^2\sigma_i^2\GC{\delta_i^2+\delta_\mref(t)^2
  +\delta_w(t)^2}+\pi^2\int S_i(\sigma)\sigma^2
  \Delta\delta_{\sigma,i}(\sigma)^2 \dd\sigma \nonumber\\
  & {}+2\pi^2\sigma_i^2\GC{\delta_i\delta_\mref(t)+\GP{\delta_i+
  \delta_\mref(t)}\delta_w(t)}.
\end{align}

This decomposes into three distinct terms:
\begin{equation}
  \left\{
    \begin{array}{r@{\hs}l}
      N_{\sigma,i} & = \pi^2\sigma_i^2\delta_i^2
      +\pi^2\int S_i(\sigma)\sigma^2
      \Delta\delta_{\sigma,i}(\sigma)^2 \dd\sigma \\[+6pt]
      N_{\delta,\mref,i}(t) & = \pi^2\sigma_i^2\GP{\delta_\mref(t)^2
      +2\delta_i\delta_\mref(t)} \\[+6pt]
      N_{\delta,w,i}(t) & = \pi^2\sigma_i^2\GP{\delta_w(t)^2+2\GP{\delta_i
      +\delta_\mref(t)}\delta_w(t)} \\
    \end{array}
  \right..
  \label{eq-N-sig-delta-t}
\end{equation}

Because $\delta_w(t)$ is zero-mean ($\langle \delta_w \rangle = 0$, $\langle \delta_w^2 \rangle = \sigma_\delta^2$), time-averaged phase contributions are:
\begin{equation}
  \left\{
    \begin{array}{r@{\hs}l}
      N_{\sigma,i} & = \pi^2\sigma_i^2\delta_i^2
      +\int\pi^2S_i(\sigma)\sigma^2
      \Delta\delta_{\sigma,i}(\sigma)^2 \dd\sigma \\[+6pt]
      \moy[T]{N_{\delta,\mref,i}} & = \pi^2\sigma_i^2
      \moy[T]{\delta_\mref(t)^2+2\delta_i\delta_\mref(t)} \\[+6pt]
      \moy[T]{N_{\delta,w,i}} & = \pi^2\sigma_i^2\sigma_\delta^2 \\
    \end{array}
  \right..
  \label{eq-N-sig-delta}
\end{equation}

This isolates the individual contributions of path delay jitter, chromatic dispersion, and setpoint offset $\delta_\mref$.

%-------------------------------------------------------------------------------
\subsubsection{Path Delay Jitter Contribution}
\label{sec-perturbation-difference}

From Equation~\eqref{eq-N-sig-delta}, the mean path delay jitter contribution across channels is:
\begin{equation}
  \moy[T]{N_{\delta,w}} = \moy[i,T]{N_{\delta,w,i}} =
  \pi^2\sigma_\mmoy^2\sigma_\delta^2.
\end{equation}

Active cophasing (S316 strain gauges enabled, LQG control) maintained residual path delay jitter at $\boldsymbol{\sigma_\delta = 0.74\text{~\bf nm RMS}}$. Time-series path delay jitter and calculated channel contributions $\moy[T]{N_{\delta,w,i}}$ are shown in Figure~\ref{fig-opd}.

\begin{figure} \centering
  \subfloat[Time-series path delay jitter $\delta_w$.]{\label{fig-opd1}
    \FIG{0.49}{false}{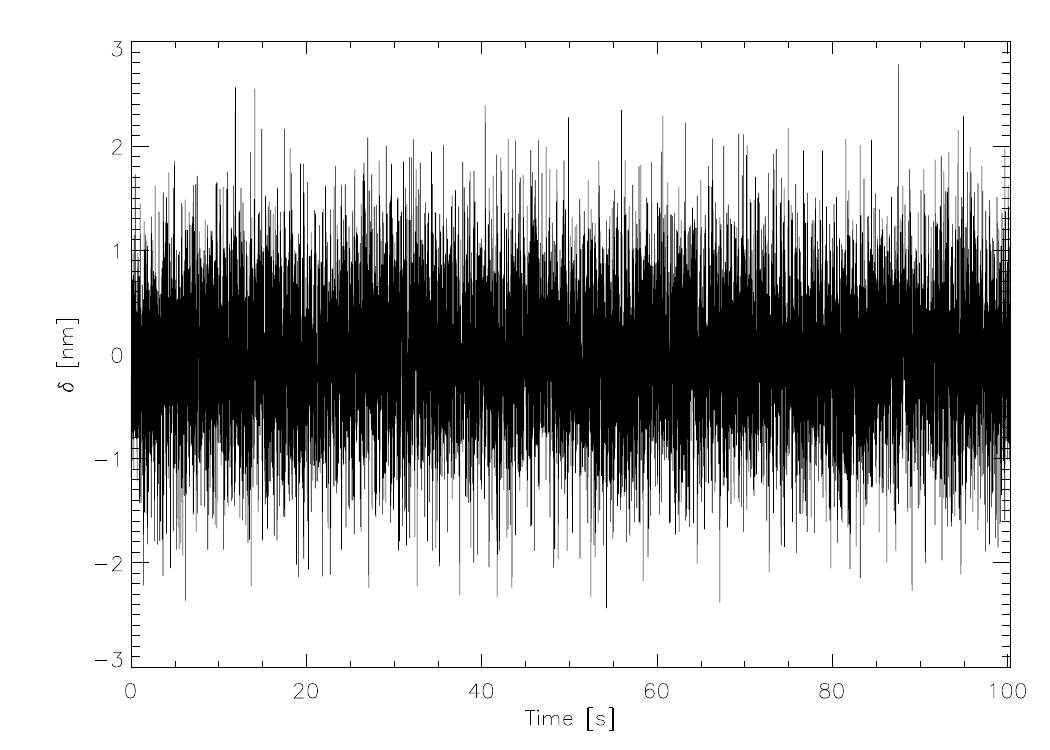}}
  \hfill\subfloat[Path delay jitter null depth contribution per channel.]{\label{fig-opd2}
    \FIG{0.49}{false}{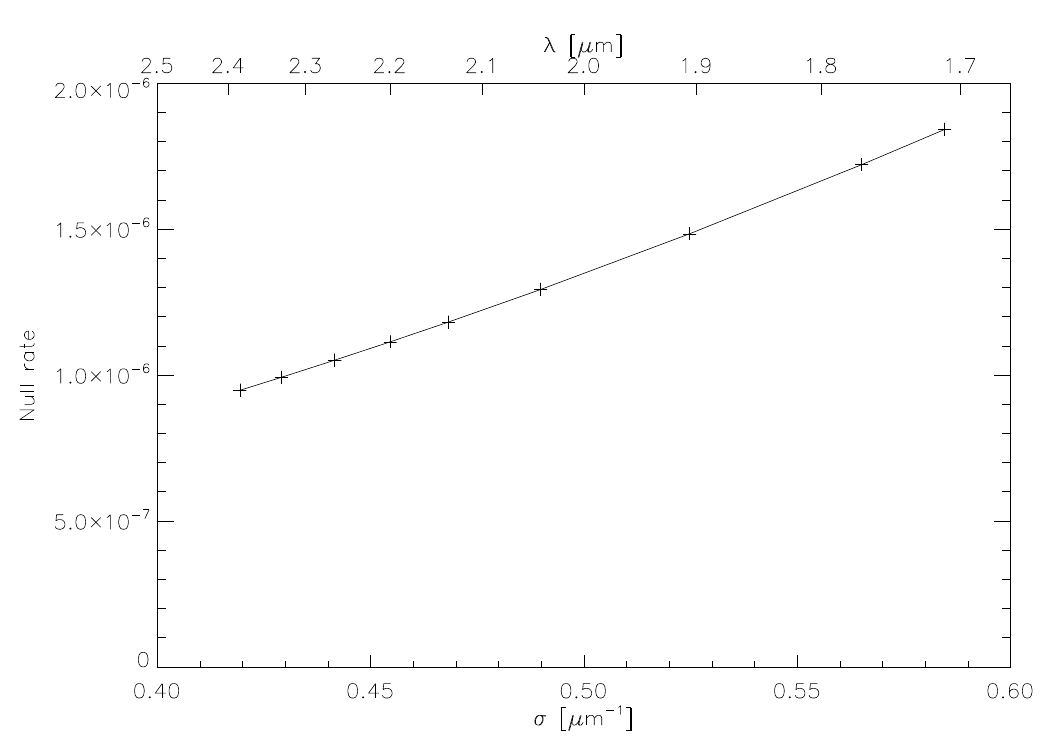}}
  \caption{Path delay jitter and corresponding channel null depth contributions.}
  \label{fig-opd}
\end{figure}

Path delay jitter leakage increases at shorter wavelengths (Figure~\ref{fig-opd2}), yielding a mean contribution of:
\begin{equation*}
  \boldsymbol{\moy[\Tc]{N_{\delta,w}} = 1.4\E{-6}}.
\end{equation*}

%-------------------------------------------------------------------------------
\subsubsection{Chromatic Dispersion Analysis}
\label{sec-analyse-chromatisme}

Setpoints optimized globally leave residual chromatic offsets $\delta_i$ across channels (Section~\ref{sec-optimisation-position}).

From Equation~\eqref{eq-N-sig-delta}, chromatic dispersion leakage per channel $N_{\sigma,i}$ is:
\begin{equation}
  N_{\sigma,i} = \GP{\pi\sigma_i\delta_i}^2+\pi^2\int S_i(\sigma)\sigma^2
  \Delta\delta_{\sigma,i}(\sigma)^2 \dd\sigma.
  \label{eq-null-lambda-i}
\end{equation}
	
Mean chromatic dispersion leakage $N_\sigma$ across channels is:
\begin{equation}
  N_\sigma = \pi^2\moy[i]{\GP{\sigma_i\delta_i}^2}+\moy[i]{\pi^2\int
  S_i(\sigma)\sigma^2\Delta\delta_{\sigma,i}(\sigma)^2 \dd\sigma}.
  \label{eq-null-lambda}
\end{equation}

Figure~\ref{fig-chromatism} plots $N_{\sigma,i}$ calculated using measured offsets $\delta_i$ (Figure~\ref{fig-chrom}).

\begin{figure} \centering
  \FIG{0.7}{false}{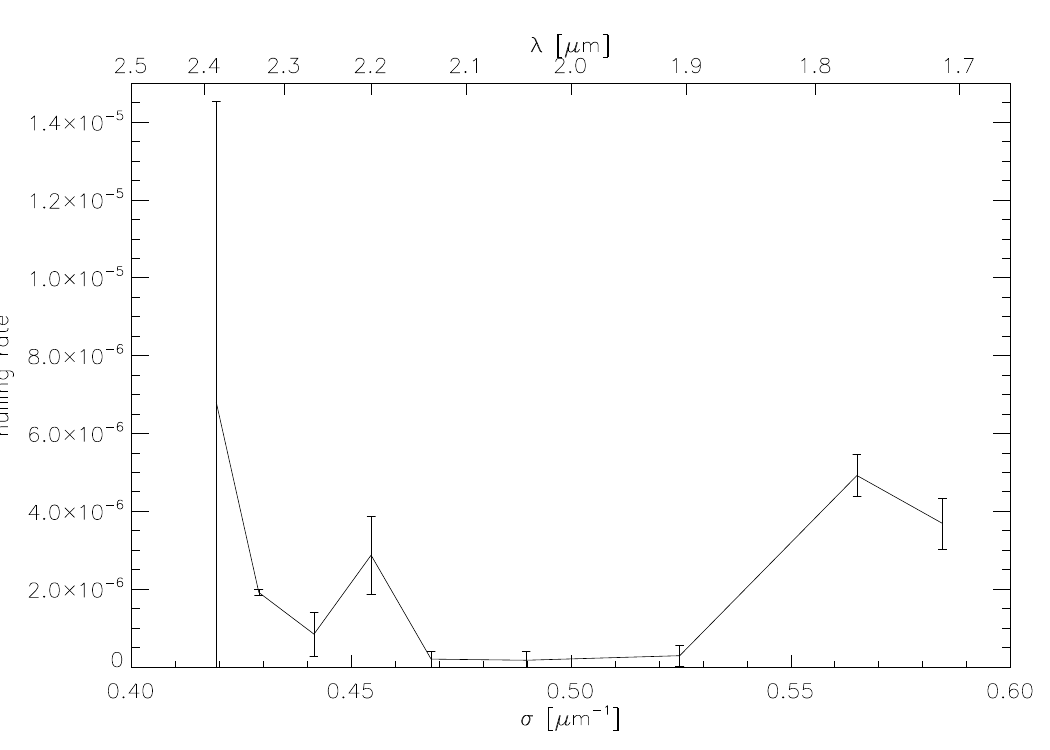}
  \caption{Chromatic dispersion null depth contribution across channels.}
  \label{fig-chromatism}
\end{figure}

Channel chromatic leakage remains below $7\E{-6}$ across all channels, yielding a mean contribution of:
\begin{equation*}
  \boldsymbol{N_\sigma = 2.4\E{-6}}.
\end{equation*}

%-------------------------------------------------------------------------------
\subsubsection{Fringe Sensor Setpoint Offset Estimation}
\label{sec-utilisation-statistique}

Because thermal drift causes setpoint offset $\delta_\mref$ to evolve, we estimate $\delta_\mref$ statistically by fitting modeled null depth probability density functions (PDF) to experimental null depth histograms.

From Equation~\eqref{eq-N-sig-delta-t}, time-varying phase contributions to mean null depth are:
\begin{equation}
  \left\{
    \begin{array}{r@{\hs}c@{\hs}l}
      N_{\delta,\mref}(t) = & \moy[i]{N_{\delta,\mref,i}}(t) & =
      \pi^2\sigma_\mmoy^2\delta_\mref(t)^2 \\[+6pt]
      N_{\delta,w}(t) = & \moy[i]{N_{\delta,w,i}}(t) & =
      \pi^2\sigma_\mmoy^2\GP{\delta_w(t)^2+2\delta_\mref(t)\delta_w(t)} \\
    \end{array}
  \right..
  \label{eq-ref-w}
\end{equation}

Summing both terms recovers Equation~\eqref{eq-N_moy}:
\begin{equation}
  N_\mmoy(t) = N_{\mmoy,0}+\pi^2\sigma_\mmoy^2\GP{\delta_\mref(t)+\delta_w(t)}^2+w_N.
\end{equation}

Fitting modeled null PDFs to experimental histograms extracts static null floor $N_{\mmoy,0}$, setpoint offset $\delta_\mref$, and measurement noise $w_N$. Figure~\ref{fig-histog} models how noise $w_N$ and offset $\delta_\mref$ alter null depth distributions.

\begin{figure} \centering
  \subfloat[Impact of measurement noise $w_N$.]{\label{fig-histog1}
    \FIG{0.49}{false}{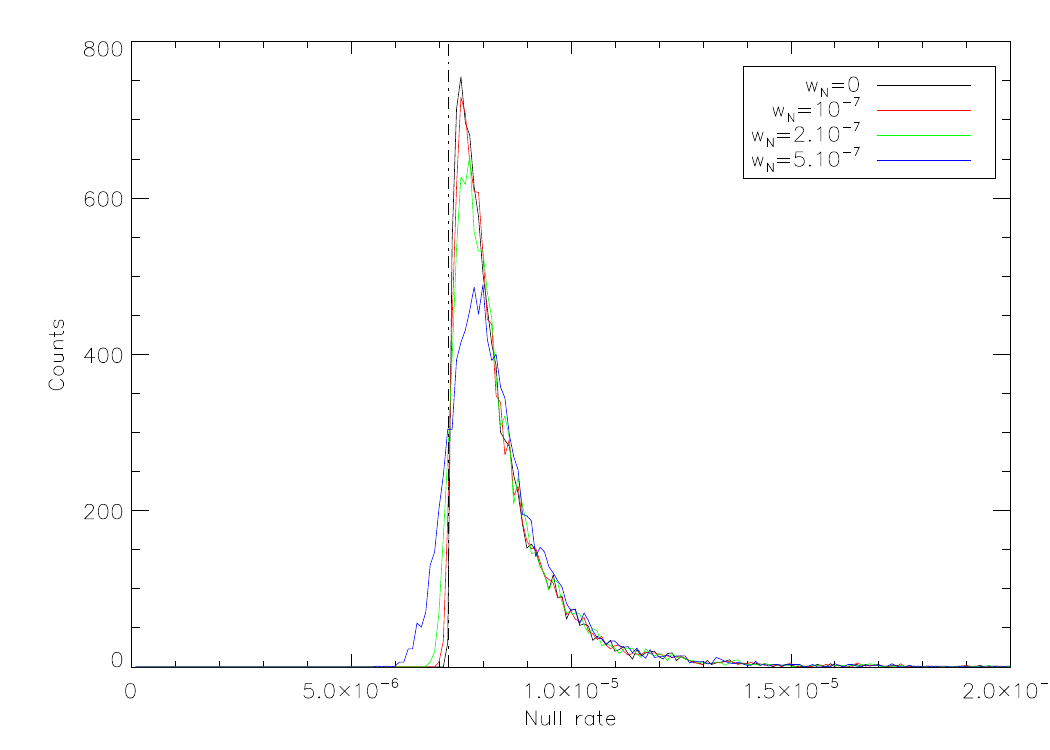}}
  \hfill\subfloat[Impact of setpoint offset $\delta_\mref$.]{\label{fig-histog2}
    \FIG{0.49}{false}{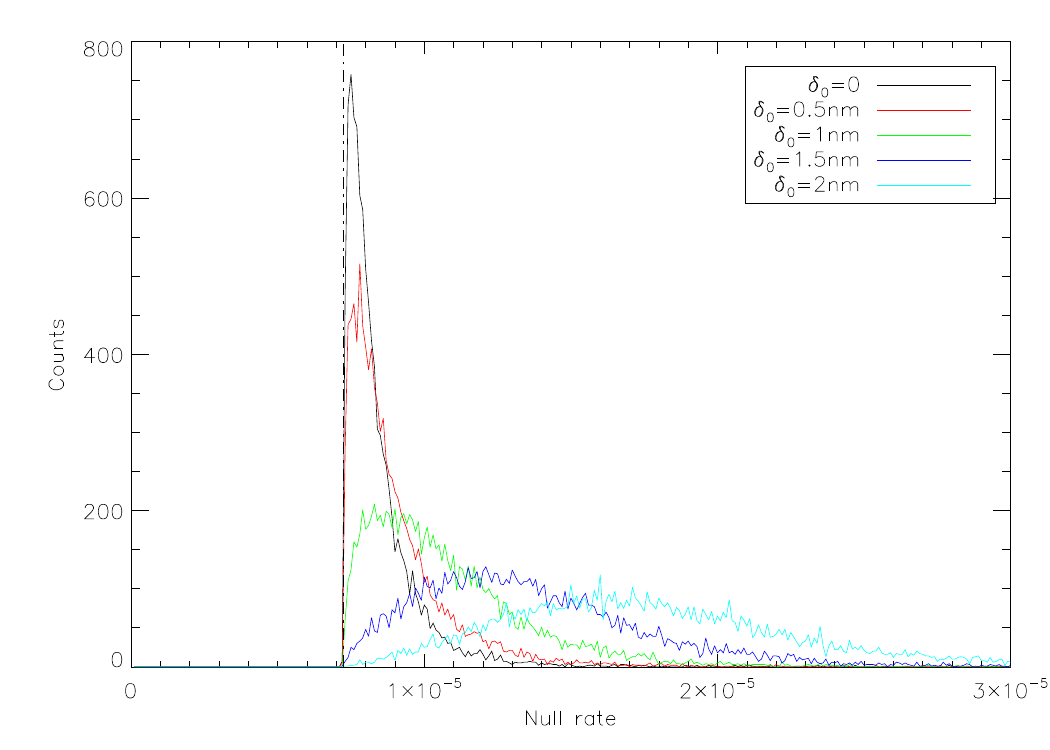}}
  \caption{Modeled null depth histograms across noise levels and setpoint offsets.}
  \label{fig-histog}
\end{figure}

For Gaussian path delay jitter $\delta_w$, the probability density function $h_\delta$ is:
\begin{equation}
  h_\delta(z_\delta) = \frac{1}{\sqrt{2\pi}\sigma_\delta}
  \exp\GC{-\frac{\GP{z_\delta-\delta_\mref}^2}{2\sigma_\delta^2}}.
\end{equation}

The corresponding null depth PDF $h_{N,\delta}$ follows a non-central chi-square distribution \cite{Hanot11}:
\begin{equation}
  h_{N,\delta}\GP{z_{N,\delta}} = \frac{1}{\sqrt{2\pi}\sigma_\delta}
  \frac{\pi\sigma_\mmoy}{\sqrt{z_{N,\delta}}}
  \exp\GC{-\frac{z_{N,\delta}+\GP{\pi\sigma_\mmoy\delta_\mref}^2}
    {2\GP{\pi\sigma_\mmoy\sigma_\delta}^2}}
  \cosh\GP{\frac{\sqrt{z_{N,\delta}}\delta_\mref}{\pi\sigma_\mmoy\sigma_\delta}}.
\label{eq-h-N}
\end{equation}

Static null $N_{\mmoy,0}$ shifts the PDF along the horizontal axis:
\begin{equation}
  h_{N}\GP{z_N} = h_{N,\delta}\GP{z_{N,\delta}-N_{\mmoy,0}}.
\end{equation}

Measurement noise $w_N$ convolves with the PDF, smoothing out the sharp low-end cutoff (Figure~\ref{fig-histog1}).

Non-zero setpoint offsets $\delta_\mref \neq 0$ shift the PDF from an exponential-like tail toward a Gaussian-like distribution (Figure~\ref{fig-histog2}) \cite{Hanot11}.

Fitting modeled PDFs (Equation~\ref{eq-h-N}) to experimental histograms (Figure~\ref{fig-histo-null}) extracts setup parameters.

\begin{figure} \centering
  \FIG{0.7}{false}{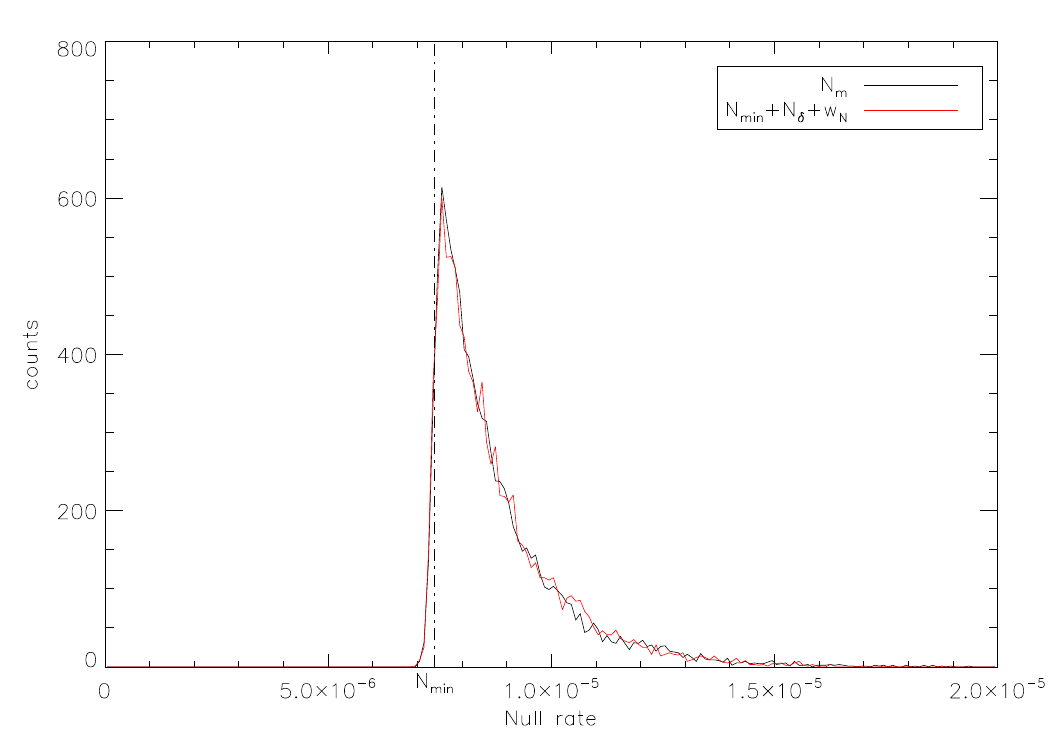}
  \caption{Experimental null depth histogram fitted with modeled PDF.}
  \label{fig-histo-null}
\end{figure}

Extracted parameters are:
\begin{equation*}
  \left\{
    \begin{array}{r@{\hs}l}
      \boldsymbol{N_{\mmoy,0}} & \boldsymbol{= 7.39\E{-6}} \\
      \boldsymbol{\delta_\mref} & \boldsymbol{= \pm 0.2\text{~\bf nm}} \\
      \boldsymbol{w_N} & \boldsymbol{= 1.2\E{-7}} \\
    \end{array}
  \right..
\end{equation*}

Because PDF fits are symmetric in $\delta_\mref$, setpoint offset direction is determined from thermal drift signs. A negative offset ($\delta_\mref = -0.2$~nm) matches room cooling trends and accounts for elevated null depths at 2.4~\mum where chromatic offsets are positive.

Evaluating $\delta_\mref = -0.2$~nm confirms path delay jitter leakage $\moy[\Tc]{N_{\delta,w}} = 1.4\E{-6}$. Setpoint offset leakage is:
\begin{equation*}
  \boldsymbol{\moy[\Tc]{N_{\delta,\mref}} = 1.0\E{-7}}.
\end{equation*}

\begin{figure} \centering
  \FIG{0.7}{false}{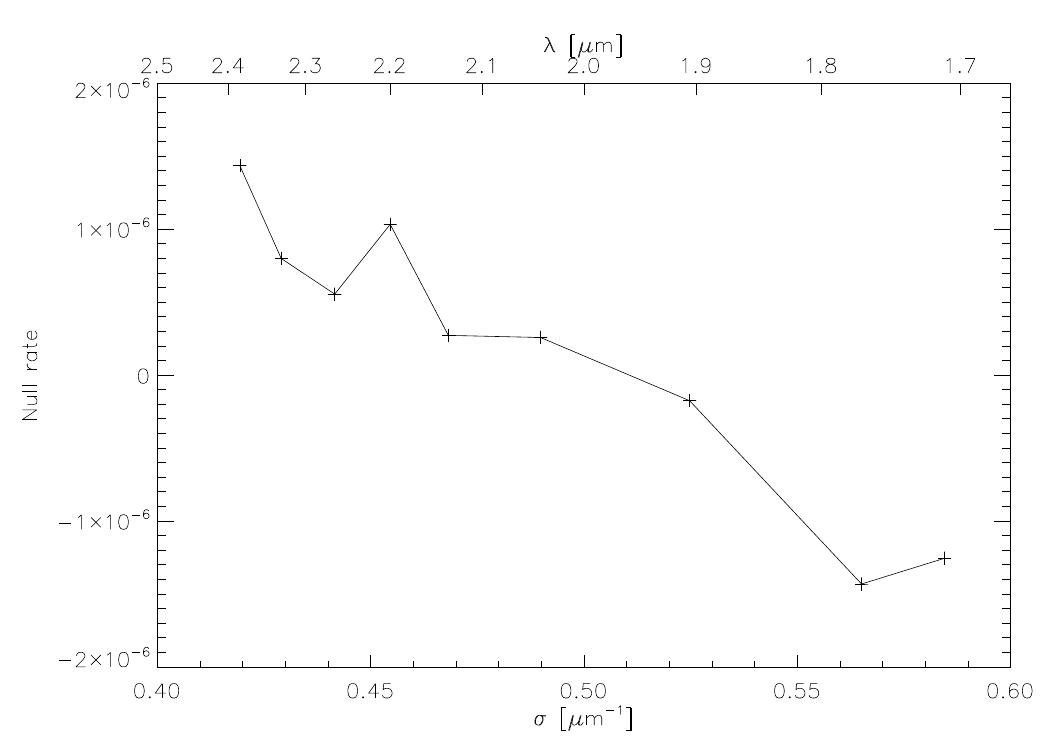}
  \caption{Setpoint offset contribution $\moy[\Tc]{N_{\delta,\mref,i}}$ across channels.}
  \label{fig-null-delta-ref}
\end{figure}

Figure~\ref{fig-null-delta-ref} plots channel setpoint leakage $\moy[\Tc]{N_{\delta,\mref,i}}$. Negative setpoint offsets lower null depths at shorter wavelengths while raising null depths near 2.4~\mum.

Path delay jitter drives short-term null stability $\sigma_{N,\delta,\tau=1\text{s}} = 10^{-7}$.

\bigskip

Comparing experimental and modeled null depth PSDs (Figure~\ref{fig-psd-null}) evaluates residual uncorrected vibration lines.

\begin{figure} \centering
  \FIG{0.7}{false}{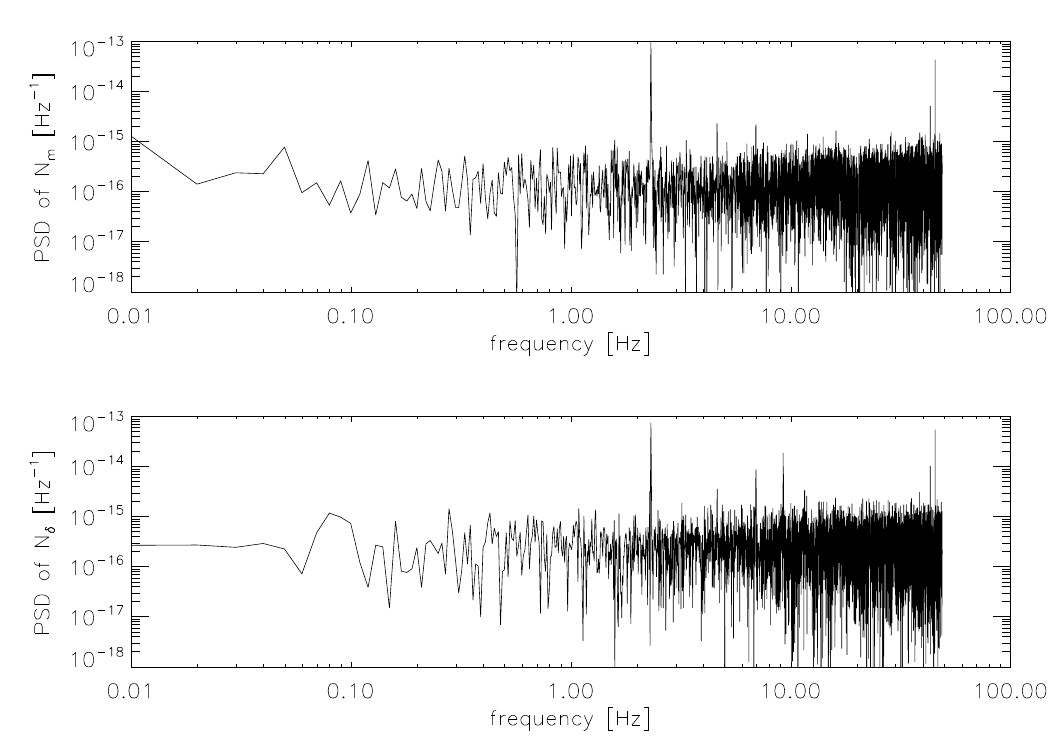}
  \caption{Experimental and modeled null depth PSDs.}
  \label{fig-psd-null}
\end{figure}

Harmonic lines in path delay $\delta$ appear at doubled frequencies ($2f$) in null depth PSDs due to quadratic response. Residual path offsets $\delta_\mref$ re-introduce fundamental lines ($f$). Dominant null depth PSD lines at 45.3~Hz and 2.3~Hz represent aliased line power harmonics (100~Hz aliased to 2.3~Hz; 150~Hz aliased to 45.3~Hz at a 97.7~Hz camera frame rate).

%¤¤¤¤¤¤¤¤¤¤¤¤¤¤¤¤¤¤¤¤¤¤¤¤¤¤¤¤¤¤¤¤¤¤¤¤¤¤¤¤¤¤¤¤¤¤¤¤¤¤¤¤¤¤¤¤¤¤¤¤¤¤¤¤¤¤¤¤¤¤¤¤¤¤¤¤¤¤¤
\subsection{Photometric Imbalance}
\label{sec-deseq-phot-1}

Photometric null leakage includes static alignment imbalance and dynamic tip/tilt tracking jitter.

From Equation~\eqref{eq-flux-angle}, single-arm throughputs $I_{\ma,i}$ and $I_{\mb,i}$ follow:
\begin{equation}
  \left\{
    \begin{array}{r@{\hs}l}
      I_{\ma,i}(t) &\simeq
      I_{\ma,i,\mmax}\GP{1-\dfrac15\GP{\pi\sigma_iD\theta_\ma(t)}^2} \\[+6pt]
      I_{\mb,i}(t) &\simeq
      I_{\mb,i,\mmax}\GP{1-\dfrac15\GP{\pi\sigma_iD\theta_\mb(t)}^2} \\
    \end{array}
  \right.,
\end{equation}
where $\theta_\ma^2 = (\alpha_\ma-\alpha_{\ma,\mref})^2+(\beta_\ma-\beta_{\ma,\mref})^2$ and $\theta_\mb^2 = (\alpha_\mb-\alpha_{\mb,\mref})^2+(\beta_\mb-\beta_{\mb,\mref})^2$.

For small tracking errors $\theta_\ma, \theta_\mb$ and $I_{\ma,i,\mmax} \approx I_{\mb,i,\mmax}$, channel flux imbalance $\varepsilon_i$ is:
\begin{equation}
  \varepsilon_i(t) = \varepsilon_{\msta,i}
  +\frac{1}{10}\GP{\pi\sigma_iD}^2\GP{\theta_\ma(t)^2-\theta_\mb(t)^2}, \qquad
  \text{with} \qquad \varepsilon_{\msta,i} =
  \dfrac{I_{\mb,i,\mmax}-I_{\ma,i,\mmax}}{I_{\mb,i,\mmax}+I_{\ma,i,\mmax}}.
\end{equation}

From Equation~\eqref{eq-N-eps}, photometric null leakage per channel $N_{\varepsilon,i}(t)$ is:
\begin{equation}
  N_{\varepsilon,i}(t) = \frac14\varepsilon_{\msta,i}^2+
  \frac{1}{400}\GP{\pi\sigma_iD}^4\GP{\theta_\ma(t)^2-\theta_\mb(t)^2}^2.
\end{equation}

Mean photometric null leakage across channels is:
\begin{equation}
  N_\varepsilon(t) = \frac14\moy[i]{\varepsilon_{\msta,i}^2}+\frac{1}{400}
  \GP{\pi D}^4\moy[i]{\sigma_i^4}\GP{\theta_\ma(t)^2-\theta_\mb(t)^2}^2.
  \label{eq-null_epsilon}
\end{equation}

Static imbalance $\varepsilon_{\msta,i}$ is measured during science calibration (Section~\ref{sec-calcul-desequilibre}). Channel static null leakage $N_{\varepsilon,\msta,i}$ is:
\begin{equation}
  N_{\varepsilon,\msta,i} = \frac14\varepsilon_{\msta,i}^2.
\end{equation}

Mean static photometric leakage $N_{\varepsilon,\msta}$ is:
\begin{equation}
  N_{\varepsilon,\msta} = \frac14\moy[i]{\varepsilon_{\msta,i}^2}.
\end{equation}

Figure~\ref{fig-epsilon} plots static photometric leakage across channels.

\begin{figure} \centering
  \FIG{.7}{false}{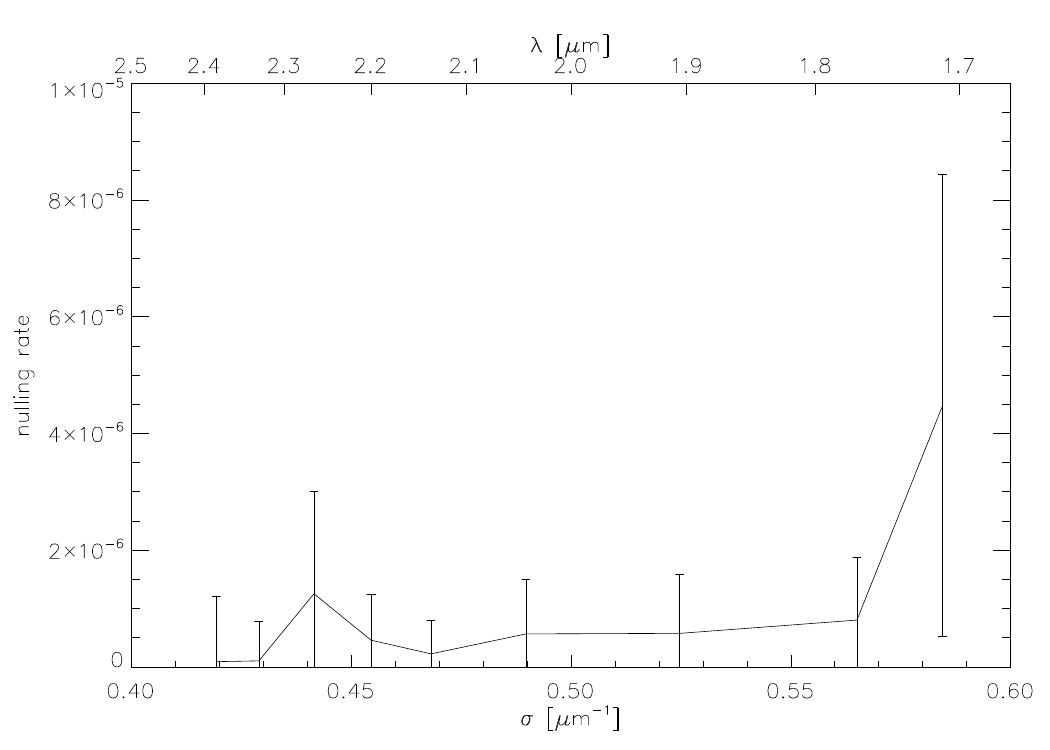}
  \caption{Static photometric imbalance null depth contribution across channels.}
  \label{fig-epsilon}
\end{figure}

Static photometric leakage remains below $1.5\E{-6}$ across channels ($4\E{-6}$ at 1.7~\mum), yielding a mean contribution of:
\begin{equation*}
  \boldsymbol{N_{\varepsilon,\msta} = 9.5\E{-7}}.
\end{equation*}

\bigskip

From Equation~\eqref{eq-null_epsilon}, dynamic tip/tilt tracking leakage $\moy[T]{N_{\varepsilon,\mdyn}(t)}$ is:
\begin{equation}
  \moy[T]{N_{\varepsilon,\mdyn}(t)} = \frac{1}{400}\GP{\pi D}^4
  \moy[i]{\sigma_i^4}\moy[T]{\GP{\theta_\ma(t)^2-\theta_\mb(t)^2}^2}.
\end{equation}

Evaluating raw moments for Gaussian tracking jitter gives:
\begin{equation}
  \left\{
    \begin{array}{r@{\hs}l}
      \moy[T]{\theta_\ma(t)^4} =& \moy[T]{\theta_\mb(t)^4} = 
      3\sigma_\theta^4 \\[+6pt]
      \moy[T]{\theta_\ma(t)^2\theta_\mb(t)^2} =& \moy[T]{\theta_\ma(t)^2}
      \moy[T]{\theta_\mb(t)^2} = \sigma_\theta^4 \\
    \end{array}
  \right..
\end{equation}

Thus, $\moy[T]{\GP{\theta_\ma(t)^2-\theta_\mb(t)^2}^2} = 4\sigma_\theta^4$, and:
\begin{equation}
  N_{\varepsilon,\mdyn} =
  \frac{1}{100}\GP{\pi D}^4\moy[i]{\sigma_i^4}\sigma_\theta^4.
  \label{eq-N-eps-dyn}
\end{equation}

Active tip/tilt tracking ($\boldsymbol{\sigma_\theta = 65\text{~\bf mas RMS}}$) yields the dynamic leakage spectrum in Figure~\ref{fig-epsilon-dyn}.

\begin{figure} \centering
  \FIG{0.7}{false}{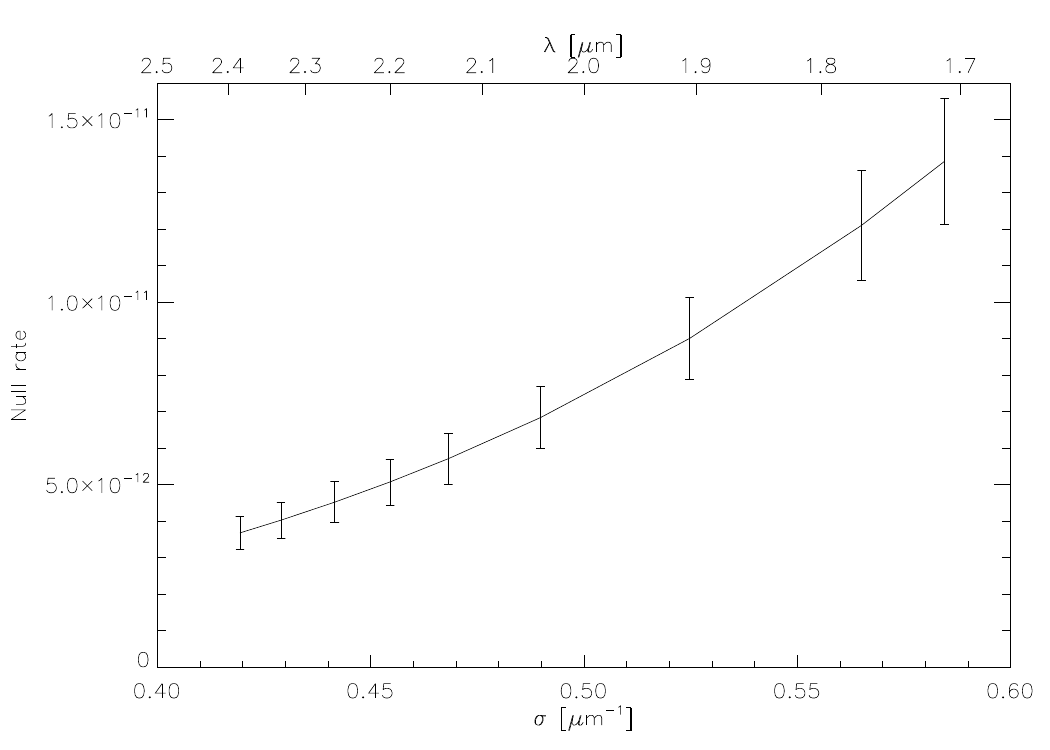}
  \caption{Dynamic tip/tilt tracking null depth contribution across channels.}
  \label{fig-epsilon-dyn}
\end{figure}

Dynamic tracking leakage exhibits a fourth-power wavelength dependency ($\sigma_i^4$), averaging:
\begin{equation*}
  \boldsymbol{\moy[\Tc]{N_{\varepsilon,\mdyn}} = 7.9\E{-12}}.
\end{equation*}

This term is negligible compared to static imbalance $N_{\varepsilon,\msta}$. Fourth-power scaling ($\sigma_\theta^4$) requires tracking errors to reach $\sigma_\theta = 2$~arcsec RMS before dynamic leakage matches static imbalance.

Short-term null stability driven by tip/tilt tracking is $\boldsymbol{\sigma_{N,\varepsilon,\tauu} = 9\E{-13}}$—negligible compared to path delay jitter.

%¤¤¤¤¤¤¤¤¤¤¤¤¤¤¤¤¤¤¤¤¤¤¤¤¤¤¤¤¤¤¤¤¤¤¤¤¤¤¤¤¤¤¤¤¤¤¤¤¤¤¤¤¤¤¤¤¤¤¤¤¤¤¤¤¤¤¤¤¤¤¤¤¤¤¤¤¤¤¤
\subsection{Polarization Analysis}
\label{sec--etude}

Polarization leakage comprises differential phase delay $\Delta\phi_\msp$ and polarization rotation misalignment $\alpha_\mrot$ (Section~\ref{sec-effets-polarisation}):
\begin{equation}
  N_\mpol = \frac{1}{16}(\Delta\phi_\msp)^2+\frac14\alpha_\mrot^2.
\end{equation}

Per-channel polarization leakage is:
\begin{equation}
  N_{\mpol,i} =
  \frac14\GP{\pi\sigma_i\Delta\delta_{\msp,i}}^2+ \frac14\alpha_{i,\mrot}^2.
\end{equation}

Because polarization components cannot be isolated directly without adding chromatic optics, polarization leakage $N_\mpol$ is estimated by subtracting modeled phase and flux terms from measured null depths:
\begin{equation}
  N_\mpol \simeq N_\mmoy-N_\phi-N_\varepsilon.
\end{equation}

Substituting $N_\mmoy = 8.8\E{-6}$, $N_\phi = 3.9\E{-6}$, and $N_\varepsilon = N_{\varepsilon,\msta} = 9.5\E{-7}$ yields an estimated polarization leakage of $\boldsymbol{N_\mpol = 3.9\E{-6}}$.

Attributing this term entirely to differential phase delay gives $\Delta\delta_\msp = 2.5$~nm; attributing it entirely to field rotation gives $\alpha_\mrot = 3.9\text{~mrad} = 14$~arcmin.

Polarization sensitivity was tested experimentally by driving angular offsets on mirror M1$\mb$ and compensating pointing via M6$\mb$. Steering mirror M1$\mb$ alters both polarization angles and spatial beam coupling on collimator M0 (owing to chromatic fiber numerical aperture shifts: $0.2$ at 1~\mum to $0.4$ at 2~\mum).

\begin{figure} \centering
  \subfloat[Zemax optical model simulation.]{\label{fig-null-vs-M1x1}
    \FIG{0.49}{false}{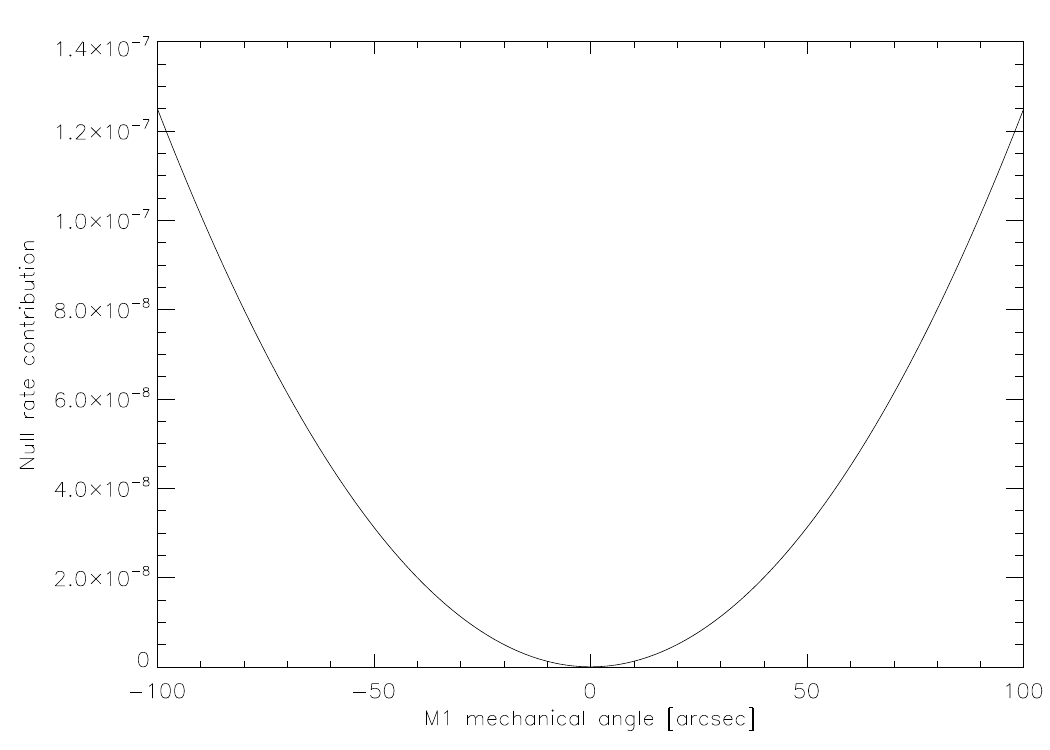}}
  \hfill\subfloat[Experimental measurements across channels.]{\label{fig-null-vs-M1x2}\FIG{0.49}{false}{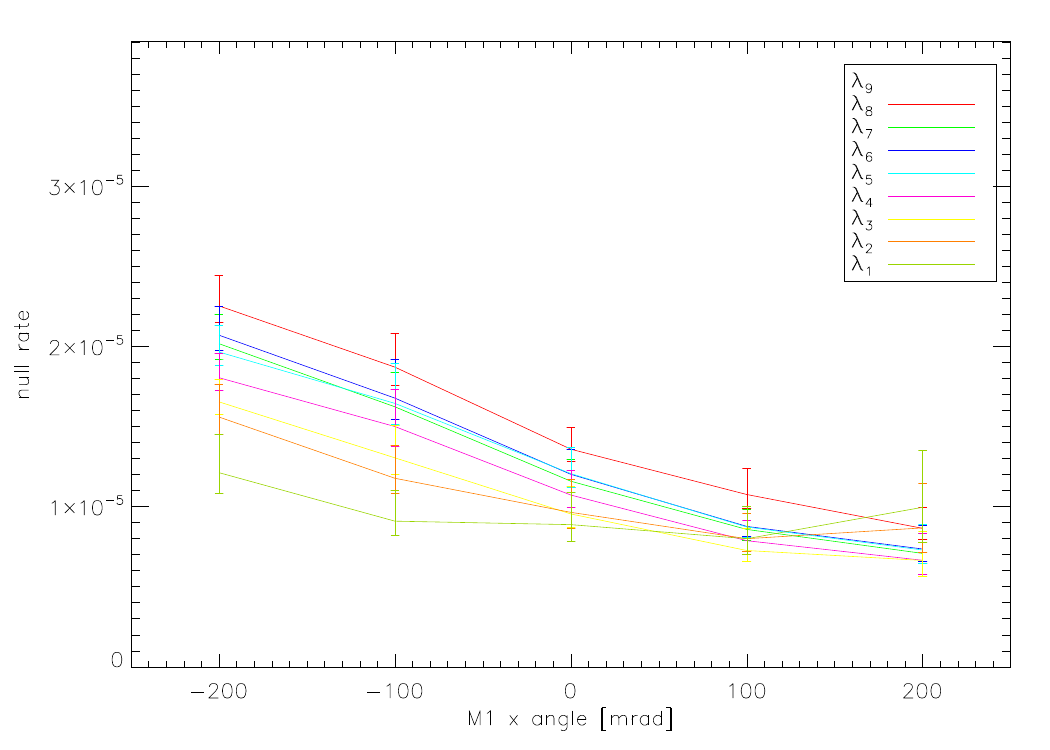}}
  \caption[Null depth versus siderostat M1$\mb$ tilt angle.]{Null depth per channel versus siderostat M1$\mb$ $x$-axis mechanical tilt angle.}
  \label{fig-null-vs-M1x}
\end{figure}

Figure~\ref{fig-null-vs-M1x1} shows Zemax modeled polarization leakage versus mechanical tilt angle $\alpha_x$ on mirror M1$\mb$ for linearly polarized light:
\begin{equation}
  N_{i,\mpol}(\alpha_x,\alpha_y) = N_{i,\mpol,0}+A_x(\alpha_x-\alpha_{x,i})^2+
  A_y(\alpha_y-\alpha_{y,i})^2.
\end{equation}

Zemax simulations give $A_x = 1.25\E{-11}$~arcsec$^{-2}$.

Figure~\ref{fig-null-vs-M1x2} plots experimental channel null depths versus $x$-axis tilt. Minimum null depth occurs at $\alpha_x \approx 250$~arcsec offset. Experimental sensitivity measures $A_x \approx 5\E{-11}$~arcsec$^{-2}$ (4$\times$ higher than Zemax models), reflecting coupled chromatic flux shifts. Adjusting M1 mirror angles provided an additional alignment degree of freedom to optimize global null depths.

%¤¤¤¤¤¤¤¤¤¤¤¤¤¤¤¤¤¤¤¤¤¤¤¤¤¤¤¤¤¤¤¤¤¤¤¤¤¤¤¤¤¤¤¤¤¤¤¤¤¤¤¤¤¤¤¤¤¤¤¤¤¤¤¤¤¤¤¤¤¤¤¤¤¤¤¤¤¤¤
\subsection{100-Second Performance Summary}
\label{sec-synthese-analyse}

Figure~\ref{fig-null-vs-sig2} compares channel null depth profiles against individual error contributions.

\begin{figure} \centering
  \FIG{0.7}{false}{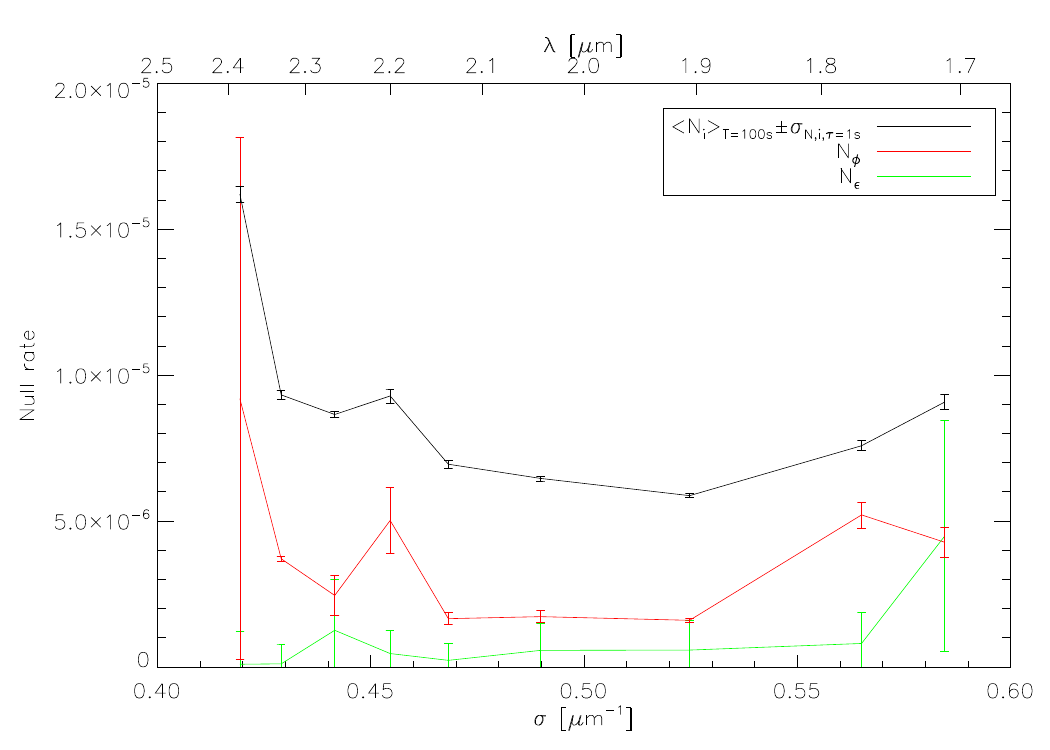}
  \caption[Breakdown of null depth error contributions across channels.]{Channel null depth profile compared against individual error contributions.}
\label{fig-null-vs-sig2}
\end{figure}

Phase and flux leakage drive null depth increases at passband edges. Local phase dispersion near 2.2~\mum stems from anti-reflection coating thickness variations on MMZ plates. Unmodeled residual leakage across channels is attributed to polarization errors.

\begin{table} \centering
  \caption[Error budget breakdown for the best 100~s null depth run.]{Error budget breakdown for the best 100~s null depth run compared against specifications.}
  \medskip
  \renewcommand{\arraystretch}{1.25} 
  \begin{minipage}{7.3cm} \centering
    \begin{tabular}[table]{lrr}
      \hline \hline Term & Requirement & Measured \\
      \hline $\moy[\Tc]{N_\mmoy}$ & $\e{-4}$ & $8.8\E{-6}$ \\
      \hline $\moy[\Tc]{N_\phi}$ & $7\E{-5}$ & $3.9\E{-6}$ \\
      $\moy[\Tc]{N_{\delta,w}}$ & \multirow{2}{*}{$\Bigg\}\ 3.5\E{-5}$} &
      $1.4\E{-6}$ \\
      $\moy[\Tc]{N_{\delta,\mref}}$ & & $\e{-7}$ \\
      $N_{\sigma}$ & $3.5\E{-5}$ & $2.4\E{-6}$ \\
      $\moy[\Tc]{N_\varepsilon}$ & $2\E{-5}$ & $9.5\E{-7}$ \\
      $N_{\varepsilon,\msta}$ & \multirow{2}{*}{$\Bigg\}$\hspace{10pt}$\
        2\E{-5}$} & $9.5\E{-7}$ \\
      $\moy[\Tc]{N_{\varepsilon,\mdyn}}$ & & $7.9\E{-12}$ \\
      $N_\mpol$\footnote{Estimated by subtracting phase and flux terms} &
      $\e{-5}$ & $3.9\E{-6}$ \\
      \hline $\sigma_{N,\tauu}$ & $1,5\E{-5}$ & $9\E{-8}$ \\
      \hline $\sigma_{N,\delta,\tauu}$ & $1.5\E{-5}$ & $\e{-7}$ \\
      $\sigma_{N,\varepsilon,\tauu}$ & $3\E{-6}$ & $9\E{-13}$ \\
      \hline \hline
    \end{tabular}
  \end{minipage}
  \label{tab-null-100s}
\end{table}

Table~\ref{tab-null-100s} summarizes the 100~s error budget breakdown. Measured error contributions surpass specification targets by 1 to 2 orders of magnitude, delivering a mean null depth 10 times deeper and a short-term stability 100 times better than baseline requirements.

%¤¤¤¤¤¤¤¤¤¤¤¤¤¤¤¤¤¤¤¤¤¤¤¤¤¤¤¤¤¤¤¤¤¤¤¤¤¤¤¤¤¤¤¤¤¤¤¤¤¤¤¤¤¤¤¤¤¤¤¤¤¤¤¤¤¤¤¤¤¤¤¤¤¤¤¤¤¤¤
\subsection{Null Depth Performance Under Injected Disturbances}
\label{sec-te-perturb}

Section~\ref{sec-lqg-persee} demonstrated LQG rejection of active path disturbances. Active LQG control maintains residual path delay jitter at 0.74~nm RMS despite piezo strain-gauge noise and mechanical bench vibration.

We evaluated nulling performance under injected formation-flying disturbances (15~nm RMS low-frequency positioning drift, 3~Hz mean reaction wheel frequency; $19.5$~nm RMS total injected jitter, Section~\ref{sec-analyse-typique}).

Active LQG control, targetting the 20 strongest vibration lines, suppressed residual path delay jitter to $\boldsymbol{\sigma_\delta = 1.17\text{~\bf nm RMS}}$. Standard integrator control left a residual jitter of $\sigma_\delta = 5.57$~nm RMS (5$\times$ higher).

This path delay reduction suppressed path-induced null leakage $\moy[\Tc]{N_{\delta,w}}$ by a factor of 23, as shown in Figure~\ref{fig-null-int-lqg}.

\begin{figure} \centering
  \subfloat[Integrator control.]{\label{fig-null-int}
    \FIG{0.49}{false}{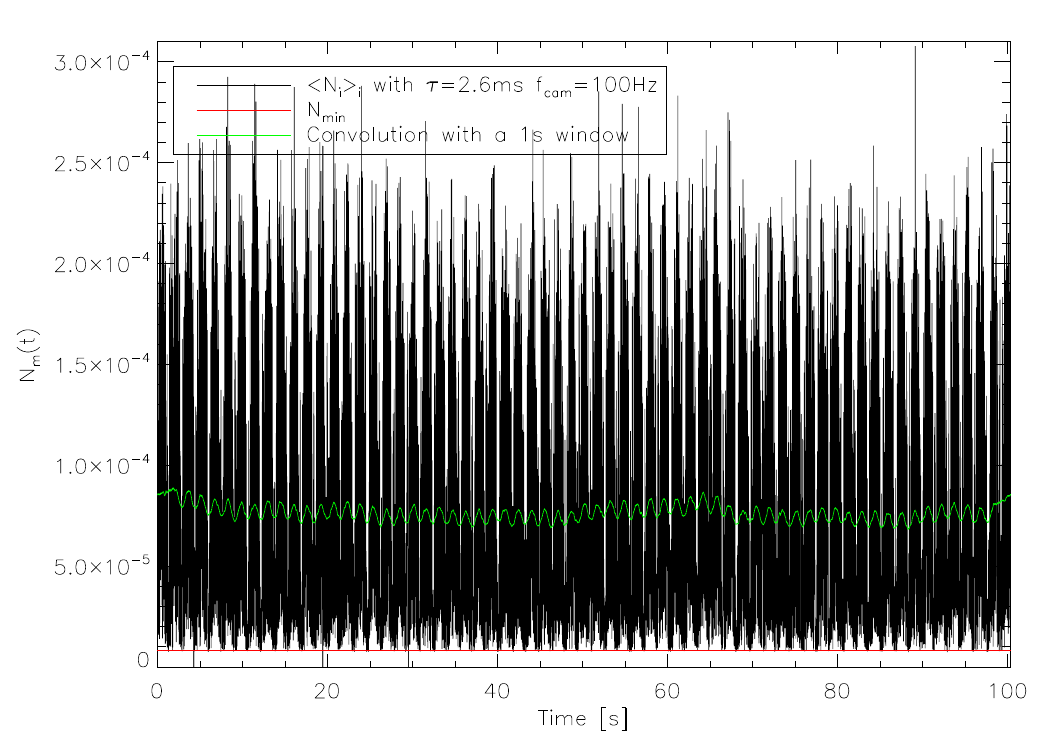}}
  \hfill\subfloat[LQG control.]{\label{fig-null-lqg}
  \FIG{0.49}{false}{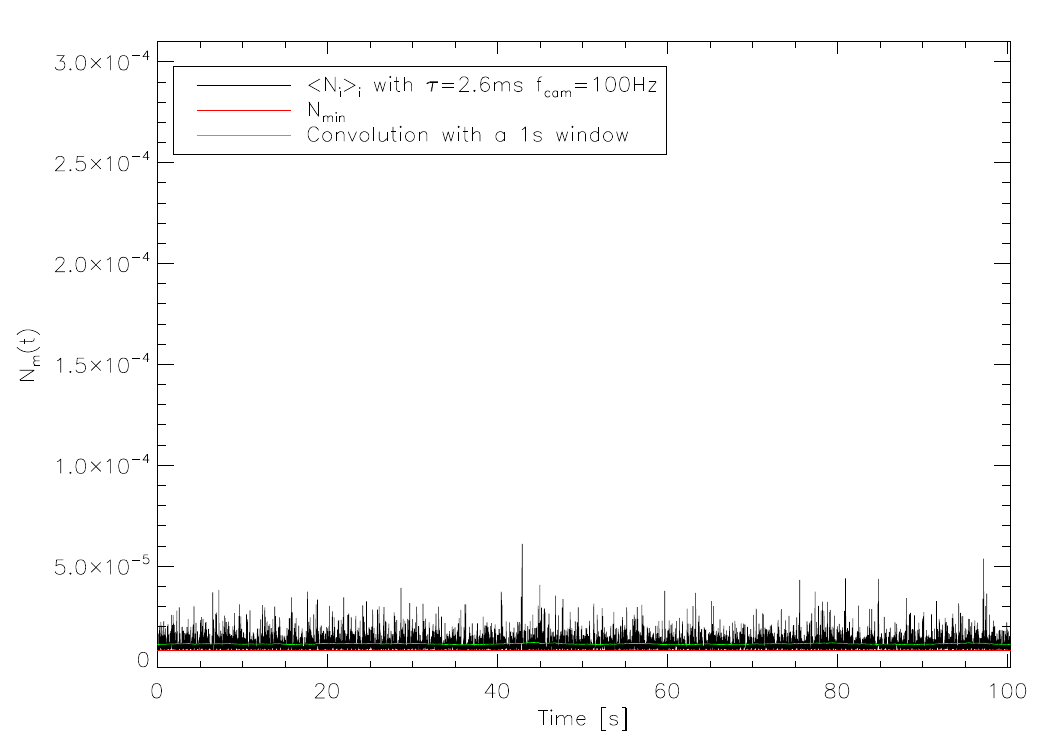}}
  \caption{Null depth time series under injected disturbances for integrator and LQG control.}
  \label{fig-null-int-lqg}
\end{figure}

Under integrator control (Figure~\ref{fig-null-int}), path jitter induces saturation spikes, yielding a mean null depth $N_\mmoy = 8.5\E{-5}$ ($\sigma_{N,\tauu} = 3\E{-6}$).

Under active LQG control (Figure~\ref{fig-null-lqg}), residual path delay jitter is suppressed, achieving a much deaper mean null depth of $\boldsymbol{\moy[\Tc]{N_\mmoy} = 1.14\E{-5}}$ and a short-term stability of $\boldsymbol{\sigma_{N,\tauu} = 3\E{-7}}$—7 times deeper and 10 times more stable than integrator control.

Comparing null depth PSDs (Figure~\ref{fig-psd-null-int-lqg}) confirms vibration suppression.

\begin{figure} \centering
  \subfloat[Integrator control.]{\label{fig-psd-null-int}
    \FIG{0.49}{false}{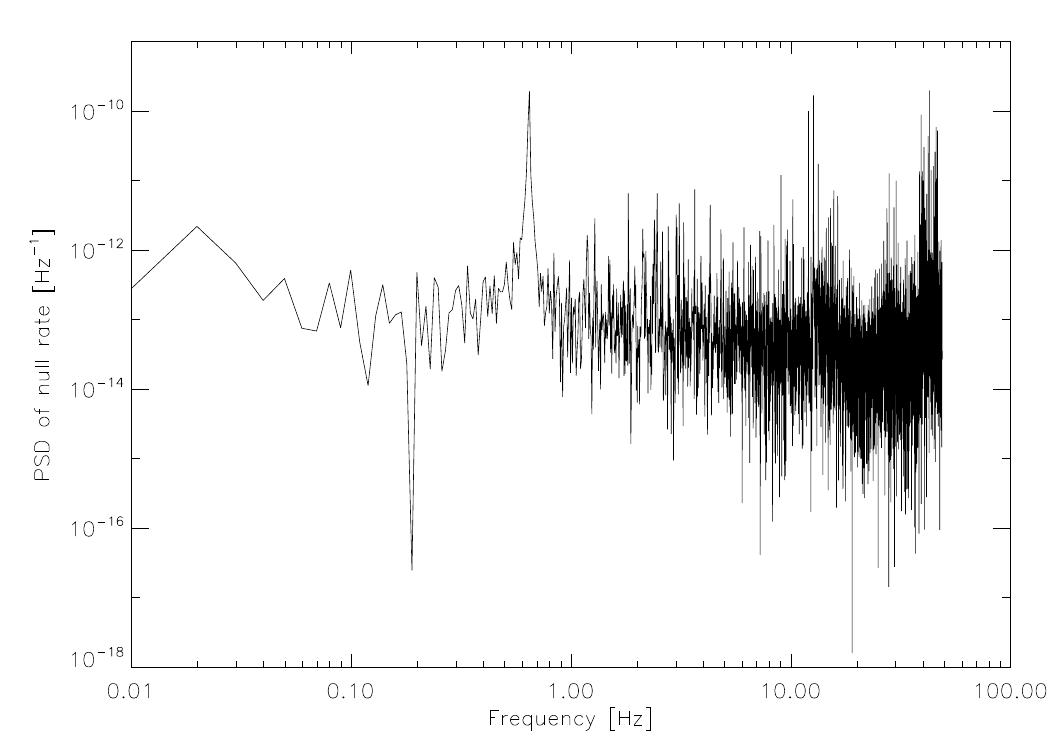}}
  \hfill\subfloat[LQG control.]{\label{fig-psd-null-lqg}
  \FIG{0.49}{false}{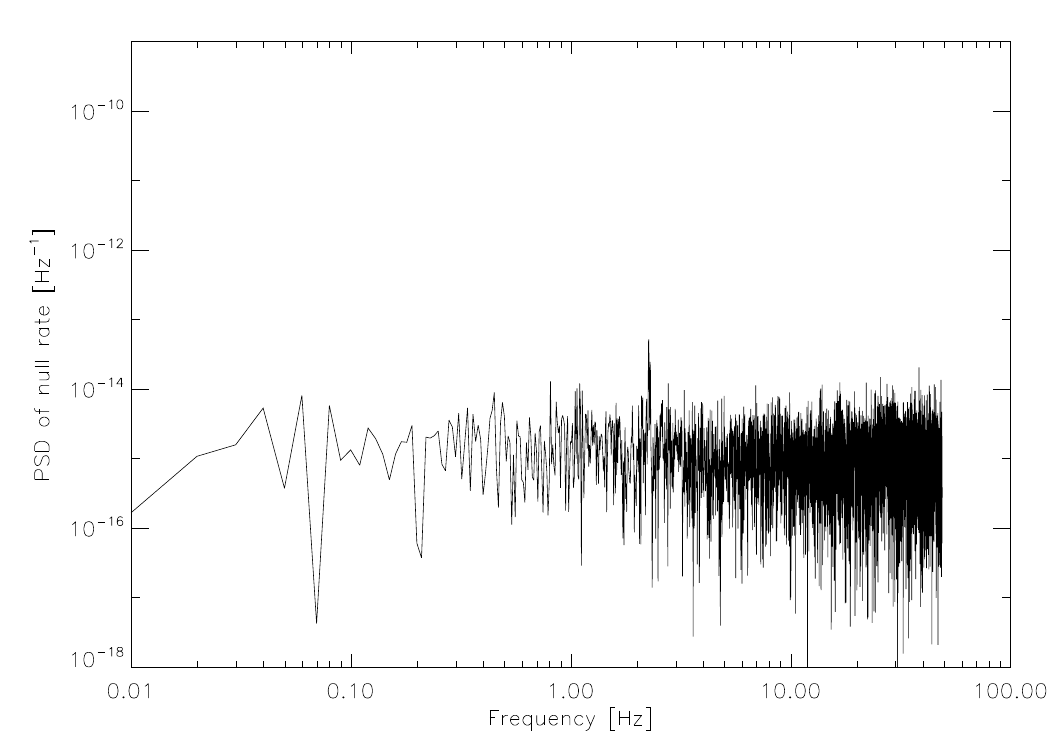}}
  \caption{Null depth PSDs under integrator and LQG control.}
  \label{fig-psd-null-int-lqg}
\end{figure}

Integrator PSDs (Figure~\ref{fig-psd-null-int}) display strong harmonic distortion spikes. Active LQG control (Figure~\ref{fig-psd-null-lqg}) suppresses these harmonic lines, leaving only a minor aliased 2.3~Hz line power residual (50~Hz aliased at 97.7~Hz camera frame rate).

\begin{table}
  \centering
  \caption{Performance summary under injected formation-flying disturbances.}
  \medskip
  \renewcommand{\arraystretch}{1.25}
  \begin{tabular}[table]{lrrr}
    \hline\hline 
    Metric & Requirement & Integrator & LQG \\
    \hline $\sigma_\delta$ [nm RMS] & 1.7 & 5.57 & 1.17 \\
    \hline $\moy[\Tc]{N_\mmoy}$ & $\e{-4}$ & $8.5\E{-5}$ & $1.14\E{-5}$ \\
    \hline $\moy[\Tc]{N_{\delta,w}}$ & $3.5\E{-5}$ & $7.3\E{-5}$ & $3.3\E{-6}$ \\
    \hline $\sigma_{N,\tauu}$ & $1.5\E{-5}$ & $3\E{-6}$ & $3\E{-7}$ \\
    \hline\hline
  \end{tabular}
  \label{tab-dis}
\end{table}

Table~\ref{tab-dis} summarizes control performance under injected flight-like disturbances. Both controllers maintain mean null depths below $10^{-4}$. However, LQG control suppresses path delay jitter below sub-nanometric levels, meeting strict mission error budgets.

%§§§§§§§§§§§§§§§§§§§§§§§§§§§§§§§§§§§§§§§§§§§§§§§§§§§§§§§§§§§§§§§§§§§§§§§§§§§§§§§
\section{Multi-Hour Long-Term Null Stability}
\label{sec-etude-preliminaire}

%¤¤¤¤¤¤¤¤¤¤¤¤¤¤¤¤¤¤¤¤¤¤¤¤¤¤¤¤¤¤¤¤¤¤¤¤¤¤¤¤¤¤¤¤¤¤¤¤¤¤¤¤¤¤¤¤¤¤¤¤¤¤¤¤¤¤¤¤¤¤¤¤¤¤¤¤¤¤¤
\subsection{Experimental Conditions}
\label{sec-conditions-mesure}

%-------------------------------------------------------------------------------
\subsubsection{Long-Term Test Overview}
\label{sec-presentation-mesure}

Long-term stability was evaluated over a 10-hour nighttime run. Air conditioning was disabled to eliminate acoustic vibration, allowing room temperature to drift slowly. Uncompensated thermal expansion shifted internal MMZ phase quadrature outside operational limits after $\sim7$~hours, bounding quantitative analysis to the first 7~hours. Room temperature drift recorded outside the bench enclosure is plotted in Figure~\ref{fig-temperature}.

\begin{figure} \centering
	\FIG{0.7}{false}{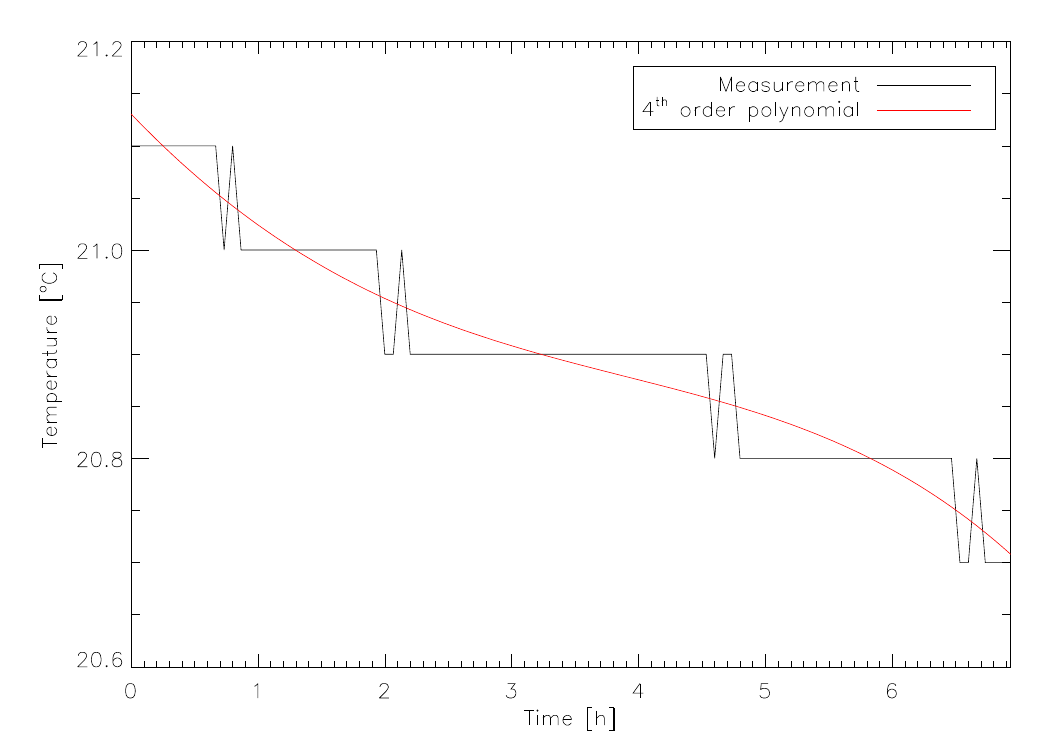}
    \caption[Room temperature drift during long-term testing.]{Room temperature drift recorded outside the bench enclosure during long-term testing.}
\label{fig-temperature}
\end{figure}

Room temperature drifted at an average rate of $-13$~\muk.s$^{-1}$ (Figure~\ref{fig-temperature}). Thermal insulation enclosed within the bench slowed internal MMZ temperature drift.

%-------------------------------------------------------------------------------
\subsubsection{Automated Calibration Workflow}
\label{sec-autom-etal}

Fringe sensor demodulation matrices were recalibrated hourly (Section~\ref{sec-etalonnage-FS}). Fine tip/tilt setpoint optimization (Section~\ref{sec-procedure-correction}) and 3-point OPD dithering (Section~\ref{sec-optimisation-position}) executed automatically prior to each 100~s science acquisition. Calibration execution times are plotted in Figure~\ref{fig-time}.

\begin{figure} \centering
  \subfloat[Execution time between tip/tilt and OPD calibration.]{\label{fig-time1}
    \FIG{0.49}{false}{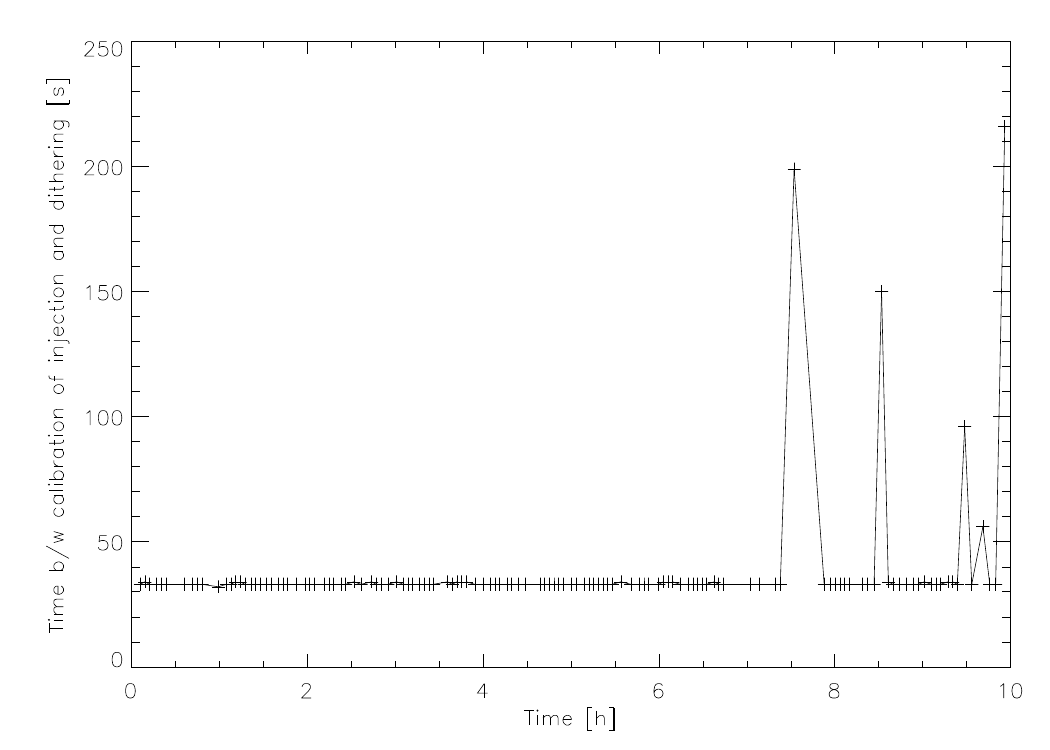}}
  \hfill\subfloat[Execution time between OPD calibration and science acquisition.]{\label{fig-time2}
  \FIG{0.49}{false}{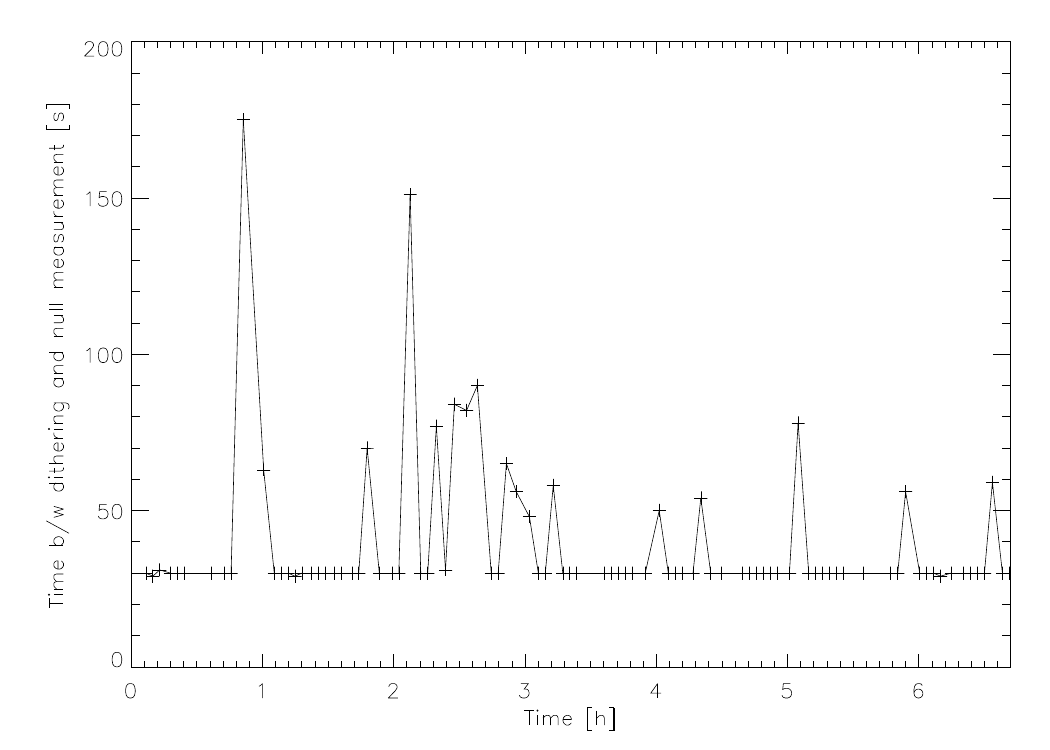}}
  \caption{Automated calibration sequence execution times.}
  \label{fig-time}
\end{figure}

Automated calibration steps executed reproducibly within 33~s during the first 7~hours (Figure~\ref{fig-time1}). Manual interventions required during later runs extended delay intervals beyond 30~s (Figure~\ref{fig-time2}), increasing thermal setpoint drift $\delta_\mref$.

%-------------------------------------------------------------------------------
\subsubsection{Control Loop Stability}
\label{sec-stabilite-boucles}

Tracking loop residuals recorded over 7 hours are shown in Figure~\ref{fig-stab-7h}.

\begin{figure} \centering
  \subfloat[Tip/tilt tracking jitter.]{\label{fig-stab-7h1}
    \FIG{0.49}{false}{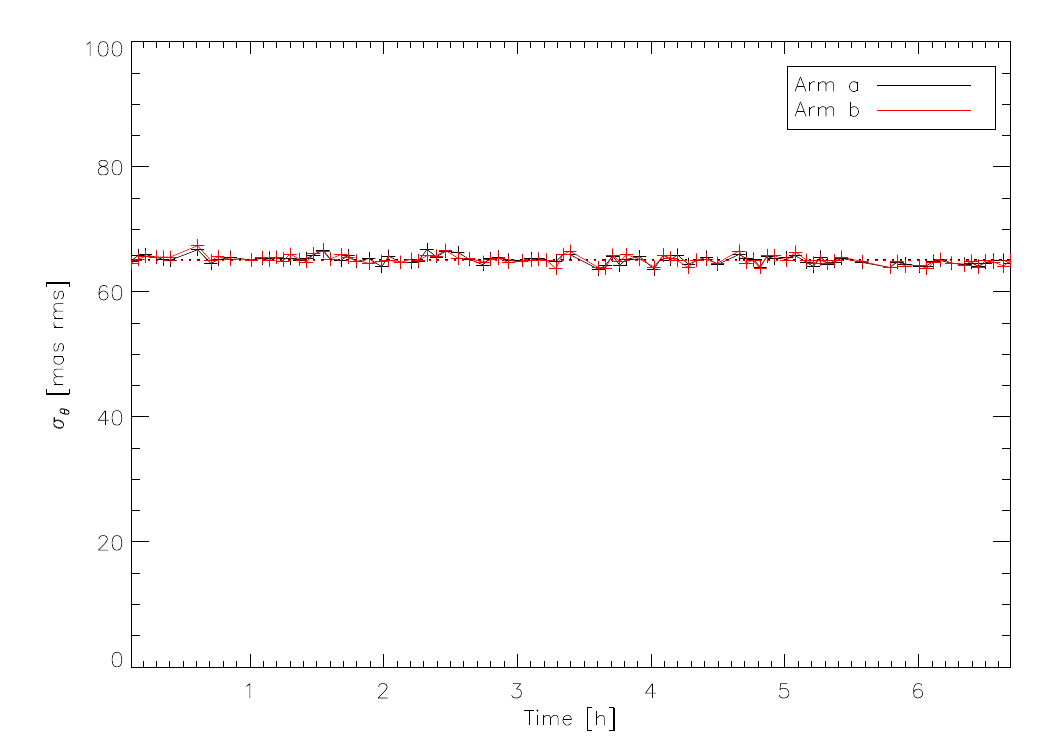}}
  \hfill\subfloat[Path delay jitter.]{\label{fig-stab-7h2}
    \FIG{0.49}{false}{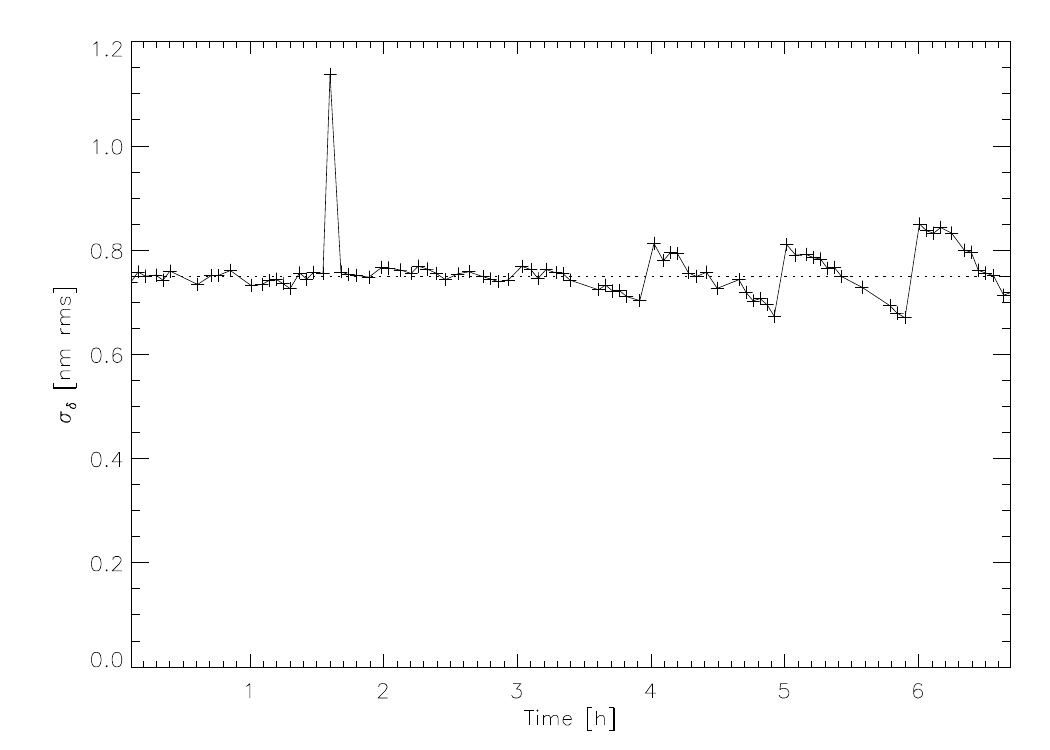}}
  \caption{Tracking loop residual stabilities recorded over 7 hours.}
  \label{fig-stab-7h}
\end{figure}

Tip/tilt tracking jitter remained stable across arms at $\sigma_\theta = 65$~mas RMS (Figure~\ref{fig-stab-7h1}). Path delay jitter (Figure~\ref{fig-stab-7h2}) remained stable at $\sigma_\delta = 0.76$~nm RMS during the first 3 hours. Thermal drift altered fringe sensor calibration matrices after 3 hours, causing apparent tracking drift.

Internal MMZ thermal drift calculated from fringe sensor phase shifts is plotted in Figure~\ref{fig-deltal}.

\begin{figure} \centering
	\FIG{0.7}{false}{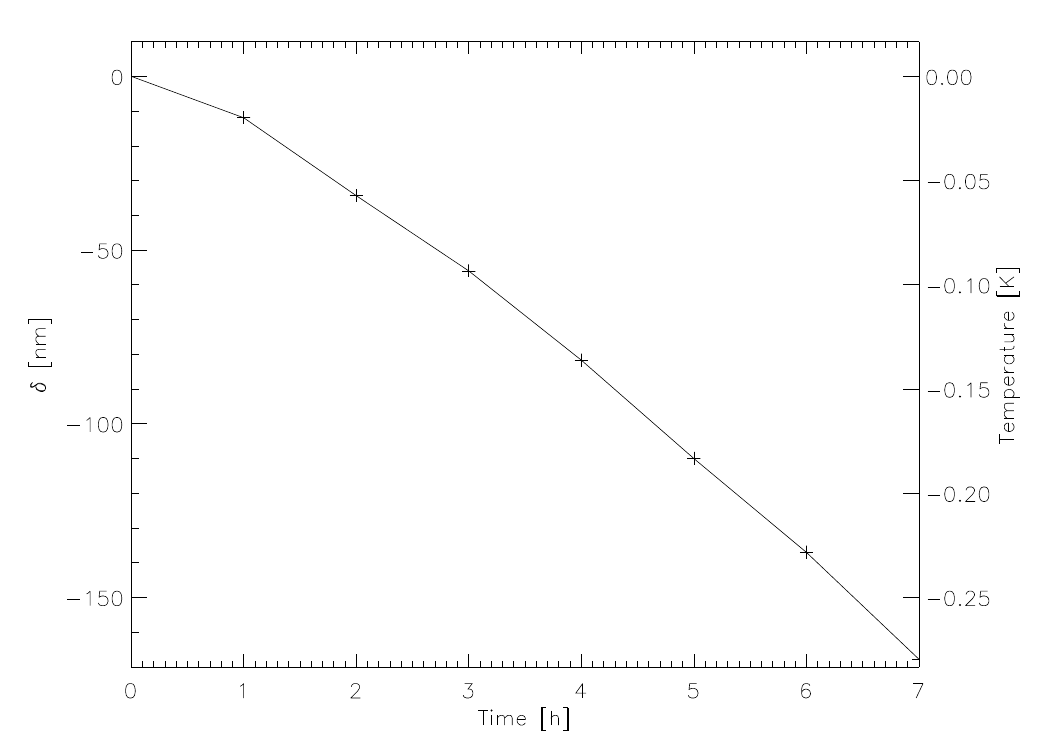}
  \caption{Internal MMZ path delay drift recorded over 7 hours.}
  \label{fig-deltal}
\end{figure}

Internal path delay drifted at a constant rate of $-11$~\muk.s$^{-1}$ ($-7$~pm.s$^{-1}$). Thermal enclosures mitigated rapid fluctuations but could not stop slow linear drift.

%¤¤¤¤¤¤¤¤¤¤¤¤¤¤¤¤¤¤¤¤¤¤¤¤¤¤¤¤¤¤¤¤¤¤¤¤¤¤¤¤¤¤¤¤¤¤¤¤¤¤¤¤¤¤¤¤¤¤¤¤¤¤¤¤¤¤¤¤¤¤¤¤¤¤¤¤¤¤¤
\subsection{Long-Term Nulling Results}
\label{sec-mesure-stabilite}

100~s null depth acquisitions recorded over 7~hours are plotted in Figure~\ref{fig-null-6h}.

\begin{figure} \centering
  \subfloat[Time series of mean null depth $N_\mmoy$.]{\label{fig-null-6h1}
    \FIG{0.49}{false}{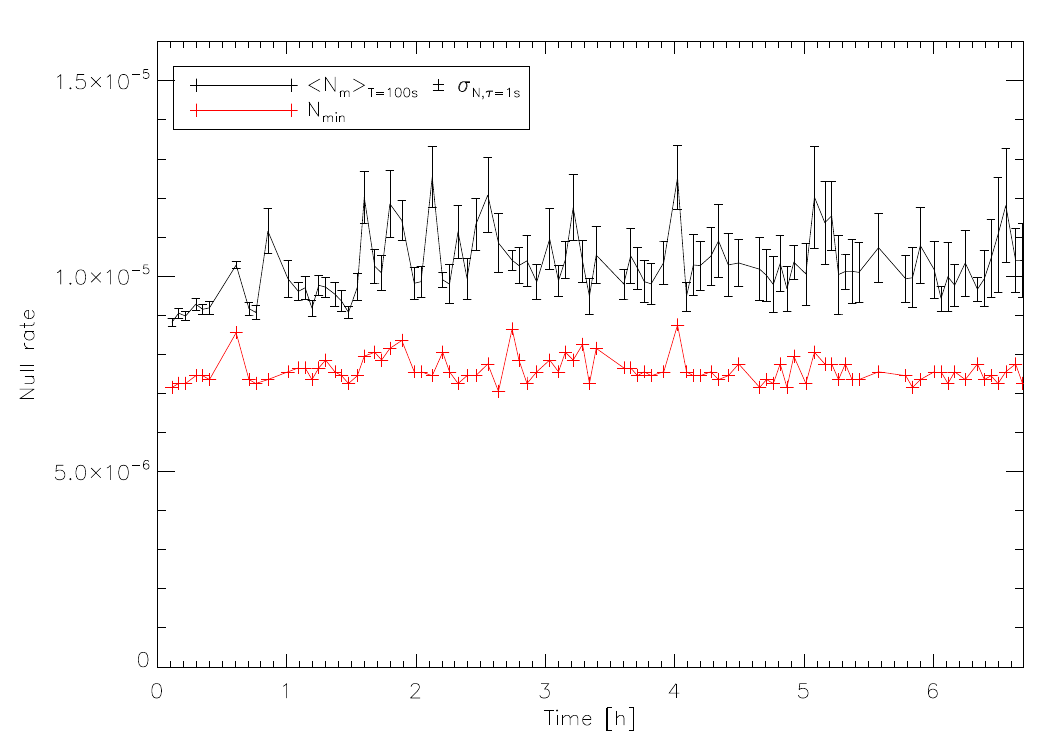}}
  \hfill\subfloat[Multi-hour mean null depth per science channel.]{\label{fig-null-6h2}
    \FIG{0.49}{false}{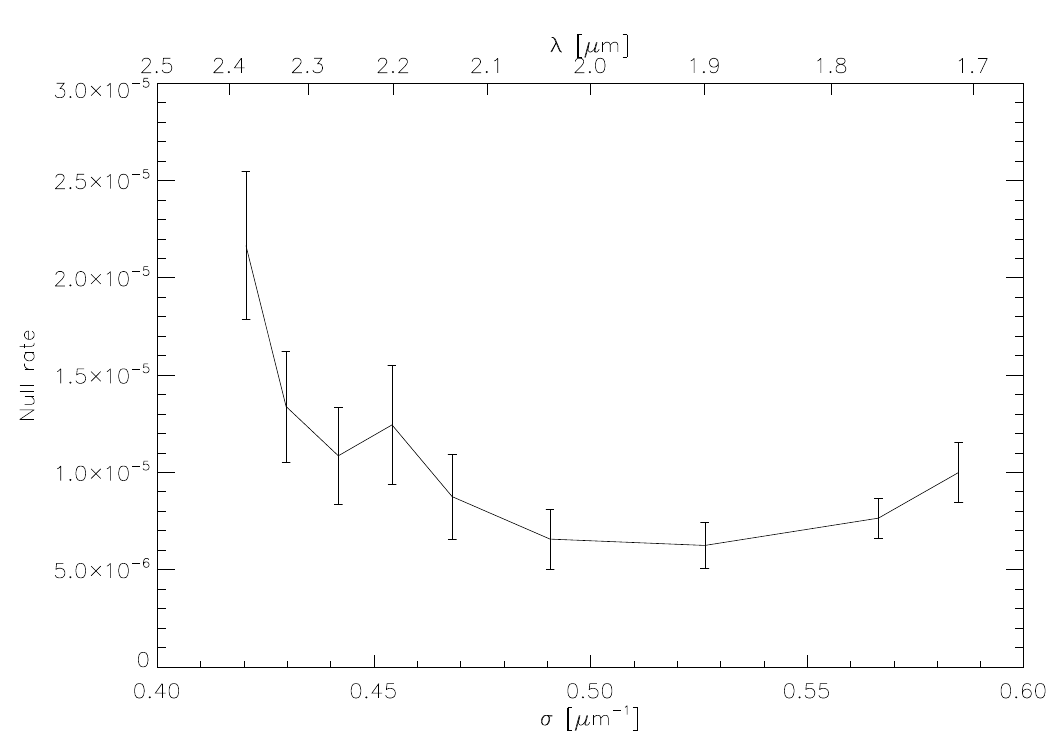}}
  \caption{Multi-hour null depth performance dataset.}
  \label{fig-null-6h}
\end{figure}

Figure~\ref{fig-null-6h1} plots 100~s mean null depths (black line) and static underlying nulls $N_{\mmoy,0}$ (red line). Multi-hour mean null depth across the $(\Delta\sigma/\sigma)_\mmoy = 37\%$ bandwidth reached:
\begin{equation*}
  \boldsymbol{\moy[\Ts]{N_\mmoy} = 1.02\E{-5}}.
\end{equation*}

Multi-hour long-term null stability across 100~s binned acquisition windows achieved:
\begin{equation*}
  \boldsymbol{\sigma_{N,\tauc} = 8\E{-7}}.
\end{equation*}

Figure~\ref{fig-null-6h2} plots time-averaged channel null depths. Null depths remain bounded between $6.2\E{-6}$ and $2.15\E{-5}$ across all 9 channels over 7~hours.

%¤¤¤¤¤¤¤¤¤¤¤¤¤¤¤¤¤¤¤¤¤¤¤¤¤¤¤¤¤¤¤¤¤¤¤¤¤¤¤¤¤¤¤¤¤¤¤¤¤¤¤¤¤¤¤¤¤¤¤¤¤¤¤¤¤¤¤¤¤¤¤¤¤¤¤¤¤¤¤
\subsection{Phase Perturbations}
\label{sec-perturbations-phase-2}

%-------------------------------------------------------------------------------
\subsubsection{Path Delay Jitter Contribution}
\label{sec-perturbation-difference-1}

Path delay tracking jitter remained stable over multi-hour runs ($\sigma_\delta = 0.76$~nm RMS). Channel leakage contributions $\moy[\Ts]{N_{\delta,w,i}}$ are plotted in Figure~\ref{fig-N-delta-w-7h}.

\begin{figure} \centering
  \FIG{0.7}{false}{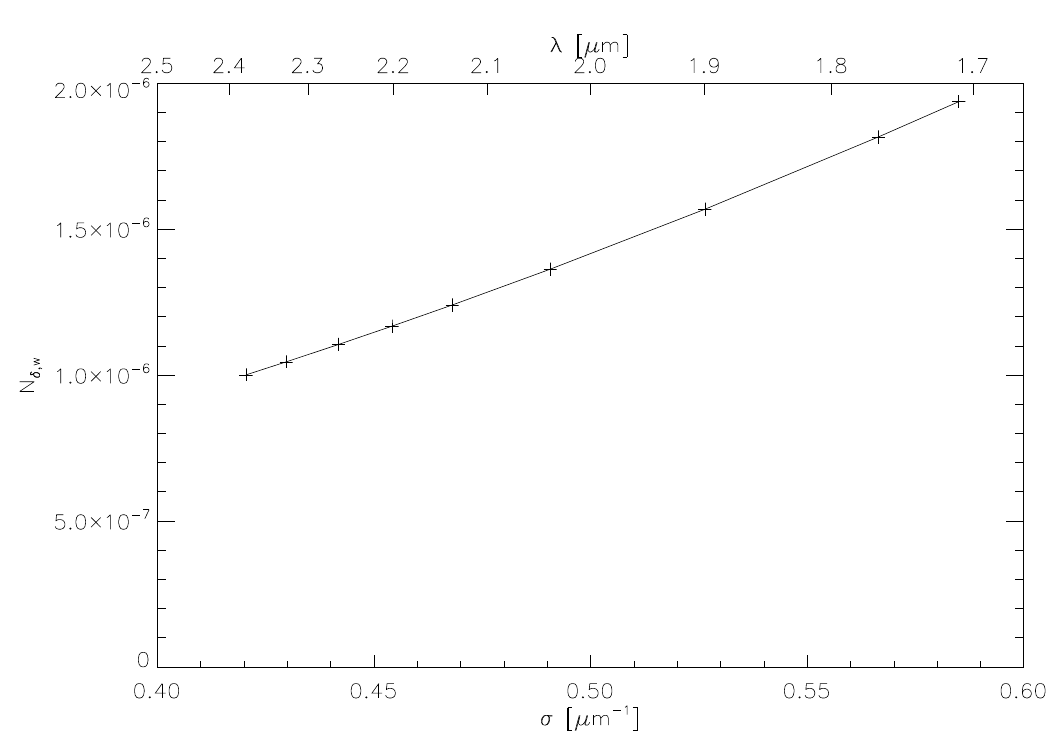}
  \caption{Multi-hour path delay jitter null depth contribution across channels.}
  \label{fig-N-delta-w-7h}
\end{figure}

The mean path delay jitter contribution across channels was:
\begin{equation*}
  \boldsymbol{\moy[\Ts]{N_{\delta,w}} = 1.4\E{-6}}.
\end{equation*}

Long-term path delay jitter stability across runs was $\boldsymbol{\sigma_{N,\delta,w,\tauc} = 2.2\E{-7}}$.

%-------------------------------------------------------------------------------
\subsubsection{Chromatic Dispersion Analysis}
\label{sec-analyse-chromatisme-1}

Chromatic dispersion profiles derived during setpoint optimization scans (Figure~\ref{fig-N-sigma-7h1}) and corresponding null leakage contributions are plotted in Figure~\ref{fig-N-sigma-7h2}.

\begin{figure} \centering
  \subfloat[Mean chromatic dispersion offsets $\delta_i$.]{\label{fig-N-sigma-7h1}
    \FIG{0.49}{false}{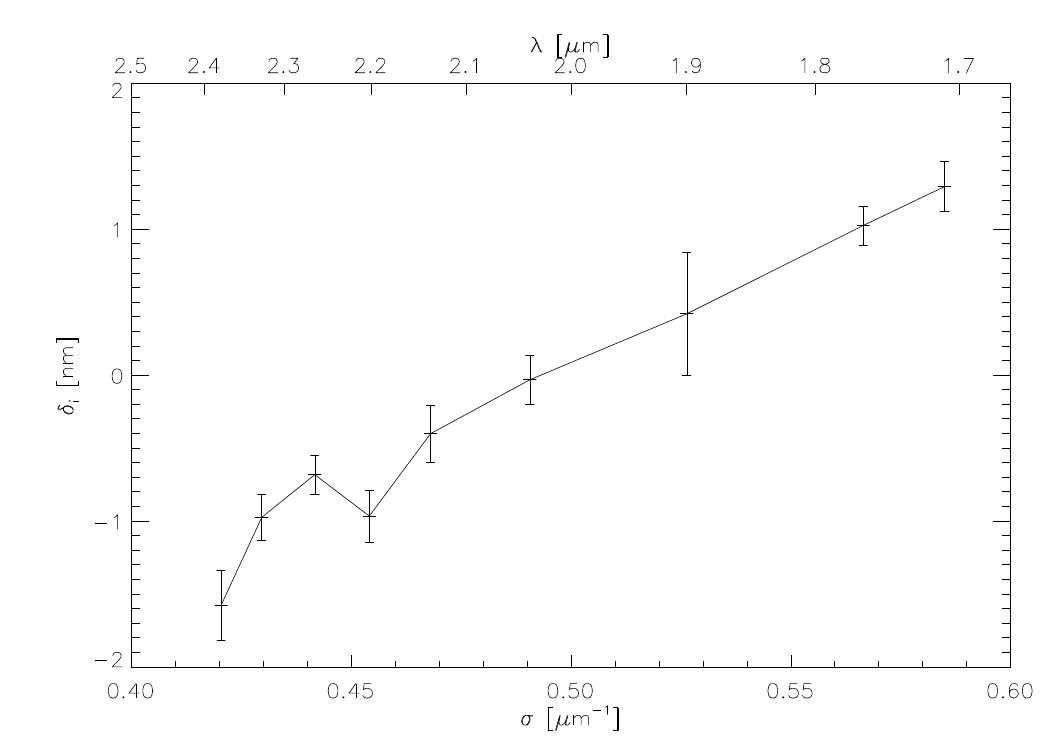}}
  \hfill\subfloat[Chromatic dispersion null depth contribution $N_{\sigma,i}$.]{\label{fig-N-sigma-7h2}
    \FIG{0.49}{false}{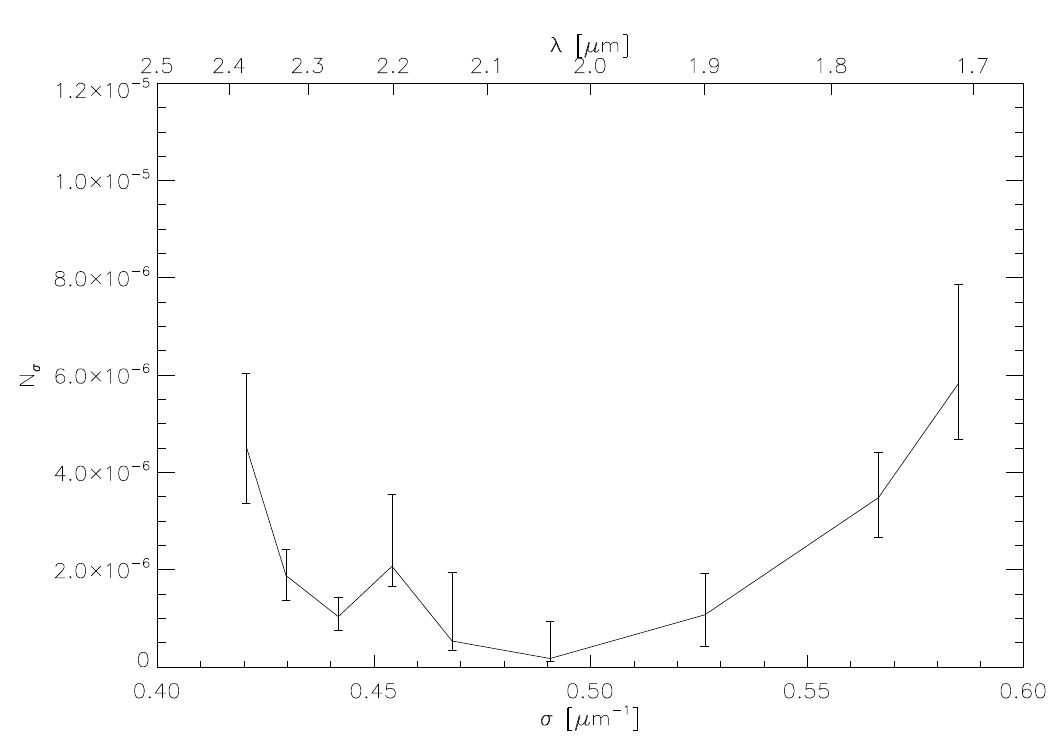}}
  \caption{Chromatic dispersion profiles and channel null depth contributions over 7 hours.}
  \label{fig-N-sigma-7h}
\end{figure}

Mean chromatic dispersion leakage across channels was:
\begin{equation*}
  \boldsymbol{N_\sigma = 2.3\E{-6}},
\end{equation*}
with multi-hour chromatic stability $\boldsymbol{\sigma_{N,\sigma,\tauc} = 7.3\E{-7}}$.

%-------------------------------------------------------------------------------
\subsubsection{Fringe Sensor Setpoint Offset Estimation}
\label{sec-estimation-biais}

Thermal drift ($-7$~pm.s$^{-1}$) over a 84~s post-calibration delay introduced a mean setpoint offset $\boldsymbol{\moy[\Ts]{\delta_\mref(t)} = -0.6\text{~\bf nm}}$. Cumulative thermal drift over time is shown in Figure~\ref{fig-der-null}.

\begin{figure} \centering
  \FIG{0.7}{false}{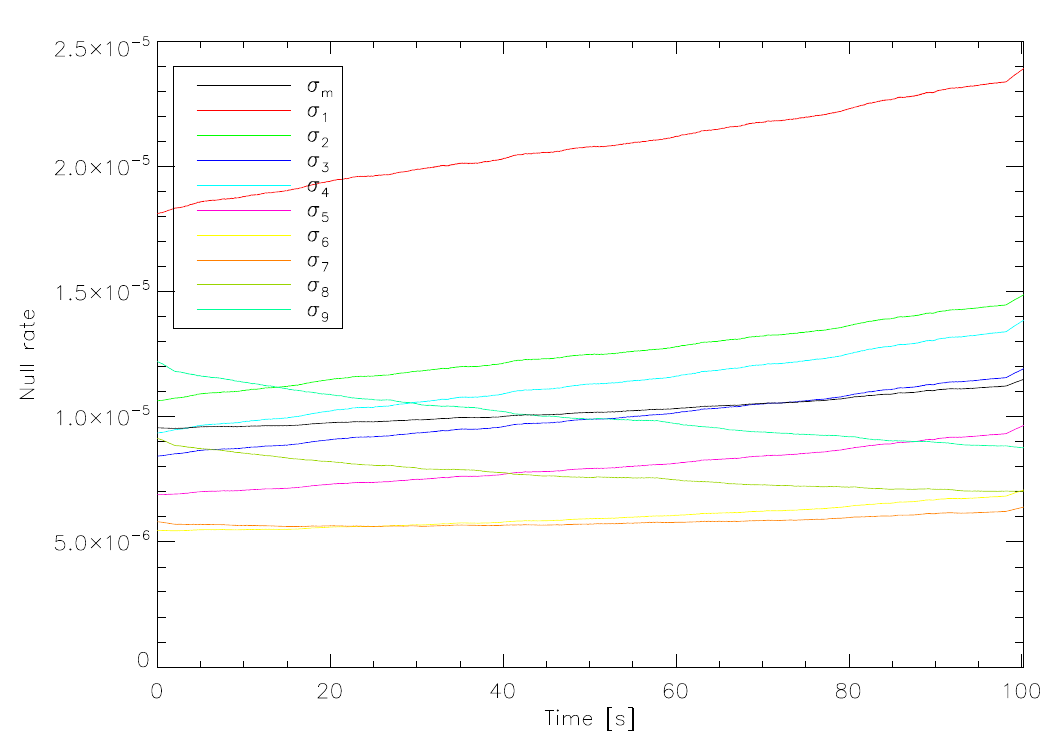}
  \caption{Mean null depth drift over time.}
  \label{fig-der-null}
\end{figure}

Negative setpoint offsets ($\delta_\mref = -0.6$~nm) degrade null depths at longer wavelengths (Figure~\ref{fig-N-delta-ref-7h}).

\begin{figure} \centering
  \FIG{0.7}{false}{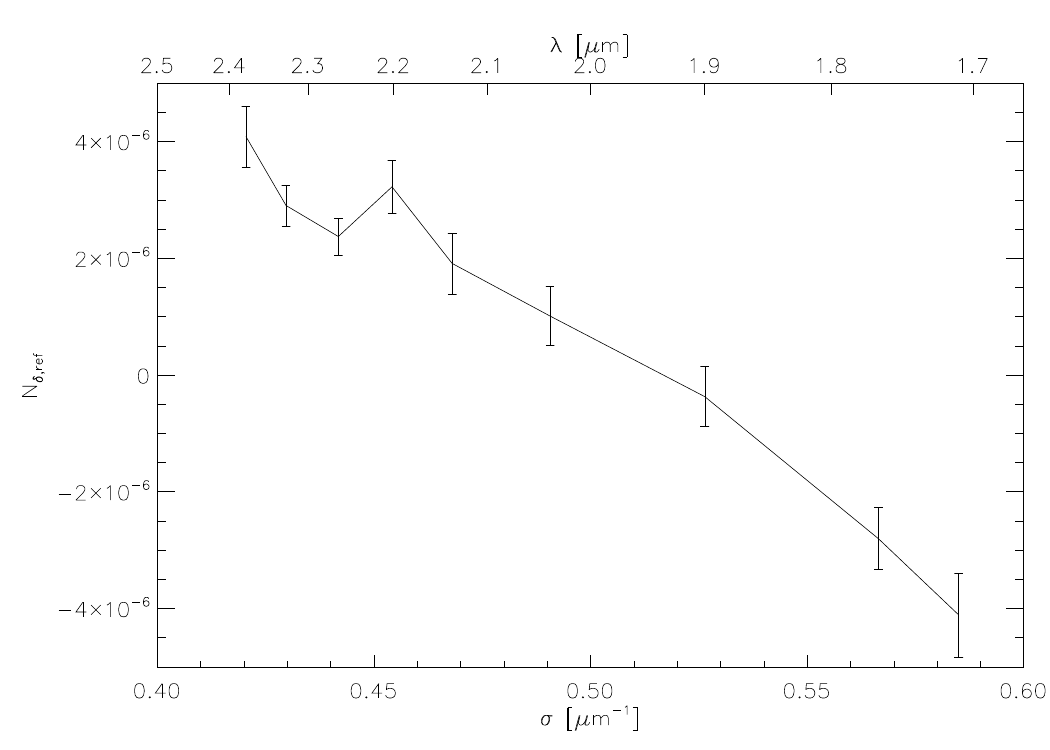}
  \caption{Multi-hour setpoint offset null depth contribution $\moy[T]{N_{\delta,\mref,i}}$.}
  \label{fig-N-delta-ref-7h}
\end{figure}

Mean setpoint offset leakage across channels was:
\begin{equation*}
  \boldsymbol{\moy[T]{N_{\delta,\mref}} = 9.1\E{-7}},
\end{equation*}
with multi-hour setpoint stability $\boldsymbol{\sigma_{N,\delta,\mref,\tauc} \approx 5\E{-7}}$.

\bigskip

Comparing experimental and modeled null depth PSDs over 7~hours (Figure~\ref{fig-psd-null}) confirms that path delay jitter drives short-term null fluctuations ($\sigma_{N,\delta,\tau=1\text{s}} = 10^{-7}$). Aliased line power harmonics (100~Hz aliased to 2.3~Hz; 150~Hz aliased to 45.3~Hz) match predicted path jitter spectra.

%¤¤¤¤¤¤¤¤¤¤¤¤¤¤¤¤¤¤¤¤¤¤¤¤¤¤¤¤¤¤¤¤¤¤¤¤¤¤¤¤¤¤¤¤¤¤¤¤¤¤¤¤¤¤¤¤¤¤¤¤¤¤¤¤¤¤¤¤¤¤¤¤¤¤¤¤¤¤¤
\subsection{Photometric Imbalance}
\label{sec-deseq-phot-2}

Measured multi-hour photometric imbalance profiles and channel leakage contributions are shown in Figure~\ref{fig-N-epsilon-7h}.

\begin{figure} \centering
  \subfloat[Mean photometric imbalance $\varepsilon_{\msta,i}$.]{\label{fig-N-epsilon-7h1}\FIG{0.49}{false}{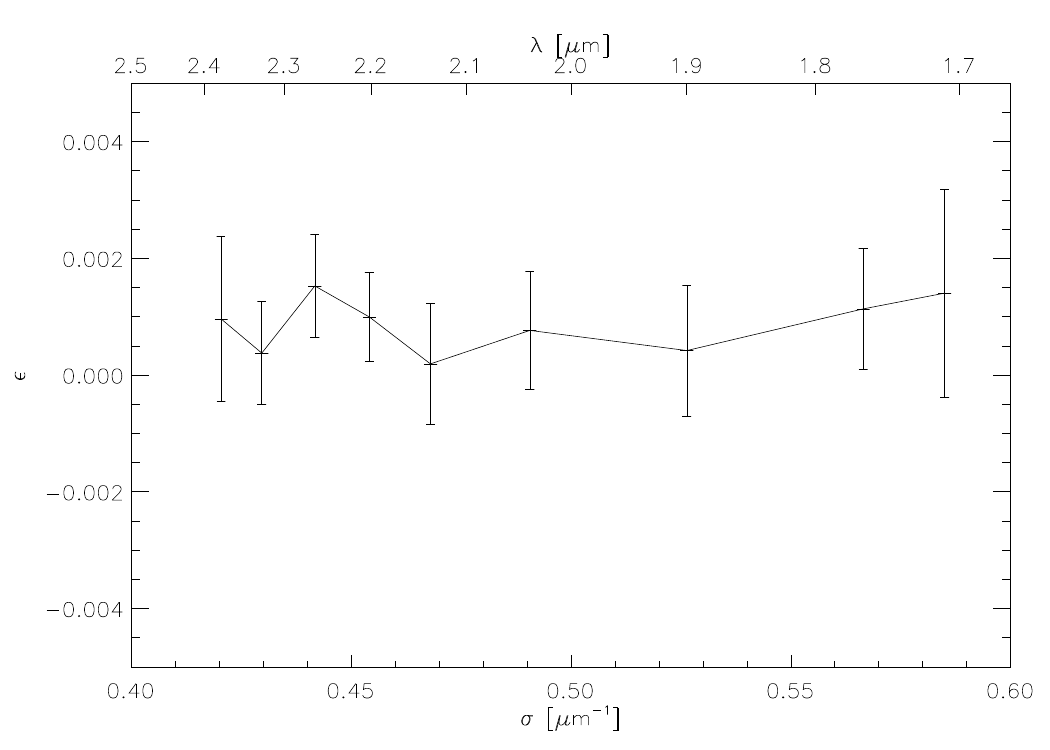}}
  \hfill\subfloat[Photometric imbalance null depth contribution $N_{\varepsilon,i}$.]{\label{fig-N-epsilon-7h2}\FIG{0.49}{false}{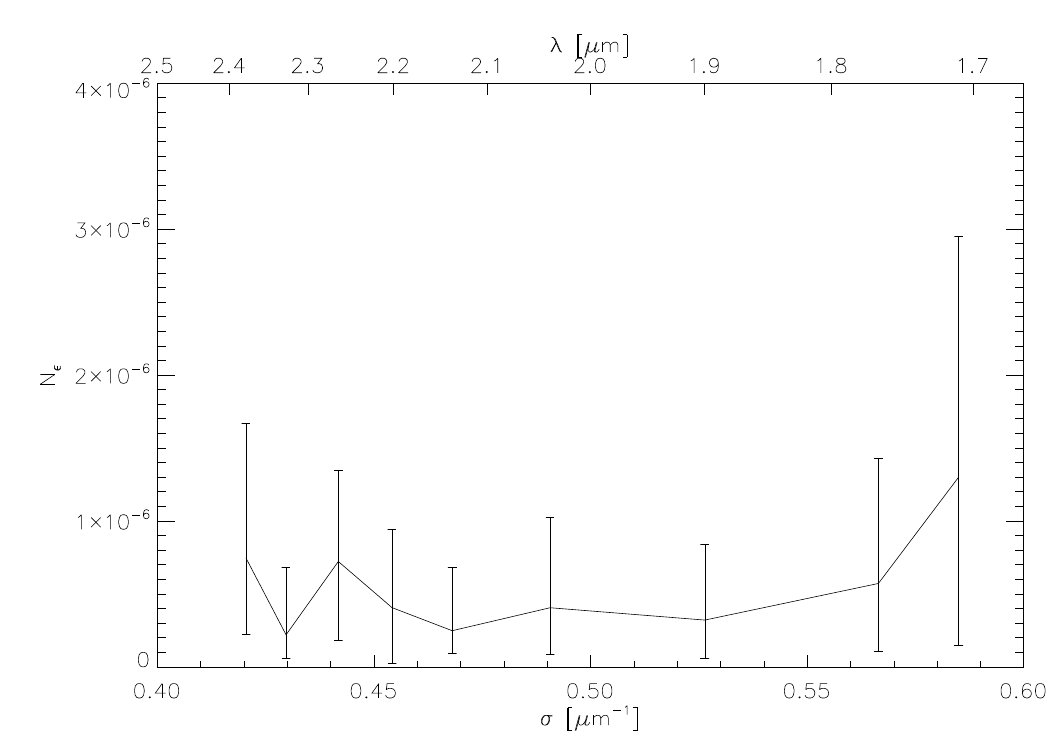}}
  \caption{Multi-hour photometric imbalance and channel null depth contributions.}
  \label{fig-N-epsilon-7h}
\end{figure}

Mean static photometric imbalance measured $\boldsymbol{\moy[i]{\varepsilon_{\msta,i}} = 0.08\%}$ (0.05\% dispersion), yielding a mean static leakage of:
\begin{equation*}
  \boldsymbol{\moy[\Ts]{N_\varepsilon} = 5.5\E{-7}},
\end{equation*}
with multi-hour photometric stability $\boldsymbol{\sigma_{N,\varepsilon,\tauc} = 5\E{-7}}$. Dynamic tip/tilt tracking leakage remained negligible ($\boldsymbol{\moy[\Tc]{N_{\varepsilon,\mdyn}} = 7.9\E{-12}}$, $\boldsymbol{\sigma_{N,\varepsilon,\tauu} = 9\E{-13}}$).

%¤¤¤¤¤¤¤¤¤¤¤¤¤¤¤¤¤¤¤¤¤¤¤¤¤¤¤¤¤¤¤¤¤¤¤¤¤¤¤¤¤¤¤¤¤¤¤¤¤¤¤¤¤¤¤¤¤¤¤¤¤¤¤¤¤¤¤¤¤¤¤¤¤¤¤¤¤¤¤
\subsection{Multi-Hour Error Budget Summary}
\label{sec-contributions}

Figure~\ref{fig-N-m2-7h} compares multi-hour channel null depth profiles against individual error contributions.

\begin{figure} \centering
  \FIG{0.7}{false}{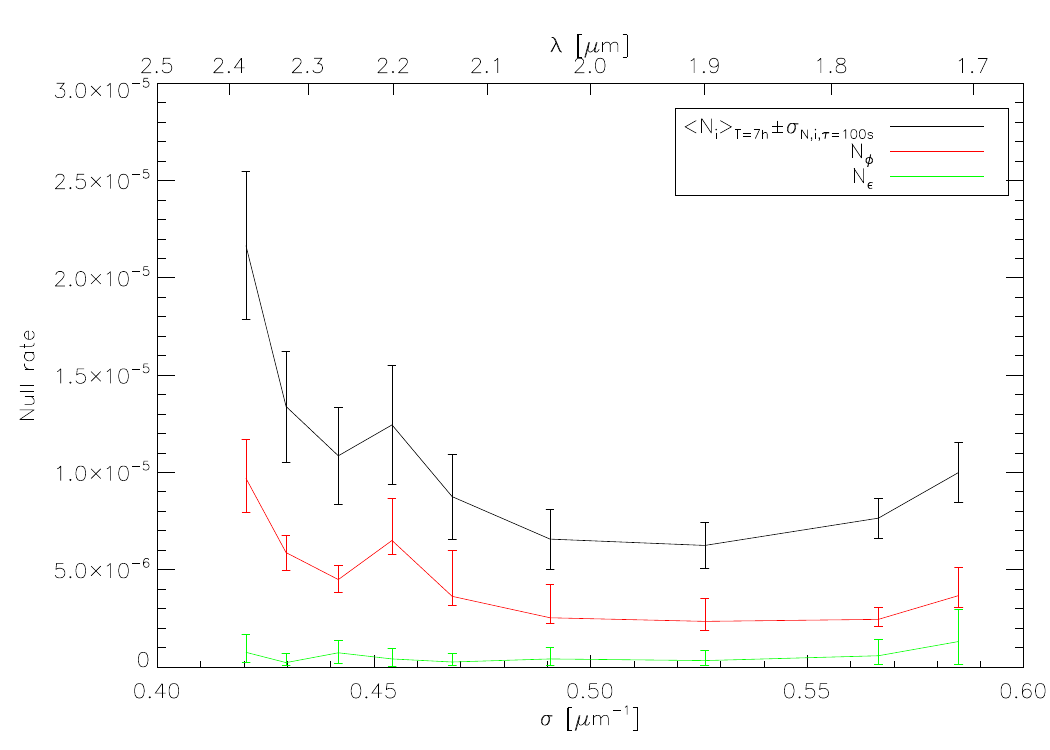}
  \caption[Multi-hour channel null depths and error contributions.]{Multi-hour channel null depths compared against individual error contributions.}
  \label{fig-N-m2-7h}
\end{figure}

Phase leakage $\moy[\Ts]{N_{\phi,i}} = \moy[\Ts]{N_{\delta,w,i}} + \moy[\Ts]{N_{\delta,\mref,i}} + N_{\sigma,i}$ drives channel variations at passband edges. Subtracting modeled phase and flux leakage leaves an estimated multi-hour polarization leakage of:
\begin{equation*}
  \boldsymbol{\moy[\Ts]{N_\mpol} = 5\E{-6}},
\end{equation*}
accounting for 50\% of total measured null depth. Flat spectral response indicates polarization rotation misalignment ($\boldsymbol{\alpha_\mrot = 4.5\text{~\bf mrad} = 15\text{~\bf arcmin}}$) as the primary polarization leakage mechanism.

\begin{table} \centering
  \caption[Multi-hour error budget summary over 7 hours.]{Multi-hour error budget summary over 7 hours compared against specifications.}
  \medskip
  \renewcommand{\arraystretch}{1.25} 
  \begin{minipage}{7.6cm} \centering
    \begin{tabular}[table]{lrr}
      \hline \hline Term & Requirement & Measured \\
      \hline $\moy[\Ts]{N_\mmoy}$ & $\e{-4}$ & $1.02\E{-5}$ \\
      \hline $\moy[\Ts]{N_\phi}$ &  $7\E{-5}$ & $4.6\E{-6}$ \\
      $\moy[\Ts]{N_{\delta,w}}$ & \multirow{2}{*}{$\Bigg\}\ 3.5\E{-5}$} &
      $1.4\E{-6}$ \\
      $\moy[\Ts]{N_{\delta,\mref}}$ & & $9.1\E{-7}$ \\
      $N_{\sigma}$ & $3.5\E{-5}$ & $2.3\E{-6}$ \\
      $\moy[\Ts]{N_\varepsilon}$ & $2\E{-5}$ & $5.5\E{-7}$ \\
      $N_{\varepsilon,\msta}$ & \multirow{2}{*}{$\Bigg\}$\hspace{10pt}$\
        2\E{-5}$} & $5.5\E{-7}$ \\
      $\moy[\Ts]{N_{\varepsilon,\mdyn}}$ & & $7.9\E{-12}$ \\
      $N_\mpol$\footnote{Estimated by subtracting phase and flux terms} &
      $\e{-5}$ & $5.0\E{-6}$ \\
      \hline $\sigma_{N,\tauu}$ & $1.5\E{-5}$ & $6\E{-7}$ \\
      \hline $\sigma_{N,\delta,\tauu}$ & $1.5\E{-5}$ & $6\E{-7}$ \\
      $\sigma_{N,\varepsilon,\tauu}$ & $3\E{-6}$ & $9\E{-13}$ \\
      \hline $\sigma_{N,\tauc}$ & $\e{-5}$ & $9\E{-7}$ \\
      \hline$\sigma_{N,\delta,w,\tauc}$ &
      \multirow{2}{*}{$\Bigg\}$\hspace{10pt}$\ 7\E{-6}$} & $2.2\E{-7}$ \\
      $\sigma_{N,\delta,\mref,\tauc}$ & & $\approx 5\E{-7}$ \\
      $\sigma_{N,\sigma,\tauc}$ & $\e{-6}$ & $7.3\E{-7}$ \\
      $\sigma_{N,\varepsilon,\tauc}$ & $\e{-6}$ & $5\E{-7}$ \\
      ${\sigma_{N,\mpol,\tauc}}^a$ & $\e{-6}$ & $\approx \e{-7}$ \\
      \hline \hline
    \end{tabular}
  \end{minipage}
  \label{tab-null-7h}
\end{table}

Table~\ref{tab-null-7h} compiles the multi-hour error budget. All measured leakage terms remain well below requirement limits. Thermal setpoint drift $\delta_\mref$ contributes 10\% of total null depth ($9.1\E{-7}$), which can be actively compensated in future software revisions. Polarization leakage represents the largest single term ($5.0\E{-6}$), which can be further optimized by fine-tuning siderostat M1 and periscope M5 alignment angles.

%§§§§§§§§§§§§§§§§§§§§§§§§§§§§§§§§§§§§§§§§§§§§§§§§§§§§§§§§§§§§§§§§§§§§§§§§§§§§§§§
\section{Conclusions}
\label{sec-conclusions-1}

We developed calibration and science data reduction pipelines to measure, optimize, and analyze broadband nulling performance on the \pe testbed. In monochromatic light, deep null depths were achieved. In broadband light, \pe demonstrated a world-record null depth of $\mathbf{8.8\E{-6}}$ over a 100~s duration across a $\mathbf{37\%}$ fractional bandwidth.

While originally designed to demonstrate $10^{-4}$ nulling depths for hot Jupiter observations, \pe's performance surpasses baseline specifications by an order of magnitude. The testbed demonstrates that simple, highly optimized optical architectures can achieve the $10^{-5}$ broadband nulling depths required for Earth-like exoplanet characterization missions.

Detailed error budget breakdowns isolate key performance drivers, providing quantitative guidance for the design of future space-based nulling interferometers.

%§§§§§§§§§§§§§§§§§§§§§§§§§§§§§§§§§§§§§§§§§§§§§§§§§§§§§§§§§§§§§§§§§§§§§§§§§§§§§§§
\chapter{Extrapolation to the Case of a Space Mission}
\label{sec-extrapolation}

\begin{flushright}
 \begin{minipage}{12cm}
   {\small {\itshape Isis and the dozen biologically active worlds
       detected by the planetary interferometer had demonstrated that life
       was not really a novelty for the Universe. It was, if not
       inevitable, at least quite widespread in the galaxy.

       But no matter how hard humanity listened with all its ears, it
       had never received the slightest intelligible signal, the slightest proof
       of non-human space travel, the slightest clue of an intersidereal
       civilization. We are spreading out into the void, thought Li. We
       call out, but no one answers us.

       We are unique.}}

   \raggedleft{{\small Robert Charles Wilson, BIOS (1999)}}
 \end{minipage}
\end{flushright}

\minitoc

\bigskip

%§§§§§§§§§§§§§§§§§§§§§§§§§§§§§§§§§§§§§§§§§§§§§§§§§§§§§§§§§§§§§§§§§§§§§§§§§§§§§§§
\section{Introduction}
\label{sec-introduction-2}

In this section, I will seek to scale the results obtained on the \pe testbed to the level of a space mission. This will allow us to provide initial specifications for the guidance, navigation, and control system, particularly in connection with formation flying (see Section~\ref{sec-objectifs}), as well as tolerances for the optical train of a space interferometer such as \peg (see Section~\ref{sec-pegase-precurseur}).

As seen in Chapter~\ref{chap-persee}, \pe aims to best simulate the optical train of \peg. However, there are certain scaling factors that are important to take into account when transposing the results obtained on the testbed.

%§§§§§§§§§§§§§§§§§§§§§§§§§§§§§§§§§§§§§§§§§§§§§§§§§§§§§§§§§§§§§§§§§§§§§§§§§§§§§§§
\section{Discussion on Stellar Leakage}
\label{sec-discussion-sur-2}

We saw in Section~\ref{sec-discussion-sur} that on \pe, the contribution of stellar leakage to the calculated null depth can be considered zero. However, in the case of \peg, the observed source is no longer spatially coherent, but possesses a non-zero angular diameter.

If we consider a hot-Jupiter type planet orbiting at $0.05$~AU around a Sun-like star \tir{a star with a diameter of $1.5\E{6}$~km, or $0.01$~AU}, the planet's orbit has a semi-major axis that is only a few times the diameter of its host star. This is indeed one of the defining characteristics of Pegasids, as seen in Section~\ref{sec-jupiters-chauds}. By comparison, the semi-major axis of Earth's orbit is 100 times the diameter of the Sun, and thus much farther out.

Thus, for Pegasids, the impact of stellar leakage on the measurement is very significant because the planet is very close to the star. Therefore, one must either maximize the transmitted planet flux by placing it near a bright fringe, which also increases the leakage term, or use a shorter baseline to reduce this leakage term, which decreases planet transmission as it then lies on the slope of an interference fringe.

By approximating the sinusoid with a quadratic function, stellar leakage is given by Equation~\eqref{eq-N-star}:
\begin{equation*}
  N_\star(\lambda) \simeq \frac{\pi^2}{16}\GP{\frac{B\theta_\star}{\lambda}}^2.
\end{equation*}

At wavelength $\lambda$, the planet transmission $T_\mpl$ is given by the equation:
\begin{equation}
  T_\mpl(\lambda) = \frac12\GC{1-\cos\GP{2\pi\frac{B}{\lambda}\theta_\mpl}}.
\end{equation}

Thus, neglecting instrumental losses, the detection contrast between the planet signal and the star's geometric residue, given by $N_\mgeo = N_\star/T_\mpl$, follows the equation:
\begin{equation}
  N_\mgeo(\lambda) \simeq \frac{\pi^2}{8}\GP{\frac{B\theta_\star}{\lambda}}^2
  \GC{1-\cos\GP{2\pi\frac{B}{\lambda}\theta_\mpl}}^{-1}.
\end{equation}

In the asymptotic limit for a small baseline $B$, this equation can be approximated by the following expression, which is independent of wavelength:
\begin{equation}
  N_\mgeo \simeq \frac{1}{16}\GP{\frac{\theta_\star}{\theta_\mpl}}^2.
  \label{eq-n-geo}
\end{equation}

We still seek to at least partially resolve the exoplanet from the star, so the baseline cannot be too small. In the scenario presented above, with a Pegasid having a semi-major axis of $0.05$~AU around a star of diameter $0.01$~AU, the ratio $\theta_\star/\theta_\mpl$ is very small, equal to 5. If the star is viewed at a distance of 20~parsecs, its angular diameter is $\theta_\star = 0.5$~mas, and the maximum separation angle between the star and the planet is $\theta_\mpl = 2.5$~mas. Figure~\ref{fig-peg-n-star} shows this contribution for different baselines.

\begin{figure} \centering
  \FIG{.7}{false}{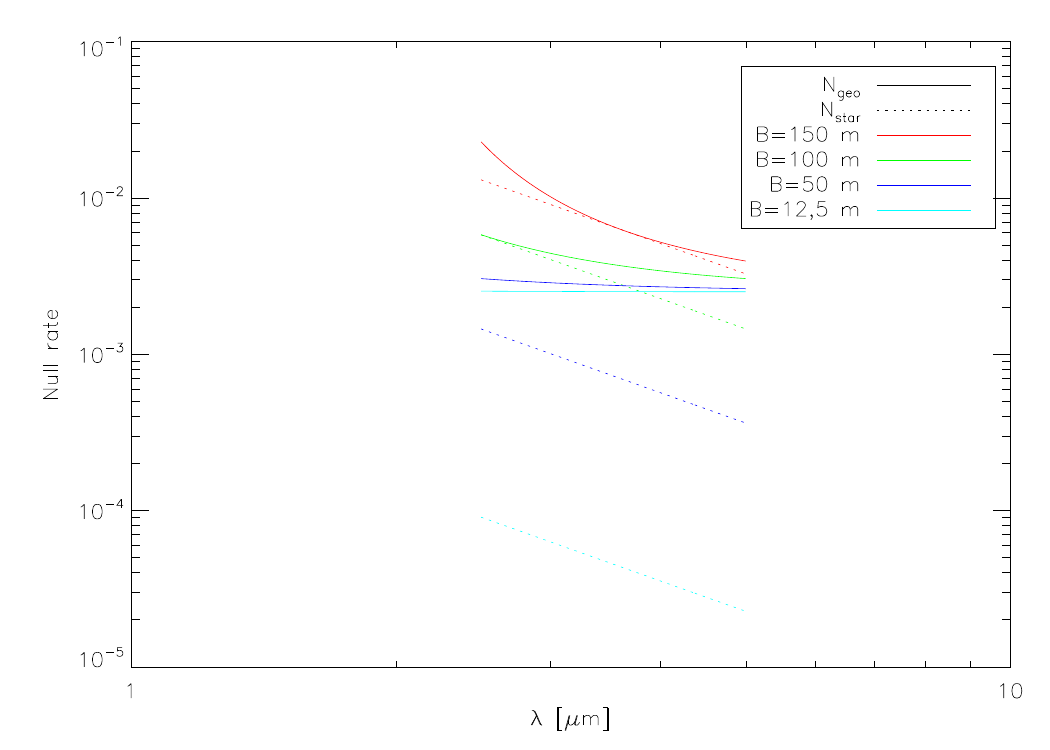}
  \caption{Contribution of geometric leakage to the null depth.}
  \label{fig-peg-n-star}
\end{figure}

On this curve, stellar leakage $N_\star$ decreases quadratically with baseline length $B$; however, as the baseline decreases, planet transmission drops as well. The two terms offset each other to a value independent of wavelength. It is clear that for \peg, placing the planet at the center of the bright fringe \tir{i.e., for a baseline of 150~m in this example} is by no means the optimal solution: stellar leakage is very high, and low planet transmission at short wavelengths degrades geometric losses even further.

We see that this contribution is at minimum given by Equation~\eqref{eq-n-geo}, whose value in this example is $N_\mgeo = 2.5\E{-3}$. For a typical baseline of 100~m, this contribution is $3.8\E{-3}$. Conversely, if we place the planet on the white fringe \tir{i.e., for a baseline $B=150$~m}, the contribution is very high, at $7.5\E{-3}$. These values are larger than the typical contrast required for Pegasid spectroscopy, which ranges between $\e{-3}$ and $\e{-4}$. However, the angular diameter of targets intended for \peg should be known with an accuracy of $0.1$\% using a visibility measurement mode; geometric leakage can therefore be subtracted with sufficient precision to achieve a signal-to-noise ratio between 5 and 30. Even if the exact diameter of the star is not known, it is possible to measure the flux in two orthogonal directions: one where the planet transmission is maximized, and one where the planet is also nulled. Subtracting the two fluxes leaves only the planet's flux.

Unlike \peg, the FKSI interferometer project (see Section~\ref{sec-autres-projets}) chose to use a shorter $12.5$~m baseline, keeping geometric leakage close to its lower bound. However, Figure~\ref{fig-peg-n-star} shows that for the case considered, baselines from 50 to 100~m only slightly degrade geometric leakage $N_\mgeo$; moreover, \peg was intended to serve as a demonstrator for more complex interferometers such as Darwin and TPF-I, described in Section~\ref{sec-darwintpf-i}. It is therefore important to use formation flying—which mandates baselines large enough to avoid spacecraft collisions—to validate this critical technology.

In the case of an Earth-like planet at 1~AU from its Sun-like star, geometric leakage is at minimum $N_\mgeo = 6.25\E{-6}$, making observation extremely difficult. Using 4 telescopes instead of 2 changes the sensitivity to stellar angular diameter from quadratic ($\theta_\star^2$) to quartic ($\theta_\star^4$); geometric leakage is thus greatly reduced.

For observing Pegasids, the choice remains open between long baselines achieved via formation flying and a $\sim$12-meter rigid boom. In both cases, cophasing will be perturbed by stabilization vibrations, as well as by drift in formation flying or flexible modes in the case of a boom.

%§§§§§§§§§§§§§§§§§§§§§§§§§§§§§§§§§§§§§§§§§§§§§§§§§§§§§§§§§§§§§§§§§§§§§§§§§§§§§§§
\section{Scaling Up to \peg}
\label{sec-mise-echelle}

%¤¤¤¤¤¤¤¤¤¤¤¤¤¤¤¤¤¤¤¤¤¤¤¤¤¤¤¤¤¤¤¤¤¤¤¤¤¤¤¤¤¤¤¤¤¤¤¤¤¤¤¤¤¤¤¤¤¤¤¤¤¤¤¤¤¤¤¤¤¤¤¤¤¤¤¤¤¤¤
\subsection{The \peg Optical Train}
\label{sec-train-optique}

The optical trains of \peg and \pe are very similar; however, there are important differences to note. Figure~\ref{fig-peg-optique} presents the layout selected for the \peg optical train during preliminary studies \cite{Leduigou06b,Ollivier07}.

\begin{figure} \centering
  \FIG{.9}{false}{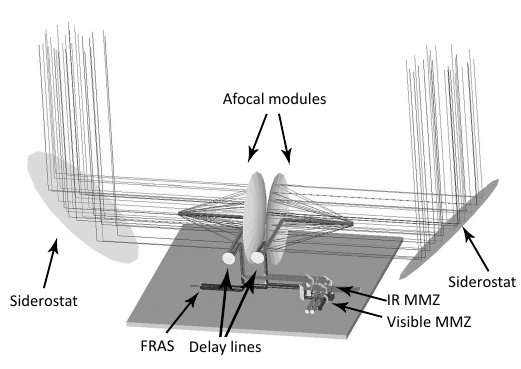}
  \caption{Layout of the \peg optical train.}
  \label{fig-peg-optique}
\end{figure}

On \peg, siderostats direct 40~cm diameter beams into two afocal modules (off-axis or otherwise), and the beams are then folded at $90\degree$ to realize, together with the siderostats, the geometric $\pi$ achromatic phase shifter. As on \pe, long-stroke delay lines correct large optical path difference drifts. Unlike \pe, the cophasing and science spectral bands are split by a dichroic plate upstream of the FRAS and MMZ. The MMZ incorporates Serabyn's MMZ design \cite{Serabyn01} and the configuration of IAS's SYNAPSE\footnote{SYmmetric Nuller for Achromatic Phase Shifter Evaluation} bench \cite{Brachet05,Chazelas07}, where the science and cophasing beams travel one above the other.

It is evident that the differential paths between the cophasing and science channels are larger than in \pe. However, an alternative configuration was conceived using a single MMZ for both channels as in \pe. This solution is reinforced by the fact that using dichroic plates to separate science and cophasing channels (consisting of dielectric multi-layer stacks) introduces additional phase shifts prior to beam combination.

%¤¤¤¤¤¤¤¤¤¤¤¤¤¤¤¤¤¤¤¤¤¤¤¤¤¤¤¤¤¤¤¤¤¤¤¤¤¤¤¤¤¤¤¤¤¤¤¤¤¤¤¤¤¤¤¤¤¤¤¤¤¤¤¤¤¤¤¤¤¤¤¤¤¤¤¤¤¤¤
\subsection{Scaling Factors}
\label{sec-facteurs-echelle}

Due to size and cost constraints, it was not possible to build the \pe testbed to the full dimensions of \peg, nor with a constant homothetic scale factor. For example, the input beams of \peg have a diameter of 40~cm, and the telescopes provide $20\times$ magnification. Thus, beams throughout the optical train downstream of the compressors have a diameter of 20~mm. On \pe, to maintain beam diameters that were neither too large at the input nor too small at the output, the magnification of the afocal modules was reduced to 3, with an input diameter of 40~mm and an output diameter of 13.3~mm. In the final setup presented in this manuscript (Section~\ref{sec-conf-finale}), I recall that the afocal modules were omitted; the magnification considered is therefore 1.

The spectral bands are also not identical, primarily due to cost considerations. Indeed, the science spectral band of \peg, originally planned between $2.5$ and 5~\mum, was shifted to $[1.65\text{--}3.3]$~\mum for \pe, with the science camera operating between $1.65$ and $2.5$~\mum, and a single-pixel detector between 3 and $3.3$~\mum. The latter band having been abandoned due to MMZ plate coating manufacturing challenges, \pe's spectral band ultimately has a smaller fractional bandwidth; similarly, spectral resolution was reduced from 60 to 10 due to photon flux limits.

For cophasing, the spectral bands are roughly similar: \peg's FRAS operates between $0.6$ and 1~\mum \tir{versus $0.8$ to 1~\mum for \pe}, and its FS operates across three spectral bands between $0.6$ and 1~\mum \tir{versus two spectral bands between $0.8$ and $1.65$~\mum for \pe}. These three spectral bands yield more precise cophasing than two, but on \pe we demonstrated that two bands suffice to significantly broaden the dynamic range, using the algorithm described in Section~\ref{sec-estim-elargi}.

The FRAS on \peg is also slightly different from \pe's. On \peg, one camera pixel corresponds to $0.18$~arcsec on the sky, or $3.6$~arcsec within the instrument, representing a spatial sampling of 2 to $2.6$~pixels per Airy disk, depending on whether the visible or $\I$ band is used. On \pe, the scale is slightly different, with 6~arcsec per pixel in the instrument, yielding a sampling of $2.2$~pixels per Airy disk.

All these scaling factors to be considered for the following analysis are summarized in Table~\ref{tab-dimensions}.

\begin{table} \centering 
  \caption{Comparison between the dimensions of \pe and \peg.}
  \medskip
  \begin{tabular}{cccc}
    \hline \hline
    Parameter & \pe & \peg & Scaling Factor \\
    \hline
    Input beam diameter [mm] & $13.3$ & 400 & 30 \\
    FRAS beam diameter [mm] & $13.3$ & 20 & $1.5$ \\
    MMZ beam diameter [mm] & 10 & 15 & $1.5$ \\
    Afocal magnification & 1 & 20 & 20 \\
    Baseline [m] & $0.05$ & 40 -- 500 & --- \\
    Science band [\mum{}] & $1.65$ -- $2.45$ & $2.5$ -- 5 & $1.5$ -- 2 \\
    Spectral resolution & 10 & 60 & 6 \\
    FS band [\mum{}] & $0.8$ -- $1.65$ & $0.8$ -- $1.5$ & 1 -- $0.9$ \\
    Number of FS channels & 2 & 3 & --- \\
    FRAS band [\mum{}] & $0.8$ -- 1 & $0.6$ -- $0.8$ & $0.75$ -- 1 \\
    Sampling on FRAS [pixels] & $2.2$ & 2 & $0.9$ \\
    \hline \hline
  \end{tabular}
  \label{tab-dimensions}
\end{table}

The FKSI space interferometer project, whose parameters are very close to \pe, has similar characteristics. Only the baseline differs significantly, as seen in Section~\ref{sec-discussion-sur-2}.

%§§§§§§§§§§§§§§§§§§§§§§§§§§§§§§§§§§§§§§§§§§§§§§§§§§§§§§§§§§§§§§§§§§§§§§§§§§§§§§§
\section{Cophasing Results}
\label{sec-resultats-cophasage}

%¤¤¤¤¤¤¤¤¤¤¤¤¤¤¤¤¤¤¤¤¤¤¤¤¤¤¤¤¤¤¤¤¤¤¤¤¤¤¤¤¤¤¤¤¤¤¤¤¤¤¤¤¤¤¤¤¤¤¤¤¤¤¤¤¤¤¤¤¤¤¤¤¤¤¤¤¤¤¤
\subsection{Fine Pointing}
\label{sec-pointage-fin}

Using piezoelectric modules to perform fine pointing significantly relaxes spacecraft pointing requirements. On \pe, the PZT modules have an optical stroke of $\pm 200$~arcsec, corresponding to $\pm 10$~arcsec on the sky for \peg. Spacecraft pointing accuracy needs only to be on the order of an arcsecond. The solution devised for \pe seems best suited for the FRAS: using annular mirrors to pick off a portion of the beam perimeter. Beamsplitters or dichroics would introduce chromatic aberrations and are more complex to qualify for spaceflight.

In Section~\ref{sec-performances-boucle}, I demonstrated \pe's ability to stabilize incident beams to 56~mas RMS in tip and tilt, representing $0.4$\% of the Airy disk. Accounting for optical scaling, this corresponds to a residual jitter of 62~mas RMS for \peg. The specification is 30~mas RMS on the sky, or 600~mas RMS inside the FRAS. Performance is thus an order of magnitude better than required.

However, this holds true for near-maximum flux on the camera. In Section~\ref{sec-sensibilite-au}, I showed that tip/tilt residual jitter is proportional to the square root of the inverse photon count, assuming a ultra-low-noise detector—as will be used in space. With a setup similar to \pe, $2\E{5}$ photons per arm per frame are required to achieve tracking better than 600~mas RMS. On \peg, each arm undergoes 11 reflections off unprotected gold mirrors (97\% reflectivity between $0.8$ and 1~\mum). Total transmission to the detector is 72\%; $2.8\E{5}$ input photons are thus required. However, \pe's FRAS camera is silicon-based, with a low quantum efficiency (~5\%) in the $\I$ band. Using a detector with an 80\% quantum efficiency reduces the required photon flux 16-fold, down to $1.8\E{4}$ photons per frame.

The collecting aperture required to capture this photon count depends on stellar magnitude. Target stars range between $m_\BV = 4$ and $m_\BV = 10$, and the magnitude of a G-type Sun-like star in the $\I$ band is $m_\I = m_\BV - 0.93$. Stellar photon flux $F$ relates to a $m_\BV = 0$ reference flux $F_0$ via:
\begin{equation}
 F = F_0 \times 10^{-\frac{m_\BV-0.93}{2.5}}.
\end{equation}

Collecting area $S$ required to capture $N_\mph$ photons at loop rate $f$ is:
\begin{equation}
  S = \frac{hc}{\lambda}\frac{N_\mph f}{F} = \frac{hc}{\lambda}\frac{N_\mph
    f}{F_0}\times 10^{\frac{m_\BV-0.93}{2.5}}.
\end{equation}

Figure~\ref{fig-surf-vs-mag} plots minimum collecting area versus stellar magnitude across frame rates for two options:
\begin{itemize}
\item Matching \pe: $\I$-band light is picked off by an annular mirror (30~cm inner, 40~cm outer diameter);
\item Visible $\BV$-band light collected via dichroics across the full 40~cm aperture.
\end{itemize}

\begin{figure} \centering
  \FIG{.7}{false}{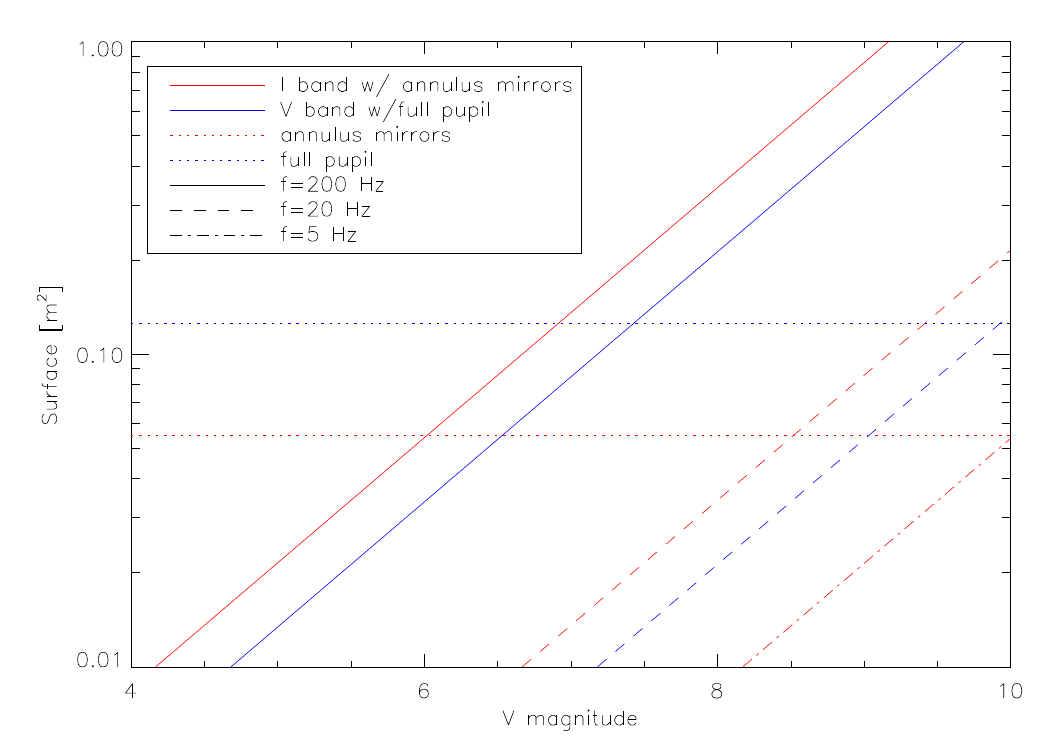}
  \caption{Minimum FRAS collecting area versus stellar magnitude.}
  \label{fig-surf-vs-mag}
\end{figure}

For \peg's core targets ($m_\BV = 4\text{--}10$), a 40~cm outer-diameter annular mirror operating at 200~Hz meets the 600~mas RMS specification only for stars brighter than $m_\BV = 6$. As shown in Section~\ref{sec-deseq-phot-3}, pointing margin exists to accommodate stars down to magnitude 7. In Section~\ref{sec-performances-boucle}, we noted that a 200~Hz loop rate is conservative relative to low-frequency disturbance spectra. Lowering the loop rate relaxes flux requirements: at 20~Hz, tracking succeeds down to magnitude $8.5$; at 5~Hz, it reaches magnitude 10. Selecting loop rates trades disturbance rejection bandwidth against sensor readout noise under a given controller (integrator vs. LQG).

Operating in the visible improves performance: at 200~Hz, tracking succeeds down to magnitude $7.5$, and at 20~Hz it covers all target magnitudes. However, this requires dichroics, which are harder to manufacture to spaceflight specs than flat pick-off mirrors.

%¤¤¤¤¤¤¤¤¤¤¤¤¤¤¤¤¤¤¤¤¤¤¤¤¤¤¤¤¤¤¤¤¤¤¤¤¤¤¤¤¤¤¤¤¤¤¤¤¤¤¤¤¤¤¤¤¤¤¤¤¤¤¤¤¤¤¤¤¤¤¤¤¤¤¤¤¤¤¤
\subsection{Optical Path Difference}
\label{sec-difference-marche}

%-------------------------------------------------------------------------------
\subsubsection{Correction of Small-Amplitude Disturbances}
\label{sec-correction-perturbations-1}

As shown in Section~\ref{sec-perf-piston}, \pe demonstrated sub-nanometric OPD control in a standard cleanroom environment over multi-meter path lengths. Testbed performance targets were fully met.

Performance reached levels 10 times better than specification under quiet conditions (electronics relocated, air conditioning/ventilation off, acoustic foam enclosure installed).

A key contribution of \pe is validating cophasing under injected formation-flying disturbances. Section~\ref{sec-analyse-typique} demonstrated that an LQG controller effectively rejects these disturbances, particularly mechanical vibrations that degrade integrator performance. Despite dynamic path disturbances exceeding 15~nm RMS (which would degrade \peg's null depth beyond $10^{-4}$), closed-loop OPD jitter was suppressed below $0.8$~nm RMS under worst-case disturbance profiles.

\pe's results show that structural resonance modes must be pushed above 150~Hz with low amplitudes to avoid degrading OPD stability and null depth. \pe's optomechanical resonances sit between 80 and 170~Hz and are sensitive to acoustic noise. While acoustic noise is absent in space, structural modes must not coincide with reaction wheel harmonics.

To guarantee $N \le 10^{-4}$ on \peg, maximum allowable path delay jitter was budgeted at $2.5$~nm RMS. Achieving $0.8$~nm RMS (3$\times$ better) reduces path delay leakage by nearly an order of magnitude. However, these results used high photon fluxes. Under photon-starved flight conditions (Section~\ref{sec-sensib-flux}), path delay jitter scales inversely with flux ($\text{variance} \propto N_\mph^{-1}$ for photon-noise limited detectors).

Achieving $2.5$~nm RMS path delay jitter requires a minimum fringe sensor signal of $\text{SNR} = 440$, corresponding to $N_\mph = \text{SNR}^2 = 1.9\E{5}$ photons.

Figure~\ref{fig-diam-vs-mag} plots the minimum collector aperture diameter required for the fringe sensor versus stellar magnitude for two target OPD jitter levels ($2.5$ and 5~nm RMS) across loop rates.

\begin{figure} \centering
  \FIG{.7}{false}{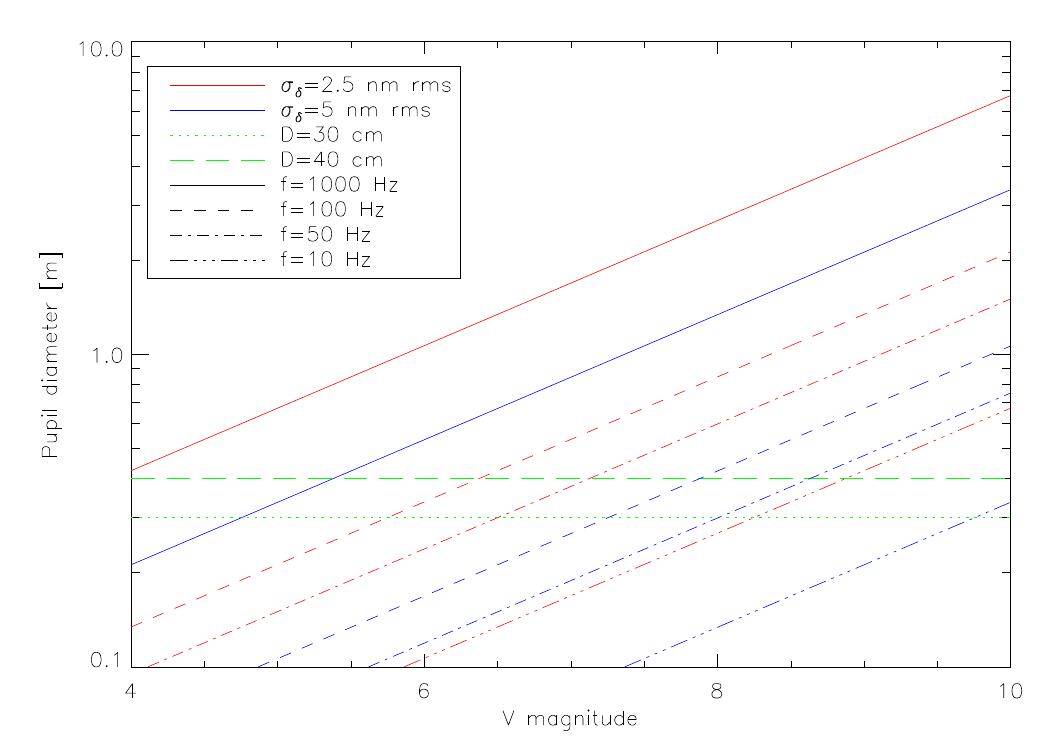}
  \caption{Minimum fringe sensor collector aperture diameter versus stellar magnitude.}
  \label{fig-diam-vs-mag}
\end{figure}

For a 1~kHz loop rate, \peg's 40~cm aperture meets requirements only for stars brighter than magnitude 4 ($5.3$ for a 5~nm RMS jitter limit). Lowering the loop rate to 100~Hz extends coverage to magnitudes $6.3$ ($2.5$~nm limit) and $7.8$ ($5$~nm limit). At 10~Hz, tracking reaches magnitudes 9 and 10.

Lowering loop rates requires that disturbance frequencies fall well below loop Nyquist limits. Selecting low reaction wheel speeds minimizes high-frequency harmonic excitation, enabling lower loop rates.

Alternatively, collector apertures could be enlarged to 64~cm (+1 magnitude) or 1~m (+2 magnitudes), trading hardware cost against target completeness. Furthermore, using dichroic splitters instead of annular mirrors for the FRAS preserves photon budget for the fringe sensor.

%-------------------------------------------------------------------------------
\subsubsection{Correction of Large-Amplitude Disturbances}
\label{sec-correction-perturbations-2}

Beyond micro-vibrations, the testbed validated coarse fringe acquisition under spacecraft drift rates up to 250~\mum.s$^{-1}$. F. Cassaing demonstrated fringe sweep detection and velocity extraction at 250~\mum.s$^{-1}$ for $\text{SNR} \ge 10$. Long-stroke delay lines ($\pm 25$~mm stroke) accommodate coarse satellite relative positioning errors ($\pm 1$~cm) derived from radio-frequency metrology.

Dual-band fringe unwrapping (Section~\ref{sec-estim-elargi}) expanded unambiguous path tracking over several micrometers (Section~\ref{sec-resultat-etalonnage}). Excessively wide metrology passbands ($\J$ band) degrade unwrapping ranges. Metrology passbands must balance photon flux against unwrapping range.

\peg's 100~nm metrology passbands yield coherence lengths of ~10~\mum, matching fringe acquisition search windows. Combining coarse radio-frequency tracking, fringe acquisition at 250~\mum.s$^{-1}$, and dual-band unwrapping enables path tracking across parabolic drift trajectories. Hierarchical control schemes (analogous to adaptive optics woofer-tweeter architectures \cite{Correia10c}) can hand off coarse delay line tracking to fine piezo stages.

%§§§§§§§§§§§§§§§§§§§§§§§§§§§§§§§§§§§§§§§§§§§§§§§§§§§§§§§§§§§§§§§§§§§§§§§§§§§§§§§
\section{Extrapolation of Nulling Performance}
\label{sec-extrapolation-te}

In Chapter~\ref{sec-performance-nulling}, \pe demonstrated deep broadband nulling ($N \le 10^{-5}$ across 37\% bandwidth) under active disturbance rejection and multi-hour stability runs. We extrapolate these results to \peg using the scaling factors from Section~\ref{sec-facteurs-echelle}.

%¤¤¤¤¤¤¤¤¤¤¤¤¤¤¤¤¤¤¤¤¤¤¤¤¤¤¤¤¤¤¤¤¤¤¤¤¤¤¤¤¤¤¤¤¤¤¤¤¤¤¤¤¤¤¤¤¤¤¤¤¤¤¤¤¤¤¤¤¤¤¤¤¤¤¤¤¤¤¤
\subsection{Phase Leakage Contributions}
\label{sec-perturbations-phase-3}

%-------------------------------------------------------------------------------
\subsubsection{Path Delay Jitter}
\label{sec-pert-ddm}

Active cophasing under flight-like disturbances maintained path delay jitter at 0.8~nm RMS ($N_\delta = 1.6\E{-6}$ on \pe). Operating at longer wavelengths ($3.75$~\mum on \peg vs. $2.0$~\mum on \pe) relaxes phase sensitivity quadratically (Figure~\ref{fig-peg-n-delta}).

\begin{figure} \centering
  \FIG{.7}{false}{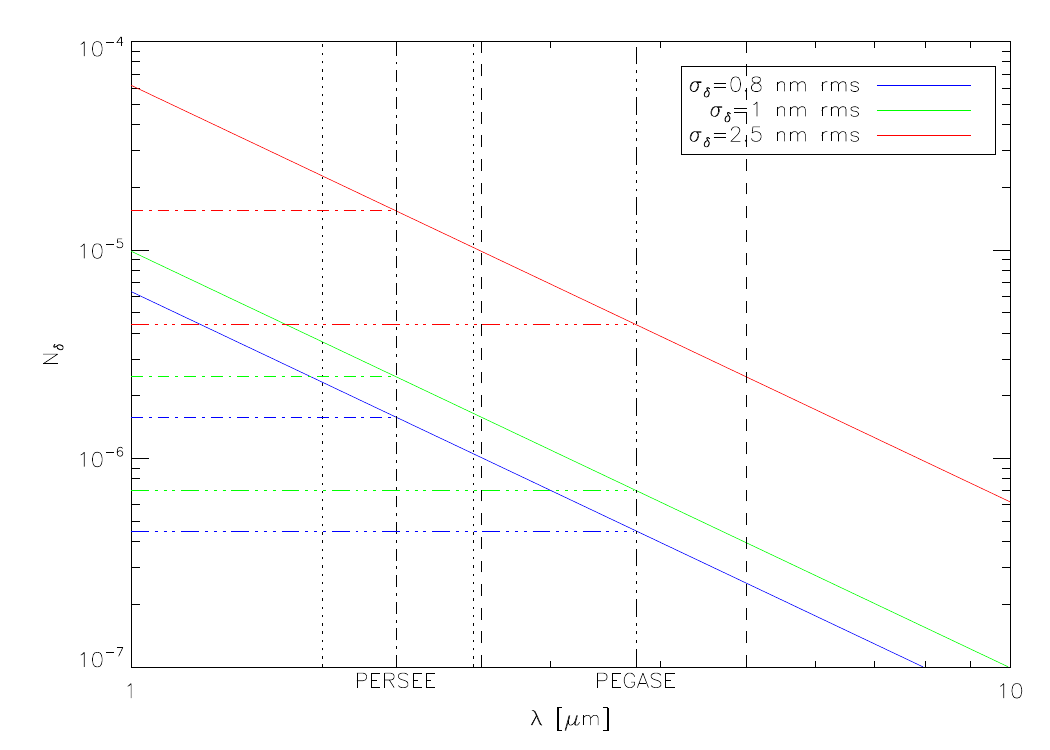}
  \caption{Path delay jitter null depth contribution versus wavelength.}
  \label{fig-peg-n-delta}
\end{figure}

At $\lambda = 3.75$~\mum, \peg is $3.5\times$ less sensitive to path delay jitter, lowering path leakage to $N_\delta = 4.6\E{-7}$ ($10^{-6}$ at $2.5$~\mum). Under a conservative 2.5~nm RMS path jitter, $N_\delta$ remains below $10^{-5}$ across the science passband. Suppressing path jitter to 0.8~nm RMS reduces path leakage tenfold relative to the $2.5$~nm baseline.

Short-term null stability driven by path delay jitter ($\sigma_{N,\delta,\tau=1\text{s}} = 3\E{-7}$ on \pe) relaxes to $9\E{-8}$ on \peg.

%-------------------------------------------------------------------------------
\subsubsection{Chromatic Dispersion}
\label{sec-correction-perturbations}

Residual chromatic dispersion ($N_\sigma = 2.3\E{-6}$ on \pe; $<5\E{-7}$ optimized) similarly relaxes by $3.5\times$ at longer wavelengths, lowering chromatic leakage on \peg to $N_\sigma = 6.5\E{-7}$.

Achieving this requires optical plate thickness uniformity and coating tolerances matching \pe's MMZ. Flight-qualified optomechanics must maintain arcsecond optical alignment through launch vibration loads without requiring complex multi-axis motorized stages.

%-------------------------------------------------------------------------------
\subsubsection{Path Offsets and Thermal Control}
\label{sec-stabilite-thermique-1}

\pe's MMZ exhibited a thermal drift sensitivity of 600~nm.K$^{-1}$ due to commercial adjustable mounts. A monolithic, optically contacted MMZ for spaceflight would reduce thermal sensitivity ten-fold to ~50~nm.K$^{-1}$.

At 50~nm.K$^{-1}$, maintaining path offsets below tolerance requires passive thermal control to $\pm 20$~mK across the optical bench—a stability achievable in space environments without requiring active L2 orbit cryogenic cooling. Slow thermal drifts can also be tracked and offset in software.

%¤¤¤¤¤¤¤¤¤¤¤¤¤¤¤¤¤¤¤¤¤¤¤¤¤¤¤¤¤¤¤¤¤¤¤¤¤¤¤¤¤¤¤¤¤¤¤¤¤¤¤¤¤¤¤¤¤¤¤¤¤¤¤¤¤¤¤¤¤¤¤¤¤¤¤¤¤¤¤
\subsection{Photometric Imbalance}
\label{sec-deseq-phot-3}

\pe achieved a static photometric imbalance of $0.08\%$ ($N_{\varepsilon,\msta} = 5.5\E{-7}$)—well below \peg's $0.2\%$ requirement.

While \pe adjusted source fiber tilts to balance Gaussian beam profiles, \peg receives uniform starlight across its primary apertures. Photometric balance on \peg can be fine-tuned using pupil spatial trimmers or slight tip/tilt offsets.

Dynamic tip/tilt tracking leakage ($8\E{-12}$ on \pe) scales as the fourth power of tracking error $(\pi D \sigma_\theta / \lambda)^4$. Even if tracking jitter increases ten-fold to 600~mas RMS, dynamic photometric leakage increases to $4\E{-8}$—remaining negligible compared to static imbalance.

Dynamic tip/tilt tracking stability ($\sigma_{N,\varepsilon,\tauu} = 9\E{-13}$) is similarly negligible.

%¤¤¤¤¤¤¤¤¤¤¤¤¤¤¤¤¤¤¤¤¤¤¤¤¤¤¤¤¤¤¤¤¤¤¤¤¤¤¤¤¤¤¤¤¤¤¤¤¤¤¤¤¤¤¤¤¤¤¤¤¤¤¤¤¤¤¤¤¤¤¤¤¤¤¤¤¤¤¤
\subsection{Polarization}
\label{sec-polarisation}

Polarization leakage ($N_\mpol = 5\E{-6}$) represents the dominant unmodeled error term on \pe. Flat spectral profiles indicate polarization rotation misalignment ($\alpha_\mrot = 15$~arcmin) as the primary contributor.

Transmissive optics (dichroics, MMZ plates) can introduce stress birefringence ($\Delta\phi_\msp$). Minimizing transmissive optics reduces polarization phase shifts.

Large siderostat M1 steering angles (several hundred arcseconds) alter polarization orientation. On \peg, siderostat tracking angles remain within a few arcseconds (50–100~arcsec at M6 after 20$\times$ afocal magnification), bounding polarization rotation errors.

%¤¤¤¤¤¤¤¤¤¤¤¤¤¤¤¤¤¤¤¤¤¤¤¤¤¤¤¤¤¤¤¤¤¤¤¤¤¤¤¤¤¤¤¤¤¤¤¤¤¤¤¤¤¤¤¤¤¤¤¤¤¤¤¤¤¤¤¤¤¤¤¤¤¤¤¤¤¤¤
\subsection{\peg Error Budget Summary}
\label{sec-bilan-te}

Table~\ref{tab-null-peg} extrapolates \pe's measured error budget to \peg under high-flux and photon-starved operating regimes.

\begin{table} \centering
  \caption{Extrapolated \peg performance based on \pe testbed measurements.}
  %\medskip
  \renewcommand{\arraystretch}{1.25}
  \begin{tabular}[table]{lrrr}
    \hline \hline Term & \pe & \multicolumn{2}{c}{\peg} \\
    \hline Flux Regime & High & High & Low \\
    \hline $N_\mgeo$ & 0 & $\approx 3\E{-3}$ & $\approx 3\E{-3}$ \\
  \hline $\moy[\Ts]{N_\mmoy}$ & $1.02\E{-5}$ & $\approx 6.7\E{-6}$ & $\approx 1.2\E{-5}$ \\
  \hline $\moy[\Ts]{N_\phi}$ & $4.6\E{-6}$ & $\approx 1.2\E{-6}$ & $\approx 6.2\E{-6}$ \\
  $\moy[\Ts]{N_{\delta,w}}$ & $1.4\E{-6}$ & $4.6\E{-7}$ & $4.5\E{-6}$ \\
  $\moy[\Tc]{N_{\delta,\mref}}$ & $9.1\E{-7}$ & $\approx \e{-7}$ & $\approx \e{-6}$ \\
  $N_{\sigma}$ & $2.3\E{-6}$ & $6.5\E{-7}$ & $6.5\E{-7}$ \\
  $\moy[\Tc]{N_\varepsilon}$ & $5.5\E{-7}$ & $5.5\E{-7}$ & $6\E{-7}$ \\
  $N_{\varepsilon,\msta}$ & $5.5\E{-7}$ & $5.5\E{-7}$ & $5.5\E{-7}$ \\
  $\moy[\Tc]{N_{\varepsilon,\mdyn}}$ & $7.9\E{-12}$ & $4\E{-12}$ & $4\E{-8}$ \\
  $N_\mpol$ & $5.0\E{-6}$ & $\approx 5\E{-6}$ & $\approx 5\E{-6}$ \\
  \hline $\sigma_{N,\tauu}$ & $6\E{-7}$ & $\approx 2\E{-7}$ & $\approx 2\E{-6}$ \\
  \hline $\sigma_{N,\delta,\tauu}$ & $6\E{-7}$ & $\approx 2\E{-7}$ & $\approx 2\E{-6}$ \\
  $\sigma_{N,\varepsilon,\tauu}$ & $9\E{-13}$ & $6\E{-13}$ & $6\E{-9}$ \\
  \hline $\sigma_{N,\tauc}$ & $9\E{-7}$ & $\approx 6\E{-7}$ & $\approx 1.5\E{-6}$ \\
  \hline$\sigma_{N,\delta,w,\tauc}$ & $2.2\E{-7}$ & $9\E{-8}$ & $9\E{-7}$ \\
  $\sigma_{N,\delta,\mref,\tauc}$ & $\approx 5\E{-7}$ & $\approx \e{-6}$ & $\approx \e{-7}$ \\
  $\sigma_{N,\sigma,\tauc}$ & $7.3\E{-7}$ & $\approx 3\E{-7}$ & $\approx 3\E{-7}$ \\
  $\sigma_{N,\varepsilon,\tauc}$ & $5\E{-7}$ & $5\E{-7}$ & $5\E{-7}$ \\
  ${\sigma_{N,\mpol,\tauc}}$ & $\approx \e{-7}$ & $\approx \e{-7}$ & $\approx \e{-7}$ \\
  \hline \hline
  \end{tabular}
  \label{tab-null-peg}
\end{table}

In high-flux regimes, longer operational wavelengths relax phase sensitivity, yielding an extrapolated instrumental null depth of $6.7\E{-6}$—15 times better than the $10^{-4}$ requirement. Polarization leakage ($5\E{-6}$) represents the primary error term.

In low-flux regimes (600~mas RMS tip/tilt jitter, 2.5~nm RMS path delay jitter), instrumental null depth increases slightly to $1.2\E{-5}$—remaining nearly an order of magnitude deeper than requirements. Margin in the error budget allows relaxing cophasing constraints or improving scientific signal-to-noise ratios.

%§§§§§§§§§§§§§§§§§§§§§§§§§§§§§§§§§§§§§§§§§§§§§§§§§§§§§§§§§§§§§§§§§§§§§§§§§§§§§§§
\section{Conclusions}

The \pe testbed validated the performance of a Bracewell nulling interferometer under flight-like disturbances and refined the error budget for missions like \peg.

Stellar leakage limits observational contrast for close-in Pegasids. Selecting shorter baselines (e.g., FKSI) minimizes stellar leakage, whereas formation-flying baselines (\peg) demonstrate multi-spacecraft operations required for future Earth-sized planet characterization missions.

\pe demonstrated nanometric cophasing control in piston and tip/tilt. Under photon-starved flight conditions, metrology photon counts bound tracking loop rates. Optimizing collector apertures and selecting lower loop rates maintains sub-nanometric path tracking across target magnitudes.

\pe demonstrated broadband nulling depths below $10^{-5}$ under active disturbance rejection. Longer operational wavelengths relax phase sensitivities on \peg, providing substantial margin in the error budget.

Metrology photon flux represents the primary design driver for \peg. Matching target catalogs to available metrology flux and optimizing loop rates confirms the technical feasibility of space-based nulling interferometry.

%§§§§§§§§§§§§§§§§§§§§§§§§§§§§§§§§§§§§§§§§§§§§§§§§§§§§§§§§§§§§§§§§§§§§§§§§§§§§§§§

\cleardoublepage
\fancyhead{}
\fancyhead[LE,RO]{\thepage}
\fancyhead[RE,LO]{\small CONCLUSION AND PERSPECTIVES}
\chapter*{Conclusion and Perspectives}
\label{sec-conclusion-fin}
\addstarredchapter{Conclusion and Perspectives}

%§§§§§§§§§§§§§§§§§§§§§§§§§§§§§§§§§§§§§§§§§§§§§§§§§§§§§§§§§§§§§§§§§§§§§§§§§§§§§§§
\section*{Context}
\label{sec-contexte}

We now know that the planets of the Solar System are far from being the only ones in the galaxy. Since the first discovery in 1995 of an exoplanet orbiting a main-sequence star, detection methods have continually evolved, enabling the discovery of over 700 planets. These techniques are now deployed in space missions: the CoRoT and Kepler satellites, for instance, are each responsible for dozens of confirmed and characterized discoveries, including the smallest ever detected around main-sequence stars, with Earth-sized and even Mars-sized diameters. However, these indirect detection techniques do not allow for the direct analysis of photons from the planet itself—for example, to discover biosignatures that could provide hints of life on its surface. Consequently, direct observation techniques have been under development for several years; these suppress the host star's flux to observe the adjacent planet. Among these techniques, nulling interferometry is highly promising, as it simultaneously achieves high angular resolution and significantly reduces the contrast ratio between the star and its planet.

Ambitious space-based nulling interferometer concepts, such as Darwin and TPF-I, were thus conceived to target Earth-like exoplanets and search for eventual traces of life. To pave the way toward this goal, simpler pathfinder concepts like the \peg interferometer aimed to demonstrate mission feasibility by testing core technologies while validating more modest scientific objectives. \peg was thus dedicated to studying Pegasids—Jupiter-sized planets orbiting extremely close to their host stars. Despite their relative simplicity, implementing these missions remains complex due to the stringent instrument precision required. It is in this context that the \pe demonstrator was developed, with the goal of proving the feasibility of achieving a deep and highly stable stellar null based on the \peg mission concept.

Funded by CNES and the Île-de-France region, \pe was developed by a consortium comprising IAS, Paris Observatory–Meudon, ONERA, Observatoire de la Côte d'Azur, and Thales Alenia Space. Its objective was to validate the requirements of the \peg mission by achieving a deep, long-term stable stellar null. To achieve this goal, an innovative cophasing system was developed to control the beams' attitude in tip/tilt and optical path difference (OPD). Following the definition phase of the bench, my PhD work aimed to integrate and characterize its various components and to bridge the gap between the cophasing control system and the scientific null depth measurement channel.

%§§§§§§§§§§§§§§§§§§§§§§§§§§§§§§§§§§§§§§§§§§§§§§§§§§§§§§§§§§§§§§§§§§§§§§§§§§§§§§§
\section*{Summary of Achievements}
\label{sec-bilan}

During my PhD, I integrated nearly all the components of the \pe testbed through a step-by-step approach, allowing me to evaluate the contribution of each added component and its interaction with the rest of the system. I developed a real-time control program to manage the feedback loops and enable live diagnostics of the testbed parameters. An initial design flaw in the source module led me to redefine this system, incorporating a higher-power source capable of feeding both the cophasing and science channels simultaneously. I also established a complete suite of calibration procedures: a Fourier transform spectroscopy mode for fine spectral characterization of the cophasing and science channels; calibration routines for the OPD and tip/tilt control loops; and a science channel calibration procedure coupled with the control loops to optimize, compute, and analyze the testbed's null depth with high precision. To optimize beam cophasing in the presence of dynamic disturbances, I also implemented and experimentally validated an optimal control law designed specifically for vibration rejection.

In the initial autocollimination configuration, I demonstrated the testbed's capability to cophase the beams with sub-nanometric precision \tir{$0.22$~nm RMS in optical path difference and 56~mas RMS in tip/tilt, corresponding to $0.4$\% of the Airy disk}. Residual path errors remained sub-nanometric in the final testbed configuration, even when injecting dynamic disturbances typical of a formation-flying mission like \peg. In the absence of injected disturbances, cophasing performance reached 0.3~nm RMS in optical path difference and maintained 56~mas RMS in tip/tilt ($0.4$\% of the Airy disk).

In the autocollimination setup, I measured very deep monochromatic null depths of $6\E{-5}$ stabilized to $3\E{-5}$. In the final configuration, thanks to accurate calibrations and precise optical alignment, I improved this result to $5.6\E{-6}$. Using the broadband supercontinuum source in this final setup, I ultimately achieved a world-record published null depth of $8.8\E{-6}$ across a very broad fractional bandwidth of 37\%; the short-term stability over a 100~s integration time was also excellent, at $9\E{-8}$. I validated deep nulling over multi-hour durations, demonstrating a mean null depth of $1.02\E{-5}$ with a long-term stability of $9\E{-7}$. Finally, I verified the coupling between cophasing control and stellar suppression by injecting and actively rejecting OPD disturbances, maintaining a deep $1.14\E{-5}$ null despite strong injected vibrations.

Overall, across both cophasing control and science measurements, I pushed testbed performance to levels an order of magnitude better than the initial requirements. I have thus demonstrated that, using this simple yet effective architecture, it is possible to achieve null depths that are not only fully compatible with observing Pegasids, but also with characterizing Earth-like exoplanets.

Furthermore, I applied my experience in vibration analysis on \pe to perform a vibration study on the \sce coronagraphic instrument for the Subaru Telescope. This analysis revealed that the instrument suffers from strong mechanical vibrations and would benefit significantly from an LQG controller. However, additional work on detector clock synchronization is required before this control law can be implemented to reject these vibrations.

%§§§§§§§§§§§§§§§§§§§§§§§§§§§§§§§§§§§§§§§§§§§§§§§§§§§§§§§§§§§§§§§§§§§§§§§§§§§§§§§
\section*{Perspectives}
\label{sec-perspectives}

Based on the complete dataset from the \pe testbed, I extrapolated our findings to a space mission like \peg. This extrapolation revealed that such a mission is feasible, provided certain critical trade-offs are taken into account. First, the fringe sensor requires a sufficiently high signal-to-noise ratio to ensure effective OPD control; this imposes a trade-off between collector aperture size and loop execution frequency. However, \pe's excellent performance demonstrated that certain constraints can be relaxed—for example, on flux imbalance between arms—to tolerate larger path delay residuals than the $2.5$~nm originally budgeted. The control loop bandwidth sets a limit on tolerable disturbance frequencies; this directly informs the specifications for reaction wheels used for satellite attitude control. Thermal stability studies on \pe's MMZ also provide realistic guidelines for the thermal control requirements of \peg's optical train, which remain very reasonable. This heavily impacts the choice of mission orbit, as a Sun-Earth L2 Lagrangian orbit—while thermally stable—is technologically more demanding and thus more costly.

One key milestone that remains to be demonstrated is coarse fringe acquisition, which is essential for a mission like \peg. The testbed has already demonstrated the ability to detect fringes sweeping at velocities of 250~\mum.s$^{-1}$; the next step is to inject the parabolic drift trajectories characteristic of the \peg mission concept and validate fringe capture using the testbed's two long-stroke delay lines. This step is scheduled for completion before the end of 2012.

As outlined in the \pe project timeline (Figure~\ref{fig-calendrier}), a new source module developed at the Observatoire de la Côte d'Azur aims to simulate the host star, its exoplanet, and its exozodiacal dust disk simultaneously. Delivered at the end of 2011, this module is scheduled for integration on the bench during 2012; subsequent tests will evaluate planetary signal extraction within the dark fringe in the presence of exozodiacal light. Other future directions have been proposed, such as building a thermally stable MMZ No. 3 or adding two additional arms to \pe to simulate a 4-aperture Darwin/TPF-I array, exploiting the testbed's high-contrast capabilities.

However, a flagship pathfinder mission like \peg remains costly. Consequently, proposals have emerged to test core functions \tir{fringe tracking and cophasing} using a balloon-borne interferometer. In addition to demonstrating the feasibility of a mission like \peg, this instrument—named BALEINE\footnote{BALlon Expérimental pour Interférométrie aNnulante d'Exosystèmes}—would fulfill the scientific objective of surveying exozodiacal dust. A NASA team has developed a similar, more ambitious balloon-borne project designed to characterize extrasolar planets and debris disks. That project, named BENI\footnote{Balloon Exoplanet Nulling Interferometer}, is a 3-aperture visible nulling interferometer; a laboratory testbed has already been integrated, achieving a deep $6.25\E{-5}$ null over a ~40\% bandwidth. The main bottleneck for both BALEINE and BENI, however, will be dynamic disturbance rejection, as atmospheric balloon environments experience much higher perturbation levels.

It will likely take several years, or even decades, before high-resolution spectra of Earth-like exoplanets can be obtained via coronagraphy or nulling interferometry. Despite persistent technological hurdles \tir{10~\mum single-mode optical fibers, formation flying, space cryogenics}, successful projects like \pe demonstrate that nulling interferometry yields exceptional results and provides solid guidelines for sizing future space missions—much as I hope \sce will do for high-contrast coronagraphy.

\cleardoublepage
\fancyhead{}
\fancyhead[LE,RO]{\thepage}
\fancyhead[RE,LO]{\small BIBLIOGRAPHY}
\bibliographystyle{authordate3}
\bibliography{these_JLozi}

%\bibliography{Acronymes,Actes,Articles,BibActex,these_JLozi}
\cleardoublepage

\cleardoublepage
\fancyhead{}
\fancyhead[LE,RO]{\thepage}
\fancyhead[RE,LO]{\small APPENDICES}
\begin{appendices}
\chapter{Communication au SPIE Astronomical telescopes and instrumentation (2010)}

Proceedings of the conference SPIE Astronomical telescopes and instrumentation, San Diego (July 2010) \cite{Lozi10}.

\newpage

\newcommand{\SPIEA}[1]{\centerline{\FIG{1.}{false,viewport=55 135 540 770,clip}{SPIE_2010_JL_#1}}}

\SPIEA{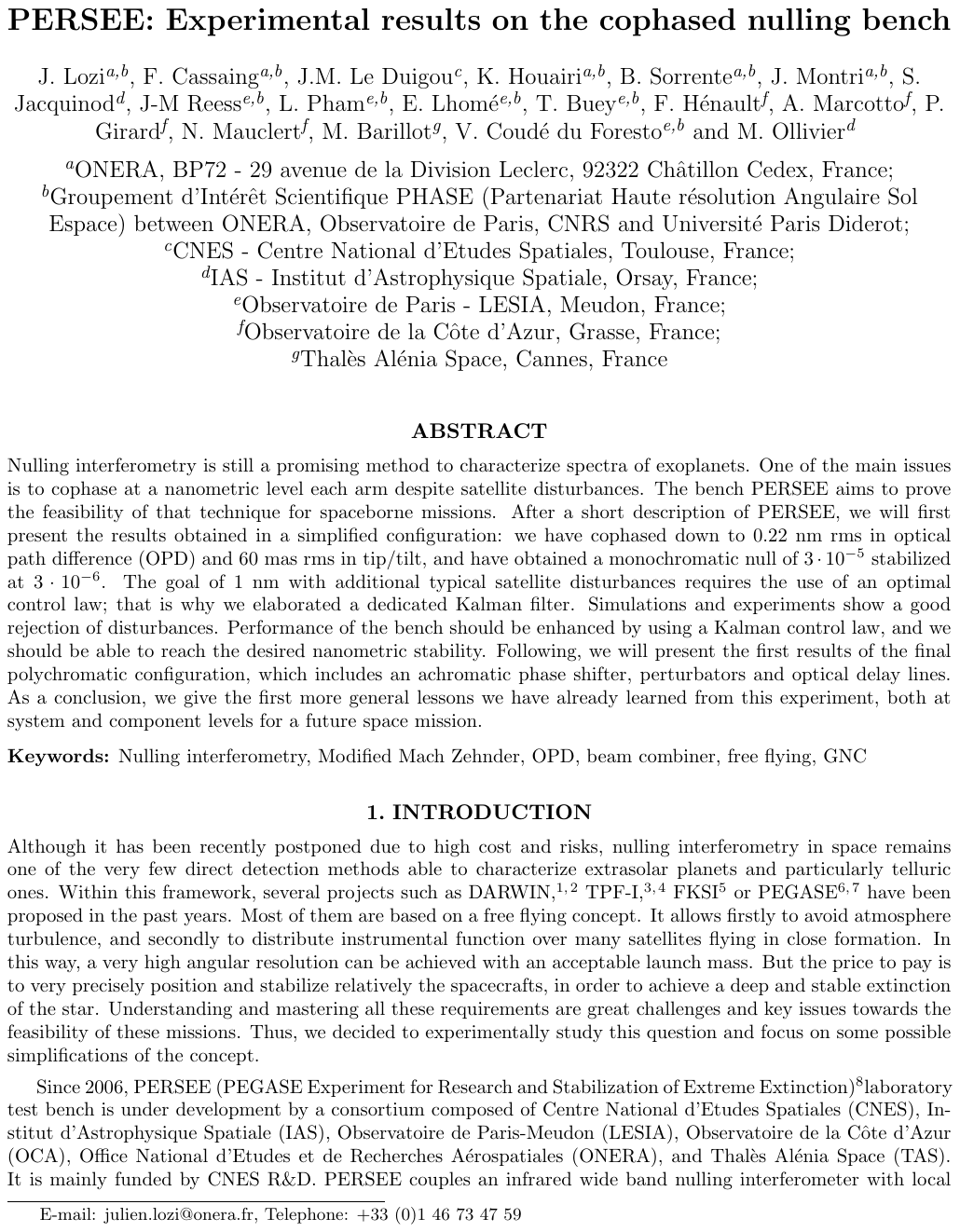}
\SPIEA{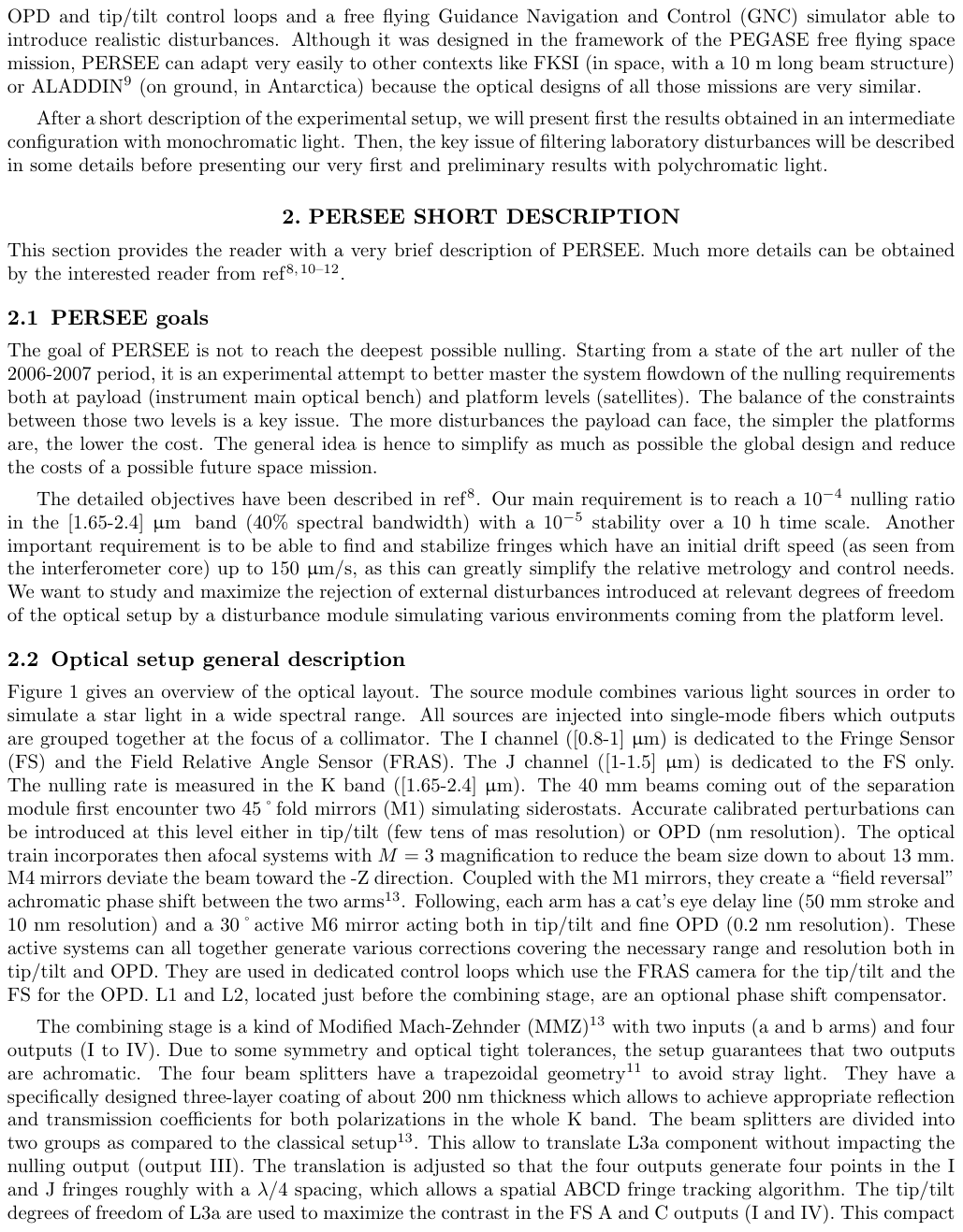}
\SPIEA{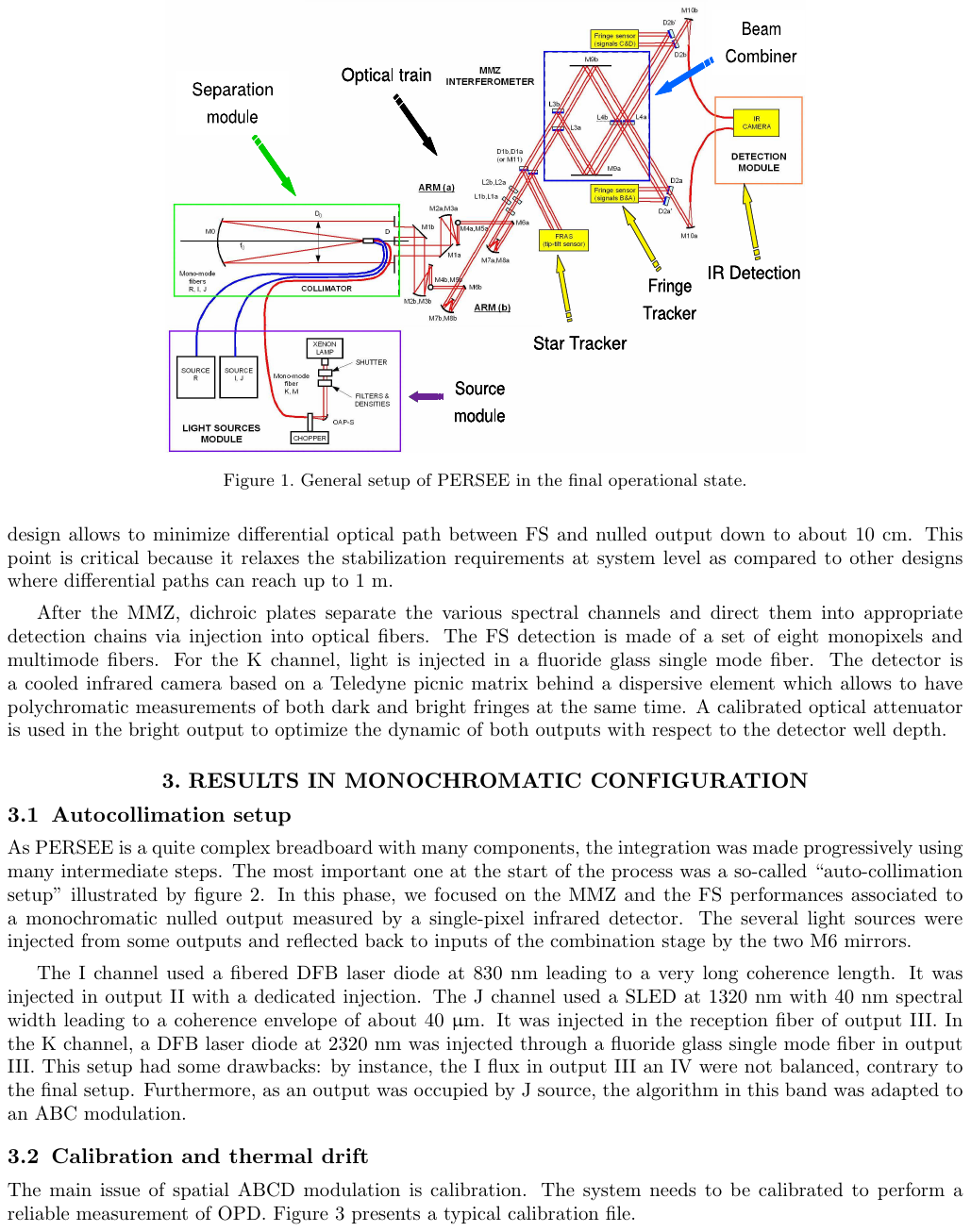}
\SPIEA{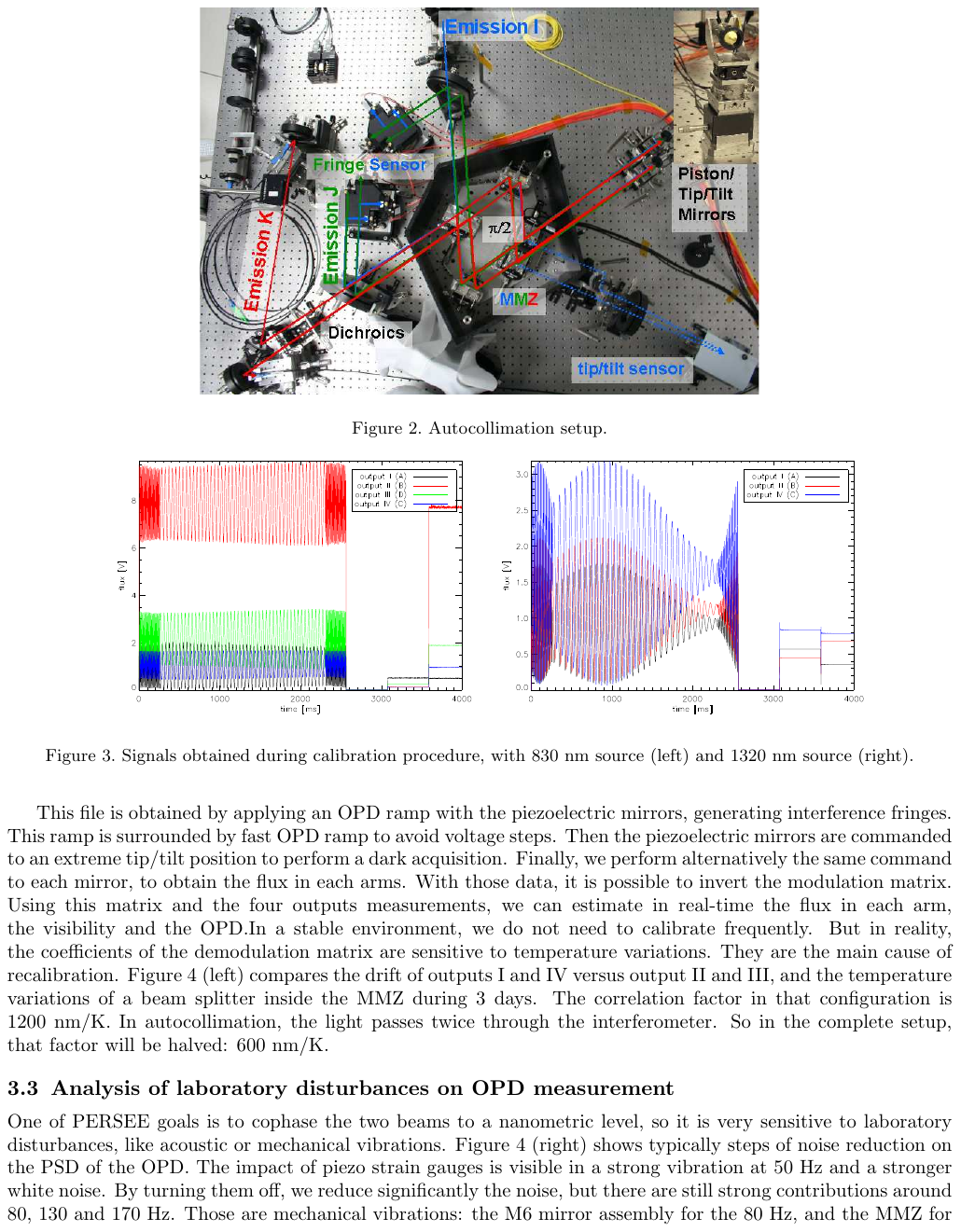}
\SPIEA{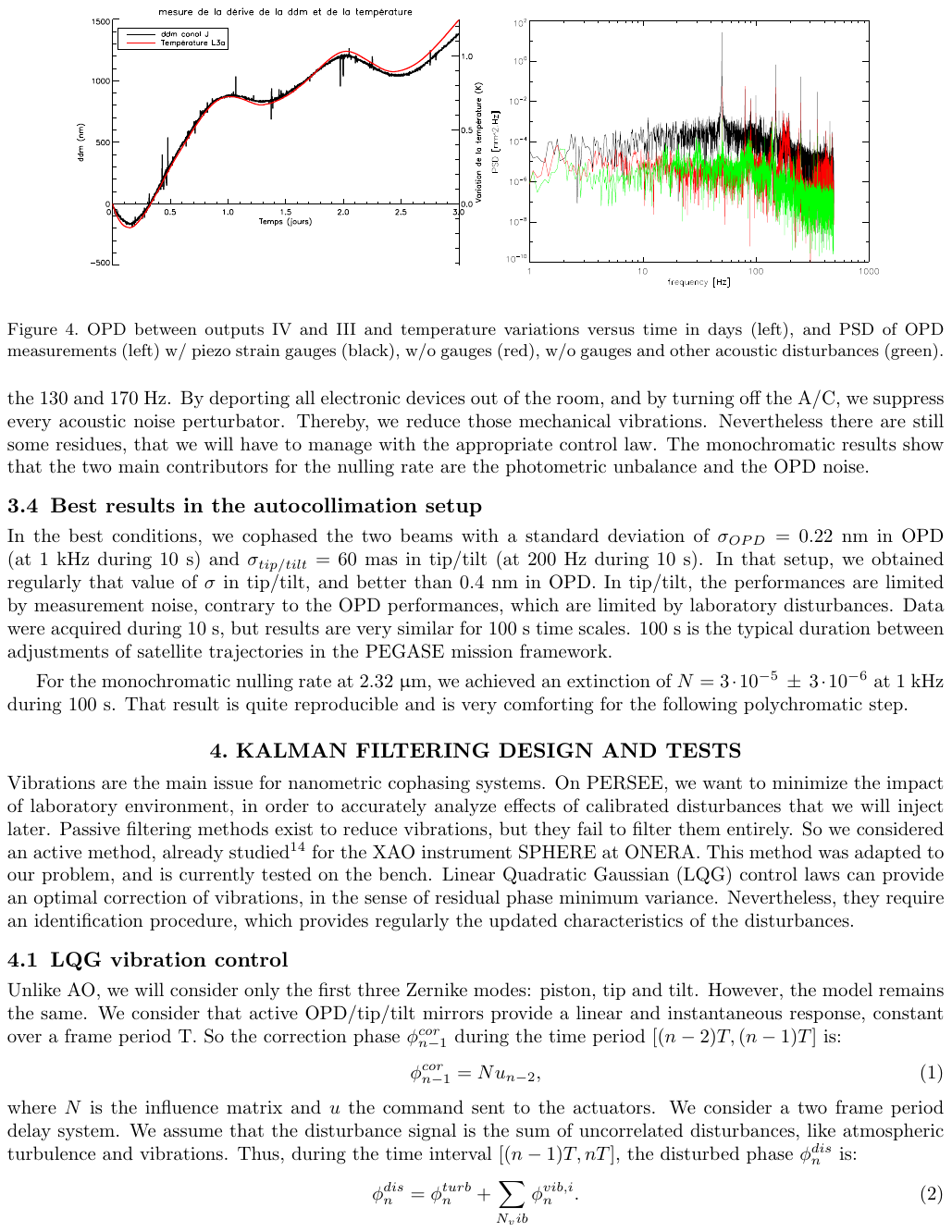}
\SPIEA{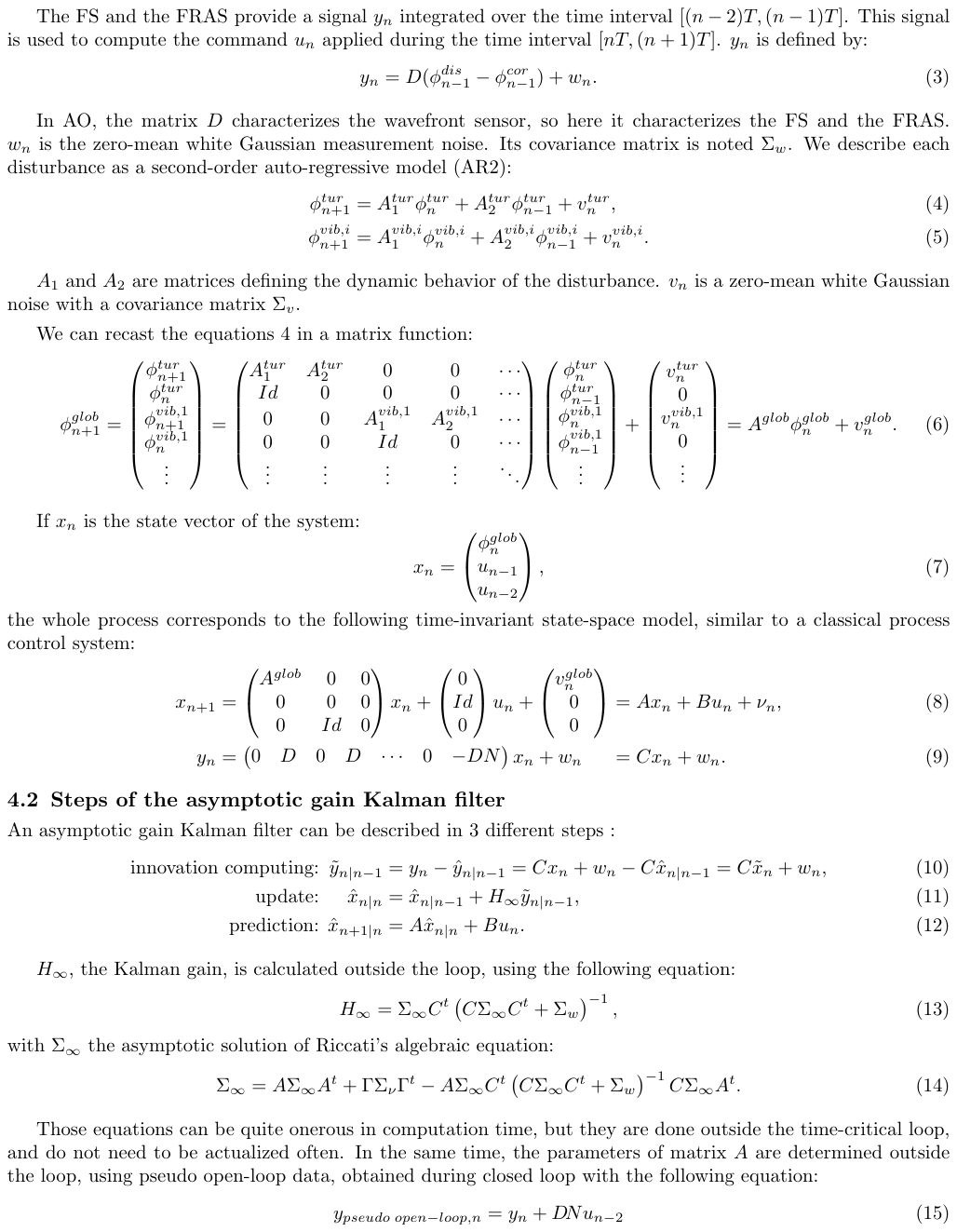}
\SPIEA{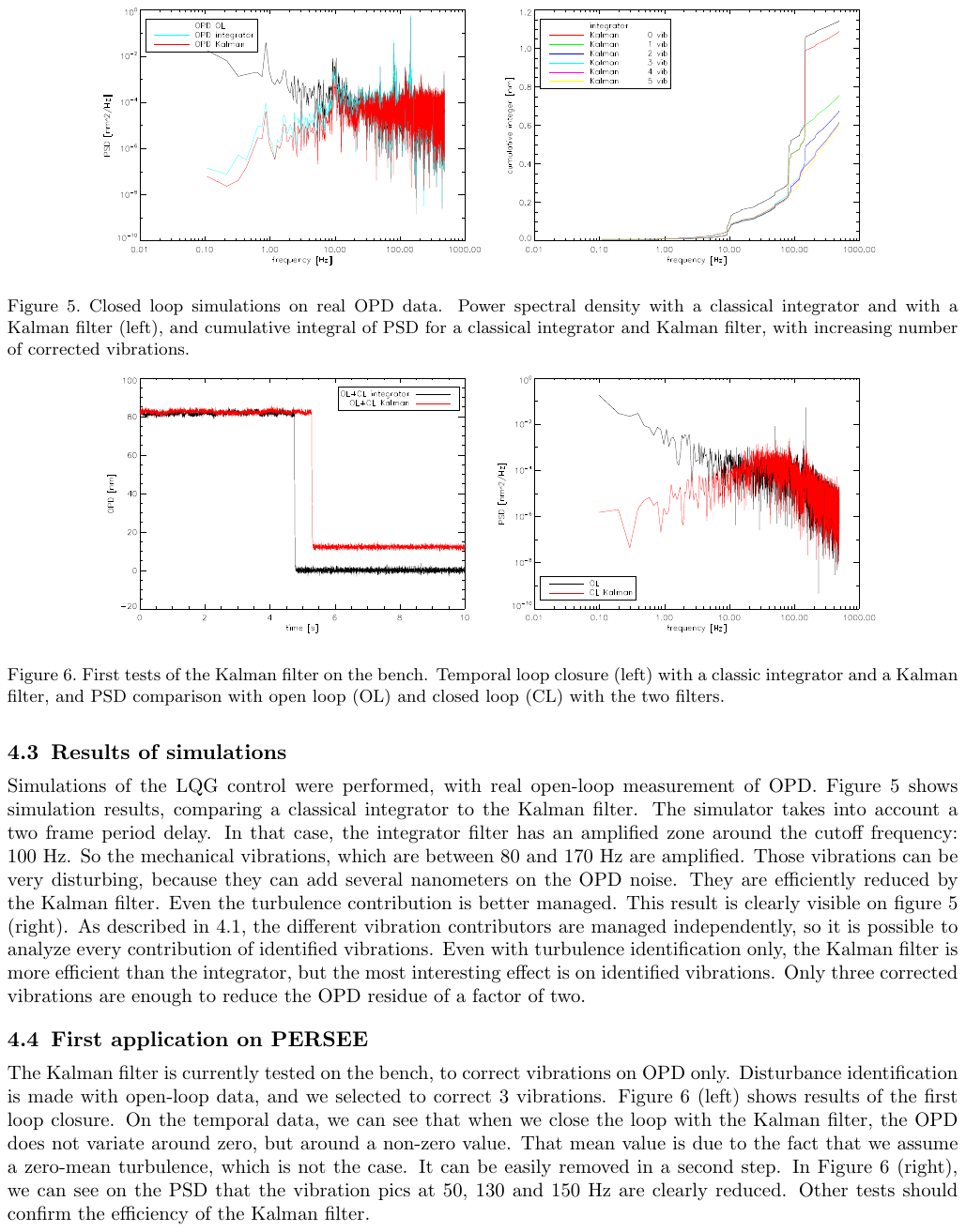}
\SPIEA{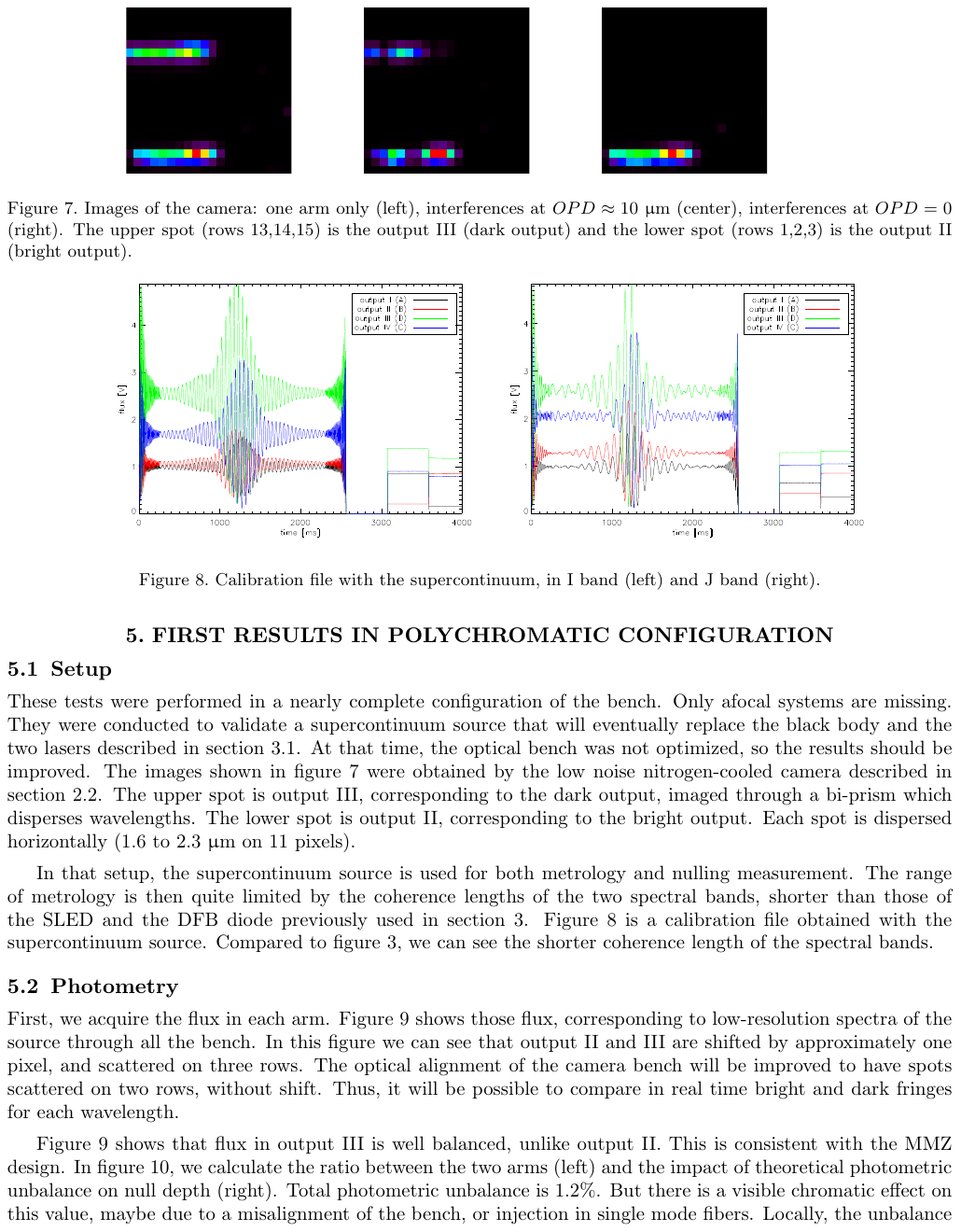}
\SPIEA{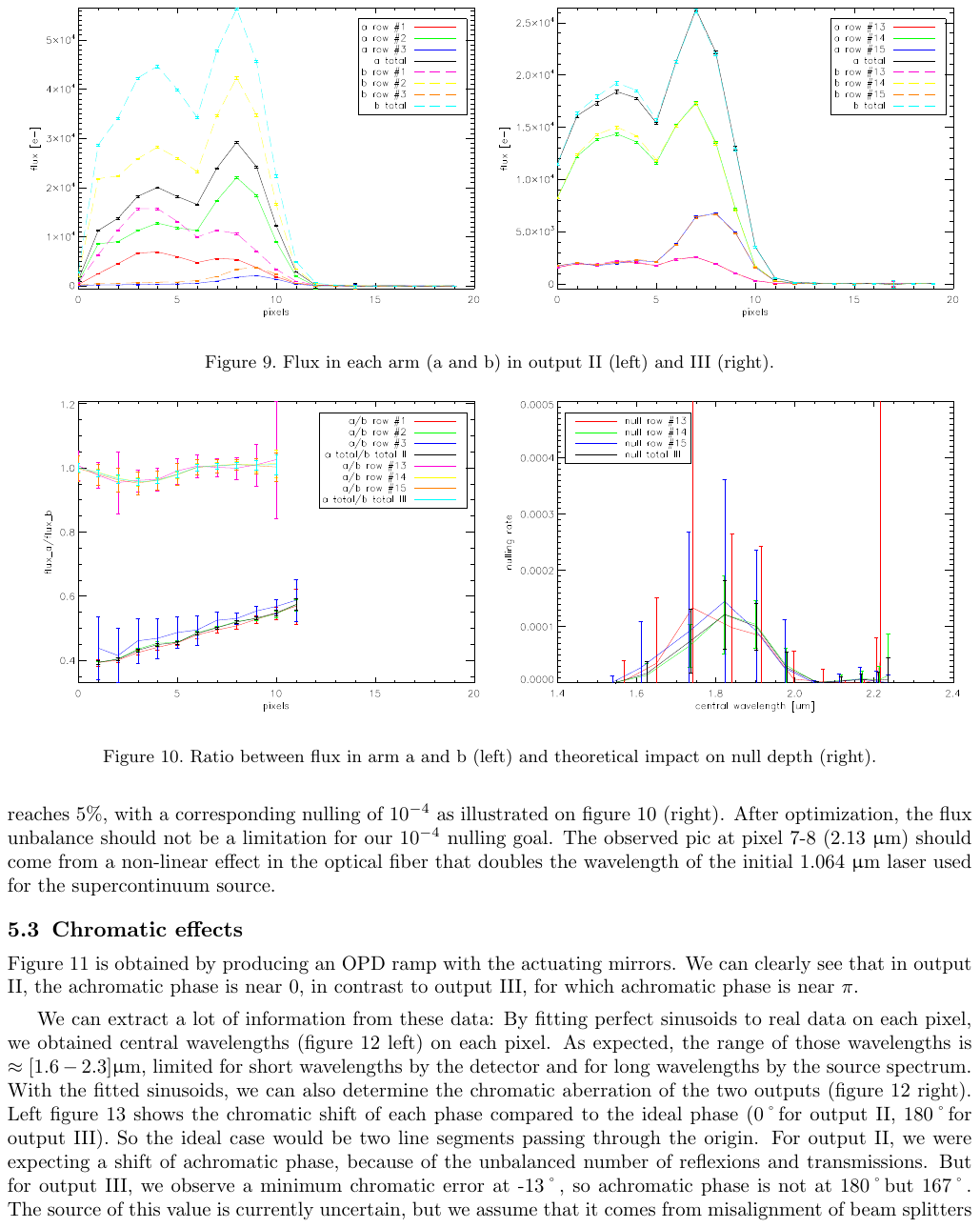}
\SPIEA{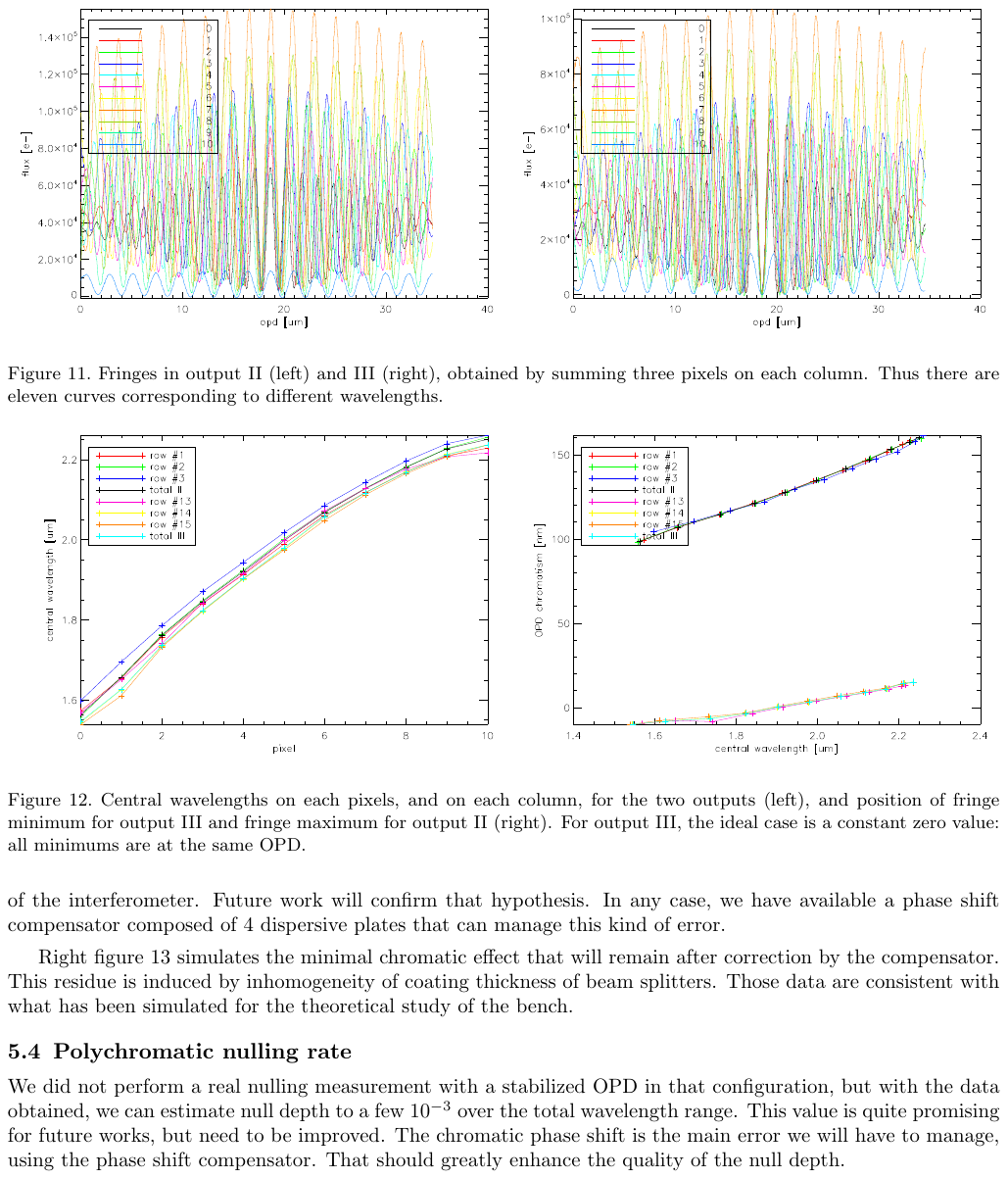}
\SPIEA{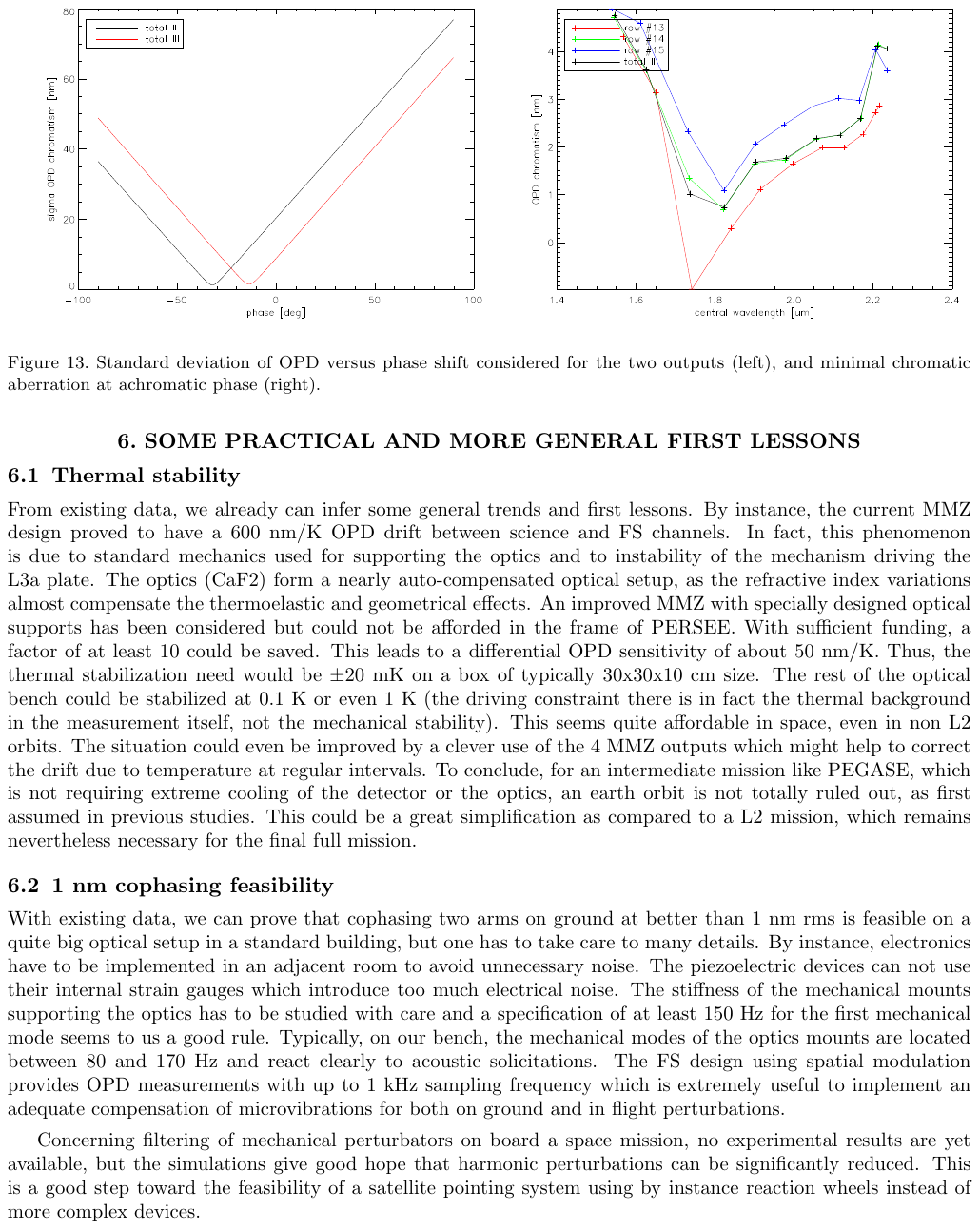}
\SPIEA{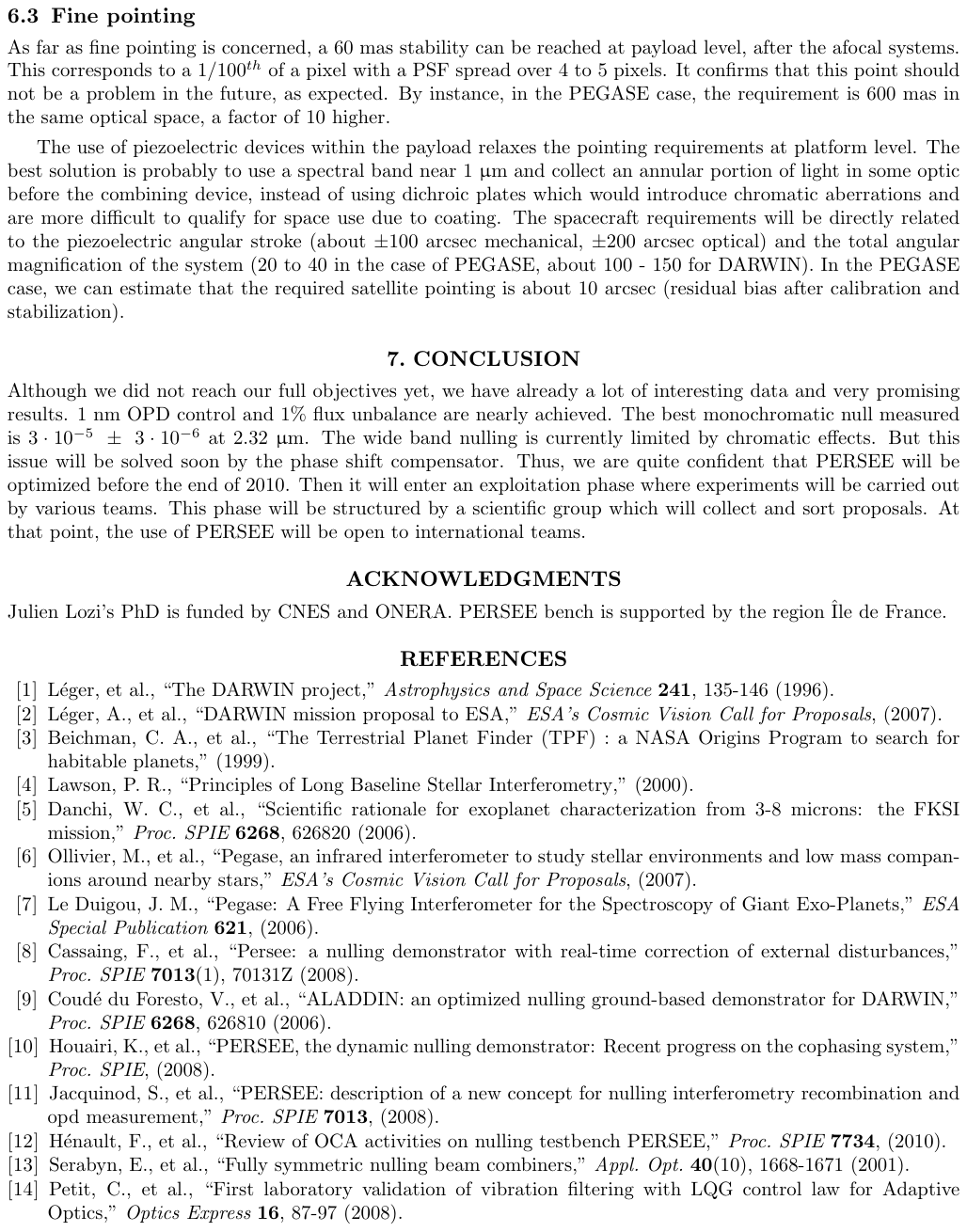}

%%% Local Variables: 
%%% mode: latex
%%% TeX-master: "../these_JL"
%%% End: 

\chapter{Communication au SPIE Optics + Photonics (2011)}

Proceedings of the conference SPIE Optics + Photonics, San Diego (August 2011) \cite{Lozi11}.

\newpage

\newcommand{\SPIEB}[1]{\centerline{\FIG{1.}{false,viewport=55 135 540 770,clip}{SPIE_2011_JL_#1}}}

\SPIEB{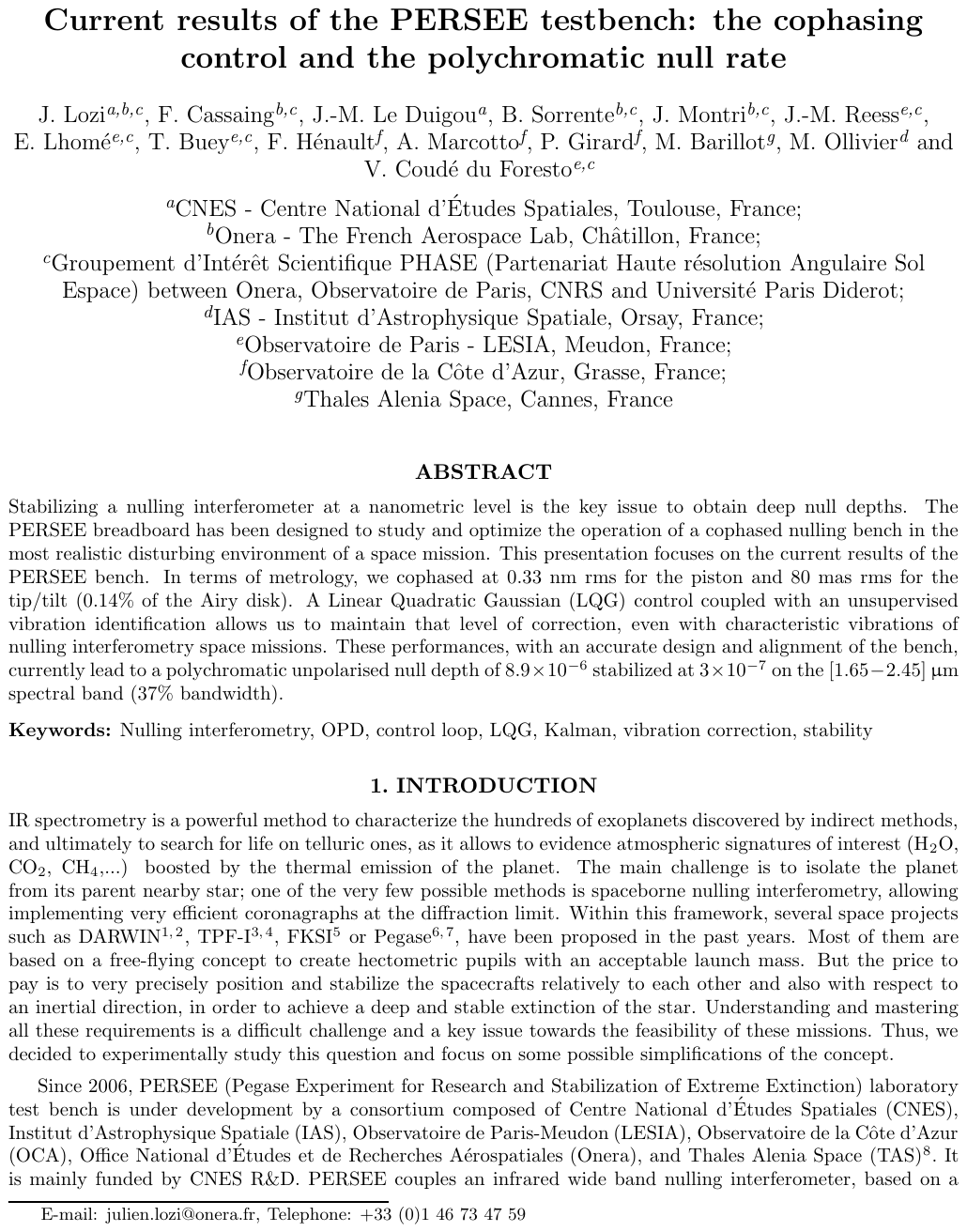}
\SPIEB{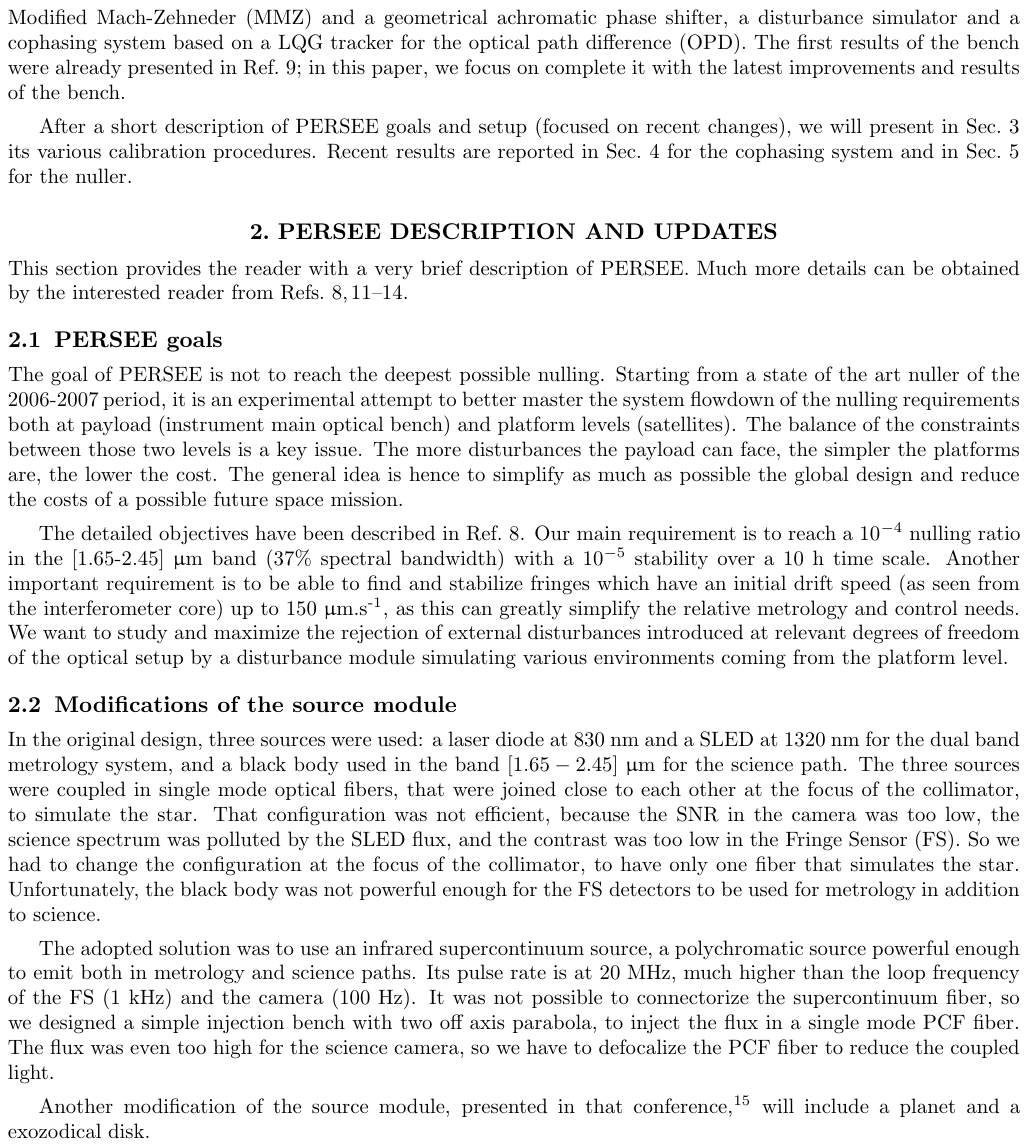}
\SPIEB{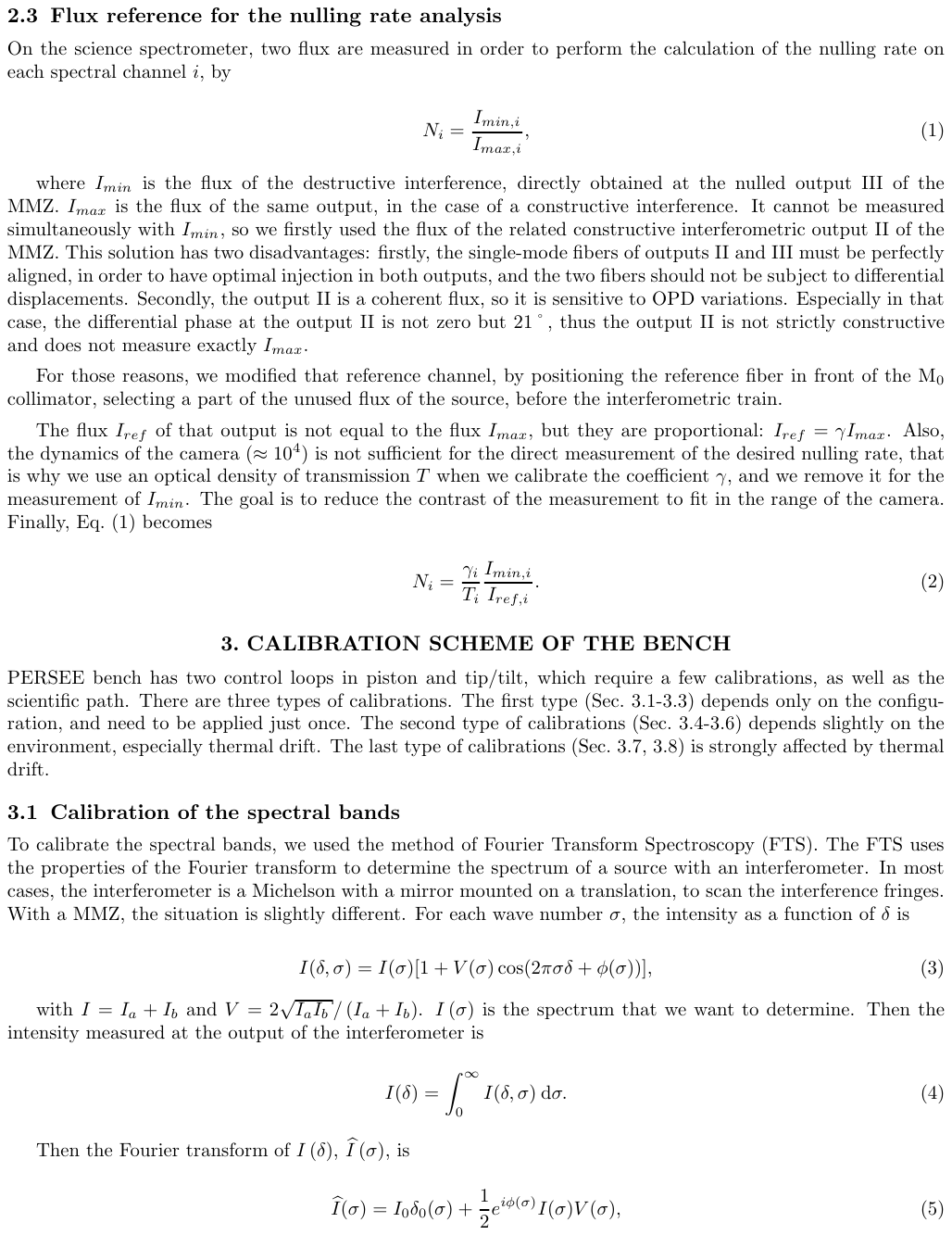}
\SPIEB{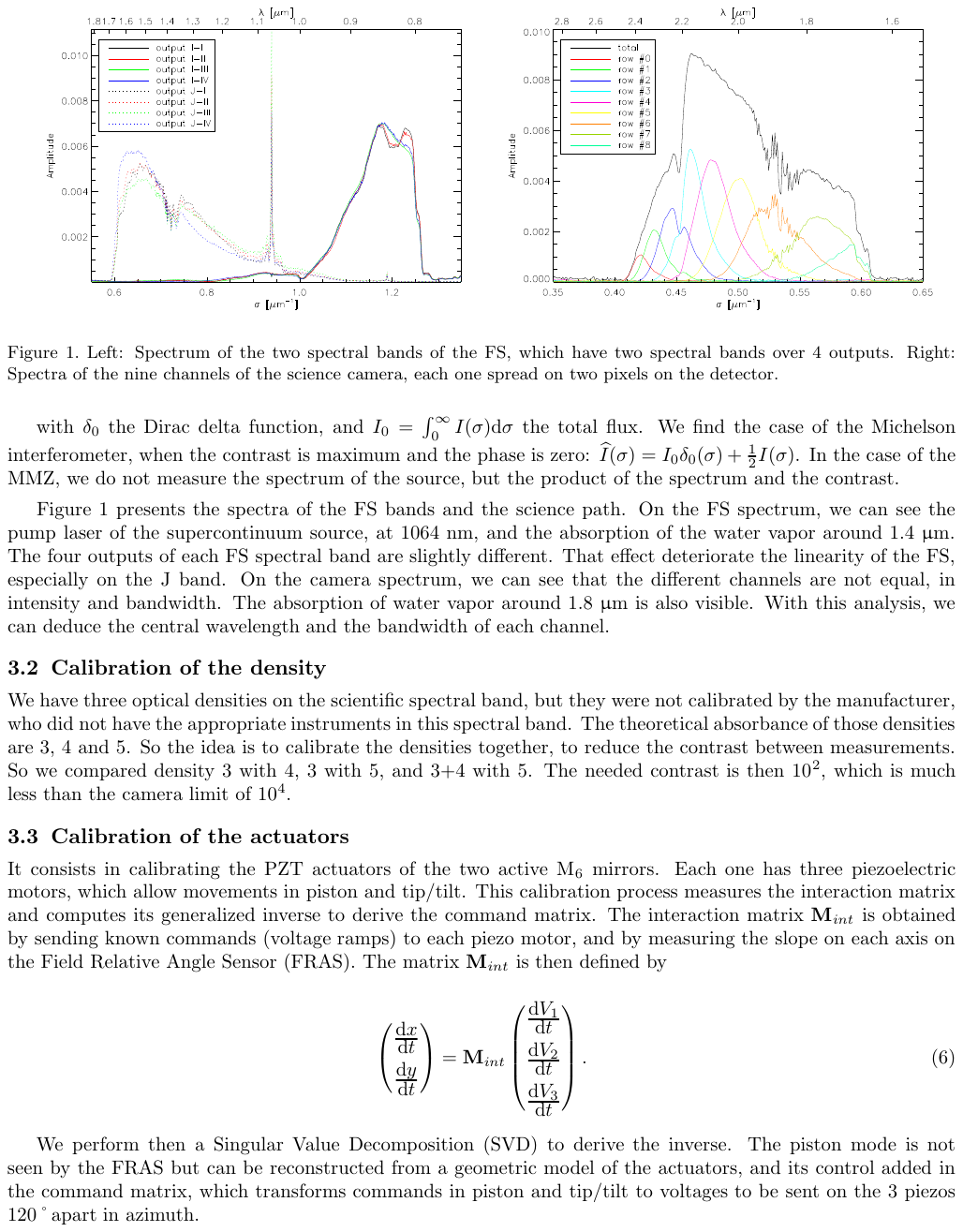}
\SPIEB{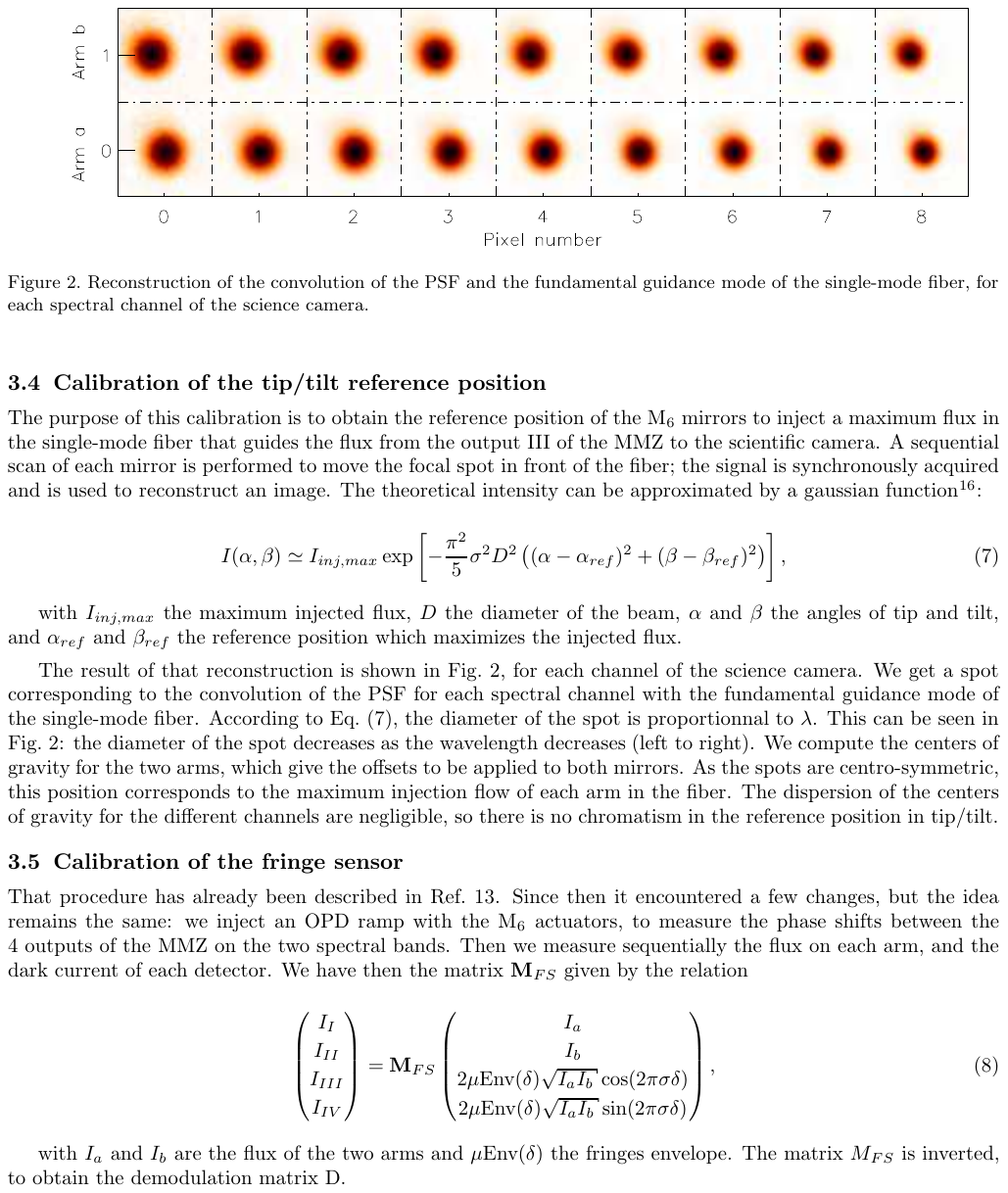}
\SPIEB{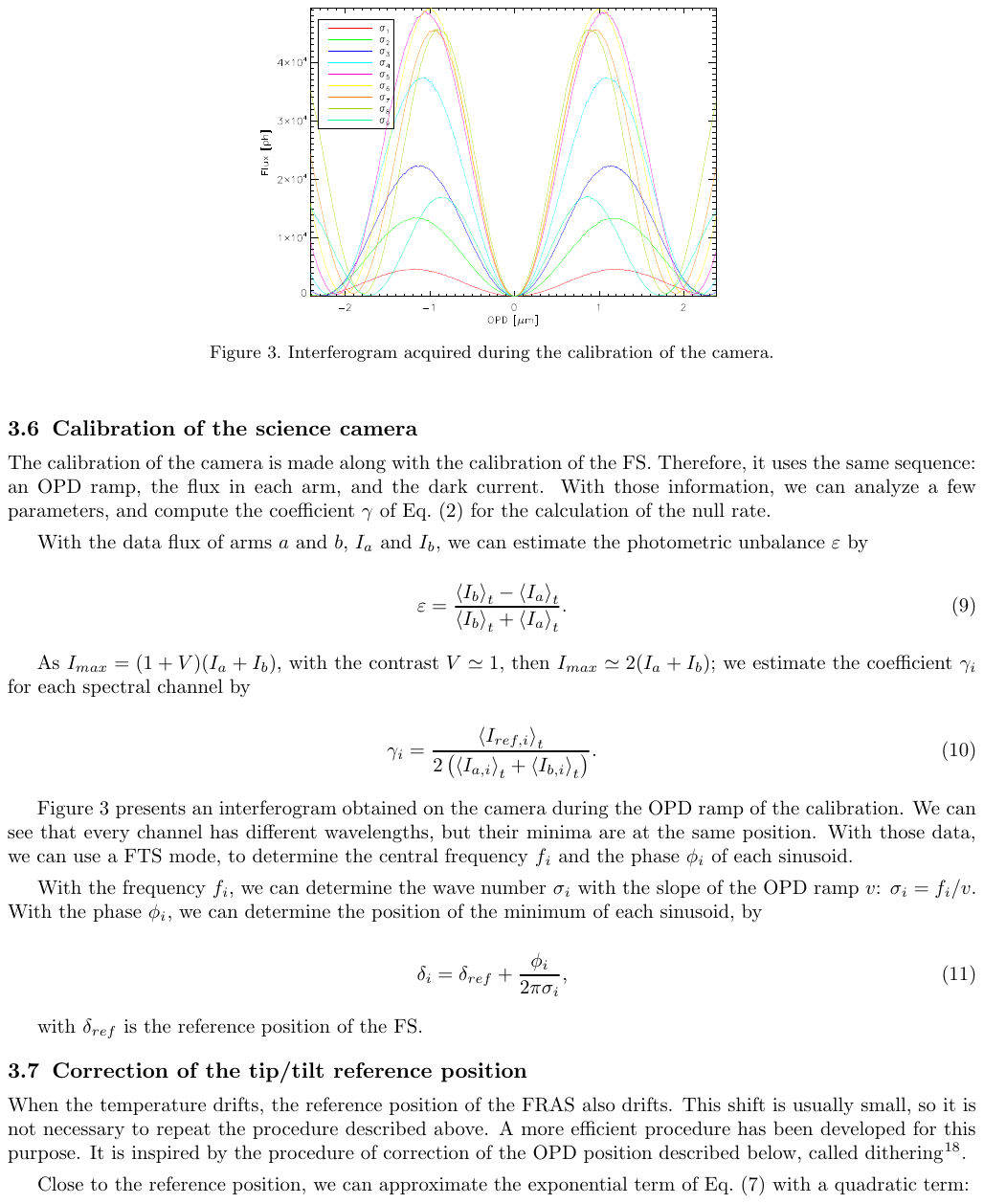}
\SPIEB{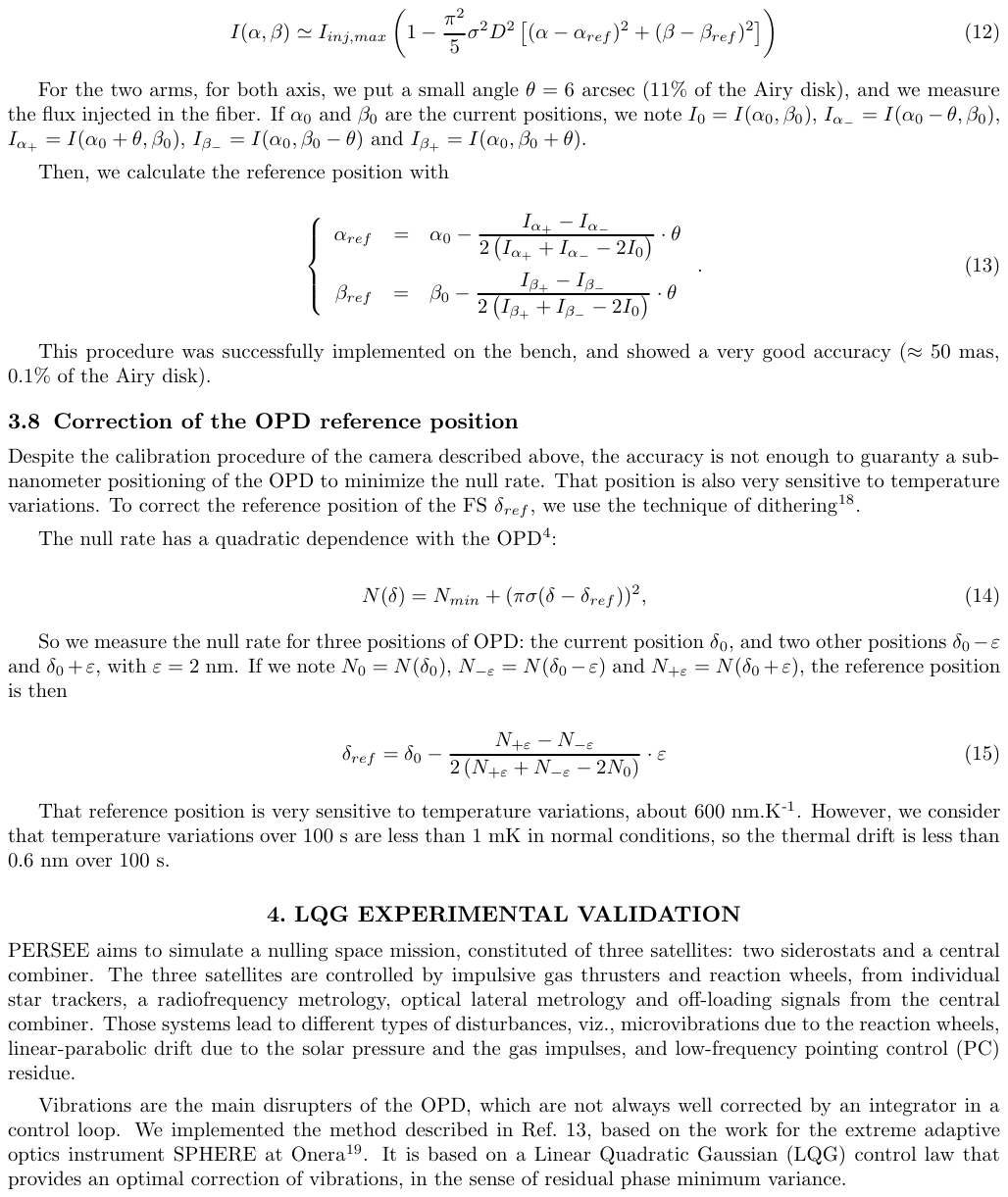}
\SPIEB{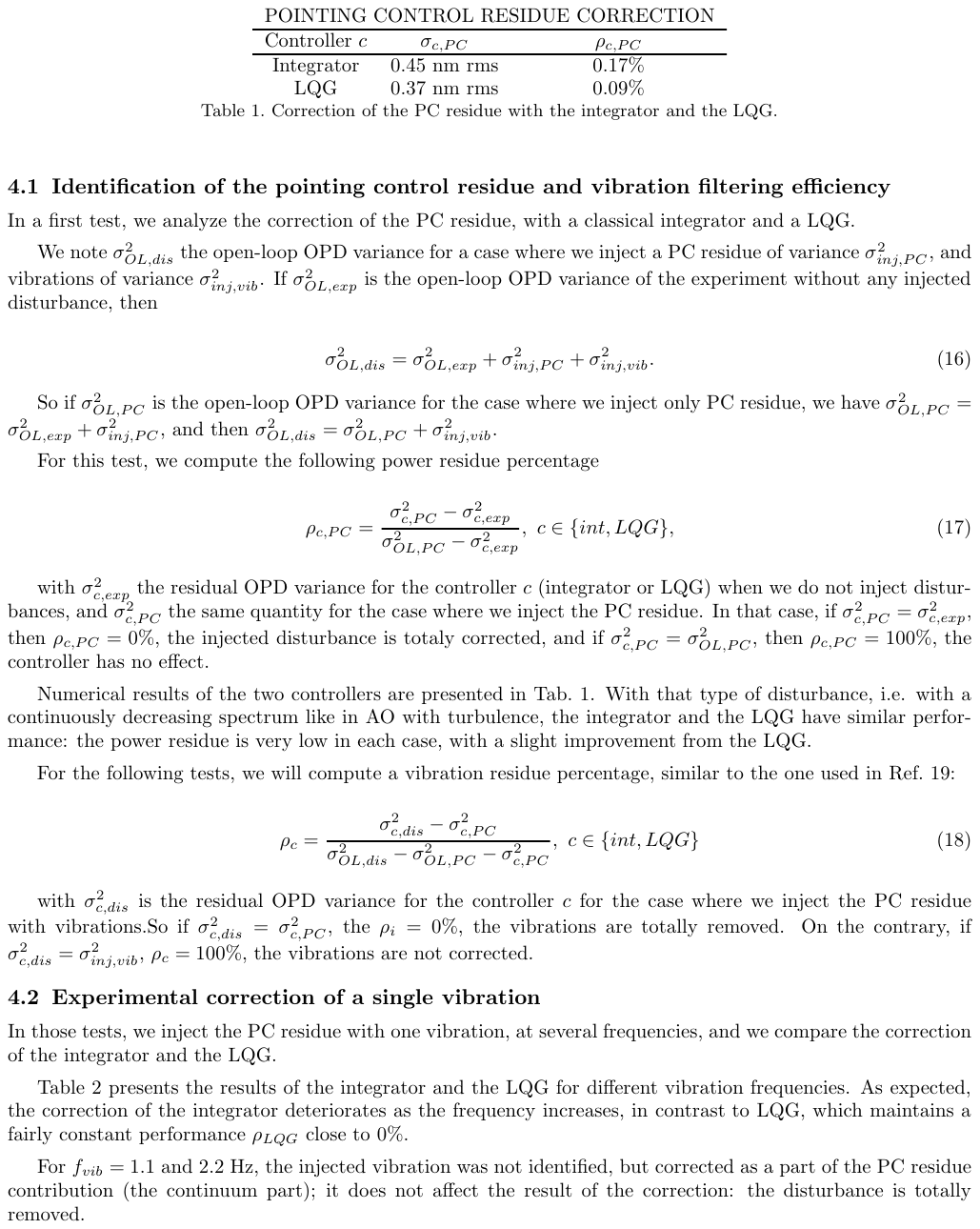}
\SPIEB{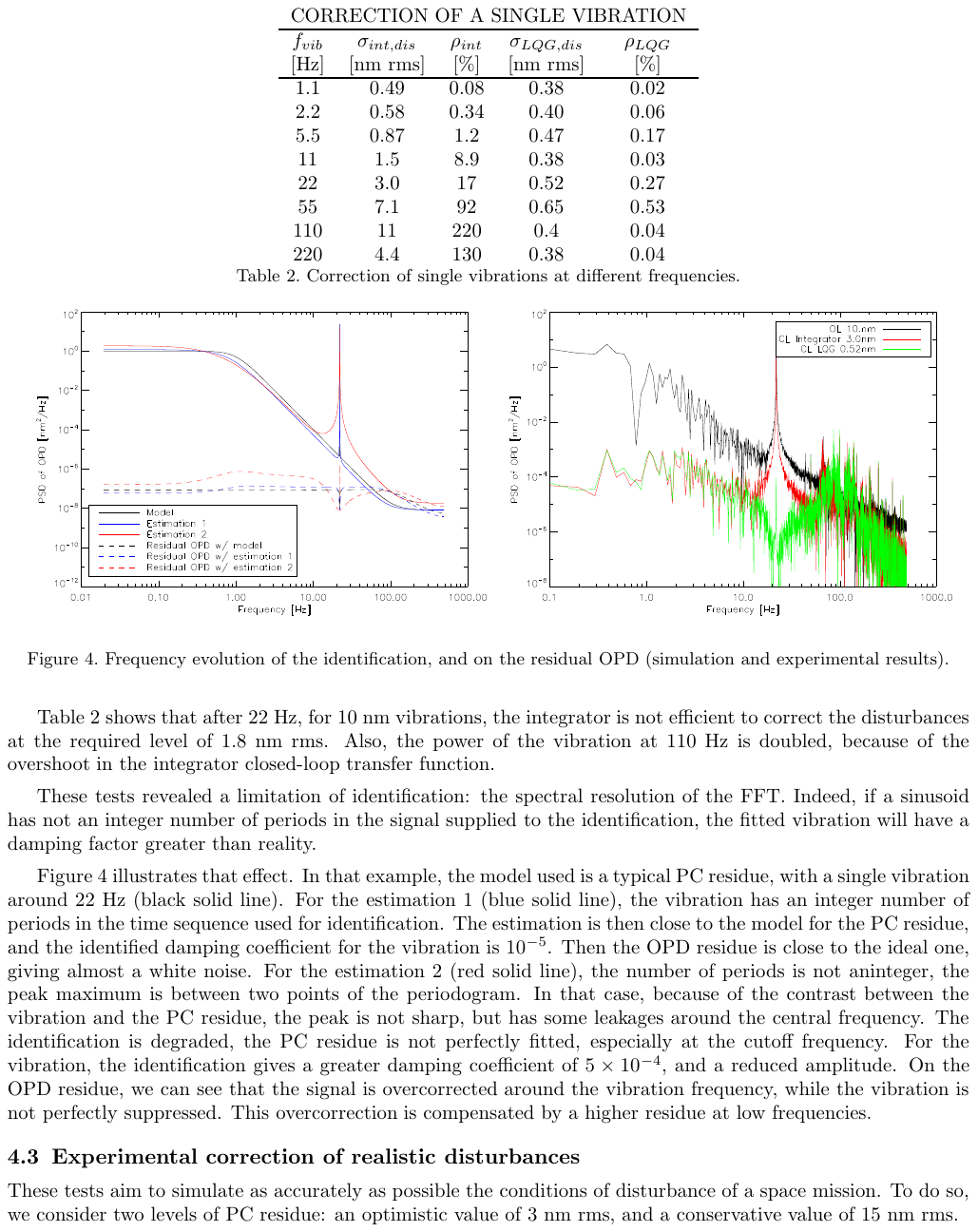}
\SPIEB{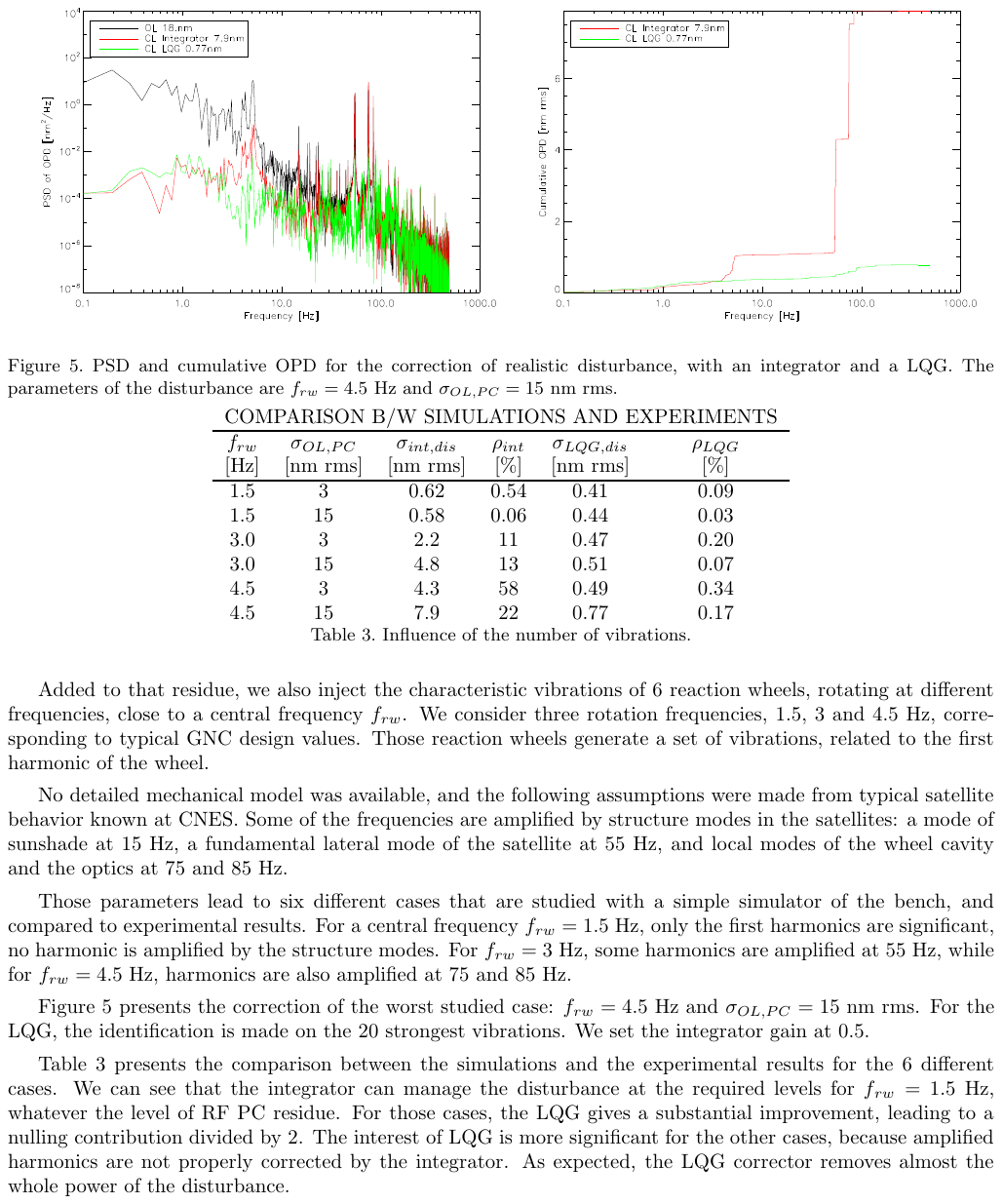}
\SPIEB{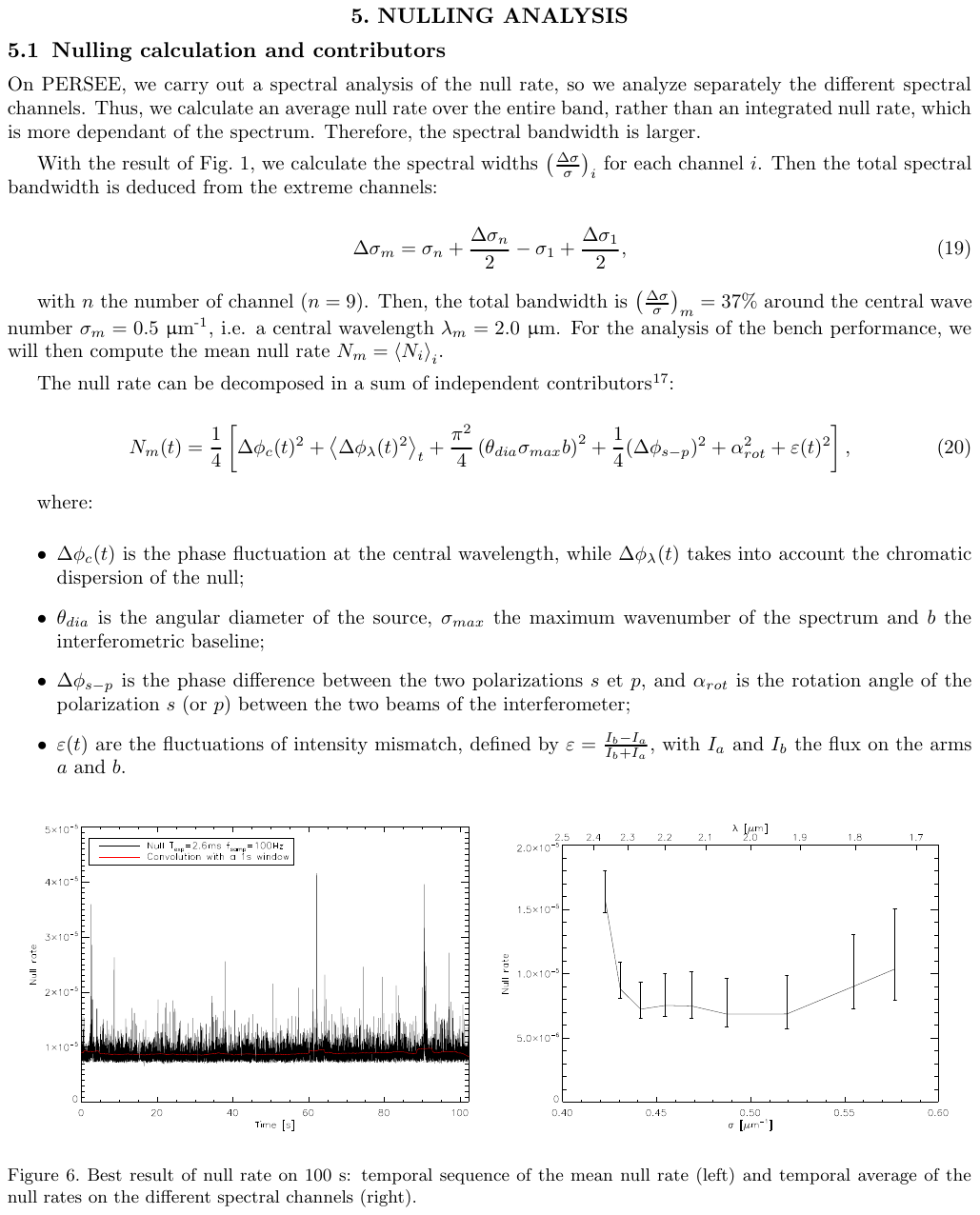}
\SPIEB{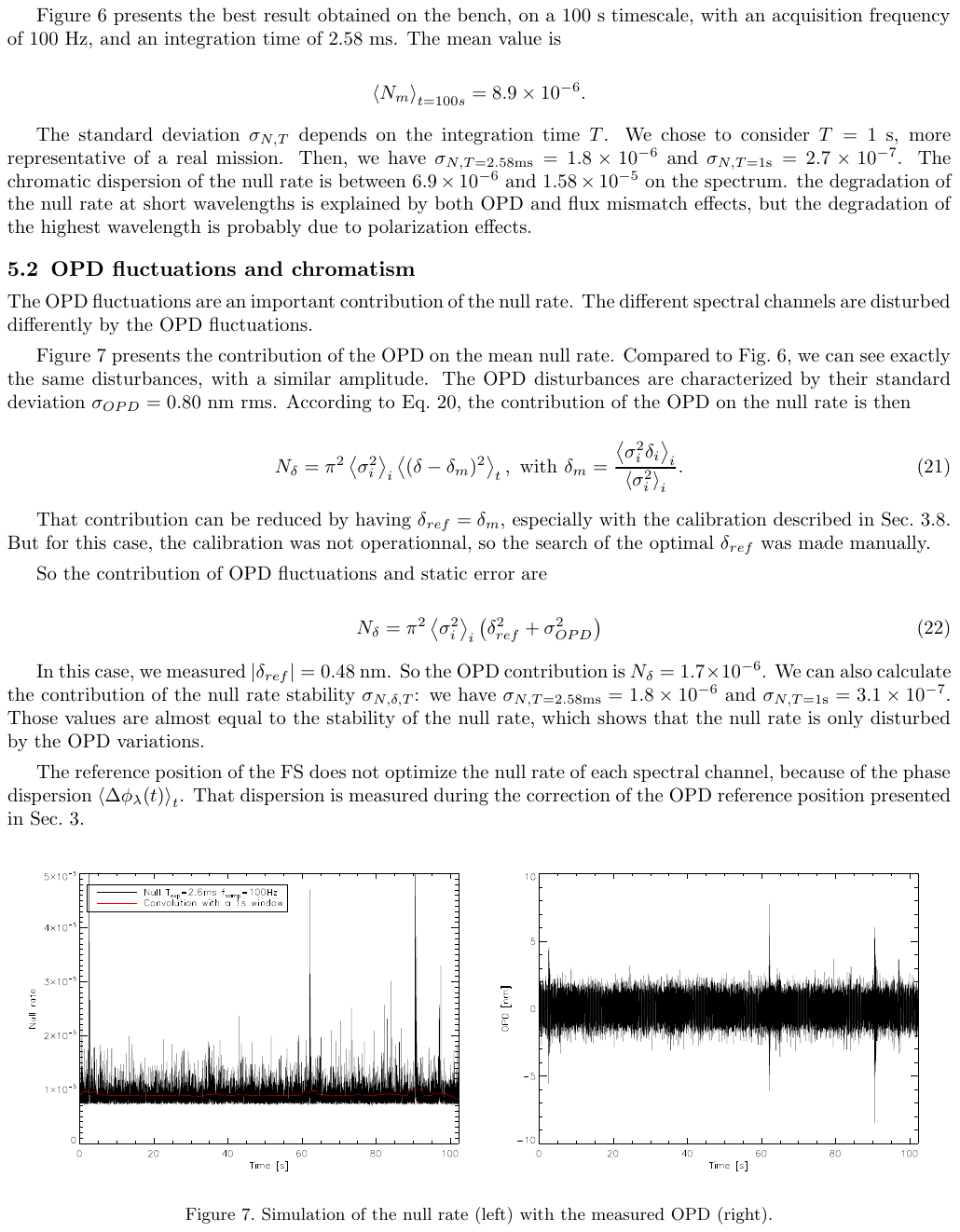}
\SPIEB{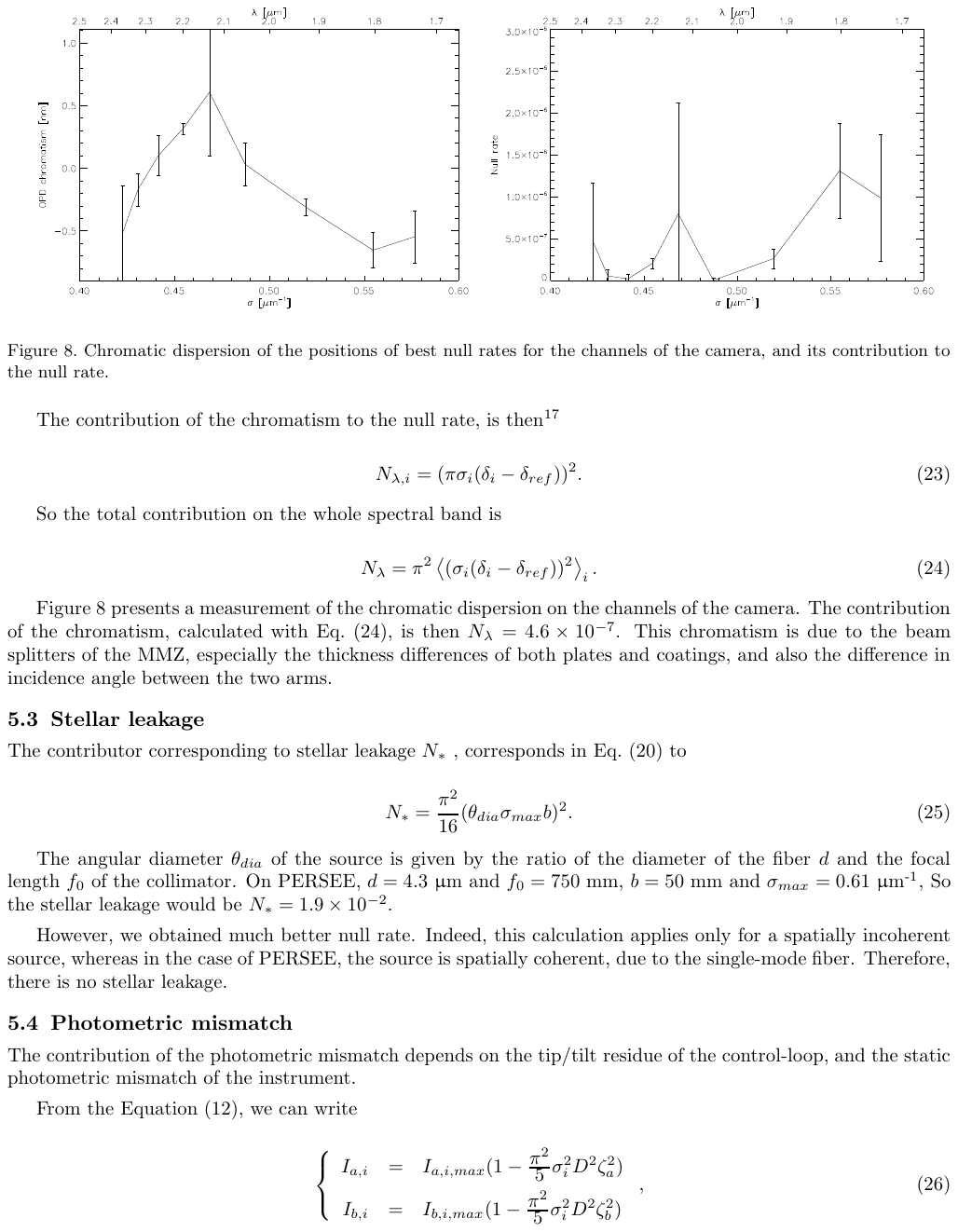}
\SPIEB{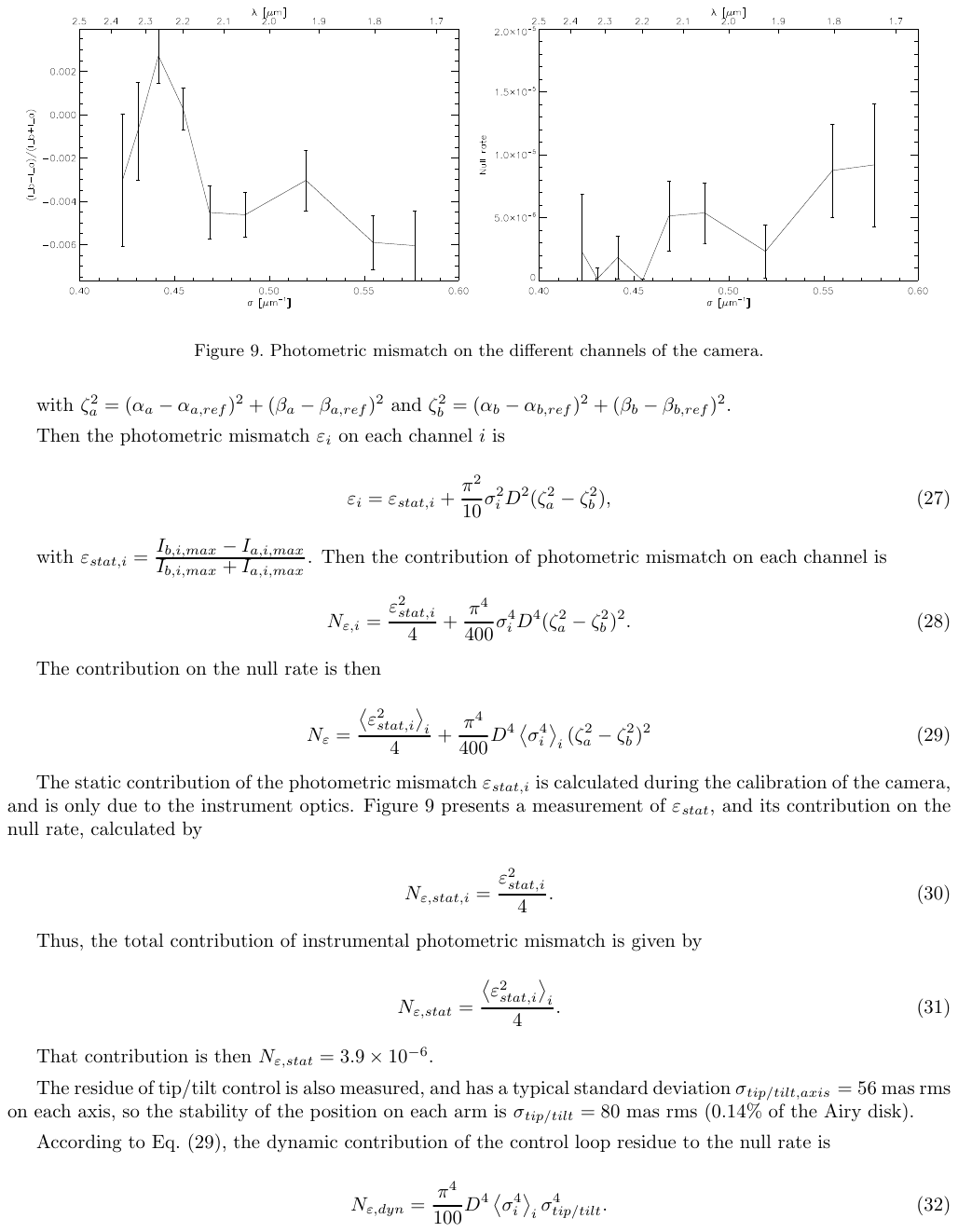}
\SPIEB{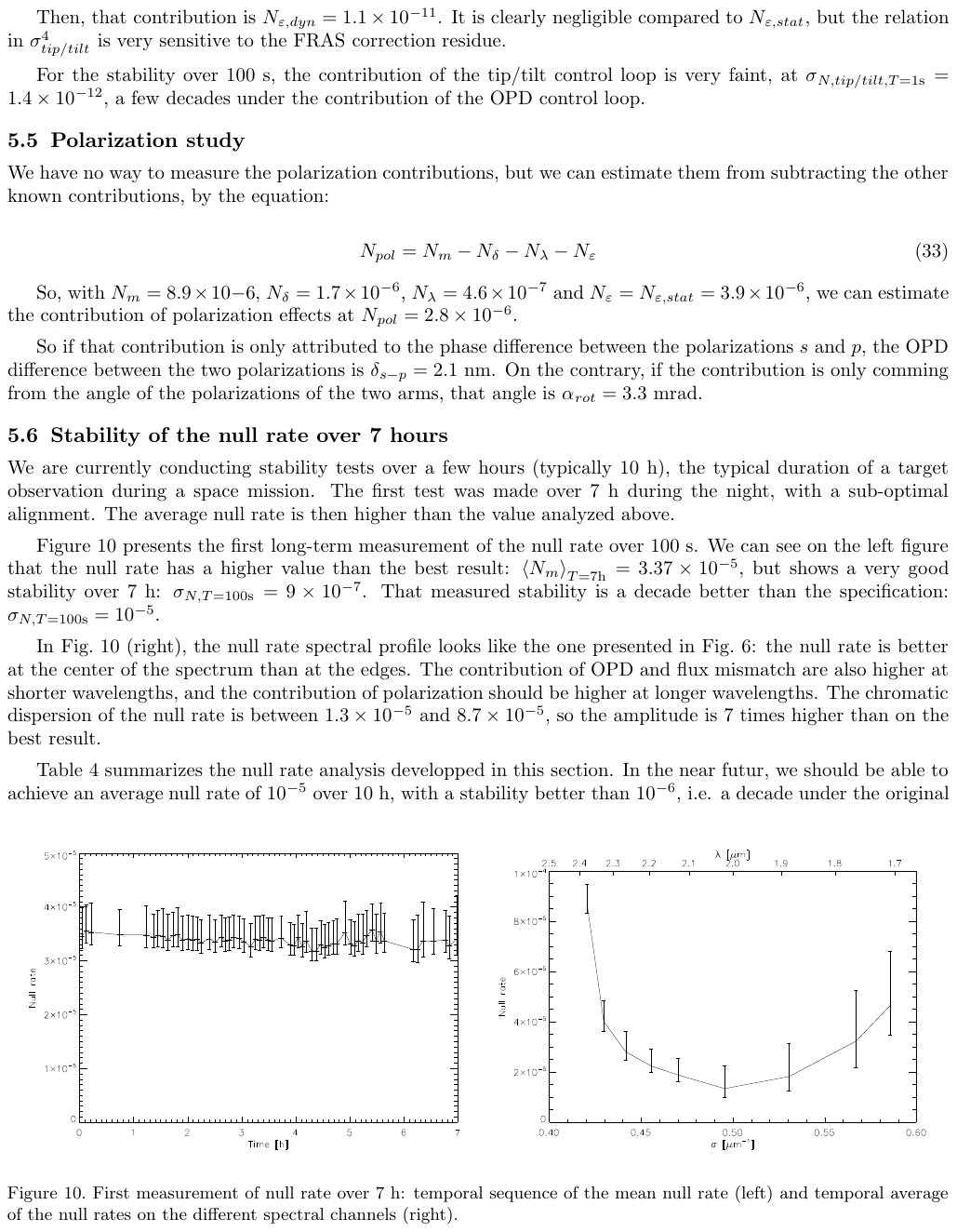}
\SPIEB{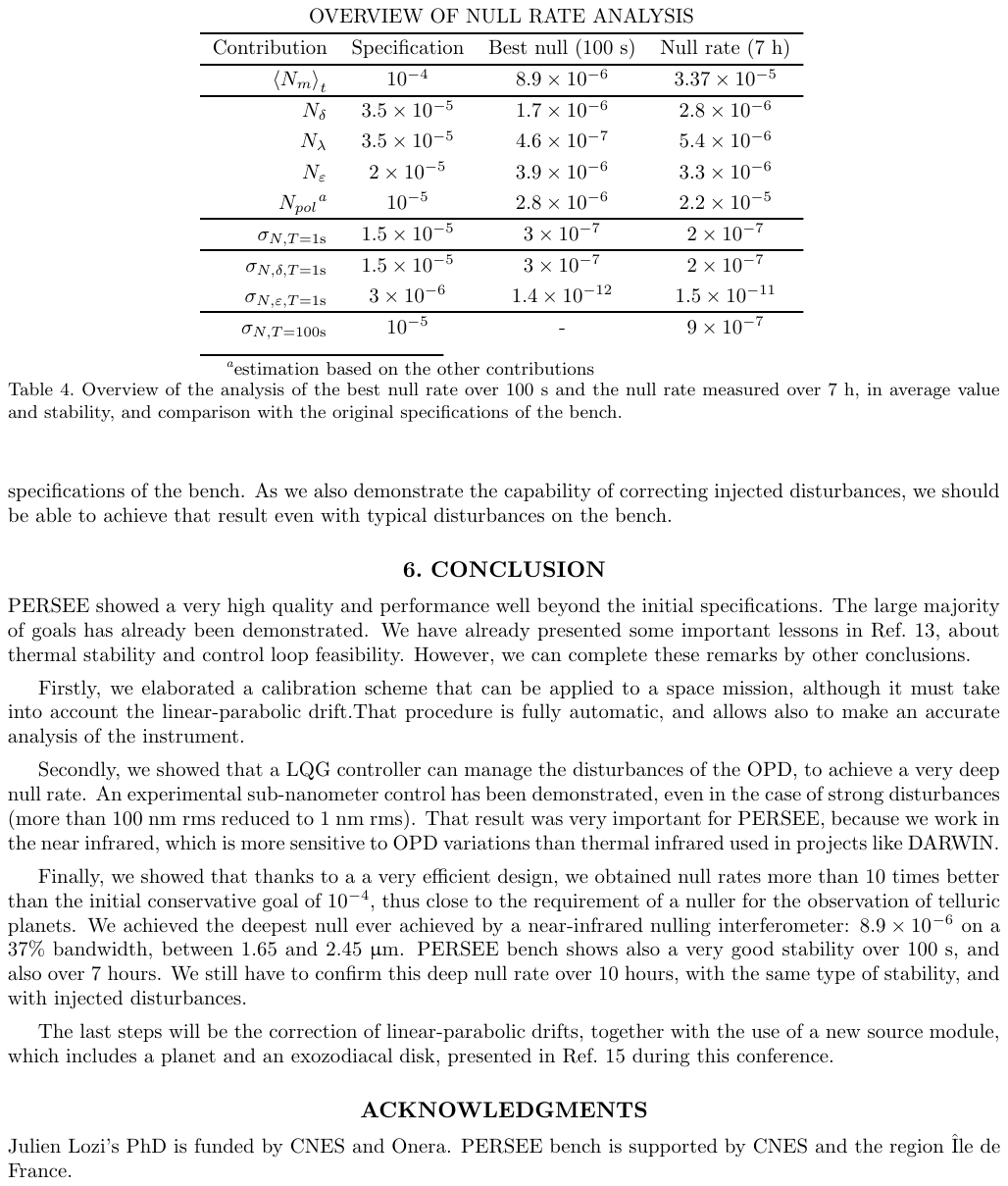}
\SPIEB{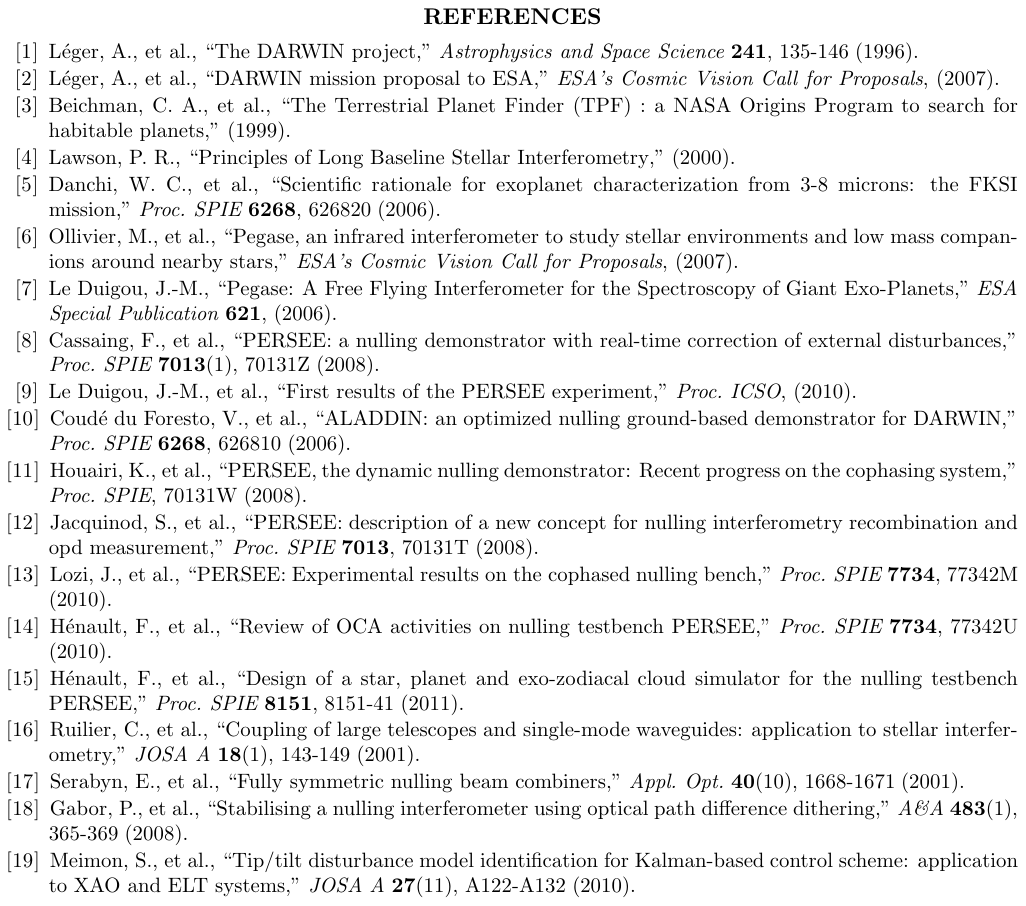}

%%% Local Variables: 
%%% mode: latex
%%% TeX-master: "../these_JL"
%%% End: 

\chapter{List of Publications}

\section*{Peer-Reviewed Journals}

{\sc \textbf{Lozi, J.}, Martinache, F., \& Guyon, O.} 2009. \newblock
Phase-Induced Amplitude Apodization on Centrally Obscured Pupils: Design and
First Laboratory Demonstration for the Subaru Telescope Pupil. \newblock {\em
  Pub.\ Astron.\ Soc.\ Pacific}, {\bf 121} (885), pp. 1232--1244.

{\sc \textbf{Lozi, J.}, Cassaing, F., Le~Duigou, J.-M., Sorrente, B., Montri,
  J., Reess, J.-M., Lhome, E., Buey, T., Henault, F., Marcotto, A., Girard,
  P., Barillot, M., Ollivier, M., \& Coude~du Foresto, V.} 2012. \newblock
{Analysis of the deep stabilized null rate of the PERSEE testbench}, in
preparation.

{\sc \textbf{Lozi, J.}, {Cassaing}, F., \& {Le~Duigou}, J.~M.} 2012. \newblock
{Linear quadratic gaussian control for interferometry: application to the
  Persee Bench}, in preparation.

\section*{Conference Proceedings}

{\sc \textbf{Lozi, J.}, Cassaing, F., Le~Duigou, J.-M., Sorrente, B., Montri,
  J., Reess, J.-M., Lhome, E., Buey, T., Henault, F., Marcotto, A., Girard,
  P., Barillot, M., Ollivier, M., \& Coude~du Foresto, V.} 2011. \newblock
Current results of the {PERSEE} testbench: the cophasing control and the
polychromatic null rate. \newblock vol. 8151. \newblock Proc.\ Soc.\
Photo-Opt.\ Instrum.\ Eng.

{\sc \textbf{Lozi, J.}, Cassaing, F., Le~Duigou, J.-M., Houairi, K., Sorrente,
  B., Montri, J., Jacquinod, S., Reess, J.-M., Pham, L., Buey, T., Hénault,
  F., Marcotto, A., Girard, P., Mauclert, N., Barillot, M., Coudé~du Foresto,
  V., \& Ollivier, M.} 2010. \newblock {PERSEE}: experimental results on the
cophased nulling bench. \newblock vol. 7734. \newblock Proc.\ Soc.\
Photo-Opt.\ Instrum.\ Eng.

{\sc H\'{e}nault, F., Girard, P., Marcotto, A., Mauclert, N., Bailet, C.,
  Lopez, B., Millour, F., Rabbia, Y., Roussel, A., Barillot, M., \textbf{Lozi,
    J.}, Cassaing, F., Houairi, K., Sorrente, B., Montri, J., Lhome, E.,
  Reess, J.-M., Pham, L., Buey, J.-T., du~Foresto, V.~C., Jacquinod, S.,
  Ollivier, M., \& Duigou, J.-M.~L.} 2011. \newblock Design of a star, planet
and exo-zodiacal cloud simulator for the nulling testbench PERSEE. \newblock
vol. 8151. \newblock Proc.\ Soc.\ Photo-Opt.\ Instrum.\ Eng.

{\sc Garrel, V., Guyon, O., Baudoz, P., Martinache, F., Stewart, P.,
  \textbf{Lozi, J.}, \& Groff, T.} 2011. \newblock The Subaru coronagraphic
extreme AO (SCExAO) system: fast visible imager. \newblock vol. 8151.
\newblock Proc.\ Soc.\ Photo-Opt.\ Instrum.\ Eng.

{\sc Le~Duigou, J.-M., \textbf{Lozi, J.}, Cassaing, F., Houairi, K., Sorrente,
  B., Montri, J., Jacquinod, S., Reess, J.-M., Pham, L., Lhomé, E., Buey, T.,
  Hénault, F., Marcotto, A., Girard, P., Mauclert, N., Barillot, M., Coudé~du
  Foresto, V., \& Ollivier, M.} 2011. \newblock Results of the PERSEE
experiment. \newblock Proc.\ SFFMT.

{\sc Le~Duigou, J.-M., \textbf{Lozi, J.}, Cassaing, F., Houairi, K., Sorrente,
  B., Montri, J., Jacquinod, S., Reess, J.-M., Pham, L., Buey, T., Hénault,
  F., Marcotto, A., Girard, P., Mauclert, N., Barillot, M., Coudé~du Foresto,
  V., \& Ollivier, M.} 2010. \newblock First results of the PERSEE experiment.
\newblock Proc.\ ICSO.

%%% Local Variables: 
%%% mode: latex
%%% TeX-master: "../these_JL"
%%% End:
\end{appendices}

\end{document}

%%% Local Variables: 
%%% mode: latex
%%% TeX-master: "these_JL"
%%% End: 